\pdfoutput=1
  \documentclass{PT1}
  \usepackage[numbers]{natbib}
  \usepackage[figuresright]{rotating}
  \usepackage{floatpag}
    \rotfloatpagestyle{empty}

\usepackage{hyperref}

 \usepackage{amssymb}
 \usepackage{amsmath}        

\usepackage{bbm}
\usepackage{bm}

 \usepackage{makeidx}
 \makeindex

\begin{document}

  \title[Theory and applications to quantum technology]
    {Quantum Atom Optics}

	\author{Tim Byrnes \\
	and \\
	Ebubechukwu O. Ilo-Okeke}


  \frontmatter
  \maketitle
  \tableofcontents
  \chapter*{Preface}
This book is a introduction to the field of quantum atom optics, the study of atomic many-body matter wave systems. In many ways the field mirrors the field of quantum optics ---  the study of light at its fundamental quantum level.  In quantum atom optics, many analogous concepts to quantum optics are encountered.  The phenomenon of Bose-Einstein condensation (BEC) can be thought of as the atomic counterpart of lasing.  Coherent states of light have analogues with spin coherent states for atoms, which can be visualized with Wigner and $Q$-functions. In this book we start from the basic concepts of many-body atomic systems, learn methods for controlling the quantum state of atoms, and understand how such atoms can be used to store quantum information and be used for various quantum technological purposes.  

Despite the many analogous concepts, there are also many differences of atoms and photons when treated at the quantum level.  Most fundamentally, atoms can be bosons or fermions and possess mass, in contrast to photons.  While lasing and BECs both have a macroscopic occupation of bosons, the mechanism that forms the BECs is ultimately at thermal equilibrium, whereas lasing is a non-equilibrium process. Coherent states of light do not conserve photon number, whereas generally atom numbers do not change shot-to-shot. Another stark difference is that atoms tend to possess much stronger interactions, whereas it is generally difficult to make photons interact strongly with each other.  This makes the details of many-body atomic systems often quite different, which often leads to rather different approaches conceptually and theoretically.  

One of the recent major developments has been the explosion of interest in the field of quantum information and technologies.  Within a span of 20 years it has turned from a niche field studied by a small community of physicists with various backgrounds in quantum optics, computer science, and foundations of quantum mechanics, to a major research field in its own right.  Much of the way of thinking in the quantum information community originates from the field of quantum optics. Now there is great excitement in how quantum systems can be utilized towards new technologies.  In this book we cover several of the promising applications that atomic systems offer, including spin and matter wave interferometry, quantum simulation, and quantum computing. 

Several excellent texts already exist in the field.  Notable are Pierre Meystre's {\it Atom Optics} and Daniel Steck's {\it Quantum and Atom Optics}.  These are both excellent comprehensive resources that cover the theory of atom optics at the fundamental level.  For BECs we refer the reader to excellent texts such as those by Pitaevskii \& Stringari, and Pethick \& Smith.  Rather than duplicate these works, we wished to provide a senior undergraduate to junior graduate level text that covers the basic principles that are necessary such that one can get up to speed with the current literature in a simple and straightforward way as possible.  While the topics that we cover inevitably are only a brief selection of the extensive achievements in the field, we hope our choices of topics reflect the current interest towards applications of such systems.  This book would suit students who wish to obtain a the necessary skills for working with many-body atomic systems, and have an interest towards quantum technology applications.

\begin{flushright}
{\it Tim Byrnes and Ebubechukwu Ilo-Okeke}
\end{flushright}

\vspace{1cm}




  \mainmatter
  \chapter[Quantum many-body systems]{Quantum many-body systems}

\label{ch:quantum}

\section{Introduction}
\label{sec:intro}

In this chapter we will give a brief overview of the mathematical formalism for describing quantum many-body systems. The types of systems that we will consider in this book will typically involve a large number of identical particles forming a composite quantum state.  Such systems are best described using the formalism of second quantized operators.  We describe how such operators can be defined starting from single particle wavefunctions, and how they can be used to build up the full Hilbert space of the composite system. We also introduce the way that interactions between atoms can be described using the formalism, illustrating this with the example of $s$-wave scattering, a fundamental type of interaction between neutral atoms.

\section{Second quantization}
\label{sec:secondquantization}\index{second quantization}

First consider the familiar case that you should already be well acquainted with from elementary quantum mechanics -- a single particle trapped in a potential $ V(\bm{x}) $.  The stationary states $ \psi_k (\bm{x}) $ are the solutions of the Schrodinger equation:
\begin{align}
H_0 (\bm{x})  \psi_k (\bm{x})  = E_k  \psi_k (\bm{x})
\label{schrodinger}
\end{align}
where 
\begin{align}
H_0 (\bm{x})  = -\frac{\hbar^2}{2m} \nabla^2 + V(\bm{x}).
\label{singleparticleham}
\end{align}
Here $ m$ is the mass of the particle, $ E_k $ are the energies of the stationary states labeled $ n $, and $ \nabla^2 $ is the Laplacian.\index{Laplacian}  As quantum mechanics teaches us, the wavefunction  tells us everything about the system that we would like to know (assuming we are working with pure states).  This approach is fine if we want to describe a single particle, but what if we have many particles, all possibly interacting with each other as we have in a quantum many-body system? 

One way is of course to simply increase the number of labels in the wavefunction and write this as $ \psi(\bm{x}_1, \bm{x}_2, \dots, \bm{x}_N )$.  Each particle has its own label, in this case we have $ N $ particles.  Then if we are dealing with bosonic particles we must impose that the wavefunction is symmetric under particle interchange, or antisymmetric in the case of fermions. 
While this approach is certainly one way and mathematically equivalent, using the language of second quantized notation tends to be much more powerful and the method of choice in modern contexts.  

In the second quantized approach, one defines an operator which creates or destroys a single particle with an implicit wavefunction.  The most intuitive way to define this is in the position basis.  The operator corresponding to the creation of one boson at the position $ \bm{x} $ is written $ a^\dagger (\bm{x}) $, where the $ \dagger $ is Hermitian conjugation. The reverse process of destroying the boson at the position $ \bm{x} $ is written $ a(\bm{x}) $. These obey bosonic commutation relations according to  \index{commutation relations}
\begin{align}
[a(\bm{x}), a^\dagger (\bm{x}') ] = \delta(\bm{x}- \bm{x}')  
\label{commutation1}
\end{align}
and
\begin{align}
[a(\bm{x}), a (\bm{x}') ] = [a^\dagger (\bm{x}), a^\dagger (\bm{x}') ] = 0  .  
\label{commutation2}
\end{align}
For fermions one can similarly define creation and destruction operators $ c^\dagger (\bm{x}) $ and $ c (\bm{x}) $, but this time obeying anticommutation relations \index{anticommutation relations}
\begin{align}
\{ c(\bm{x}), c^\dagger (\bm{x}') \}= \delta(\bm{x}- \bm{x}') 
\label{fermioncomm1}
\end{align}
and
\begin{align}
\{ c(\bm{x}), c (\bm{x}') \}= \{ c^\dagger(\bm{x}), c^\dagger (\bm{x}') \} = 0 .
\label{fermioncomm2}
\end{align}

\begin{figure}[t]
\centerline{\includegraphics[width=\textwidth]{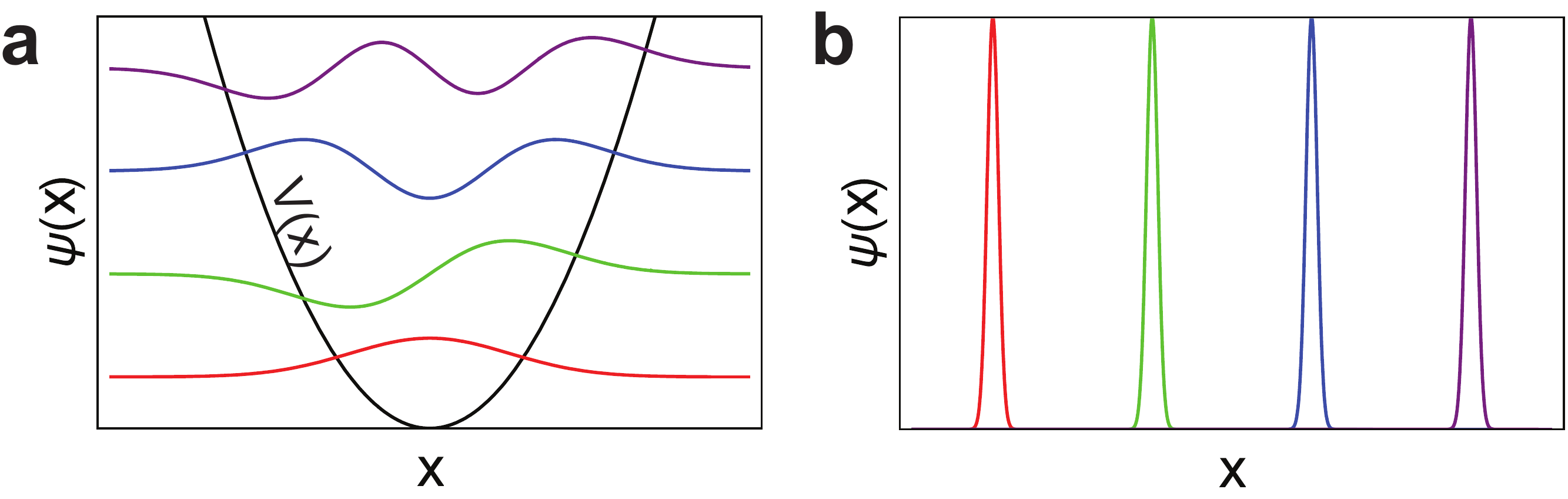}}
\caption{Basis wavefunctions.  (a) Wavefunctions $ \psi_k ( \bm{x}) $ of a quantum harmonic oscillator. (b) Position basis wavefunction take the form of delta functions.  }
\label{fig1-1}
\end{figure}

The bosonic operator $ a(\bm{x}) $ (and also the fermion operator $ c(\bm{x}) $) is only one of an infinity of ways this can be defined.  The reason for this is the same as why there are an infinity of ways of defining a basis.  The position basis is just the most intuitive way of defining a basis, where the particle is located at the position $ \bm{x} $.  We could equally have used a Fourier basis or the complete set of states $ \psi_k (\bm{x}) $ (see Fig. \ref{fig1-1}).  This can be written explicitly in the following way: \index{basis transformation}
\begin{align}
a_k = \int d \bm{x} \psi_k^* (\bm{x}) a (\bm{x}) 
\label{basisan}
\end{align}
and
\begin{align}
a_k^\dagger = \int d \bm{x} \psi_k (\bm{x}) a^\dagger (\bm{x})  .
\label{adaggerdef}
\end{align}
The very neat property of these newly defined operators is that they still have the same commutation relations that we are familiar with of bosons
\begin{align}
[a_k , a_l^\dagger ] = \delta_{kl}
\label{commu1}
\end{align}
and
\begin{align}
[a_k , a_l ] = [a_k^\dagger , a_l^\dagger ] = 0 .
\label{commu2}
\end{align}
We see that regardless of the basis, bosons always have the same properties. This is one of the reasons why the labels $ \bm{x} $ or $ n,m $ are omitted, and sometimes even fail to precisely say what the implicit wavefunction of the boson is. It is however worth remembering that there is always an implicit wavefunction when defining bosons.

\begin{exerciselist}[Exercise]
\item  \label{q1-1}
Verify (\ref{commu1}) and (\ref{commu2}).
\item \label{q1-2}
Show that the fermion version of (\ref{basisan}), 
\begin{align}
c_k = \int d \bm{x} \psi_k^* (\bm{x}) c (\bm{x})  ,
\label{fermionwavefunc}
\end{align}
satisfies fermion anticommutation relations \index{anticommutation relations}
\begin{align}
 \{ c_k, c_l^\dagger \} & = \delta_{kl}, \nonumber \\
 \{ c_k, c_l \} &  = \{ c_k^\dagger, c_l^\dagger \} = 0 . \label{fermionanticommu}
\end{align}
\end{exerciselist}

\section{Fock states}
\label{sec:fockstates}

We now have the right mathematical tools to write down a state involving any number of particles. Let's first start with the 
example that we began the discussion in the previous section -- a single particle in a particular stationary state of the Schrodinger equation (\ref{schrodinger}).  
Our starting point is the {\it vacuum} state, which has absolutely no particles in it at all, which is denoted $ | 0 \rangle $. \index{vacuum} The vacuum state is defined 
as the state such that destroying a particle from this gives
\begin{align}
a_k | 0 \rangle = 0 .  
\label{vacuumstate}
\end{align}
Then one particle with wavefunction $ \psi_k (\bm{x}) $ is written
\begin{align}
| \psi_k \rangle & = a_k^\dagger | 0 \rangle \nonumber \\
& = \int d \bm{x} \psi_k (\bm{x}) a^\dagger (\bm{x})  | 0 \rangle \nonumber \\
& = \int d \bm{x} \psi_k (\bm{x}) | \bm{x} \rangle .
\end{align}
In the second line above, we see how the wavefunction explicitly comes into the definition of the state, where we used (\ref{adaggerdef}).  In the third line, we defined delta-function localized position eigenstates as
\begin{align}
| \bm{x} \rangle = a^\dagger (\bm{x})  | 0 \rangle .
\end{align}
We see that the states can be expanded according to any basis, in the usual way that quantum mechanics allows us to.

Extending this to more than one particle in the way that you can probably already guess: for each particle in the state $ \psi_k (\bm{x}) $, we apply a creation operator $  a_k^\dagger $.  There is one small complication which is how the normalization factors are defined. Suppose that there are $n $ particles that occupy, say, the ground state. Since we are only talking about the ground state here, let's temporarily drop the $ k $-label which specifies the state and write 
\begin{align}
a \equiv a_0  .
\end{align}
Such a state with a particular number of atoms in a state is called a Fock (or number) state, which we would write as
\begin{align}
|n \rangle = \frac{1}{\sqrt{{\cal N}_n}} (a^\dagger)^n | 0 \rangle 
\label{fockstatephoton} .
\end{align}
Working out the normalization factor is an example of a very routine type of calculation when working with bosonic operators, so is recommended if you haven't done it before.  The essential steps are  can be done by repeatedly applying the commutation relations (\ref{commu1}).  Since $ \langle n | n \rangle = 1 $, the normalization factor must be
\begin{align}
{\cal N}_n = \langle 0 | \underbrace{a a \dots a}_{n\text{ of these}} \underbrace{a^\dagger a^\dagger \dots a^\dagger}_{n\text{ of these}} | 0 \rangle .
\end{align}
The aim is then to take advantage of (\ref{vacuumstate}) by commuting all the $ a $'s to hit the vacuum state ket on the right.  Similarly, we can equally commute all the $ a^\dagger $'s to the left to hit the vacuum state bra on the left and use the Hermitian conjugate of (\ref{vacuumstate}):
\begin{align}
\langle 0 | a_k^\dagger = 0  .  
\end{align}
For example, taking one of the $ a $'s all the way to the right gives
\begin{align}
{\cal N}_n = n \langle 0 | \underbrace{a a \dots a}_{n-1\text{ of these}} \underbrace{a^\dagger a^\dagger \dots a^\dagger}_{n-1\text{ of these}} | 0 \rangle .
\end{align}
Repeating this process we find that $ {\cal N}_n = n! $, and so the properly normalized Fock state is \index{Fock states}
\begin{align}
|n \rangle = \frac{1}{\sqrt{n!}} (a^\dagger)^n | 0 \rangle  .
\label{normalizedfock}
\end{align}
From the commutation relations it follows that the Fock states are orthonormal\index{Fock states}
\begin{align}
\langle n' |n \rangle = \delta_{n n'} .
\label{orthonormalfock}
\end{align}

Taking into account of this normalization factor, we can use similar methods to work out that applying a creation or destruction operator gives numerical coefficients in addition to increasing or decreasing the number of bosons \index{creation and destruction operators}
\begin{align}
a |n \rangle = \sqrt{n}  |n-1 \rangle 
\label{lowering}
\end{align}
and
\begin{align}
a^\dagger |n \rangle = \sqrt{n+1}  |n+1 \rangle .
\label{raising}
\end{align}

For fermions, it is simpler.  We can perform the same logic for fermion operators.  Define
\begin{align}
c \equiv c_0 
\end{align}
where we have a fermion in the ground state of the states as defined in (\ref{fermionwavefunc}). Then the only Fock states we can define are the vacuum state $ | 0 \rangle $, which satisfies\index{Fock states}\index{vacuum}
\begin{align}
c_k | 0 \rangle = 0 
\end{align}
and the one particle fermion Fock state
\begin{align}
|1 \rangle_{\text{f}} = c^\dagger | 0 \rangle ,
\label{fermionfock}
\end{align}
where we have labeled the fermion Fock states by a subscript\index{Fock states} $ \text{f} $. We could try and write down a fermion version of (\ref{normalizedfock}), but from (\ref{fermionanticommu}) we would find that any state with $ n\ge 2 $ gives zero.  This is the Pauli exclusion principle\index{Pauli exclusion principle} at work --- no two fermions can occupy the same state. The creation and destruction operator then simply shifts between these two states
\begin{align}
c |1 \rangle_{\text{f}} & = | 0 \rangle \nonumber \\
c^\dagger | 0 \rangle & = |1 \rangle_{\text{f}} .
\label{fermioncreationdest}
\end{align}

\begin{exerciselist}[Exercise]
\item \label{q1-3}
Work through the steps to verify (\ref{normalizedfock}) and (\ref{orthonormalfock}).
\item \label{q1-4}
Verify (\ref{lowering}) and (\ref{raising}). 
\item \label{q1-5}
(a) Show explicitly that the only fermion Fock states are (\ref{fermionfock}) and the vacuum.  
(b) Verify (\ref{fermioncreationdest}) using the fermion anticommutation relations (\ref{fermionanticommu}).  \index{anticommutation relations}
\end{exerciselist}

\section{Multi-mode Fock states}
\label{sec:multifockstates}

\begin{figure}[t]
\centerline{\includegraphics[width=\textwidth]{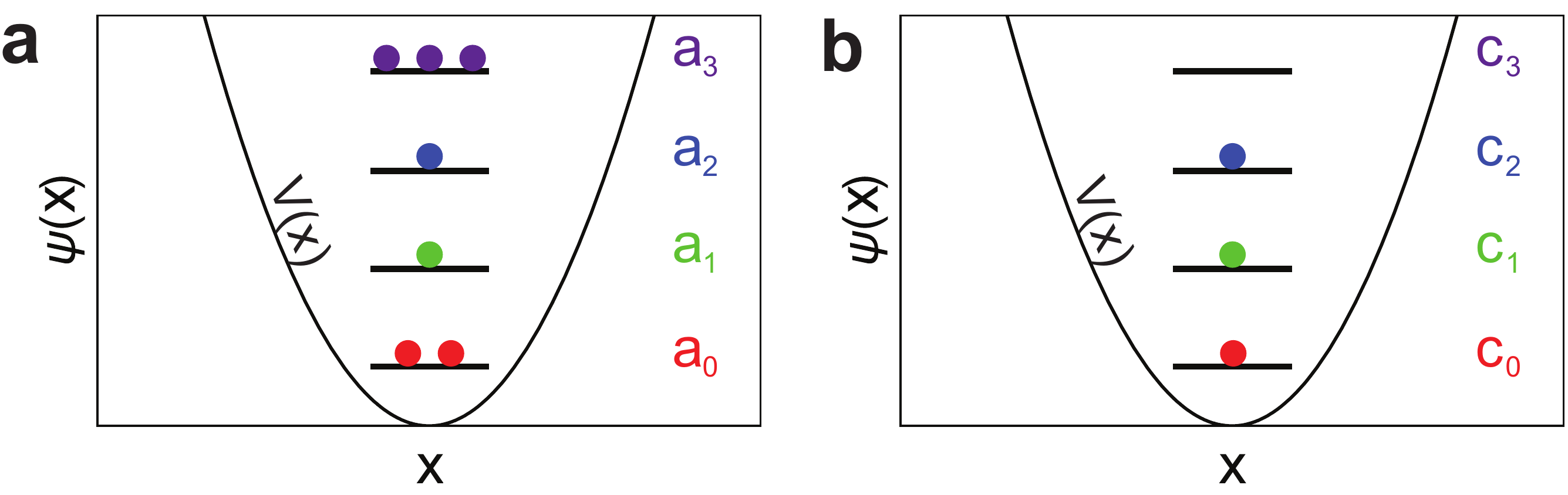}}
\caption{Generalized Fock states for (a) bosons and (b) fermions. The case shown in (b) corresponds to lowest energy configuration possible for 3 fermions (Fermi sea).  The lowest energy configuration for bosons would correspond to all the particles in the ground state, hence (a) is an excited state for the multiparticle system.   }
\label{fig1-2}
\end{figure}

The Fock states\index{Fock states} that we have written in the previous section are fine if all the particles are in the ground state.  This might be the case at zero temperature, or indeed the situation in a Bose-Einstein condensate as we will be discussing in later chapters.  But more typically particles will occupy all energy levels, so we need to take into account of more than just the ground state.   In this case we can simply just do the same thing as above, for each of the eigenstates labeled by $ k $.  The generalized Fock state is then for bosons
\begin{align}
| n_0, n_1, \dots, n_k, \dots \rangle = \prod_{k} \frac{(a^\dagger_k)^{n_k} }{\sqrt{n_k !}} | 0 \rangle
\label{generalizedbosonstate}
\end{align}
and for fermions it is similarly
\begin{align}
| n_0, n_1, \dots, n_k, \dots \rangle_{\text{f}} = \prod_{k} (c^\dagger_k)^{n_k}  | 0 \rangle .
\end{align}
Notice we don't need the normalization factors for the fermions because we only ever have $ n_k \in \{0,1 \} $.

For a single particle state, obviously the energy of the state is simply given by the eigenvalue of the Schrodinger equation $ E_k $.  If there are multiple particles involved, how do we write the total energy of a Fock state\index{Fock states} such as that written above? First we should generalize the single-particle Schrodinger equation to a multi-particle one, which can be done by simply writing
\begin{align}
{\cal H}_0 = \int d \bm{x} a^\dagger (\bm{x}) H_0(\bm{x}) a (\bm{x}) 
\label{multiham}
\end{align}
where $ H_0(\bm{x}) $  is given by (\ref{singleparticleham}).  For fermions, this takes the same form, except that the bosonic operators are changed to fermionic ones.  We now need the inverse relations of (\ref{basisan}) and (\ref{adaggerdef}), given by 
\begin{align}
a (\bm{x}) = \sum_k  \psi_k(\bm{x})  a_k 
\label{basisaninv}
\end{align}
and
\begin{align}
a^\dagger (\bm{x}) = \sum_k  \psi_k^* (\bm{x})  a_k^\dagger  .
\label{adaggerdefinv}
\end{align}
Substituting this into (\ref{multiham}), we obtain
\begin{align}
{\cal H}_0  = \sum_k E_k  N_k 
\label{hamiltoniandiagonal}
\end{align}
where 
\begin{align}
N_k  \equiv a_k^\dagger  a_k  .
\label{numberoperator}
\end{align}

The operators (\ref{numberoperator}) are {\it number operators} \index{number operators} which have Fock states\index{Fock states} as their eigenstates:
\begin{align}
N_k | n_0, n_1, \dots, n_k, \dots \rangle = n_k | n_0, n_1, \dots, n_k, \dots \rangle .
\label{numbereigenstate}
\end{align}
The multi-particle Fock state\index{Fock states} is thus a eigenstate of the Hamiltonian (\ref{hamiltoniandiagonal})
\begin{align}
{\cal H}_0  | n_0, n_1, \dots, n_k, \dots \rangle = E_{\text{tot}} | n_0, n_1, \dots, n_k, \dots \rangle ,
\end{align}
where the total energy is
\begin{align}
E_{\text{tot}}  = \sum_k E_k n_k .
\label{totalsingleenergy}
\end{align}
We thus have the intuitive result that the total energy of the multi-particle Hamiltonian is the sum of the energies of each particle.  

We can picture such Fock states as shown in Fig. \ref{fig1-2}. Each single particle quantum level is labeled by an index $ k $, and this is occupied by a number $ n_k $, which tells us how many particles are occupied in that level.  The total energy of the system is just the sum of the energies of the individual particles.  For bosons, we are allowed to fill each level more than once so the minimum energy is if all the particle are in the ground state.  In this case we would have 
\begin{align}
E_{\text{tot}} = N E_0 \hspace{2cm} \text{(minimum, bosons)}
\end{align}
where $ N $ is the total number of bosons in the system.  For fermions, each level cannot be filled more than once, so the minimum energy will be
\begin{align}
E_{\text{tot}} = \sum_{k=0}^{N-1} E_k . \hspace{2cm} \text{ (minimum, fermions)}
\end{align}

\begin{exerciselist}[Exercise]
\item \label{q1-6}
(a) Derive (\ref{basisaninv}) by multiplying it by $ \psi_k (\bm{x}') $ and summing (\ref{basisan}) over $ k $, using the completeness relation.
(b) Derive (\ref{basisan}) by multiplying it by $ \psi_l^* (\bm{x}) $ and integrating (\ref{basisaninv}) over $ x $.  
\item \label{q1-7}
Verify that (\ref{hamiltoniandiagonal}) starting from (\ref{multiham}).  
\item \label{q1-8}
Verify that (\ref{numbereigenstate}) starting from the bosonic commutation relations.\index{commutation relations}  
\end{exerciselist}

\section{Interactions}
\label{sec:interactions}

Up to this point we have not included any interactions between the particles. The energy of the whole system was therefore determined entirely by the sum of the single particle energies, as seen in 
(\ref{totalsingleenergy}). This is usually called the single particle or {\it non-interacting limit}.  \index{single particle limit} More realistically, there are interactions present between the particles, due to the presence of 
some forces --- naturally occurring or otherwise --- that particles feel between each other.  Suppose that the energy between two particles can be written $ U(\bm{x},\bm{y}) $, where the location of the first particle is $ \bm{x} $ and the second is $ \bm{y} $.  The two-particle time-independent Schrodinger equation would be written in this case
\begin{align}
\left[ H_0( \bm{x}) + H_0( \bm{y}) \right] \psi(\bm{x},\bm{y}) + U(\bm{x},\bm{y})  \psi(\bm{x},\bm{y}) = E \psi(\bm{x},\bm{y}) .
\end{align}
The first term on the left hand side is the single particle Hamiltonian for the two particle case, as given in (\ref{multiham}).  
The second is the interaction term between the particles. There is only one term because the interaction corresponds to a pair of particles

\begin{figure}[t]
\centerline{\includegraphics[width=\textwidth]{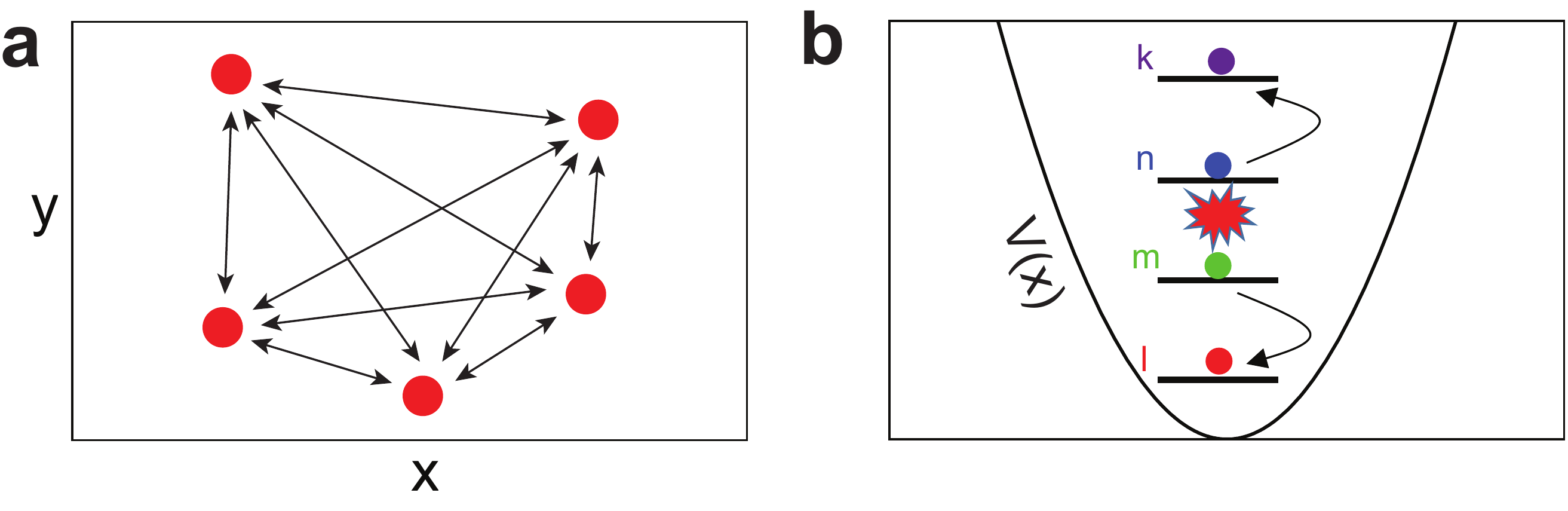}}
\caption{Interactions between particles shown in (a) the position basis, and (b) single particle eigenstate basis. In (a), for $ N $ particles there are $ N (N-1)/2 $ pairs of interactions. In (b), two particles initially in levels labeled by $ n $ and $ m $ scatter to other levels $ k $ and $ l $. }
\label{fig1-3}
\end{figure}

To generalize this to any number of particles, we write it in the form 
\begin{align}
{\cal H}  = {\cal H}_0  +  {\cal H}_I.
\label{multiintham}
\end{align}
where the interaction Hamiltonian is
\begin{align}
 {\cal H}_I & = \frac{1}{2} \int dx dy a^\dagger (\bm{x}) a^\dagger (\bm{y}) U(\bm{x},\bm{y})  a (\bm{y}) a(\bm{x})  \nonumber \\
 & = \frac{1}{2} \int dx dy U(\bm{x},\bm{y}) n(\bm{x}) (n (\bm{y})-1) ,
\label{interactionhamch1}
\end{align}
where we have used the commutation relations (\ref{commutation1}) and (\ref{commutation2}).\index{commutation relations}  Eq. (\ref{interactionhamch1}) has the simple interpretation of counting all the pairs of particles between the particles (see Fig. \ref{fig1-3}).  For a $ N $-particle system, there will be $ N (N-1)/2 $ pairs of interactions, which are potentially dependent on the positions of the particles $ U(\bm{x},\bm{y}) $.

As was done in the previous sections, it is possible to write the interaction Hamiltonian $  {\cal H}_I  $ in terms of the eigenstates of the single particle Hamiltonian.  This is done by simply substituting (\ref{basisaninv}) and (\ref{adaggerdefinv}) into (\ref{interactionhamch1}).  This gives
\begin{align}
 {\cal H}_I & =  \sum_{klmn } U_{klmn} a^\dagger_{k}  a^\dagger_{l} a_{m} a_{n} 
\label{interactionmanybodyham}
\end{align}
where
\begin{align}
 U_{klmn}  = \frac{1}{2}\int d \bm{x} d \bm{y} \psi_{k}^* (\bm{x})  \psi_{l}^*  (\bm{y})  U(\bm{x},\bm{y})   \psi_{m} (\bm{x})  \psi_{n}  (\bm{y})  .
\label{interactionmatrix}
\end{align}
In this basis we find that the interactions are not diagonal.  This is in contrast to (\ref{interactionhamch1}) which consists entirely of number operators.  This means that for particles that are spatially localized as shown in Fig. \ref{fig1-3}(a), the interaction Hamiltonian  $  {\cal H}_I $ leaves the state unaffected.  In the basis of the eigenstates of the single particle Hamiltonian, the interaction Hamiltonian $  {\cal H}_I $ can cause some scattering between states.  Specifically, a particle in level $ n $ can feel the effect of another particle in level $ m $, and both particles can be scattered to different levels $ k $ and $ l $ (see Fig. \ref{fig1-3}(b)).

In many cases it is possible to simplify the form of the interaction.  For example, typically we can safely say that the
interaction only depends upon the relative position between the particles.  This is a reasonable assumption in most experimental situations. The most common type of interaction for Bose-Einstein condensates (BECs) is the $ s $-wave interaction, which is a type of contact interaction
\begin{align}
U(\bm{x},\bm{y}) = U_0 \delta (\bm{x} - \bm{y}),
\label{swaveintch1}
\end{align}
where the interaction energy is
\begin{align}
U_0  = \frac{4 \pi \hbar^2 a_s}{m} .
\label{gintdef}
\end{align}
Here $ a_s $ is the scattering length\index{scattering length}, an experimentally measured quantity.  In this case the interaction matrix elements take the form (\ref{interactionmatrix})
\begin{align}
g_{klmn}  = \frac{2 \pi \hbar^2 a_s}{m} \int d \bm{x} \psi_{k}^* (\bm{x})  \psi_{l}^*  (\bm{x})  \psi_{m} (\bm{x})  \psi_{n}  (\bm{x})  .
\label{interactionmatrix2}
\end{align}
Specifically, for the atoms in the ground state the interaction is
\begin{align}
g_{0000}  = \frac{2 \pi \hbar^2 a_s}{m} \int d \bm{x} |\psi_{0} (\bm{x}) |^4  .
\end{align}
As we will see in the next chapter, in a BEC many of the atoms occupy the ground state and this will be the dominant interaction energy.  

For the general interaction  (\ref{interactionmatrix2}), very often some of the matrix elements will be zero automatically.  To see an example of this, consider the simple case where $ V(\bm{x}) = 0 $.  In this case the eigenstates are simply plane waves
\begin{align}
\psi_{\bm{k}} (\bm{x}) = \frac{e^{i \bm{k} \cdot \bm{x}}}{ \sqrt{V} },
\end{align}
where $ V$ is a suitable normalization factor.  Substituting into (\ref{interactionmatrix2}) we find that 
\begin{align}
 g_{\bm{k} \bm{l} \bm{m} \bm{n}}  = \frac{2 \pi \hbar^2 a_s}{m V } \delta (\bm{n} + \bm{m} - \bm{k} - \bm{l}) .
\label{swaveexpression}
\end{align}
In this case the labels $ k, l, m, n $ have the interpretation of momentum, and the delta function is a statement of the conservation of momentum between the initial states and the final states. This occurs due to the translational invariance of the potential\index{translational invariance} (\ref{swaveintch1}).   While realistic potentials are not exactly $ V(\bm{x}) = 0 $, approximate relations for the conservation of momentum are obeyed in practice.

\section{References and further reading}

\begin{itemize}
\item Sec. \ref{sec:secondquantization}: For further details about quantum mechanics and second quantization see the textbooks \cite{sakurai1995modern,altland2010condensed,cohen1977quantum}.
\item Secs. \ref{sec:fockstates},\ref{sec:multifockstates}:  Fock states and their algebra in a quantum optics setting are further described in  \cite{scully1999quantum,gerry2005introductory,walls2007quantum}.
\item Sec. \ref{sec:interactions}: Further details about interactions in Bose-Einstein condensates are provided in \cite{pethick2002bose,pitaevskii2016bose,yamamoto1999mesoscopic,landau2013quantum,gibble1995direct}. 
\end{itemize}

  \chapter[Bose-Einstein condensation]{Bose-Einstein condensation}

\label{ch:bec}

\section{Introduction}
\label{sec:intro2}

The previous chapter introduced the formalism for treating many-particle indistinguishable quantum systems.  This can describe the wavefunction of any system of bosonic or fermionic particles, possibly interacting with each other.  In this chapter, we introduce how such a system can undergo Bose-Einstein condensation (BEC).  The essential feature of a Bose-Einstein condensate (BEC) is the macroscopic occupation of the ground state.  The fact that such a state occurs is not completely obvious from the point of view of statistical mechanics -- one might naively expect that there is a exponential decay of the probability of occupation following a Boltzmann distribution\index{Boltzmann!distribution} $ p_n \propto \exp( - E_n/ k_B T ) $, where $E_n $ is the energy of the state, $ k_B $ is the Boltzmann constant,\index{Boltzmann!constant} and $ T $ is the temperature.  We first discuss the original argument by Bose and Einstein showing why such a macroscopic occupation might occur.  We then show how macroscopic occupation of the ground state occurs in a grand canonical ensemble and give key results for the condensation temperature and the fraction of atoms in the ground state. We then discuss the effect of interactions on the energy-momentum dispersion relation, which gives rise to the Bogoliubov dispersion relation\index{dispersion!Bogoliubov }.  This is the key to understanding superfluidity\index{superfluidity}, one of the most astounding features of a quantum fluid.  

Bose-Einstein condensation is a expansive subject and the purpose of this chapter is introduce the minimal amount of background such that one can understand the more modern applications of such systems.  For a more detailed discussion of the physics of BECs we refer the reader to excellent texts such as those by Pitaevskii and Stringari \cite{pitaevskii2016bose} and Pethick and Smith \cite{pethick2002bose}.

\section{Bose and Einstein's original argument}
\label{sec:einstein}


To see why macroscopic occupation of the ground state occurs in a system of bosons, we first examine a simple model of $ N $ non-interacting two level particles. To be specific, we can imagine that there is a gas of bosonic atoms and the system is cooled down enough such that the spatial degrees of freedom are all the same for all the particles. As the spatial wavefunction is all the same, all the atoms are indistinguishable\index{indistinguishable particles} from each other, with the exception of the internal states.  The two levels are then internal states of the atom, which have an energy $ E_0 $ and $ E_1 $.  Due to the indistinguishable\index{indistinguishable particles} nature of the atoms, it is appropriate to denote them by bosonic creation operators $ a_0^\dagger $ and $ a_1^\dagger $ for the two states.   As we did in Sec. \ref{sec:fockstates}, we can write the Fock states corresponding to this system as \index{indistinguishable particles}
\begin{align}
| k \rangle = \frac{(a_0^\dagger)^{N-k} (a_1^\dagger)^{k} }{\sqrt{  (N-k)!k!}} | 0 \rangle ,
\label{bosontwolevel}
\end{align}
where $ k $ is the number of atoms in the state $E_1 $, and the number of atoms in the other state with energy $ E_0 $ is $ N - k $ because we have $ N $ atoms in total.   It is easy to see that there are a total of $ N+1 $ states, because $ k $ can be any integer between $ 0 $ and $ N $.  

For comparison, let's also look at the case when we don't have indistinguishable bosons.  Physically this might correspond to a situation where the system is not quite as cold as the previous case, such that the spatial wavefunctions of the atoms are not the same as each other.  In this case, the particles are distinguishable\index{distinguishable particles} by virtue of some other property of the atoms, such as their velocity or position. Note that the distinguishability\index{distinguishable particles} has to be merely in principle, no measurement has to be made to ensure this.  Let's again look at all the possible states.  Since the particles are distinguishable\index{distinguishable particles}, we can label the states in the $j$th particle by $ | \sigma_j \rangle_j $, where $ \sigma_j = 0, 1 $ are 
the two states with energy $ E_0 $ and $ E_1 $ respectively. The state of the whole system is then \index{distinguishable particles}
\begin{align}
| \sigma_1 \sigma_2 \dots \sigma_N \rangle = | \sigma_1 \rangle_1 \otimes | \sigma_2 \rangle_2 \otimes
\dots \otimes | \sigma_N \rangle_N .
\label{disttwolevel}
\end{align}
Counting the total number of different states, we see a different result to what we saw in (\ref{bosontwolevel}).  Since each atom can be in one of two states, the total number of combinations is $ 2^N $. 

Let's now look at the energy spectrum for the two cases.  For the indistinguishable\index{indistinguishable particles} bosonic case, we can simply use (\ref{totalsingleenergy}) to obtain
\begin{align}
E_{\text{indis}}  = E_0 (N-k ) + E_1 k= N E_0  +  k \Delta E
\label{energybos}
\end{align}
where we have defined $ \Delta E = E_1 - E_0 $.  For the distinguishable\index{distinguishable particles} case, we have
\begin{align}
E_{\text{distin}}  & = \sum_{j=1}^N E_0 (1- \sigma_j) + E_1 \sigma_j \nonumber \\
& = N E_0  + k \Delta E 
\label{energydist}
\end{align}
where we defined as $ k = \sum_{j=1}^N \sigma_j $, which counts the total number of atoms in the excited state $ E_1 $.

\begin{figure}[t]
\centerline{\includegraphics[width=\textwidth]{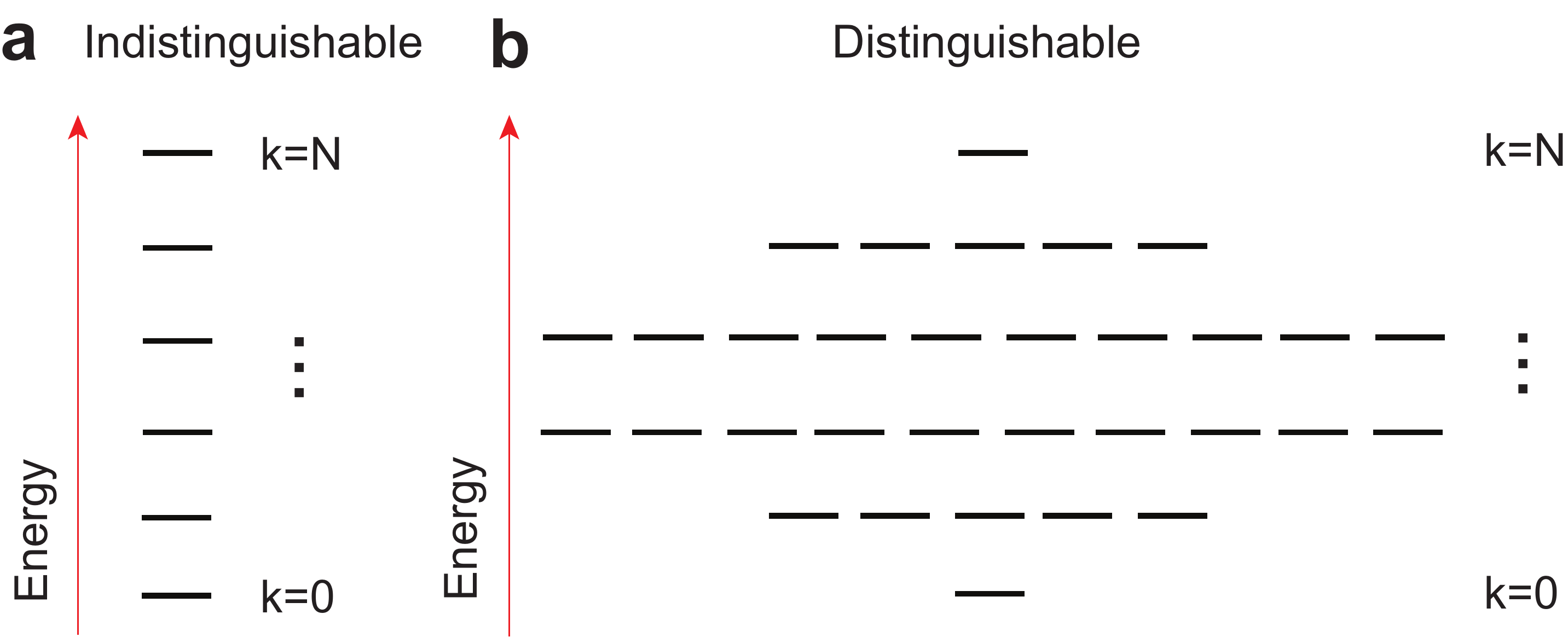}}
\caption{Number of levels for (a) indistinguishable\index{indistinguishable particles} bosonic and (b) distinguishable\index{distinguishable particles} two-level systems. We show the case $ N = 5 $.  }
\label{fig2-1}
\end{figure}

The expressions (\ref{energybos}) and (\ref{energydist}) are exactly the same, which is not surprising considering in both cases we have a set of $ N $ two-level atoms with the same energy structure. 
While both have a variable $ k $ which is the total number of atoms in the excited state, they have a difference in terms of the degeneracy of the states (\ref{bosontwolevel}) and (\ref{disttwolevel}).  In Fig. \ref{fig2-1}, we see the energy spectrum for the whole system.  While the indistinguishable\index{indistinguishable particles} case only has one possible state for each $ k $, the distinguishable\index{distinguishable particles} has various combinations of spin configurations (\ref{disttwolevel}) that have the same $ k $.  For example, for $ N = 5 $, there are 5 states with $ k = 1 $:
\begin{align}
|00001 \rangle, |00010 \rangle,|00100 \rangle, |01000 \rangle, |10000 \rangle
\label{N5stateexample}
\end{align}
In general, one can argue from simple combinatorics that the number of states in each sector $ k $ is $ \binom{N}{k} $, so that we can label the state (\ref{disttwolevel}) equivalently as 
\begin{align}
|k, d_k \rangle = | \sigma_1 \sigma_2 \dots \sigma_N \rangle 
\label{distinfock2}
\end{align}
where $ d_k \in [1, \binom{N}{k}] $.

One may wonder whether there is some kind of correspondence of the states between the distinguishable\index{distinguishable particles} and indistinguishable cases\index{indistinguishable particles}.  The key feature of the bosonic system is that the wavefunction is symmetric under particle interchange.  This restriction that indistinguishability\index{indistinguishable particles} imposes greatly reduces the available states that the system can occupy.  For example, none of the states (\ref{N5stateexample}) are symmetric under particle interchange --- for each state interchanging the ordering of the $ \sigma_j $ will give a completely different state.  There is however a state that is symmetric under interchange
\begin{align}
\frac{1}{\sqrt{5}} \left( |00001 \rangle +|00010 \rangle + |00100 \rangle + |01000 \rangle + |10000 \rangle
\right) .
\label{N5symmetric}
\end{align}
For this state, interchanging the order of the spins still gives the same state. More generally, any state of the form
\begin{align}
\sum_{d_k=1}^{\binom{N}{k}} |k, d_k \rangle
\label{generalsymmetric}
\end{align}
is symmetric under particle interchange.  This is the only state in the space of states $ |k, d_k \rangle$ that has this property --- all the phases in the sum (\ref{N5symmetric}) must be the same.  This is thus the state that is equivalent to the bosonic Fock state $ | k \rangle $ in (\ref{bosontwolevel}).  \index{Fock states}

This difference in the number of states ultimately is the origin of Bose-Einstein condensation.  To see this, let us find the probability of picking a particular atom, and finding it in the ground state for the two cases.  Starting with the distinguishable case,\index{distinguishable particles} first find the probability for the $j$th atom to be in the ground state using Boltzmann statistics:\index{Boltzmann!statistics}
\begin{align}
p_j = \frac{e^{-\beta E_0}}{e^{-\beta E_0} + e^{-\beta E_1}} = \frac{1}{1+ e^{-\beta \Delta E}} ,
\label{pjprobability}
\end{align}
where $ \beta = 1/ k_B T $, $ k_B $ is the Boltzmann constant\index{Boltzmann!constant}, and $ T $ is the temperature.  Here the state of the remaining $ N - 1 $ atoms do not have any effect on the $ j $th atom.  
Therefore, if we then measure all $ N $ atoms, of course the fraction of atoms that are in the ground state will be
\begin{align}
f_{\text{distin}} = \frac{1}{1+ e^{-\beta \Delta E}}
\label{fracdis}
\end{align}
In this case, we see that the number of atoms $ N $ doesn't affect the probability because the particles are independent.

In the indistinguishable\index{distinguishable particles} boson case, there is no way of simply picking out the $j$th atom as we did above, since the atoms cannot be distinguished. We can however talk about the statistics of the whole system in a well-defined way, since we know the energy structure as shown in Fig. \ref{fig2-1}. The probability of the state  (\ref{bosontwolevel}) can be obtained again using Boltzmann statistics\index{Boltzmann!statistics}:
\begin{align}
p_k & = \frac{e^{-\beta(NE_0 + k \Delta E)}}{\sum_{k=0}^N e^{-\beta(NE_0 + k \Delta E)}} \nonumber \\
& \approx e^{-\beta k \Delta E} ( 1 - e^{-\beta \Delta E} )
\label{pkdistribution}
\end{align}
where we assumed that $ N $ is large for simplicity.  This is the distribution of the whole system, but what we are interested in is the probability of {\it one} atom to be  in the ground state.  To calculate this, for each state $ |k \rangle  $, let us count the number of atoms are in the ground state, which is just equal to $ N - k $.  On average, one would thus find that
\begin{align}
\langle n_0 \rangle & = \sum_{k=0}^N p_k (N-k) \nonumber \\
& = N - \frac{e^{-\beta \Delta E}}{1 - e^{-\beta \Delta E}} .
\end{align}
The fraction of atoms in the ground state is then 
\begin{align}
f_{\text{indis}} = \frac{\langle n_0 \rangle}{N} = 1 - \frac{1}{N} \frac{e^{-\beta \Delta E}}{1 - e^{-\beta \Delta E}} .
\label{fracindis}
\end{align}
In the limit of $ N \rightarrow \infty $, we see that the fraction $ f $ approaches 1.  This means that regardless of the temperature $ \beta $ and energy separation $ \Delta E $, the fraction of atoms in the ground state is unity!

The two different results from the fraction of atoms in the ground state (\ref{fracdis}) and (\ref{fracindis}) originates from the different way of counting states for distinguishable\index{distinguishable particles} and indistinguishable particles\index{indistinguishable particles}. As we see from Fig. \ref{fig2-1}, the main difference is that for the distinguishable case, there are many more levels away from the ground state.  In the distinguishable case, due to the a state with $ k $ excited states having $ \binom{N}{k} $ possibilities, this strongly biases the population towards excited states with $k \approx N/2 $.  Thus even though the Boltzmann factor\index{Boltzmann!factor} $ \propto e^{-\beta E} $ exponentially biases the probability towards lower energies, these two effects effectively cancel each other.  On the other hand the indistinguishable\index{indistinguishable particles} case does not have the  handicap of the combinatorics.  In the limit of $ N \rightarrow \infty $, the ladder in Fig. \ref{fig2-1}(a) extends indefinitely, and the Boltzmann factor ensures that the population is biased towards the bottom.  This makes it much more likely that the ground state is occupied for each atom.

\begin{exerciselist}[Exercise]
\item \label{q2-1}
Show using a combinatorial argument that for $ N $ distinguishable two-level atoms the total number of states with $ k $ atoms in one of the excited states is $ \binom{N}{k} $ and hence verify \index{distinguishable particles} (\ref{distinfock2}).  
\item \label{q2-2}
(a) Verify that (\ref{N5symmetric}) is symmetric under particle interchange.  (b) Show explicitly, by evaluating the fidelity or otherwise, that the state (\ref{N5symmetric}) with a minus sign on one of the terms is not symmetric under particle interchange.   (c) Verify that (\ref{generalsymmetric}) is symmetric under particle interchange.
\item \label{q2-3}
For the full system with $ N $ distinguishable particles, the\index{distinguishable particles} probability that the state occupies the state $ \sigma_j = 0 $ should be equal to $ p_{00 \dots 0} = \prod_{j=1}^N p_j $ in (\ref{pjprobability}).  This should agree with directly constructing the partition function using (\ref{distinfock2}) and (\ref{energydist}).  Check that these two methods give the same answer. 
\item \label{q2-4}
(a) Verify (\ref{pkdistribution}) is true for $ N \rightarrow \infty $.   (b) What is the corresponding result for any $ N $? 
\end{exerciselist}

\section{Bose-Einstein condensation for a grand canonical ensemble}
\label{sec:grand}


In the previous section, we saw how the indistinguishable nature of the particles could affect the occupation of the ground state for a collection of two-level atoms.  While this is possibly the simplest example of the macroscopic occupation of the ground state, it if far from a realistic model of Bose-Einstein condensation (BEC) of atoms.  In the previous section we did not examine the movement of the atoms at all.  In fact, what happens more typically in a BEC experiment is that a collection of atoms are all cooled down to low enough temperatures such that they condense into their ground states. The energy levels that we are talking about here are the motional degrees of freedom, which we can take to be momentum states. \index{distinguishable particles} 

In this section, we will derive BEC for a collection of non-interacting atoms in three dimensions. The argument has a few subtleties, so let's tread carefully and work through it step by step.  The argument works in the grand canonical ensemble formulation of statistical mechanics.  We will review a few things in the context of bosons, and then move onto the main argument showing BEC.

\subsection{Bose-Einstein distribution and chemical potential}

Recall the situation that we had in Fig. \ref{fig1-2}(a): a potential $ V(\bm{x}) $ gave rise to a set of levels labeled by $ k $, each of which could be multiply occupied by an integer $ n_k $.  The total energy is given by (\ref{totalsingleenergy}).  According to the Boltzmann statistics, the probability of a particular Fock state configuration being occupied is\index{Boltzmann!statistics}\index{Fock states}
\begin{align}
p(n_0,n_1,\dots,n_k, \dots) = \frac{\exp( -\beta \sum_k (E_k-\mu) n_k ) }{Z},
\end{align}
where $ Z $ is the partition function, \index{partition function}  which plays the role of the normalization factor for the probability distribution.  We have also offset the energy by the chemical potential $ \mu $, \index{chemical potential} which can for now be viewed as shifting the definition of the zero-point in the energy per particle.  The partition function can be evaluated by summing over all possible configurations
\begin{align}
Z & = \sum_{n_0} \sum_{n_1} \dots \sum_{n_k} \dots \exp( -\beta \sum_k (E_k-\mu) n_k ) \nonumber \\
& = \sum_{n_0} \exp( -\beta (E_0-\mu) n_0 ) \sum_{n_1} \exp( -\beta (E_1-\mu) n_1 ) \dots
\label{partitionfunc}
\end{align}
Normally one would like to consider a fixed number of particles, such that $ \sum_k n_k = N $.  This unfortunately put some restrictions on the summations in (\ref{partitionfunc}), which makes it difficult to evaluate mathematically.  Instead, let's for now consider the unrestricted case, and impose the constraint later.  This is the so-called grand canonical distribution\index{grand canonical distribution}, where the particle number is not fixed, and particles can enter or leave the system, very much in the same way as energy being exchanged between the system and reservoir.  Evaluating each of the sums by a geometric series, we can write the probability as
\begin{align}
p(n_0,n_1,\dots,n_k, \dots) =  \prod_k p_k (n_k)
\end{align}
where
\begin{align}
p_k (n_k) = e^{-\beta (E_k-\mu) n_k}(1-e^{-\beta(E_k-\mu)})
\end{align}
is the probability that the $ k $th level is occupied by $ n_k $ atoms.  The average number of atoms in this level
\begin{align}
\bar{n}_k (\mu)  = \sum_k p_k n_k = \frac{1}{e^{\beta(E_k - \mu)} - 1}
\label{boseeinsteindist}
\end{align}
gives the well-known Bose-Einstein distribution.  

Since we are using the grand canonical distribution\index{grand canonical distribution}, the number of particles in a given level --- and hence the whole system --- fluctuates.  The number of particles that are in the system is controlled by the chemical potential $ \mu $, as can be seen from the Bose-Einstein distribution (\ref{boseeinsteindist}).  When the chemical potential\index{chemical potential} tends to infinity $ \mu \rightarrow  - \infty $, we can see from (\ref{boseeinsteindist}) that the number of particles in all the levels become zero
\begin{align}
\lim_{\mu \rightarrow - \infty} \bar{n}_k (\mu)  = 0 .
\end{align}
The total number of particles in the whole system is also zero in this limit.  For the ground state, as the chemical potential\index{chemical potential} approaches $ E_0 $ from below, the population diverges
\begin{align}
\lim_{\mu \rightarrow E_0^- } \bar{n}_0 (\mu)  = \infty .
\label{groundchemical}
\end{align}
We can thus say that the valid range of the chemical potential is
\begin{align}
-\infty < \mu < E_0 ,
\label{rangemu}
\end{align}
since otherwise we will have a negative population on the state with energy $ E_0 $.  Between these two extremes, the total particle number
\begin{align}
\langle N \rangle = \sum_k \bar{n}_k (\mu)  
\end{align}
can be swept from 0 to infinity.  If we would like to talk about a specific particle number $ N_0 $, then we can do this by first finding the chemical potential associated with $ \langle N \rangle  = N_0 $, and then take this to be the chemical potential associated with the system. \index{chemical potential}

\begin{figure}[t]
\centerline{\includegraphics[width=\textwidth]{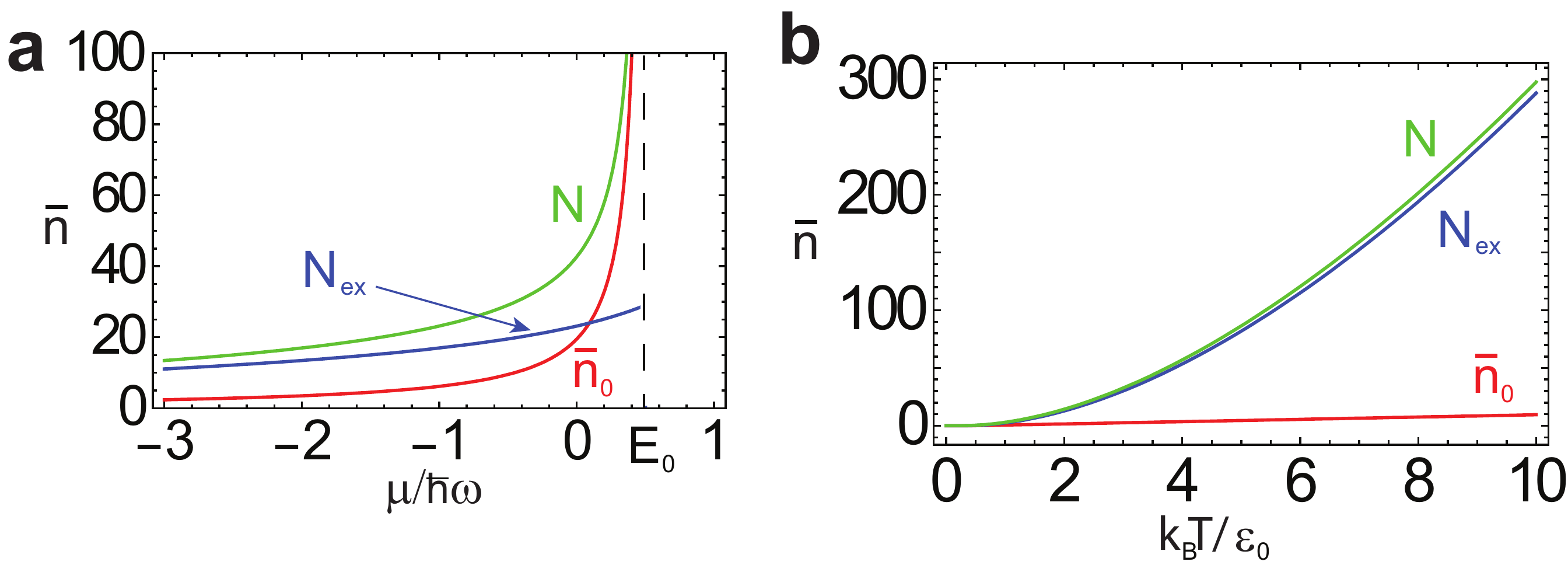}}
\caption{Number of particles occupying the ground state $ \bar{n}_0 $, all the remaining excited states $ N_{\text{ex}} $, and the total particle number $ N = \bar{n}_0 +  N_{\text{ex}}  $ according to the Bose-Einstein distribution for (a) a one-dimensional harmonic oscillator with fixed temperature $ \beta =  \hbar \omega $; (b) three-dimensional free particle with fixed chemical potential\index{chemical potential} $ \mu = -\epsilon_0 $.   }
\label{fig2-2}
\end{figure}

The fact that the population in the ground state (\ref{groundchemical}) diverges is not particularly surprising, since at this point the chemical potential $ \mu = E_0 $.  But there is something interesting that happens to the remaining levels that is particularly relevant to BEC.  If we assume that $E_1 - E_0  > 0 $, that is, that there is a non-zero energy difference between the ground state and first excited state, then there is a maximum to the number of particles that the excited states can contain. This is by virtue of the restricted range that the chemical potential can take (\ref{rangemu}).   This maximum is given by
\begin{align}
N_{\text{ex}}^{\max} = \sum_{k=1}^\infty \bar{n}_k (\mu = E_0)  = \sum_{k=1}^\infty \frac{1}{e^{\beta(E_k - E_0)} - 1} .
\label{excitedstatepop}
\end{align}
The specific maximum number depends upon the particular distribution of energies.  
In Fig. \ref{fig2-2}(a) we show an example for the case of a one-dimensional harmonic oscillator distribution  $ E_k = \hbar \omega ( k + 1/2) $.  We see that the ground state population diverges as expected from (\ref{groundchemical}), while the population in all the remaining states reaches a maximum according to (\ref{excitedstatepop}).  As $ \mu \rightarrow E_0^- $, the total population $ N $ is dominated by the ground state population $ \bar{n}_0 $.

\begin{exerciselist}[Exercise]
\item \label{q2-5}
Plot the number of particles occupying the ground state $ \bar{n}_0 $, all the remaining excited states $ N_{\text{ex}} $, and the total particle number $ N = \bar{n}_0 +  N_{\text{ex}}  $  versus the chemical potential for three-dimensional harmonic oscillator, where the bosons have also a spin degree of freedom $ \sigma = \pm 1 $.  The total energy can be modeled by  $ E_k = \hbar \omega ( k_x + k_y + k_z + 3/2) + \mu_B B \sigma $, where $ \mu_B $ is the Bohr magneton\index{Bohr magneton} (not to be confused with the chemical potential)\index{chemical potential}, and $ B $ is the magnetic field.  Describe the different cases of $ B = 0 $, $ B > 0 $, and $ B < 0 $.
\end{exerciselist}

\subsection{Bose-Einstein condensation}

In the previous section we derived the Bose-Einstein distribution, which tells us the average number of bosons for a given energy level $ E_k $ for a grand canonical ensemble\index{grand canonical ensemble}.  As we see from Fig. \ref{fig2-2}, as the chemical potential\index{chemical potential} approaches $ \mu \rightarrow E_0^- $, the entire population of the bosons resides in the ground state:
\begin{align}
\lim_{\mu \rightarrow E_0^-} \frac{\bar{n}_0}{N} = 1 .
\label{mulimit}
\end{align}
If we consider the signature of Bose-Einstein condensation to be a macroscopic occupation of the ground state, this certainly seems to be what is happening.  There is one problem here, which is that this is not exactly what typical experiments do.  As the chemical potential\index{chemical potential} is increased, what we are doing is letting more and more particles into the system, and thereby increasing the density.  But a more typical situation is that we cool the system to lower and lower temperatures, and at a certain point there is a critical point where macroscopic occupation of the ground state occurs.  Eq. (\ref{mulimit}) doesn't show any sign of a critical point, so it appears that we haven't quite derived the effect we are seeking yet. 

To see Bose-Einstein condensation, it turns out that we must look in three dimensions.  As a simple example, let us look at the case of a particle in three dimensions within a box of dimensions $ a \times a \times a $. Choosing periodic boundary conditions, the spectrum is 
\begin{align}
E_{k_x k_y k_z} = \epsilon_0  (k_x^2 + k_y^2 + k_z^2),
\end{align}
where $ \epsilon_0  = \frac{h^2}{2 m a^2} $ is the energy scale of the Hamiltonian, and $ k_{x,y,z} \in \{0,\pm 1, \pm 2, \dots \} $.  Periodic boundary conditions are chosen for convenience since this gives a unique ground state with $ k_x = k_y = k_z = 0 $ at zero energy, but does not affect the overall results.  The ground state occupation number\index{occupation number} according to (\ref{boseeinsteindist}) is
\begin{align}
\bar{n}_0 = \frac{1}{e^{-\beta \mu} - 1} .
\end{align}
Meanwhile, the population in the excited states is
\begin{align}
N_{\text{ex}} & =\int d^3 k \frac{1}{e^{\beta( \epsilon_0   k_r^2 - \mu)} - 1} \nonumber \\
& = (\frac{\pi}{ \epsilon_0  \beta})^{3/2} \text{Li}_{3/2} (e^{\beta \mu}),
\label{nexcitedthree}
\end{align}
where $ \text{Li}_n (z) $ is the polylogarithm function and $ k_r^2 = k_x^2 + k_y^2 + k_z^2$.  One may be concerned here that in (\ref{nexcitedthree}) we have included the contribution of the ground state as we have not made any restriction to the integral, as there should be in (\ref{excitedstatepop}).  There is however nothing to worry about since the ground state automatically has no contribution because in spherical coordinates $ d^3 k = k_r^2 \sin k_\theta d k_r d k_\theta d k_\phi $ and for the ground state its weight in the integral is zero.  

Now let us see whether we see the macroscopic occupation effect as the temperature is lowered. If we calculate the ratio of the ground state to the total population $ N = \bar{n}_0 + N_{\text{ex}} $, then we find that 
\begin{align}
\lim_{T \rightarrow 0 } \frac{\bar{n}_0}{N} = 1 ,
\end{align}
so we do see that all the particles occupy the ground state as the temperature is lowered. 
In Fig. \ref{fig2-2}(b) we plot the occupations for the ground state, and all the excited states.  We see that the basic effect of lowering the temperature is to reduce the population of the excited state. However we don't see any evidence of a critical temperature where there is a threshold where BEC occurs. 

What went wrong? A hint of this can be found in Fig. \ref{fig2-2}(b), where we see that changing the temperature for a fixed chemical potential\index{chemical potential} really has the effect of changing the total population $ N $ quite dramatically.  What more realistically happens in an experiment is that the particle number should be fixed to a constant, as the temperature is lowered.  All this happened because we chose to work in the grand canonical ensemble formalism where particles can freely enter and leave the system. \index{grand canonical ensemble}

\begin{figure}[t]
\centerline{\includegraphics[width=\textwidth]{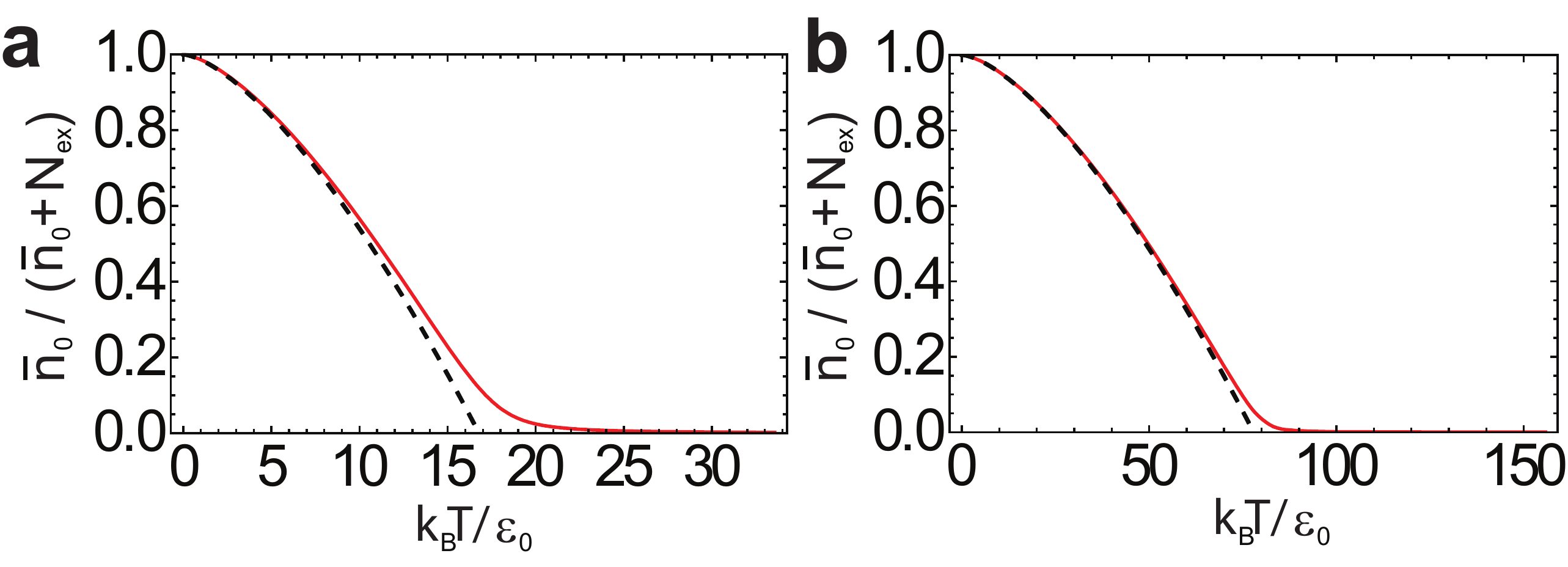}}
\caption{The condensate fraction as a function for temperature for a fixed total number of particles (a) $ N =10^3 $ corresponding to $k_B T_c/\epsilon_0 = 16.8 $; (b) $ N = 10^4 $  corresponding to $k_B T_c/\epsilon_0 = 77.9 $. Solid lines show a numerical solution of (\ref{chemfind}) for the chemical potential\index{chemical potential}, the dashed lines show the asymptotic result for an infinite number of particles.  }
\label{fig2-3}
\end{figure}

Let's try again but this time we vary of the chemical potential with the temperature such that the total number of particles $ N $ is fixed.  That is, for a particular temperature $ T $ we find the solution with respect to $ \mu(T) $ of the equation
\begin{align}
\bar{n}_0 +N_{\text{ex}} = \frac{1}{e^{-\beta \mu(T) } - 1} +  (\frac{\pi}{ \epsilon_0  \beta})^{3/2} \text{Li}_{3/2} (e^{\beta \mu(T) })  = N .
\label{chemfind}
\end{align}
Once we have found our function $ \mu(T) $  for a fixed $ N $, we can then compare the number of particles in the ground state to the total number $ \bar{n}_0/N $.  The results are shown in Fig. \ref{fig2-3}.  This time we do see a sharp transition where the number of particles in the ground state starts to pick up strongly at a particular temperature. As the total number of particles increases, this transition gets sharper and sharper.  As before, at $ T = 0 $ the condensate fraction is 1, but most importantly there is  a range of temperatures below which there is a macroscopic occupation of the ground state.  This is Bose-Einstein condensation.

Can we estimate what the critical temperature is? We can do this by returning to Fig. \ref{fig2-2}(a) and noticing the crucial property that $ N_{\text{ex}} $ has a maximum even when the chemical potential\index{chemical potential} is at its largest point $ \mu = E_0 $. Also note from Fig. \ref{fig2-2}(b) that the occupation numbers\index{occupation number} generally decrease with temperature.  Suppose then that there are a total number of particles that happens to be much bigger than $ N_{\text{ex}} $.  Then we would surely have a macroscopic population in the ground state since $ \bar{n}_0  = N- N_{\text{ex}} $.  The temperature that the ground state population starts to increase strongly can then be estimated by setting $ \mu \approx E_0 =0$ and finding when $ N = N_{\text{ex}}  $.  This gives \index{critical temperature for BEC}
\begin{align}
k_B T_{\text{c}} = \frac{h^2}{2 \pi m} \left( \frac{n}{\text{Li}_{3/2} (1)} \right)^{2/3}
\label{criticaltemp}
\end{align}
where $ n = N/a^3 $ is the density of the particles, and $ \text{Li}_{3/2} (1) = 2.612 $.  This can be conveniently rewritten in terms of the thermal de Broglie wavelength \index{de Broglie wavelength}
\begin{align}
\lambda_T = \frac{h}{\sqrt{2 \pi m k_B T}},
\end{align}
giving
\begin{align}
n \lambda_{T_c}^3 = 2.612 .
\label{debrogliebeccrit}
\end{align}
Since the thermal de Broglie wavelength is the average de Broglie wavelength of particles moving at a particular temperature, this has the physical interpretation that BEC occurs when the matter wave wavelengths start to be comparable to the interparticle distances.  We derived the BEC criterion for the particular case of bosons obeying a harmonic oscillator potential.  For other geometries we would generally expect similar results to (\ref{debrogliebeccrit}), with possibly a different constant on the right hand side. \index{ de Broglie wavelength}

We can find an approximate dependence to the ground state population in Fig. \ref{fig2-3} by again noting that for this temperature range the chemical potential will be close to the ground state $ \mu \approx E_0 =0$.  The excited state population (\ref{nexcitedthree}) will therefore be
\begin{align}
N_{\text{ex}} & \approx N \left( \frac{T}{T_c} \right)^{3/2} ,
\end{align}
where we have used our expression for the critical temperature (\ref{criticaltemp}).  The ground state is then according to $ \bar{n}_0  = N - N_{\text{ex}} $:
\begin{align}
\bar{n}_0 (T) =  N \left( 1 - \left( \frac{T}{T_{\text{c}}} \right)^{3/2} \right) .  
\label{groundstatetemp}
\end{align}
In practice the approximation works rather well as can be seen in Fig. \ref{fig2-3} by the dashed lines. Strictly speaking (\ref{groundstatetemp}) is only valid in the limit of infinite density, so the sharp transition is smoothed out in practice.

\section{Low-energy excited states}
\label{sec:excitedstates}

We have seen that a bosonic gas at low enough temperatures will have a macroscopic occupation of the ground state.  Up to this point we have not included the effect of interactions as we have discussed in  Sec. \ref{sec:interactions}.  Typically the energy due to interactions has a much lower energy than the kinetic energy.  Hence to lowest order the state of the gas can be approximated by all the bosons occupying the lowest energy state of the Hamiltonian (\ref{multiham}). In this section we show the effect of introducing interactions between bosons.  We shall see that this has a dramatic effect on the dispersion relation\index{dispersion! relation} of the bosons, and is a key reason why BECs have spectacular properties such as superfluidity\index{superfluidity}. 

Let us start by writing the full Hamiltonian of the interacting boson gas, assuming the most common situation of a $s$-wave interaction
\begin{align}
{\cal H}  & = \sum_{\bm{k}} E_{\bm{k}} a^\dagger_{\bm{k}}  a_{\bm{k}}  + 
 \frac{U_0}{2 V}   \sum_{\bm{k} \bm{k}' \bm{q}} a^\dagger_{\bm{k}'+\bm{q}}  a^\dagger_{\bm{k}-\bm{q}} a_{\bm{k}} a_{\bm{k}'} 
\label{interactingtotalham}
\end{align}
where we have substituted (\ref{swaveexpression}) into (\ref{interactionhamch1}) and $ U_0 $ is the interaction energy as defined in (\ref{gintdef}).  We now would like to find the solution of (\ref{interactingtotalham}) in the regime where there is macroscopic occupation of the ground state $ a_0 $ and the interactions are weak.  Generally a Hamiltonian that is quadratic in the bosonic annihilation and creation operators $ a_{\bm{k}} $ and $ a^\dagger_{\bm{k}} $ can be solved by a bosonic transformation, but since (\ref{interactingtotalham}) is fourth order, it cannot be solved.  However, using the fact that we expect most of the bosons to occupy the ground state $ a_0 $, we can approximate the above Hamiltonian to second order in bosonic operators, such that it can be solved. 

%
We approximate the interaction part of the Hamiltonian (\ref{interactingtotalham}) by keeping terms only involving the macroscopically occupied  state $ a_0 $. There are a total of 7 such terms.   The first such term amounts to setting $ \bm{k} = \bm{k}' = \bm{q} = 0 $.  The other 6 terms involve setting two of the momentum labels to zero, and the other two to non-zero terms.  Due to momentum conservation there are no terms with one or three $ a_0 $'s.  This gives
\begin{align}
{\cal H}_{\text{B}}  & = \sum_{\bm{k}} E_{\bm{k}} a^\dagger_{\bm{k}}  a_{\bm{k}}  + 
 \frac{U_0}{2 V}  a^\dagger_0 a^\dagger_0  a_0 a_0  + 
\sum_{\bm{k} \ne 0 } \Big[ 4 a^\dagger_0 a^\dagger_{\bm{k}} a_{\bm{k}} a_0 + a^\dagger_{\bm{k}} a^\dagger_{-\bm{k}} a_0 a_0  + a^\dagger_{0}  a^\dagger_{0} a_{-\bm{k}} a_{\bm{k}}  \Big] .
\end{align}
Due to the large population in the ground state we may replace $ a_0 \rightarrow e^{i\varphi} \sqrt{N_0} $, where $ N_0 $ is the number of atoms in the ground state. Here $ \varphi $ is the phase which may be present since $ a_0 $ is a non-Hermitian operator, and thus can have a complex expectation value.  We then have
\begin{align}
{\cal H}_{\text{B}}  & = \sum_{\bm{k}} E_{\bm{k}} a^\dagger_{\bm{k}}  a_{\bm{k}}  + 
 \frac{U_0}{2 V} N_0^2  + 
\frac{U_0}{2 V} \sum_{\bm{k} \ne 0 } \Big[ 4 N_0  a^\dagger_{\bm{k}} a_{\bm{k}} 
+ N_0 e^{2i\varphi}  a^\dagger_{\bm{k}} a^\dagger_{-\bm{k}}  + N_0 e^{-2i\varphi}  a_{-\bm{k}} a_{\bm{k}}  \Big] .
\end{align}
The number of atoms in the ground state is related to the total number of atoms by 
\begin{align}
N_0 = N - \sum_{\bm{k} \ne 0 }  a^\dagger_{\bm{k}} a_{\bm{k}}  .
\end{align}
Substituting this into  $ {\cal H}_{\text{B}}  $ and keeping only quadratic terms in bosonic operators we have
\begin{align}
{\cal H}_{\text{B}}  & = E_0 N +  \frac{U_0 nN }{2}   
+ \sum_{\bm{k} \ne 0} (E_{\bm{k}}-E_0) a^\dagger_{\bm{k}}  a_{\bm{k}}  + 
\frac{U_0 n}{2} \sum_{\bm{k} \ne 0 } \Big[ 2  a^\dagger_{\bm{k}} a_{\bm{k}} 
+ e^{2i\varphi}  a^\dagger_{\bm{k}} a^\dagger_{-\bm{k}}  +  e^{-2i\varphi}  a_{-\bm{k}} a_{\bm{k}}  \Big] .
\label{bogham}
\end{align}
where we have defined the density $ n = N/V $.  

Now that the Hamiltonian has been approximated to quadratic powers of bosonic operators, it can be diagonalized.  The Bogoliubov transformation\index{Bogoliubov transformation} defines new bosonic operators $ b_{\bm{k}} $ and are related to the existing ones according to
\begin{align}
a_{\bm{k}} & = e^{i\varphi} (\cosh \xi_{\bm{k}} b_{\bm{k}} + \sinh \xi_{\bm{k}} b_{-\bm{k}}^\dagger )\nonumber \\
a_{-\bm{k}}^\dagger & = e^{-i\varphi} ( \sinh \xi_{\bm{k}} b_{\bm{k}}  + \cosh \xi_{\bm{k}} b_{-\bm{k}}^\dagger ).
\label{bogoliubovtrans}
\end{align}
The coefficients of the new bosonic operators are chosen such as the operators $ a_{\bm{k}} $ always satisfy bosonic commutations as they should. We have assumed that the coefficients are symmetric $ \xi_{\bm{k}} = \xi_{-\bm{k}} $.  This ensures that the new operators $ b_{\bm{k}} $ satisfy bosonic commutation relations.  Substituting (\ref{bogoliubovtrans}) into (\ref{bogham}) we obtain\index{commutation relations}
\begin{align}
& {\cal H}_{\text{B}}  =  E_0 N + \frac{U_0 nN }{2}  + \sum_{\bm{k} \ne 0} \Big\{  \Big[  
\delta E_{\bm{k}} \cosh 2 \xi_{\bm{k}}   + U_0 n \sinh 2 \xi_{\bm{k}} \Big] b_{\bm{k}}^\dagger b_{\bm{k}}   \nonumber \\
& \Big[  
\frac{\delta E_{\bm{k}}}{2} \sinh 2 \xi_{\bm{k}}  +  \frac{gn}{2}  \cosh 2 \xi_{\bm{k}}   \Big] ( b_{\bm{k}} b_{-\bm{k}} + b_{\bm{k}}^\dagger b_{-\bm{k}}^\dagger )  + \delta E_{\bm{k}}  \sinh^2 \xi_{\bm{k}} + \frac{U_0 n}{2}   \sinh 2 \xi_{\bm{k}}  \Big\} ,
\end{align}
where $ \delta E_{\bm{k}} = E_{\bm{k}}-E_0 + U_0 n $. To diagonalize this Hamiltonian, we require setting the off-diagonal terms proportional to $  b_{\bm{k}} b_{-\bm{k}} + b_{\bm{k}}^\dagger b_{-\bm{k}}^\dagger  $ to zero.  The condition for this is
\begin{align}
\coth 2 \xi_{\bm{k}} = -\frac{\delta E_{\bm{k}}}{U_0 n} .
\end{align}
Solving for the coefficients themselves, we take the solutions with signs
\begin{align}
\cosh \xi_{\bm{k}} & = \sqrt{\frac{\delta E_{\bm{k}}}{2 \epsilon_{\bm{k}} } + \frac{1}{2} } \nonumber \\
\sinh \xi_{\bm{k}} & = -\sqrt{\frac{\delta E_{\bm{k}}}{2 \epsilon_{\bm{k}} } - \frac{1}{2} } ,
\end{align}
where 
\begin{align}
\epsilon_{\bm{k}} = \sqrt{(E_{\bm{k}}-E_0 )(E_{\bm{k}}-E_0  + 2 U_0 n)} .
\label{bogdispersion}
\end{align}
The diagonalized form of the Hamiltonian then reads
\begin{align}
{\cal H}_{\text{B}}  =  E_0 N + \frac{U_0  nN }{2 }  
+ \sum_{\bm{k} \ne 0} \Big\{ \epsilon_{\bm{k}} b_{\bm{k}}^\dagger b_{\bm{k}}  + [ \frac{\epsilon_{\bm{k}} -  \delta E_{\bm{k}} }{2} ]  \Big\} .
\label{bogdiag}
\end{align}

Now that the Hamiltonian is diagonalized, we can better understand the effect of the interactions. What (\ref{bogdiag}) tells us is that the effect of the interactions is to make a new type of bosonic particle, called a {\it bogoliubovon}, which is annihilated by the operators $ b_{\bm{k}} $.  \index{bogoliubovon} Since $ b_{\bm{k}}^\dagger b_{\bm{k}} $ is just a number operator for these particles, the eigenstates of (\ref{bogdiag}) are simply the Fock states with respect to\index{bogoliubovon} bogoliubovons:\index{Fock states}
\begin{align}
| \bm{n}_{\bm{k}}  \rangle_b = \prod_{\bm{k}} \frac{ (b_{\bm{k}}^\dagger)^{n_{\bm{k}}} }{\sqrt{ n_{\bm{k} !}}}  | \tilde{0} \rangle
\end{align}
where we used the same form as (\ref{generalizedbosonstate}).  The difference to  (\ref{generalizedbosonstate}) is that the ground state $ | \tilde{0} \rangle $ in this case is not the state with no atoms as it was before.  It can explicitly be written as
\begin{align}
| \tilde{0} \rangle = \frac{1}{\cosh \xi_{\bm{k}}} \exp ( e^{2i \varphi} \tanh \xi_{\bm{k}} a_{\bm{k}}^\dagger
a_{-\bm{k}}^\dagger  ) | 0 \rangle  ,
\label{bogvacuum}
\end{align}
which satisfies 
\begin{align}
 b_{\bm{k}} | \tilde{0} \rangle = 0 .
\label{bogonvac}
\end{align}
The lowest energy state has no bogoliubovons\index{bogoliubovons} and has an energy 
\begin{align}
{\cal H}_{\text{B}} | \tilde{0} \rangle = \left[ E_0 N + \frac{U_0 nN }{2 }  
+ \sum_{\bm{k} \ne 0}  \frac{\epsilon_{\bm{k}} -  \delta E_{\bm{k}} }{2}   \right] | \tilde{0} \rangle .
\end{align}
The state (\ref{bogvacuum}) in terms of the original $ a_{\bm{k}} $ operators has a non-zero population in the excited state as can be found by evaluating
\begin{align}
\langle \tilde{0} | a_{\bm{k}}^\dagger a_{\bm{k}} |  \tilde{0} \rangle =  \sinh^2 \xi_{\bm{k}} .
\label{averageak}
\end{align}
We can interpret this as the interactions causing some of the non-zero momenta states to be excited, thereby lowering the energy of the whole system.  

\begin{figure}[t]
\centerline{\includegraphics[width=0.5\textwidth]{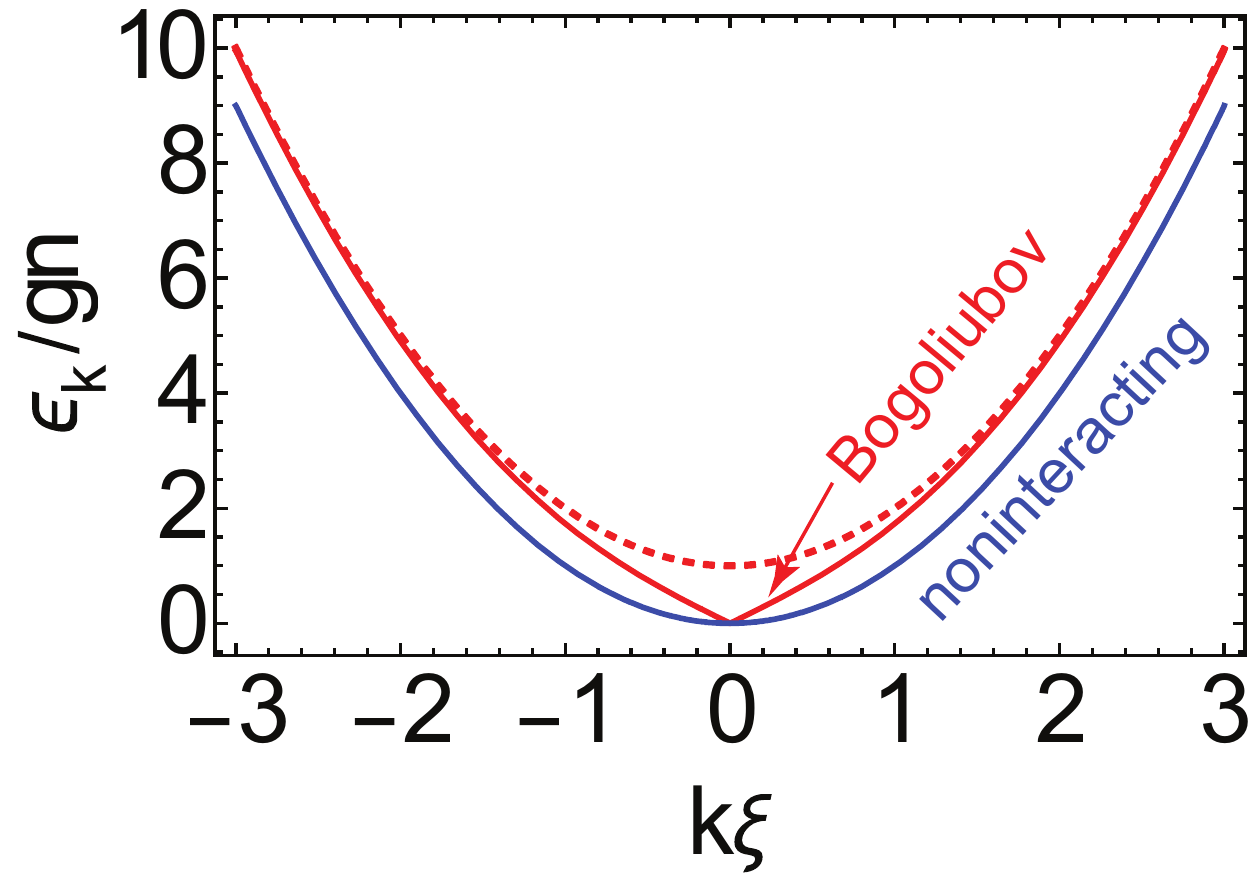}}
\caption{The Bogoliubov dispersion\index{dispersion!Bogoliubov }  $\epsilon_{\bm{k}} $ in (\ref{bogdispersion}). The momentum is in units of the healing length $ \xi =\hbar/\sqrt{2m U_0 n} $ and the energy is in units of the interaction strength $ U_0 n $. The original non-interacting energy dispersion \index{dispersion!energy}$ E_{\bm{k}}-E_0  = \frac{\hbar^2 k^2}{2m} $, and the limiting case for large $ k $ is also shown for comparison.\index{healing length}  }
\label{fig2-4}
\end{figure}

The elementary excitations have an energy dispersion\index{dispersion! energy} according to (\ref{bogdispersion}) and is shown in Fig. \ref{fig2-4}.  Comparing it to the standard parabolic dispersion\index{dispersion!parabolic},  the Bogoliubov dispersion\index{dispersion!Bogoliubov } exhibits a linear dispersion\index{dispersion!linear} at low momenta.  Expanding the dispersion (\ref{bogdispersion}) for small $ k $ we can write
\begin{align}
\epsilon_{\bm{k}} & \approx \sqrt{2 U_0  n(E_{\bm{k}}-E_0 )}  \hspace{1cm} (|k\xi | < 1)   \nonumber \\
& = \sqrt{2} U_0 n \xi |k|  ,
\label{linearbogo}
\end{align}
where in the second line we took the original dispersion\index{dispersion} to be $ E_{\bm{k}}-E_0 = \frac{\hbar^2 k^2}{2m} $, 
and $ \xi = \hbar/\sqrt{2m U_0 n} $ is the healing length of the BEC which will be discussed more in Sec. \ref{sec:healinglength}.  At larger momenta, the Bogoliubov dispersion again has a parabolic form, which can be approximated by\index{dispersion!Bogoliubov }\index{healing length}
\begin{align}
\epsilon_{\bm{k}}&  \approx E_{\bm{k}}-E_0  + U_0 n  \hspace{1cm} (|k \xi | \gg 1) .
\label{bogdispersionapprox}
\end{align}
The parabolic dispersion is offset by a constant amount $ U_0 n $.  The approximation works well for momenta $ k \xi \gg 1 $. \index{dispersion!parabolic}

\begin{exerciselist}[Exercise]
\item \label{q2-6}
Verify that the $ b_{\bm{k}} $ operators satisfy bosonic commutation relations $ [b_{\bm{k}},b_{\bm{k}'}^\dagger ] = \delta_{\bm{k} \bm{k}'} $. Hint: First invert the relation (\ref{bogoliubovtrans}) such that that $ b_{\bm{k}} $ operators are written in terms of the $ a_{\bm{k}}, a_{-\bm{k}}^\dagger $ operators. This is best done by multiplying the two equations by a suitable coefficient and taking advantage of the relation $ \cosh^2 x - \sinh^2 x = 1 $.  
\item \label{q2-7}
Show that (\ref{bogvacuum}) is the vacuum for the Bogoliubov operators\index{Bogoliubov operators} $ b_{\bm{k}} $.  Hint: Expand (\ref{bogvacuum}) as a Taylor series and apply the operator $ b_{\bm{k}} $ written in terms of $ a_{\bm{k}}, a_{-\bm{k}}^\dagger $ operators found in the previous question.  
\item \label{q2-8}
Verify that the average number of $ a_{\bm{k}} $ particles is (\ref{averageak}). Try this two ways: First, evaluate it by substituting the state (\ref{bogvacuum}) into  (\ref{averageak}) and evaluating the Taylor expanded sum.  Second, transform the $ a_{\bm{k}} $ operators to $ b_{\bm{k}} $  operators and use the fact that (\ref{bogonvac}).
\end{exerciselist}

\section{Superfluidity}\index{superfluidity}
\label{sec:superfluidity}


\begin{figure}[t]
\centerline{\includegraphics[width=\textwidth]{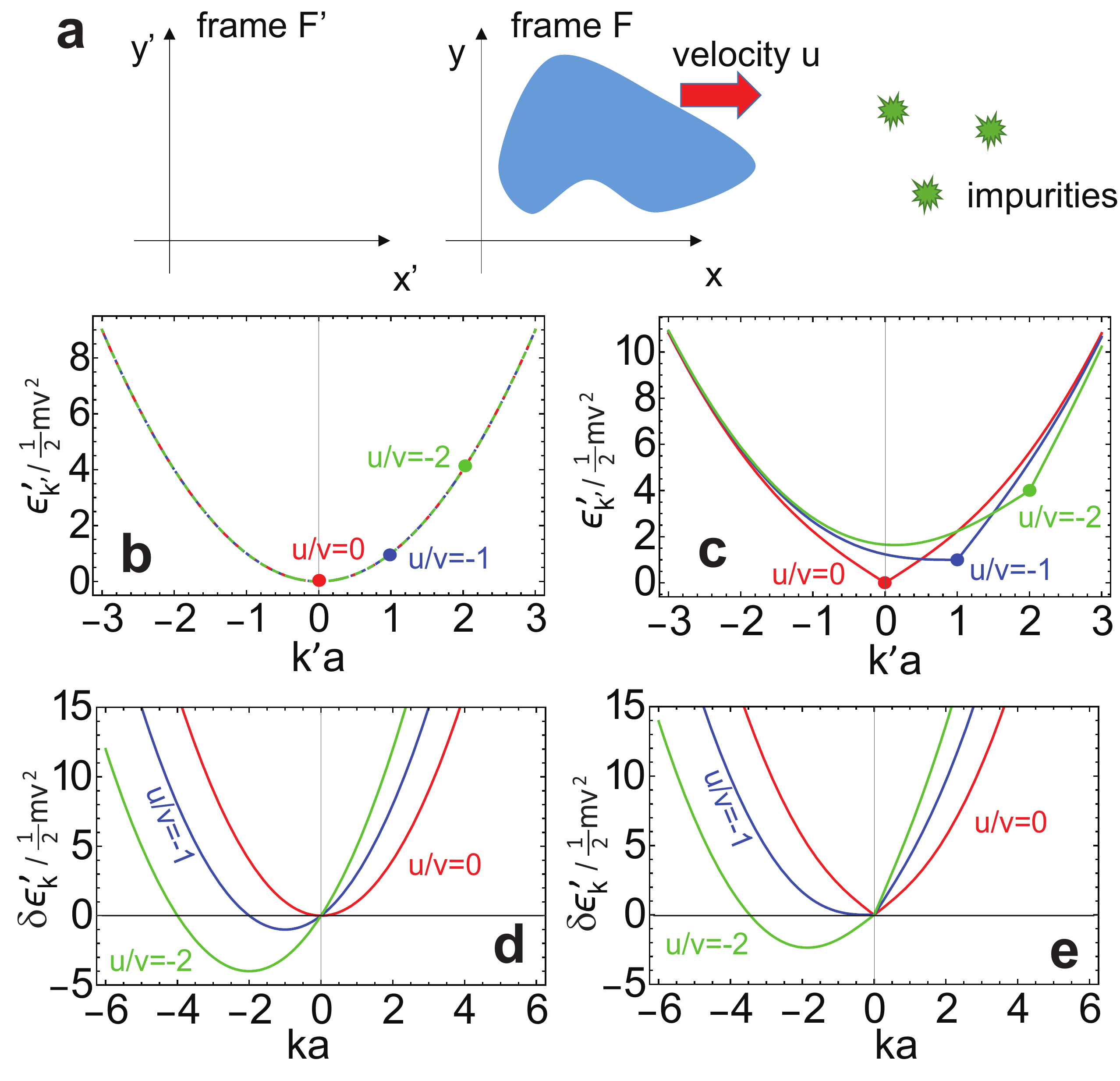}}
\caption{Landau's argument for superfluidity\index{superfluidity}.  (a) Coordinate reference frames as defined in the text.  Dispersion relations after a Galilean transformation\index{Galilean transformation} for various velocities $ u $ and a (b) non-interacting standard dispersion with $ U_0 n = 0 $; (c) interacting Bogoliubov dispersion\index{dispersion!Bogoliubov} with $ U_0 n=mv^2$.  The excitation energies defined by (\ref{landauinequality}) for the (d) non-interacting and (e) interacting case with the same parameters as (b) and (c) respectively.  The length scale as set to be $ a =\hbar/mv $, and the energy scale is $ \frac{1}{2}mv^2 $.  The dots indicate the momentum of the zero momentum state $ k = 0 $ with respect to the frame $ F $.  }
\label{fig2-5}
\end{figure}

The Bogoliubov distribution\index{Bogoliubov distribution} derived in the last section gives rise to one of the most remarkable effects seen in BECs: superfluidity.  As the name suggests, superfluidity is the phenomenon  of a fluid possessing zero viscosity, such that it may flow without resistance.  For example, when arranged in a circular geometry, a superfluid can keep flowing around the loop indefinitely.  In practice, once the fluid exceeds a particular velocity, the superfluid\index{superfluid} loses its properties and can no longer flow with zero viscosity.  In this section we derive Landau's criterion for this velocity, which gives insight to the reason why superfluidity occurs. \index{Landau's criterion}

Consider a BEC which is moving with velocity $ \bm{u} $ in the $x$-direction. Such a configuration might be prepared by first preparing a BEC such that the atoms condense into a zero-momentum state, then momentum is added to all the atoms together such that the whole system is moving with velocity $ u $.  For simplicity, we assume that the temperature is zero, so there are no thermal excitations, and the initial state is exactly in the ground state. Now consider two coordinate frames, one in the laboratory frame $ F' $ and another in the moving frame $ F $, which is also moving in the $ x$-direction with velocity $ u $. Thus with respect to frame $ F $, the laboratory frame $ F' $ is moving with velocity $ \bm{v} = - \bm{u} $ (see Fig. \ref{fig2-5}(a))  The coordinates between the two frames are related according to
\begin{align}
\bm{x}' = \bm{x} + \bm{u} t  ,
\end{align}
where $ \bm{x}, \bm{x}' $ are the coordinates in the frames $ F, F' $ respectively.

Given that in the frame $ F $ the dispersion relation\index{dispersion!relation} is of a form $ \epsilon_{\bm{k}} $, let us work out what the excitation spectrum is in the coordinates of $ F' $.  From general considerations of coordinate transformations, we know (see the Box in this section) that this must be
\begin{align}
\bm{k}' & = \bm{k} + \frac{m\bm{u}}{\hbar} \label{ktransform} \\
 \epsilon_{\bm{k}}' & = \epsilon_{\bm{k}} + \hbar \bm{k}\cdot \bm{u} + \frac{1}{2} m u^2 ,
\label{energytransform}
\end{align}
where $  \epsilon_{\bm{k}}' $ is the energy dispersion\index{dispersion!energy} in frame $ F' $, in terms of the momentum variables $ \bm{k} $ of frame $ F $.  

First let us consider the case of a regular dispersion\index{dispersion} without interactions $ U_0 = 0 $.  In this case the dispersion relation\index{dispersion!relation} in frame $ F $ is 
\begin{align}
\epsilon_{\bm{k}} = \frac{\hbar^2 k^2}{2m} .  
\end{align}
Substituting this relation into (\ref{ktransform}) and (\ref{energytransform}), we can find the dispersion relation in frame $ F' $ with the new coordinates which are
\begin{align}
\epsilon_{\bm{k}'}' = \frac{\hbar^2 (k')^2}{2m} ,  
\end{align}
which takes exactly the same form.  In this frame the atoms are all moving with momentum
\begin{align}
\bm{k}' =  \frac{m\bm{u}}{\hbar} .
\label{becmomentum}
\end{align}
In Fig. \ref{fig2-5}(b) we plot the dispersion of the non-interacting case.  \index{dispersion}

 Now suppose that the BEC encounters a few impurities which have fixed positions in the laboratory frame $ F' $, and the BEC must flow around these.  In a normal fluid ($ U_0 = 0 $), we expect that such impurities will create excitations such that the atoms  no longer all have the momenta given by (\ref{becmomentum}).  The collision process should scatter many of the atoms to the opposite momenta (e.g. $\bm{k}' <0 $ for $ u > 0 $, which also implies $ \bm{k} < 0 $) so that eventually the fluid will slow down and finally stop. 

Let us look at the energy cost of creating such an excitation in the frame $ F' $.  We compare the energy of no excitation (i.e. keeping a particular atom at $\bm{k} = 0 $) and exciting it to some $ |\bm{k}|>0 $.  This energy difference, or excitation energy, is
\begin{align}
\delta \epsilon_{\bm{k}}= \epsilon_{\bm{k}}' -  \epsilon_{0}' = \epsilon_{\bm{k}} - \epsilon_{0} + \hbar \bm{k} \cdot \bm{u}  .
\label{landauinequality}
\end{align}
A plot of this for the non-interacting case is shown in Fig.  \ref{fig2-5}(d) (note that the horizontal axis is with respect to $ \bm{k} $, not $ \bm{k}' $).  We see that the excitation energy becomes negative for a range of momenta with $ \bm{k} <0 $ as expected.  Since the excitation energy is negative, the system is susceptible to create many excitations with these momenta, which eventually slows and stops the BEC. 

If we now repeat the argument for the interacting case, we see completely different behavior.  In Fig. \ref{fig2-5}(c) the dispersion relation \index{dispersion!relation}for the Bogoliubov dispersion\index{dispersion!Bogoliubov} derived in the previous section is shown.  In this case, the dispersion is not invariant like in the non-interacting case. In the excitation energy plotted in Fig. \ref{fig2-5}(e), we observe that for the $ u/v = - 1 $ curve the excitation energy remains positive for all $ \bm{k} $.  This means that the system is not susceptible to the creation of many excitations. Without the creation of excitations, the atoms moving with velocity $ \bm{u} $ are not scattered, and the movement of the atoms are unhindered. For the  $ u/v = -2 $ curve however, we see that the excitation energy is negative, and the system is again susceptible to the creation of excitations which eventually slow the system down.  \index{dispersion}

The criterion for superfluidity is then given by the condition that (\ref{landauinequality}) is positive for all momenta:\index{superfluidity}
\begin{align}
\epsilon_{\bm{k}} - \epsilon_{0}  + \hbar \bm{k} \cdot \bm{u} > 0 .  
\label{landaufun}
\end{align}
As the velocity of the fluid $ \bm{u} $ is increased,  superfluidity  eventually breaks down, as illustrated above. Let us find what this critical velocity is that marks this boundary.  Setting the left hand side of (\ref{landaufun}) to zero and rearranging, we have 
\begin{align}
u_c = \min_{\bm{k}} \frac{\epsilon_{\bm{k}}- \epsilon_{0}  }{\hbar \bm{k}}
\label{landaucriterion}
\end{align}
where we have taken the minimum of all the momenta that satisfy the boundary condition, in case there are multiple solutions.  This is the famous {\it Landau's criterion}. \index{Landau's criterion}  This predicts the maximum velocity that a fluid can flow in the superfluid phase.

\begin{framed}
{\centering \bf Galilean transformation of the Schrodinger equation \\\index{Galilean transformation}
}
\bigskip

Consider the standard Schrodinger equation in a frame $ F $ defined with 
coordinates $ \bm{x} $ and time $ t $:
\begin{align}
i\hbar  \frac{\partial  \psi (\bm{x},t)}{\partial t} = \left[ -\frac{\hbar^2}{2m} \nabla_{\bm{x}}^2  + V(\bm{x}) \right] \psi (\bm{x},t) ,
\end{align}
where $ \nabla_{\bm{x}} $ indicates the gradient with respect to the variable $ \bm{x} $.  
Now consider another frame $ F' $ that is moving with velocity $ \bm{v} $ with constant velocity.  
The coordinates in this frame are defined by $ \bm{x}' = \bm{x} -\bm{v} t $.  Making the transformation, in  the new coordinates
the Schrodinger equation is written
\begin{align}
i\hbar \frac{\partial  \psi (\bm{x}'+\bm{v}t,t)}{\partial t} &  -i \hbar \bm{v} \cdot \nabla_{\bm{x}'}  \psi (\bm{x}'+\bm{v}t,t) = \nonumber \\
&  \left[ -\frac{\hbar^2}{2m} \nabla_{\bm{x}'}^2  + V(\bm{x}'+\bm{v}t ) \right] \psi (\bm{x}'+\bm{v}t,t),
\label{schrodingerfpframe}
\end{align}
where we used
\begin{align}
\psi (\bm{x},t)& = \psi (\bm{x}'+\bm{v}t,t) \nonumber \\
\frac{\partial  \psi (\bm{x},t)}{\partial t}  & = \frac{\partial  \psi (\bm{x}'+\bm{v}t,t)}{\partial t}  - \bm{v} \cdot \nabla_{\bm{x}'}  \psi (\bm{x}'+\bm{v}t,t) \nonumber \\
\nabla_{\bm{x}}\psi (\bm{x},t) & = \nabla_{\bm{x}'} \psi (\bm{x}'+\bm{v}t,t) .
\end{align}

In the new frame $ F' $, the particle must also obey the Schrodinger equation, which must take the form
\begin{align}
i\hbar \frac{\partial  \psi' (\bm{x}',t)}{\partial t} = \left[ -\frac{\hbar^2}{2m} \nabla_{\bm{x}'}^2  + V'(\bm{x}') \right] \psi' (\bm{x}',t).
\label{transschro}
\end{align}
The same physical wavefunction  $ \psi' (\bm{x}',t) $ in the frame $ F'$  will take a different form to that in the original frame $ F $ since it is described with different variables.  For example, the momentum and energy will not be measured to be the same values.  We can relate the same wavefunction in the two coordinates using the transformation
\begin{align}
\psi (\bm{x}'+\bm{v}t,t) = e^{i( m \bm{v} \cdot \bm{x}' + m v^2 t/2)/\hbar} \psi' (\bm{x}',t) .
\label{transformationgalileo}
\end{align}
One can verify that substitution of (\ref{transformationgalileo}) into (\ref{schrodingerfpframe}) yields (\ref{transschro}), where we have defined $  V'(\bm{x}') =  V(\bm{x}'+\bm{v}t ) $.

For example, consider a particle in a plane wave in the frame $ F $ with the wavefunction
\begin{align}
\psi (\bm{x},t) = e^{i(\bm{p} \cdot \bm{x}-  Et)/\hbar} .
\end{align}
Using the formula (\ref{transformationgalileo}), we find that the wavefunction in the frame $ F'$ is 
\begin{align}
\psi' (\bm{x}',t) = e^{i[(\bm{p}-m\bm{v}) \cdot \bm{x}'- (E - \bm{p}\cdot \bm{v} + m v^2/2) t] /\hbar} . 
\end{align}
Hence the momentum $\bm{p}' $ and energy $ E' $ in the frame $F ' $ is related to the original values as
\begin{align}
\bm{p}' & = \bm{p}-m\bm{v} \nonumber \\
E' & = E - \bm{p}\cdot \bm{v} + \frac{1}{2} m v^2 .
\end{align}
\end{framed}

\begin{exerciselist}[Exercise]
\item \label{q2-9}
 Verify that  substitution of (\ref{transformationgalileo}) into (\ref{schrodingerfpframe}) yields (\ref{transschro}).
\item \label{q2-10}
For the Bogoliubov dispersion find the maximum velocity that a superfluid can flow according to Landau's criterion assuming the linear dispersion (\ref{linearbogo}).  \index{dispersion!Bogoliubov}
\end{exerciselist}\index{Landau's criterion}

\section{References and further reading}

\begin{itemize}
\item Sec. \ref{sec:intro2}: To learn more about Bose-Einstein condensates in general, see the textbooks \cite{pitaevskii2016bose,pethick2002bose}.
\item Sec. \ref{sec:einstein}: The original papers detailing Bose and Einstein's original argument \cite{bose1924plancks,einstein1924quantentheorie}.
\item Sec. \ref{sec:grand}: The first experiments showing Bose-Einstein condensation are \cite{anderson1995observation,davis1995bose}.  Other early experiments are reported in \cite{fried1998bose,jochim2003bose,zwierlein2003observation,griesmaier2005bose,kraft2009bose}.  Reviews and books relating to the topic are given in \cite{anglin2002bose,pitaevskii2016bose,pethick2002bose,griffin1996bose,dalfovo1999theory,leggett2001bose}.  
\item Sec. \ref{sec:excitedstates}: Bogoliubov's original theory of low-energy excited states of a BEC \cite{bogolyubov1947theory}.  Experimental observation \cite{steinhauer2002excitation}.  Review articles and books on the topic \cite{pitaevskii2016bose,leggett2001bose,pethick2002bose}.
\item Sec. \ref{sec:superfluidity}: The original theories of superfluidity \cite{landau1941theory,bogolyubov1947theory,ginzburg1958theory}. Experimental observation \cite{marago2000observation}. Review articles and books on the topic \cite{pitaevskii2016bose,khalatnikov2018introduction,tilley2019superfluidity,leggett1999superfluidity}.  
\end{itemize}

  \chapter[The Order Parameter and Gross-Pitaevskii equation]{The Order Parameter and Gross-Pitaevskii equation}

\label{ch:order}

\section{Introduction}

We have seen that a BEC can be described by a quantum many-body state of bosons, with a macroscopic occupation of the ground state. While this a mathematically complete framework to describe the system, it is also rather difficult to visualize since it is inherently involves many particles.  In many situations it is useful to approximately capture the essential physics, without having all the details that describe the system.  In this respect the order parameter, and its equation of motion, the Gross-Pitaevskii equation, is a very popular framework to describe a BEC, as it gives a simple way to visualize the state.    In this chapter we describe these and related concepts such as the healing length, vortices, solitons, and hydrodynamic equations. \index{healing length} \index{vortices} \index{solitons} \index{order parameter} \index{hydrodynamic equation}

\section{Order parameter}
\label{sec:orderparameter}

In the previous chapters we established the essential feature of a BEC, that there is a macroscopic occupation of the ground state.  What is the wavefunction of this state? Assuming there are no interactions between the bosons, we can easily write this down using the methods that we worked out in Chapter \ref{ch:quantum}. Using the notation of (\ref{generalizedbosonstate}) we have
\begin{align}
|\text{BEC} (N) \rangle  = | N, 0, \dots \rangle =  \frac{(a^\dagger_0)^{N} }{\sqrt{N !}} | 0 \rangle .
\label{becwavefunc}
\end{align}
Simple enough! Of course this is an idealization in that we have assumed that all the particles occupy the ground state, which is equivalent to zero temperature.  As we have seen in Sec. \ref{sec:excitedstates}, more realistically there will be some fraction of the bosons occupying the excited states, although most of the bosons will be in the ground state. 

Recall that the meaning of $ a_0 $ was actually in terms of a single particle wavefunction which were eigenstates of the potential $ V(\bm{x}) $ (see Eq. \ref{adaggerdef}).  Thus in terms of position space, the BEC wavefunction is
\begin{align}
|\text{BEC} (N) \rangle  = \frac{1}{\sqrt{N !}} \left( \int d \bm{x} \psi_0 (\bm{x}) a^\dagger (\bm{x}) 
\right)^N | 0 \rangle .
\label{becwavefunctionpsi}
\end{align}
This is obviously $ N $ bosons, all with the wavefunction $ \psi_0 (\bm{x}) $.  Since all the bosons have the same wavefunction, it is tempting to define a single macroscopic wavefunction $ \Psi (\bm{x})  $, which captures the wavefunction of the whole BEC.  This is the idea of the {\it order parameter}, \index{order parameter} which in this case would be 
\begin{align}
 \Psi (\bm{x}) = \sqrt{N} \psi_0 (\bm{x}) .
\label{orderparameterexample}
\end{align}
By convention, this is normalized such that when the order parameter is integrated we obtain the total number of particles in the ground state
\begin{align}
\int d \bm{x} |  \Psi (\bm{x}) |^2 = N  .
\end{align}

We note that the idea of the order parameter\index{order parameter} only makes sense when a large number of particles are all occupying the same state, which is not usually what happens in a typical macroscopic state. Taking the example of a general thermal state above the BEC critical temperature, we would expect bosons occupying all kinds of energy states in Fig. \ref{fig1-2}.  This means that all the particles would have different wavefunctions $ \psi_k (\bm{x}) $, and there would not be any kind of common wavefunction that we can define for the particles.  But in this case, since all (or at least most) of the particles have the same wavefunction, we can think of a common giant macroscopic wavefunction for the BEC. 

For the case that we wrote above, it was clear already what the answer should be, it is the same wavefunction as the underlying particles.  But what if we had a more complicated state, perhaps under more realistic conditions where the particles interact with each other? What would be better is if we could calculate the order parameter given some arbitrary state $ | \psi \rangle $.  Although the order parameter is a natural idea, it is not completely obvious what the general definition should be. Suppose we tried the most obvious thing, which is to measure the average position of the particles:
\begin{align}
\langle \psi | a^\dagger (\bm{x}) a (\bm{x}) | \psi \rangle = | \psi_0 (\bm{x}) |^2 .
\end{align}
We could take the square root of this to obtain the magnitude of the wavefunction, but we have completely lost all the phase component.  Since the phase is an important part of the wavefunction in general, this doesn't really work as a good definition of the order parameter. 

To keep the phase information, suppose we define it instead as
\begin{align}
\Psi(\bm{x}) = \langle \psi(N-1) | a (\bm{x})  | \psi(N) \rangle 
\label{orderparameterdef}
\end{align}
where $  | \psi(N) \rangle  $ is the state of the BEC with $ N $ particles, and $  | \psi(N-1) \rangle  $ is the state
of the BEC with $ N -1 $ particles.  This is of course not a regular expectation value, since usually we use the same state for both the bra and ket sides of the expectation value.  This is necessary however, as $ a (\bm{x}) $ will reduce the number of bosons by one, and otherwise we will immediately obtain zero.  What this means is that to explicitly work out the order parameter\index{order parameter}, we have to not only know the wavefunction of the state, but also the  version of the wavefunction with one less particle. This is not necessarily completely trivial to obtain, which makes the definition not completely practical in every case. Nevertheless, for simple cases you can check that this gives the desired result including the phase (see Ex. \ref{q2-6}). The benefit of defining the order parameter in this way is that this will work for a more general case, even if the many-body state is rather complicated. 

An equivalent way to view this is by using the expansion of the boson operators (\ref{basisaninv}) in terms of a complete set of states 
\begin{align}
a (\bm{x}) =  \psi_0(\bm{x})  a_0 + \sum_{k\ne 0}  \psi_k(\bm{x})  a_k  .
\label{expansion}
\end{align}
If we assume a BEC, then we have a large population of 
bosons, all occupying the same state as in (\ref{becwavefunc}). Applying  (\ref{expansion}) to (\ref{becwavefunc}) we obtain
\begin{align}
a (\bm{x}) |\text{BEC} (N) \rangle = \sqrt{N} \psi_0(\bm{x})   |\text{BEC} (N-1) \rangle .
\end{align}
If the state with $ N $ and $ N -1 $ particles is not considerably different, and anticipating that every $ a_0 $ will give a factor of $ \sqrt{N} $ we can approximately write
\begin{align}
a (\bm{x}) \approx \Psi(\bm{x}) + \sum_{k\ne 0}  \psi_k(\bm{x})  a_k .
\label{bogoliubovapprox}
\end{align}
This is called the Bogoliubov approximation, and it amounts to treating the macroscopic mode $ a_0 $ classically (i.e. ignoring the commutation relations of the bosonic operators).  \index{Bogoliubov approximation}

In the approximation (\ref{bogoliubovapprox}) we used the same wavefunctions $ \psi_k(\bm{x}) $, which in the context of Chapter \ref{ch:quantum}, were the eigenstates of the potential $ V(\bm{x}) $. But there is no rule to say that we have to use this set of states, we could equally take another set of complete states. As we will see in the next section, the presence of interactions and other effects can modify the order parameter, meaning that the expansion (\ref{expansion}) can be in terms of another basis that is not necessarily the same as the eigenstates of the potential $ V(\bm{x}) $.

\begin{exerciselist}[Exercise]
\item \label{q3-1}
(a) Find $[  a (\bm{x}), a^\dagger_0 ] $.  
(b) Using your result in (a), verify (\ref{orderparameterexample}) by substituting (\ref{becwavefunc}) into (\ref{orderparameterdef}).  
\end{exerciselist}

\section{The Gross-Pitaevskii equation}
\label{sec:gpequation}

In the previous section we defined the order parameter\index{order parameter} of the BEC, which gives an effective single particle wavefunction of the system.  While we wrote a definition in terms of the entire $N$-particle macroscopic wavefunction of the whole system, it would useful to derive a self-contained equation of motion for the order parameter itself.  We could then completely bypass working out the $N$-particle wavefunction of the whole system, which can be difficult to find. 

We start with the total many-particle Hamiltonian (\ref{multiintham}) which we repeat here for convenience
\begin{align}
{\cal H} =  \int dx a^\dagger (\bm{x}) \left[ -\frac{\hbar^2}{2m} \nabla^2 + V(\bm{x})  \right] a (\bm{x}) 
+ \frac{U_0}{2} \int dx n(\bm{x}) (n (\bm{x})-1)  ,
\label{manyparticleham}
\end{align}
where we have used the $s$-wave scattering interaction (\ref{swaveintch1}) with $ U_0 = \frac{4 \pi \hbar^2 a_s }{m} $. 
The Heisenberg equation of motion for the operator $ a (\bm{x}) $ can be written\index{Heisenberg equation}
\begin{align}
i \hbar \frac{\partial a (\bm{x},t)}{\partial t} & = [a (\bm{x},t), {\cal H}] \nonumber \\
& =
\int dx \left[ -\frac{\hbar^2}{2m} \nabla^2 + V(\bm{x}) + U_0 n(\bm{x},t) \right]  a (\bm{x},t) .
\label{heisenbergax}
\end{align}

Now let us suppose that we have a state of the form (\ref{becwavefunc}) where all $N $ bosons are occupying  
the same state.   Writing the macroscopically occupied state as
\begin{align}
a_0 = \frac{1}{\sqrt{N}} \int dx \Psi^* (\bm{x}) a(\bm{x}) ,
\end{align}
where $ \Psi (\bm{x}) $ is the order parameter\index{order parameter} that we are trying to find an equation of motion for.  The factor of $\frac{1}{\sqrt{N}} $ is there because by convention the order parameter is normalized as  $ \int dx  |\Psi (\bm{x}) |^2 = N $.  Using the prescription (\ref{orderparameterdef})  by applying $ | \psi(N) \rangle $ on the right and 
$ | \psi(N-1) \rangle $ on the left for the left hand side and the first two terms in (\ref{heisenbergax}) we can immediately obtain an expression involving the order parameter.  For the interaction term, we can evaluate
\begin{align}
\langle \psi(N-1) | n(\bm{x}) a(x) | \psi(N) \rangle = |\Psi (\bm{x}) |^2 \Psi (\bm{x}) .
\end{align}
We can then obtain the Gross-Pitaevskii equation
\begin{align}
i \hbar \frac{\partial \Psi(\bm{x},t)}{\partial t} =
\left( -\frac{\hbar^2}{2m} \nabla^2 + V(\bm{x}) +  U_0 |\Psi (\bm{x}) |^2 \right) \Psi (\bm{x},t) .
\label{gpequation}
\end{align}
This has a similar form to the Schrodinger equation, except with the presence of a non-linear interaction term proportional to $ U_0 $. 

Several approximations have been introduced in deriving (\ref{gpequation}).  First, we started with a BEC wavefunction of the form (\ref{becwavefunc}), which assumes that all the particles are in the ground state.  This obviously neglects thermal effects so corresponds to a zero temperature approximation.  Furthermore, (\ref{becwavefunc}) does not properly account for interactions in the sense that it is of a form where all the bosons are in a simple product state. By assuming a wavefunction of the form (\ref{becwavefunc}), these are only included at the level of mean field theory.  In addition, in (\ref{manyparticleham}) an $s$-wave scattering interaction\index{s-wave scattering interaction} was included, which is an approximation to a more realistic interatomic potential $ U(\bm{x},\bm{y}) $ between particles.  This amounts to the assumption that the order parameter varies slowly over the distances in the range of the interatomic potential, hence for phenomena with distances shorter than the scattering length it is not valid to use the Gross-Pitaevskii equation.  \index{order parameter}

\begin{exerciselist}[Exercise]
\item \label{q3-2}
Verify (\ref{heisenbergax}) using the commutation relations for the bosonic operators.
\end{exerciselist}

\section{Ground state solutions of the Gross-Pitaevskii equation}
\label{sec:examplesgp}

It is instructive to find the solution of the Gross-Pitaevskii equation for some simple examples to get a feel for how it works.  Unfortunately, solving for the solution of the Gross-Pitaevskii equation is technically not quite as easy as the Schrodinger equation, due to the presence of the non-linear term.  Only very specific cases have exact solutions, which we will discuss in later sections.  In many cases we must turn to numerical methods to obtain a solution.

\subsection{Stationary solutions}

The Gross-Pitaevskii equation possess stationary solutions \index{stationary solutions} much in the same way that the Schrodinger equation possess solutions that do not evolve in time up to a phase factor.  Here we will be concerned with stationary solutions of the lowest energy, which correspond to the BEC state.  As such we will look for solutions of 
\begin{align}
\mu \Psi(\bm{x}) =
\left( -\frac{\hbar^2}{2m} \nabla^2 + V(\bm{x}) +  U_0 |\Psi (\bm{x}) |^2 \right) \Psi (\bm{x}) .
\label{gpequationindependent}
\end{align}
where $\mu $ is the chemical potential\index{chemical potential}.  Note that we use the term ``chemical potential'' rather than ``energy'' here, even though they play a rather similar role.  The chemical potential is defined as
\begin{align}
\mu = \frac{\partial E}{\partial N }
\end{align}
where $ E $ is the total energy of the whole system.  Thus the chemical potential is the energy needed to add an extra particle into the system.  It is roughly speaking the energy per particle, rather than the energy of the whole system together.  Once the stationary solution is found, the time evolution works in exactly the same way as the Schrodinger equation
\begin{align}
\Psi(\bm{x},t) = \Psi(\bm{x}) e^{-i \mu t/\hbar} .
\end{align}

\subsection{No confining potential}

The simplest analytical case is when there is no confining potential at all $ V(\bm{x}) = 0 $.  In this case, the stationary solutions can be written
\begin{align}
 \Psi (\bm{x}) = \sqrt{\frac{N}{V}} e^{ i \bm{k} \cdot \bm{x}} 
\label{zeropotentialgp}
\end{align}
where $ N $ is the number of atoms, $ V $ is the volume (assumed to extend to infinity), and $ \bm{k} $ is a wavenumber that can be chosen arbitrarily, and the chemical potential is\index{chemical potential}
\begin{align}
\mu = \frac{\hbar^2 k^2}{2m} + U_0 n_0 ,
\label{flatpotentialchemical}
\end{align}
where we define the average density to be 
\begin{align}
n_0 = \frac{N}{V} .
\end{align}
In this case, the solution of the Gross-Pitaevskii equation is identical to the Schrodinger equation, with the exception of the extra interaction term in the energy. In the same way as the Schrodinger equation, the solutions (\ref{zeropotentialgp}) are stationary solutions in the sense that the time evolution does not change the order parameter\index{order parameter} up to a global phase.  This means that in principle all the $ \bm{k} $  are stable solutions in terms of the Gross-Pitaevskii dynamics.  However, in the context of a BEC, usually the most relevant solution is the lowest energy one since this is the state that is thermodynamically stable according to the discussion in Chapter \ref{ch:bec}.  Thus we would typically say that the solution is just the  $ \bm{k} =0$ case:
\begin{align}
 \Psi (\bm{x}) = \sqrt{n_0}  . 
\end{align}
This is not to say that the $ \bm{k} >0$ solutions are impossible, they would simply be described as not being in equilibrium.  Such states would have to prepared in a special way such as to favor the formation of macroscopic occupation at non-zero momentum.

\subsection{Infinite potential well}

For a less trivial example, we need to include a non-uniform potential $ V(\bm{x}) $.  The obvious similarity to the Schrodinger equation suggests that we should look at the same classic example that is studied in any introductory quantum mechanics course: the one dimensional infinite potential well, defined as
\begin{align}
V(x) = \left\{
\begin{array}{cc}
0 & 0 \le x \le L \\
\infty & \text{otherwise}
\end{array}
\right.
\end{align}
where $ L $ is the width of the well.  This is actually not particularly realistic for a BEC for several reasons --- BECs do not strictly even exist in one dimension, and producing a sharp potential that goes from zero to infinity abruptly is virtually impossible in the lab.  Nevertheless, the point of this section is to get a feel for the similarities and differences to the standard Schrodinger equation.

Numerically, the Gross-Pitaevskii equation can be solved by discretizing space on a lattice, and time evolving the order parameter step by step in time.  We make the approximations
\begin{align}
\Psi(x_n,t_m) & \approx \Psi_{n,m} \nonumber \\
\frac{\partial^2 \Psi(x_n,t_m)}{\partial x^2} & \approx \frac{\Psi_{n-1,m} - 2 \Psi_{n,m} + \Psi_{n+1,m}}{\Delta x^2} \nonumber \\
\frac{\partial \Psi(x_n,t_m)}{\partial t}  & \approx \frac{\Psi_{n,m+1} - \Psi_{n,m} }{\Delta t} 
\end{align}
where $ x_n = n \Delta x $, $ t_m = m \Delta t $. Substituting this into (\ref{gpequation}), one obtains
\begin{align}
\Psi_n' = \Psi_n - \frac{i \Delta t }{\hbar} \left[ - \frac{\hbar^2}{2m \Delta x^2 } ( \Psi_{n-1} - 2 \Psi_{n} + \Psi_{n+1} )
+ V(x_n) \Psi_n + g |\Psi_n|^2 \Psi_n \right]
\label{gpequationdiscrete}
\end{align}
where we have written $ \Psi'(x_n)  = \Psi_{n,m+1} $ and $ \Psi(x_n)  = \Psi_{n,m} $ which makes it clear that the order parameter\index{order parameter} can be written as a recursion relation in a vector containing the spatial distribution of the order parameter at a particular time.  In practice one only needs to keep the order parameter at one time instant, and it can be discarded once the next time evolution is evaluated.  

The discrete Gross-Pitaevskii equation (\ref{gpequationdiscrete}) time evolves the order parameter.  In the same way as the Schrodinger equation does not converge to a solution by simple time evolution since it is unitary, neither does the Gross-Pitaevskii equation.  Thus if we put a random order parameter in (\ref{gpequationdiscrete}) and time evolve it, it would not give the solution for the lowest energy state associated with a BEC.  A simple trick allows us to obtain the lowest energy solution, which works rather well in practice.  Making the replacement $ t \rightarrow -i \tau $ changes the unitary dynamics of Hamiltonian evolution to non-unitary dissipation:
\begin{align}
e^{-iHt/\hbar} \rightarrow e^{- H \tau /\hbar} .
\end{align}
Expanding in the eigenstates of $ H = \sum_n E_n | E_n \rangle \langle E_n | $,  this exponentially dampens high energy states, and the lowest energy state (i.e. the ground state) will have the largest relative amplitude.  Thus by simply removing the factor of $ i $ in (\ref{gpequationdiscrete}), we can switch to imaginary time evolution and drive the order parameter\index{order parameter} towards the lowest energy solution!

\begin{figure}[t]
\centerline{\includegraphics[width=\textwidth]{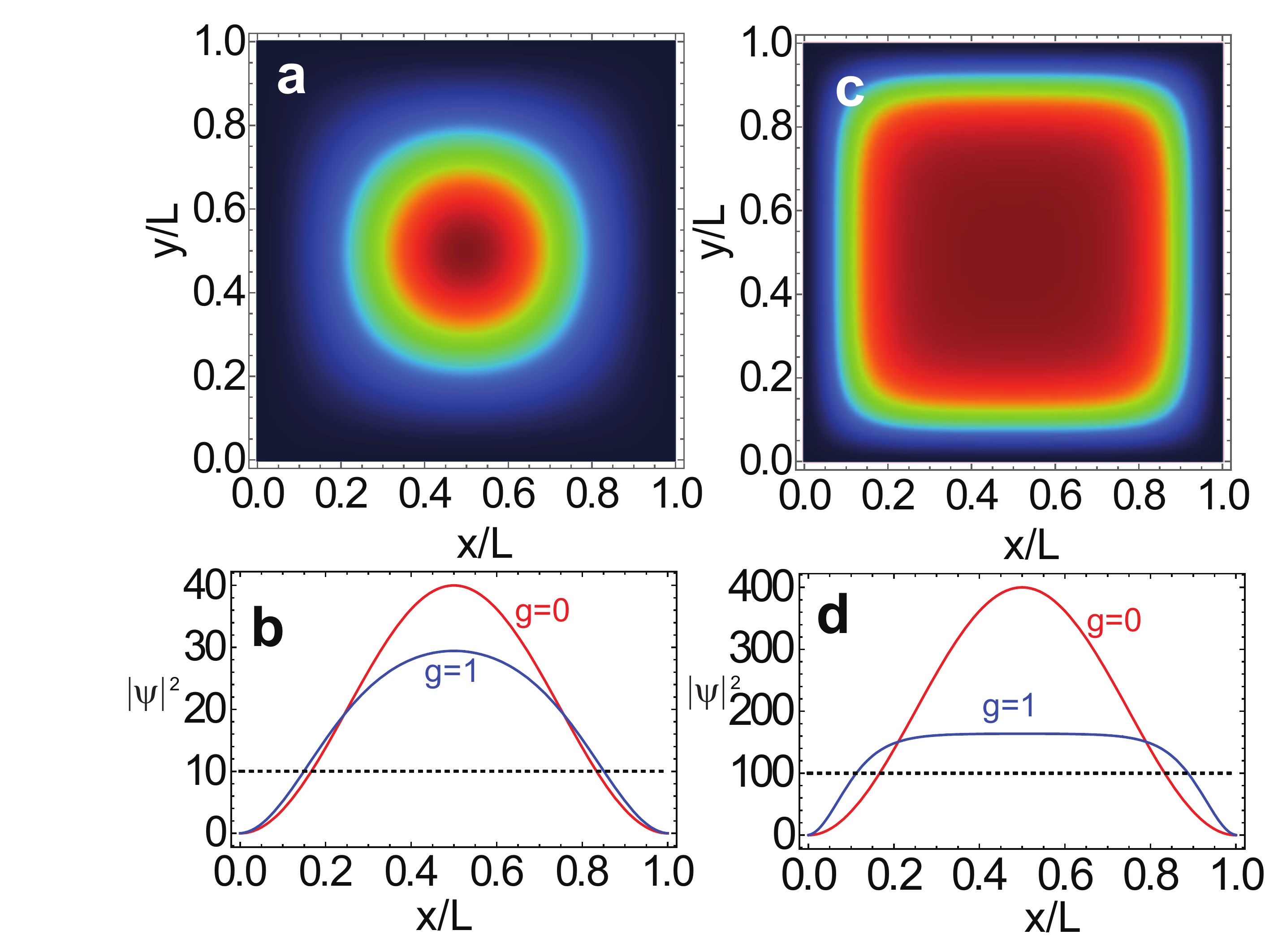}}
\caption{Solution of Gross-Pitaevskii equation for a two dimensional infinite square well with interaction $ U_0 $ as marked. Parameters used are (a) $ N/L^2 = 10 $ with $ U_0 /E_0 = 0 $; (b) $ N/L^2 = 10 $ and interactions as marked for $ y = 0 $; (c) 
$ N/L^2 = 100 $ with $ U_0 /E_0 = 1 $; (d) $ N/L^2 = 100 $ and interactions as marked for $ y = 0 $.  Solutions is numerically evolved under imaginary time for $ t = 0.1 \tau $.  The initial condition are (\ref{exactg0case}) for each case.  
Parameters used are $ L = 1 $ where the units are $ E_0 = \frac{\hbar^2}{2m L^2} $, $ \tau = \hbar/E_0 $.  Dotted line shows the Thomas-Fermi approximation. \index{Thomas-Fermi approximation} }
\label{fig3-1}
\end{figure}

Figure \ref{fig3-1} show the numerical evolution of the Gross-Pitaevskii equation for the two dimensional infinite well of dimension $ L \times L $.  For zero interaction $ U_0=0 $, the Gross-Pitaevskii equation is exactly equivalent to the Schrodinger equation, hence the spatial part of the order parameter takes the form
\begin{align}
 \Psi (\bm{x}) = \sqrt{\frac{4N}{L^2}} \sin ( \frac{\pi x}{L} ) \sin ( \frac{\pi y}{L} ) .
\label{exactg0case}
\end{align}
Evolving this state under imaginary time evolution gives the solution including interactions.  From Fig. \ref{fig3-1}(b) we see that the wavefunction becomes broadened out when interactions are included.  One can interpret this to be the effect of the particles minimizing energy by spreading themselves out more than the case without interactions.  As the interactions are increased, the order parameter is more evenly distributed within the allowed space.  When the density is increased, the effect of the interactions becomes stronger, as can be seen in Fig. \ref{fig3-1}(c).  This is because the higher density allows for more opportunity to interact with each other.

\subsection{Thomas-Fermi approximation}
\index{Thomas-Fermi approximation}

Due to the quadratic nature of the interaction term $  U_0 |\Psi (\bm{x}) |^2 $, and remembering that the order parameter\index{order parameter} has a magnitude $ \sim \sqrt{N} $, we can see that for high densities the interaction term tends to dominate more and more.  In fact in the limit of very high density, the kinetic energy term plays a smaller and smaller role.  The time-independent Gross-Pitaevskii equation (\ref{gpequationindependent}) can then approximated by
\begin{align}
\mu \Psi(\bm{x}) =
\left( V(\bm{x}) +  U_0 |\Psi (\bm{x}) |^2 \right) \Psi (\bm{x}) ,
\label{gpequationindependent2}
\end{align}
which has the solution
\begin{align}
\Psi(\bm{x}) = \sqrt{\frac{\mu - V(\bm{x})}{U_0 }} .
\end{align}
This is called the Thomas-Fermi limit.   \index{Thomas-Fermi limit}

Figure \ref{fig3-1}(b)(d) shows comparisons of the Thomas-Fermi limit versus exact stationary solutions.  In Fig. \ref{fig3-1}(b) we see the results for the low-density case.  In this case we see that the Thomas-Fermi approximation does not work very well since the kinetic energy term still plays an important role.  As the density is increased in Fig. \ref{fig3-1}(d), the Thomas-Fermi approximation\index{Thomas-Fermi approximation} increasingly becomes a better approximation, with the numerically evaluated distribution flattening out.  In the limit of large density or high interactions, the Thomas-Fermi approximation gives a more accurate expression for the ground state distribution.

\subsection{Healing length}
\label{sec:healinglength}\index{healing length}

We saw from Fig. \ref{fig3-1} that as the density is increased, this has the effect of spreading the order parameter\index{order parameter} more evenly, due to the repulsive interactions.  This would be true also if the interactions were increased and the density is kept constant, since the last term in the Gross-Pitaevskii equation (\ref{gpequationindependent}) is the product of the interaction and the density.  In Fig. \ref{fig3-1}(c)(d) we see that at the edges of the well the order parameter is zero according to the boundary conditions, and approaches a relatively flat region particularly for the higher densities.  We can estimate the distance that this transition occurs, which is commonly called the {\it healing length}. \index{healing length}

Let us assume that the infinite potential well has a very large area, and we consider just the region near the walls, away from any corners --- take for example the middle of the left wall in Fig. \ref{fig3-1}(c).  We expect that as we move away from the walls the BEC will have a similar behavior to the zero potential case $ V(\bm{x}) = 0 $ where the effect of the boundary is not noticeable.  As we saw in (\ref{flatpotentialchemical}), in this case for the ground state $ \bm{k} = 0 $ the chemical potential\index{chemical potential} is $ \mu = U_0 n_0 $.  
Substituting this into the stationary state (\ref{gpequationindependent}) we obtain the equation
\begin{align}
 \frac{\hbar^2}{2m} \frac{d^2 \Psi}{dx^2} =   U_0 (|\Psi (x) |^2 - n_0 ) \Psi (x) .
\label{healinglengthgpe}
\end{align}
We would like to solve this for the boundary condition that $ \Psi (0)=0 $. One can easily verify that a solution of this equation with this boundary condition takes a form
\begin{align}
\Psi (x) = \sqrt{n_0} \tanh \left( \frac{x}{\sqrt{2} \xi} \right)
\label{healinglengthsolution}
\end{align}
where the healing length\index{healing length} is defined as
\begin{align}
\xi = \frac{\hbar}{\sqrt{2m U_0 n_0}} .
\label{healinglengthdef}
\end{align}
Physically, we can interpret this to be the length when the kinetic energy $ \frac{\hbar^2}{2 m \xi^2} $ is equal to 
to the interaction energy $ U_0 n_0 $.  Calculating the healing length for Fig. \ref{fig3-1}(a)(c), we get 
$ \xi/L = \sqrt{\frac{E_0}{U_0 n_0}} $ to be 0.32 and 0.1 respectively, which matches with the numerical calculation.

\begin{exerciselist}[Exercise]
\item \label{q3-3}
Using your favorite programming language, code the recursive equation (\ref{gpequationdiscrete}) and implement the discrete Gross-Pitaevskii equation for the potential $ V(x) = x^2 $.  Compare the dynamics including the factor of $  i $ (corresponding to real time dynamics) and removing the $ i $ (imaginary time dynamics).  Experiment to see the various solutions for different parameters. \item \label{q3-4}
Check that (\ref{healinglengthsolution}) is actually a solution of the equation (\ref{healinglengthgpe}).  
\end{exerciselist}

\section{Hydrodynamic equations}\index{hydrodynamic equation}

\label{sec:hydrodynamicequations}

From our discussion in Sec. \ref{sec:superfluidity} we have seen that BECs can be thought of being a fluid with special properties such as dissipationless flow\index{disspationless flow}.  The GP equation can be rewritten in a way that makes the analogy with fluids explicit, by writing it as a hydrodynamic equation.  The first step is to parametrize the order parameter\index{order parameter} according to
\begin{align}
\Psi (\bm{x}) = \sqrt{n(\bm{x})} e^{i \phi(\bm{x})}
\label{fluidparameterize}
\end{align}
where 
\begin{align}
n(\bm{x}) & =  | \Psi (\bm{x}) |^2 \nonumber  \\
\phi(\bm{x}) & = \arg ( \Psi (\bm{x}) ) 
\label{densityphasedef}
\end{align}
is a spatially dependent density and phase of the order parameter respectively.  To describe the flow of the fluid, we can define the current
\begin{align}
\bm{j} (\bm{x}) & = -\frac{i \hbar}{2m} \left(  \Psi^* (\bm{x}) \bm{\nabla}  \Psi (\bm{x}) - \Psi (\bm{x}) \bm{\nabla}  \Psi^* (\bm{x}) \right)  \label{currentdef} \\
& = n(\bm{x}) \frac{\hbar}{m} \bm{\nabla}  \phi(\bm{x}) 
\end{align}
where in the second line we used the parametrization (\ref{fluidparameterize}). 
 You may wonder why the current is written in the form (\ref{currentdef}). A simple way to understand this is that it can be equivalently written $ \bm{j} (\bm{x})  = \frac{1}{2m}( \Psi^* (\bm{x}) \bm{p} \Psi (\bm{x}) - \Psi (\bm{x}) \bm{p} \Psi^* (\bm{x}) ) $ where $  \bm{p} = -i\hbar  \bm{\nabla}  $ is the momentum operator. Therefore the average current is
\begin{align}
\langle \bm{j} (\bm{x}) \rangle = \frac{\langle \bm{p} \rangle}{m} .
\label{averagej}
\end{align}
It is tempting to say that classically $ \bm{p} = m \bm{v} $ and hence the right hand side of (\ref{averagej}) is a velocity.  This would be correct if $ \Psi (\bm{x})  $ was a single particle wavefunction.  However, since $ \Psi (\bm{x})  $ is an order parameter\index{order parameter} that is normalized to $ N $ rather than $1$, $ \langle \bm{j} (\bm{x}) \rangle  $ corresponds to a total current of all the particles combined.  To obtain the local velocity, we must divide (\ref{currentdef}) by the local density, giving \index{velocity}
\begin{align}
\bm{v} (\bm{x}) & = \frac{\hbar}{m} \bm{\nabla}  \phi(\bm{x}) 
\label{localvelocity}
\end{align}
Since for any scalar field $ f $ the curl of the gradient is zero $ \bm{\nabla} \times (\bm{\nabla} f )= 0 $, we then have
\begin{align}
\bm{\nabla}  \times \bm{v} (\bm{x}) = 0 .
\end{align}
This has the physical interpretation that a superfluid\index{superfluid} is irrotational.  This means that when traversing a closed loop, the particles do not experience a net rotation.  For example, irrotational flow\index{irrotational flow} is like a carriage in a Ferris wheel, which keeps the same direction with respect to the Earth as it goes around a loop.  This is in contrast to rotational flow\index{rotational flow} where a person sitting in a carriage on a roller coaster loop, who would rotate once with each revolution. 

By multiplying (\ref{gpequation}) by $\Psi^* $ and subtracting the complex conjugate we obtain
\begin{align}
\bm{\nabla} \cdot \bm{j} (\bm{x})  = - \frac{\partial n  (\bm{x})}{\partial t} 
\label{continuityequation}
\end{align}
which is called the continuity equation\index{continuity equation}.  This is a statement of the conservation of the amount of fluid. If there is fluid flow out of a volume (the divergence of the current is positive --- the left hand side), then this must be accompanied by a decrease in the amount of fluid within the volume (the right hand side).  We can obtain another equation involving the phase by substituting (\ref{fluidparameterize}) into the Gross-Pitaevskii equation (\ref{gpequation}):
\begin{align}
\hbar \frac{\partial \phi (\bm{x})}{\partial t} + \frac{m}{2}  \bm{v}^2 (\bm{x}) + V(\bm{x}) + U_0 n (\bm{x}) 
- \frac{\hbar^2 (\nabla^2 \sqrt{n (\bm{x})})}{2m \sqrt{n (\bm{x})}} = 0
\label{phasehydroequation}
\end{align}
The two equations (\ref{continuityequation}) and (\ref{phasehydroequation}) are an equivalent set of equations that has the same mathematical content as the Gross-Pitaevskii equation (\ref{gpequation}).

\begin{exerciselist}[Exercise]
\item \label{q3-5}
Derive equations (\ref{continuityequation}) and  (\ref{phasehydroequation}) using the Gross-Pitaevskii equation.  
\end{exerciselist}

\section{Excited state solutions of the Gross-Pitaevskii equation}
\label{sec:examplesgpexcited}

The previous section examined the ground state solutions of the Gross-Pitaevskii equation.  It is also possible to
examine particular excited states using the same equations.  It is important to understand that the types of states that the 
Gross-Pitaevskii equation considers are special types of states where there is a {\it macroscopic occupation} of all the bosons in 
the system.  We argued this from the point of view that Bose-Einstein condensation should occur at low enough temperatures,
so that for the ground state this is a reasonable picture to have.  In general for excited states the assumption of macroscopic
occupation is generally not true --- a proper treatment should be described more along the lines of Chapter \ref{ch:quantum}, where the full quantum many-body wavefunction is evaluated.  Thus the excited states of the Gross-Pitaevskii equation only describe specifically those states which are macroscopically populated.  You may then be worried whether examining such states would have any relevance at all to states that are seen in the lab.  One way to get around this is to specifically arrange the system such that the excited state of interest is the ground state of a different Hamiltonian ---  this is what is done for the case of vortices \index{vortices}as will be explained in the next section.   It turns out that such states can also be prepared under non-equilibrium situations, so that the excited state solutions of the Gross-Pitaevskii equation are quite relevant in practice.  We will examine two examples of such excited states: vortices and solitons.\index{soliton}

\subsection{Vortices}
\label{sec:vortices}

If you are given a cup of water that initially has zero flow everywhere (e.g. there are no convection currents\index{convection currents} inside, or any other kind flow within the cup), one of the simplest ways to get it moving is to stir it.  A BEC is no different, and it is possible to create states where there is a current moving circularly, i.e. vortex flow.   Unlike a regular fluid, where eventually frictional forces cause the vortices to slow down and disappear, in a superfluid\index{superfluid} it is possible to have vortices that never dissipate energy.  These configurations are also solutions of the Gross-Pitaevskii equation, and have interesting properties which are a simple example of a topological state. 

To find these solutions, let us revisit the stationary Gross-Pitaevskii equation (\ref{gpequationindependent}) and examine it in polar coordinates for two-dimensions.  For the potential, it is natural to consider a radially symmetric case $  V(\bm{x}) = V(r) $ since we are looking for vortices.  The equation then reads
\begin{align}
\mu \Psi(r,\theta) =
-\frac{\hbar^2}{2m} \left( \frac{\partial^2 \Psi}{\partial r^2} + \frac{1}{r} \frac{\partial \Psi}{\partial r} \right) +
\frac{1}{r^2} \frac{\partial^2 \Psi}{\partial \theta^2}  +
 V(r) \Psi(r,\theta)  + U_0 |\Psi (r,\theta) |^2 \Psi (r,\theta) .
\end{align}
This equation can be solved using similar techniques to that used when finding the eigenstates of the hydrogen atom \index{hydrogen atom}or harmonic oscillator in polar coordinates.  The primary difference here is that there is the non-linear interaction term.  Assuming a separable form of the order parameter $ \Psi (r,\theta) = R(r) Y(\theta) $ we obtain
\begin{align}
 -\frac{\hbar^2}{2m} \frac{1}{R} \left( r^2 \frac{\partial^2 R}{\partial r^2} + r \frac{\partial R}{\partial r} \right) + V(r) - r^2 \mu + U_0 r^2 | R(r) |^2 |Y(\theta)|^2 = - \frac{1}{Y} \frac{\partial^2 Y}{\partial \theta^2} .
\label{separationvariables}
\end{align}
This would be separable if the last term on the left hand side did not have the $  |Y(\theta)|^2 $ term.  This looks like our separation of variables trick didn't work, but if it so happens that $ |Y(\theta)|^2 $ is a constant then it might still work.  Continuing with the separation of variables we would then say that the left and right hand side of are equal to the same constant. Setting this to $ l^2 $ we have the equation 
\begin{align}
\frac{\partial^2 Y}{\partial \theta^2} =  l^2 Y(\theta),
\end{align}
which has solutions
\begin{align}
Y(\theta) = C_1 e^{il\theta} + C_2 e^{-il\theta} .
\end{align}
As for the hydrogen atom\index{hydrogen atom}, we require that $ Y(\theta) = Y(\theta + 2 \pi) $ since this refers to the same physical spatial location.  This means that $ l $ must be an integer.  Furthermore, in order to have $ |Y(\theta)|^2 $  to be a constant, we should have either $ C_1=0 $ or $ C_2=0 $.  This satisfies the separability problem (\ref{separationvariables}) which then allows the radial solution to be solved.  The solution is then
\begin{align}
\Psi (r,\theta) = R(r) e^{il\theta}.
\label{vortexsolution}
\end{align}

We can check whether the solution of this form is vortex-like at this stage.  Assuming that $ R(r) $ is real, we can substitute it into (\ref{localvelocity}) to obtain
\begin{align}
\bm{v} (r,\theta) = \frac{\hbar l}{mr}  \hat{\theta}
\label{vortexvelocity}
\end{align}
where $ \hat{\theta} $ is the unit vector in the angular direction.  This has zero component in the radial direction $ \hat{r} $ hence shows that the velocities are pure rotations around the origin. This certainly looks like a vortex as long as we have $ l \ne 0  $.  Unlike classical rotations, the velocities are quantized and only take discrete values at a given radius. 

\begin{figure}[t]
\centerline{\includegraphics[width=\textwidth]{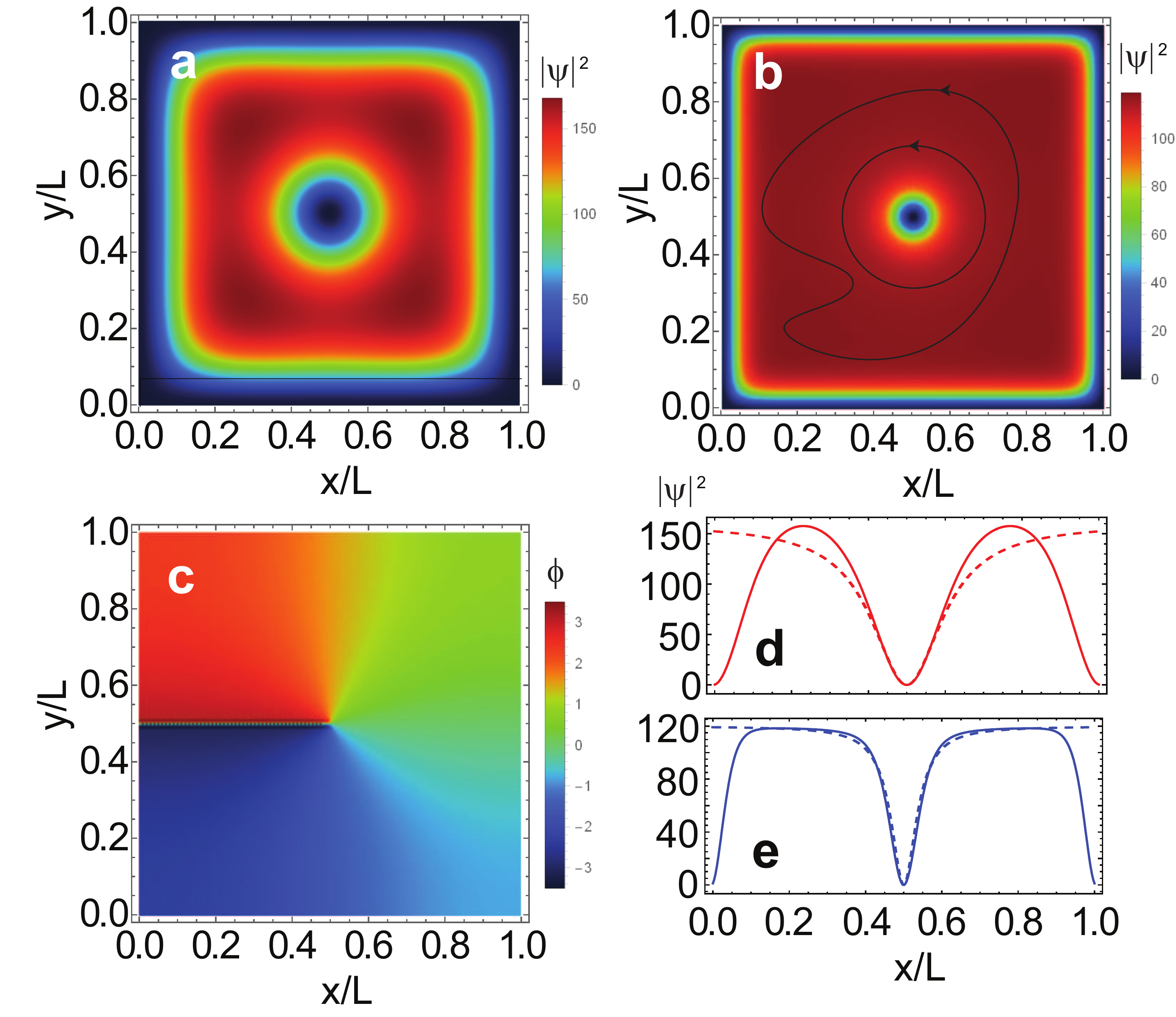}}
\caption{Vortex solutions of the Gross-Pitaevskii equation for a two dimensional infinite square well with $ N/L^2 = 100 $ (a)(d) $ U_0/E_0 = 1 $; (b)(e) $U_0/E_0 = 10 $.  (c) The phase distribution of the solution is the same for both parameters. Solutions are numerically evolved under imaginary time for $ t = 0.1 \tau $.  The initial condition is Eq. (\ref{exactg0case}) multiplied by $ \exp(i \theta') $ where $ \tan \theta' = (x-L/2)/(y-L/2) $ for each case.  
Parameters used are $ L = 1 $ where the units are $ E_0 = \frac{\hbar^2}{2m L^2} $, $ \tau = \hbar/E_0 $.  Dotted line shows the Thomas-Fermi approximation.  Solid line in (b) shows two examples of contours in the integral (\ref{quantizationvortex}). }
\label{fig3-2}\index{Thomas-Fermi approximation}
\end{figure}

To obtain the vortex solution (\ref{vortexsolution}) we assumed radially symmetry, but this is not necessary in general.  In Fig. \ref{fig3-2} we show the solutions to the same infinite potential well that we looked at in Fig. \ref{fig3-1}, but with a single vortex in them.  To obtain these, we start with the initial condition (\ref{exactg0case}) but multiplied by the vortex phase $ e^{i\theta} $ offset to the center of the trap.  Starting with such a vortex-like solution, by evolving in imaginary time we obtain the stable solutions seen in Fig. \ref{fig3-2}.  The phase distribution\index{phase distribution} is virtually unchanged from the initial condition, as can be seen in Fig.  \ref{fig3-2}(c).  The most obvious difference between Fig. \ref{fig3-1} and Fig. \ref{fig3-2} is then the appearance of a density dip at the location of the vortex, called a vortex core\index{vortex core}.   We can see that increasing the interaction reduces the size of the vortex core, and it is always similar to the healing length\index{healing length} $\xi $ from the edges of the well.  This is because exactly at the vortex core the density drops to zero, and a similar argument can be applied to Sec. \ref{sec:healinglength} to get the distribution in the bulk part of the BEC.  Using (\ref{healinglengthsolution}) gives a reasonable estimate of the distribution near the vortex. Another commonly used distribution which works rather well is found using Pad{\'e} approximants \index{Pad{\'e} approximants} 
\begin{align}
 R(r) = \sqrt{n_0} \frac{r/\xi}{\sqrt{ (r/\xi)^2 + 2}} ,
 \label{padeapproxvortex}
\end{align}
where $ n_0 $ is in this case the density far away from the vortex. This is plotted in Fig.  \ref{fig3-2}(d)(e).  We see that it works quite well in reproducing the density variation near the vortex core.

Returning to (\ref{vortexvelocity}), we see there is another difference to classical rotation.  We know from classical rotation of a rigid object that $ \bm{v} = r \omega  \hat{\theta} $.  The proportionality to $ r $ is simply because further away from the origin the particles have to move faster to keep up with the rotation.  For the BEC we see that it is exactly the opposite of this, the velocity is {\it inversely} proportional to $  r $, such that the particles move faster near the origin.  This occurs because in (\ref{localvelocity}) the velocity is proportional to the rate of change of the phase.  We can see this from Fig. \ref{fig3-2}(c): the phase variation near the vortex core goes to a singular point. This is one of the reasons that the density at this point must drop to zero.  Otherwise we would have particles moving at infinite velocity, costing infinite energy!

If we calculate the line integral of the velocity around a vortex we find that
\begin{align}
\oint \bm{v} \cdot d \bm{l} = 2 \pi l \frac{\hbar}{m} ,
\label{quantizationvortex}
\end{align}
where for a circular contour of radius $ r $ we can take $ d \bm{l} = r d \theta  \hat{\theta} $.  We see that the quantization occurs because of the precise cancellation of the factor of $ r $ from the line integral and velocity.  This seems like a remarkable coincidence and seems to be dependent upon the somewhat arbitrary choice of a perfectly circular contour and a vortex in a radially symmetric potential.  However, this is actually true no matter what choice of contour or type of potential.  As long as the contour encloses the vortex core, one obtains the same result on the right hand side of (\ref{quantizationvortex}).  In Fig. \ref{fig3-2}(b) we show two examples of contours that would give exactly the same result.  Since the result of (\ref{quantizationvortex}) is only dependent upon the topological properties of the contour relative to the vortex, the $ l$ is sometimes referred to as a topological charge. \index{topological charge}

But what gives rise to this remarkable property? If we look at the same contours with respect to the phase in Fig. \ref{fig3-2}(c) we get a better idea of where this comes from. The key aspect of a vortex is that there is a phase singularity\index{phase singularity} at the vortex core\index{vortex core}, and the phase evolves by a multiple of $ 2 \pi $ as you go around the vortex in a loop.  It has to be a multiple of $ 2 \pi $ because of the single-valuedness of $ Y(\theta) $ --  otherwise the wavefunction would be discontinuous.  Actually the phase is discontinuous about the vortex core, but by making the density go to zero, the discontinuity is avoided.  Thus (\ref{quantizationvortex}) is really a statement that the phase of the order parameter around a vortex evolves by a factor of $ 2 \pi $ before joining back together.  This is obvious from the point of view of the definition of the velocity (\ref{localvelocity}) which is the derivative of the phase.  So if we add up all the phase changes around a vortex then this has to add to a multiple of $ 2 \pi $.  What is interesting is that this can be related to a physical quantity such as the velocity.  Since quantum mechanically we have quite a different notion of velocity (i.e. the derivative of the phase) in comparison to classical physics ---  because we deal with waves which possess a phase --- we end up with a relation that looks quite unusual.\index{order parameter}

\begin{figure}[t]
\centerline{\includegraphics[width=\textwidth]{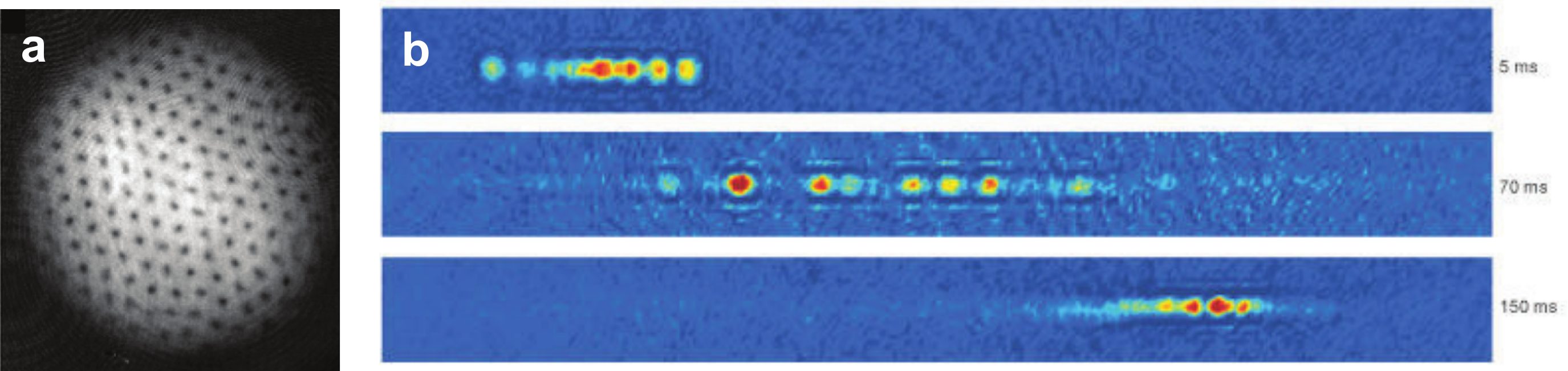}}
\caption{Experimental realization of excited states in Bose-Einstein condensates.  (a) Vortex array produced by stirring BECs, reproduced from \cite{PhysRevLett.87.210402}.  (b) Bright solitons in BECs, reproduced from  \cite{strecker2002formation}.} \index{soliton}
\label{fig3-3}
\end{figure}

\subsection{Solitons}
\label{sec:solitons}\index{soliton}

When a wave packet is evolved in time with the Schrodinger equation, we know that it starts to spread out, or disperse. Taking the potential $ V(x) = 0 $, and considering the one-dimensional case for simplicity, we have 
\begin{align}
i \hbar \frac{\partial \psi}{\partial t} = - \frac{\hbar^2}{2m} \frac{ \partial^2 \psi}{\partial x^2} .
\end{align}
Starting with a Gaussian wave packet moving with velocity $ v $, the probability density evolves as
\begin{align}
| \psi (x,t) |^2 = \frac{\sigma}{\sqrt{\pi (\sigma^4 + \hbar^2 t^2/m^2)}} 
\exp \left[ - \frac{\sigma^2 ( x - v t)^2}{\sigma^4 + \hbar^2 t^2/m^2} \right]
\end{align}
where $ \sigma $ is the initial spread of the Gaussian.  We can see that the wave packet spreads because the denominator in the exponential increases with time, which controls how spread out the Gaussian is.  The amplitude of the Gaussian also decreases, to preserve normalization. 

The origin of this behavior is that the kinetic energy term in the Hamiltonian has a diffusive effect, where the eigenstates are delocalized plane waves.  In the presence of a confining potential $ V(x) $, this acts as a counter to the spreading effect of the kinetic energy.  But in this case, there is nothing to stop the spreading, and hence the wavefunction keeps on broadening. 

Now turning to the Gross-Pitaevskii equation, we have another way to prevent the wave packets from spreading --- the interaction term.  If we had repulsive interactions, we can see that having a wave packet configuration is not a good idea energetically since the particles are bunched together.  So what might work better is if the interactions are attractive.  Considering one dimension and setting $ V(x) = 0 $ in (\ref{gpequationindependent}) we have the solution
\begin{align}
\Psi_\text{B}(x) = \sqrt{\frac{2\mu_\text{B}}{U_0}} \text{sech} \left[ \frac{x-vt}{\xi_{\text{B}}} \right] e^{-i \mu_\text{B}^0 t/\hbar + i m v x/\hbar},
\label{brightsoliton}
\end{align}
where 
\begin{align}
\xi_\text{B} & = \frac{\hbar}{\sqrt{2 m | \mu_\text{B} |}} ,\nonumber \\
\mu_\text{B} & = \mu_\text{B}^0 - \frac{1}{2} m v^2 \nonumber, \\
\mu_\text{B}^0 & = \frac{1}{2} U_0 n_0 , 
\end{align}
and $ \mu_\text{B}^0 $ and $ n_0 $ is the chemical potential and density at the position of the soliton for $ v = 0 $ respectively. Here we have $ U_0 < 0 $ and therefore $ \mu > 0 $.  The special thing about this solution is that unlike the Gaussian form that we considered initially, the shape of the wave packet doesn't change with time.  \index{soliton!bright}

For repulsive interactions, another type of soliton can be produced which has the form of a density dip, instead of a ``mound'' shape that we saw above.  This has a form
\begin{align}
\Psi_\text{D}(x) = \sqrt{n_0} \left( \frac{iv}{c} + \sqrt{1-\frac{v^2}{c^2}} \tanh \left[ \frac{x-vt}{\xi_{\text{D}}} \right]  \right)e^{-i \mu_\text{D} t/\hbar},
\label{darksoliton}
\end{align}
where
\begin{align}
\xi_{\text{D}} & = \frac{\hbar}{\sqrt{m U_0 n_0( 1- v^2/c^2)}} \nonumber,\\
\mu_\text{D} & = U_0 n_0 \nonumber, \\
c & = \sqrt{\frac{U_0 n_0}{m}} .
\end{align}
Here $ n_0 $ is the density far away from the soliton, and $ c $ is the sound velocity.  Because this is the form of a density dip, they are called ``dark solitons'', in contrast to (\ref{brightsoliton}) which are called ``bright solitons''.  \index{soliton!dark} \index{soliton!bright}

While these solution works perfectly well in one dimension, in higher dimensions they are unstable because they correspond to an infinitely long wave packet.  Due to the attractive interactions, such long structures are not stable and small perturbations can destabilize the solitons. In the case of attractive interactions, it is energetically more favorable to form smaller, more clumped structures.   In practice, they can be produced if the trapping potential is quasi-one dimensional, such as a long channel, where they have been observed experimentally.  In Fig. \ref{fig3-2} we see an example of solitons formed in a long one-dimensional trap.  We see that the solitons are stable and can propagate for long distances.

\begin{exerciselist}[Exercise]
\item \label{q3-6}
Substitute (\ref{brightsoliton}) into  (\ref{gpequation}) and verify that it is a solution.  
\item \label{q3-7}
Substitute (\ref{darksoliton}) into  (\ref{gpequation}) and verify that it is a solution. 
\end{exerciselist}

\section{References and further reading}

\begin{itemize}
\item Sec. \ref{sec:orderparameter}: Early review on the
topic of order parameters and off-diagonal long range order \cite{yang1962concept}.  Textbooks detailing the concept further \cite{pitaevskii2016bose,pethick2002bose}.
\item Sec. \ref{sec:gpequation}: Original references on the Gross-Pitaeveskii equation \cite{gross1961structure,pitaevskii1961vortex}. Further theoretical works expanding on the approach \cite{rogel2013gross,erdHos2007rigorous,erdHos2010derivation,kolomeisky2000low,salasnich2002effective}.
\item Sec. \ref{sec:examplesgp}: Time-dependent solutions of the Gross-Pitaevskii equation \cite{bao2003numerical,perez1997dynamics,muruganandam2009fortran}.
\item Sec. \ref{sec:hydrodynamicequations}: For an introduction to hydrodynamic equations of the Gross-Pitaevskii equation \cite{pitaevskii2016bose,rogel2013gross}. The stochastic Gross-Pitaevskii equation is derived in \cite{gardiner2002stochastic}.  Theoretical analyses of the hydrodynamics are given in  \cite{kolomeisky2000low,o2004exact}. Observation in exciton-polariton BECs is given in \cite{amo2011polariton}.  
\item Sec. \ref{sec:vortices}: Experimental observation of vortices in BECs \cite{matthews1999vortices,madison2000vortex,hodby2001vortex,weiler2008spontaneous}.  Experimental observation of vortex lattices \cite{abo2001observation,schweikhard2004vortex,engels2002nonequilibrium}.  Experimental observation of solitons \cite{strecker2002formation}. Theoretical analyses of vortices \cite{simula2006thermal,parker2005emergence,ruostekoski2001creating}.  Review articles and books discussing vortices \cite{fetter1969,fetter2001vortices,pitaevskii2016bose}.   Eq. (\ref{padeapproxvortex}) is taken from \cite{fetter1969}.
\item Sec. \ref{sec:solitons}: Experimental observation of solitons \cite{burger1999dark,denschlag2000generating,anderson2001watching,strecker2002formation,khaykovich2002formation,eiermann2004bright,marchant2013controlled}.  Theoretical analyses of solitons in BECs \cite{perez1998bose,busch2000motion,busch2001dark,drummond1998coherent,ostrovskaya2003matter,theocharis2005matter}.  Review articles and books discussing solitons further \cite{griffiths2018introduction,pitaevskii2016bose}.  
\end{itemize}

	\chapter[Spin dynamics of atoms]{Spin dynamics of atoms}

\label{ch:spin}

\section{Introduction}
\label{sec:spinintro}

Up to this point we have only considered the motional degrees of the bosons.  In fact the atoms used to form Bose-Einstein condensates have a rich internal spin structure. This gives another degree of freedom that the many-body quantum state can occupy.  In this chapter we examine the spin dynamics that will affect the quantum state of the system.  This will include processes that are naturally present in a typical setup involving trapped atoms, such as the Zeeman energy shift, collisional interactions, spontaneous emission, and loss.  We also consider processes where the spin dynamics can be actively manipulated in the laboratory, using electromagnetic transitions between energy levels and Feshbach resonances.\index{Feshbach resonance}\index{Zeeman energy shift}\index{spontaneous emission}\index{loss}

\section{Spin degrees of freedom}
\label{sec:spindegrees}

Many different types of atoms have been used in the context of atom trapping and cooling.  In the context of BECs, the alkali atoms are a particularly popular choice, which have a simple electronic structure and amenable for laser cooling.  In this section we will discuss the spin structure of such atoms, taking the example of rubidium, one of the most commonly used atom for BECs.  While other atoms have different details in their atomic structure, they can be considered variations of what is described in this section.  

\begin{figure}[t]
\centerline{\includegraphics[width=\textwidth]{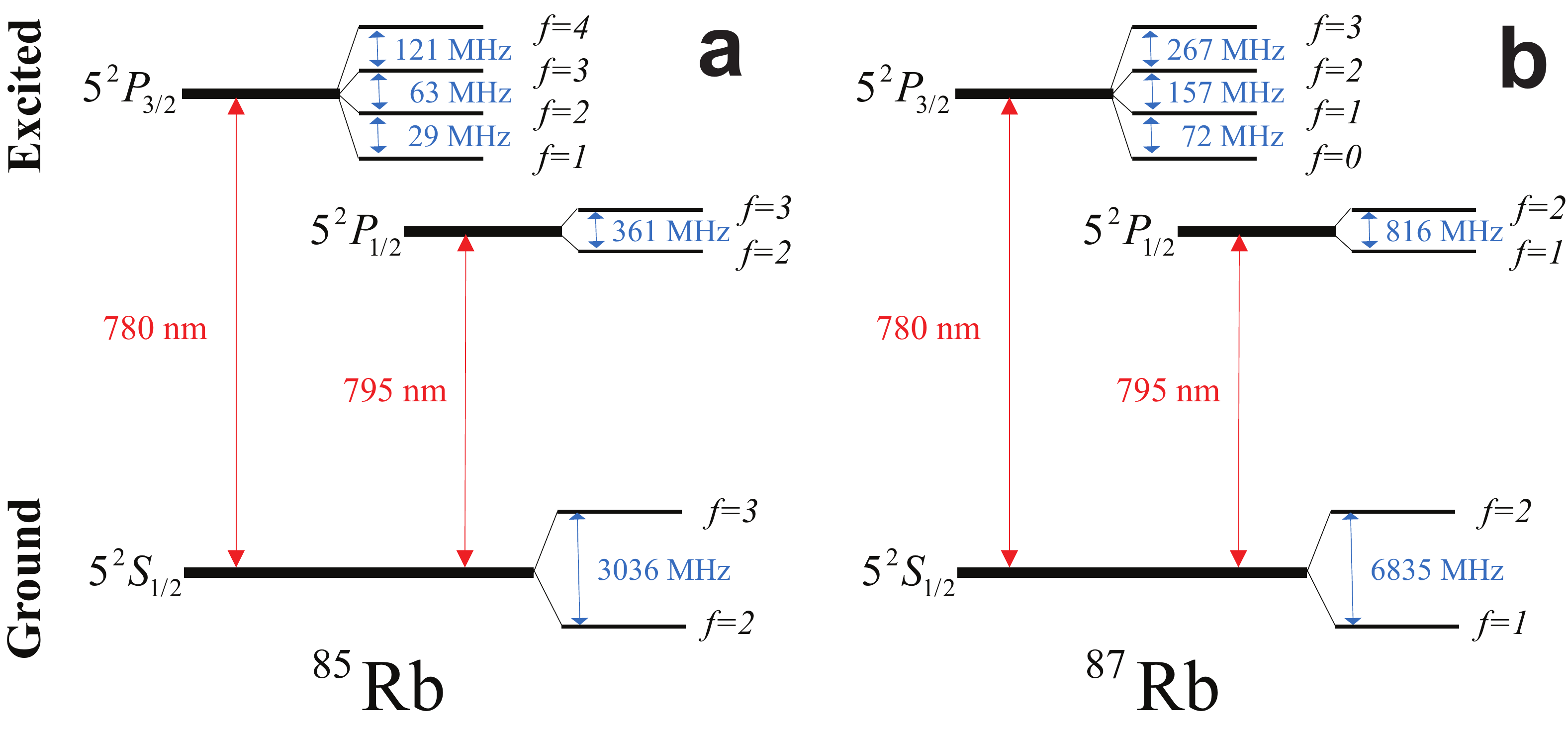}}
\caption{Energy level structure for (a) $^{85} \mbox{Rb} $ and (b) $ ^{87} \mbox{Rb} $.  The total spin $ F $ are marked next to each level. The transition frequencies for the $ \text{D}_2 $ and $ \text{D}_1 $ transitions are given in terms of the wavelength, the hyperfine splittings are given in terms of frequencies. }
\label{fig4-1}
\end{figure}

Figure \ref{fig4-1} shows the internal state structure for two isotopes of rubidium, $^{87} \mbox{Rb} $ and $ ^{85} \mbox{Rb} $.  The levels of atoms are typically denoted using the hyperfine spin  $ \bm{F} $ which consists of the total angular momentum of the electrons $ \bm{J} $ and the nuclear spin $ \bm{I}  $
\begin{align}
\bm{F} = \bm{J} + \bm{I}
\end{align}
The electron angular momentum is itself decomposed according to
\begin{align}
\bm{J} = \bm{L} + \bm{S} ,
\end{align}
where $ \bm{L} $ is the orbital angular momentum  and $ \bm{S} $ is the electronic spin.  As usual, the total angular momentum quantum numbers are eigenvalues of the total spin according to 
\begin{align}
\bm{F}^2 | f, m_f \rangle & = f(f+1) \hbar^2 | f, m_f \rangle \nonumber \\
F_z | f, m_f \rangle & = m_f \hbar | f, m_f \rangle .
\end{align}
Since $\bm{F}^2$ and $F_z$ commute  $ [\bm{F}^2,F_z] = 0  $\index{commutation relations}, in the above we defined the simultaneous eigenstates of $ \bm{F}^2 $ and $F_z $, where $ m_f $ is the magnetic quantum number.  The magnetic quantum number takes a range $ m_f \in \{-f,-f+1,\dots, f\} $, having $ 2f+1 $ values.  For example in Fig. \ref{fig4-1}(b), for $^{87} \mbox{Rb} $, there are in fact three quantum states for $ f = 1 $, and five for $ f = 2 $.  

These levels can be split with the addition of a magnetic field due to the Zeeman effect.  To take this into account, we need to add an extra term to the Hamiltonian (\ref{singleparticleham}).  The single particle Hamiltonian taking into account of the first order Zeeman effect is\index{Zeeman effect}
\begin{align}
H_0 (\bm{x})  = -\frac{\hbar^2}{2m} \nabla^2 + V(\bm{x})  + g \mu_B B m_f  ,
\label{singleparticlehamzeeman}
\end{align}
where $ \mu_\text{B} = e\hbar/2 m_\text{e} $ is the Bohr magneton and $ e $ the elementary charge, $ m_\text{e} $ is the mass of an electron, and $ B $ is the magnetic field.   The Land{\'e} $g$-factor is given by \index{Land{\'e} $g$-factor}
\begin{align}
g = \left( \frac{f(f+1) + j(j+1) - i(i+1)}{2 f (f+1)} \right) \left( \frac{3}{2} + \frac{ s(s+1) - l (l+1)}{2 j (j+1)} \right) .
\end{align}
The spin structure of several types of atoms that are used for Bose-Einstein condensates are given in Table \ref{tab4-1}.

\begin{table}[t]
\processtable
{Spin structure of several types of atoms that are used for Bose-Einstein condensation. \label{tab4-1} }
{
\begin{tabular}{cccccc}
Isotope & $ s $  & $ l $ & $ j $ & $ i $ & $ f $ \\ 
\hline
$ ^{1} \text{H} $ & $ 1/2 $ & $0$ & $ 1/2 $ & $ 1/2 $ & $ 0,1 $ \\
$ ^{7} \text{Li} $,$ ^{23} \text{Na} $, $ ^{39} \text{K} $, $ ^{41} \text{K} $, $ ^{87} \text{Rb} $  & $ 1/2 $ & $0$ & $ 1/2 $ & $ 3/2 $ & $ 1,2 $ \\ 
 $ ^{85} \text{Rb} $  & $ 1/2 $ & $0$ & $ 1/2 $ & $ 5/2 $ & $ 2,3 $ \\ 
 $ ^{52} \text{Cr} $  & $ 3 $ & $0$ & $ 3 $ & $ 0 $ & $ 3 $ \\ 
 $ ^{133} \text{Cs} $  & $ 1/2 $ & $0$ & $ 1/2 $ & $ 7/2 $ & $ 3,4 $ \\ 
 $ ^{164} \text{Dy} $  & $ 2 $ & $6$ & $ 8 $ & $ 0 $ & $ 8 $ \\ 
 $ ^{168} \text{Dy} $  & $ 1 $ & $5$ & $ 6 $ & $ 0 $ & $ 6 $ \\ 
\end{tabular}}
\end{table}

In order to account for the spin quantum numbers of the atoms, we need an additional label on the bosonic operators as defined in (\ref{adaggerdef}).  Including the spin we have
\begin{align}
a_{k \sigma} = \int d \bm{x} \psi_{k \sigma}^* (\bm{x}) a_\sigma (\bm{x}) 
\end{align}
where $ \sigma $ is a label for the spin states which correspond to $ | f, m_f \rangle $.  The commutation relations then follow
\begin{align}
[a_{k \sigma} , a_{l \sigma'}^\dagger ] & = \delta_{kl} \delta_{ \sigma  \sigma'} \nonumber \\
[ a_{k \sigma}, a_{l \sigma'} ] & = [ a_{k \sigma}^\dagger , a_{l \sigma'}^\dagger ] = 0  .
\end{align}
In the case of a Bose-Einstein condensate, as we discussed in Chapter \ref{ch:bec}, the bosons occupy the same spatial state.  In this case we can implicitly assume that all the spatial quantum numbers $ k $ are the same.  Then only the spin degrees of freedom are relevant and we drop the label $ k $ and leave the spin label $ \sigma $
\begin{align}
a_{k \sigma} \rightarrow a_{\sigma} .
\end{align}
Bose-Einstein condensates with such a spin degree of freedom are called {\it spinor Bose-Einstein condensates}, which we examine in more detail in the next chapter.  \index{spinor Bose-Einstein condensates}
We can now write down the many-body Hamiltonian including spin degrees of freedom, in the same way that we did in (\ref{hamiltoniandiagonal}).  This is
\begin{align}
{\cal H}_0  = \sum_\sigma (E_0 + g \mu_B B \sigma) a^\dagger_\sigma a_\sigma  ,
\label{hamiltoniandiagonal2}
\end{align}
where $ E_0 $ is the ground state energy as given in (\ref{schrodinger}) and we have assumed that all the atoms are in the ground state.

\section{Interaction between spins}
\label{sec:interaction}

In the previous section we only wrote down the single particle Hamiltonian.  For the interaction term, we extend (\ref{interactionmanybodyham}) so that it includes spin degrees of freedom.  The most general way to write this is
\begin{align}
{\cal H}_I  = \frac{1}{2} \sum_{\sigma_1 \sigma_2 \sigma_1' \sigma_2'}  g_{\sigma_1' \sigma_2' \sigma_1 \sigma_2} 
a^\dagger_{\sigma_1'} a^\dagger_{\sigma_2'} a_{\sigma_1} a_{\sigma_2}
\label{interactionham}
\end{align}
where the matrix elements are
\begin{align}
g_{\sigma_1' \sigma_2' \sigma_1 \sigma_2}    =  \int d \bm{x} d \bm{y} \psi_{\sigma_1'}^* (\bm{x})  \psi_{\sigma_2'}^*  (\bm{y})  U(\bm{x},\bm{y})   \psi_{\sigma_1} (\bm{x})  \psi_{\sigma_2}  (\bm{y})  .
\label{interactionmatrix3}
\end{align}
We have omitted the labels for the spatial degrees of freedom because we have assumed that all the atoms are in the ground state.  Specifically the wavefunctions are given by
\begin{align}
\psi_{\sigma} (\bm{x}) \equiv \psi_{0 \sigma} (\bm{x})  ,
\end{align}
where $ \sigma $ labels a particular hyperfine spin state $ | f, m_f \rangle $.  

In (\ref{swaveintch1}) we implied that the $s$-wave scattering was only dependent on the relative spatial positions of the bosons.  In fact this is not entirely true, there is a dependence on the total spin of the interacting atoms.  When two atoms collide, only the outer electrons contribute to the interaction because the nuclear spin is deep within the atom and are generally unaffected to a good approximation.  As such, the interaction of the spin is related to the total electron spin of the two atoms $ \bm{J}_1 + \bm{J}_2 $.  The total spin is calculated by angular momentum addition of the interacting atoms. For example, for two $^{87} \text{Rb} $ atoms with  $ j = 1/2 $, the total spin of the atoms can be either $ j_{\text{tot}} = 0,1 $ (see Table \ref{tab4-1}).  The interaction does not change the total spin $  j_{\text{tot}} $.  We thus write the interaction in general as \index{s-wave scattering}
\begin{align}
U(\bm{x},\bm{y}) = \frac{4 \pi \hbar^2 }{m}  \delta (\bm{x} - \bm{y}) \sum_{j_{\text{tot}}}  a_s^{(j_{\text{tot}})} P_{j_{\text{tot}}}
\label{swaveint}
\end{align}
where 
\begin{align}
P_{j_{\text{tot}}} = \sum_{m_{j_{\text{tot}}}= - j_{\text{tot}}}^{j_{\text{tot}}} | j_{\text{tot}} , m_{j_{\text{tot}}} \rangle \langle j_{\text{tot}} , m |
\end{align}
is the projection operator for all the states that have a total spin $ j_{\text{tot}}  $.

For alkali atoms, the orbital angular momentum is zero $ l = 0 $ and only the single spin $ s = 1/2 $ contributes, giving $  \bm{J} = \bm{S} $ in this case.  In scattering theory, a particular combination of initial states that transition to a set of final states is called a {\it channel}.  \index{channel}  The two output total angular momenta  $ j_{\text{tot}} = 0,1 $ are thus called the {\it singlet} and {\it triplet  channels} respectively.  \index{singlet channel} \index{triplet channel} The projection operators can in this case be written
\begin{align}
P_{0}  & = \frac{1}{4} - \bm{S}_1 \cdot \bm{S}_2 \nonumber \\
P_{1}  & = \frac{3}{4} + \bm{S}_1 \cdot \bm{S}_2
\label{projectionalkali}
\end{align}
Thus in this case the interaction (\ref{swaveint}) can be written
\begin{align}
U(\bm{x},\bm{y}) = \frac{4 \pi \hbar^2 }{m}  \delta (\bm{x} - \bm{y}) \left[ 
\frac{a_s^{(0)} + 3 a_s^{(1)} }{4} + (a_s^{(1)} - a_s^{(0)}) \bm{S}_1 \cdot \bm{S}_2  \right].
\label{alkaliinteraction}
\end{align}

Substituting (\ref{swaveint}) into (\ref{interactionmatrix3}) we obtain
\begin{align}
g_{\sigma_1' \sigma_2' \sigma_1 \sigma_2}   = \frac{4 \pi \hbar^2 }{m}  \int d \bm{x} | \psi_{0} (\bm{x}) |^4 
 \sum_{j_{\text{tot}}}   a_s^{(j_{\text{tot}})}
\sum_{\sigma_1 \sigma_2 \sigma_1' \sigma_2'} \langle \sigma_1' | \langle \sigma_2' |  P_{j_{\text{tot}}}  |\sigma_1 \rangle |\sigma_2 \rangle .
\end{align}
In the case of alkali atoms, using (\ref{alkaliinteraction}) we have
\begin{align}
g_{\sigma_1' \sigma_2' \sigma_1 \sigma_2}   = & \frac{4 \pi \hbar^2 }{m}  \int d \bm{x} | \psi_{0} (\bm{x}) |^4 \nonumber \\
& \times \left[ \frac{a_s^{(0)} + 3 a_s^{(1)} }{4} \delta_{\sigma_1 \sigma_1'}  \delta_{\sigma_2 \sigma_2'}  
+ (a_s^{(1)} - a_s^{(0)}) \langle \sigma_1' | \langle \sigma_2' |  \bm{S}_1 \cdot \bm{S}_2  |\sigma_1 \rangle |\sigma_2 \rangle \right].
\label{alkaliscattering}
\end{align}
As was the case in Sec. \ref{sec:interactions}, we see that some of the matrix elements of $ g_{\sigma_1' \sigma_2' \sigma_1 \sigma_2}  $ are zero because of symmetries present in the interaction. These can be found by decomposing the total spins $ |f, m_f \rangle $ in terms of the electronic and nuclear spins according to Clebsch-Gordan coefficients \index{Clebsch-Gordan coefficient}
\begin{align}
| \sigma \rangle = |f, m_f \rangle = \sum_{m_j=-j}^j \sum_{m_i=-i}^i  \langle j, m_j, i, m_i | f, m \rangle | j, m_j, i, m_i  \rangle .
\end{align}
Some combinations of initial and final spins are zero by virtue of being in different total spin sectors. For example, the total $ z $-component of the spins before and after the interaction cannot change for both of the terms in (\ref{alkaliscattering})
\begin{align}
m_{f1}' + m_{f2}' = m_{f1} + m_{f2} .
\end{align}
Some scattering lengths for typical atoms are shown in Table \ref{tab4-2}.


\begin{table}[t]
\processtable
{Scattering lengths of alkali atoms.\label{tab4-2}}
{\addtolength\tabcolsep{2pt}
\begin{tabular}{ccc}
Atom & $  a_s^{(0)} $  & $  a_s^{(1)} $ \\ 
\hline
$^7 \text{Li} $ & 34 & -27.6  \\
$^{23} \text{Li} $ & 19 & 65 \\
$^{41} \text{K} $ & 85 & 65 \\
$^{85} \text{Rb} $ & 2400 & -400 \\
$^{87} \text{Rb} $ & 90 & 106 \\
$^{133} \text{Cs} $ & -208 & -350 \\
\end{tabular}}
\end{table}

\begin{exerciselist}[Exercise]
\item \label{q4-1}
Evaluate the $s$-wave interaction for $ ^{87} \text{Rb} $ for the scattering processes (a) $ | f= 1, m= -1 \rangle | f= 2, m= 1 \rangle \rightarrow 
 | f= 1, m= 0 \rangle | f= 2, m= 0 \rangle $; (b)  $ | f= 2, m= 2 \rangle | f= 2, m= 0 \rangle \rightarrow 
 | f= 1, m= 1 \rangle | f= 1, m= 1 \rangle $; and (c) $ | f= 1, m= 1 \rangle | f= 2, m= 1 \rangle \rightarrow 
 | f= 1, m= -1 \rangle | f= 2, m= -1 \rangle $.  
\item \label{q4-2}
(a) Verify that (\ref{projectionalkali}) are the projection operators for two spin-$1/2$ atoms by directly multiplying the 
singlet state $ (|\uparrow \rangle | \downarrow \rangle - |\downarrow \rangle | \uparrow \rangle)/\sqrt{2} $ and triplet states 
$ (|\uparrow \rangle | \downarrow \rangle + |\downarrow \rangle | \uparrow \rangle)/\sqrt{2}, |\uparrow \rangle | \uparrow \rangle, |\downarrow \rangle | \downarrow \rangle $.  (b) Derive (\ref{projectionalkali}) using only the fact that the singlet and triplet states are eigenstates of the total spin operator $ (\bm{S}_1 + \bm{S}_2)^2 $.  \index{s-wave scattering interaction} \index{triplet state} \index{singlet state}
\end{exerciselist}

\section{Electromagnetic transitions between spin states}
\label{sec:transitions}

In the last two sections we have described the Hamiltonian that is present for the spin degrees of freedom.  Specifically, (\ref{hamiltoniandiagonal2}) described the diagonal energy of atoms including the Zeeman shift,\index{Zeeman shift} and (\ref{interactionham}) described the interactions between atoms.  In the absence of any applied fields, these describe the coherent spin dynamics of the BEC.  However, often in experiments it is desirable and interesting to manipulate the spins by applying external electromagnetic fields to create transitions between spin levels.  This can be in the form of optical, microwave, or radio frequency radiation, and depends entirely upon the energy difference between the states in question.  For example, in Fig. \ref{fig4-1}, the transition 
$ 5^2 \text{S}_{1/2} \leftrightarrow 5^2 \text{P}_{3/2}, 5^2 \text{P}_{1/2} $ are in the optical frequency range with a wavelength of 780 nm and 795 nm respectively.  Meanwhile, the energy difference between the hyperfine ground states are microwave frequencies with wavelength $ \sim 0.1 $ mm.  Energy levels split by a Zeeman shift are typically in the radio frequency range.\index{Zeeman shift}  

To obtain the Hamiltonian for the interaction of the Bose-Einstein condensate with the electromagnetic field we first look at how it interacts with a single atom.  Specifically, we first look at the interaction of a single atom localized in space. To illustrate the procedure we first consider the hydrogen atom, which only contains a single electron and gives a virtually exact way of obtaining the transition. For more complex atoms, in principle all the electrons interact with the field, but for commonly used atoms such as alkali atoms, to a good approximation, the outer electron alone determines the state of the whole atom.  In this way we can determine the interaction of an atom with the electromagnetic field by looking at a single electron. 

The interaction of an electron with an electromagnetic field is given in general by
\begin{align}
H_e = \frac{1}{2m_e} 
\left( - i \hbar \bm{\nabla}- e \bm{A} (\bm{x},t ) \right)^2 + e U (\bm{x},t)  + V_e (\bm{x}) 
\label{generalham}
\end{align}
where $ m_e $ is the electron mass, and $ V_e (\bm{x}) $ is the binding energy\index{binding energy} for the electron within the atom. The vector and scalar potentials\index{scalar potential} for the electromagnetic field are $ \bm{A} (\bm{x},t ) $ and $ U (\bm{x},t) $, respectively.  Let us start in the radiation gauge, where we take\index{radiation gauge}\index{vector potential}
\begin{align}
U (\bm{x},t) = 0 .
\end{align}
This may always be done since scalar and vector potentials can be defined by subtracting a scalar function $ \chi(\bm{x},t ) $\index{scalar function}
\begin{align}
U' (\bm{x},t ) & =  U (\bm{x},t ) - \frac{\partial  \chi}{\partial  t} \nonumber \\
\bm{A}'(\bm{x},t ) & = \bm{A}(\bm{x},t ) + \bm{\nabla} \chi(\bm{x},t )  . \label{gaugetransform}
\end{align}
For an electromagnetic wave, the vector potential can be taken to 
be 
\begin{align}
\bm{A} (\bm{x},t ) = \bm{A}_0 e^{i( \bm{k} \cdot \bm{x} - \omega t) } ,
\end{align} 
where $ \bm{A}_0 $ is the amplitude of the vector potential, $ \bm{k} $ is the wavenumber and $ \omega $ is the angular frequency.  The wavelength of electromagnetic radiation is typically 
much larger than size of the atom.  For example, the radius of a Rb atom is $ 2.5 \times 10^{-10} $ m, in comparison to the $\text{D}_2$ transition with a wavelength of $ 7.8 \times 10^{-7} $ m.  \index{vector potential}
The spatial variation is therefore negligible and the vector potential may be approximated as  
\begin{align}
\bm{A} (\bm{x},t ) \approx \bm{A}_0 e^{i (\bm{k} \cdot \bm{x}_0 - \omega t )} = \bm{A}_0 (t ) ,
\end{align}
where $ \bm{x}_0 $ is the location of the atom. 
This approximation is called the {\it electric dipole approximation}.  \index{electric dipole! approximation}
The interaction Hamiltonian is then
\begin{align}
H_e = \frac{1}{2m_e} 
\left( - i \hbar \bm{\nabla}- e \bm{A}_0 (t) \right)^2  + V_e (\bm{x})  .
\label{vectorham}
\end{align}
Since the vector potential now has no spatial dependence, we can go further and remove this from the Hamiltonian by taking
\begin{align}
\chi(\bm{x},t) = - \bm{x} \cdot \bm{A}_0 (t)  .
\end{align}
This means according to (\ref{gaugetransform}) that the scalar potential is non-zero, giving the interaction Hamiltonian \index{scalar potential} 
\begin{align}
H_e = H_e^{(0)} + H_e^{(1)} ,
\label{dipoleham}
\end{align}
where the unperturbed Hamiltonian is
\begin{align}
H_e^{(0)}  = -\frac{\hbar^2}{2m_e} \nabla^2 + V_e (\bm{x}) 
\end{align}
and the electric dipole Hamiltonian is\index{electric dipole!Hamiltonian}
\begin{align}
H_e^{(1)} = - e \bm{x} \cdot \bm{E}_0 (t) .
\label{edmham}
\end{align}
Here we have used $ \bm{E} = - \frac{\partial \bm{A}}{\partial t} $.  The Hamiltonian (\ref{edmham}) is the desired term which causes transitions between energy levels of the atoms.

Note that the gauge transformation (\ref{gaugetransform}) actually also requires transformation of the wavefunction as well, according to
\begin{align}
\psi' (\bm{x},t ) = e^{i e \chi(\bm{x},t )/ \hbar } \psi (\bm{x},t ) .
\label{wavefunctiongauge}
\end{align}
Thus since the Hamiltonians (\ref{generalham}), (\ref{vectorham}), and (\ref{dipoleham}) are all in different gauges, the wavefunctions will differ
by phase definitions according to (\ref{wavefunctiongauge}).

Let's now take the example of the hydrogen atom to illustrate the use of (\ref{dipoleham}).  Suppose that we start with the electromagnetic field turned off, so that $  \bm{E}_0 = 0 $.  In this case
the eigenstates of (\ref{dipoleham}) are just the solutions of the Schrodinger equation with a potential $ V_e (\bm{x}) = - \frac{e}{4 \pi \epsilon_0 r } $.  The ground and first excited state the wavefunctions are
\begin{align}
\psi_{100} (r, \theta, \phi) & = \frac{1}{\sqrt{\pi a_B^3}} e^{-r/a_B} \nonumber \\
\psi_{200} (r, \theta, \phi) & = \frac{1}{\sqrt{8 \pi  a_B^3}} (1- \frac{r}{2a_B}) e^{-r/a} ,
\end{align}
where $ a_B $ is the Bohr radius.  \index{Bohr radius}
Suppose the electromagnetic field is relatively weak and is of the appropriate frequency such that the transitions only occur between these two atomic states.  We will see what kind of conditions are required for this assumption later.  The electronic wavefunction can then be restricted to superpositions of the two levels 
\begin{align}
\psi( \bm{x} ) = c_1 \psi_{100} ( \bm{x} ) + c_2 \psi_{200} ( \bm{x} ) 
\end{align}
where $ c_1, c_2 $ are complex coefficients such that $ |c_1 |^2 +  |c_2 |^2 = 1 $.  In this two-dimensional space, we can write the Hamiltonian in matrix form, with matrix elements 
\begin{align}
H_e (n,m) = \int d^3 \bm{x} \psi_{n00}^* ( \bm{x} ) H_e \psi_{m00} ( \bm{x} ) 
\end{align} 
The matrix corresponding to the unperturbed component is 
\begin{align}
H_e^{(0)}  = 
\left(
\begin{array}{cc}
\hbar \omega_1 & 0 \\
0 & \hbar \omega_2 \\
\end{array}
\right) ,
\label{unperturbed}
\end{align}
where $ \hbar \omega_n $ is the energy of the states $ \psi_{n00} $. For the electric field, let us assume a form
\begin{align}
H_e^{(1)} = - e x E_0 \cos  \omega_0 t
\end{align}
where we have taken the electric field to be in the $ x $-direction and $ E_0 $ is the amplitude. The electric dipole Hamiltonian is\index{electric dipole!Hamiltonian} then 
\begin{align}
H_e^{(1)} = 
\left(
\begin{array}{cc}
0 & \hbar \Omega^* \cos  \omega_0 t \\
\hbar \Omega \cos  \omega_0 t  & 0 \\
\end{array}
\right) ,
\label{edmham1}
\end{align}
where we have defined the Rabi frequency \index{Rabi frequency}
\begin{align}
\hbar \Omega = - e E_0 \int d^3 \bm{x} \psi_{200}^* ( \bm{x} ) x \psi_{100} ( \bm{x} ) .
\label{rabifrequency}
\end{align} 
The diagonal elements of (\ref{edmham1}) are zero due to the fact that
\begin{align}
\int d^3 \bm{x} | \psi_{n00} ( \bm{x} ) |^2 x = 0 .
\end{align}
This follows from the fact that due to the spherical symmetry of the atom, the wavefunctions are all odd or even functions $ \psi_{nlm} ( -\bm{x} ) =(-1)^l \psi_{nlm} ( \bm{x} ) $.  Since (\ref{edmham}) is an odd function, thus ensures that all diagonal components evaluate to zero. 

We can remove the time dependence of the Hamiltonian by working in the interaction picture. To do this we separate the Hamiltonian into two parts according to \index{index picture}
\begin{align}
H_e & = {H_e^{(0)}}' + {H_e^{(1)}}' \nonumber \\
{H_e^{(0)}}' & = \hbar \omega_1 | 1 \rangle \langle 1 | +  \hbar (\omega_1+\omega_0) | 2 \rangle \langle 2 |  \nonumber \\
{H_e^{(1)}}' & = \hbar \Delta | 2 \rangle \langle 2 | + \hbar \Omega \cos  \omega_0 t | 2 \rangle \langle 1 | 
+ \hbar \Omega^* \cos  \omega_0 t | 1 \rangle \langle 2 | 
\end{align}
where we have switched to bra-ket notation, $ \Delta = \omega_2 - \omega_1 -\omega_0  $ and the state $ | n \rangle $ corresponds to the wavefunction $ \psi_{n00} $.  
An operator $ O $ in the interaction picture is related to that in the Schrodinger picture according to\index{Schrodinger picture}
\begin{align}
[O]_\text{I} = {U^{(0)}_e}^{\dagger} O U^{(0)}_e ,  
\end{align}
where $ U^{(0)}_e $ is the time evolution operator for $ {H_e^{(0)}}' $.  
We denote interaction picture operators by $ [ \dots ]_\text{I} $ and leave Schrodinger picture operators unlabeled.  The time evolution operator can be evaluated to give
\begin{align}
U^{(0)}_e (t) = e^{-i {H_e^{(0)}}' t / \hbar}  = e^{-i \omega_1 t} | 1 \rangle \langle 1 | + e^{-i (\omega_1+\omega_0) t} | 2 \rangle \langle 2 | .
\label{timeevolution}
\end{align}
We see that each of the states pick up a time evolving phase related to their energies.  The states then evolve according to only the 
interaction Hamiltonian
\begin{align}
[{H_e^{(1)}}']_\text{I} =  {U_e^{(0)}}^{\dagger} {H_e^{(1)}}' U_e^{(0)} = & \frac{(e^{2 i \omega_0 t} + 1 )\hbar \Omega}{2}   
| 2 \rangle \langle 1 | 
+ \frac{(e^{-2 i  \omega_0 t} + 1 ) \hbar \Omega^* }{2}    | 1 \rangle \langle 2 | \nonumber \\
& + \hbar \Delta | 2 \rangle \langle 2 |  .
\label{interactionham1}
\end{align}

We see that in the interaction picture there are two terms which contribute to each off-diagonal matrix element.  The term that has no time dependence is an energy conserving term since the energy difference between the atomic levels $ \hbar (\omega_{2} - \omega_{1}) $ is approximately matched by the energy of the photon $ \hbar \omega_0 $.  Physically this corresponds to the absorption of a photon by the atom and a transition to a higher energy state, or emission of a photon by the atom and a transition to a lower energy state.  The other term does not conserve energy as it corresponds to the absorption of a photon {\it and} a transition to a lower energy state, or the reverse.  A common step at this point is to neglect such energy non-conserving terms, which amounts to the {\it rotating wave-approximation}. \index{rotating wave approximation}  The name comes from neglecting terms in (\ref{interactionham1}) that involve phases $ e^{i \omega t } $ where the $ \omega t  $ is a large number within the relevant timescale such that the average integrates to zero. We finally have under this approximation 
\begin{align}
[H_e^{(1)} ]_\text{I} = \hbar \left(
\begin{array}{cc}
0 & \Omega^*/2  \\
\Omega/2   & \Delta \\
\end{array}
\right) .
\label{interactionham2}
\end{align}  

In the above example we considered two simple states of the hydrogen atom for the sake of a simple example.  More generally the transitions that the electric dipole Hamiltonian\index{electric dipole!Hamiltonian} allows depends upon selection rules\index{selection rule} which depend on the matrix element as was calculated in (\ref{rabifrequency}).  There are generally two considerations which determine whether a given transition is zero or non-zero.  The first the parity effect\index{parity effect} as we discussed above.  Since $ \psi_{nlm} ( -\bm{x} ) =(-1)^l \psi_{nlm} ( \bm{x} ) $, this means that if the initial and final states are both the same parity then the transition matrix element evaluates to zero since (\ref{edmham}) is odd.  The other consideration is due to angular momentum conservation.  To see this, first write
\begin{align}
\bm{x} \cdot \bm{E} = \sqrt{\frac{4 \pi}{3}} \left(
E_z Y_{10} + \frac{-E_x + i E_y}{\sqrt{2}} Y_{11} + 
\frac{E_x + i E_y}{\sqrt{2}} Y_{1-1} \right) ,
\end{align}
where $ Y_{10} = \sqrt{\frac{3}{4 \pi}} \cos \theta $, $ Y_{1\pm 1} = \mp  \sqrt{\frac{3}{8 \pi}} \sin \theta e^{\pm i \phi} $ are  the spherical harmonics.  Then by the rules of angular momentum addition, the angular momentum of the final and initial states must obey $ l' = l-1, l, l+1 $ respectively. The purely spatial form of the electric dipole moment\index{electric dipole!moment} means that the spin is unaffected and hence $ s' = s $.  Between the parity and angular momentum addition rules, we thus only have transitions such that $ l' = l \pm 1 $.  In terms of the total angular momentum quantum numbers, for hydrogen we have the selection rules\index{selection rule}
\begin{align}
\Delta j & = 0, \pm 1 \nonumber \\
\Delta m & = 0, \pm 1 .
\label{selectionrules}
\end{align}
For more complex atoms involving more than one electron, similar rules apply.  The parity selection rule is not an exact symmetry for more complex atoms, but the angular momentum addition rules still apply, hence we again obtain (\ref{selectionrules}).  In this case $ \Delta l = 0, \pm 1 $ is allowed, except for the case $ l' = l = 0 $, but  $ s' = s $ again.  

In the above discussion we were only concerned with a single atom.  We can straightforwardly write down the Hamiltonian for the many atom case, such as in a BEC.  The transitions simply occur for each atom individually, and we have 
\begin{align}
{\cal H}_\Omega = \frac{\hbar \Omega  e^{i \bm{k} \cdot \bm{x}}}{2} a^\dagger_2 (\bm{x})  a_1  (\bm{x})
+   \frac{\hbar \Omega^*  e^{-i \bm{k} \cdot \bm{x}}}{2} a^\dagger_1 (\bm{x}) a_2 (\bm{x})
+ \hbar \Delta a^\dagger_2 (\bm{x})  a_2  (\bm{x})
\label{transitionham}
\end{align}
One difference to the single atom case is that the transition matrix now takes a spatial dependence due to the phase of the electromagnetic field.  In a BEC, atoms can be delocalized over distances that are larger than optical wavelengths, hence it becomes important to take into account of the spatial dependence. As a result of the phase dependence, the atoms experience a momentum shift equal to $ \bm{k} $ according to momentum conservation when a photon is absorbed or emitted.  We will show an example of this in Sec. \ref{sec:spinorbitcoupling}.

\begin{exerciselist}[Exercise]
\item \label{q4-3}
Show that (\ref{timeevolution}) is true by expanding the exponential operator, and using the fact that the identity operator is $ I = |1 \rangle \langle 1 | + |2 \rangle \langle 2 |  $.  
\item \label{q4-4}
Verify (\ref{interactionham1}). 
\end{exerciselist}

\section{The ac Stark shift}
\label{sec:acstark}

In the previous section we saw that light will in general cause a transition to an excited state of the atom according to (\ref{interactionham2}) for a single atom and (\ref{transitionham}) for the multi-atom case. We now consider the case where the light field has a large detuning with respect to the transition $ \Delta \gg | \Omega | $. Diagonalizing (\ref{interactionham2}), the eigenstates and energies are 
\begin{align}
|\pm \rangle & = \frac{1}{{\cal N}_\pm} \left[ (\Delta \pm \sqrt{\Delta^2 + |\Omega|^2})  |1 \rangle - \Omega | 2 \rangle \right],  \nonumber \\
E_{\pm} & = \frac{1}{2} \left( \Delta \mp  \sqrt{\Delta^2 + |\Omega|^2} \right) ,
\label{acstarkdiagonalized}
\end{align} 
where $ | 1 \rangle $ is a ground state of the atom and $ | 2 \rangle $ is an excited state and $ {\cal N}_\pm $ is a suitable normalization factor.  The frequency of the laser is chosen such that it is detuned from the excited state, such that $ |\Omega| \ll \Delta $.  We can thus expand the square roots to give
\begin{align}
| + \rangle & \approx  | 1 \rangle - \frac{\Omega}{2 \Delta} |2 \rangle \\
| - \rangle & \approx  | 2 \rangle + \frac{\Omega^*}{2 \Delta} |2 \rangle \\
E_+ &  \approx - \frac{ |\Omega|^2}{4 \Delta} \label{acstarkenergy} \\
E_- & \approx \Delta + \frac{ |\Omega|^2}{4 \Delta} .
\end{align}
The energetically lower state $ |+ \rangle $ is to a good approximation the same as the state $ | 1 \rangle $, except that it is shifted lower by an energy $ - \frac{ |\Omega|^2}{4 \Delta}  $.  The energy shift is called the {\it ac Stark shift}.  \index{ac Stark shift} 

We can then write down an effective Hamiltonian that describes the ac Stark shift for the many atom case straightforwardly. Working in the diagonal basis (\ref{acstarkdiagonalized}), the Hamiltonian  (\ref{transitionham}) reads
\begin{align}
{\cal H}_\Omega & = \hbar \Big[ E_+ b_{+}^\dagger b_{+} + E_- b_{-}^\dagger b_{-}   \Big] \nonumber \\
& \approx  \hbar \Big[ - \frac{ |\Omega|^2}{4 \Delta} b_{+}^\dagger b_{+}  +
\left( \Delta + \frac{ |\Omega|^2}{4 \Delta}  \right) b_{-}^\dagger b_{-}  \Big]
\end{align}
where $ b_{\pm} $ are associated with the states $ | \pm \rangle  $. Since the energy level of the diagonalized states have corrections to the original state, it often interpreted as a second order effect in perturbation theory, where the off-diagonal terms in (\ref{interactionham2}) are the perturbative terms, which modify the energy of the state $ | 1 \rangle $. \index{perturbation theory}

The ac Stark shift \index{ac Stark shift}is a very common effect that is exploited in many situations in atomic physics.  For instance, in optical lattices a light field at a frequency detuned with the atomic resonance is prepared with a spatially varying intensity.  Since the light field intensity is $ \propto  |\Omega|^2 $,  an energy potential (\ref{acstarkenergy}) proportional to the intensity of the light at that point in space is produced.  It is also the basis for atom trapping methods such as that using optical dipole traps.  In the treatment of the previous section, the light was considered to be a classical field, which is valid as long as the intensities are high. However in some applications such as non-destructive measurements\index{non-destructive measurement} it is also important to take into account of the quantum nature of light.  The quantum version of the ac Stark shift is used as the basis of techniques such as optical imaging of atoms and entanglement generation\index{entanglement generation}.

\section{Feshbach resonances}\index{Feshbach resonance}
\label{sec:feshbach}

\begin{figure}[t]
\centerline{\includegraphics[width=\textwidth]{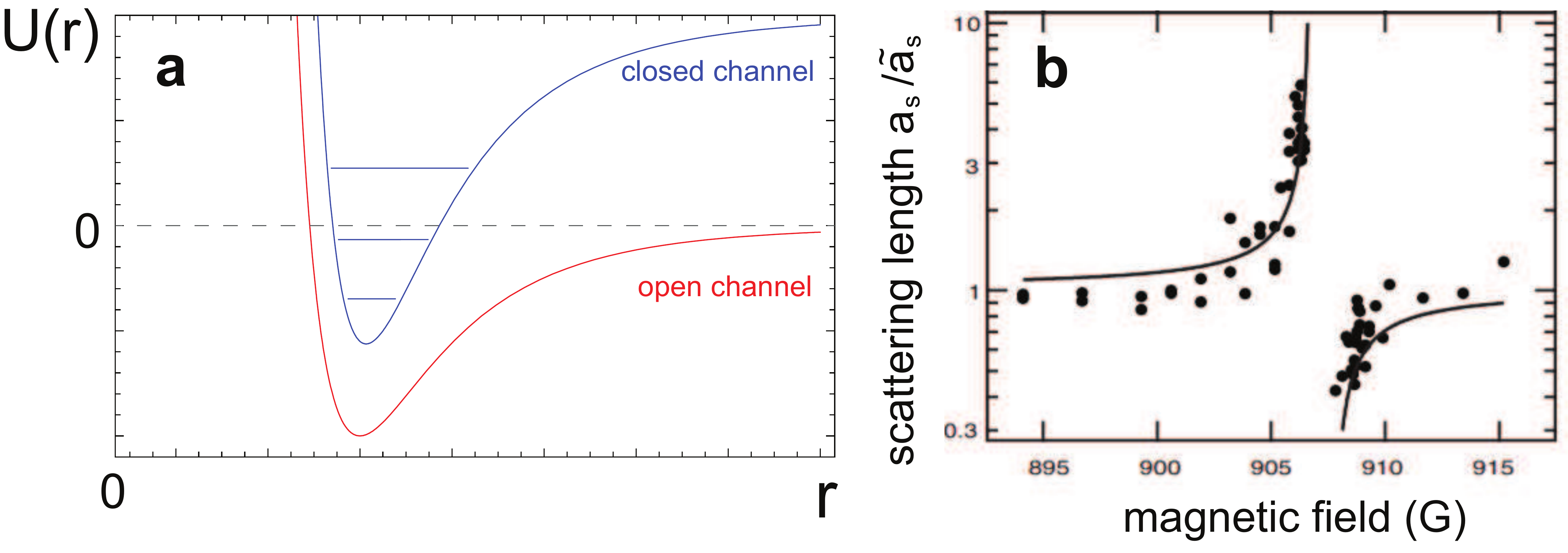}}
\caption{(a) Schematic of the process involved in a Feshbach resonance. (b) Experimental data showing the scattering length variation with magnetic field in a Feshbach resonance in a gas of sodium atoms from  \cite{inouye1998observation}.    }\index{Feshbach resonance}
\label{fig4-3}
\end{figure}

We now discuss another effect that originates from a second order transition, which affects the interactions between the atoms.   We consider a scattering picture, such that there is a particular initial state of two atoms that are initially far away from each other, which undergo some interaction at close range, then scatter to leave the atoms at distant locations. Depending upon the initial and final states, there are various scattering processes, or channels, which have different forms of interatomic potential.  Two such channels are shown in Fig. \ref{fig4-3}(a).  An open channel\index{channel!open} refers to a type of interaction which is allowed by energy conservation, such that at large interatomic distances the energy matches that of the free atoms.  Meanwhile, a closed channel\index{channel!closed} refers to a type of interaction where there is an energy barrier between the initial and final states. \index{channel}

Although the closed channel\index{channel!closed} may be highly off-resonant when the atoms are highly separated, at close distances the potential may possess some bound molecular states which have a similar energy. The atoms in the open channel\index{channel!open} cannot directly scatter to the closed channel, because by definition there are no continuum states in closed channels.  This means that the first order correction in perturbation theory is zero.  However, the atoms in the open channel can be affected by a second order process in a similar way to that seen in the last section.  This affects the interatomic potential of the open channel, which in turn affects the scattering length.  The scattering length is modified according to the form\index{scattering length}
\begin{align}
\frac{4 \pi \hbar^2}{m} a_s = \frac{4 \pi \hbar^2}{m} \tilde{a}_s + \sum_n \frac{ | \langle \psi_n | V | \psi_0 \rangle |^2}{E - E_n}  ,
\label{feshbachform}
\end{align}
where $ a_s, \tilde{a}_s $ are the modified and non-resonant scattering length respectively, $ | \psi_0 \rangle $ are the incoming and outgoing spherical waves,  $ | \psi_n \rangle $ are the closed channel bound state with energies $ E_n $, $ V $ is the Hamiltonian which causes the transitions between the open and closed channel states, and $ E $ is the energy of the particles in the open channel.  

In order to adjust the energy of the bound states to match the initial states of the atoms, a magnetic field is typically used to tune the energy via the Zeeman effect\index{Zeeman effect}.  A measurement of the scattering length across a Feshbach resonance is\index{Feshbach resonance} shown in Fig. \ref{fig4-3}(b).  We see the typical form of the scattering length which follows the form of (\ref{feshbachform}).  Depending upon the sign of $ E-E_n $, the additional term to the background $ \tilde{a}_s $ can be positive or negative.  This means that it is possible to tune the atomic interactions such that they are either attractive or repulsive, by applying a magnetic field.  Other ways of achieving the coupling between the closed and open channels is possible also using optical methods, which also produces an effective tuning of the interactions.

\section{Spontaneous emission}
\label{sec:spontaneous}\index{spontaneous emission}

The spin processes described up to now are all coherent processes\index{coherent process}.  Coherent processes are describable by a Hamiltonian, which implies that they are reversible, since they are described by the Schrodinger equation.  By changing the Hamiltonian from $ {\cal H} \rightarrow -{\cal H} $ one can always reverse the dynamics, and ``undo'' the time evolution $ e^{- i {\cal H} t/\hbar} $.  Not all process are however reversible.  The example that we examine in this section is a prime example: spontaneous emission\index{spontaneous emission}.  This is the process where an atom relaxes from a high to low energy state.  The reverse process of starting in a low energy state and ending up in a high energy state usually does not happen unless the atom is made to do so, perhaps by illuminating it with laser light as we saw in Sec. \ref{sec:transitions}. Such irreversible processes are called incoherent processes, and cannot be described by a Hamiltonian evolution.  \index{incoherent process}

But first, what is the physical origin of this difference? Isn't the universe ultimately described by a huge wavefunction that evolves according to a Hamiltonian? The reason for the irreversibility can be seen in the spontaneous emission case because the relaxation from high to low energy state is tied to the emission of a photon.  But unless the photon is reflected back in some way, it usually flies away and doesn't come back.  The reverse process on the other hand requires a photon for the atom to go from a low to high energy state.  Therefore if we don't specifically arrange for a photon to be near the atom, it cannot undergo the reverse process. 

\index{open system} \index{closed system} 
The key difference between the irreversible and reversible cases comes down to whether we consider the system as being an open or closed system (not to be confused by the open and closed channels of the previous section!).   In a closed system, we cannot possibly have any situation where the photon ``flies away and doesn't come back''.  In contrast, in an open system we can have the system coupled to many (usually an continuum, i.e. an infinity) of degrees of freedom, called the {\it bath} or {\it environment}.  \index{bath} \index{environment} \index{open system} \index{closed system} Then there is the opportunity for the photon escaping into the environment and never coming back.  To obtain the irreversible dynamics one then obtains an effective equation for the system alone.  This is called the {\it master equation}, and is a time evolution equation in the density matrix, rather than the wavefunction of the system.  This is capable of describing both the coherent and incoherent dynamics of the system.\index{master equation}

In this book we shall not go through the explicit derivation of the master equation\index{master equation}. Often the derivation does not give much insight into the dynamics anyway, one usually starts at the master equation and uses that as a starting point to examine the dynamics.  Fortunately, there is a simple prescription for most incoherent processes that one can write down the master equation without explicitly working through the derivation, as we explain below.  We will show just the start and end points of the master equation derivation, and examine a few simple consequences. \index{incoherent process}

\begin{figure}[t]
\centerline{\includegraphics[width=\textwidth]{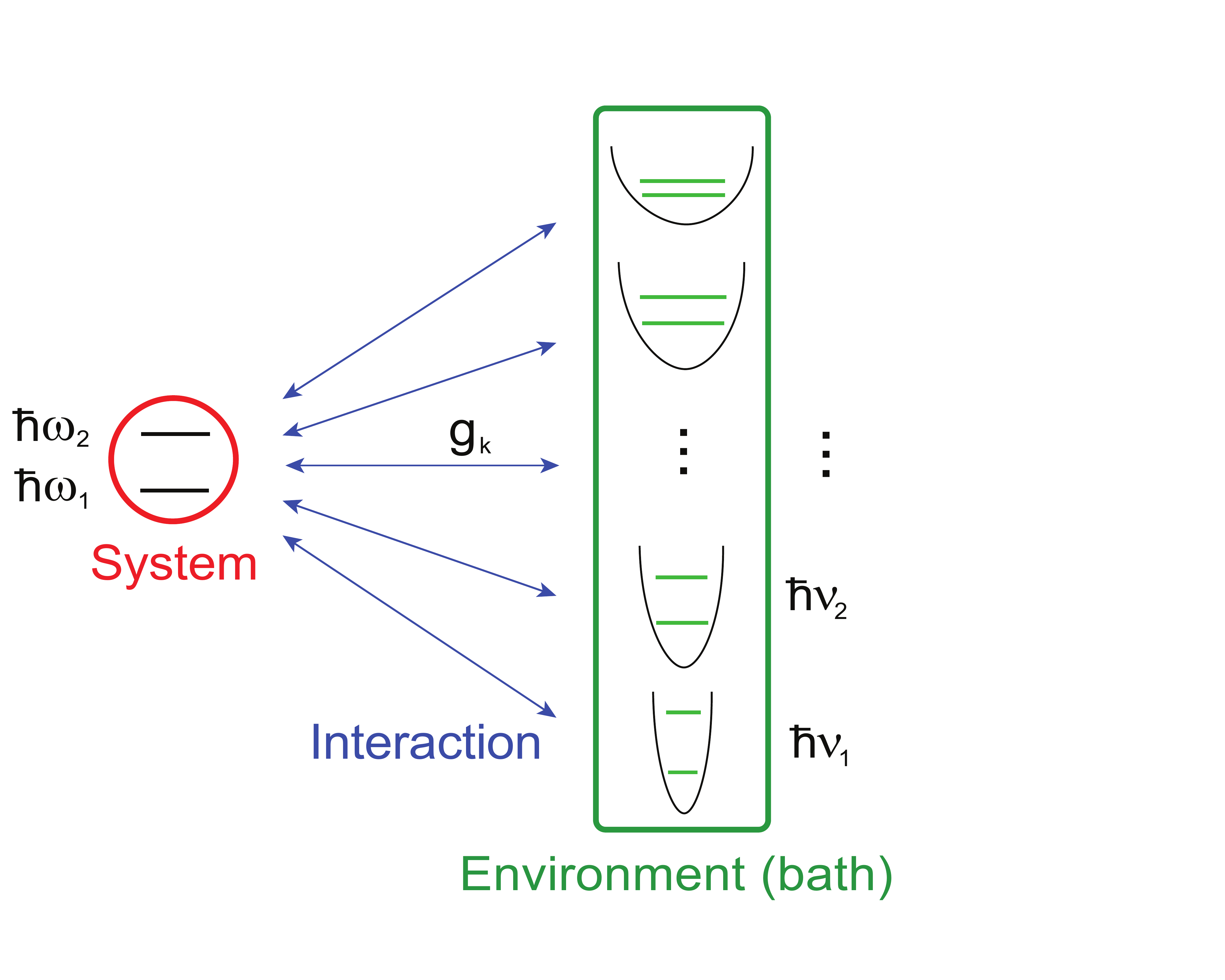}}
\caption{Schematic of model to describe spontaneous emission given in the Hamiltonian (\ref{spontaneousemissionham}).  Two atomic levels comprise the system and are coupled to a large number of electromagnetic modes, which form the environment.\index{spontaneous emission}   }
\label{fig4-2}
\end{figure}

For spontaneous emission, the model to describe the system (S) and environment (E) is
\begin{align}
H_\text{spon} & = H_\text{S} + H_\text{E} + H_\text{I} \label{spontaneousemissionham} \\
H_\text{S} & = \frac{\hbar (\omega_2-\omega_1)}{2} \sigma^z + \frac{\hbar (\omega_2+\omega_1)}{2} I \nonumber \\
H_\text{E} & = \hbar \sum_k  \nu_k  b^\dagger_k b_k  \nonumber \\
H_\text{I} & = \hbar \sum_k \left( g_k \sigma^+ b_k + g_k^* \sigma^- b_k^\dagger \right) ,
\end{align}
where we have used the following definitions
\begin{align}
\sigma^z & = | 2\rangle \langle 2 | - | 1\rangle \langle 1 | \nonumber \\
\sigma^- & = | 1\rangle \langle 2 | = (\sigma^+)^\dagger \nonumber \\
I & = | 2\rangle \langle 2 | + | 1\rangle \langle 1 |
\end{align}
for the two atomic levels $ | 1\rangle , |2 \rangle $ with energies $ \hbar \omega_{1},  \hbar \omega_2 $ respectively, and a continuum of electromagnetic modes $ b_k $ with energy $ \hbar \nu_k $.  The interaction (I) between the system and environment has a coupling $ g_k $.  Schematically the coupling takes a form that is shown in Fig. \ref{fig4-2}.

At this point everything is described by a Hamiltonian, so there is no irreversible dynamics in the system.  The irreversibility comes about because one would like to obtain an effective equation that only involves the system in question, which is the atom.  Treating an infinity of modes is needless to say rather inconvenient if one would like to perform simple calculations to capture the dynamics of the atom. The final result that is obtained, after a standard derivation under the Markovian assumption and the initial state of the environment being a vacuum state for all the modes, yields\index{Markovian approximation}
\begin{align}
\frac{ d \rho}{dt} = \Gamma {\cal L} ( \sigma^-, \rho)
\label{lindbladmaster}
\end{align}
where the Lindblad operator for an arbitrary operator $ X $ is defined as\index{Lindblad operator}
\begin{align}
{\cal L} ( X, \rho) \equiv  X \rho X^\dagger - \frac{1}{2} (  X^\dagger  X \rho +  \rho X^\dagger  X  ) .
\label{lindblad}
\end{align}
The constant $ \Gamma $ is a decay rate 
\begin{align}
 \Gamma  = \frac{  (\omega_2 - \omega_1)^3 | \langle 1 | e \bm{x} | 2 \rangle |^2}{3 \pi \epsilon_0 \hbar c^3},
\end{align}
where $ \epsilon_0 $ is the permittivity of free space and $ c $ is the speed of light. The equation (\ref{lindbladmaster}) describes the irreversible dynamics due to spontaneous emission.  \index{spontaneous emission}

In (\ref{lindbladmaster}), the only effect that we have taken account of is spontaneous emission --- there are no other dynamics for the system alone, or other types of baths causing different types of dynamics.  In the presence of other dynamics, one can simply include them by adding the corresponding terms to the master equation.  Thus the effects of spontaneous emission can be included by simply adding the Lindblad operator\index{Lindblad operator} (\ref{lindblad}) to the master equation\index{master equation}, with a suitable decay coefficient which determines its rate. Generally as a rule of thumb the operator $ X $ is the same operator that appears in the interaction Hamiltonian $ H_\text{I} $.  We will see below that simply by changing the operator $ X $ all types of other incoherent processes\index{incoherent process} can be easily included.

Let us solve (\ref{lindbladmaster}) to show explicitly that it indeed captures the spontaneous emission process.  The way to solve a matrix equation is to look at each matrix element and see how it evolves with time.  Since we are looking at a two level atom in this example, the density matrix is $ 2 \times 2 $, and there are four matrix elements that we must find the evolution of.  We can do this by multiplying (\ref{lindbladmaster}) by $ \langle \sigma | $ from the left and $ | \sigma' \rangle $ from the right.  Writing this out, we obtain the four coupled equations
\begin{align}
\frac{d \rho_{11}}{dt} & = \Gamma \rho_{22} \nonumber \\
\frac{d \rho_{22}}{dt} & = -\Gamma \rho_{11} \nonumber \\
\frac{d \rho_{12}}{dt} & = - \frac{\Gamma}{2} \rho_{12} \nonumber \\
\frac{d \rho_{21}}{dt} & = - \frac{\Gamma}{2} \rho_{21} ,
\end{align}
where we have defined $ \rho_{\sigma \sigma'} = \langle \sigma | \rho | \sigma' \rangle $. These can be solved explicitly to give the time evolution
\begin{align}
\rho(t) = \left(
\begin{array}{cc}
1 - \rho_{22}(0) e^{-\Gamma t} & \rho_{12} (0) e^{-\Gamma t/2} \\
\rho_{21} (0) e^{-\Gamma t/2} & \rho_{22} (0) e^{-\Gamma t}
\end{array}
\right) ,
\label{spontaneousdensity}
\end{align}
where $ \rho_{\sigma \sigma'}(0) $ are the initial matrix elements of the density matrix. 

We see that (\ref{spontaneousdensity}) has exactly the expected behavior.  The probability of the excited state $\rho_{22} $ decays exponentially from its initial value.  The loss of $ \rho_{22} $ is the gain of $ \rho_{11} $, hence it increases in probability accordingly. The off-diagonal elements of the density matrix are associated with coherence\index{coherence} between the levels $ | 1 \rangle $ and $ |2 \rangle $.  For example an equal superposition state between the two states $ ( | 1 \rangle + | 2 \rangle )/\sqrt{2} $ has a density matrix with elements $ \rho_{\sigma \sigma'} =1/2 $ everywhere.  We see that the off-diagonal terms also are damped, so that spontaneous emission\index{spontaneous emission} also reduces coherence between the states, albeit at half the rate of the diagonal terms.

Generalizing to the many atom case is again straightforward.  If we consider the annihilation operator for the two levels $ | 1,2 \rangle $ as $ a_{1,2} $ respectively, then all that is required in this case is to change the operator in the Lindblad operator\index{Lindblad operator} to $ X = a^\dagger_1 a_2 $ such that the master equation reads\index{master equation}
\begin{align}
\frac{ d \rho}{dt} & = \Gamma {\cal L} ( a^\dagger_1 a_2 , \rho) \nonumber \\
& = \frac{\Gamma}{2} \left( 2 a^\dagger_1 a_2 \rho a_1 a_2^\dagger - n_2 (n_1 + 1) \rho -  \rho n_2 (n_1 + 1) \right)
\label{bosemaster}
\end{align}
where $ n = a^\dagger a $ is the number operator for the two levels as labeled.  

Let us consider a simple example of the dynamics for the many atom case.  Suppose that we start in a state with $ N $ atoms occupying the
level denoted by $ a_2 $.  We can derive an equation that describes the number of atoms under the presence of spontaneous emission\index{spontaneous emission} from the master equation above.\index{master equation}  Multiplying (\ref{bosemaster}) by $ n_2 = a_2^\dagger a_2 $, and taking the trace we obtain the equation
\begin{align}
\frac{ d \langle n_2 \rangle }{dt} = - \Gamma (N+1) \langle n_2 \rangle + \Gamma \langle (n_2)^2 \rangle.
\end{align}
To exactly solve for the dynamics we would at this point need the evolution equation for $ \langle (n_2)^2 \rangle $, which can be obtained by multiplying (\ref{bosemaster}) by $ n_2^2 $ and taking the trace.  One approach is to make the approximation that 
$ \langle (n_2)^2 \rangle \approx \langle (n_2) \rangle^2  $, which amounts to saying that the variance of $ n_2 $ is rather small, which is usually obeyed if $ N \gg 1 $.  Another way to view this is that it is a mean-field approximation\index{mean-field approximation}
\begin{align}
\langle X Y \rangle & = \langle (\langle X \rangle + \delta X ) (\langle Y \rangle+ \delta Y ) \rangle  \nonumber \\
& = \langle X \rangle \langle Y \rangle + \langle X \rangle \langle \delta Y  \rangle +  \langle \delta X \rangle \langle Y  \rangle + 
\langle \delta X \delta Y \rangle \label{meanfield1} \\
& \approx \langle X \rangle \langle Y \rangle
\end{align}
where $ \delta X = X - \langle X \rangle $ are operators that offset by the mean, and we have used $ \langle \delta X \rangle = 0 $, and discarded the last term in (\ref{meanfield1}) since it is a product of two small quantities.  We can then approximate
\begin{align}
\frac{ d \langle n_2 \rangle }{dt} = - \Gamma (N+1) \langle n_2 \rangle + \Gamma \langle n_2 \rangle^2 ,
\end{align}
which can be solved analytically to obtain the dynamics
\begin{align}
 \langle n_2 \rangle = \frac{N(N+1)}{N + e^{(N+1) \Gamma t}} .
\label{superradiance}
\end{align}
We can see that the timescale of the exponential decay is $ \sim \frac{1}{(N+1) \Gamma} $.  This means that the more particles that occupy the system, the faster the decay rate is.  This is an example of the collective bosonic enhancement\index{bosonic enhancement} effect, and is an example of superradiance. \index{superradiance} The factor of $ N+1 $ often arises and originates from the way bosonic creation and annihilation operators work.  Adding an extra boson to a system which already has $ N $ bosons gives a factor of $ \sqrt{N+1} $:
\begin{align}
a^\dagger | N \rangle = a^\dagger \frac{ (a^\dagger)^N}{\sqrt{N!}} | 0 \rangle = \sqrt{N+1} | N+1 \rangle .
\end{align}
Thus matrix elements have this bosonic factor, and probabilities then have factors which are increased by  $ N+1$. \index{bosonic enhancement}

\begin{exerciselist}[Exercise]
\item \label{q4-5}
Solve (\ref{lindbladmaster}) and verify (\ref{spontaneousdensity}).  Keep in mind that for a density matrix the diagonal terms always add to 1, so that $ \rho_{11} + \rho_{22} = 1 $ at any time. 
\item \label{q4-6}
Work out the matrix elements of (\ref{bosemaster}) for a Fock state\index{Fock states}
\begin{align}
| N-k, k \rangle = \frac{(a_1^\dagger)^{N-k} (a_2^\dagger)^{k} }{\sqrt{(N-k)! k! }}
\end{align}
where the total number of atoms is $ N $.  Assuming that the system starts in the state $ | 1, N-1 \rangle $, find the dynamics of the system thereafter. Compare the dynamics to the single atom case and (\ref{superradiance}).  Does it decay faster, the same, or slower due to the presence of many atoms in the lower energy state? 
\end{exerciselist}

\section{Atom loss}
\label{sec:loss}\index{loss}

Another source of decoherence that acts on many-atom systems is loss.  The atoms in a trapped BEC are not perfectly trapped and generally they will be lost through various processes.  Various mechanisms exist for atoms to be lost from the trap.  The simplest is one-body loss\index{loss!one-body}, where an atom is lost from the trap since it exceeds the energy barrier imposed by the trap.  This can occur due to the background interaction with untrapped atoms or absorption of stray photons.  This is described by a master equation \index{master equation}
\begin{align}
\frac{ d \rho}{dt} & = \gamma_1{\cal L} ( a , \rho) \nonumber \\
& = \frac{\gamma_1}{2} \left( 2  a \rho a^\dagger - n \rho - \rho n  \right)
\label{bosemaster1loss}
\end{align}
where the rate of the one-body loss is $ \gamma_1 $, and the bosonic operators for a particular atomic species is given by $ a $. For the one-body loss term alone, it is easy to show that the atom number exponentially decays, by multiplying (\ref{bosemaster1loss}) by $ n $ and taking the trace.  This gives
\begin{align}
\frac{ d \langle n_l \rangle  }{dt} = \gamma_1 \langle n_l \rangle 
\end{align}
which has the solution
\begin{align}
\langle n \rangle = N e^{- \gamma_1 t } ,
\end{align}
where $N $ is the initial number of atoms. 

Atoms can also be lost from the BEC by the interactions between the atoms. In a two-body loss process\index{loss!two-body}, two atoms scatter and change the internal state of the atoms.  The dominant process is spin-exchange\index{spin-exchange}, and occurs in a similar way to that described in Sec. \ref{sec:interaction}.  Two-body loss is described according to the master equation\index{master equation}
\begin{align}
\frac{ d \rho}{dt} & = \gamma_2 {\cal L} (  a^2 , \rho) \nonumber \\
& = \frac{\gamma_2}{2} \left( 2  a^2 \rho (a^\dagger)^2 - (a^\dagger)^2 a^2 \rho - \rho (a^\dagger)^2 a^2  \right) ,
\label{bosemaster2loss}
\end{align}
where $ \gamma_2 $ is the two-body loss rate. 

For three-body loss\index{loss!three-body}, the relevant process is molecule formation.  Two atoms can be unstable towards the formation of a molecule, for example in Rb it is energetically favorable to form a $ \text{Rb}_2 $ molecule. The binding energy\index{binding energy} of the molecule must be carried away by a third atom, in order to conserve energy and momentum.  The energy of the molecule and the third atom have larger energies than the trapping potential, which results in the loss of three atoms.  Three-body loss is described according to the master equation
\begin{align}
\frac{ d \rho}{dt} & = \gamma_3 {\cal L} (  a^3 , \rho) \nonumber \\
& = \frac{\gamma_3}{2} \sum_j \left( 2  a^3 \rho (a^\dagger)^3 - (a^\dagger)^3 a^3 \rho - \rho (a^\dagger)^3 a^3  \right) ,
\label{bosemaster3loss}
\end{align}
where $ \gamma_3 $ is the three-body loss rate.

\section{Quantum jump method}
\label{sec:quantumjump}\index{quantum jump method}

In the previous two sections we introduced methods for simulating the time dynamics of several incoherent processes\index{incoherent process}.  These are generally described by master equations which take the general form 
\begin{align}
\frac{ d \rho}{dt} = -\frac{i}{\hbar} [H_0, \rho] +  \sum_j  \Gamma_j {\cal L} ( X_j, \rho) ,
\label{lindbladmaster2}
\end{align}
where $ X_j $ is the Lindblad operator\index{Lindblad operator} for the particular process,  $  \Gamma_j  $ is its rate, and $ H_0 $ is a Hamiltonian describing the coherent processes\index{coherent process}.    For processes involving several incoherent processes\index{incoherent process}, the Lindblad terms simply sum together.  A straightforward way to solve the master equations is to take matrix elements of both sides of (\ref{lindbladmaster2}), such that it is a set of coupled first order differential equations
\begin{align}
\frac{ d \rho_{nm} }{dt} = - \frac{i}{\hbar} H_{nm} +   \sum_j \Gamma_j L_{nm}(X_j,\rho)
\label{lindbladmaster3}
\end{align}
where 
\begin{align}
\rho_{nm} & = \langle n | \rho | m \rangle  \\
H_{nm} & = \langle n | [ \rho, H_0 ] | m \rangle = \sum_{n'} \left[ \langle n | H_0 | n ' \rangle \rho_{n' m} - \langle n ' | H_0  | m \rangle \rho_{n n'} \right]  \\
L_{nm}(X_j,\rho)&  = \langle n | {\cal L} ( X_j, \rho)| m \rangle  \nonumber \\
& =  \langle n | \left[ X \rho X^\dagger - \frac{1}{2} (  X^\dagger  X \rho +  \rho X^\dagger  X  )  \right] | m \rangle ,  
\end{align}
and $ |n \rangle, |m \rangle $ are an orthogonal set of states. For the coherent Hamiltonian evolution $ H_0 $, we introduced a complete set of states $ I = \sum_{n'} | n ' \rangle \langle n ' | $.  The set of equations (\ref{lindbladmaster3}) involve the square of the dimension of the Hilbert space, and is quite a numerically intensive task. An alternative method, called the {\it quantum jump method} \index{quantum jump method} gives equivalent results which can be more efficient in terms of calculation in many circumstances, since it can be calculated using only states which have the dimension of the Hilbert space.  

The procedure for the quantum jump method is as follows. Suppose at time $ t $ the state is in the state $ | \psi(t) \rangle $. One then evolves the state at some small time  $ \delta t $ later following the algorithm:
\begin{enumerate}
\item Calculate the quantity 
\begin{align}
\delta p_j   =\langle \psi(t)  | X_j^\dagger X_j |  \psi(t)  \rangle \Gamma_j \delta t ,   
\end{align}
and define $ \delta p = \sum_j \delta p_j $.  
\item Draw a random number $ \epsilon $ from the range $ [0,1] $.  
\item If $ \epsilon > \delta p $, then evolve the state according to
\begin{align}
| \psi(t+\delta t ) \rangle = \frac{1}{\sqrt{1-\delta p}} \left( 1 - \frac{i}{\hbar} H_{\text{eff}} \delta t \right) | \psi(t) \rangle
\end{align}
where
\begin{align}
H_{\text{eff}} = H_0 - \frac{i \hbar \Gamma_j }{2} \sum_j X_j^\dagger X_j .  
\end{align}
\item If $ \epsilon \le \delta p $, randomly choose one of the jumps labeled by $ j $ according to the probability $ \delta p_j /\delta p $, and perform the time evolution
\begin{align}
| \psi(t+\delta t ) \rangle = \sqrt{ \frac{ \Gamma_j \delta t }{ \delta p_j}} X_j | \psi(t) \rangle .
\end{align}
\item Go to step 1) and repeat until the desired time evolution is achieved.   
\end{enumerate}

The above procedure obviously has a stochastic nature, and will give a different time evolution for each run. Let us label the trajectory for a single run as $ | \psi^{(k)} (t ) \rangle $.   Running the time evolution $ M $ times, then the statistical mixture\index{statistical mixture} of these random runs approaches the true time evolved density matrix:
\begin{align}
\frac{1}{M} \sum_{k=1}^M | \psi^{(k)} (t ) \rangle \langle   \psi^{(k)} (t ) | \rightarrow \rho(t) .
\end{align}
As long as $ M $ is sufficiently large, the the time evolution of any pure state can be found. For an initially mixed state, one first writes this a mixture of pure states
\begin{align}
\rho(0) = \sum_l P_l | \psi_l (0) \rangle \langle \psi_l (0) | .  
\end{align}
Randomly choosing the initial state with probability $ P_l $, one performs the state steps as before to find the ensemble average.

\begin{exerciselist}[Exercise]
\item \label{q4-7}
Solve (\ref{lindbladmaster}) using the quantum jump method and compare it to the exact solution (\ref{spontaneousdensity}).  Do you obtain the same results?
\end{exerciselist}

\section{References and further reading}
\label{sec:ch4refs}

\begin{itemize}
\item Sec. \ref{sec:spindegrees}: For a more in depth introduction on the spin degrees of freedom of atoms see the textbooks \cite{foot2005atomic,bransden2003physics}.  For more details on the atomic structure of $ ^{87}\text{Rb}$ \cite{steck2001rubidium}.
\item Sec. \ref{sec:interaction}: Introductions discussing atomic interactions can be found in 
 \cite{pethick2004,krane1987introductory,bransden2003physics,dalibard1999collisional}.
\item Sec. \ref{sec:transitions}: Textbooks discussing electromagnetic transitions between spin states  \cite{foot2005atomic,fox2006quantum,steck2007quantum,scully1999quantum,walls2007quantum,meystre2001atom}.
\item Sec. \ref{sec:acstark}: The original reference introducing the ac Stark effect \cite{autler1955stark}. For further textbook level introductions see   \cite{pethick2004,meystre2001atom,fox2006quantum}.
 Experimental demonstration of atom trapping \cite{raab1987trapping}.  Reviews relating to atom trapping are given in  \cite{phillips1985laser,cohen1992laser,metcalf1994cooling,adams1997laser,ashkin1997optical,phillips1998nobel,wieman1999atom,grimm2000optical,ashkin2006optical,metcalf2007laser}. Treating the ac Stark shift quantum mechanically (the optical fields are treated mechanically) gives the related quantum nondemolition Hamiltonian \cite{happer1967off}. Theoretical works examining this are given in \cite{kuzmich1998atomic,molmer1999twin,kuzmich2000atomic,ilo2014theory}.
\item Sec. \ref{sec:feshbach}: First experimental works observing Feshbach resonance \cite{inouye1998observation,courteille1998observation}. Further experiments using Feshbach resonances \cite{cornish2000stable,donley2001dynamics,zwierlein2004condensation,jochim2003bose,durr2004observation}.   Reviews and books further discussing Feshbach resonances \cite{chin2010feshbach,pethick2002bose,kohler2006production,timmermans1999feshbach}. For the full derivation of  (\ref{feshbachform}) see \cite{pethick2002bose}. 
\item Sec. \ref{sec:spontaneous}: Early 
theories of spontaneous emission \cite{dirac1927quantum,weisskopf1930z}.  For further textbook level introductions on spontaneous emission see \cite{gerry2005introductory,scully1999quantum,schlosshauer2007decoherence,steck2007quantum,pethick2002bose,fox2006quantum,nielsen2000,meystre2001atom}.
\item Sec. \ref{sec:loss}: For textbooks and reviews discussing atom loss see \cite{scully1999quantum,walls2007quantum,gerry2005introductory,grimm2000optical,foot2005atomic,pethick2004,phillips1998nobel,kordas2015dissipative}.  Specific discussions of the role of atom loss is discussed in 
\cite{treutlein2008coherent,byrnes2015macroscopic,soding1999three}. 
\item Sec. \ref{sec:quantumjump}    
For reviews and books discussing the quantum jump method see \cite{plenio1998quantum,wiseman2009quantum,carmichael2009open,kordas2015dissipative}.  Experimental observation of quantum jumps were reported in \cite{sauter1986observation,bergquist1986observation,benson1994quantum,gleyzes2007quantum}. Theoretical works developing the topic are performed in \cite{dalibard1992wave,dum1992monte,hegerfeldt1993ensemble,molmer1993monte}.
\end{itemize}

  \chapter[Spinor Bose-Einstein condensates]{Spinor Bose-Einstein condensates}

\label{ch:spinorbec}

\section{Introduction}
\label{sec:introrepresentations}

We have seen in the previous chapter that the atoms constituting a BEC typically have a rich spin structure.  For the case of $^{87}\text{Rb} $, there are 8 spin states (the $ F =  1,2 $ states) which all have an energy which are relatively close.  The energies of these states are close enough such that they can be considered to be approximately degenerate, and are collectively called the ``ground states''.  By external manipulation of the spin states, various spin states of the BEC can be produced.  In this chapter we will describe these states and methods of how to characterize them.  We will examine several archetypal states such as spin coherent states and squeezed states. 
Many of the techniques in this chapter are borrowed from quantum optics, where useful methods to visualize states such as the $ P$-, $Q$-, and Wigner functions were first developed.  These can also be defined for the spin case, which serve as a powerful visualization of the quantum state. We also show a gallery of typical states for a spinor BEC, and how they appear for various quasiprobability distributions.  \index{coherent state!spin} \index{squeezed state} \index{P-function} \index{Q-function} \index{Wigner function}

\section{Spin coherent states}
\label{sec:spincoherentstates}

The simplest spin system that we can consider is that involving two levels.  The annihilation operators for the two levels will be labeled by $ a, b $, and we assume that there are a total of $ N $ atoms. The general wavefunction of a state can then be written in terms of the Fock states, which we define as\index{Fock states}
\begin{align}
|k \rangle = \frac{ (a^\dagger)^{k} (b^\dagger)^{N-k}}{\sqrt{k! (N-k)! }} | 0 \rangle ,
\label{fockstates}
\end{align}
where we only label the number of atoms in the state $ a $ in the ket because we always assume a total number of atoms as being $ N $. Recall that the meaning of (\ref{fockstates}) is that (see Sec. \ref{sec:spindegrees}), that the spatial degrees of freedom are all taken to be in the same state for both the levels $ a, b $.  This means that the only degrees of freedom that are left are the spin, and thus the different states arise from different combinations of occupations of the two levels. The total number operator 
\begin{align}
{\cal N } = a^\dagger a + b^\dagger b ,
\label{numberoperator2}
\end{align}
satisfies 
\begin{align}
{\cal N } |k \rangle  = N |k \rangle .
\end{align}
The general wavefunction of a state for $ N $ particles is then 
\begin{align}
| \psi \rangle = \sum_{k=0}^N \psi_k | k \rangle .
\label{genearlspin}
\end{align}
This state is an eigenstate of the total number operator
\begin{align}
{\cal N } | \psi \rangle = N | \psi \rangle .
\label{neigenstate}
\end{align}

In a Bose-Einstein condensate, we typically start with all the atoms occupying the same quantum state.  For example, this might be the energetically lowest hyperfine state such as that given in Fig. \ref{fig4-1}.  Taking this to be the state $ a $, the state would be written
\begin{align}
| k = N \rangle = \frac{  (a^\dagger)^{N}}{\sqrt{N!}} | 0 \rangle
\label{initialstatebec}.
\end{align}
If all the bosons were in the other state then the state would be
\begin{align}
| k = 0 \rangle = \frac{  (b^\dagger)^{N}}{\sqrt{N!}} | 0 \rangle .
\label{initialstatebec2}
\end{align}
It is then natural to consider the general class of states where the $ a $ and $ b $ states are in an arbitrary superposition. 
\begin{align}
| \alpha, \beta \rangle \rangle =  \frac{1}{\sqrt{N!}} ( \alpha a^\dagger + \beta b^\dagger )^N | 0 \rangle,
\label{scs}
\end{align}
where $ \alpha, \beta $ are coefficients such that $ | \alpha |^2 + | \beta |^2 =1 $.  This class of states is called a {\it spin coherent state}\index{coherent state!spin}  and plays a central role in manipulations of BECs.  Spin coherent states are aptly named in the sense that they have analogous properties to coherent states of light.\index{coherent state!optical}  We will see how such states arise naturally from a physical process in the next section, but for now let us take a mathematical point of view and explore its properties.

\begin{figure}[t]
\centerline{\includegraphics[width=\textwidth]{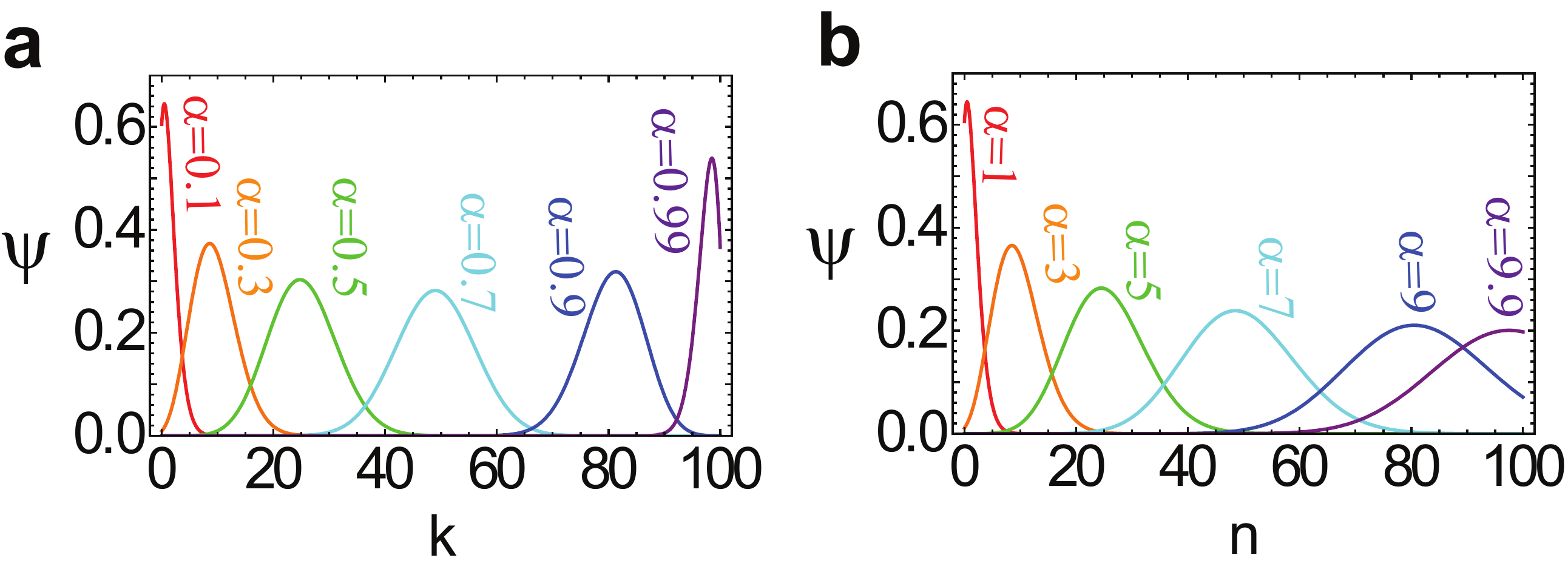}}
\caption{Amplitude of Fock states for (a) spin coherent states and (b) optical coherent states.  For the spin coherent state we take a value of $ \beta = \sqrt{1-\alpha^2} $ and $ N = 100 $. }\index{Fock states}\index{coherent state!spin} \index{coherent state!optical}
\label{fig5-1}
\end{figure}

Firstly, we can write the spin coherent state in the Fock representation (\ref{genearlspin}) by expanding the brackets in (\ref{scs}) with a binomial series
\begin{align}
| \alpha, \beta \rangle \rangle & =  \frac{1}{\sqrt{N!}} \sum_{k=0}^N {N \choose k} (\alpha a^\dagger)^{k} (\beta b^\dagger)^{N-k} | 0 \rangle \nonumber \\
& = \sum_{k=0}^N \sqrt{{N \choose k}} \alpha^k \beta^{N-k} | k \rangle .
\label{scsexpansion}
\end{align}
where we used (\ref{fockstates}).  We see that the spin coherent state wavefunction takes the form of (\ref{genearlspin}) so that indeed it is a state of fixed atom number $ N $. Eq. (\ref{scsexpansion}) has an analogous role to the expansion of an optical coherent state\index{coherent state!optical} in terms of Fock states\index{Fock states}.  Recall that 
\begin{align}
| \alpha \rangle & = e^{-| \alpha |^2/2} e^{\alpha a^\dagger} | 0 \rangle \nonumber \\
& = e^{-| \alpha |^2/2} \sum_{n=0}^\infty \frac{\alpha^n}{\sqrt{n!}} | n \rangle
\label{coherentfock}
\end{align}
where for this equation $ \alpha $ is any complex number and the optical Fock states are $ | n \rangle = \frac{(a^\dagger)^n}{\sqrt{n!}} | 0 \rangle $.  The functional form of (\ref{scsexpansion}) and (\ref{coherentfock}) have rather similar forms as can be seen in Fig. \ref{fig5-1}. Since $|\alpha|^2 + |\beta|^2 = 1$, we  see almost an identical distribution of the coefficients for $|\alpha| \ll1$ and $n\ll N$.  The reason for this can be seen by observing that one can approximate the amplitudes of the spin coherent states as
\begin{align}
\psi_k = \sqrt{\frac{N!}{k! (N-k)!}} \alpha^k \beta^{N-k} \approx \frac{(\sqrt{N} \alpha)^k}{\sqrt{k!}}e^{-N\frac{|\alpha|^2}{2}}
\label{psikexpansion},
\end{align}
since $ \exp[(N-k)\ln \beta ] = \exp[(N-k)\ln\sqrt{1-|\alpha|^2} ]\approx \exp[-N|\alpha|^2/2] $ and $ N!/(N-k)! \approx N^k $ .  Thus by rescaling the $ \alpha $ by a factor of $ \sqrt{N} $ one can approximate the same coefficients as in (\ref{coherentfock}).  From Fig. \ref{fig5-1} we see that the approximation starts to fail when $ \alpha \approx \beta $.  Optical coherent states\index{coherent state!optical} have a distribution where the average photon number and variance is equal to $ |\alpha|^2 $. The spin coherent state\index{coherent state!spin} distribution is symmetrical such that when $ \alpha \rightarrow 1 $, it again narrows.  This is due to the finite Hilbert space of the spins, whereas the photonic Hilbert space is infinite.

\begin{exerciselist}[Exercise]
\item \label{q5-1}
Verify (\ref{psikexpansion}).   
\end{exerciselist}

\section{The Schwinger boson representation}
\label{sec:schwingerboson}\index{Schwinger boson operator}

We now examine some observables of the spin coherent state (\ref{scs}).  For optical coherent states, the natural observables are the position and momentum, defined in terms of bosonic operators as \index{position operator} \index{momentum operator} \index{coherent state!spin} \index{coherent state!optical}
\begin{align}
X & = \sqrt{\frac{\hbar}{2 \omega}} ( a + a^\dagger) \nonumber, \\
P & = -i\sqrt{\frac{\hbar \omega}{2}}   ( a - a^\dagger),
\label{positionmomentumdim}
\end{align}
which have commutation relations\index{commutation relations} as $ [X,P] = i \hbar $ and $ \hbar \omega $ is the photon energy. Typically it is more convenient to work in dimensionless units, defining instead
\begin{align}
x & = \frac{1}{\sqrt{2}} ( a + a^\dagger) \nonumber ,\\
p & = -i\frac{1}{\sqrt{2}}   ( a - a^\dagger) ,
\label{positionmomentum}
\end{align}
which have the commutation relation $ [x,p] = i $. 

In our case, due to the two-level nature we are working with, it is more natural to define operators that have a form that closely resembles Pauli spin operators. We define the dimensionless total spin operators as\index{Pauli operators}
\begin{align}
S_x & = a^\dagger b + b^\dagger a \nonumber ,\\
S_y & = -i a^\dagger b + i b^\dagger a \nonumber, \\
S_z & = a^\dagger a - b^\dagger b .
\label{schwingerboson}
\end{align}
These operators are also called the {\it Schwinger boson operators}. \index{Schwinger boson operators}
These operators have many of the same properties as Pauli spin operators, such as
\begin{align}
[ S_j, S_k ] = 2 i \epsilon_{jkl} S_l
\label{commutationschwinger}
\end{align}
where $ j,k,l \in \{ x,y,z \} $ and  $ \epsilon_{jkl} $ is the Levi-Civita antisymmetric tensor\index{Levi-Civita antisymmetric tensor}.  Although the {\it commutation} relations of the operators are identical to Pauli operators, they do not have identical properties to them.  For example, $ (S_j)^2 \ne I $, where $ I $ is the identity operator, and the anticommutator is $ \{ S_i, S_j \} \ne 0 $, unlike Pauli operators.  We note that when $ N = 1 $, the spin operators reduce to Pauli operators\index{Pauli operators}
\begin{align}
\sigma_x & = |a \rangle \langle b |  + |b \rangle \langle a |  \nonumber, \\
\sigma_y & = -i |a \rangle \langle b |  + i |b \rangle \langle a |, \nonumber \\
\sigma_z & =|a \rangle \langle a |  - |b \rangle \langle b | .
\label{paulioperators}
\end{align}

Raising and lowering operators are defined according to \index{raising and lowering operators} \index{ladder operators}
\begin{align}
S_+ & = \frac{S_x + i S_y}{2} = a^\dagger b \nonumber, \\
S_- & = \frac{S_x - i S_y}{2} = b^\dagger a .
\label{raisinglowering}
\end{align}
These operators have the effect of changing the Fock state by one unit\index{Fock states}
\begin{align}
S_+ |k \rangle & = \sqrt{(k+1)(N-k)} | k + 1\rangle \nonumber, \\
S_- |k \rangle & = \sqrt{k(N-k+1)} | k - 1\rangle .
\label{ladderfock}
\end{align}
The raising and lowering operators can be related to the $ S_x, S_y $ spin operators by
\begin{align}
S_x & = S_+ + S_- ,\nonumber \\
S_y & = -i S_+ + i S_- .
\label{inverseladder}
\end{align}
Operating the spin operators on the Fock states gives
\begin{align}
S_x | k \rangle & = \sqrt{(k+1)(N-k)} | k + 1\rangle + \sqrt{k(N-k+1)} | k - 1\rangle \nonumber, \\
S_y | k \rangle & = -i \sqrt{(k+1)(N-k)} | k + 1\rangle + i \sqrt{k(N-k+1)} | k - 1\rangle \nonumber, \\
S_z | k \rangle & = (2k -N) | k \rangle .
\label{fockstateop}
\end{align}
The Fock states are eigenstates of the $ S_z $ operator.  

What are the eigenstates of the $ S_x $ and $ S_y $ operators? These are also Fock states, but where the bosons are defined in different basis.  If we define bosonic operators
\begin{align}
c & = \frac{a+b}{\sqrt{2}} ,\nonumber \\
d & = \frac{a-b}{\sqrt{2}}, 
\end{align}
the eigenstates of the $ S_x $ operator take the form
\begin{align}
|k \rangle_x & = \frac{ (c^\dagger)^{k} (d^\dagger)^{N-k}}{\sqrt{k! (N-k)! }} | 0 \rangle \nonumber \\
& = \frac{ (a^\dagger+ b^\dagger)^{k} (a^\dagger- b^\dagger)^{N-k}}{\sqrt{2^N k! (N-k)! }} | 0 \rangle 
\label{fockstatesx}
\end{align}
and the  eigenstates of the $ S_y $ operator are
\begin{align}
|k \rangle_y &  = \frac{ (a^\dagger+ i b^\dagger)^{k} (i a^\dagger + b^\dagger)^{N-k}}{\sqrt{2^N k! (N-k)! }} | 0 \rangle  .
\label{fockstatesy}
\end{align}
The eigenvalue equation then are written
\begin{align}
S_x |k \rangle_x & = (2k -N ) |k \rangle_x \nonumber \\
S_y |k \rangle_y & = (2k -N ) |k \rangle_y .
\label{sxyeigenvalueeq}
\end{align}

The Schwinger boson operators conserve the total number of particles.  The particle number is thus a conserved quantity which arises because\index{Schwinger boson operators}
\begin{align}
[{\cal N}, S_x ] = [{\cal N}, S_y ] = [{\cal N}, S_z ] = 0 .
\label{spinnumbercommutation}
\end{align}
The eigenstates of the spin operators and total number operators thus have simultaneous eigenstates.  This is apparent from (\ref{neigenstate}), which shows that states of this form are eigenstates of the total number operator, and the fact that the eigenstates (\ref{fockstatesx}) and (\ref{fockstatesy}) can be expanded in Fock states (\ref{fockstates}).  \index{Fock states}

\begin{exerciselist}[Exercise]
\item \label{q5-2}
Verify (\ref{commutationschwinger}).  Why isn't $ (S_j)^2 \ne I $ and  $ \{ S_i, S_j \} \ne 0 $?  If $ N = 1 $, do any of your conclusions change?
\item \label{q5-2b}
Show that (\ref{fockstatesx}) and (\ref{fockstatesy}) satisfy the eigenvalue equations (\ref{sxyeigenvalueeq}).  
\item \label{q5-2c}
Expand (\ref{fockstatesx}) and (\ref{fockstatesy}) and show that it can be written in the form of (\ref{neigenstate}).  
\end{exerciselist}

\section{Spin coherent state expectation values}
\label{sec:spincohexp}\index{coherent state!spin}

Let us now evaluate the expectation values of the Schwinger boson operators (\ref{schwingerboson}) with respect to the spin coherent states (\ref{scs}).  First turning to the expectation value of $ S_x $, we require evaluation of the expression\index{Schwinger boson operators}
\begin{align}
\langle S_x \rangle & = \langle \langle \alpha, \beta | S_x | \alpha, \beta \rangle \rangle \nonumber \\
& = \frac{1}{N!} \langle 0 | ( \alpha^* a + \beta^* b )^N S_x ( \alpha a^\dagger + \beta b^\dagger )^N | 0 \rangle .
\end{align}
To evaluate this our strategy will be to commute the $ S_x $ operator one by one through the $ ( \alpha a^\dagger + \beta b^\dagger ) $ product to the far right hand side until it is immediately preceding $ | 0 \rangle $.  Then using the fact that $ S_x | 0 \rangle = 0 $ we can remove the operator from the expression. The commutator that we will need is
\begin{align}
[S_x,  \alpha a^\dagger + \beta b^\dagger ] = \beta a^\dagger + \alpha b^\dagger .
\end{align}
Every time the $ S_x $ is passed through, we obtain the above factor. Hence we obtain
\begin{align}
\langle S_x \rangle = \frac{1}{(N-1)!} \langle 0 | ( \alpha^* a + \beta^* b )^N (\beta a^\dagger + \alpha b^\dagger)
( \alpha a^\dagger + \beta b^\dagger )^{N-1} | 0 \rangle .
\end{align}
Next, we will commute the $ (\beta a^\dagger + \alpha b^\dagger)  $ factor one by one through all the terms on the left so that it is immediately after the $ \langle 0 | $.  Then using the fact that $ \langle 0 | (\beta a^\dagger + \alpha b^\dagger)  = 0 $ we can eliminate this factor. The commutation relation that we will need is
\begin{align}
[ \alpha^* a + \beta^* b,  \beta a^\dagger + \alpha b^\dagger] & = \alpha^* \beta + \beta^* \alpha ,
\label{commrel1}
\end{align}
which gives
\begin{align}
\langle S_x \rangle = N ( \alpha^* \beta + \beta^* \alpha) \frac{1}{(N-1)!} \langle 0 | ( \alpha^* a + \beta^* b )^{N-1} ( \alpha a^\dagger + \beta b^\dagger )^{N-1} | 0 \rangle .
\end{align}
Nothing that the remaining expectation value is simply the normalization condition for spin coherent state with $ N - 1 $ particles $ \langle \langle \alpha, \beta | \alpha, \beta \rangle \rangle_{N-1} = 1 $, we have
\begin{align}
\langle S_x \rangle = N ( \alpha^* \beta + \beta^* \alpha) .
\label{expectationvaluesx}
\end{align}
We can work out the expectation values of the other operators in a similar way, which yield
\begin{align}
\langle S_y \rangle & = N ( -i \alpha^* \beta + i \beta^* \alpha), \nonumber \\
\langle S_z \rangle & = N ( |\alpha|^2 - |\beta|^2 ) .
\label{expectationvaluesyz}
\end{align}

\begin{figure}[t]
\centerline{\includegraphics[width=\textwidth]{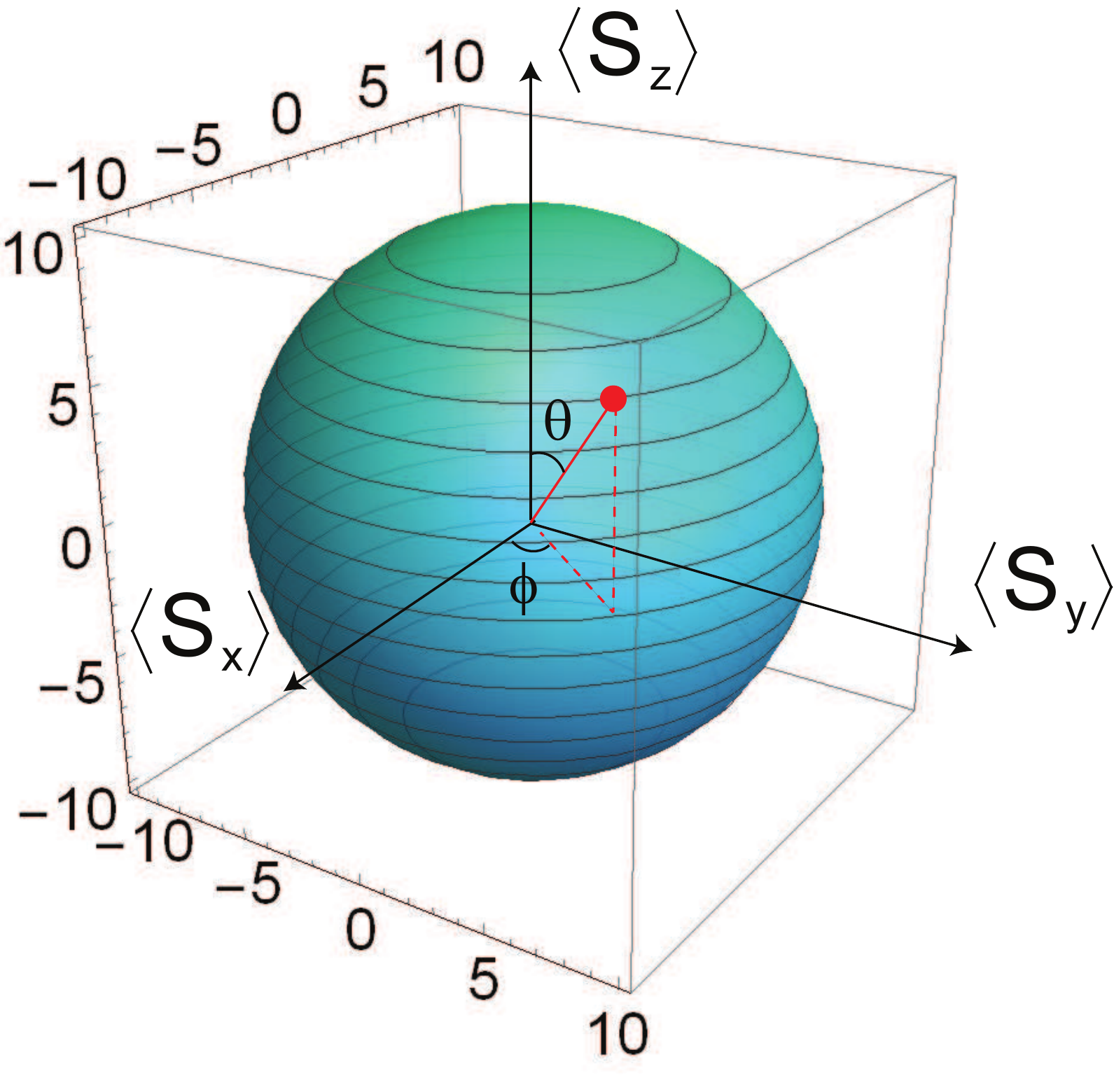}}
\caption{The Bloch sphere representation of a spin coherent state for $ N = 10 $.   }\index{Bloch sphere,}
\label{fig5-2}
\end{figure}

Comparing the expectation values of (\ref{expectationvaluesx}) and (\ref{expectationvaluesyz}) we see that they are identical to the expectation values of Pauli operators (\ref{paulioperators}) for a state
\begin{align}
\alpha | a \rangle  + \beta | b \rangle ,
\label{qubitstate}
\end{align}
up to a multiplicative factor of $ N $. 
For standard qubits, the above quantum state can be visualized on the Bloch sphere using the parametrization\index{Bloch sphere}
\begin{align}
\alpha & = \cos \frac{\theta}{2} e^{-i\phi/2} \nonumber, \\
\beta & = \sin \frac{\theta}{2} e^{i\phi/2},
\label{blochpara}
\end{align}
where the angles $ \theta, \phi $ give a position on the unit sphere.  Using this parametrization the expectation values take a form
\begin{align}
\langle S_x \rangle & = N \sin \theta \cos \phi \nonumber ,\\
\langle S_y \rangle & = N \sin \theta \sin \phi \nonumber ,\\
\langle S_z \rangle & = N \cos \theta .
\label{scsexpectations}
\end{align}
It is easy to see that the expectation values for the spin coherent state exactly coincides with the usual $ x, y, z $ coordinates on a sphere of radius $ N $\index{coherent state!spin}
\begin{align}
\langle S_x \rangle^2 + \langle S_y \rangle^2  + \langle S_z \rangle^2 = N^2 .  
\label{sphereequation}
\end{align}
Figure \ref{fig5-2} shows the Bloch sphere representation of a spin coherent state with $ N = 10 $.  The set of all spin coherent states (\ref{scs}) form a sphere in the expectation values $ (\langle S_x \rangle , \langle S_y \rangle, \langle S_z \rangle ) $.  

Eq. (\ref{sphereequation}), which is only true for spin coherent states,  should not be confused with the total spin operator, which in the language of group theory is a Casimir operator for SU(2). This is \index{Casimir operator}
\begin{align}
\bm{S}^2 & = S_x^2 +  S_y^2 + S_z^2 \nonumber \\
& = {\cal N} ( {\cal N} + 2 ) \label{totalspin}
\end{align}
where we have used (\ref{numberoperator2}).  Since this can be written in terms of the number operator $ {\cal N} $ this means that any state with fixed number as given in (\ref{genearlspin}) is an eigenstate with eigenvalue $ N (N+2) $.  This is the same as the usual expression for the eigenvalue of total spin operator $ s(s+1) $ if we note that our dimensionless spin operators have an extra factor of 2 throughout.

\begin{exerciselist}[Exercise]
\item \label{q5-3}
Evaluate (\ref{expectationvaluesyz}) using the same method as (\ref{expectationvaluesx}). You will need to first evaluate 
$ [S_{y,z},  \alpha a^\dagger + \beta b^\dagger ] $ and the similar relations to (\ref{commrel1}).  
\item \label{q5-4}
Check that the expectation value of the Pauli operators for the state (\ref{qubitstate}) give the same values as\index{Pauli operators} (\ref{expectationvaluesx}) and (\ref{expectationvaluesyz})  up to a factor of $ N $. 
\item \label{q5-4b}
Substitute the expressions for the spin operators into (\ref{totalspin}) and show that all the off-diagonal terms cancel to give the expression in terms of the number operator. 
\end{exerciselist}

\section{Preparation of a spin coherent state}
\label{sec:prepscs}\index{coherent state!spin}

Previously we defined the spin coherent state without describing how it can be prepared physically.  How would one prepare a state such as (\ref{scs})?  Suppose we then apply some electromagnetic radiation of a suitable frequency such that transitions now take place between the levels $ a $ and $ b $, such as that described in (\ref{transitionham}).  Taking the detuning to be zero we have the Hamiltonian
\begin{align}
{\cal H}_x = \frac{\hbar \Omega}{2} \left( a^\dagger b + b^\dagger a \right) .
\end{align}
In terms of the Schwinger boson operators, the transition Hamiltonian takes a form\index{Schwinger boson operators}
\begin{align}
{\cal H}_x = \frac{\hbar \Omega}{2} S_x .
\label{spintransitionham}
\end{align}
It is reasonable to assume that (\ref{initialstatebec}) can be created through the process of Bose-Einstein condensation, since it is the lowest energy state. If the transition Hamiltonian is applied for a time $ t $, the new state is
\begin{align}
| \psi(t) \rangle = e^{- i {\cal H}_x t / \hbar } | k = N \rangle  = e^{- i S_x \Omega t / 2} | k = N \rangle .
\label{rotatedspin}
\end{align}
For Pauli operators, evaluating the time evolution operator is straightforward because we can take advantage of the relation\index{Pauli operators}
\begin{align}
e^{- i \sigma_x \Omega t / 2} = \cos ( \Omega t / 2) - i \sigma_x  \sin ( \Omega t / 2)  ,
\label{exponentialsigmax}
\end{align}
which was possible because $ \sigma_x^2 = I $.  Unfortunately we cannot do the same here so we must use a different method.  The way to evaluate the time evolution operator is to diagonalize the $ S_x $ operator by performing the transformation
\begin{align}
a & = \frac{c+d}{\sqrt{2}} \nonumber ,\\
b & =  \frac{c-d}{\sqrt{2}} .
\label{bosontransform}
\end{align}
The spin operators (\ref{schwingerboson}) then take the form
\begin{align}
S_x & = c^\dagger c - d^\dagger d \nonumber ,\\
S_y & = i c^\dagger d - i d^\dagger c \nonumber, \\
S_z & = c^\dagger d +  d^\dagger c  .
\label{schwingerboson2}
\end{align}
We see that in this representation $ S_x $ looks like how the $ S_z $ operator appeared, in diagonal form. The state then can be written
\begin{align}
| \psi(t) \rangle =  \frac{1}{\sqrt{2^N N!}} e^{- i n_c \Omega t / 2} e^{ i n_d \Omega t / 2} (c^\dagger + d^\dagger)^N | 0 \rangle,
\label{transitionrotation1}
\end{align}
where $ n_c = c^\dagger c $ and $ n_d = d^\dagger d $. This can be evaluated noting that for an arbitrary function $ f $ 
\begin{align}
f(n_c) c^\dagger & = \left( f(0) + f'(0) n_c + \frac{f''(0)}{2} (n_c)^2 + \dots \right) c^\dagger \nonumber \\
& = c^\dagger f(n_c+ 1 ) ,
\end{align}
where the function was expanded as a Taylor series and we used
\begin{align}
n_c c^\dagger= c^\dagger (n_c + 1 ),
\end{align}
and similarly for $ n_d $.  Our strategy in evaluating (\ref{transitionrotation1}) will be to commute the exponentials to the right since they can be removed once they act on the vacuum according to  $ n_{c,d} | 0 \rangle = 0 $.  Moving the exponentials through one factor of $ (c^\dagger + d^\dagger) $ we find
\begin{align}
 e^{- i n_c \Omega t / 2} e^{ i n_d \Omega t / 2} (c^\dagger + d^\dagger) = ( e^{- i \Omega t / 2} c^\dagger + e^{ i \Omega t / 2} d^\dagger) e^{- i n_c \Omega t / 2} e^{ i n_d \Omega t / 2} .
\end{align}
After commuting exponentials through all $ N $ factors of $ (c^\dagger + d^\dagger) $, we obtain
\begin{align}
| \psi(t) \rangle =  \frac{1}{\sqrt{2^N N!}}  ( e^{- i \Omega t / 2} c^\dagger + e^{ i \Omega t / 2} d^\dagger)^N | 0 \rangle ,
\label{finaltransitionpre}
\end{align}
Reverting back to the original bosonic variables by inverting (\ref{bosontransform}), we have 
\begin{align}
| \psi(t) \rangle & =   \frac{1}{\sqrt{N!}} ( \cos (\Omega t / 2) a^\dagger - i \sin (\Omega t / 2) b^\dagger )^ N | 0 \rangle \nonumber \\
& = | \cos (\Omega t / 2),  -i \sin (\Omega t / 2) \rangle \rangle
\label{finaltransition}.
\end{align}
This is the form of a spin coherent state\index{coherent state!spin} (\ref{scs}) where $ \alpha = \cos (\Omega t / 2)  $ and $ \beta = -i \sin (\Omega t / 2)  $.  By choosing the time $ t $ appropriately, the magnitudes of $ \alpha, \beta $ can be chosen arbitrarily.  From (\ref{expectationvaluesx}), (\ref{expectationvaluesyz}), and Fig. \ref{fig5-2} we see that the electromagnetic radiation can produce states which are along the circle
\begin{align}
\langle S_y \rangle^2 + \langle S_z \rangle^2 = N^2,
\end{align}
which is the great circle with axis along the $ S_x $-direction. 

But what about the other points on the Bloch sphere?\index{Bloch sphere} To obtain spin coherent states on an arbitrary location of the Bloch sphere we must perform an additional operation with a different axis of rotation.  This time, suppose we apply a magnetic field such that the levels experience a Zeeman effect\index{Zeeman effect}.  Then according to (\ref{hamiltoniandiagonal}) we will have the Hamiltonian
\begin{align}
{\cal H}_z & = (E_0 + g \mu_B \sigma_a) a^\dagger a + (E_0 + g \mu_B \sigma_b) b^\dagger b  \nonumber \\
& = E_0' {\cal N } +  \frac{\hbar \Delta}{2} S_z,
\label{hzeroham}
\end{align}
where we have defined
\begin{align}
E_0' & = E_0 + g \mu_B \frac{\sigma_a + \sigma_b}{2}  \nonumber, \\
\hbar \Delta & =  g \mu_B (\sigma_a - \sigma_b),
\end{align}
and used (\ref{numberoperator2}) and (\ref{schwingerboson}).  As we have seen from (\ref{neigenstate}), the spin coherent state is an eigenstate of the total number operator.  Thus the application of $ {\cal N } $ only creates a global phase on the state which is physically unobservable.  The only important part of the Hamiltonian is then the $ S_z $ component which as we will evaluate below, creates a rotation around the $ S_z $ axis.  

Applying (\ref{hzeroham}) to (\ref{finaltransition}) we then would like to evaluate
\begin{align}
| \psi(t,\tau) \rangle & = e^{-i {\cal H}_z \tau /\hbar } | \psi(t) \rangle \nonumber \\
& = e^{-i E_0' N \tau / \hbar } e^{-i \frac{\Delta}{2} S_z \tau} \sum_{k=0}^N \sqrt{ N \choose k} 
\cos^k ( \Omega t/2) i^{N-k} \sin^{N-k}  ( \Omega t/2) | k \rangle,
\end{align}
where we used (\ref{scsexpansion}) to expand the spin coherent state and the number operator $ \cal N $ was evaluated to its eigenvalue $ N $.  The exponential can be evaluated noting that the Fock states \index{Fock states}are eigenstates of the $ S_z $ operator (\ref{fockstateop}).  The exponential of $ S_z $ can then be evaluated by expanding it as a Taylor series\index{Taylor series}
\begin{align}
e^{-i \frac{\Delta}{2} S_z \tau}  | k \rangle & = \left( 1 - i \frac{\Delta}{2} S_z \tau + (i \frac{\Delta}{2} S_z \tau)^2/2 + \dots \right)  | k \rangle  \nonumber \\
& = \left( 1 - i \frac{\Delta}{2} (2k -N) \tau + (i \frac{ \Delta}{2} (2k -N) \tau)^2/2 + \dots \right)  | k \rangle  \nonumber \\
& = e^{-i \frac{\Delta}{2} (2k - N) \tau}  | k \rangle .
\end{align}
We thus obtain 
\begin{align}
| \psi(t,\tau) \rangle & = e^{-i E_0' N \tau/ \hbar } e^{-i \frac{\Delta}{2} S_z \tau} \sum_{k=0}^N \sqrt{ N \choose k } 
[e^{-i \frac{\Delta}{2}\tau}   \cos ( \Omega t/2) ]^k  [ i e^{i \frac{\Delta}{2}\tau}  \sin  ( \Omega t/2) ]^{N-k}  | k \rangle  \nonumber \\
& = e^{-i E_0' N \tau/\hbar } | e^{-i \frac{\Delta}{2}\tau} \cos ( \Omega t/2) , i e^{i \frac{\Delta}{2}\tau}  \sin  ( \Omega t/2) \rangle \rangle .
\label{psittau}
\end{align}
Comparing this to (\ref{blochpara}) we see that an arbitrary position on the Bloch sphere can be attained if we associate\index{Bloch sphere}
\begin{align}
\theta & = \Omega t \nonumber, \\
\phi & = \Delta \tau + \frac{\pi}{2} , 
\end{align}
up to a physically irrelevant global phase factor. In summary, we have seen that an arbitrary state on the Bloch sphere can be attained by two successive rotations, first around the $ S_x $ axis, then around the $ S_z $ axis:
\begin{align}
e^{-i \phi S_z/2} e^{-i \theta S_x/2}  | 1,0\rangle \rangle = 
| e^{-i \phi/2} \cos ( \theta/2) , -i e^{i \phi/2 }  \sin  ( \theta/2) \rangle \rangle
\end{align}
where we used the fact that 
\begin{align}
|k = N \rangle & = | 1,0 \rangle \rangle \nonumber, \\
|k = 0 \rangle & = | 0,1 \rangle \rangle .
\end{align}

It is not a coincidence that the axis of the great circle for the first rotation is along the same operator that appeared in (\ref{spintransitionham}), and the axis for the second rotation was the same as that in (\ref{hzeroham}).  In fact in general, for a Hamiltonian containing a sum of $ S_x, S_y, S_z $ operators, the axis of rotation will be along the unit vector $ \bm{n} = (n_x, n_y, n_z ) $ given by the coefficients of these operators. The general rotation is
\begin{align}
e^{-i \theta \bm{n} \cdot \bm{S} } | \alpha, \beta \rangle \rangle = | \alpha', \beta' \rangle \rangle,
\label{generalrotation}
\end{align}
where
\begin{align}
\left(
\begin{array}{c}
\alpha' \\
\beta' 
\end{array}
\right)
=
\left(
\begin{array}{cc}
\cos \theta - i n_z \sin \theta & - (n_y+i n_x ) \sin \theta\\
(n_y - i n_x) \sin \theta  & \cos \theta + i n_z \sin \theta\\
\end{array}
\right)
\left(
\begin{array}{c}
\alpha \\
\beta 
\end{array}
\right)
\label{singlequbitrotation},
\end{align}
and
\begin{align}
\bm{S} = (S_x, S_y, S_z )  .
\end{align}
Thus by a suitable combination of spin operators any point can be rotated to any other point on the Bloch sphere via a great circle rotation. Special cases of this when the spins are aligned along the $x,y,z$ axes are given by\index{Bloch sphere}
\begin{align}
e^{-i \theta S_x/2 }  | \alpha, \beta \rangle \rangle & = | \alpha \cos (\theta/2)  -i \beta \sin (\theta/2) , \beta \cos (\theta/2)  -i \alpha \sin (\theta/2)   \rangle \rangle \nonumber, \\
e^{-i \theta S_y/2 }  | \alpha, \beta \rangle \rangle & = |\alpha  \cos (\theta/2) - \beta \sin (\theta/2) , \beta  \cos (\theta/2) + \alpha \sin (\theta/2)  \rangle \rangle \nonumber, \\
e^{-i \theta S_z/2 }  | \alpha, \beta \rangle \rangle & = |e^{-i \theta/2} \alpha , e^{i \theta/2} \beta \rangle \rangle  .
\label{spinrotationsxyz}
\end{align}
These correspond to a rotation around the $S_x$, $S_y $, and $S_z $ axis by an angle $ \theta $ respectively.

\begin{exerciselist}[Exercise]
\item \label{q5-5}
Evaluate (\ref{finaltransition}) from (\ref{transitionrotation1}) in a different way.  First expand the factor $ (c^\dagger + d^\dagger)^N $ using a binomial expansion and obtain a superposition in terms of Fock states\index{Fock states}
\begin{align}
|k \rangle' = \frac{ (c^\dagger)^{N-k} (d^\dagger)^{k}}{\sqrt{(N-k)! k!}} | 0 \rangle
\end{align}
Then using the fact that $ e^{i \theta n_c} |k \rangle'  = e^{i \theta (N-k) }  |k \rangle' $ and 
 $ e^{i \theta n_d} |k \rangle'  = e^{i \theta k }  |k \rangle' $ evaluate the exponentials, then refactorize the expression to obtain (\ref{finaltransitionpre}). 
\item \label{q5-6}
Verify that (\ref{generalrotation}) with (\ref{singlequbitrotation}) produces the correct spin coherent state.  Hint: make a linear transformation of the $ a, b $ operators such that unit vector $ \bm{n} $ is rotated to $ (0,0,1) $ in the new basis.  Then you can use the same method as that used in (\ref{psittau}) to evaluate the $ S_z $ rotation.  Reverting back to the original basis then gives the final result. 
\item \label{q5-6b}
Substitute the parametrization (\ref{blochpara}) into (\ref{spinrotationsxyz}) and show that original spin coherent state is rotated by an angle $ \theta $. 
\end{exerciselist}

\section{Uncertainty relations}
\label{sec:uncertainty}\index{uncertainty relation}

One of the central properties of quantum mechanics is the Heisenberg uncertainty relation, which in the most familiar form states that \index{Heisenberg uncertainty relation}
\begin{align}
\sigma_X \sigma_P  \ge \frac{\hbar}{2}
\label{heisenberguncertainty},
\end{align}
where $ \sigma_X, \sigma_P $ are the standard deviations of the position and momentum operators (\ref{positionmomentumdim}) respectively.  For a coherent state we can evaluate the standard deviations to be
\begin{align}
\sigma_X & = \sqrt{\langle X^2 \rangle - \langle X \rangle^2} =  \sqrt{\frac{\hbar}{2 \omega}} \nonumber, \\
\sigma_P & = \sqrt{\langle P^2 \rangle - \langle P \rangle^2} =  \sqrt{\frac{\hbar \omega}{2 }} ,
\label{coherentstateuncertainty}
\end{align}
which satisfies the relation (\ref{heisenberguncertainty}). 

The uncertainty relation as written in (\ref{heisenberguncertainty}) is in fact a special case of an inequality that can be derived for two arbitrary operators.  As we have seen in the previous sections, the main operators 
that we will be dealing with are spin operators which have a different commutation relation to the position and momentum operators. For two arbitrary operators $ A, B $ the {\it Schrodinger uncertainty relation} holds
\index{Schrodinger uncertainty relation}
\begin{align}
\sigma_A^2 \sigma_B^2 \ge \left| \frac{1}{2}\langle\{A,B\}\rangle-\langle A \rangle \langle B \rangle \right|^2+\left| \frac{1}{2i}\langle[A,B]\rangle \right|^2 .
\label{schrodingeruncertainty}
\end{align}
Dropping the first term on the right hand side of the inequality then coincides with the {\it Robertson uncertainty relation}. \index{Robertson uncertainty relation}  Evaluating the Schrodinger uncertainty relation for a coherent state\index{coherent state} again achieves a minimum uncertainty state where the left and right hand sides are equal.

\begin{framed}
{\centering \bf The Schrodinger uncertainty relation \\
}\index{Schrodinger uncertainty relation}
\bigskip

To prove the Schrodinger uncertainty relation, define the auxiliary states
\begin{align}
|f_A \rangle & = (A-\langle A \rangle) |\Psi\rangle , \nonumber \\
|f_B \rangle & = (B-\langle B \rangle) |\Psi\rangle ,
\end{align}
which then can be used to write the variances as
\begin{align}
\sigma_A^2 & =  \langle\Psi|(A-\langle A \rangle)^2|\Psi \rangle =\langle f_A| f_A \rangle \nonumber, \\
\sigma_B^2 & =  \langle\Psi|(B-\langle B \rangle)^2|\Psi \rangle =\langle f_B| f_B \rangle .
\label{sigmarelation}
\end{align}
Applying the Cauchy-Schwarz inequality, we have\index{Cauchy-Schwarz inequality}
 \begin{align}
\langle{f_A}|f_A\rangle \langle{f_B}|f_B\rangle \geq |\langle{f_A}|f_B\rangle|^2   .
 \end{align}
Using the fact that the modulus squared is the square of the real and imaginary parts of a complex number, we can write this in an equivalent form
\begin{align}
\sigma_A^2 \sigma_B^2 \geq \left( \frac{ \langle f_A | f_B \rangle + \langle f_B | f_A \rangle}{2} \right)^2 
+ \left( \frac{ \langle f_A | f_B \rangle - \langle f_B | f_A \rangle}{2i} \right)^2 ,
\end{align}
where we used (\ref{sigmarelation}).  Then using
\begin{align}
\langle f_A | f_B \rangle  = \langle A B \rangle - \langle A \rangle \langle B \rangle  ,
\end{align}
we obtain (\ref{schrodingeruncertainty}).  
\end{framed}

Since optical coherent states\index{coherent state!optical} are minimum uncertainty states, it is reasonable to expect that spin coherent states\index{coherent state!spin} are also minimum uncertainty states.  Let us directly evaluate the Schrodinger uncertainty relation\index{Schrodinger uncertainty relation} now to see whether this is so. For convenience, we take $ A = S_z $ and $ B = S_x $ and evaluate all the terms in (\ref{schrodingeruncertainty}) for the spin coherent state (\ref{scs}).  

On the left hand side we have the variances of the spin operators which can be evaluated to be
\begin{align}
\sigma_{S_z}^2 & = \langle S_z^2 \rangle - \langle S_z \rangle^2 \nonumber \\
& = 4 N |\alpha \beta |^2  = N \sin^2 \theta,
\label{szvariance}
\end{align}
where we used the parametrization (\ref{blochpara}).  For the other spin operators the variances are
\begin{align}
\sigma_{S_x}^2 & = \langle S_x^2 \rangle - \langle S_x \rangle^2  \nonumber \\
& = N ( 1- (\alpha \beta^* + \beta \alpha^*)^2) = N ( 1- \cos^2 \phi \sin^2 \theta) \nonumber, \\
\sigma_{S_y}^2 & =  \langle S_y^2 \rangle - \langle S_y \rangle^2  \nonumber \\
& =  N ( 1- (i\alpha \beta^* -i \beta \alpha^*)^2) = N ( 1- \sin^2 \phi \sin^2 \theta) .
\label{sxyvariance}
\end{align}
The above clearly shows that the sum of the variances evaluate to 
\begin{align}
\sigma_{S_x}^2 + \sigma_{S_y}^2 + \sigma_{S_z}^2 = 2N, 
\end{align}
which can equally be seen by combining (\ref{sphereequation}) and (\ref{totalspin}). 

On the right hand side, the anticommutator can be written
\begin{align}
\langle \{ S_z, S_x \} \rangle = 2 \langle S_x S_z \rangle + 2 i \langle S_y \rangle .
\end{align}
Evaluating the first correlation we obtain (see problem \ref{q5-9} for a workthrough)
\begin{align}
\langle S_x S_z \rangle = -i \langle S_y \rangle + \frac{N-1}{N} \langle S_x \rangle \langle S_z \rangle .  
\label{sxszcorrelation}
\end{align}
For the commutator, we can evaluate 
\begin{align}
\langle [S_x, S_z] \rangle = 2i \langle S_y \rangle.
\end{align}
Putting all this together we find that the left and right sides are equal as expected, yielding
\begin{align}
\sigma_{S_z}^2 \sigma_{S_x}^2 & = \left| \frac{1}{2}\langle\{S_z,S_x\}\rangle-\langle S_z \rangle \langle S_x \rangle \right|^2+\left| \frac{1}{2i}\langle[S_z,S_x]\rangle \right|^2  \nonumber \\
& = N^2 \sin^2 \theta  (1  - \cos^2 \phi \sin^2 \theta) .
\end{align}
Thus all spin coherent states are minimum uncertainty states in the same way as optical coherent states.  \index{coherent state!optical} \index{coherent state!spin}

\begin{exerciselist}[Exercise]
\item \label{q5-7}
Evaluate the Schrodinger uncertainty relation\index{Schrodinger uncertainty relation} for an optical coherent state (\ref{coherentfock}) and the position and momentum operators (\ref{positionmomentumdim}). 
\item \label{q5-8}
(a) Find the variance of $ S_z $ giving the result (\ref{szvariance}).  (b) Find the variance of $S_x $ to give the result (\ref{sxyvariance}).  Hint: The simplest way to do this is by reusing the result that you found in (a).  Perform a basis transformation on the spin coherent state to make $S_x$ take the form of $ S_z $ using
\begin{align}
a_x & = \frac{1}{\sqrt{2}}(a+b), \nonumber \\
b_x  & = \frac{1}{\sqrt{2}}(a-b), 
\label{axtransform}
\end{align}
Put this in the definition of the spin coherent state\index{coherent state!spin} (\ref{scs}) to transform it, and then use the new coefficients in the formula (\ref{szvariance}). 
\item \label{q5-9}
Verify (\ref{sxszcorrelation}).  Hint: First evaluate the commutator relations\index{commutation relations}
\begin{align}
[S_z, \alpha a^\dagger + \beta b^\dagger] \nonumber \\
[S_x, \alpha a^\dagger + \beta b^\dagger]  .
\end{align}
Use this result to commute first $S_z$ to the right side of the expectation value, then do the same for $ S_x $.  Then simplify the expression using the fact that 
\begin{align}
[\alpha^* a + \beta^* b, \alpha a^\dagger + \beta b^\dagger] = 1
\end{align}
Finally use the expressions (\ref{expectationvaluesx}) and (\ref{expectationvaluesyz}) to write the correlator in the form (\ref{sxszcorrelation}). 
\end{exerciselist}

\section{Squeezed states}
\label{sec:squeezedstates}

\subsection{Squeezed optical states}
\label{sec:squeezedoptical}\index{squeezed state}
\index{squeezed state!optical}

In the previous sections we saw that spin coherent states are minimum uncertainty states and play an analogous role to optical coherent states.  Optical coherent states have an uncertainty that is equal in both position and momentum operators.  In terms of dimensionless variables we have\index{coherent state!spin} \index{coherent state!optical}
\begin{align}
\sigma_x = \sigma_p = \frac{1}{\sqrt{2}},
\end{align}
where we used (\ref{coherentstateuncertainty}) and (\ref{positionmomentum}).  Another type of state which is important in the context of quantum optics are squeezed states, where uncertainties in one direction are suppressed at the expense of the other.  The optical squeezed vacuum state is \index{squeezed state!optical squeezed vacuum} 
\begin{align}
|r e^{i \Theta} \rangle =  \exp \left[ - \frac{r e^{-i \Theta}}{2} a^2 + \frac{r e^{i \Theta}}{2}  (a^\dagger)^2 \right] | 0 \rangle, 
\end{align}
where $ r $ is the squeezing parameter and $  \Theta $ specifies the direction of the squeezing. This has uncertainties for $ \Theta = 0 $
\begin{align}
\sigma_x & = \frac{e^{r} }{\sqrt{2}} \nonumber, \\
\sigma_p & = \frac{e^{-r} }{\sqrt{2}},
\end{align}
which is a squeezed state for the momentum. This is also a minimum uncertainty state as the Heisenberg uncertainty relation is satisfied as an equality $ \sigma_x \sigma_p = 1/2 $.  A different choice of $ \Theta $ would produce squeezing in a different direction. In this section we will explore two types of squeezed states for spin ensembles.  \index{squeezed state} \index{ensembles}

\subsection{One-axis twisting squeezed states}
\label{sec:oneaxistwisting}
\index{one-axis one-spin squeezed states}
\index{squeezed state}

A similar type of squeezed state for a spin coherent state can be produced by the {\it one-axis twisting} Hamiltonian\index{squeezed state} \index{coherent state!spin} \index{one-axis twisting Hamiltonian}
\begin{align}
H_{\text{1A1S}} = \hbar \kappa (S_z)^2 ,
\label{squeezingham}
\end{align}
where $ \kappa $ is the interaction constant. Here 
1A1S stands for ``one-axis one-spin'', in anticipation of two-spin generalizations that we will encounter in Chapter \ref{ch:entanglement}. This type of Hamiltonian arises from a diagonal interaction between the bosons.  This corresponds to a particular form of the interaction Hamiltonian (\ref{interactionham}).  For example, an interaction of the form
\begin{align}
{\cal H}_I = \frac{g_{aa}}{2} n_a^2  + g_{ab} n_a n_b + \frac{g_{bb}}{2} n_b^2,
\end{align}
contains the interaction (\ref{squeezingham}), where $ g_{aa},g_{bb} $ is the intraspecies interaction and $ g_{ab} $ is the interspecies interaction.  To see this explicitly, we may use the relations
\begin{align}
n_a & = a^\dagger a = \frac{{\cal N}+S_z}{2}, \nonumber \\
n_b & = b^\dagger b = \frac{ {\cal N}-S_z}{2},
\end{align}
to obtain
\begin{align}
{\cal H}_I =  \frac{{\cal N}^2}{8} ( g_{aa} + g_{bb} + 2 g_{ab}) + \frac{\cal N}{4}(g_{aa} -g_{bb}) S_z + \frac{1}{8}(g_{aa} - 2 g_{ab} +  g_{bb}) (S_z)^2  .
\end{align}
The first term contributes to a physically irrelevant global phase, and the second produces a $ S_z $ rotation if $ g_{aa} \ne g_{bb} $, and the third term produces squeezing.    

Let us now see what effect of the squeezing is on a spin coherent state.  From (\ref{szvariance}) and (\ref{sxyvariance}) we can see that the variance depends upon the location of the state on the Bloch sphere.  Let us choose an initial spin coherent state with parameters $ \theta = \pi/2, \phi = 0 $, which corresponds to spins fully polarized in the $ S_x $ direction.  Applying the squeezing operation we obtain \index{coherent state!spin} \index{Bloch sphere}
\begin{align}
e^{-i (S_z)^2 \tau } | \frac{1}{\sqrt{2}}, \frac{1}{\sqrt{2}} \rangle \rangle = 
\frac{1}{\sqrt{2^N}} \sum_{k=0}^N \sqrt{N \choose k} e^{-i (2k-N)^2 \tau} |k \rangle
\label{squeezedszstate},
\end{align}
where $ \tau = \kappa t $ is the dimensionless squeezing parameter.  To see the effect of the squeezing we can calculate the variance of the spin in a generalized direction
\begin{align}
\sigma_{\bm{n}}^2 \equiv \langle ( \bm{n} \cdot \bm{S})^2 \rangle - \langle \bm{n} \cdot \bm{S} \rangle^2,
\end{align}
where
\begin{align}
\bm{n} = (\sin \theta \cos \phi, \sin \theta \sin \phi, \cos \theta) .
\end{align}
It will turn out that the largest variation in the variance occurs when $ \phi = \pi/2 $, which gives a spin in the direction
\begin{align}
\bm{n} \cdot \bm{S} = \sin \theta S_y +  \cos \theta S_z .
\end{align}
The reason for this is that the spin coherent state is originally polarized in the $S_x $ direction, hence for small $ \tau $ it remains an approximate eigenstate with small variance. For spins in the $ S_y$-$S_z$ plane the variance is
\begin{align}
\sigma_\theta^2 = \sin^2 \theta \sigma_{S_y}^2 + \cos^2 \theta \sigma_{S_z}^2 + \sin \theta \cos \theta ( \langle S_y S_z + S_z S_y \rangle - 2 \langle S_y \rangle \langle S_z \rangle  ) .
\label{sigmatheta}
\end{align}
The various quantities can be evaluated as
\begin{align}
\langle S_y^2 \rangle & = \frac{N(N+1)}{2} - \frac{N(N-1)}{2} \cos^{N-2} 8 \tau \cos 16 \tau,  \nonumber\\
\langle S_y \rangle  & = N \cos^{N-1} 4 \tau  \sin 4 \tau,  \nonumber\\
\langle S_z^2 \rangle & = N,  \nonumber\\
\langle S_z \rangle & = 0,  \nonumber\\
 \langle S_y S_z + S_z S_y \rangle & = N(N+2) \cos^{N-1} 4 \tau \sin 8 \tau .
\label{sigmathetaexpressions}
\end{align}

\begin{figure}[t]
\centerline{\includegraphics[width=\textwidth]{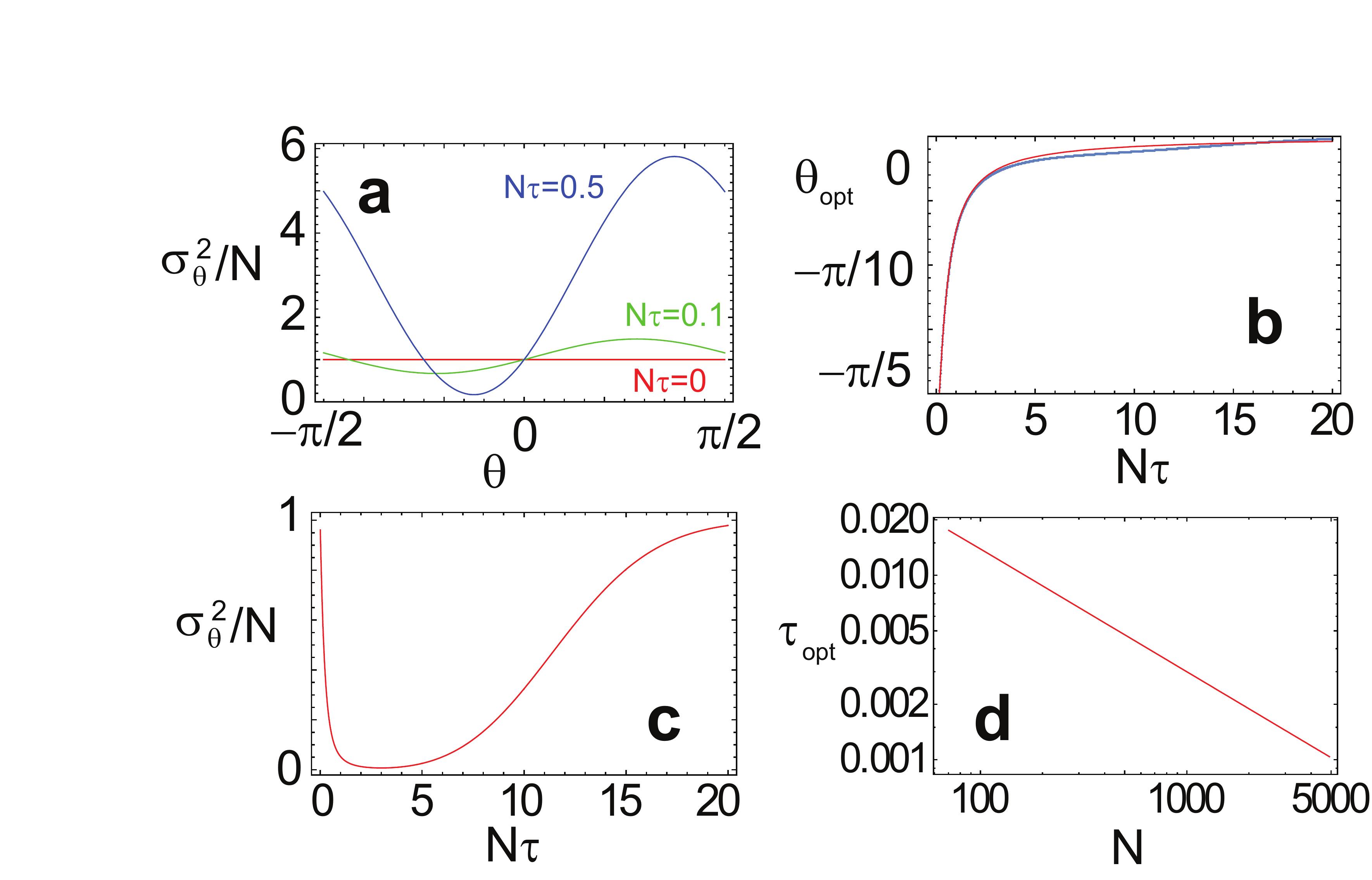}}
\caption{(a) The variance of the one-axis squeezed state (\ref{squeezedszstate}) given by the expression (\ref{sigmatheta}). (b) The angle for which the minimum uncertainty occurs.  The points are an exact minimization of (\ref{sigmatheta}) and the line is the approximate expression (\ref{minimumangle}). (c)  Log-log plot of variance of the squeezed state (\ref{squeezedszstate}) for the optimum angle $ \theta_{\text{opt}} $ that gives the minimum variance.  (d) The time $ \tau_{\text{opt}} $ that gives the minimum squeezing for a given $ N $ plotted on a log-log scale.  $N = 1000 $ is used for all calculations.  }
\label{fig5-3}
\end{figure}

Putting these together, we obtain a dependence that is plotted in Fig. \ref{fig5-3}(a). For $ \tau = 0 $ (i.e. no squeezing) there is no dependence on the angle and we have
\begin{align}
\sigma_\theta^2 = \sigma_{S_y}^2 = \sigma_{S_z}^2 = N  .
\label{variancescs}
\end{align}
This is again analogous to an unsqueezed optical coherent state where the variance is the same in all directions.  As the squeezing is increased $ \tau > 0 $, the variance depends upon the spin direction, and can become smaller than (\ref{variancescs}).  
The angle at which the variance is minimized can be found by solving for $ \frac{d \sigma_\theta}{d \theta} = 0 $.  To obtain an approximate expression for the minimum variance, we can first expand $ \frac{d \sigma_\theta}{d \theta} $  to second order in $ \tau $, which yields
\begin{align}
\theta_{\text{opt}} \approx - \frac{1}{2} \tan^{-1} ( \frac{1}{2 N \tau})
\label{minimumangle},
\end{align}
where we assumed $ N \gg 1 $.  Fig. \ref{fig5-3}(b) compares the exact minimum angle with the above formula.  We see that the formula agrees extremely well for $ N \gg 1 $.  For very small $ \tau \ll 1 $ the angle starts at $ \theta_{\text{opt}} = \pi/4 $, and approaches $ \theta_{\text{opt}} = 0 $ for large $ \tau $.  

In Fig. \ref{fig5-3}(c) we show the variance (\ref{sigmatheta}) at the minimum angle $ \theta_{\text{opt}} $. For example, for $ N =1000 $ atoms the minimum value of squeezing achieved in this case is $ \sigma_\theta^2 \approx 7 $, much smaller than the unsqueezed value. The optimal squeezing time can be found for the curve Fig. \ref{fig5-3}(c), and the results are plotted in Fig. \ref{fig5-3}(d) for various $ N $.  The log-log plot reveals a power law behavior to the optimal time, which depends as
\begin{align}
\tau_{\text{opt}} \approx \frac{0.3}{N^{2/3}} .
\end{align}
Kitagawa and Ueda (\citeyear{kitagawa1993squeezed}) found that the optimal squeezing that can be attained scales as
\begin{align}
\frac{\sigma_\theta}{N} \propto  \frac{1}{N^{2/3}} . \hspace{1cm} \text{(1 axis twisting)}
\label{oneaxislimit}
\end{align}

We can now compare the variances with and without squeezing. For the original unsqueezed state, we can see from (\ref{variancescs}) that the standard deviation scales as 
\begin{align}
\frac{\sigma_\theta}{N} \propto \frac{1}{\sqrt{N}}, \hspace{1cm} \text{(standard quantum limit)}
\label{sqllimit}
\end{align}
which is a typical result that is characteristic of systems where the noise is dominated by quantum mechanics (as opposed to technical noise in the experiment), and is called the {\it standard quantum limit}. \index{standard quantum limit} 
Since (\ref{oneaxislimit}) decreases with $ N $ faster than (\ref{sqllimit}), this is an indication of the squeezing effect of the one-axis twisting Hamiltonian.\index{one-axis twisting Hamiltonian}

\begin{exerciselist}[Exercise]
\item \label{q5-10}
Evaluate the expression (\ref{sigmatheta}) including the expectation values (\ref{sigmathetaexpressions}).  Hint: For these expectation values try evaluating them by using the expanded form (\ref{squeezedszstate}).  For the $ S_z $ expectation values you will see that the time dependence immediately disappears, hence you can use the results of the previous section.  For the expectation values involving $ S_y $ you should still have the time dependence.  Use the fact that 
\begin{align}
S_y | k \rangle = -i \sqrt{(k+1)(N-k)} | k + 1 \rangle + i \sqrt{k(N-k+1)} | k - 1 \rangle
\end{align}
and 
\begin{align}
\sqrt{{N \choose k }} = \sqrt{\frac{N!}{k!(N-k)!}}
\end{align}
to simplify the square root factors.    
\end{exerciselist}

\subsection{Two-axis countertwisting squeezed states}
\label{sec:twoaxiscounter}
\index{two-axis one-spin squeezed states}
\index{squeezed state}

We have seen that the noise fluctuations in the one-axis twisting Hamiltonian has a scaling with $ N $ that is less than the standard quantum limit, and is hence a demonstration of squeezing. 
There are however other types of Hamiltonians that can produce a higher level of squeezing.  In many systems it has been observed that the limits of squeezing that can be obtained scales as  $ \sigma_\theta/N  \propto 1/N $ and is commonly called the {\it Heisenberg limit}.  This is the typical best scaling of a quantum mechanical system in a single collective state such that noise fluctuations are further suppressed.  \index{Heisenberg limit}  In this section we introduce the {\it two-axis countertwisting} squeezed state, which attains this type of scaling.  \index{squeezed state!two-axis countertwisting }

The  two-axis countertwisting Hamiltonian is given by
\begin{align}
H_{\text{2A1S}} = \frac{\hbar \kappa}{2} [ (S_x)^2 - (S_y)^2 ] = \hbar \kappa [ (S_+)^2 + (S_-)^2 ],
\label{tacsqueezingham}
\end{align}
where $ \kappa $ is the interaction constant. Here 2A1S stands for ``two-axis one-spin'' in anticipation of two-spin generalizations we will encounter in Chapter \ref{ch:entanglement}. Realizing the two-axis countertwisting Hamiltonian experimentally is not quite as straightforward as the one-axis version.  One way that it can be produced is by taking advantage of the Suzuki-Trotter expansion\index{Suzuki-Trotter expansion}
\begin{align}
e^{(A+B)} = \lim_{n \rightarrow \infty} \left( e^{\frac{A}{n}} e^{\frac{B}{n}} \right)^n .
\end{align}
If one wishes to perform a time evolution with the Hamiltonian (\ref{tacsqueezingham}), then we can perform a decomposition 
\begin{align}
e^{-i H_{\text{2A1S}} t /\hbar } \approx \left( e^{-i (S_x)^2  \tau/2n } e^{i (S_y)^2 \tau/2n } \right)^n ,
\end{align}
where $ \tau =  \kappa t $ again. The above terms can be realized using a combination of one-axis squeezing\index{one-axis twisting Hamiltonian} terms and rotations according to \cite{liu2011spin}
\begin{align}
e^{-i (S_x)^2 \tau } & = e^{- i S_y \pi/4} e^{-i (S_z)^2 \tau } e^{ i S_y \pi/4} \nonumber \\
e^{i (S_y)^2 \tau } & = e^{- i S_x \pi/4} e^{-i (S_z)^2 \tau } e^{ i S_x \pi/4} .
\end{align}

Let us now see the effect of the two-axis countertwisting Hamiltonian\index{two-axis countertwisting Hamiltonian} on a spin coherent state, which we choose to be the state at $ \theta = 0 $ this time.  We define the state\index{coherent state!spin}
\begin{align}
|\psi_{\text{2A1S}} (\tau) \rangle = e^{-i [ (S_x)^2 - (S_y)^2 ] \tau } | 1, 0 \rangle \rangle  .
\label{tacstate}
\end{align}
Unlike the one-axis twisting Hamiltonian for which we can easily write down the exact wavefunction at any time, the time evolution for the two-axis countertwisting must in general be evaluated numerically.  One nice feature of the two-axis countertwisting state is that the direction of the squeezing and anti-squeezing occurs for the variables 
\begin{align}
\tilde{S}^x & = \frac{S^x + S^y}{\sqrt{2}} \nonumber \\
\tilde{S}^y & = \frac{S^y - S^x}{\sqrt{2}} 
\label{tildevariables}
\end{align}
respectively, hence no optimization needs to be performed as was required for the one-axis case. 

\begin{figure}[t]
\centerline{\includegraphics[width=\textwidth]{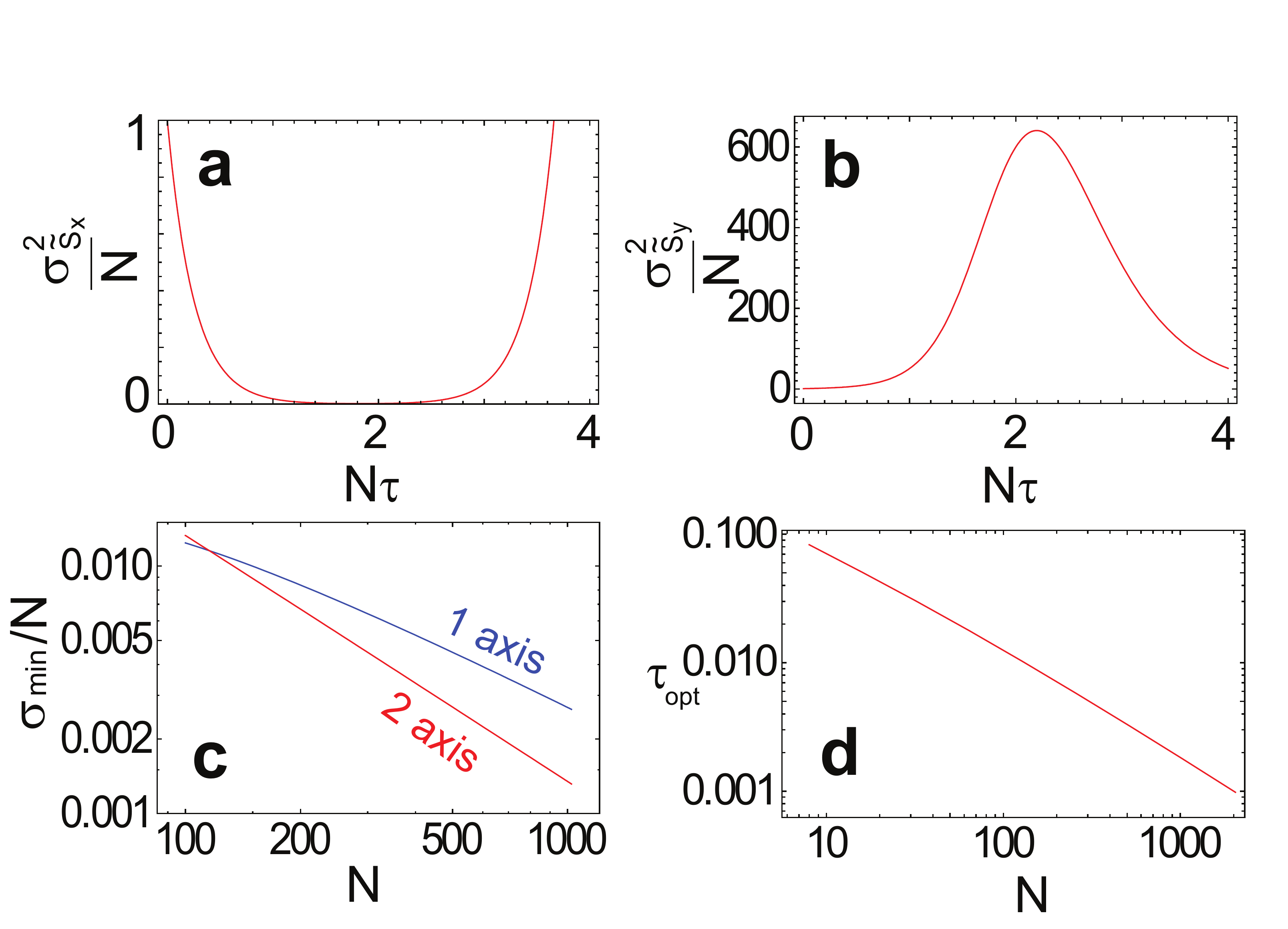}}
\caption{The variance of the two-axis countertwisting squeezed state (\ref{tacstate}) for the spin (a)  $ \tilde{S}^x $ and (b)  $ \tilde{S}^y $ as defined in (\ref{tildevariables}) for $N = 1000 $. (c) The minimum standard deviations for the one-axis and two-axis squeezed states. (d) The time $ \tau_{\text{opt}} $ that gives the minimum squeezing for a given $ N $ plotted on a log-log scale. }\index{squeezed state!two-axis countertwisting }
\label{fig5-4}
\end{figure}

Figure \ref{fig5-4}(a) shows the variance of the operator $ \tilde{S}^x $.  Comparing to the one-axis twisting squeezed states\index{squeezed state!one-axis twisting} (Fig. \ref{fig5-3}(c)), we see that the standard deviation reaches a smaller value for the same $ N = 1000 $, giving a value $ \sigma_{\tilde{S}^x } \approx 2 $. Figure \ref{fig5-4}(a) shows the variance of the operator $\tilde{S}^y  $ in the orthogonal direction, which displays antisqueezing.  In Fig. \ref{fig5-4}(c) we show the scaling of standard deviations for the one-axis and two-axis squeezed states.  The two-axis squeezed states indeed have a scaling which is close to the Heisenberg scaling of  $ \sigma_{\tilde{S}^x }/N \propto 1/N  $.  We finally show the optimal times to reach the minimum squeezing in \ref{fig5-4}(d) for the two-axis countertwisting Hamiltonian, which scales as $ \tau_{\text{opt}} \propto 1/N $.  \index{two-axis countertwisting Hamiltonian}

\section{Entanglement in spin ensembles}
\label{sec:entanglementsqueezing}\index{ensembles}

We saw in the previous section that it is possible to reduce the variance of spin operators such that they are below the value in a spin coherent state. The reason for this can be understood as follows.  Initially the state is prepared in a spin coherent state\index{coherent state!spin} (\ref{scs}), which can be viewed as an ensemble of non-interacting spins.  When the squeezing operation is introduced, the spins interact, and a collective quantum state is formed.  Since the state is now a single, non-separable quantum object, suitable observables can take a reduced value since it is a single quantum system.  
This means that due to the interactions between the bosons, we might expect that there is entanglement between the bosons.  Several ways of detecting this have been proposed for spin ensemble states.  In this section we list two entanglement witnesses that serve this purpose.

\subsection{Spin squeezing criterion}
\label{sec:wineland}

To quantify the degree of squeezing, it is useful to introduce a parameter
\begin{align}
\xi^2 = \frac{N \sigma_{\theta_{\min}}^2}{\langle S_x \rangle^2} ,
\end{align}
where $ \sigma_{\theta_{\min}}^2 $ is the variance as defined in (\ref{sigmatheta}) along the minimal angular direction.  For zero squeezing $ \tau = 0 $, the variance is $ \sigma_{\theta_{\min}}^2 = N $ as given in (\ref{variancescs}), and the initial state is polarized in the $ S_x $ direction, such that $ \langle S_x \rangle = N $. The unsqueezed initial state thus takes a value $ \xi = 1 $.  

S{\o}rensen, Duan, Cirac, and Zoller found that this squeezing parameter can be used to signal entanglement.  The criterion takes the simple form
\begin{align}
\xi^2 < 1 \hspace{1cm} \text{(for entangled states)} .
\label{spinsqueezingcrit}
\end{align}
Thus any squeezed state also signals the presence of entanglement. If the criterion (\ref{spinsqueezingcrit}) is satisfied, it guarantees that the state is entangled.  However, there are some entangled states which have $ \xi^2 \ge 1 $, hence this is a {\it sufficient} criterion for entanglement, but not {\it necessary}.

\subsection{Optimal spin squeezing inequalities}
\label{sec:toth}

A more sensitive entanglement criterion than the spin squeezing criterion was developed by Toth, Knapp, G{\"u}hne, and Briegel, only involving the first and second moments of the spin operators.  Violation of any of the following inequalities implies entanglement:
\begin{align}
\langle S_x^2 \rangle + \langle S_y^2 \rangle + \langle S_z^2 \rangle &  \le N(N+2) & \hspace{1cm} \text{(for separable states)}, \label{toth1} \\
\sigma_x^2 + \sigma_y^2  + \sigma_z^2  & \ge 2N & \hspace{1cm} \text{(for separable states)}, \label{toth2} \\
\langle S_i^2 \rangle + \langle S_j^2 \rangle - 2N &  \le (N-1) \sigma_k^2 & \hspace{1cm} \text{(for separable states)}, \label{toth3}  \\
(N-1)( \sigma_i^2 + \sigma_j^2 ) &  \ge \langle S_k^2 \rangle + N(N-2) & \hspace{1cm} \text{(for separable states)}, \label{toth4} 
\end{align}
where $ i, j, k $ take all the possible permutations of $ x, y, z $.  These criteria can detect a large range of entangled states, such as Fock states, and ground states of an anti-ferromagnetic Heisenberg chain. \index{Fock states}

\begin{exerciselist}[Exercise]
\item \label{q5-11}
Evaluate the criteria (\ref{toth1})-(\ref{toth4}) for a Fock state (\ref{fockstates}) with $ k = N/2 $.  Is there any difference if $ k = 0 $ or $ k = N $? Why?
\end{exerciselist}

\section{The Holstein-Primakoff transformation}
\label{sec:holstein}

The spin operators $ S_{x,y,z} $ involve two bosonic operators $ a,b $ in the Schwinger boson representation.  As we saw in\index{Schwinger boson operators} (\ref{spinnumbercommutation}) the spin operators $ S_{x,y,z} $ conserve the total boson number.  Suppose that we restrict the total number of bosons to $ N $, such that we consider the space of states
\begin{align}
{\cal N } | \psi \rangle  = N | \psi \rangle
\end{align}
where the wavefunction takes the form (\ref{genearlspin}).  In this case we may write
\begin{align}
b^\dagger b  = N - a^\dagger a  .
\label{bequation}
\end{align}
This suggests that we may try and eliminate the $ b $ operators and only deal with the $ a $ operators alone.  

For the $ S_z $ operator this is straightforwardly done by substituting (\ref{bequation}) into the definition (\ref{schwingerboson})
\begin{align}
S_z = 2 a^\dagger a - N .
\label{holsteinz}
\end{align}
Defining the Fock states with only the $ a $ operators\index{Fock states}
\begin{align}
| k \rangle = \frac{ (a^\dagger)^{k} }{\sqrt{k!}} | 0 \rangle ,
\label{fockstatesphoton}
\end{align}
we observe that we obtain the same relation as (\ref{fockstateop}).  

To find a relation for the remaining off-diagonal operators, it is easiest to look at the raising and lowering operators as defined in (\ref{raisinglowering}).  Looking at the action of these on the Fock states (\ref{ladderfock}), we can see that $ S_{\pm} $ has the effect of shifting the number of $ a $ bosons, hence should be similar to $ a $ and $ a^\dagger $ respectively. Applying these operators to (\ref{fockstatesphoton}) have similar bosonic factors of $ \sqrt{k} $ and $ \sqrt{k+1} $, except that there is another bosonic factor arising from the $ b $ operator.  To account for these we can add appropriate number operators\index{Fock states}
\begin{align}
S_+ & = a^\dagger \sqrt{N - a^\dagger a} \nonumber, \\
S_- & = \sqrt{N - a^\dagger a} a ,
\label{holstein}
\end{align}
which gives the same relations as (\ref{ladderfock}), with the use of the Fock states (\ref{fockstatesphoton}).  The relations for the other spin operators can be obtained simply by substitution into (\ref{inverseladder}).  \index{Holstein-Primakoff transformation} 

The relations (\ref{holstein}) and (\ref{holsteinz}) are together taken to be the {\it Holstein-Primakoff transformation}, which relates spin operators to bosonic operators.  The price to be paid for making the transformation is the non-linear square root factor, which gives an infinite series in terms its expansion as a Taylor series.  Keeping just the leading order gives the approximate relation\index{Holstein-Primakoff transformation}
\begin{align}
S_+ & \approx \sqrt{N}  a^\dagger \nonumber \\
S_- & \approx \sqrt{N} a .
\label{holsteinapprox2}
\end{align}
This approximation is often used in the context of large spins $ N \gg 1 $, where the leading order is often enough to capture the basic physics.

\begin{exerciselist}[Exercise]
\item \label{q5-12b}
Verify that the Holstein-Primakoff transformations (\ref{holstein}) acting on the Fock states  (\ref{fockstatesphoton}) gives the same relations as (\ref{ladderfock}).  \index{Fock states}
\end{exerciselist}

\section{Equivalence between bosons and spin ensembles}
\label{sec:equivalencebosonspin}\index{ensembles}

So far we have only considered BECs consisting of indistinguishable bosons that populate several levels.  In fact under certain conditions it is \index{indistinguishable particles}
possible to have ensembles of spins which are not indistinguishable to give the same types of states. For example, instead of a BEC where all the atoms occupy the same spatial quantum state, a hot gas of atoms where condensation has not occurred can be put into a spin coherent state\index{coherent state!spin} (\ref{scs}) or a squeezed state \index{squeezed state} (\ref{squeezedszstate}). \index{distinguishable particles} \index{ensembles}

How is this possible? After all the dimension of the Hilbert space for indistinguishable and distinguishable particles do not even match. Recall from Fig. \ref{fig2-1} that for $ N $ two-level indistinguishable bosons the dimension of the Hilbert space is $ N+1 $, but for $N $ two-level  distinguishable particles the dimension is $ 2^N $.  But as we discussed in Sec. \ref{sec:einstein}, there are the set of states for the spin ensemble which have analogous properties to the indistinguishable bosons.  Specifically, the states (\ref{generalsymmetric}), which we repeat here
\begin{align}
|k\rangle^{\text{(ens)}} = \sum_{d_k=1}^{\binom{N}{k}} |k, d_k \rangle
\label{symmetricfock},
\end{align}
are the analogue of the Fock states for the indistinguishable bosons (\ref{fockstates}).  These states are symmetric under particle interchange, which is a general property of any state of a BEC.  Then the analogue of the general state (\ref{genearlspin}) is then 
\begin{align}
| \psi \rangle^{\text{(ens)}} = \sum_{k=0}^N \psi_k | k \rangle^{\text{(ens)}} .
\label{spingeneralstate}
\end{align}
The remaining states (i.e. $ 2^N - (N + 1) $ of them) are then unpopulated.

How can we ensure that the remaining unsymmetric states are unpopulated?  This can be guaranteed if the initial state and the Hamiltonian used to prepare the given state are both symmetric.  Suppose we are following the procedure to prepare the spin coherent state as in Sec. \ref{sec:prepscs}.  There we first prepared the state in $ | k = N \rangle $, which for a spin ensemble corresponds to 
\begin{align}
|k=N\rangle^{\text{(ens)}} = \prod_{j=1}^N | a \rangle_j = | a a \dots a \rangle .
\label{extremalstate}
\end{align}
We have taken the state $ | 1 \rangle \rightarrow | a \rangle $ and $ | 0 \rangle \rightarrow | b \rangle $ in the language of Sec. \ref{sec:einstein}.  This is obviously symmetric under particle interchange.  Next we apply symmetric total spin operators for the ensembles
\begin{align}
S^{\text{(ens)}}_x & = \sum_{j=1}^N \sigma_x^{(j)} \nonumber \\
S^{\text{(ens)}}_y & = \sum_{j=1}^N \sigma_y^{(j)} \nonumber \\
S^{\text{(ens)}}_z & = \sum_{j=1}^N \sigma_z^{(j)} ,
\label{ensemblespin}
\end{align}
where the spin operator for the $ j$th spin is defined in Eq. (\ref{paulioperators}).  Applying the $ S^{\text{(ens)}}_x $ operation in this case is simple because all the Pauli operators\index{Pauli operators} in (\ref{ensemblespin}) commute.  The resulting state is
\begin{align}
e^{-i S^{\text{(ens)}}_x \Omega t/2} |k=N\rangle^{\text{(ens)}} & = \prod_{j=1}^N  e^{-i \sigma_x^{(j)} \Omega t/2} | a \rangle_j \nonumber \\ 
& = \prod_{j=1}^N \left[ \cos (\Omega t/2) | a \rangle_j - i  \sin (\Omega t/2) | b \rangle_j  \right],
\label{rotatedensemble}
\end{align}
where we used (\ref{exponentialsigmax}).  This state is symmetric under particle interchange since all the spins are in the same state.  The general spin coherent state for ensembles is thus written\index{coherent state!spin} \index{ensemble}
\begin{align}
| \alpha, \beta \rangle \rangle^{\text{(ens)}}  =  \prod_{j=1}^N \left[ \alpha | a \rangle_j +  \beta  | b \rangle_j  \right] .
\label{scsensemble}
\end{align}
The equivalence between the total spin operators (\ref{ensemblespin}) and the Schwinger boson operators (\ref{schwingerboson}) is the result of the {\it Jordan map}, which historically was used by Schwinger to derive the theory of quantum angular momentum in a simpler way. \index{Jordan map}  \index{Schwinger boson operators} 

The example above can be generalized to any operation involving symmetric Hamiltonians.  As long as the initial state is symmetric (typically a state such as (\ref{extremalstate})) and the applied Hamiltonians are symmetric such as those involving (\ref{ensemblespin}), the final state is also symmetric and can be written in the form (\ref{genearlspin}).

We note that the symmetric Fock states (\ref{symmetricfock}) are eigenstates of the total spin operator\index{Fock states}
\begin{align}
(\bm{S}^{\text{(ens)}})^2 |k\rangle^{\text{(ens)}} = N(N+2) |k\rangle^{\text{(ens)}}
\label{totalspin2}
\end{align}
where 
\begin{align}
(\bm{S}^{\text{(ens)}})^2 = (S^{\text{(ens)}}_x)^2 + (S^{\text{(ens)}}_y)^2 + (S^{\text{(ens)}}_z)^2 .
\end{align}
This is the equivalent result of (\ref{totalspin}) for the spin ensemble. This has the interpretation that the symmetric Fock states are maximum spin eigenstates for $ N $ spins.  The remaining $ 2^N - (N+1) $ states will have a spin magnitude that is smaller than this value.

\begin{exerciselist}[Exercise]
\item \label{q5-13}
Show that the state (\ref{scsensemble}) can be expanded into the form (\ref{genearlspin}) where the Fock states only involve the symmetric Fock states (\ref{symmetricfock}).  
\item \label{q5-14}
Verify  (\ref{totalspin2}) for a symmetric Fock state (\ref{symmetricfock}).  
\end{exerciselist}

\section{Quasiprobability distributions}
\label{sec:quasiprobability}
\index{quasiprobability distribution}

The order parameter and Gross-Pitaevskii equation gives an effective way to understand the spatial degrees of freedom of a Bose-Einstein condensate. The usefulness of this approach is that it makes an effective single particle representation which is easily visualized.  In fact the Bose-Einstein condensate has an extremely complicated many-body wavefunction as we explored in Chapter \ref{ch:quantum}, so the order parameter is a great simplification to the full wavefunction.  However, due to the weakly interacting nature of a typical BEC, it often contains much of the important information of interest.    In the same way, the spin degrees of freedom alone can give a complex wavefunction, which is not always easily graspable.  Methods to visualize the wavefunction are useful to understand in a simple way the nature of the quantum state. 

In quantum optics, phase space representations are an indispensable tool to visualize quantum states of light.  The reason is that the quantum state of a single mode of light has an infinite Hilbert space, and hence potentially an infinite number of parameters to specify it.  Quasiprobability distributions such as the $Q$- and Wigner functions allow for a simple plot that can be interpreted in terms of the amplitude and phase of the light.  For spins, we have a similar situation.  Typically the number of spins $ N $ is large, and the full quantum state is then specified within a large Hilbert space.  What is desirable is to again convert an arbitrary quantum state to a distribution specified by phase space. For the spin case, the natural phase space is the Bloch sphere, as already discussed in Fig. \ref{fig5-2}. Thus in contrast to optical quantum states which have a distribution on a two-dimensional plane, spin states are plotted on a sphere.  We can then define the spin versions of the $Q$-, and Wigner functions.\index{Q-function} \index{Wigner function} \index{Bloch sphere} 

In all the quasiprobability distributions that we discuss here, we only discuss the visualization of a general bosonic state (\ref{genearlspin}) and the equivalent general state with spin ensembles (\ref{spingeneralstate}). For the bosonic states, the state (\ref{genearlspin}) is completely general, but (\ref{spingeneralstate}) is a subclass of a more general state within the full $ 2^N $ Hilbert space.  We also note that we do not discuss the $ P $-function because it is more difficult than the analogous optical case to give a simple and consistent way of calculating it. For this reason the  $Q$- and Wigner functions are more commonly used, and for most applications these are sufficient to visualize the state.  \index{P-function}

\subsection{$Q$-function}

The $Q$-function is defined in terms of the overlap of the state in question with a spin coherent state:\index{Q-function} \index{coherent state!spin}
\begin{align}
Q(\theta, \phi ) = \frac{N+1}{4\pi} \langle \langle \cos \frac{\theta}{2} e^{-i \phi/2}, 
\sin \frac{\theta}{2} e^{i \phi/2} | \rho | \cos \frac{\theta}{2} e^{-i \phi/2}, 
\sin \frac{\theta}{2} e^{i \phi/2} \rangle \rangle ,
\label{qfuncdefinition}
\end{align}
where the factor of $ \frac{N+1}{4\pi} $ is needed for the normalization. We note that the same normalization factor appears in the completeness relation for spin coherent states
\begin{align}
\frac{N+1}{4\pi} \int \sin \theta d \theta d \phi | \cos \frac{\theta}{2} e^{-i \phi/2}, \sin \frac{\theta}{2} e^{i \phi/2} \rangle \rangle \langle \langle \cos \frac{\theta}{2} e^{-i \phi/2}, \sin \frac{\theta}{2} e^{i \phi/2}  | = I ,
\end{align}
where $ I $ is the identity operator.  We then have for any state 
\begin{align}
\int Q(\theta, \phi ) \sin \theta d\theta d\phi = 1 .
\end{align}
The $Q$-function can also be interpreted as the probability of making a measurement in the spin coherent state basis, where the outcome is the spin coherent state with parameters $ \theta, \phi $.

\subsection{Wigner function}
\label{sec:wignerreps}

The Wigner function is defined as\index{Wigner function}
\begin{align}
W(\theta, \phi) = \sum_{l=0}^{2j} \sum_{m=-l}^{l} \rho_{lm} Y_{lm} (\theta, \phi), \label{wigner}
\end{align}
where $Y_{lm} (\theta, \phi) $ are the spherical harmonic functions.  Here, $\rho_{lm}$ is defined as 
\begin{align}
\rho_{lm}= \sum_{m_1=-j}^{j} \sum_{m_2=-j}^{j} (-1)^{j-m_1-m} 
\langle j m_1; j -m_2 | l m \rangle \langle j m_1 | \rho | j m_2 \rangle,
\end{align}
where $ \langle j_1 m_1; j_2 m_2 | J M\rangle  $ is a Clebsch-Gordan coefficient\index{Clebsch-Gordan coefficient} for combining two angular momentum eigenstates $ |j_1 m_1 \rangle $ and $ |j_2 m_2 \rangle $ to $ | J M \rangle $.  The angular momentum eigenstates are equivalent to the Fock states  (\ref{fockstates}), which are also eigenstates of the $ S_z $ operator.   We temporarily use a different notation for these states\index{Fock states}
since this is more natural in terms of the above formula:
\begin{align}
| j m \rangle = | k = j + m \rangle = \frac{(a^\dagger)^{j+m} (b^\dagger)^{j-m}}{\sqrt{(j+m)!(j-m)!}} | 0 \rangle ,
\label{fockdickeconversion}
\end{align}
which are also called Dicke states. \index{Dicke states}
In the above, 
\begin{align}
j = \frac{N}{2}
\end{align}
throughout. Since the Clebsch-Gordan coefficients are zero unless $ m = m_1 - m_2  $, one of the summations may be removed to give the formula\index{Clebsch-Gordan coefficient}
\begin{align}
W(\theta, \phi) = & \sum_{l=0}^{2j} \sum_{m_1=-j}^{j} \sum_{m_2=-j}^{j} (-1)^{j-2 m_1+m_2} 
\langle j m_1; j -m_2 | l m_1 - m_2 \rangle  Y_{l m_1-m_2} (\theta, \phi) \nonumber \\
& \times \langle j m_1 | \rho | j m_2 \rangle  .
\label{wignerfunctionone}
\end{align}

\begin{figure}[t]
\centerline{\includegraphics[width=\textwidth]{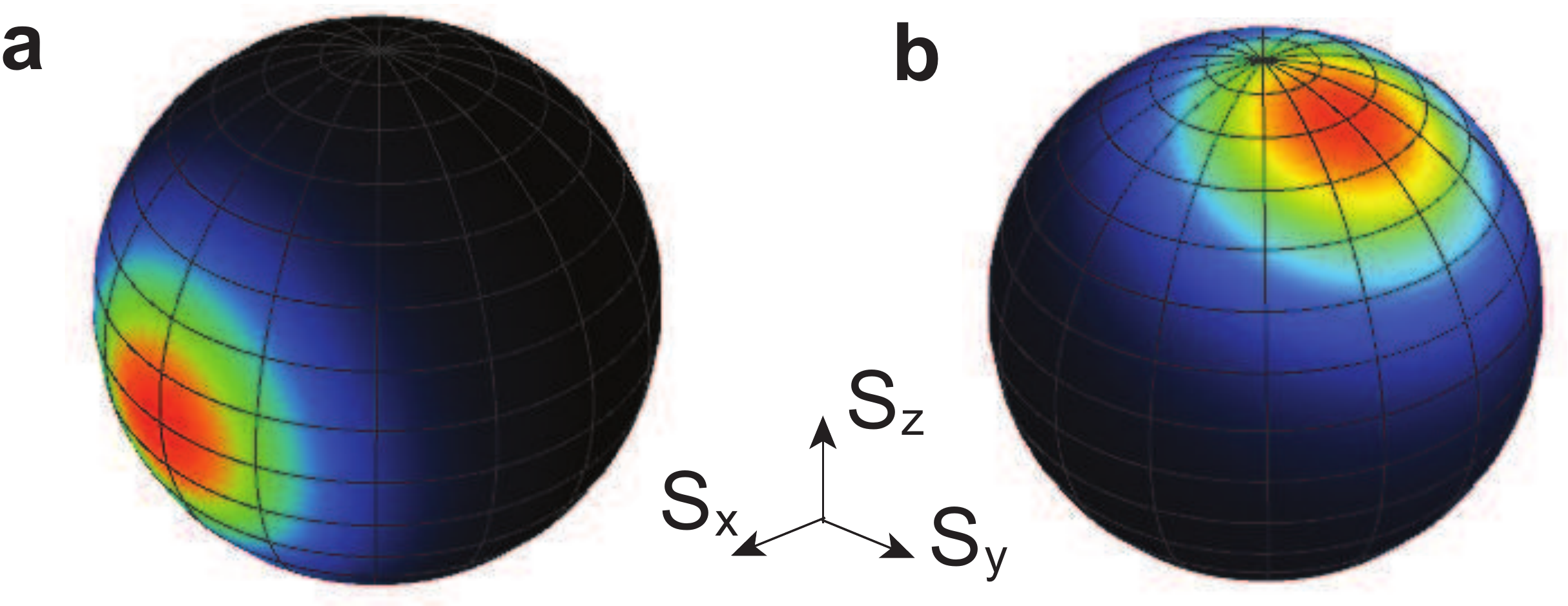}}
\caption{  $ Q $-distributions for spin coherent states (\ref{scs}) of two-component Bose-Einstein condensates with $ N=10 $ atoms. (a) $ \theta = \pi/2 $, $ \phi = 0 $, (b) $ \theta = \pi/8 $, $ \phi = \pi/2 $ for the parametrization (\ref{blochpara}). }\index{coherent state!spin} \index{Q-function}
\label{fig5-5}
\end{figure}

\subsection{Examples}

We now show several examples of quasiprobability distributions of spinor BEC states. Both the $Q$- and Wigner functions have a  distribution that is defined in terms of angular variables $ \theta, \phi $ and the natural way to plot this is on the surface of a sphere, much as we represented the spin coherent state on the Bloch sphere Fig. \ref{fig5-2}.  \index{Bloch sphere} \index{coherent state!spin} \index{Q-function} \index{Wigner function}

\subsubsection{Spin coherent states}

Figure \ref{fig5-5} shows the $Q$-distribution for two examples of a spin coherent state.  In both cases the distributions are of a Gaussian form, centered at different locations on the sphere.  The Gaussian nature of the $Q$-functions can be easily deduced using (\ref{expansionoverlap}), which shows that the overlap of two spin coherent states are of this form. For example, for the spin coherent state centered at $ \theta_0 , \phi_0 $, the $Q$-distribution is\index{Q-function} \index{coherent state!spin}
\begin{align}
Q(\theta, \phi) & = \frac{N+1}{4 \pi} \left| \langle \langle \cos \frac{\theta_0}{2} e^{-i \phi_0/2}, 
\sin \frac{\theta_0}{2} e^{i \phi_0/2} | \cos \frac{\theta}{2} e^{-i \phi/2}, 
\sin \frac{\theta}{2} e^{i \phi/2} \rangle \rangle \right|^2 \nonumber \\
& =  \frac{N+1}{4 \pi} \cos^{2N} \left( \frac{\Delta \Theta}{2} \right) \nonumber \\
& \approx  \frac{N+1}{4 \pi} e^{-N (\Delta \Theta)^2/4}
\label{qfuncscs}
\end{align}
where
\begin{align}
\cos \Delta \Theta = \cos \theta \cos \theta_0 + \sin \theta \sin \theta_0 \cos (\phi - \phi_0)  .
\label{greatcircleangle}
\end{align}
From (\ref{qfuncscs}) it is clear that for larger $ N $, the distributions become narrower.  

\begin{figure}[t]
\centerline{\includegraphics[width=\textwidth]{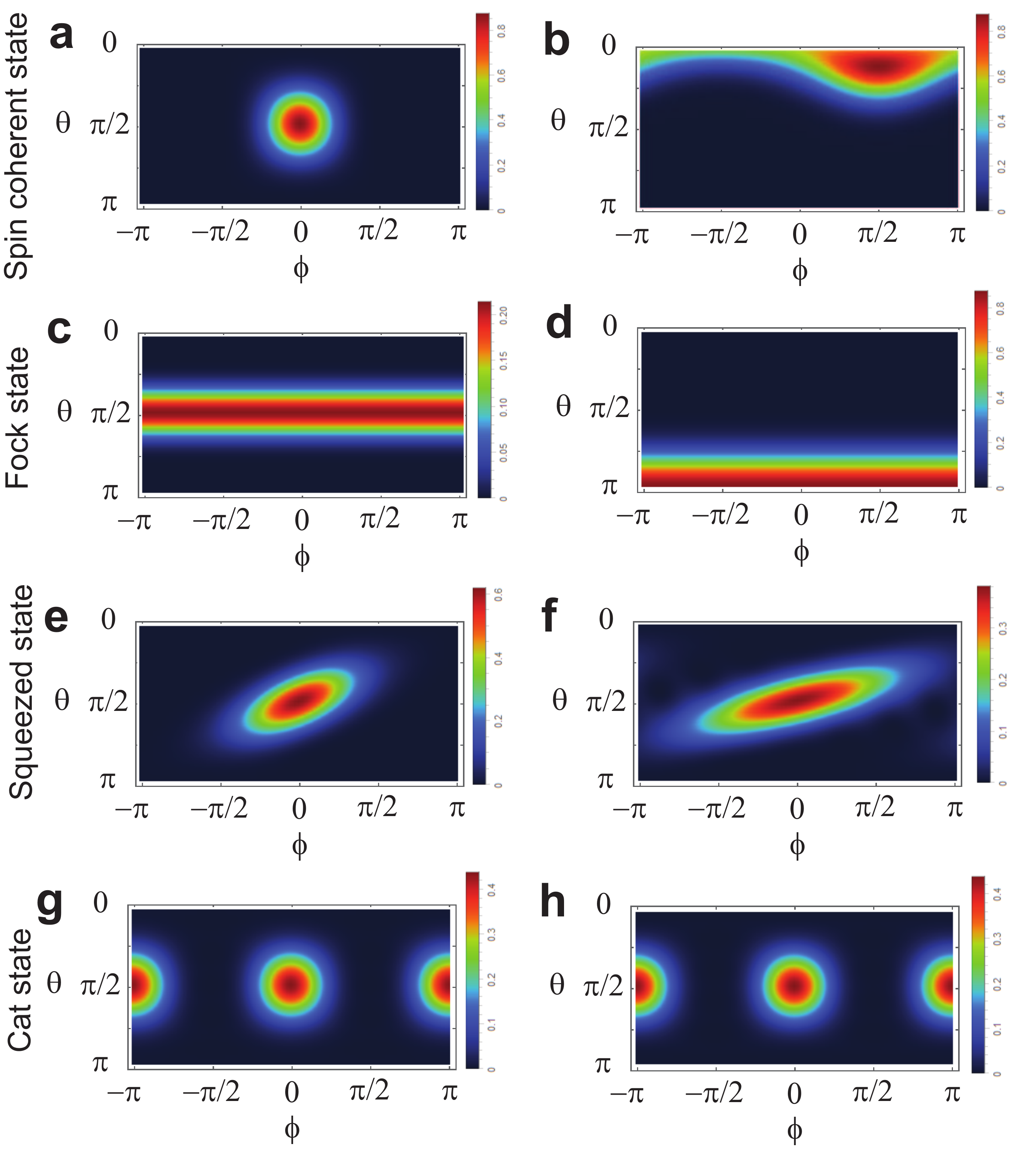}}
\caption{  $ Q $-distributions for various states of a two-component Bose-Einstein condensates with $ N=10 $ atoms. States shown are: spin coherent state (\ref{scs}) (a) $ \theta = \pi/2 $, $ \phi = 0 $, (b) $ \theta = \pi/8 $, $ \phi = \pi/2 $ for the parametrization (\ref{blochpara}); Fock state (\ref{fockstates}) (c) $ k = N/2 $, (d) $ k = 0 $; one-axis twisting squeezed state (\ref{squeezedszstate}) (e) $ \tau = 1/2N $, (f) $ \tau = 1/N $; Schrodinger cat state (\ref{catstate}) (g) $ \sigma = +1 $ (h) $ \sigma = - 1 $.  }\index{Fock states} \index{Q-function} \index{squeezed state!one-axis twisting}
\label{fig5-6}
\end{figure}

Although plotting the quasiprobability distribution on sphere as shown in Fig. \ref{fig5-5} is the most natural, it is not convenient for obvious reasons -- you cannot see the other side of the sphere without rotating it.  A simple solution to this is to flatten the sphere as you do with a world map, which inevitably distorts the distribution.  Fig. \ref{fig5-6}(a)(b) shows the same spin coherent states\index{coherent state!spin} as shown in Fig. \ref{fig5-5}  using the Mercator projection\index{Mercator projection}.  Near the center of the projection (Fig. \ref{fig5-6}(a)) the distribution appears Gaussian but near the poles the distribution is heavily distorted (Fig. \ref{fig5-6}(b)).  We must thus be careful to interpret the distribution in the correct way when using such projected maps.

\begin{figure}[t]
\centerline{\includegraphics[width=\textwidth]{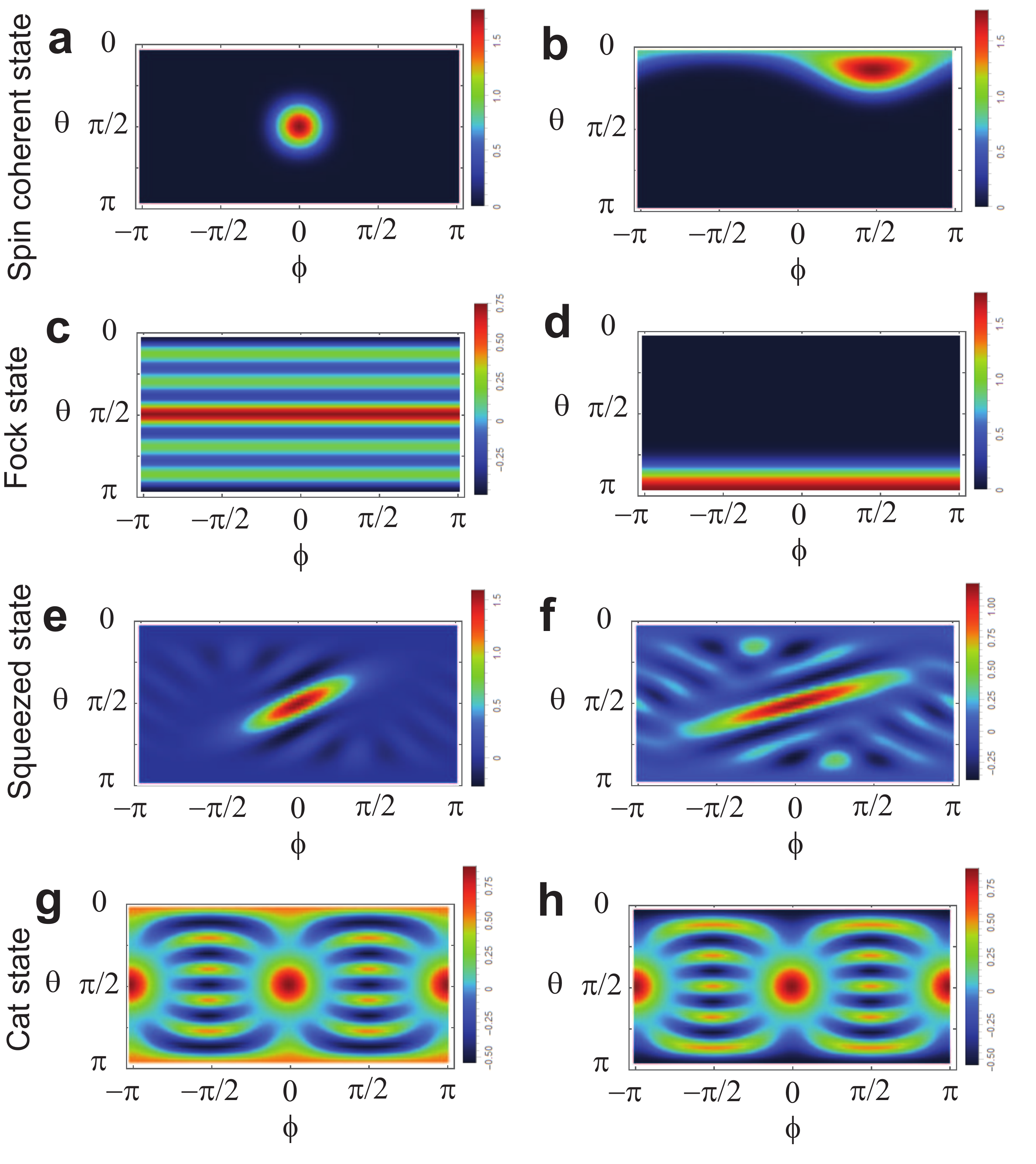}}
\caption{Wigner distributions for various states of a two-component Bose-Einstein condensates with $ N=10 $ atoms. States show are: spin coherent state (\ref{scs}) (a) $ \theta = \pi/2 $, $ \phi = 0 $, (b) $ \theta = \pi/8 $, $ \phi = \pi/2 $ for the parametrization (\ref{blochpara}); Fock state (\ref{fockstates}) (c) $ k = N/2 $, (d) $ k = 0 $; one-axis twisting squeezed state (\ref{squeezedszstate}) (e) $ \tau = 1/2N $, (f) $ \tau = 1/N $; Schrodinger cat state (\ref{catstate}) (g) $ \sigma = +1 $ (h) $ \sigma = - 1 $.  }\index{squeezed state!one-axis twisting} \index{Wigner function} \index{Fock states}\index{coherent state!spin}
\label{fig5-7}
\end{figure}

Figure \ref{fig5-7}(a)(b) shows the Wigner distributions for the same spin coherent states\index{Wigner function}\index{coherent state!spin}.  We again see a similar Gaussian-shaped distribution as given by the $Q$-functions.  A simplified expression for the Wigner function of a general spin coherent state can obtained by noticing that it must be the same as that of a displaced spin coherent state starting from the poles.  Recall from Sec. \ref{sec:eqivfockscs} that there are two special spin coherent states which are also Fock states\index{Fock states}.  For these cases the evaluation of the Wigner function simplifies and displacing the distribution one obtains
\begin{align}
W(\theta, \phi) & = N! \sum_{l=0}^N \sqrt{\frac{2l+1}{(N-l)! (l+1+N)!}} Y_{l0} ( \Delta \Theta,0) ,
\label{wignerscs}
\end{align}
where the spherical harmonic function in this case is given by
\begin{align}
Y_{l0} ( \theta, \phi) = \sqrt{\frac{2l+1}{4 \pi}} P_l ( \cos \theta)
\label{sphericallzero}
\end{align}
and $ P_l(x) $ is the Legendre polynomial\index{Legendre polynomials}. We see generally the same type of distribution but the Wigner functions have a narrower distribution.  This is also true for optical $Q$- and Wigner functions which share this same characteristic. \index{Q-function} \index{Wigner function}

\subsubsection{Fock states}

The $ Q $-function of a Fock state $ | k \rangle $ is\index{Q-function} \index{Fock states} \index{Q-function}
\begin{align}
Q(\theta, \phi ) & = \frac{N+1}{4\pi} | \langle k | \cos \frac{\theta}{2} e^{-i \phi/2}, 
\sin \frac{\theta}{2} e^{i \phi/2} \rangle \rangle |^2 \nonumber \\
& =  \frac{N+1}{4\pi} {N \choose k} \cos^{2k} \frac{\theta}{2} \sin^{2N-2k} \frac{\theta}{2}  .
\label{qfunctionfock}
\end{align}
We see that this is independent of the angle $ \phi $ as shown in Fig. \ref{fig5-6}(c)(d).  The $Q$-distribution is peaked at the location
\begin{align}
\cos \theta =\frac{2k - N}{N} .
\label{peakqfunc}
\end{align}
which can be found by setting $ \frac{dQ}{d\theta} = 0 $.  This is a rather intuitive result since the eigenstate of a Fock state is 
\begin{align}
S_z | k \rangle = (2k - N ) | k \rangle .
\label{szeigenstate}
\end{align}
Meanwhile, the expectation value of $ S_z $ for a spin coherent state is given by (\ref{scsexpectations}).  Since the $ Q$-function is nothing but the overlap with various spin coherent states, the maximum occurs when these two expressions coincide. 

For the Wigner function, we can evaluate the expression (\ref{wignerfunctionone}) for a general Fock state to obtain\index{Wigner function} \index{Fock  states}
\begin{align}
W(\theta, \phi) = & \sum_{l=0}^{N}  (-1)^{N-k} 
\langle \frac{N}{2}  k-\frac{N}{2} ; \frac{N}{2}  \frac{N}{2}-k | l 0 \rangle  Y_{l 0} (\theta, \phi) .
\label{wignerfunctionfock}
\end{align}
This again has no $ \phi $ dependence because it only involves spherical harmonic functions (\ref{sphericallzero}).  In this case the Wigner function \index{Wigner function}has significant differences to that of the $ Q $-function\index{Q-function}, as can be observed from Fig. \ref{fig5-7}(c)(d).  The main distinguishing feature is the appearance of several ripples around the largest peak, which is shared with the $ Q $-function.  Looking closer at the scale we note that the distribution has negative regions.  The possibility of becoming negative is one large difference between the $ Q $- and Wigner functions.  The appearance of negative regions is often associated with non-classical behavior.  In contrast, the Wigner functions for the spin coherent states were positive.  A common interpretation for this is that spin coherent states have a quasi-classical nature, particularly when $ N $ becomes large.  However, we note that spin coherent states always have a quantum nature in that they are minimum uncertainty states, as given by (\ref{variancescs}).  \index{coherent state!spin}

\subsubsection{Squeezed states}\index{squeezed state}

The $Q$-function for the squeezed state (\ref{squeezedszstate}) can be evaluated to be\index{Q-function}
\begin{align}
Q(\theta, \phi ) & = \frac{N+1}{2^{N+2}\pi} \left| \sum_{k=0}^N   {N \choose k}  e^{i(N-2k)(\phi/2 + (N-2k)\tau)} \cos^{k} \frac{\theta}{2} \sin^{N-k} 
\frac{\theta}{2}  \right|^2 ,
\end{align}
which is plotted in Fig. \ref{fig5-6} for two squeezing times.  We see that the Gaussian distribution becomes stretched in a diagonal direction.  For longer squeezing times the distribution gets stretched to a greater degree, and the angle tends towards a smaller gradient in the $ \theta$-$\phi$ plane. 

An intuitive explanation for why the distribution becomes elongated in a diagonal direction can be obtained by rewriting the form of the squeezing interaction as
\begin{align}
e^{-i (S_z)^2 \tau} = \sum_{k=0}^N e^{-i(2k-N)\tau S_z} | k \rangle \langle k |  .
\end{align}
Recalling from (\ref{spinrotationsxyz}) that application of an operator $ e^{-i \theta S_z} $ produces a rotation around the $S_z $ axis, we can interpret the squeezing term to be a rotation of the state around the $ S_z $ axis, by an angle $2(2k-N)\tau $.  The factor $ (2k-N) $ can be either positive or negative, depending on the Fock state in question.  Since the Fock states are represented by a position (\ref{peakqfunc}) on the Bloch sphere\index{Bloch sphere}, the lower hemisphere rotates in a clockwise direction, while the upper hemisphere rotates in a anti-clockwise direction. \index{Fock states} 

Comparing to the discussion of Sec. \ref{sec:squeezedstates}, we see a clearer picture of the effects of the squeezing.  Previously we analyzed the variances in various spin directions and found that there was a particular direction where it was at a minimum, as seen in Fig. \ref{fig5-3}(a) and  (\ref{minimumangle}).  We can attribute this to the elongation of the distribution, which minimizes the variance in direction perpendicular to the stretching. According to (\ref{minimumangle}) the angle of elongation starts in the $45^\circ $ direction and rotates to $0^\circ$ in $ \theta$-$\phi$ plane.

The Wigner functions\index{Wigner function} for the squeezed states are shown in Fig. \ref{fig5-7}(e)(f).  The main diagonal feature is similar to the $Q$-function,\index{Q-function} except that it has a narrower distribution, as was observed for the spin coherent and Fock states.  The major difference is that the remaining parts of the distribution are less even than the $ Q$-function case.  These arise primarily because of the relatively small number of atoms $ N = 10 $ that are used in these plots.  Such small $ N $ Wigner functions tend to always have fluctuations which arise due to the sum of a relatively small number of spherical harmonic functions.  An interesting feature of Fig. \ref{fig5-7}(e)(f) is the presence of negative regions of the Wigner functions, which are associated with non-classical behavior.  The negative regions tend to increase with $ \tau $, showing that the squeezing operation produces increasingly non-classical states.

\subsubsection{Schrodinger cat states}\index{Schrodinger cat state}

The final state we examine are Schrodinger cat states.  These are defined as
\begin{align}
|\text{cat}^\pm \rangle = \frac{1}{\sqrt{2}} \left( | \cos \frac{\theta_0}{2} e^{-i \phi_0/2}, 
\sin \frac{\theta_0}{2} e^{i \phi_0/2}  \rangle \rangle \pm | - \sin \frac{\theta_0}{2} e^{i \phi_0/2} , \cos \frac{\theta_0}{2} e^{-i \phi_0/2}  \rangle \rangle \right) ,
\label{catstate}
\end{align}
where the two states in the superposition are two spin coherent states\index{coherent state!spin} at antipodal points and are orthogonal (\ref{antipodalscs}).  For the special case of $ \alpha = \beta = 1/\sqrt{2} $, the superposition causes either the even or odd Fock states to completely cancel\index{Fock state}
\begin{align}
|\text{cat}^\pm ( \alpha = \beta = \frac{1}{\sqrt{2}}) \rangle = \frac{1}{\sqrt{2^{N-1}}} \sum_{k\in \text{even, odd} } \sqrt{{N \choose k}} | k \rangle ,
\end{align}
where the even terms are for the $ + $ superposition and odd for the $ - $ superposition.  

The $Q$-functions \index{Q-function}for both types of Schrodinger cat states is given by 
\begin{align}
Q(\theta,\phi) & = \frac{N+1}{8 \pi} \left[  \cos^{2N} \left( \frac{\Theta}{2} \right) +  \cos^{2N} \left( \frac{\Theta'}{2} \right) \pm 2  \cos^{N} \left( \frac{\Theta}{2} \right) \cos^{N} \left( \frac{\Theta'}{2} \right) \right] \nonumber \\
& \approx \frac{N+1}{8 \pi} \left[  \cos^{2N} \left( \frac{\Theta}{2} \right) +  \cos^{2N} \left( \frac{\Theta'}{2} \right)  \right] 
\label{qfunccatstate}
\end{align}
where
\begin{align}
\cos \Delta \Theta & = \cos \theta \cos \theta_0 + \sin \theta \sin \theta_0 \cos (\phi - \phi_0)  \nonumber \\
\cos \Delta \Theta' & = - \cos \theta \cos \theta_0 -  \sin \theta \sin \theta_0 \cos (\phi + \phi_0) .
\label{greatcircleangle2}
\end{align}
Here, $ \Delta \Theta, \Delta \Theta' $ is the great circle angle from $ (\theta,\phi) $ to $ (\theta_0, \phi_0), (\theta_0+ \pi, -\phi_0)  $ respectively. In the second line of (\ref{qfunccatstate}) we have dropped the last term since this is small for $ N \gg 1 $.  The reason this term is small is that the cross term is a product of two overlaps between the point $ (\theta,\phi) $ and two antipodal points.  This term is thus exponentially suppressed according to (\ref{expansionoverlap}), regardless of the choice of $ (\theta,\phi) $.

A numerical plot of the $ Q $-function\index{Q-function} is shown in Fig. \ref{fig5-6}(g)(h).  We see that both distributions for the even and odd cat states are identical, according to the approximation given in (\ref{qfunccatstate}). The dominant structure is simply a combination of two spin coherent state\index{coherent state!spin} distributions, as predicted by (\ref{qfunccatstate}).  The Wigner function\index{Wigner function}, as shown in Fig.  \ref{fig5-7}(g)(h) shows a very different distribution.  In addition to the peaks associated with the spin coherent states, strong fringes appear connecting the two peaks.  The fringes show strong negative regions, indicating the highly non-classical\index{non-classical state} nature of the state.  The two types of cat states also show a complementary structure, where the negative regions of one are positive in the other.  For optical Wigner functions, superpositions of coherent states\index{coherent state} also exhibit a similar structure, with fringes appearing connecting the coherent states. In this case, the Wigner function provides more information about the structure of the state, distinguishing between the type of quantum state.

\begin{exerciselist}[Exercise]
\item \label{q5-15}
Verify (\ref{qfunctionfock}) and show that the peak of the distribution occurs at the angle (\ref{peakqfunc}). 
\item \label{q5-16}
Verify (\ref{wignerscs}).  First evaluate (\ref{wignerfunctionone}) for the case 
\begin{align}
\rho = |k = N \rangle \langle k = N | = | 1,0 \rangle \rangle \langle \langle 1,0 | .
\end{align} 
Then use the fact that the great circle angle between $ (\theta,\phi) $  and  $ (\theta_0,\phi_0) $ is (\ref{greatcircleangle}).  
\end{exerciselist}

\section{Other properties of spin coherent states and Fock states}

Here we list several other important relations between spin coherent states and Fock states.  \index{Fock states}

\subsection{Overlap of two spin coherent states}
\label{sec:overlapscs}

Like optical coherent states, the spin coherent states are in general not orthogonal.  The overlap of two spin coherent states can be evaluated to be\index{coherent state!optical} \index{coherent state!spin}
\begin{align}
\langle \langle \cos \frac{\theta'}{2} e^{-i \phi'/2}, 
\sin \frac{\theta'}{2} e^{i \phi'/2} |\cos \frac{\theta}{2} e^{-i \phi/2}, 
\sin \frac{\theta}{2} e^{i \phi/2} \rangle \rangle = \cos^N \left( \frac{\Delta \Theta}{2} \right) ,
\label{overlapscs}
\end{align}
where
\begin{align}
\cos \Delta \Theta = \cos \theta \cos \theta' +  \sin \theta \sin \theta' \cos (\phi - \phi')
\end{align}
is the central angle (i.e. the angle at the center of the sphere for the great-circle distance) between the two spin coherent states.  For $ N \gg 1 $, the powers of cosine can be well-approximated by a Gaussian and hence we have
\begin{align}
\langle \langle \cos \frac{\theta'}{2} e^{-i \phi'/2}, 
\sin \frac{\theta'}{2} e^{i \phi'/2} |\cos \frac{\theta}{2} e^{-i \phi/2}, 
\sin \frac{\theta}{2} e^{i \phi/2} \rangle \rangle \approx e^{-N (\Delta \Theta)^2 / 8 } .
\label{expansionoverlap}
\end{align}

For a given spin coherent state, there is always another spin coherent state which is exactly orthogonal to it.  From (\ref{overlapscs}) one can easily deduce that this is when $ \Delta \Theta = \pi $, which is the state that is at the antipodal point to it.  That is, if 
\begin{align}
\theta' & = \theta + \pi \nonumber \\
\phi' & = - \phi
\end{align}
then
\begin{align}
\langle \langle -\beta^*, \alpha^* | \alpha, \beta \rangle \rangle = 0 .
\label{antipodalscs}
\end{align}

\subsection{Equivalence of Fock states and spin coherent states}
\label{sec:eqivfockscs}

In general, the Fock states (\ref{fockstates}) are a completely different class of state to spin coherent states (\ref{scs}).  However, there are two states where the two classes coincide.  These are the extremal Fock states\index{Fock states} \index{coherent state!spin}
\begin{align}
|k = 0 \rangle & = \frac{(b^\dagger)^N}{\sqrt{N!}} = | 0,1 \rangle \rangle \nonumber \\
|k = N \rangle & = \frac{(a^\dagger)^N}{\sqrt{N!}} = | 1,0 \rangle \rangle  .
\end{align}
In both these cases, all the bosons in the same state, which is essentially the definition of a spin coherent state.

\subsection{Transformation between Fock states in different basis}
\label{sec:kxkzconv}

The Fock states (\ref{fockstates}) are eigenstates of the $ S_z $ operator.  For a more general spin operator \index{Fock states}
\begin{align}
\bm{n} \cdot \bm{S} = \sin \theta \cos \phi  S_x + \sin \theta \sin \phi S_y + \cos \theta  S_z
\end{align}
where $ \bm{n} = (\sin \theta \cos \phi , \sin \theta \sin \phi , \cos \theta ) $ is a unit vector, the Fock states are defined as
\begin{align}
|k \rangle_{\bm{n}} =  \frac{ (c^\dagger)^{k} (d^\dagger)^{N-k}}{\sqrt{k! (N-k)! }} | 0 \rangle ,
\label{krotatedfock}
\end{align}
where
\begin{align}
c & = \cos \frac{\theta}{2} e^{-i \phi/2}  a + \sin \frac{\theta}{2} e^{i \phi/2}   b \nonumber \\
d & =\sin \frac{\theta}{2} e^{-i \phi/2} a - \cos \frac{\theta}{2} e^{i \phi/2}  b .
\end{align}
The Fock states satisfy 
\begin{align}
\bm{n} \cdot \bm{S} |k \rangle_{\bm{n}} = ( 2k - N) |k \rangle_{\bm{n}}
\end{align}
in the same way as we have seen already for the special cases of $ \bm{n}=(1,0,0) $ and $ \bm{n}=(0,1,0) $ in (\ref{sxyeigenvalueeq}). 

In order to convert the Fock state $ |k \rangle_{\bm{n}} $ to the $ S_z $ basis operators $ |k \rangle $, one may use the result
\begin{align}
\langle k' | e^{-i S_y \theta/2} | k \rangle = & \sqrt{ k! (N-k)! k'! (N-k')!}  \nonumber \\
& \times \sum_{n=\max(k'-k,0)}^{\min(k',N-k)}  \frac{(-1)^n \cos^{ k'- k + N - 2n} (\theta/2) \sin^{2n + k - k'} (\theta/2) }{(k'-n)!(N-k-n)!n!(k-k'+n)!} . \label{kexpansiongroup}
\end{align}
We may then write
\begin{align}
|k \rangle_{\bm{n}} & =  e^{-i S_z \phi/2} e^{-i S_y \theta/2} | k \rangle  \nonumber \\
& =  e^{-i S_z \phi/2} \left(\sum_{k'} | k' \rangle \langle k '| \right)  e^{-i S_y \theta/2} | k \rangle  \nonumber \\
& = \sum_{k'} e^{-i (2k'-N) \phi/2} \langle k '|  e^{-i S_y \theta/2} | k \rangle   | k' \rangle 
\label{expansionkykz}
\end{align}
where the terms on the expansion are given by (\ref{kexpansiongroup}).  For example, for eigenstates of the $ S_x $ operator where $ \theta = \pi/2, \phi=0 $, the Fock states are
\begin{align}
|k \rangle_x = &  \frac{1}{\sqrt{2^N}} \sum_{k'=0}^N \sqrt{ k! (N-k)! k'! (N-k')!}  \nonumber \\
& \times \sum_{n=\max(k'-k,0)}^{\min(k',N-k)}  \frac{(-1)^n  }{(k'-n)!(N-k-n)!n!(k-k'+n)!} | k'\rangle  .  
\label{kxeigenstates}
\end{align}
For the eigenstates of the $ S_y $ operator, we use parameters $ \theta = \pi/2, \phi=\pi/2 $, and obtain a similar expression to (\ref{kxeigenstates}) but with an extra factor of $ e^{i(N-2k')\pi/4} $ in the sum. 
We note that for $ k \ll  N/2 $ or $ k \gg N/2 $, the overlap $ | \langle k' |k \rangle_x  |^2 $ is well-approximated by the modulus squared of the quantum harmonic oscillator eigenstates.

\begin{exerciselist}[Exercise]
\item \label{q5-17}
Set $ \phi = \phi' $ and confirm that the formula (\ref{overlapscs}) gives the correct result.  
\item \label{q5-18}
Expand the first three terms of (\ref{overlapscs}) as a Taylor series and confirm that it agrees with the Taylor expansion of (\ref{expansionoverlap}).  \index{Taylor series}
\end{exerciselist}

\section{Summary}

In this chapter, we have seen that there is a close analogy between optical states and spin states.  We summarize the correspondences in Table \ref{tab5-1}.  The spin coherent states play an analogous role to coherent states and are minimum uncertainty states.  These form a set of non-orthogonal states with a Gaussian overlap with respect to the great circle distance between two spin coherent states.  Due to three operators, rather than two for the optical case, the phase space is a two dimensional space with a spherical topology.  Analogous quasiprobability distributions such as the $Q$- and Wigner functions can be defined for the spin case equally, in the same way as the optical case.

\begin{table}[t]
\processtable
{Equivalencies between optics and spins. \label{tab5-1} }
{
\begin{tabular}{ccc}
Quantity & Optics & Spin \\
\hline
Coherent state & $ | \alpha \rangle = e^{-|\alpha|^2/2} e^{\alpha a^\dagger} | 0 \rangle $ & $ | \alpha, \beta \rangle \rangle = \frac{1}{\sqrt{N!}} ( \alpha a^\dagger + \beta b^\dagger)^N | 0 \rangle $ \\
Fock state & $ | n \rangle = \frac{1}{\sqrt{n!}} (a^\dagger)^n | 0 \rangle $ & $ | k \rangle =  \frac{1}{\sqrt{k!(N-k)!}} (a^\dagger)^k (b^\dagger)^{N-k} | 0 \rangle $ \\
Overlap of coherent states & $ |\langle \alpha' | \alpha \rangle|^2 = e^{-|\alpha - \alpha'|^2} $ &  $ |\langle \langle \alpha', \beta' | \alpha, \beta \rangle \rangle |^2 = \cos^{2N} (\Delta \Theta/2) $ \\
& & $ \approx e^{-N (\Delta \Theta)^2/4}  $ \\
Observables & $X = \frac{a+a^\dagger}{\sqrt{2}} $ & $S_x = a^\dagger b + b^\dagger a $\\
	& $ P = \frac{a+a^\dagger}{\sqrt{2}}$  & $S_y = -i a^\dagger b + i b^\dagger a $ \\
	&   &  $S_z = a^\dagger a - b^\dagger b $  \\
Phase space & 2-dimensional plane $ (X,P) $ & 3-dimensional sphere $ (S_x, S_y, S_z ) $ \\
Squeezing Hamiltonian & $e^{-i \Theta} a^2 + e^{i \Theta} (a^\dagger)^2 $ & $(S_z)^2$ \\
Displacement Hamiltonian & $e^{-i \Theta} a + e^{i \Theta} a^\dagger$ & $\bm{n} \cdot \bm{S} $\\
Quasiprobability distributions & $P$-, $Q$-, Wigner functions & $P$-, $Q$-, Wigner functions \\
\end{tabular}}
\end{table}

\section{References and further reading}

\begin{itemize}
\item Sec. \ref{sec:spincoherentstates}:
Theoretical works discussing the mathematics of spin coherent states \cite{radcliffe1971some,arecchi1972atomic,holtz1974coherent}. Edited volume on coherent states \cite{bo1985coherent}.  Book on coherent states taking a group theoretical perspective \cite{perelomov2012generalized}. Effect of interactions in spinor Rb BECs \cite{klausen2001nature}.   
\item Sec. \ref{sec:schwingerboson}: Original work introducing the Schwinger boson representation \cite{jordan1935zusammenhang}.  Edited volume by Schwinger on angular momentum \cite{schwinger1965quantum}.  Textbooks further discussing Schwinger boson representation \cite{sakurai1995modern,auerbach2012interacting}.
\item Sec. \ref{sec:spincohexp}: Original work discussing spin coherent state expectation values  \cite{radcliffe1971some}.  Further discussion can be found in the textbook \cite{auerbach2012interacting}.  
\item Sec. \ref{sec:prepscs}: Experimental preparation of a spin coherent state has been performed in \cite{harber2002effect,treutlein2004coherence,bohi2009}. The original work for Ramsey interference \cite{ramsey1950molecular}.  Textbook level discussion of spin transitions and interference \cite{nielsen2000}. 
\item Sec. \ref{sec:uncertainty}: Original work on the Robertson uncertainty relation \cite{robertson1929uncertainty}.  Heisenberg's original paper on the uncertainty relation \cite{heisenberg1985anschaulichen}.  Textbooks further discussing the topic  \cite{scully1999quantum,walls2007quantum,gerry2005introductory}.  Original works showing that atomic spin coherent state have  minimal uncertainty \cite{arecchi1972atomic,bacry1978physical}. Review article focusing on atomic spin squeezing \cite{gross2012}.  
\item Sec. \ref{sec:squeezedoptical}: Experimental observation of a  optical squeezed state \cite{slusher1985observation}. Using cold atoms to squeeze light \cite{lambrecht1996squeezing}.  Early review article on optical squeezed states \cite{walls1983squeezed}.  Textbooks and review articles discussing squeezed optical states  \cite{braunstein2005,scully1999quantum,walls2007quantum,gerry2005introductory}. 
\item Secs. \ref{sec:oneaxistwisting},\ref{sec:twoaxiscounter}:
Original theoretical works introducing spin squeezing \cite{kitagawa1993squeezed,wineland1992spin,wineland1994}.  
 Experimental demonstrations of spin squeezing \cite{hald1999spin,kuzmich2000generation,orzel2001squeezed,appel2009mesoscopic,takano2009spin,esteve2008,riedel2010,leroux2010implementation}. Further theoretical approaches to spin squeezing \cite{kuzmich1997spin,sorensen2001many,thomsen2002spin,bouchoule2002spin}.  Review articles on spin squeezing \cite{ma2011quantum,gross2012}.  Theoretical scheme to turn one-axis squeezing to two-axis squeezing \cite{liu2011spin}.  
\item Sec. \ref{sec:entanglementsqueezing}: Theoretical works relating entanglement in spin ensembles to squeezing  \cite{sorensen2001many,sorensen2001entanglement,wineland1994,wang2003spin,korbicz2005spin,toth2007optimal,toth2009spin,toth2014quantum,dalton2017quantum}. Other theoretical works investigating entanglement in BECs \cite{pu2000creating,helmerson2001creating,duan2002quantum,micheli2003many,hines2003entanglement,he2011einstein}.  Review article on entanglement detection \cite{guhne2009entanglement}. Experimental observation of  entanglement \cite{hald1999spin,esteve2008,gaetan2009observation,kunkel2017,lange2017,fadel2017}.  
\item Sec. \ref{sec:holstein}: Original paper introducing the Holstein-Primakoff transformation \cite{holstein1940field}.  Textbooks discussing the transformation \cite{caspers1989spin,kittel1987quantum}.  Review article for atomic ensemble systems that uses the transformation \cite{hammerer2010quantum}.
\item Sec. \ref{sec:equivalencebosonspin}: Original works discussing the equivalence between bosons and spin ensembles \cite{jordan1935zusammenhang,schwinger1965quantum}. Textbook level discussion of the topic \cite{sakurai1995modern}. 
\item Sec. \ref{sec:quasiprobability}: Original reference introducing the spin $Q$-function \cite{lee1984q} and the spin Wigner function \cite{dowling1994wigner}. Other definitions of quasiprobability distributions for spins \cite{scully1994spin,tilma2016wigner}. 
\item Sec. \ref{sec:overlapscs}: Original references discussing spin coherent states \cite{radcliffe1971some,arecchi1972atomic}. 
\item Sec. \ref{sec:kxkzconv}: Eq. (\ref{kexpansiongroup}) can be found in Chapter 6 of \cite{thompson2008angular}.  Other textbooks that discuss the transformation \cite{brink1968angular,sakurai1995modern}.
\end{itemize}

	\chapter{Diffraction of atoms using standing wave light \label{ch:atomdiffraction}}

%
\section{Introduction \label{sec:chp5:intro}}

Diffraction\index{diffraction} of waves like sound waves, water waves and light waves is characterized by the spreading or ``flaring''of waves after passing through an aperture. For diffraction to occur, the size of the aperture must be comparable to the wavelength of  incident wave. A classic diffraction experiment was performed by Thomas Young\index{Young's double slit experiment} in 1801 where a monochromatic beam of light is coherently split into two distinct beams after passing through two slits. In the region beyond the aperture, the split beams diffract and produce interference fringes on a screen placed a large distance away in comparison with the distance between the slits. 
As with other waves mentioned above, matter waves\index{matter wave} like electrons and neutrons also diffract after passing through an aperture. For instance in electron beam diffraction, highly energetic electrons  that are incident on a periodic crystalline structure diffract in a predictable manner prescribed by Bragg's law. \index{Bragg's law} \index{electron beam diffraction}

Beyond particles such as photons, electrons and neutrons, Young's double slit experiment\index{Young's double slit experiment} have been performed with atoms. In 1991, a highly excited Helium atom was interfered after passing through two slits in gold foil. In contrast to the gold foil technique, most modern experiments use apertures made from light crystals to diffract atoms.  This will be the focus of this chapter. We begin our discussion in the next section by reviewing the diffraction of light and matter waves. Next we describe the interaction with detuned off-resonant light and solve the resulting dynamics under Bragg conditions and Raman transitions\index{Raman transition}. Finally we present some experiments that use this technique and describe other techniques used to diffract atoms. 

\begin{figure}[t]
\begin{center}
\includegraphics[angle=0,width= 0.7\textwidth]{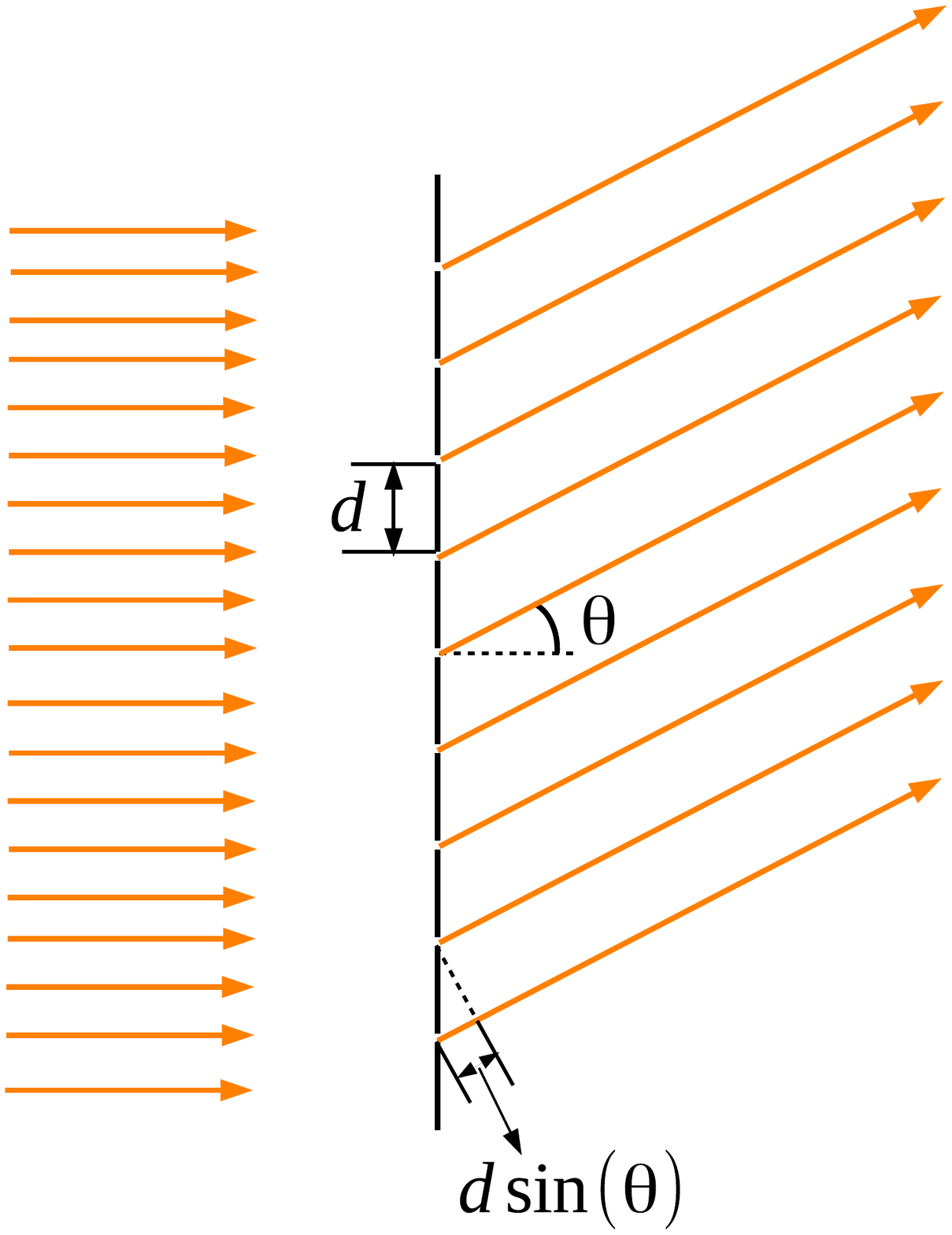}
\caption{The diffraction of light by a grating. Each slit on the grating is a source of light wave and is at a distance $d$ from one another. The wave from each slit is nearly parallel to each other, and the path difference of the light waves from any adjacent slits is $  d\sin\theta$.   }\label{fig:chp5-1}
\end{center}
\end{figure}

\begin{figure}[t]
	\begin{center}
		\includegraphics[angle=0,width= 0.7\textwidth]{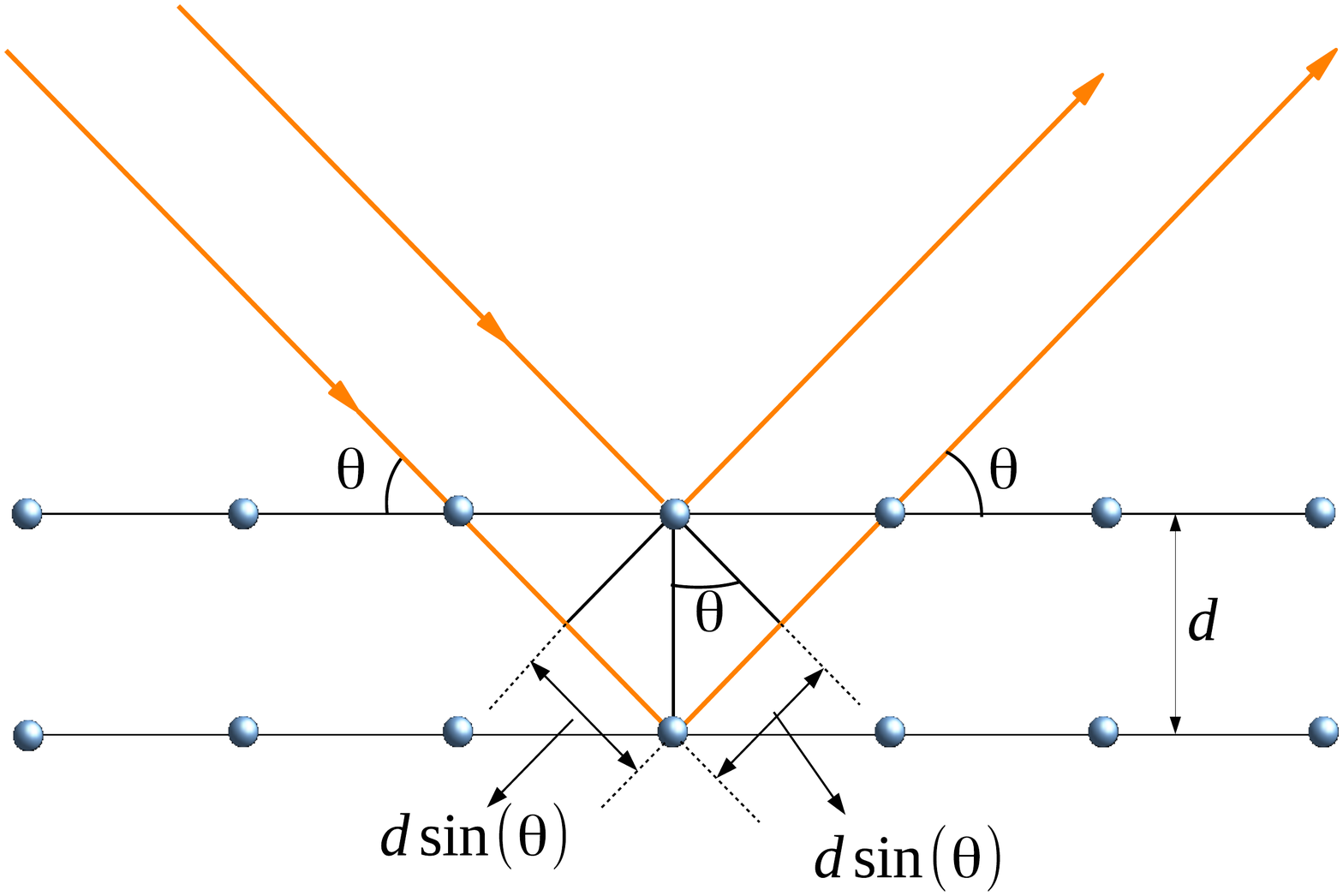}
		\caption{The diffraction of matter wave by a crystal. Each plane consists of atoms arranged in a regular fashion, and the planes are separated by a distance $d$. For waves incident on the crystal at angle $\theta$ to the plane, waves from  the bottom plane travel extra distance of $2d\sin\theta$ relative to those from the upper plane.}\label{fig:chp5-2}
	\end{center}
\end{figure}


%
\section{Theory of diffraction \label{sec:chp5:theory}\index{diffraction}}
Diffraction gratings consist of many slits that are spaced at constant interval of distance $d$ between the slits. Gratings are often used to study different light sources. When light passes through a slit, it diffracts in accordance with Huygens' wave theory, where each  slit acts as source of light waves that produces a secondary wavefront. The wavefronts as they advance spread and give rise to diffraction. Thus it is possible for a light wave from one slit to interfere with a light wave from another slit. The final interference pattern depends on the directional angle measured from the horizontal at which the rays are examined. To see this consider a grating illuminated by a light wave of wavelength $\lambda$ as shown in Fig.~\ref{fig:chp5-1}. Each slit on the grating is a source of light wave that emits a ray of light that are mutually parallel to each other.  They are also in phase as they leave the slit. As the rays propagate further towards a point on a screen, it is observed that the path difference between the rays depends on the angle at which a ray makes with the horizontal. For any two adjacent rays, the path difference between them is $d\sin\theta$. At the screen placed a large distance from the slit, bright bands are observed when all the waves arriving at the point on the screen are in phase. This is realized if the path difference is equal to a wavelength or integral multiples of the wavelength. Therefore the maximum in a bright band is  observed if the path difference is
\begin{equation}
\label{eq:chp501}
d\sin\theta = m\lambda,
\end{equation}
where $m = 0, \pm1, \pm2 \pm3 ,\cdots$ is the  diffraction order. To observe diffraction from the grating, the separation distance $d$ between the slits should be large in comparison with the wavelength $\lambda$ of light. Otherwise, the slits would appear as the same source and no diffraction pattern would be observed. Because of this requirement, a diffraction grating made for one wave of particular wavelength may not be suitable  in the study of another wave of very different wavelength.  

Possibly the simplest way to realize matter wave\index{matter wave} interference is to observe the diffraction of electrons in a crystal lattice.  The slits in the crystal --- for instance NaCl --- are made of atoms, which are arranged in a specific order. When matter wave\index{matter wave} are incident on a crystal, the atoms forming the crystal reflect the matter wave at regular intervals. It is found that the intensity of emerging matter wave is very strong in a particular direction, which corresponds to constructive interference. For example, suppose that the incident matter wave makes an angle $\theta$ with the planes of the crystal as shown in Fig.~\ref{fig:chp5-2}. Each plane is at constant spacing $d$, and we consider two planes as shown in Fig.~\ref{fig:chp5-2}. Matter waves reflected from the upper plane travel a smaller distance compared with those reflected from the bottom plane. Waves from the two planes would reinforce at a point on the screen if the path difference $2d\sin\theta$ traveled by the waves is an integer multiple of the wavelength $\lambda$ of the matter waves. Thus a maximum exists within a constructive interference\index{interference} if 
\begin{equation}
\label{eq:chp502}
2d\sin\theta = m\lambda,
\end{equation}
where $m = 1, 2, 3, \cdots$.

%
\section{Ultra-cold atom interaction with standing light wave \label{sec:chp5:atomlight}}
Material gratings such as crystals described in Sec.~\ref{sec:chp5:theory}, work well to perform diffraction of particles such as electrons and neutrons, but are less suited to diffraction\index{diffraction} of atoms. This is partly because atoms have a large mass compared to that of electrons and neutrons,  and results in a much smaller de Broglie wavelength \index{de Broglie wavelength} for a given velocity. Also due to the charge neutrality of atoms, producing an energetic atom with high velocity was almost impossible until high-energy Helium were produced via high impact electron excitation. But even at that, the challenge still remained on how to diffract a stationary atom or an atom with negligible velocity as found in ultra-cold atoms. The answer came with advances in techniques for atomic manipulation and better understanding of atom-light interactions\index{atom-light interactions} that is used to impart momentum on an atom.   


Photons possess both energy and momentum. A photon in a light wave with wave vector $\kappa_l$ has a momentum $\hbar\kappa_l$. When an atom absorbs a photon, it absorbs its energy and momentum, and the excited atom will move in the direction of the absorbed photon. To see how the momentum transfer is used to provide directed motion for the atoms, consider a two-level atom which has a mirror to the right of it. Light traveling from left to right is reflected by the mirror. The reflected light traveling to the left and the incident light traveling to the right superpose to form a standing wave. Thus, the atom is bathed by streams of photons in the light wave traveling to the right and left. If the atom absorbs a photon from the light traveling to the right, it receives a momentum kick to the right and excited atom would start moving to the right with momentum $\hbar\kappa_l$. The excited atom can interact with either of the light waves that will result in a stimulated emission. If the excited atom interacts with the light wave traveling to the right, it will emit a photon with momentum $\hbar\kappa_l$, and receives a momentum kick to the left. As such its net momentum after stimulated emission is zero. On the other hand, if the excited atom was to interact with light traveling to the left, it emits a photon that will be traveling to the left and receives a kick to the right. Thus, the net momentum of the atom after de-excitation would be $2\hbar\kappa_l$ and the atom will be seen moving to the right. Similar intuitive picture can be built for atom moving to the left after the absorption-stimulated emission cycle. Higher order diffraction can be achieved by repeating the whole process for each order increase in steps of $2\hbar\kappa_l$. Spontaneous emission\index{spontaneous emission} is possible if the excited atom interacts instead with the vacuum mode of light. Of course, the atom will still receive a momentum kick in this case but in a random direction that is not useful in the context of diffraction. We will however examine the absorption-spontaneous emission cycle employed in the cooling of atoms.

\subsection{Evolution of internal states of the atom \label{sec:sec:chp5:evolution}}

\begin{figure}[t]
	\begin{center}
		\includegraphics[angle=0,width= 0.3\textwidth]{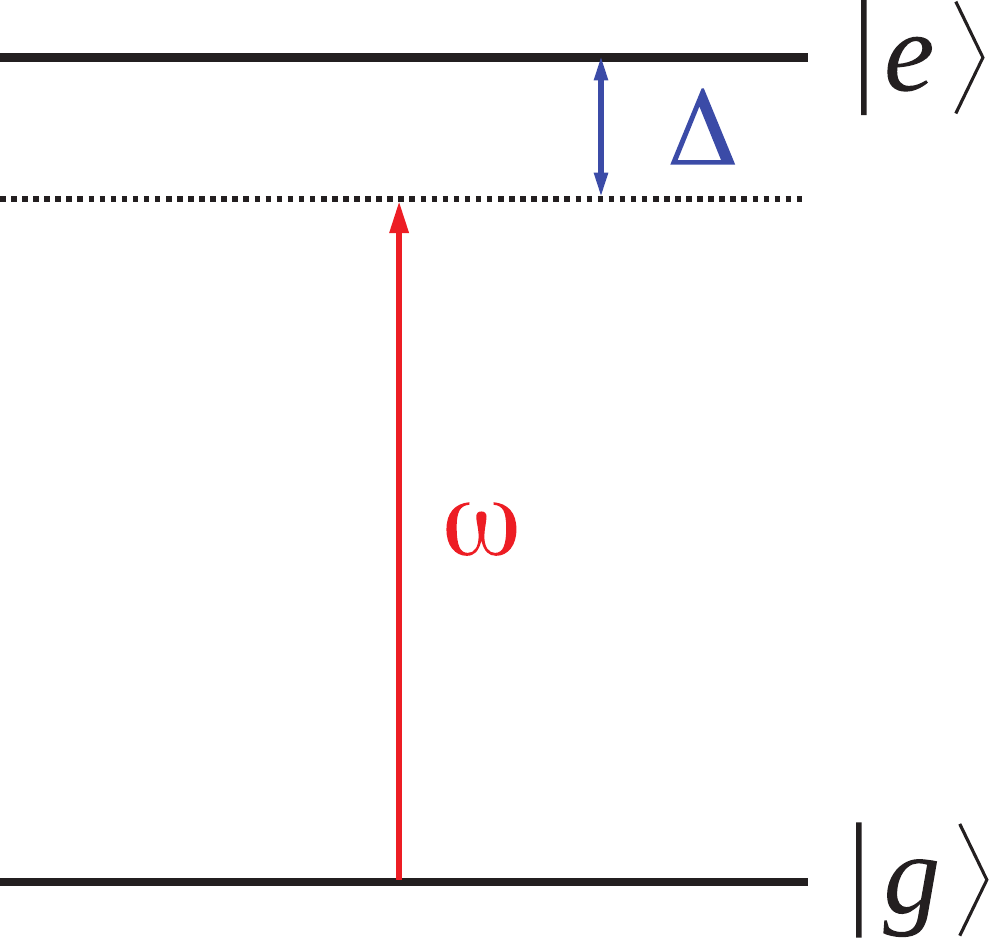}
		\caption{Two-level atom with ground state $\lvert g\rangle $ and excited state $\lvert e \rangle$ is coupled by a laser field of frequency $\omega$ that is detuned from resonance by frequency $\Delta$ as defined in the text.}\label{fig:chp5-3}
	\end{center}
\end{figure}

The Hamiltonian for a two-level stationary atom interacting with standing wave light as described in the previous section Sec.~\ref{sec:chp5:atomlight} is under the dipole approximation\index{dipole approximation} (see Sec. \ref{sec:transitions})
\begin{equation}
\label{eq:chp503}
H = \hbar\omega_g \lvert g\rangle \langle g\rvert + \hbar\omega_e \lvert e \rangle \lvert e\rangle  - \mathbf{d}\cdot\mathbf{E},
\end{equation}
where $\mathbf{d}= \mathbf{d}_{ge} \left[\lvert g \rangle\langle e \rvert + \lvert e \rangle\langle g \rvert \right]$  is the dipole moment of the atom assumed to be real, $\mathbf{d}_{ge} = e \langle g\rvert \mathbf{x} \lvert e \rangle$,   and  $\omega_{g,e}$ are frequencies of the ground  $\lvert g\rangle $ and  excited $\lvert e\rangle$ of the atom, respectively. The classical standing wave laser field $\mathbf{E}$ experienced by the atom is 
\begin{equation}
\label{eq:chp504}
\mathbf{E} = \mathbf{E}_0(xt)\cos(\omega t + \phi_L(x)),
\end{equation}
where $\mathbf{E}_0 $ is the amplitude of laser standing wave, $\omega$ is the frequency of the laser field and $\phi_L(x)$ is the phase of  laser beam. The light field couples two internal states of the atom through dipole interactions as shown in Fig.~\ref{fig:chp5-3}. The time evolution of the internal states of the atom 
\begin{equation}
\label{eq:chp505}
\lvert \psi(t)\rangle = a_g(t)\lvert g\rangle + a_e(t) \lvert e \rangle,
\end{equation}
is governed by the Schrodinger equation
\begin{equation}
\label{eq:chp506}
i\hbar \frac{d}{dt}\lvert\psi(t)\rangle = H\lvert\psi(t)\rangle,
\end{equation}
and the amplitudes $a_{g,e}$ are normalized  to unity, $|a_g(t)|^2 + |a_e(t)|^2 = 1$.  Substituting the state vector (\ref{eq:chp505}) into (\ref{eq:chp506}) gives coupled differential equations for  the amplitudes
\begin{align}
\label{eq:chp507}
i\hbar \dot{a}_g(t) & = \hbar\omega_g a_g(t) + \hbar \Omega_{ge} \cos(\omega t + \phi_L)a_e(t),\\
i\hbar \dot{a}_e(t) & = \hbar \Omega_{ge} \cos(\omega t + \phi_L)a_g(t) + \hbar\omega_e a_e(t), \nonumber
\end{align}
where 
\begin{equation}
\label{eq:chp508}
\Omega_{ge} = -\frac{ \mathbf{d}_{ge}\cdot\mathbf{E}_0(x,t)}{\hbar}.
\end{equation}
 
The cosine function can be decomposed as $\cos(\omega t + \phi_L) = ( e^{i (\omega t + \phi_L)} +  e^{-i (\omega t + \phi_L)})/2 $, which contains both fast and slow oscillatory terms. For instance, the field component $e^{-i\omega t}$ oscillates rapidly that its net effect on the ground state $\lvert g \rangle$ is zero, and vice versa. Transforming into the rotating frame\index{rotating wave approximation} according to
\begin{eqnarray}
\label{eq:chp509}
a_g(t) & = c_g(t)e^{-i\omega_g t - i\Delta t/2},\\
a_e(t) & = c_e(t)e^{-i\omega_e t + i\Delta t/2}\nonumber,
\end{eqnarray}
where $\Delta = (\omega_e - \omega_g - \omega)$, and neglecting terms that oscillate rapidly, (\ref{eq:chp507}) becomes
\begin{align}
\label{eq:chp510}
i\dot{c}_g = -\frac{\Delta}{2} c_g + \frac{\Omega_{ge}e^{i\phi_L}}{2} c_e,\\
i\dot{c}_e = \frac{\Omega_{ge}e^{-i\phi_L}}{2} c_g + \frac{\Delta}{2} c_e. \nonumber
\end{align}

To solve (\ref{eq:chp510}), we assume $\Omega_{ge}(x,t)$ is constant when the light wave is interacting with the atom.  This is true to a good approximation in most atom diffraction experiments. Defining the following parameters
\begin{equation}
\label{eq:chp511}
\tan\theta = \frac{\Omega_{ge}}{\Delta},\quad \sin\theta = \frac{\Omega_{ge}}{\Omega_r},\quad \cos\theta = \frac{\Delta}{\Omega_r}, 
\end{equation} 
where $\Omega_r = \sqrt{\Delta^2 + \Omega_{ge}^2}$ , and $0 < \theta < \pi$, the solution of  amplitudes for an atom that is initially in the ground state are found to be
\begin{align}
\label{eq:chp512}
a_g & = e^{-i(\omega_g + \Delta/2)t }\left(\cos^2\theta/2 e^{i\Omega_rt/2} + \sin^2\theta/2 e^{-i\Omega_rt/2}\right),\\
a_e  &= \frac{\sin\theta}{2} e^{-i(\omega_e - \Delta/2)t}\left(e^{-i(\Omega_rt/2 +\phi_L)} - e^{i(\Omega_rt/2 -\phi_L )}\right) . \nonumber
\end{align}
The energies $E_{g\pm}$ associated with the ground state in the presence of light are
\begin{align}
\label{eq:chp513}
E_{g \pm} & = \hbar\left[\omega_g + \frac{\Delta}{2} \pm \frac 12 \sqrt{\Delta^2 + |\Omega_{ge}|^2}\right] .
\end{align} 

The detuning~\index{detuning} $\Delta$ can be controlled using the frequency $\omega$ of laser light. For large positive detuning\index{detuning!red-detuned} $\Delta > 0$, $\theta$ is approximately zero and the state vector of the system becomes $\lvert \psi\rangle \approx e^{-iE_{g-}t/\hbar}\lvert g \rangle$, where $E_{g-} \approx \hbar\left(\omega_g - \tfrac{1}{4}\tfrac{|\Omega_{ge}|^2}{|\Delta|}\right)$. Similarly, for very large negative detuning\index{detuning!blue-detuned} $\Delta <0$, $\theta$ is roughly $\pi$ and the state vector of the system is given as $\lvert\psi\rangle \approx e^{-iE_{g+}t/\hbar}\lvert g\rangle$ where $E_{g+} \approx \hbar\left(\omega_g + \frac{1}{4}\frac{|\Omega_{ge}|^2}{|\Delta|}\right)$. Notice that in either detuning considered, the atoms are always found in the ground state $\lvert g\rangle$ while the 
excited state $\lvert e\rangle$ is negligibly occupied. The overall effect of large detuning is to raise or lower the ground state energy of the atoms by a constant amount $\frac{1}{4}\frac{|\Omega_{ge}|^2}{|\Delta|}$. It also produces a periodic potential to the atoms since $\Omega_{ge} \sim \mathbf{d}_{ge}\cdot\mathbf{E}_0(t)\cos(\kappa_lx)$, which the ground state follows adiabatically. The discussion above is carried out without damping. Nevertheless it captures the essential physics of the system even in presence of decay like spontaneous emission\index{spontaneous emission} that leads to damping. For a treatment that includes damping and decay channels see references and further reading at the end of this chapter.

\begin{exerciselist} [Exercise]
\item \label{q6-1}
By solving (\ref{eq:chp507}) given that initially the atom is in the ground state $a_g =1$, show that the solution is (\ref{eq:chp512}), and verify that the energies of the ground state $E_{g,\pm}$ are as given in (\ref{eq:chp513}).
\item \label{q6-2}
For zero detuning $\omega = \omega_e-\omega_g$, zero phases of the laser $\phi_L = 0$, and pulse duration $\Omega_r t=\pi/2$, show that the initial state $\lvert\psi(0)\rangle=\lvert g\rangle $ is transformed to the state $\lvert \psi_{\pi/2}\rangle = \{\lvert g\rangle - i\lvert e\rangle\}/\sqrt{2}$ up to some global phase factors. 
\item \label{q6-3}If the pulse duration the problem above is $\pi$ instead, what is the effect of the laser pulse on the initial state $\lvert\psi(0)\rangle=\lvert g\rangle $? What would be the effect of having two $\pi$ pulses act on the initial state $\lvert\psi(0)\rangle=\lvert g\rangle $? 
\end{exerciselist}

\subsection{Forces on atom and potential due to light field \label{sec:sec:chp5:Forces}\index{light potential}}

It is evident from the above discussion that there is a force acting on the atom due to the light which can be estimated from the Hamiltonian, and must be in accordance with Newton's first law. The starting point to relate the quantum dynamics of the system with classical mechanics is to find the evolution of the quantum operators $p, x$ in the Hamiltonian (\ref{eq:chp503}). The Heisenberg\index{Heisenberg equation of motion} equation of motion for the dynamical operators $x, p$ of  (\ref{eq:chp503}) are $i\hbar\dot{x} = [x,H]$, $i\hbar\dot{p} = [p,H]$, where the dot denotes time derivative,  and $[A,B]$ is the usual commutator.  Averaging the Heisenberg equation of motion for the operators over the atomic wave function gives (Ehrenfest equations) \index{Ehrenfest equations}
\begin{align}
\label{eq:chp514}
\langle \dot{x} \rangle & = 0,\\
\langle \dot{p} \rangle & = \langle\mathbf{d} \rangle\cdot\frac{\partial}{\partial x}\mathbf{E}(x,t)\nonumber.
\end{align}

The first equation in (\ref{eq:chp514}) shows that the time evolution  for  center of  atom wave $\langle \dot{x}\rangle$ is zero. This suggests three distinct possibilities for the atomic motion. The first case would be that the atom is stationary and so would be its wave packet. In this case the center of the wave packet would not move for all time. This possibility is ruled out by the second equation of (\ref{eq:chp514}). Another scenario would be that the atom makes a random motion about some point, such that the net classical trajectory averages to zero. A third possibility, which is not intuitive classically, is for the center of wave packet to be in a linearly superposed directed motion in opposite directions. As such, the atom is in an superposition of being at two places $-x$ and $+x$ at the same time such that on averaging the velocity of the atom's center of wave packet, one obtains zero. This ability to put the atoms in a directed motion in opposite directions will  be exploited in the splitting of a wave packet of atom. 

When the spatial derivative is applied to the field as prescribed in the second equation of (\ref{eq:chp514}), it  gives
\begin{equation}
\label{eq:chp515}
\frac{\partial}{\partial x} \mathbf{E}(x,t) = \left(\frac{\partial\mathbf{E}_0(x,t)}{\partial x} \right) \cos(\omega t + \phi_L(x)) - \mathbf{E}_0(x,t)\sin(\omega t + \phi_L(x))\frac{\partial \phi_L(x) }{\partial x}.
\end{equation} 
Thus the force $\langle \dot{p} \rangle$ is composed of two terms: the dissipative force $F_\mathrm{diss}$ and the reactive force $F_\mathrm{react}$. The dissipative force $F_{\mathrm{diss}}$ is proportional to the gradient of light phase $\frac{\partial \phi_L(x)  }{\partial x}$, and is due to the spontaneous emission\index{spontaneous emission} of light by the atom. Spontaneous emission\index{spontaneous emission} is possible when there is a finite population of atoms in the excited state. An atom with a probability $|a_e(t)|^2 \approx \tfrac{|\Omega_{ge}|^2}{\Delta^2 +|\Omega_{ge}|^2}$ of being in excited state and with natural linewidth $\Gamma$ will decay spontaneously to ground state at a rate $\Gamma |a_e(t)|^2 \approx \Gamma\tfrac{|\Omega_{ge}|^2}{\Delta^2 +|\Omega_{ge}|^2} $. If the momentum of a photon lost in the spontaneous decay\index{spontaneous emission!decay} process is $\tfrac{\partial \phi_L}{\partial x} = \hbar\kappa_l$, then the dissipative force\index{dissipative force} experienced by the atom is $F_\mathrm{diss} \sim \Gamma\tfrac{|\Omega_{ge}|^2}{\Delta^2 + |\Omega_{ge}|^2}\hbar\kappa_l$. This type of force is used in the cooling of atoms\index{cooling of atoms}. However, for the purposes of directed motion of the atom, this process is not wanted but cannot be completely eliminated. It can however, be controlled using  light fields that are strongly detuned from atomic resonance transitions, $\Delta \gg 0$ and $|\Omega_{ge}|/\Delta \ll 1$. 

The reactive force $F_\mathrm{react}$\index{light potential!reactive force} is given by the term proportional to the gradient of field amplitude. To estimate this force, we will ignore the contributions from the gradient of the field phase as discussed in the preceding paragraph. Hence we consider a standing wave field of the form $\mathbf{E}(x,t ) = \mathbf{E}_0(x,t)\cos\omega t$, with spatial derivative
\begin{equation}
\label{eq:chp516}
\frac{\partial}{\partial x}\mathbf{E}(x,t) = \left(\frac{\partial\mathbf{E}_0(x,t)}{\partial x}\right)\cos\omega t.
\end{equation}
The average dipole moment $\langle \mathbf{d} \rangle = \langle \psi \lvert \mathbf{d}\rvert\psi\rangle$ within the rotating frame defined in (\ref{eq:chp509}) gives 
\begin{equation}
\label{eq:chp517}
\langle \mathbf{d} \rangle = \mathbf{d}_{ge}\sin\theta\cos\theta\left[ \cos(\Omega_r t)  - 1\right].
\end{equation} 
This choice for calculating the dipole is a matter of convenience. Nevertheless, it gives the leading order term necessary to describe the effects of the reactive force. If the bare atom states $a_{g,e}$ are used instead of the rotating frame state $c_{g,e}$, the term not captured in this approximation is of the order $\left| \tfrac{\Omega_{ge}}{\Delta}\right |^3$, which is negligible in regime of interest, $\Delta \gg 0$, and $ |\Omega_{ge}|/\Delta \ll 1$. From  (\ref{eq:chp517}), the dipole oscillates at frequency $\Omega_r/2$, so the atom responds to the field at frequencies lower than or equal $\Omega_r/2$. For oscillations greater than $\Omega_r/2$, the atom will not have felt the effect of the field appreciably. Substituting (\ref{eq:chp516}) and (\ref{eq:chp517}) into (\ref{eq:chp514}) for $\langle\dot{p}\rangle$ and averaging over one period of the fastest oscillation $T = 2\pi/(\Omega_r + \omega)$ in the limit $(\Omega_r -\omega)T$, $\omega T\ll 1$ gives  the reactive force 
\begin{equation}
\label{eq:chp518}
F_\mathrm{react} = \frac{\hbar\Delta}{4}\frac{1}{\Delta^2 + \Omega_{ge}^2} \frac{\partial}{\partial x}( |\Omega_{ge}|^2),
\end{equation}
where the definition of $\Omega_{ge}$, (\ref{eq:chp508}) has been used, and intensity of light is proportional to $|\Omega_{ge}|^2$. We see that the reactive force $F_\mathrm{react}$ depends on the detuning of laser light.  For red-detuned light\index{detuning!red-detuned} $\Delta >0$,  the atom is attracted to  region of high light intensity. Similarly, for blue-detuned light\index{detuning!blue-detuned} $\Delta <0$, the atom is attracted to region of low light intensity, and will be repelled from region of high light intensity. This points to the force $F_\mathrm{react} = -\nabla V$ being  conservative, thus derivable from a potential $V$ given by 
\begin{equation}
\label{eq:chp519}
V = -\frac{\hbar \Delta}{4} \ln \left[1 + \left(\frac{\Omega_{ge}}{\Delta}\right)^2\right].
\end{equation}
For large detuning, $V \approx -\tfrac{\hbar \Delta}{4}  \tfrac{ |\Omega_{ge}|^2 }{\Delta^2}$, which is exactly the shift in the energy of atom calculated before in Sec~\ref{sec:sec:chp5:evolution}. For $\Delta >0$, the atom will be attracted to intensity maximum which appears as a minimum of the potential $V$. Such potential minima can be used for trapping atoms.

\begin{exerciselist} [Exercise]
    \item \label{q6-4}
    Let $\lvert\psi\rangle = c_g \lvert g\rangle + c_e\lvert e \rangle$ where $c_{g,e}$ are the solutions of (\ref{eq:chp510}), prove (\ref{eq:chp517}). Show that if the bare state solutions $a_{g,e}$ are used, there would be an additional term of order $\lvert \Omega_{g,e}\rvert^2/\lvert\Delta\rvert^2$.
    \item \label{q6-5}
    Prove (\ref{eq:chp518}) and obtain the potential $V$ (\ref{eq:chp519}). Also, obtain the potential $V$ in the large detuning limit, $\Delta > 0,|\Omega_{g,e}|$.
\end{exerciselist}


%
\section{Bragg diffraction by standing wave light \label{sec:chp5:squarewave}\index{diffraction!Atom diffraction}}
The diffraction of atomic beams can be classified into two regimes, the Raman-Nath\index{diffraction!Raman-Nath diffraction} regime where only the momentum is conserved,  and the Bragg regime where both the momentum and energy conservation are met. As such an atomic beam in Raman-Nath diffraction can be diffracted into many orders, while in Bragg diffraction\index{diffraction!Bragg diffraction}, only few diffraction orders are possible. In diffraction experiments, the momentum distribution of the atom shows a few narrow peaks. In this section we examine Bragg diffraction using a square-wave pulse. Using a square wave pulse removes the requirement of resonance and adiabaticity for the laser pulses. Instead, emphasis is placed on the proper timing of the square pulses in order to achieve the desired goal.

As shown in the previous sections, an atom  in the presence of an off-resonant standing wave of light experiences a periodic force and follows the light adiabatically. In the limit of large detuning and assuming the atom has no initial momentum, the Hamiltonian governing the dynamics of an atom is 
\begin{equation}
\label{eq:chp520}
i\hbar\frac{d}{dt}\psi(x,t) = \left[-\frac{\hbar}{2m} \frac{d^2}{dx^2} + \tilde{\Omega} (t)\cos\left(2\kappa_l x\right) \right] \psi(x,t), 
\end{equation}
where $\psi(x,t)$ is the ground state wave function of the atom, $\tilde{\Omega} (t)$ is the time dependent amplitude of the applied laser pulses.  Here the optical potential $\tilde{\Omega}(t)\cos(2\kappa_lx)$ presents a grating of periodicity $\lambda_l/2$ to the atom's wave function, where $\kappa_l$ and $\lambda_l$ are the wavenumber and wavelength of the laser beam, respectively. The periodicity of the grating has a characteristic width of $2\hbar\kappa_l$ in momentum space, and fulfills the Bragg condition\index{Bragg condition} 
\begin{equation}
\label{eq:chp521}
p = 2n\hbar\kappa_l,
\end{equation}
where $n$ is the diffraction order and takes integer values only, and $p$ is the momentum of the atom. Introducing the recoil frequency $\omega_\mathrm{rec} = \hbar(2\kappa_l)^2/(2m)$, dimensionless coordinate $\nu = 2\kappa x$ and  time $\tau = 2\omega_\mathrm{rec} t $, (\ref{eq:chp520}) is written in dimensionless form 
\begin{equation}
\label{eq:chp522}
i\frac{\partial}{\partial \tau} \psi(\nu,\tau) = \left[-\frac 12 \frac{\partial^2}{\partial \nu^2} + \Omega(\tau)\cos\nu \right]\psi(\nu,\tau),
\end{equation}
where $\Omega(\tau) = \tilde{\Omega}(t)/(2\omega_\mathrm{rec})$. The above equation belongs to a special class of differential equations called the Mathieu equations and its solutions cannot be written in terms of elementary functions in general. However, approximate solutions can be found if the properties of the equation are understood. \index{Mathieu equations}

In Fourier space, if the width of an atom's wave function is much smaller than the  length of the grating wavevector, then $\psi(\nu,\tau)$ consists of a series of narrow peaks. It is therefore convenient to expand $\psi(\nu,\tau)$ as 
%
\begin{equation}
\label{eq:chp523}
\psi(\nu,\tau) = \sum_{n=-\infty}^{\infty}\phi_n(\tau)e^{in\nu},
\end{equation}
where $\phi_n(t)$ are the time varying amplitudes. Substituting (\ref{eq:chp523}) into the Schrodinger equation, (\ref{eq:chp522}) gives 
\begin{equation}
\label{eq:chp524}
i\dot{\phi}_n =  \frac{n^2}{2}\phi_n + \left(\phi_{n-1} + \phi_{n+1}\right)\frac{\Omega(\tau)}{2}.
\end{equation}
%
Equation (\ref{eq:chp524}) comprises an infinite set of coupled differential equations. In the Raman-Nath diffraction\index{diffraction!Raman-Nath diffraction} regime, the interaction time between the atoms and the laser pulses is short, resulting in the kinetic energy of the atoms being small in comparison with the optical potential energy presented by the laser pulses (second term of (\ref{eq:chp524})). Hence the kinetic energy term is neglected in solving (\ref{eq:chp524}), and the resulting solution is given in terms of Bessel functions.  

In the Bragg limit, the kinetic energy term is no longer negligible. For some given diffraction order $L$, only diffraction orders $n =0,\pm 1,\pm 2, \cdots,\pm(L-1) $ less than $L$ are excited. Thus the series can be truncated if $L^2 \gg \Omega(\tau)$. For the lowest order diffraction $L=2$, we have $n=0,\pm1$, leading to the following three coupled differential equations \index{Bragg diffraction}
\begin{equation}
\label{eq:chp525}
i\left(\begin{array}{c}
\dot{\phi}_{-1}\\
\dot{\phi}_0\\
\dot{\phi}_1
\end{array}
\right) = \frac{1}{2}\left(\begin{array}{ccc}
1 & \Omega(\tau) &0\\
\Omega(\tau) & 0 & \Omega(\tau)\\
0 & \Omega(\tau) & 1
\end{array}\right)\left(\begin{array}{c}
\phi_{-1}\\
\phi_0\\
\phi_1
\end{array}\right).
\end{equation}
The states $\phi_{\pm 1}$ correspond to atomic packets moving with momenta $\pm1$ in Fourier space, while $\phi_0$ corresponds to atomic wave packet with zero momentum in the Fourier space. The symmetric nature of the couplings in  (\ref{eq:chp525}) means that we can eliminate one variable using the transformation $\eta_{\pm} = \tfrac{\phi_1 \pm \phi_{-1}}{\sqrt2}$, reducing it to two coupled differential equations and an uncoupled differential equation:
\begin{align}
\label{eq:chp526}
i\left(\begin{array}{c}
\dot{\phi_0}\\
\dot{\eta_+} 
\end{array}\right)& = \frac{1}{2}\left(\begin{array}{cc}
0 & \sqrt{2}\Omega(\tau)\\
\sqrt{2}\Omega(\tau) & 1
\end{array}\right) \left(\begin{array}{c}
\phi_0\\ \eta_+
\end{array}\right),\\
\label{eq:chp527}
i\dot{\eta_-} &= \frac{1}{2}\eta_-.
\end{align} 
The transformation $\eta_-$, which does not couple to $\phi_0$, has a solution for the initial condition $\eta_-(0)$
\begin{equation}
\label{eq:chp528}
\eta_- (\tau) = \eta_-(0)e^{-i\tau/2}.
\end{equation}  
This is the state that is called a {\it dark state} \index{dark state} and is often encountered in three-level systems. To solve (\ref{eq:chp526}), take the function $\Omega(\tau)$ a square pulse of the form $\Omega(\tau) =  \Theta(\tau) - \Theta(\tau + \tau_1)  $, where $\Theta(\tau)$ is the Heaviside step function, and the amplitude is $\Omega_0$. The solution to (\ref{eq:chp526}) for the initial condition $\phi_0(0)$, $\eta_+(0)$ is 
\begin{equation}
\label{eq:chp529}
\left(\begin{array}{c}
\phi_0(\tau)\\
\eta_+(\tau)
\end{array}\right) = e^{-i\frac{\tau}{4}}\left(\begin{array}{cc}
M_{11} & M_{12}\\
M_{12}& M_{11}^*
\end{array}\right)\left(\begin{array}{c}
\phi_0(0)\\\eta_+(0)
\end{array}\right),
\end{equation}
where 
\begin{align}
\label{eq:chp530}
M_{11} &= \cos\left(\frac{\sqrt{1 + 8\Omega_0^2}}{4}\tau\right) - \frac{i}{\sqrt{1 + 8\Omega_0^2}}\sin\left(\frac{\sqrt{1+8\Omega_0^2}}{4}\tau\right),\\
\label{eq:chp531}
M_{12}& =\frac{ 2\sqrt{2}\Omega_0 i}{\sqrt{1+ 8\Omega_0^2}}\sin\left(\frac{\sqrt{1+8\Omega^2}}{4}\tau\right).
\end{align}
Equation (\ref{eq:chp529}) gives the evolution of the atoms' wave packet in the presence of laser compound pulses for different initial conditions. A particular case of interest in many experiments using cold atoms is the coherent population transfer initially at zeroth order $n=0$ to first order $n=\pm1$,  corresponding to atoms moving left and right of the atomic wave packet with momentum $2\hbar\kappa$. This can be realized using two compound laser pulses with pulse amplitude $\Omega_0 = 1/\sqrt{8}$ each with a duration of $\tau_d =\sqrt{2}\pi$. The first and the second pulses are separated by a period of free evolution lasting for a time $\tau_\mathrm{free} = 2\pi$  during which the laser pulses are switched off, and clouds are allowed to rephase. The evolution matrix for the splitting sequence is thus described as
\begin{equation}
\label{eq:chp532}
U_{0\leftrightarrow +} = e^{-i\frac{\pi}{\sqrt{2}}}\left[\begin{array}{cc}
0 & -1 \\
-1 & 0
\end{array}\right].
\end{equation} 
The expression (\ref{eq:chp532}) shows that the total population in $\phi_0$ is transferred to the state $\eta_+$ at the end of the second pulse as described. This transfer of population to $\eta_+$ happens in such a way that $\phi_+ $ and $\phi_-$ are equally populated. This is because according to (\ref{eq:chp527}) and (\ref{eq:chp528}), the evolution of the population difference has to be zero at all times if initially $\eta_-(0) = 0$.
 
The analytic expression for the solution of (\ref{eq:chp525}) can be obtained directly by assuming that the light amplitude $\Omega(\tau)$ is a square wave as done in the preceding paragraph. The solutions obtained in this case is
\begin{equation}
\label{eq:chp533}
\left(\begin{array}{c}
\phi_{-1}(\tau)\\
\phi_0(\tau)\\
\phi_1(\tau)
\end{array}\right) = e^{-i\tau/4}\left(\begin{array}{ccc}
A_{11} & A_{12} & A_{13}\\
A_{12} & A_{22} & A_{12}\\
A_{13} & A_{12} & A_{11}
\end{array}\right)\left(\begin{array}{c}
\phi_{-1}(0)\\
\phi_0(0)\\
\phi_1(0)
\end{array}\right),
\end{equation} 
where 
\begin{align}
\label{eq:chp534}
A_{11} & = \frac 12\left[\cos\left(\frac{\sqrt{1 + 8\Omega_0^2}\tau}{4}\right) + e^{-i\tau/4} - \frac{i}{\sqrt{1 + 8\Omega_0^2}} \sin\left(\frac{\sqrt{1 + 8\Omega_0^2}\tau}{4} \right)\right],\\
\label{eq:chp535}
A_{12} & = -2i\frac{\Omega_0}{\sqrt{1+ 8\Omega_0^2}}\sin\left(\frac{\sqrt{1 + 8\Omega_0^2}\tau}{4} \right),\\
\label{eq:chp536}
A_{13} & = \frac 12\left[\cos\left(\frac{\sqrt{1 + 8\Omega_0^2}\tau}{4}\right) - e^{-i\tau/4} - \frac{i}{\sqrt{1 + 8\Omega_0^2}} \sin\left(\frac{\sqrt{1 + 8\Omega_0^2}\tau}{4} \right)\right],\\
\label{eq:chp537}
A_{22} & = \cos\left(\frac{\sqrt{1 + 8\Omega_0^2}\tau}{4}\right) + \frac{i}{\sqrt{1 + 8\Omega_0^2}} \sin\left(\frac{\sqrt{1 + 8\Omega_0^2}\tau}{4} \right).
\end{align}
Evolving the initial state $[\phi_{-1}, \,\phi_0\,, \phi_1]^T = [0\, ,\, 1\, ,0]^T$ under same parameters considered in the preceding section gives the same result. In fact, it is easily seen that after Bragg pulses are applied in the sequence as described, two wave packets of equal amplitude in a superposition of mode-entangled states emerge.

\begin{exerciselist} [Exercise]
    \item \label{q6-6}
    By solving (\ref{eq:chp525}), obtain the solution (\ref{eq:chp533}) for an atom that is initially stationary $\phi_0 = 1$, $\phi_{\pm} = 0$.  
    \item \label{q6-7}
    In diffraction experiment with atoms, it is desired to put  atoms in zero-momentum state $\phi_0 = 1$  into a superposition of state with atoms moving either to the left or to the right, $\phi_0 \rightarrow (\phi_{+1} + \phi_{-1})/\sqrt{2}$, and none in the stationary state $\phi_0 =0$. Using the results of Exercise~\ref{q6-6}, find the splitting matrix that achieves this. Can the splitting be achieved by a single pulse? Compare the obtained result with answer to Exercise~\ref{q6-2}
    \item \label{q6-8}
    Laser pulses can be used as mirrors whereby they reverse the motion of atoms in the moving state $\phi_\pm \rightarrow \phi_\mp$. Find the reflection matrix.
\end{exerciselist}

%
\section{Bragg diffraction by Raman pulses \label{sec:chp5:ramanpulse}}
\index{Bragg diffraction}

\begin{figure}[t]
	\begin{center}
		\includegraphics[angle=0,width= 0.7\textwidth]{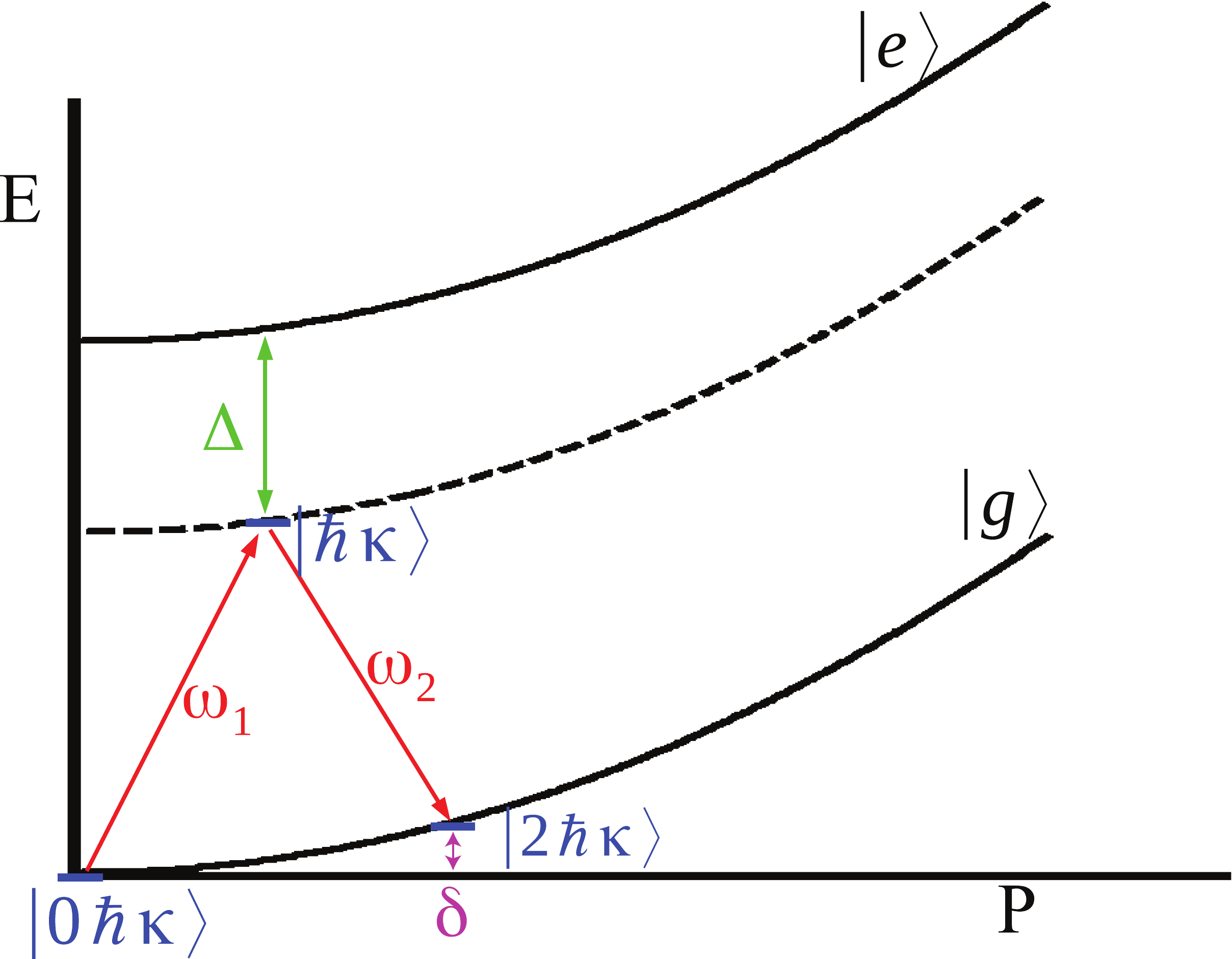}
		\caption{The interaction of an atom with a detuned light of frequencies $\omega_1$ and $\omega_2$. In (a), the frequencies couple same ground state to an excited state such that at the end of a cycle the atom returns to its ground state. In (b), the atom returning to the ground state no longer has the same momentum.}\label{fig:chp5-4}
	\end{center}
\end{figure}

Another scheme for diffraction of matter waves uses the Raman pulses as shown in Fig. \ref{fig:chp5-4}. In this technique two linearly polarized laser pulses with frequencies $\omega_1 = \omega_0$ and $\omega_2 = \omega_0 + \delta$ detuned from the atomic resonance are incident on an atom. The frequency difference $\delta$ is chosen to be an integer multiple of the recoil frequency $2\hbar\kappa_l$, where $\kappa_l$ is the wave number  of the laser. An atom exposed to linearly polarized light makes a transition to the excited state allowed by the dipole transition rule. Since linearly polarized light consists of light with $\sigma^+$ and $\sigma^-$ polarization, the two lasers provide channels for excitation and de-excitation of the atoms from virtual states. For instance, an atom absorbing a photon from light of any of frequency and de-excites via stimulated emission into the same frequency gains no energy. Thus its momentum change after the round trip is zero. However, an atom absorbing a photon from frequency $\omega_2$ and de-exciting via stimulated emission into light with frequency $\omega_1$ gains a net energy $\hbar\delta$. The net energy $\hbar\delta$ gained by the atom changes the kinetic energy of the atom. Because the light at frequencies $\omega_1$ and $\omega_2$ both couple to the same ground state, there is no change in the internal state of the atom at the end of the diffraction. Since an atom that is excited via $\omega_1$ may return to the ground state via emission through the same frequency only because of energy conservation, the observed atomic motion is only in one direction when compared to the method of Sec.~\ref{sec:chp5:squarewave}. This diffraction technique contrasts with a Raman transition where both the internal and motional state of an atom is different from what it was at the beginning of the process. Because the transitions in momentum space resemble Raman transitions, it is also called a Raman transition in momentum space. \index{Raman transition}

%
\subsection{Evolution of the ground state momentum family\label{sec:sec:chp5:familyofstates}}
Consider atoms interacting with a linearly polarized beam as described above. When the lateral dimensions of the atoms are comparable or greater than the wavelength of light,  each atom experiences a different relative phase. The external motional state of the atoms are no longer negligible. The atomic wave function is then described as a tensor product of the atom's internal energy state, and external motional state represented by momentum state. As such the internal state of the atoms is associated with many different momenta.  To see this, consider the Hamiltonian of a two-level atom 
\begin{equation}
\label{eq:chp538}
H = \frac{\mathbf{p}^2}{2m} + \hbar\omega_e \lvert e\rangle\langle e\rvert + \hbar\omega_g \lvert g\rangle\langle g\rvert - \mathbf{d}\cdot\mathbf{E}(z,t),
\end{equation}
where $\mathbf{p}$ operates on the momentum part of the atom's wave function, $\mathbf{d}$ is the electric dipole moment, and the external field $\mathbf{E}(z,t)$ on the atom is 
\begin{align}
\label{eq:chp539}
\mathbf{E}(z,t) = \frac 12 \left[\mathbf{E}_1e^{i(k_1z-i\omega_1t) } + \mathbf{E}_2 e^{i(k_2z + \omega_2t)}+ \mathrm{c.c.}\right]. 
\end{align}
Here, the  incident laser beams have wave vectors  $k_{1,2}$, frequencies $\omega_{1,2}$, and c.c. stands for complex conjugate. The operator $e^{\pm ik_mz}$ ($m = 1,2$) acts on the atom's momentum state $\lvert p\rangle$, and shifts it as 
\begin{equation}
\label{eq:chp540}
e^{\pm ik_m z} = \int dp\, \lvert p\rangle\langle p \mp \hbar k_m\rvert.
\end{equation}
Therefore an atom absorbing or emitting a photon of wave vector $k_m$ from the laser has its momentum changed by an amount $\hbar k_m$. Writing the basis state as a tensor product of internal energy and momentum $\lvert g,p\rangle$ and $\lvert e,p\rangle$, any state can be expanded in terms of these states as
\begin{equation}
\label{eq:chp541}
\lvert\psi(t)\rangle = \int dp\, \left[a_0(p,t)\lvert g,p\rangle + a_1(p,t)\lvert e,p\rangle\right].
\end{equation}
The state $\lvert g,p\rangle$ is the ground state of an atom that has momentum $p$ while $\lvert e,p\rangle$ is the excited state on atom that has momentum $p$.

The evolution of the state $\lvert\psi(t)\rangle$ is governed by the Schrodinger equation $i\hbar \tfrac{d}{dt}\lvert\psi(t)\rangle = H \lvert\psi(t)\rangle$. Substituting (\ref{eq:chp541})  and (\ref{eq:chp538}) in the Schrodinger equation, and equating coefficients in same basis $\lvert g,p\rangle,\, \lvert e,p\rangle$ give the following coupled differential equations 
\begin{align}
\label{eq:chp542}
i\dot{c}_1(p) & =\frac 12 [\Omega_1^*e^{-i(\omega_1 -\omega_{eg} + \omega_{-k_1})t} c_0(p-\hbar k_1) + \Omega^*_2e^{-i(\omega_2 -\omega_{eg} +\omega_{k_2})t} c_0(p + \hbar k_2)],\\
\label{eq:chp543}
i\dot{c}_0(p) & = \frac 12 [\Omega_2e^{i(\omega_2 -\omega_{eg} - \omega_{-k_2})t} c_1(p-\hbar k_2) +  \Omega_1e^{i(\omega_1 -\omega_{eg} -\omega_{k_1})t} c_1(p + \hbar k_1)],
\end{align}
where $\omega_{eg} = \omega_e - \omega_g$ and the Rabi frequencies are $\Omega_i = -\langle g\lvert \mathbf{d}\cdot\mathbf{E}_i\rvert e\rangle/\hbar$, $i=1,2$. Note for brevity in (\ref{eq:chp542}) and (\ref{eq:chp543}), $c_0(p,t)$ is written as $c_0(p)$. The variables $a_0(p,t),\,a_1(p,t)$ are related to $c_0(p,t), \,c_1(p,t)$ as
\begin{align}
\label{eq:chp544}
a_0(p,t) &= c_{0}(p,t) \exp\left[-i\left(\frac{p^2}{2\hbar m} + \omega_g\right)t \right]\nonumber,\\
a_0(p\pm\hbar k,t)& = c_{0}(p\pm\hbar k,t) \exp\left[-i\left(\frac{(p \pm \hbar k)^2}{2\hbar m} + \omega_g\right) t\right],\\
a_1(p,t) &= c_{1}(p,t) \exp\left[-i\left(\frac{p^2}{2\hbar m} + \omega_e\right)t\right]\nonumber,\\
a_1(p\pm\hbar k,t)& = c_{1}(p\pm\hbar k,t) \exp\left[-i\left(\frac{(p \pm \hbar k)^2}{2\hbar m} + \omega_e\right)t \right]\nonumber.
\end{align}
Terms that are fast rotating at frequencies $\omega_i + \omega_{eg}$ ($i=1,2$) were neglected in arriving at (\ref{eq:chp542})  and (\ref{eq:chp543}). The transformation (\ref{eq:chp544}) and the neglect of fast terms at frequency $\omega_i + \omega_{eg}$  is known as rotating wave approximation. \index{rotating wave approximation}

Equations (\ref{eq:chp542})  and (\ref{eq:chp543}) do not lend to easy analytical treatment because  the coefficients depend on time. We begin our analysis by examining (\ref{eq:chp542}), which describes the coupling between the set of momenta $\lvert g,p-\hbar k_1\rangle$, $\lvert e,p\rangle$ and $\lvert g,p +\hbar k_1\rangle$. The two ground states $\lvert g,p-\hbar k_1\rangle$ and $\lvert g,p+\hbar k_2\rangle$ are coupled to the excited state  $\lvert e,p\rangle$. The coupling between the ground state $\lvert g,p-\hbar k_1\rangle$ and the excited state $\lvert e,p\rangle$ is mediated at frequencies $\omega_1$. Similarly the coupling between the  excited state $\lvert e,p\rangle$ state and the ground state $\lvert g,p+\hbar k_2\rangle$ is mediated at frequency $\omega_2$.  The detunings of laser light from the atomic resonance transition for the two transitions at wave vector  of magnitude $k_{1},k_2$  are given by
\begin{align}
\label{eq:chp545}
\Delta_{{-k_1}} & = \omega_1 - \omega_{eg} + \frac{\hbar {k}^2_1}{2\hbar m} - \frac{{p}{k_1}}{m},\\
\Delta_{k_2} & = \omega_2 - \omega_{eg} + \frac{\hbar {k}_2^2 }{2\hbar m} + \frac{{p}{k_2}}{m}.\nonumber
\end{align}
In comparison with (\ref{eq:chp510}), the above detunings have additional terms. There is an additional  photon-recoil frequency $ \tfrac{\hbar {k}^2 }{2 m}$ term which corresponds to the kinetic energy of an atom with momentum $\hbar k$, and a Doppler shift frequency $\frac{pk}{m}$. 

\index{adiabatic elimination}
It is then possible to eliminate the  excited state adiabatically from the coupled differential equations (\ref{eq:chp542}) and (\ref{eq:chp543}), if the Rabi frequencies $\Omega_{1,2}$ are small in comparison with the detuning, and the coefficients $c_0(p+\hbar k_2,t)$ and $c_0(p -\hbar k_1,t)$ in (\ref{eq:chp542}) are slow varying functions of time, within the interval specified by the detuning (\ref{eq:chp545}). Integrating (\ref{eq:chp542}) gives
\begin{equation}
\label{eq:chp546}
c_1(p,t) = \frac 12\left[\frac{\Omega_1}{\Delta_{{-k_1}}} e^{-i\Delta_{{-k_1}}t}c_0(p-\hbar k_1,t) + \frac{\Omega_2^*}{\Delta_{k_2}}e^{-i\Delta_{k_2} t} c_0(p + \hbar k_2,t)\right].
\end{equation}
Substituting  (\ref{eq:chp546}) in (\ref{eq:chp543}) gives a three term recursive differential equation for the evolution of the ground state
\begin{align}
\label{eq:chp547}
i\dot{c}_0(p,t) &= \left[\frac{|\Omega_1|^2}{4\Delta_{{-k_1}}}e^{-i\frac{\hbar k_1^2}{m}t}  + \frac{|\Omega_2|^2}{4\Delta_{{k_2}}}e^{-i\frac{\hbar k_2^2}{m}t}\right]c_0(p,t) +\\
 &\frac{\Omega_2\Omega_1^*}{4\Delta_{{-k_1}}} e^{-i\delta_1t} c_0(p -\hbar K,t) +  \frac{\Omega_1\Omega_2^*}{4\Delta_{{k_2}}} e^{i\delta_2t} c_0(p +\hbar K,t)\nonumber,
\end{align}
where $K = k_1 + k_2$, and 
\begin{align}
\label{eq:chp548}
\delta_1 &= \omega_1 - \omega_2 +\frac{\hbar(k_1^2 + k_2^2)}{2m} - \frac{pK}{m} ,\\
\label{eq:chp549}
\delta_2 &= \omega_1 - \omega_2 -\frac{\hbar(k_1^2 + k_2^2)}{2m} - \frac{pK}{m}.
\end{align}

Equation (\ref{eq:chp547}) shows coupling between three different ground state momentum families $\lvert g,p\rangle$, $\lvert g,p \pm \hbar K \rangle$ at frequencies  $\delta_{1,2}$. Since the frequencies $\delta_{1,2}$ are different, only one  of the  $\lvert g,p \pm \hbar K \rangle$ family is coupled to the $\lvert g,p\rangle$ at resonance. To see this, if the lasers are detuned such that $\omega_1 - \omega_2 = \tfrac{p'K}{\hbar} + \tfrac{\hbar(k^2_1+ k^2_2)}{2m}$, $\delta_1 = \tfrac{\hbar(k^2_1 + k^2_2)}{m} + \tfrac{(p'-p)}{m}K$, and $\delta_2 = \tfrac{(p'-p)}{m}K$, then the frequency $\delta_1$ oscillates fast compared to $\delta_2$. At resonance, which is a special case where $p=p'$, the phases of the  $c_0(p,t)$ and $c_0(p - \hbar K)$ terms in (\ref{eq:chp547}) are roughly the same order of magnitude and their contribution to the evolution of  $\lvert g,p\rangle$ averages to zero. Then, only the state $\lvert g,p + \hbar K\rangle$ contributes to the evolution of $\lvert g,p\rangle$. Using similar analysis for evolution of the $\lvert g,p + \hbar K\rangle$ state, one arrives at the following effective coupled differential equations
\begin{align}
\label{eq:chp550}
i\dot{c}_0(p,t) & = \frac{\Omega_1\Omega^*_2}{4\Delta_{{k_2}}} c_0(p+\hbar K,t),\\
i\dot{c}_0(p + \hbar K, t) & = \frac{\Omega_2\Omega_1^*}{4\Delta_{-k_1}} c_0(p,t).\nonumber
\end{align}
The solution of (\ref{eq:chp550}) for the initial condition $c_0(p,t=0)$, $c_0(p+\hbar K,t=0)$ is 
\begin{equation}
\label{eq:chp551}
\left(\begin{array}{c}
c_0(p,t)\\
c_0(p+\hbar K,t)
\end{array}\right) = \left(\begin{array}{cc}
\cos\left(\dfrac{\Omega t}{2}  \right) & -\dfrac{-iW}{\Omega}\sin\left(\dfrac{\Omega t}{2} \right)\\
-\dfrac{-iW}{\Omega}\sin\left(\dfrac{\Omega t}{2} \right) & \cos\left(\dfrac{\Omega t}{2}  \right)
\end{array}\right) \left(\begin{array}{c}
c_0(p,0)\\
c_0(p+\hbar K,0)
\end{array}
\right),
\end{equation}
where  $W$ and the effective Rabi frequency $\Omega$ are
\begin{align}
\label{eq:chp552}
\Omega^2 & = \frac{1}{4} \frac{|\Omega_1 |^2}{\Delta_{-k_1}}   \frac{|\Omega_2 |^2}{\Delta_{k_2}},\\
\label{eq:chp553}
W & = \frac{\Omega_1\Omega_2^*}{2\Delta_{k_2}}.
\end{align}
Equation (\ref{eq:chp551}) gives the amplitudes of finding atoms in the states $\lvert g,p\rangle$ and $\lvert g,p + \hbar K\rangle$ moving with momenta $p$ and $p + \hbar K$, respectively. Since the detuning depends sensitively on the momentum $p$, only a fraction of atoms that meet the resonance condition  are transferred to  $\lvert g,p+\hbar K\rangle$.  For $c_0(p,0)=1$ and $c_0(p +\hbar K,0)= 0$, the probabilities of finding  atoms in the state $\lvert g,p\rangle$ and $\lvert g,p+\hbar K\rangle$ are proportional to 
\begin{align}
\label{eq:chp554}
\left|c_0(p )\right|^2 &= \cos^2\left(\frac{\Omega t}{2}\right),\\
\label{eq:chp555}
\left|c_0(p +\hbar K)\right|^2 &= \left|\frac{W}{\Omega}\right|^2\sin^2\left(\frac{\Omega t}{2}\right),
\end{align}
respectively. As inferred from (\ref{eq:chp554}), for $\Omega t=\tfrac{\pi}{2}$ half of the population initially in the ground state $\lvert g,p\rangle$ moving with momentum $p$ is transfered to ground state momentum family $\lvert g,p -\hbar K\rangle$, $\lvert g,p+\hbar K\rangle$, and $\lvert g,p+ 2\hbar K\rangle$ each moving with momentum $p-\hbar K$, $p + \hbar K$ and $p + 2\hbar K$, respectively. However, of the population  transfered to the momentum state families, only the fraction $W^2/\Omega^2 = \frac{\Delta_{-k_1}}{\Delta_{k_2}}$ is found in the state $ \lvert p +\hbar K\rangle$. This is because  $W^2/\Omega^2 = \frac{\Delta_{-k_1}}{\Delta_{k_2}}$ depends on the momentum of atoms, from (\ref{eq:chp545}). Unless this ratio is unity, the ground state momentum family $\lvert g,p-\hbar K\rangle$, $\lvert g,p+ 2\hbar K\rangle$  other than the intended $\lvert g,p+\hbar K\rangle$ are likely to be populated with finite probability. For a special case where $p = 0$ and $\omega_2 \approx \omega_1$, the ratio of laser detuning is roughly unity,  and one-half of the atomic population is transferred to the target state $\lvert g,p + n\hbar K \rangle$, with atoms in that state having momentum $p + n\hbar K$, ($n = \pm1,\,  \pm2,\,\cdots$). This is routinely achieved with atomic BEC samples where the momentum distribution of the atoms is centered at  $p=0$. Similar analysis can extended to the reflection of atoms using Bragg optical pulses.

\begin{exerciselist} [Exercise]
    \item \label{q6-9}
    Substitute  (\ref{eq:chp540}) and (\ref{eq:chp541}) in the Schrodinger equation and obtain the coupled differential equations (\ref{eq:chp542}) and (\ref{eq:chp543}).
    \item \label{q6-10}
    Apply the rotating wave approximation and adiabatic elimination to the results of the Exercise~\ref{q6-9}, and obtain the differential equation for the evolution of the ground state probability amplitudes $c_0(p,t)$, and $c_0(p-\hbar K,t)$, see for example (\ref{eq:chp547}).     
    \item \label{q6-11}
    Solve the coupled differential equations of Exercise~\ref{q6-10} and use the solutions to find the splitting matrix that takes atoms initially in the state$\lvert g,p\rangle$ to the state $ (\lvert g,p\rangle + \lvert g,p-\hbar K\rangle)/\sqrt{2}$; $\lvert g,p\rangle \rightarrow (\lvert g,p\rangle + \lvert g,p-\hbar K\rangle)/\sqrt{2}$. What can you say about the reflection matrix? 
\end{exerciselist}

\section{References and further reading}
\begin{itemize}
    \item Sec.\ref{sec:chp5:theory}: For a general introduction to the theory of diffraction see the following texts~\cite{young2012, goodman1996}.
    \item Sec.\ref{sec:chp5:atomlight}: The first experiment demonstrating the diffraction of an atom, which used a mechanical grating is detailed here~\cite{carnal1991}.  For an introduction to interactions between atoms and radiation~\cite{dalibard1985, allen1975, shore1990, cohen1998, metcalf1999}. The text by Ref.~\cite{metcalf1999} is a very good resource on the topic of trapping and manipulating atoms using light.
    \item Sec.~\ref{sec:chp5:squarewave}: Several theory proposals~\cite{marte1992, gupta2001, wu2005, stickney2007} and experimental realizations are given in Refs. \cite{martin1988, rasel1995, oberthaler1999, wang2005, garcia2006, hughes2007, burke2009}.
    \item Sec.~\ref{sec:chp5:ramanpulse}: Bragg diffraction of atoms using  Raman pulses~\cite{kozuma1999, torii2000, sadgrove2007, kasevich1991, moler1992, peters1999, peters2001, james2000, brion2007, horikoshi2007}.
	\item For a more detailed analysis of a two-level or multi-level atom interacting with light employing a semi-classical approach including the 
	effects of spontaneous emission may be found in the books \cite{allen1975, shore1990}.
	\item For a quantum treatment of atom interaction with light see \cite{cohen1998, gardiner2004, fray2004atomic}.  
	\item Effect of the spontaneous emission on the Bragg diffraction is discussed in the book~\cite{kazantsev1990}.
\end{itemize}

\chapter{Atom Interferometry \label{ch:atominterferometry}}

\section{Introduction\label{sec:chp6:intro}}
An interferometer is a measuring device that uses waves for its operation. It works on the principle of interference of waves; any object (not limited to physical objects) that can alter the path of the  waves introduces a phase shift that leads to interference effects. The information about the object can then be estimated from the interference pattern. The most common type of interferometer uses light wave for its operation. Examples of interferometers are the Michelson interferometer, the Mach-Zehnder interferometer and the Fabry-Perot interferometer. The Michelson interferometer played a significant role in the understanding of light at the turn of 19\textsuperscript{th} century. \index{interferometer!Michelson interferometer}\index{interferometer!Mach-Zehnder interferometer}\index{interferometer!Fabry-Perot interferometer}\index{interferometer}

Matter-waves too can be used in the operation of an interferometer. Typical examples include neutron and atomic BEC interferometers. As already discussed in the last chapter, atoms can be manipulated using light to perform diffraction and other interference effects. The analogue of optical elements such as beam-splitters and mirrors are made of light. These elements are then used to split and diffract atomic BECs in an atom interferometer. This feat has been demonstrated in a number of BEC experiments that will be discussed in this chapter. We begin our discussion in the next section by first looking at optical interferometers. This is followed by the discussion of atomic BEC interferometry.

\section{Optical Interferometry\label{sec:chp6:opticalinterferometer}}

\begin{figure}[t]
	\begin{center}
		\includegraphics[angle=0,width= \textwidth]{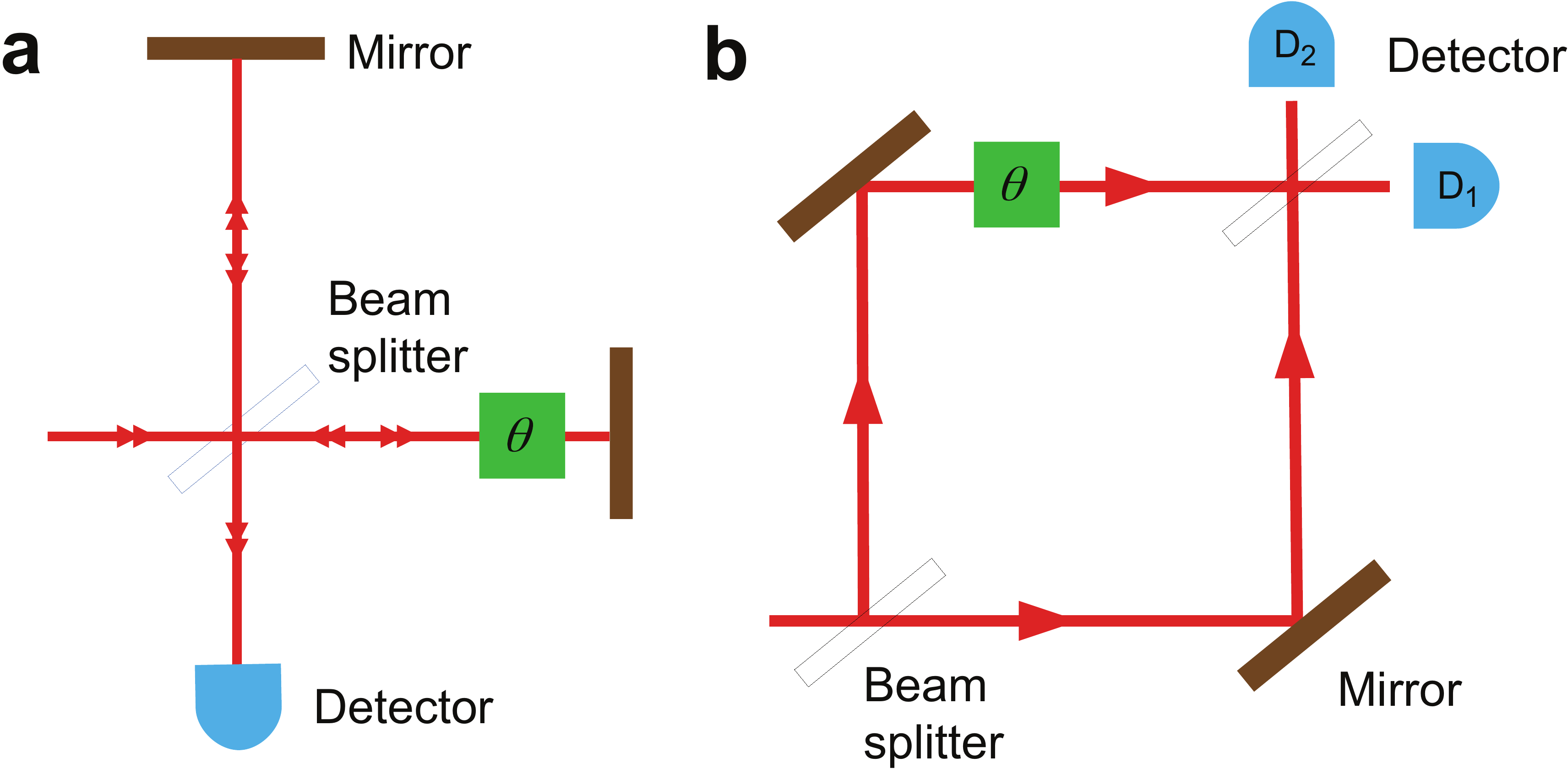}
		\caption{A schematic representation of a (a) Michelson interferometer and  (b)  Mach-Zehnder interferometer. The shaded box $\theta$ represents the relative phase between the two arms of the interferometer.}\label{fig:chp6-1}
	\end{center}
\end{figure}

We first describe two types of optical interferometers: the Michelson and Mach-Zehnder interferometers. Their principle of operation is closely related to the atom BEC interferometers described in the preceding sections. A Michelson interferometer is shown in Fig~\ref{fig:chp6-1}(a). In a Michelson interferometer, coherent light enters a beam splitter and splits it into two parts that are allowed to evolve along different paths while being guided by mirrors. The split beams are then recombined by a beam splitter again to produce interference. The splitting and recombination of the light beam takes place at the same location, and the outgoing light is detected to observe the interference effect.   \index{interferometer!Michelson interferometer}

In a Mach-Zehnder interferometer as shown in Fig~\ref{fig:chp6-1}(b), the splitting of the incident  light beam and the recombination of the split light beams occur at different locations. If light of a particular phase is fed into the Mach-Zehnder interferometer, and there is a relative phase $\theta$ between the two arms as shown,  it will be partly observed at detector D\textsubscript{1} with probability $P_1 = \tfrac{1 + \cos\theta}{2}$, while the remaining will be observed at detector D\textsubscript{2} with probability $P_2 = \tfrac{1 - \cos\theta}{2}$. Notice that if the paths traversed by the split beams are identical $\theta = 0$, only detector D\textsubscript{1} clicks and the light is not split between the two detectors. \index{interferometer!Mach-Zehnder interferometer}

 In general this result can be written as 
\begin{equation}
\label{eq:chp601}
P = \frac{1 + {\cal V}\cos\theta}{2}
\end{equation} 
where ${\cal V}$ is the visibility \index{visibility}of the interference fringes and is usually defined with respect to the intensity maximum and minimum as follows
\begin{equation}
\label{eq:chp602}
{\cal V} = \frac{I_{\textrm{max}} - I_{\textrm{min}}}{I_{\textrm{max}} + I_{\textrm{min}}}.
\end{equation}
The visibility can be considered a measure of the coherence of the interfering beams.  For $ {\cal V} = 1$ corresponding to perfect visibility as discussed above, one is able to measure accurately the interference signal. For ${\cal V} < 1$, an interference signal can still be observed. However, if  ${\cal V}\ll 1$  it becomes difficult  to observe anything (for ${\cal V} = 0$ nothing is observed).

\section[BEC interferometry]{BEC interferometry\label{sec:chp6:BEC interferometry}}

As discussed in previous sections, in a BEC the matter waves of the bosons are coherent, forming a macroscopic quantum wave.  For light, the formation of coherent light is the key to observing interference effects, as otherwise the final interference is a mixture of different phases, and no consistent interference pattern is obtained.  We thus expect that it is possible to obtain similar interference effects as those discussed in the previous section with BECs. 

One of the most notable experiments showing matter wave interferometry is the  demonstration in 1997 of interference between two independently  prepared atomic condensates. It is striking because the atomic waves\index{atom wave} that were interfered did not share the same coherence initially. However, the interference\index{interference} was observed to vary from shot-to-shot (i.e. each run of the experiment). Since then, a number of atomic BEC interferometry experiments have been demonstrated in trapped-atom and guided-wave interferometers. In trapped atom interferometers, a condensate in a potential well is divided into two condensates by deforming the well into a double-potential well. The split condensates are allowed to evolve before the double well is switched off allowing the atomic BEC cloud to expand and interfere. 

Interference of atomic BECs  have also been demonstrated using guided-wave atom\index{interference} interferometers where a condensate in trapping potential is split into two condensates using light. The arms of the potential well act as a guiding path for the split condensates. During the interferometric cycle, reflective laser pulses may be used to reverse the momentum of the split atomic condensates. At the end of the interferometric cycle, a light beam identical to the one used in the splitting of an atomic BEC cloud is used to recombine the split clouds.  \index{interferometer!guided-wave atom interferometer}

The  contrast\index{visibility} observed in both the trapped-atom and guided-wave interferometers is degraded by atom-atom interaction within each split condensate. The quantum state of atomic BEC can be represented by a number state consisting of $ N$  atoms. When this state is split into two condensates with $ n_1 $ and $ n_2 $ atoms in each condensate, the presence of atom-atom interactions causes each number state to evolve at different rates, and result in the accumulation of a relative time dependent phase between the BECs. As a result, the interference fringes are lost at recombination. This effect is called phase diffusion\index{phase diffusion}. In addition, the spatially dependent phase, induced by the atom-atom interactions and the confining harmonic trap\index{harmonic trap}, degrades the observed contrast of guided-wave interferometers. In the rest of this chapter, we will analyze phase diffusion effects on guided-wave atom interferometry where the standing-wave light pulses of Sec.~\ref{sec:chp5:squarewave} and Sec.\ref{sec:chp5:ramanpulse} are used in the operation of interferometer.\index{interferometer!guided-wave atom interferometer}



\subsection{Splitting of trapped atomic condensate \label{sec:sec:chp6:BECsplitting}}
Consider an atomic BEC cloud with wavefunction $\psi_0 (\bm{x}) $ at rest in a confining harmonic potential\index{harmonic trap}. Applying the splitting laser pulses as discussed in Sec.~\ref{sec:chp5:squarewave},  the initial cloud at rest $\psi_0 (\bm{x}) $ evolves to mode-entangled states $\psi_\pm (\bm{x}) $ moving in opposite directions, where each state is a linear superposition of a number state with $ N$  atoms.  The clouds are allowed to evolve for a duration $T$ after which they are subjected to recombination laser pulses that are identical to the splitting laser pulses. After applying the recombination pulses, atoms in general populate all the three modes $\psi_0 (\bm{x}) $, $\psi_\pm (\bm{x}) $ as a result of of the relative phase accumulated by the split clouds during their evolution. 

The many-body Hamiltonian describing the atomic BEC in the presence of an external potential $V$ is 
\begin{equation}
\label{eq:chp603}
H (t) = \int d^2\bm{x}\, \Psi^\dagger \left[-\frac{\hbar^2}{2m}\nabla^2 + V + \frac{U_0}{2} \Psi^\dagger \Psi \right] \Psi,
\end{equation}
where $m$ is the atomic mass, $U_0 = 4\pi\hbar^2 a_s m^{-1}$ is the strength of the two-body interaction within the condensate, $a_s$ is the $s$-wave scattering length, $\Psi^\dagger$ is the creation field operator, which at a given time $t$ creates an atom at position $\bm{x}$. Performing a transformation in a similar way to that done in (\ref{basisan}), let $b^\dagger_0$, $b^\dagger_{+1}$ and $b^\dagger_{-1}$ be the operators which, upon acting on vacuum state create an atom belonging to a cloud at rest, and moving to the right and left, respectively. They also satisfy the bosonic commutation relation $\left[b_i,b_j^\dagger\right] = \delta_{ij}$, $\left[b_k,b_k\right] = \left[b_k^\dagger,b_k^\dagger\right] = 0$, where $i,j,k = 0,\pm$. The field operator $ \Psi $ may be expanded in harmonics moving to the left and right as 
\begin{equation}
\label{eq:chp604}
\Psi (\bm{x},t) = b_{+1} \psi_{+1}(\bm{x},t)  + b_{-1}\psi_{-1}(\bm{x},t),
\end{equation}
where $\psi_\pm(\bm{x},t) $ are the wavefunctions of the BEC moving to right and left, respectively. They are solutions of the Gross-Pitaevskii equations, and are normalized to unity, $\int d\bm{x}\, |\psi_\pm(\bm{x},t)|^2 = 1$.\index{Gross-Pitaevskii equation}

Substituting (\ref{eq:chp604}) into (\ref{eq:chp603}) gives the following Hamiltonian
\begin{equation}
\label{eq:chp605}
H_{\mathrm{eff}} = \frac{W}{2} (n_+ - n_-) + \frac{g}{2} \left[{\cal N}^2 + (n_+ - n_-)^2 - 2{\cal N}\right]
\end{equation}
where $n_{\pm1} = b^\dagger_{\pm1}b_{\pm1}$, $ {\cal N} = n_{+1} + n_{-1}$, $W = \varepsilon_+ - \varepsilon_-$ is the relative environment-introduced energy shift between the right- and  left- propagating clouds $\varepsilon_\pm  = \int d^3\bm{x} \psi_{\pm1}^* \left[-\frac{\hbar^2}{2M} \nabla^2 + V\right] \psi_{\pm1}$, and 
\begin{align}
\label{eq:chp606}
g  = \frac{U_0}{2}\int d^3\bm{x}\, |\psi_{+1}|^4 =  \frac{U_0}{2}\int d^3\bm{x}\, |\psi_{-1}|^4,
\end{align}
is the coefficient characterizing the strength of atom-atom interaction within each cloud. In arriving at (\ref{eq:chp605}), a term that introduces constant energy shift has been omitted. 

The initial state vector of the condensate, before the splitting laser pulses are applied, is well-described by a number state with fixed  atom number $ N $
\begin{equation}
\label{eq:chp607}
\lvert \Psi_{\mathrm{ini}} \rangle = \frac{(b^\dagger_0)^N}{\sqrt{N!}}\lvert 0\rangle,
\end{equation}
where $\lvert 0 \rangle$ is the vacuum state. The splitting and recombination pulses couple the bosonic operators $b_0^\dagger$, $b_{\pm1}^\dagger$ as described in (\ref{eq:chp529}) or (\ref{eq:chp533})
\begin{align}
\label{eq:chp608}
b^\dagger_{+1} & \rightarrow -\frac{b^\dagger_{+1}}{2} + \frac{e^{i\pi/\sqrt{2}}}{\sqrt{2}}b_0^\dagger + \frac{b^\dagger_{-1}}{2},\nonumber\\
b_0^\dagger & \rightarrow \frac{b^\dagger_{+1}}{2} + \frac{b^\dagger_{-1}}{2},\\
b^\dagger_{-1} & \rightarrow \frac{b^\dagger_{+1}}{2} + \frac{e^{i\pi/\sqrt{2}}}{\sqrt{2}}b^\dagger_0 - \frac{b^\dagger_{-1}}{2}.\nonumber
\end{align}
The state vector (\ref{eq:chp607}) after the splitting pulses is applied becomes
\begin{align}
\label{eq:chp609}
\lvert \Psi_\mathrm{split}\rangle&  = \frac{(b_{+1}^\dagger + b^\dagger_{-1})^N}{\sqrt{2^N N!}}\lvert 0\rangle,\nonumber\\
& = \frac{1}{2^{N/2}\sqrt{N!}}\sum_{n= 0}^{N}\binom{N}{n} \left(b^\dagger_{+1}\right)^n\left(b^\dagger_{-1}\right)^{N - n}\lvert 0 \rangle ,
\end{align}
where $\binom{N}{n} =\tfrac{N!}{n!(N-n)!}$ is the binomial coefficient. \index{binomial}

The state vector $\lvert \Psi(t)\rangle$  at any other time after the splitting pulses have been applied is obtained from the time evolution of $\lvert \Psi_{\mathrm{split}}\rangle$ under the Hamiltonian (\ref{eq:chp605})
\begin{equation}
\label{eq:chp610}
\lvert \Psi(t) \rangle = e^{-i/\hbar\int H_{\mathrm{eff}}\,dt }\, \lvert \Psi_{\mathrm{split}} \rangle.
\end{equation}  
and has the form 
\begin{align}
\label{eq:chp611}
\lvert \Psi_\mathrm{split}(t)\rangle & = \frac{1}{2^{N/2}\sqrt{N!}}\sum_{n= 0}^{N}\binom{N}{n} e^{-i\frac{\theta}{2}(2 n - N) -i\frac{\varphi}{2}[2n^2 + 2 (n - N)^2] } \left(b^\dagger_{+1}\right)^n\left(b^\dagger_{-1}\right)^{N - n}\lvert 0 \rangle,
\end{align}
where 
\begin{equation}
\label{eq:chp612}
\theta = \frac{1}{\hbar}\int_{0}^{t} d\tau\, W
\end{equation}
is the accumulated phase difference between the left and right atomic clouds due to the environment, and 
\begin{equation}
\label{eq:chp613}
\varphi = \frac{1}{\hbar} \int_{0}^{t} d\tau\, g
\end{equation}
is the accumulated nonlinear phase per atom due to inter-atomic interactions within the each atomic cloud. In arriving at (\ref{eq:chp611}), the global phase term $\exp(iN\varphi)$ was neglected.

At the end of the interferometric cycle $t = T$, the recombination pulses act on $\lvert \Psi(T)\rangle$ in accordance with (\ref{eq:chp608}), and the state afterward is 
\begin{align}
\label{eq:chp614}
\lvert \Psi_{\mathrm{rec}} \rangle & = \frac{1}{\sqrt{2^N N!}}\sum_{n= 0}^{N} \binom{N}{n}e^{-i\left[\frac{\theta}{2}(2n - N) + \varphi (n^2 + (n - N)^2)\right]} \left(-\frac{b^\dagger_{+1}}{2} +  \frac{e^{i\pi/\sqrt{2}}b^\dagger_0}{\sqrt{2}} + \frac{b^\dagger_{-1}}{2} \right)^n\nonumber\\
& \times \, \left(\frac{b^\dagger_{+1}}{2} +  \frac{e^{i\pi/\sqrt{2}}b^\dagger_0}{\sqrt{2}} - \frac{b^\dagger_{-1}}{2} \right)^{N-n} \lvert 0 \rangle .
\end{align}

\begin{exerciselist} [Exercise]
    \item \label{q7-1}
    Using the results of Exercise~\ref{q6-7} verify (\ref{eq:chp608}).
    \item \label{q7-2}
    Verify (\ref{eq:chp611}).     
\end{exerciselist}

\subsection[Probability density]{Probability density\label{sec:sec:chp6:ProbabilityDensity}}
After recombination, the particle numbers for the clouds at rest $n_0$, moving left $n_-$ and moving right  $n_+$ are measured.  The state describing this result is given by
\begin{equation}
\label{eq:chp615}
\lvert n_+, n_-, n_0\rangle = \frac{\left(b^\dagger_{+1}\right)^{n_+}}{\sqrt{n_+!}}\frac{\left(b^\dagger_{-1}\right)^{n_-}}{\sqrt{n_-!}}\frac{\left(b^\dagger_{0}\right)^{n_0}}{\sqrt{n_0!}}\lvert 0 \rangle.
\end{equation}
The probability of measuring atoms in the state $\lvert n_+, n_-, n_0\rangle$ after recombination is given by the modulus square of the probability amplitude $\langle n_+, n_-, n_0 \rvert \Psi_{\mathrm{rec}}\rangle$. Using (\ref{eq:chp614}) and (\ref{eq:chp615}) the probability amplitude becomes
\begin{align}
\label{eq:chp616}
\langle n_+, n_-, n_0 \rvert \Psi_{\mathrm{rec}}\rangle & = \frac{1}{\sqrt{2^N N!}}\sum_{n= 0}^{N} \binom{N}{n}e^{-i\left[\frac{\theta}{2}(2n - N) + \varphi (n^2 + (n - N)^2)\right]}\nonumber\\
& \times \langle 0 \lvert \frac{\left(b^\dagger_{0}\right)^{n_0}}{\sqrt{n_0!}}\frac{\left(b^\dagger_{-1}\right)^{n_-}}{\sqrt{n_-!}}\frac{\left(b^\dagger_{+1}\right)^{n_+}}{\sqrt{n_+!}}\left(-\frac{b^\dagger_{+1}}{2} +  \frac{e^{i\pi/\sqrt{2}}b^\dagger_0}{\sqrt{2}} + \frac{b^\dagger_{-1}}{2} \right)^n\nonumber\\
& \times  \left(\frac{b^\dagger_{+1}}{2} +  \frac{e^{i\pi/\sqrt{2}}b^\dagger_0}{\sqrt{2}} - \frac{b^\dagger_{-1}}{2} \right)^{N-n}\lvert 0\rangle,
\end{align}
where
\begin{align*}
\left(-\frac{b^\dagger_{+1}}{2} +  \frac{e^{i\pi/\sqrt{2}}b^\dagger_0}{\sqrt{2}} + \frac{b^\dagger_{-1}}{2} \right)^n
  \left(\frac{b^\dagger_{+1}}{2} +  \frac{e^{i\pi/\sqrt{2}}b^\dagger_0}{\sqrt{2}} - \frac{b^\dagger_{-1}}{2} \right)^{N-n}  =\nonumber \\
  \sum_{j=0}^{n}\sum_{k=0}^{N-n} \binom{n}{j}\binom{N-n}{k}
  \left(-1\right)^{n-j}\left(\frac{b^\dagger_{0}e^{i\frac{\pi}{\sqrt{2}}}}{\sqrt{2}} \right)^{j+k}  \left(\frac{b^\dagger_{+1} - b^\dagger_{-1}}{2} \right)^{N-k-j}.
\end{align*}

\subsubsection[Probability in absence on atom-atom interactions]{Probability in absence on atom-atom interactions \label{sec:sec:sec:chp6:ProbabilityXiZero}}
Here we consider the case where there is no atom-atom interactions  among  the atoms  in the split clouds during evolution. In this case $\varphi = 0$, and the probability amplitude (\ref{eq:chp616}) takes the form
\begin{equation}
\label{eq:chp617}
\langle n_+, n_-, n_0 \rvert \Psi_{\mathrm{rec}}\rangle = \sqrt{\frac{N!}{2^N n_+! n_-! n_0!}}(-1)^{n_-} \left(-i\sin\frac{\theta}{2}\right)^{N-n_0}\left(\sqrt{2}\cos\frac{\theta}{2}\right)^{n_0},
\end{equation}
and the probability density $P(n_+,n_-,n_0) = |\langle n_+, n_-, n_0 \rvert \Psi_{\mathrm{rec}}\rangle|^2$ is given by
\begin{equation}
\label{eq:chp618}
P(n_+,n_-,n_0) = \frac{1}{2^N}\frac{N!}{n_+! n_-! n_0!}\left(\sin^2\frac{\theta}{2}\right)^{N- n_0}\left(2\cos^2\frac{\theta}{2}\right)^{n_0}. 
\end{equation}
Equation (\ref{eq:chp618}) is a binomial distribution and can be written as a product of two probability density functions\index{binomial}
\begin{equation}
\label{eq:chp619}
P(n_+,n_-,n_0) = P_\pm (n_+, n_-) P(n_0),
\end{equation}
where
\begin{equation}
\label{eq:chp620}
P_\pm (n_+, n_-) = \frac{(N - n_0)!}{2^{N-n_0}n_+!n_-!},
\end{equation}
and 
\begin{equation}
\label{eq:chp621}
P(n_0) = \frac{N!}{n_0!(N - n_0)!}\left(\sin^2\frac{\theta}{2}\right)^{N- n_0}\left(\cos^2\frac{\theta}{2}\right)^{n_0}.
\end{equation}

The function $P_\pm$ describes the probability of observing $n_+$ and $n_-$ atoms in the right and left moving clouds respectively for a fixed number of atoms in the cloud at rest. This function is independent of the phase angle $\theta$ and is normalized to unity. The function $P$ is the probability of observing $n_0$ atoms in cloud at rest. It is normalized to unity and depends on the phase angle $\theta$ introduced by the environment. 

For very large population of atoms ${N}\gg 1$, the factorials may be approximated using Stirling's formula $n! = \sqrt{2\pi n}n^ne^{-n}$, and the probability densities that correspond to $P $ and $P_\pm$ become\index{Stirling's approximation}\index{Probability}
\begin{equation}
\label{eq:chp622}
P(n_0) = \frac{2}{\sqrt{2\pi N}\sin\theta}\exp\left[-\frac{2}{N} \frac{(n_0 - N\cos^2(\theta/2))^2}{\sin^2\theta} \right],
\end{equation}
and 
\begin{equation}
\label{eq:chp623}
P_\pm(n_+, n_-) = \sqrt{\frac{2}{\pi (n_+ + n_-)}}\exp\left[-\frac{2}{n_+ + n_-} \left(\frac{n_+ - n_-}{2}\right)^2 \right],
\end{equation}
where $n_+ + n_- \gg 1$.

\begin{figure}[t]
	\begin{center}
		\includegraphics[angle=0,width= \textwidth]{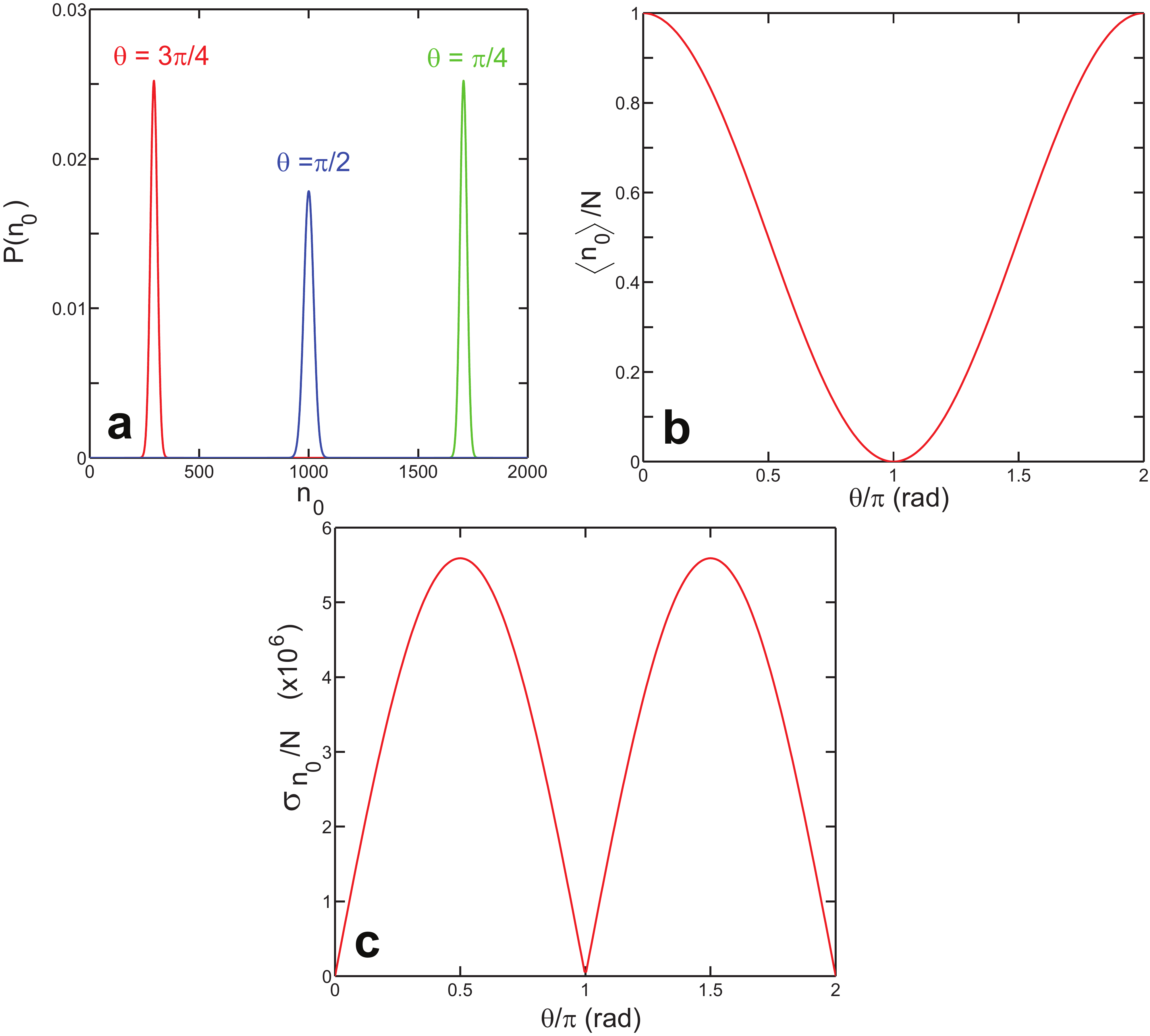}
		\caption{(a) The probability density $P(n_0)$ as a function of  $n_0$ at three different values of $\theta$. (b) The relative mean value $\langle n_0 \rangle/N $ as a function of $\theta$. (c) The relative standard deviation $\sigma_{n_0}/N$ as a function of  $\theta$. }\label{fig:chp6-3}
	\end{center}
\end{figure}

Both the probability densities $P_\pm$ and $P$ are Gaussian. For a fixed value of $n_0$ atoms in the stationary cloud, the peak of the probability function $P_\pm$ is located at $(N- n_0)/2$, with an average values of $n_+$ and $n_-$ given by
\begin{equation}
\label{eq:chp624}
\langle n_+ \rangle = \langle n_- \rangle = \frac{1}{2} (N - n_0),
\end{equation}
and standard deviations
\begin{equation}
\label{eq:chp625}
\sigma_{n_+} = \sigma_{n_-} = \frac{1}{2} \sqrt{N - n_0}.
\end{equation}
The number of atoms in the right and left clouds are anti-correlated according to
\begin{equation}
\label{eq:chp626}
\mathrm{cov}(n_+,n_-) = \langle n_+ n_- \rangle  - \langle n_+ \rangle  \langle n_- \rangle = -\frac{1}{4} (N - n_0).
\end{equation}\index{Probability!covariance}

The maximum of the probability density function $P(n_0)$ is located at $n_0 = N\cos^2(\theta/2)$. Since $n_0 $ takes values in the interval $[0, N]$, $\theta$ takes values in the interval $ 0 < \theta < \pi$. The end points $\theta = 0$ and $\theta = \pi$ are excluded because the probability density (\ref{eq:chp622}) is not defined at the end points. However, the values of $P(n_0)$ at the end points may be obtained using (\ref{eq:chp621}). This gives $P(n_0) = 1$ for $\theta = 0, \pi$.  The probability density $P(n_0 = N)= 1$ for $\theta = 0$ means that all the atoms are in the cloud at rest after recombination, and $P(n_0 = 0) = 1 $ for $\theta = \pi$ implies that no atom is observed in the cloud at rest after recombination; all the atoms are found in the clouds moving to the left and right after recombination.

Figure \ref{fig:chp6-3}(a) shows a plot of the probability density (\ref{eq:chp622}) at three different values of $\theta$. The width of each peak on the graph scales roughly as $\sqrt{(N\sin^2\theta)/4}$ so that the relative width of the distribution scales roughly as $ \sqrt{\sin^2\theta/(4N)}$. Because of the dependence of the width of the distribution function on $\theta$, the width of the probability density is largest at $\theta = \pi/2$ and vanishes at $\theta = 0, \pi$. The changing values of $\theta$ move the peak of the probability density $P(n_0)$ from $n_0 = N$ corresponding to the situation where more atoms are in the stationary cloud towards $n_0 = 0$ that corresponds to situations where less number of atoms are in stationary cloud. \index{Probability}

The mean value and variance of the probability density $P(n_0)$ are 
\begin{equation}
\label{eq:chp627}
\langle n_0 \rangle = N\cos^2\frac{\theta}{2},
\end{equation}
and 
\begin{equation}
\label{eq:chp628}
\sigma^2_{n_0}  = N\cos^2\frac{\theta}{2}\sin^2\frac{\theta}{2}.
\end{equation}\index{Probability!mean}\index{Probability!variance}
Figures~\ref{fig:chp6-3}(b)(c) show the plots of the relative mean value and relative standard deviation, respectively. In Fig.~\ref{fig:chp6-3}(b), the contrast is unity and the visibility as defined in (\ref{eq:chp602}) is maximum, taking the value of unity. Thus for non-interacting condensates, full fringes would be observed in every run of the experiment. The error associated in counting the number of atoms in the stationary cloud shows sinusoidal oscillations with a periodicity of $\pi$ as shown in Fig. \ref{fig:chp6-3}(c). At $\theta = 0, m\pi$ (where $m$ is any integer value), the standard deviation is zero and corresponds to situations where all the atoms are known with absolute certainty to be either in the cloud at rest or in the moving clouds. At this point, the width of the probability density vanishes as previously described above. Even values of $m$ correspond to the case when all the atoms are in the cloud at rest, while odd values of $m$ correspond to the case when all the atoms are in the moving clouds. The standard deviation is maximum at odd multiples of $ \pi/2$ as shown in Fig. \ref{fig:chp6-3}(c).  This can also be seen from the width of the distribution at $\theta = \pi/2$ in Fig. \ref{fig:chp6-3}(a), which occurs when equal population of atoms are found in the moving clouds and the cloud at rest.\index{Probability!mean}\index{Probability!variance}

\subsubsection[Probability in presence of atom-atom interactions]{Probability in presence of atom-atom interactions \label{sec:sec:sec:chp6:ProbabilityXiNotZero}}
Calculating the probability density when $\varphi \neq 0$ seems a daunting task. However, due to orthogonality of the number states, all terms in (\ref{eq:chp616}) vanish except for the term $j+k =n_0$, giving\index{Probability}
\begin{align}
\label{eq:chp629}
\langle n_{+1}, n_{-1}, n_0\rvert\Psi_{\mathrm{rec}}\rangle = \sqrt{\frac{N!n_0!}{2^{(3N - n_0)} n_{+1}!n_{-1}!}  } (N-n_0)! e^{in_0\frac{\pi}{\sqrt{2}}} (-1)^{n_{-1}}\nonumber\\
\times\sum_{n=0}^{N}e^{-i\theta(n - N/2) - i\varphi[n^2 + (n -N)^2]} S(n_0,n),
\end{align}
where
\begin{equation}
\label{eq:chp630}
S(n_0,n) = \sum_{j=\max(0,n_0+n-N)}^{\min(n,n_0)}\frac{(-1)^{n-j}}{j!(n-j)!(n_0 -j)!(N-n - n_0 + j)!}.
\end{equation}
Comparing (\ref{eq:chp629}) for $\varphi = 0$ and (\ref{eq:chp617}) shows that $S(n_0,n)$ may be obtained from a Fourier transform of the trigonometric functions in (\ref{eq:chp617})
\begin{equation}
\label{eq:chp631}
S(n_0,n) = \frac{1}{2\pi}\frac{2^N}{(N-n_0)! n_0!}\int_{0}^{2\pi}d\theta\,e^{i(n-N/2)\theta} \left(\cos\frac{\theta}{2}\right)^{n_0}
\left(-i\sin\frac{\theta}{2}\right)^{N-n_0} .
\end{equation}
Substituting (\ref{eq:chp631}) into (\ref{eq:chp629}), the probability 
\begin{align}
P(n_+, n_-, n_0) = \lvert \langle n_{+1}, n_{-1}, n_0\rvert\Psi_{\mathrm{rec}}\rangle\rvert^2
\end{align}
of observing $n_-$ atoms in the clouds moving to the left, $n_+$ atoms in the cloud moving to the right, and $n_0$ atoms in the stationary cloud after recombination  may be written as a product of two functions $P(n_+, n_-, n_0) = P_\pm (n_+, n_-) P_0(n_0,\theta,\varphi)$, where $P_\pm$ defined in (\ref{eq:chp620}) gives the probability of finding $n_+$ and $n_-$ atoms in the clouds that is moving to the right and left, respectively, after recombination.  The probability density function $ P_0(n_0,\theta,\varphi)$ describes the probability of finding $n_0$ atoms in the stationary clouds after recombination, 
\begin{equation}
\label{eq:chp632}
P_0(n_0,\theta,\varphi) = \frac{N!}{n_0!(N- n_0!)} \lvert f(n_0,\theta,\varphi) \rvert^2.
\end{equation}
The  function $f(n_0,\theta,\varphi)$ is 
\begin{align}
\label{eq:chp633}
f(n_0,\theta,\varphi) & = \frac{e^{-iN^2\varphi/2}}{\sqrt{1 - 2iN\varphi}} \left[(N-n_0) \ln\sqrt{1 - \frac{n_0}{N}} + n_0\ln\sqrt{\frac{n_0}{N}}\right]\nonumber\\
&\times \left(e^{-\eta_-^2} + (-1)^{N-n_0}e^{-\eta_+^2} \right),
\end{align}
where
\begin{equation}
\label{eq:chp634}
\eta_{\pm} = \frac{N\left(\arccos\sqrt{\frac{n_0}{N}} \pm \frac{\theta}{2}  \right)^2}{1 - 2iN\varphi}.
\end{equation}
Unlike the probability density function $P_\pm$, $ P_0 $ has a non-trivial dependence on its arguments, and  contains parameters $\varphi$, $\theta$ that affect the fringes observed in the interferometer. The next section is devoted to understanding the the dependence of $ P_0 $ on its arguments. The dependence of $P_\pm$ on its arguments has been described previously in Sec.~\ref{sec:sec:sec:chp6:ProbabilityXiZero}.\index{Probability}

\begin{exerciselist} [Exercise]
    \item \label{q7-3}
    Show that powers of operators $b_0^\dagger$ $b_{+1}^\dagger$ and $b^\dagger_{-1}$ in (\ref{eq:chp616}) can be expanded as written.
    \item \label{q7-4}
    For $\varphi=0$, verify that the probability (\ref{eq:chp616}) takes the form (\ref{eq:chp617}). Show that (\ref{eq:chp617}) can be decomposed into two probability density functions (\ref{eq:chp620}) and (\ref{eq:chp621}).
    \item \label{q7-5}
    Assuming $N\gg1$ and using the Stirling's approximation prove (\ref{eq:chp621}) and (\ref{eq:chp623}). What is the assumption on $n_0$ or  $n_+ + n_-$?
    \item \label{q7-6}
    Verify (\ref{eq:chp624}), (\ref{eq:chp625}) and (\ref{eq:chp626}). 
    \item \label{q7-7}
    Verify (\ref{eq:chp627}) and (\ref{eq:chp628}).
    \item \label{q7-8}
    Show that the Fourier transform of (\ref{eq:chp630}) is (\ref{eq:chp631}).
\end{exerciselist}

\begin{figure}[t]
	\begin{center}
		\includegraphics[angle=0,width= \textwidth]{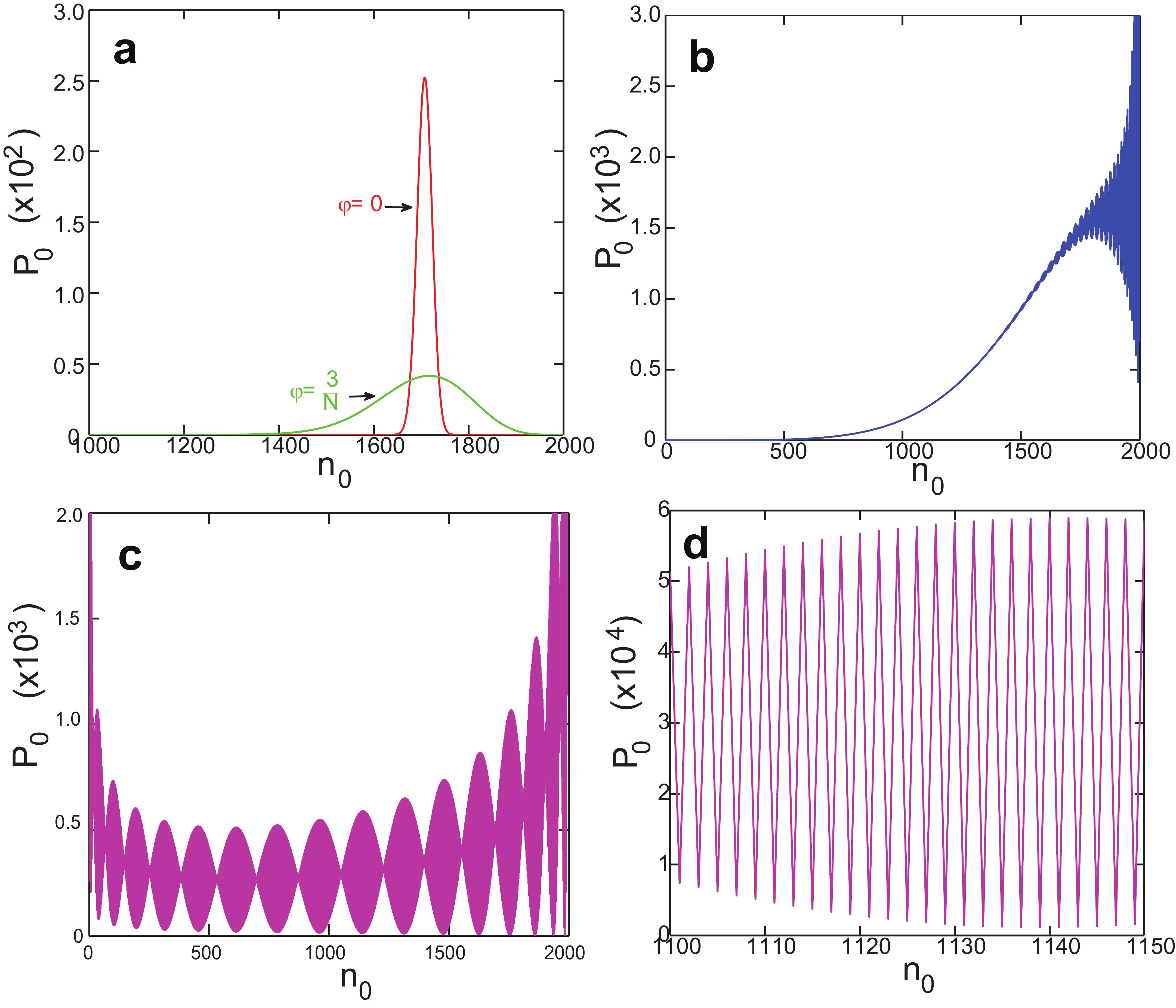}
		\caption{(a) The probability density function $P_0(n_0,\theta,\varphi)$ as function of  $n_0$ for $\varphi=0$ and $\varphi = 3/N$. (b) The probability density function $P_0(n_0,\theta,\varphi)$ as a function of $n_0$ for $\varphi=0.2/\sqrt{N}$. (c) The probability density function $P_0(n_0,\theta,\varphi)$ as a function of $n_0$ for $\varphi= 1/\sqrt{N}$. (d) An enlargement of the plot in (c) showing the fast-scale spatial oscillations of the probability density function $P_0(n_0,\theta,\varphi)$. For all plots $\theta = \pi/4$ and $N= 2000$. }\label{fig:chp6-6}\index{Probability}
	\end{center}
\end{figure}

\subsection[Features of the probability density]{Features of the probability density\label{sec:sec:chp6:ProbabilityFeatures}}
In the limit of very large number of atoms, $N,n_0\gg 1$, the factorial function can be approximated by Stirling's approximation. The probability density function  $ P_0(n_0,\theta,\varphi)$ is proportional to the modulus square of two terms
\begin{equation}
\label{eq:chp635}
P_0(n_0,\theta,\varphi) = \frac{1}{\sqrt{(1 + 4N^2\varphi^2)}}\sqrt{\frac{N}{2\pi n_0 (N-n_0)}}\big| e^{-\eta_-^2} + (-1)^{N-n_0}e^{-\eta_+^2}\big|^2,
\end{equation}
where $\eta_{\pm}$ is given in (\ref{eq:chp634}). The relative phase difference between the two terms of $P_0(n_0,\theta,\varphi)$ changes rapidly with $n_0$ due the multiplier $(-1)^{N-n_0}$. Thus, the interference terms do not contribute to the averages and are  neglected in calculating the mean $\langle n_0\rangle = \int_0^N dn_0 \, n_0 P_0(n_0,\theta,\varphi)$ and standard deviation of the probability density function. Evaluating the integral gives\index{interference}\index{Probability!mean}\index{Probability!variance}
\begin{equation}
\label{eq:chp636}
\langle n_0 \rangle = \frac{N}{2}\left[1 + \exp\left(-\frac{1 + 4N^2\varphi^2}{2N}\right)\cos\theta\right].
\end{equation}
Similarly, the variance is
\begin{equation}
\label{eq:chp637}
\sigma^2_{n_0} = \frac{N^2}{2}\left[\frac{1}{4} + \frac{\exp\left(-2\frac{1 + 4N^2\varphi^2}{N}\right)\cos2\theta}{4}  - \frac{\exp\left(-\frac{1 + 4N^2\varphi^2}{N}\right)\cos^2\theta}{2}\right].
\end{equation}

These results are understood by studying the dependence of the function $P_0(n_0, \theta,\varphi)$ on the number of atoms $n_0$ for different values of the strength of two-body atom interactions $\varphi$. At relatively small values of $\varphi $, such that $\varphi \ll 1/\sqrt{N}$, the term $\exp(-\eta_-)$ in (\ref{eq:chp635}) dominates the other. The probability is then of the form of a Gaussian \index{Probability}
\begin{equation}
\label{eq:chp638} 
P_0(n_0,\theta,\varphi) \approx \frac{1}{\sqrt{1 + 4N^2\varphi^2}}\sqrt{\frac{N}{2\pi n_0 (N-n_0)}} \exp\left[-\frac{2N\left(\theta/2 - \arccos\sqrt{n_0/N} \right)^2}{1 + 4N^2\varphi^2}\right],
\end{equation} 
with a maximum located at $n_0 = N\cos^2\theta/2$. This situation is shown in Fig. \ref{fig:chp6-6}(a). The two curves in the figure are plots of the probability density $P_0(n_0,\theta,\varphi)$ given by (\ref{eq:chp638}) versus $n_0$ for two different values of two-body atomic interaction strength $\varphi$. Both curves correspond to the same value of angle $\theta$. The noticeable feature of Fig.~\ref{fig:chp6-6}(a) is the increase in the width of the probability distribution with $\varphi$. This behavior is explained by (\ref{eq:chp637}), which in the limit $\varphi \ll 1/\sqrt{N}$ reduces to
\begin{equation}
\label{eq:chp639}
\sigma_{n_0} = \frac{\sqrt{N}}{2}\sqrt{1 + 4N^2\varphi^2}\sin\theta.
\end{equation}\index{Probability!variance}
For very small values of $\varphi \ll 1/N$, the influence of interatomic interactions on the operations of the atom interferometer is negligible. The relative standard deviation of the number of atoms in the central cloud is inversely proportional to the square root of the total number of atoms in the system: $ \sigma_{n_0}  \propto 1/\sqrt{N}$. For $1/N \ll \varphi \ll 1/\sqrt{N} $, the width of the distribution grows linearly with the increases in $\varphi$. The mean value of $n_0$  for $\varphi \ll 1/\sqrt{N}$ reasonably corresponds to the position of the peak. Equation~(\ref{eq:chp636}) for $\langle n_0 \rangle$ in this limit gives 
\begin{equation}
\label{eq:chp640}
\langle n_0 \rangle \approx \frac{N}{2}\left(1 + \cos\theta\right).
\end{equation} 
As is seen, $n_0$ depends on $\theta $ but not on $\varphi$.\index{Probability!mean}

For large values of $\varphi \approx 1/\sqrt{N}$, the two terms $\exp(-\eta_-)$ and $\exp(-\eta_+)$ in (\ref{eq:chp635}) are now comparable in magnitude. The width of the probability density $P_0(n_0,\theta,\varphi)$ becomes of the order of the total number $N$ of atoms in the systems. The transition to this limit is shown in Fig.~\ref{fig:chp6-6}(b) and Fig.~\ref{fig:chp6-6}(c). The solid regions not resolved in Fig.~\ref{fig:chp6-6}(b) and Fig.~\ref{fig:chp6-6}(c) correspond to rapid spatial oscillations with period 2. These oscillations are clearly seen in Fig.~\ref{fig:chp6-6}(d), which shows part of Fig~\ref{fig:chp6-6}(c) for a narrow range of values of $n_0$. The oscillations are caused by the interference between the two terms in (\ref{eq:chp635}). As the magnitude of $\varphi$ approaches  $1/\sqrt{N} $, the $\exp(-\eta_-)$ and $\exp(-\eta_+)$ terms become comparable in magnitude. However, because of the nearly $\pi$-phase change between the two terms every time $n_0$ changes by one due to the factor $(-1)^{N - n_0}$, the two terms constructively add in phase or out of phase when one steps through different values of $n_0$. Along with rapid oscillations, both Fig.~\ref{fig:chp6-6}(b) and Fig.~\ref{fig:chp6-6}(c) exhibit oscillations of the envelopes at a much longer timescale which are more pronounced for large values of the interaction strength. These oscillations are due to the fact that the relative phase of the terms $\exp(-\eta_-)$ and $\exp(-\eta_+)$ in (\ref{eq:chp635}) changes with $n_0$. The nodes in Fig.~\ref{fig:chp6-6}(c) correspond to the value of this relative phase being equal to $0$ or a $\pi$ and the antinodes have the phase shifted by $\pm\pi/2$.

\begin{figure}[t]
	\begin{center}
		\includegraphics[angle=0,width= \textwidth]{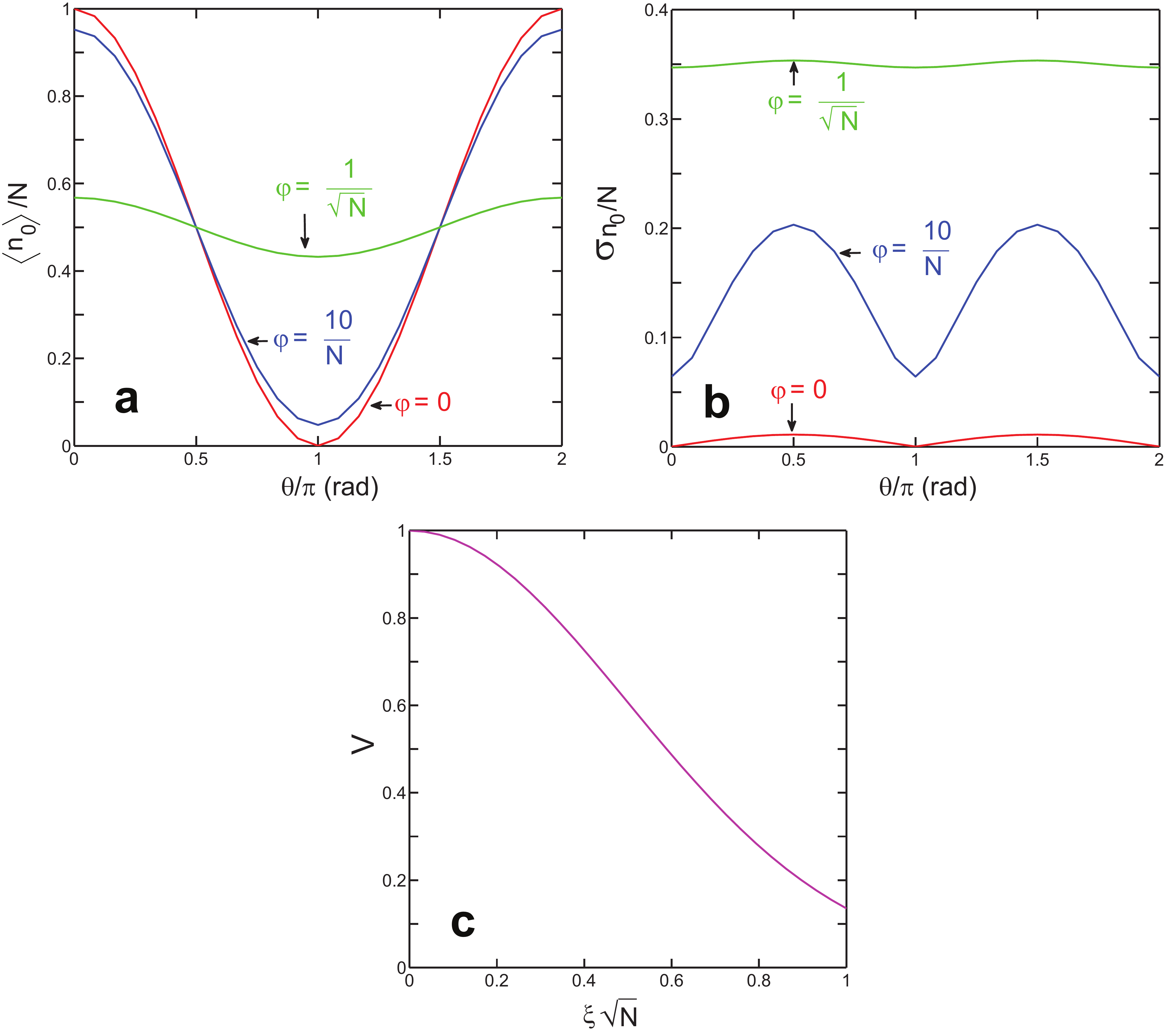}
		\caption{(a) The normalized mean value of the number of atoms in the central cloud $\langle n_0\rangle/N$ as a function of $\theta$. (b) The normalized standard deviation $\sigma_{n_0} /N$ as functions of $\theta$. (c) Interference fringe contrast ${\cal V}$ as a function of the interatomic in interactions $\varphi\sqrt{N}$. All plots use $ N = 2000$.}\label{fig:chp6-10}
	\end{center}
\end{figure}\index{Probability!mean}\index{Probability!variance}

Figures~\ref{fig:chp6-6}(b)(c) indicate that the probability $P_0(n_0,\theta,\varphi)$ and, as a consequence $\langle n_0\rangle$  and $\sigma_{n_0}$, become less sensitive to changes in the environment-introduced angle $\theta$. This fact is illustrated in Figs.~\ref{fig:chp6-10}(a)(b) showing the average value of the number of atoms $\langle n_0\rangle$ in the central cloud and the standard deviation $\sigma_{n_0}$ versus $\theta$  as given by (\ref{eq:chp636}) and (\ref{eq:chp637}), respectively. Figure \ref{fig:chp6-10}(a) demonstrates that increased interatomic interactions eventually lead to the loss of contrast of interference fringes. Additionally, larger interatomic interactions cause large shot-to-shot fluctuations in the number of atoms in each of the three ports, as seen in Fig.~\ref{fig:chp6-10}(b). The loss of interference fringe contrast can be quantified by writing (\ref{eq:chp636}) as 
\begin{equation}
\label{eq:chp641} 
\langle n_0 \rangle  = \frac{N}{2}\left(1 + {\cal V}\cos\theta \right),
\end{equation}
where the contrast ${\cal V}$ is 
\begin{equation}
\label{eq:chp642}
{\cal V} = \exp\left(- \frac{1+ 4N^2\varphi^2}{N}\right).
\end{equation}\index{visibility}
Figure \ref{fig:chp6-10}(c) shows the fringe contrast (\ref{eq:chp642}) as a function of $\varphi$ and demonstrates that the values of $\varphi$ approaching $1/\sqrt{N}$ results in the washing out of interference fringes.

\begin{exerciselist} [Exercise]
    \item \label{q7-9}
    Using (\ref{eq:chp631}) in (\ref{eq:chp629}) show that the probability of atoms in the stationary cloud may be written as (\ref{eq:chp636}), and state all your assumptions. Hint first use the steepest decent method to get (\ref{eq:chp633}) and apply Stirling's approximation to simplify your result.
    \item \label{q7-10}
    Calculate the variance and expectation value of $n_0$. Will there be any changes to expectation value and variance of atoms in the moving clouds change?  
    \item \label{q7-11}
    Show that in the limit $\varphi\sqrt{N} \ll 1$, the variance and expectation value of $n_0$ reduces to (\ref{eq:chp640}) and (\ref{eq:chp641}), respectively.
\end{exerciselist}

\subsection{Controlling nonlinear phase per atom \label{sec:sec:chp6:discussion}}
As shown in the previous section, limited interference fringes were observed when the nonlinear phase per atom $\varphi$ due to two-body interactions is about the order of $1/\sqrt{N}$. In order to quantify the effect due to two-body interactions within the condensate, the phase $\varphi$ is calculated in terms of experimental parameters. We consider experiments that are conducted in parabolic traps with confining potential of the form
\begin{equation}
\label{eq:chp643}
V = \frac{M}{2}\left(\omega_x^2x^2 + \omega_y^2y^2 + \omega_z^2z^2\right).
\end{equation}\index{harmonic trap}
The density profiles of the moving clouds are well-described by the Thomas-Fermi approximation \index{Thomas-Fermi approximation}
\begin{equation}
\label{eq:chp644}
\lvert \psi_\pm\rvert^2 = \frac{\mu_n}{U_0 n}\left(1 - \frac{x^2}{R^2_x} - \frac{y^2}{R^2_y} - \frac{z^2}{R^2_z}\right), \quad \lvert \psi_\pm\rvert^2 \geq 0, 
\end{equation}
where $R_i$ is the radial sizes of the cloud for the $i$th dimension, $\mu_n$ is the chemical potential of the BEC cloud with $n$ atoms, and $U_0 = 4\pi \hbar^2 a_{sc}/M$ is the strength of the two-body interactions within the cloud. 

Immediately after the splitting pulses are applied, the density profiles of the moving clouds are the same as that of the initial BEC cloud containing $N$ atoms, and is in equilibrium in the confining potential given by (\ref{eq:chp643}). After the splitting, each moving cloud contains on average $N/2$ atoms. The repulsive nonlinearity is no longer balanced by the confining potential and the radii of both clouds start to oscillate. The maximum size of the oscillating clouds is the equilibrium size corresponding to $N$ atoms and the minimum size lies below the equilibrium size corresponding to $N/2$ atoms. As an estimate, the radial size $R_i$ of the condensate is taken to be the equilibrium size of a cloud with $N/2$ atoms. Evaluating (\ref{eq:chp613}) we obtain the nonlinear phase per atom $\varphi$  for the duration $T$ of interferometric cycle
\begin{equation}
\label{eq:chp645}
\varphi = \frac{2}{7} \frac{\mu_n T}{n\hbar},
\end{equation}
where $\mu_n = 2^{-3/5} \mu$, $\mu$ being the equilibrium chemical potential~\citep{baym1996,dalfovo1999}\index{chemical potential}
\begin{equation}
\label{eq:chp646}
\mu = \frac{\hbar \bar{\omega}}{2} \left(15 N\frac{a_{sc}}{\bar{a}}\right)^{2/5},
\end{equation}
$\bar{\omega} = (\omega_x\omega_y\omega_z)^{1/3}$, $\bar{a} = \sqrt{\hbar/(M\bar{\omega})}$.

The relative importance of interatomic interaction effects on the operation of interferometer is determined by the parameter $P = \varphi\sqrt{N} \ll 1$, 
\begin{equation}
\label{eq:chp647}
P = 0.64\left(\frac{a_{sc}}{\bar{a}}\right)^{2/5}\bar{\omega} T N^{-1/10}.
\end{equation}
Figure~\ref{fig:chp6-10}(c) shows that the contrast of the interference fringes decreases with $P$. The condition of good contrast is in the regime  $P < 1/2$; for $P = 0.5$, the contrast ${\cal V}= 0.6$.

Equation (\ref{eq:chp647}) shows that the $P\propto T\bar{\omega}^{6/5} N^{-1/10}$ The dependence of $P$ on the total number $N$ of atoms in the BEC cloud is  very weak. Hence, the parameter $P$ is primarily dependent on the duration $T$ of the interferometric cycle and averaged frequency $\bar{\omega}$ of the trap. Equation (\ref{eq:chp647}) is handy in quantifying the amount of nonlinear phase per atom that would be present in an experiment. For example, consider an experiment where a condensate consisting of about $10^5$ \textsuperscript{87}Rb atoms are used to perform interferometry, say in a Michelson geometry with transverse and longitudinal frequencies of the trap being $177$~Hz and $5$~Hz, respectively. If the interferometry time is $10$ns and given the $s$-wave scattering length  to be $ 5.2\times 10^{-9}$m, one obtains  $P\approx 1.6 \times 10^{-2}$. This value of $P$ is very small in comparison to unity. As such, the interatomic interactions will not limit the visibility of the interference fringes obtained in the experiment.

It is of importance to quantify the amount of error one makes in estimating parameters of the measurement, such as the $\theta$ parameter in this chapter. One would expect that the large number of atoms in BEC would easily allow one to estimate $\theta$ errors with scaling better than the standard quantum limit, $1/\sqrt{N}$. However, this is not case as one can see from the probability density distribution before and after the interferometry. This is not surprising since the initial input state is a linear superposition of atomic coherent state which in the limit of large $N$ is Gaussian. Indeed, the self-interaction within each atomic condensate during propagation belongs to the family of correlations that produce squeezing---the one-axis twisting squeezed states of Sec.~\ref{sec:oneaxistwisting}. Incidentally, the measured observables $n_\pm$ or $n_0$ of interest in these experiments commute with the generator of the correlations. To observe the squeezing, one would have to find ways to calculate averages of operators that do not commute with the generators of the correlations in the condensate. Most importantly is that the self-interaction which is responsible for the nonlinear evolution would only lead to phase diffusion\index{phase diffusion}. We will be studying ways to utilize squeezing contained in quantum states for improved estimation in the next chapter.

\begin{exerciselist} [Exercise]
    \item \label{q7-12}
    One of the principal conditions required to arrive at (\ref{eq:chp636}) is that $\varphi\sqrt{N}\ll 1$. For $\varphi\sqrt{N} = 1$ verify  (\ref{eq:chp647}).
    \item \label{q7-13}
    A trap holding \textsuperscript{$87$}Rb atom condensate containing $3\times 10^3$ atoms was used in an interferometry experiment.  The trap has radial frequency fixed at $60$ Hz while the axial frequency of the trap was varied. For propagation time $T$ of $60$ ms the axial frequency was $17$ Hz, while for propagation time of $97$ ms the axial frequency was $10.29$ Hz. Comment on the effect of nonlinear phase per atom in each experiment.
\end{exerciselist}

\section{References and further reading}

\begin{itemize}
    \item Sec.~\ref{sec:chp6:opticalinterferometer}: For a general introduction to interferometry especially in optics see \cite{hecht2002, mandel1995}. 
    \item Sec.~\ref{sec:chp6:BEC interferometry}: Experiments describing BEC interferometry in trapped and guided-wave interferometers  detailing the effect of two-body interactions \cite{andrews1997b, shin2004, schumm2005, jo2007, wang2005, horikoshi2006, garcia2006, horikoshi2007, burke2008}. Theoretical works describing BEC atom interferometry and  the effects of two-body interactions \cite{javanainen1997, ilo-okeke2010, olshanii2005, stickney2007, kafle2011, ilo2014theory, ilo-okeke2016}. A theory paper giving the scattering coefficients of an atomic BEC \cite{julienne1997}.  For more review articles on the subject see~\cite{berman1997, godun2001, cronin2009}. For texts describing the steepest descent method and asymptotic evaluation of integrals see \cite{bender1978, arfken2005}.
    \item Other interferometers involving for instance fermions and other methods for splitting the atom BEC~\cite{roati2004atom, hall2007condensate, schellekens2005hanbury, jeltes2007comparison, rosi2014precision, lesanovsky2007time, fortagh2007magnetic}. 
    \item Applications of interferometers to atomtronic devices and gyroscopes \cite{seaman2007atomtronics, amico2017focus, cassettari2000beam, dumke2002interferometer, roach1995realization, deng1999four, hinds2001two, gustavson1997precision, gustavson2000rotation, dowling1998correlated}.  
    \item Other applications to precision measurements \cite{weld2009spin, blatt2008new, obrecht2007measurement}. 
\end{itemize}

\chapter[Atom interferometry beyond the standard quantum limit]{Atom interferometry beyond the standard quantum limit \label{ch:squeezedstates}}
\index{atom interferometry}
\section{Introduction\label{sec:chp7:intro}}
As pointed out in the previous chapter, the error in interferometry depends sensitively on the input state that is used in the measurement. The types of input states used in the interferometer of Chapter~\ref{ch:atominterferometry} are linear superposition of atoms in different modes. Using such states composed of $N$ independent particles in interferometry results in phase estimation that cannot be better than the standard quantum limit scaling as $1/\sqrt{N}$. However, non-linear interactions among the particles can modify the linear superposed states  such that the particles become correlated. Such correlations give rise non-classical states such as twin-Fock states\index{non-classical state! twin-Fock state}, NOON states \index{non-classical state!N00N state}, and squeezed states\index{non-classical state!squeezed state}. These non-classical states\index{non-classical state} have special feature in their distribution where the width of the distribution in one of the degrees of freedom  is less than square-root of $N$. An interferometer using these non-classical states at the input realizes phase estimation error that is better than the standard quantum limit scaling as $1/\sqrt{N}$\index{standard quantum limit}. 

In optics, passing coherent light through nonlinear materials allow for the generation of non-classical states such as the squeezed and twin-Fock states. Similarly, by using collisions, dipole interactions, or optical field modes, non-classical states are generated in condensed atoms like Bose-Einstein condensates (BECs) and atomic ensembles. The collisions in atomic systems, such as in two-component\index{two-component Bose-Einstein condensates} BECs, dipole interactions in Rydberg atoms, \index{Rydberg atom} or spin-1 (or spinor) condensates\index{spinor Bose-Einstein condensates}, produce non-linear interactions that have similar form as those found in their optics counterpart. These types of states have been extensively studied and have been suggested to provide improvements in rotation sensing, and the precision of atomic clocks. We have already examples of squeezed states in BECs in Sec. \ref{sec:squeezedstates}.  In this chapter, we further study the generation of squeezed states in two-component atomic condensates, and its characterization in an interferometric context.

\section[Two-component atomic Bose-Einstein condensates]{Two-component atomic Bose-Einstein condensates\label{sec:chp7:two-component}}\index{two-component Bose-Einstein condensates}
Multicomponent atomic condensates, such as two-component BECs or spinor condensates consist of atoms with different degrees of freedom.  The most common way of realizing this is using internal states of the atom, such as the hyperfine ground states as seen in Sec. \ref{sec:spindegrees}. When the components are put in a linear superposition, the nonlinear two-body atom interactions cause fluctuations in the relative number and phase between the components. The net effect is that a non-classical state emerges. In this section, we derive the non-classical state generated by collisions in a two-component atomic condensate. \index{two-component Bose-Einstein condensates}

\subsection[Two-mode model Hamiltonian]{Two-mode model Hamiltonian\label{sec:sec:chp7:two-mode}}\index{Two-mode model}
Consider a condensate consisting of different atomic species labeled by $a,b$, that are in the same trap.  In the case of \textsuperscript{87}Rb, the two species may be realized by the hyperfine ground states $\lvert F =1, m_f = -1\rangle$ and $\lvert F = 2, m_f = 1\rangle$  for instance. This can be achieved by population transfer after initially preparing the condensate in one of the atomic states. As discussed in Sec. \ref{sec:interaction}, the atoms interact experiencing inter- and intra-species collisions. Assuming that the inter-species collisional interactions are energetically insufficient to cause spin flips from one species to the other, the number of atoms in any species is then constant and there is no weak link\index{weak link} or Josephson oscillations between the species.\index{Josephson oscillations} The many-body Hamiltonian governing their dynamics during the nonlinear interactions is
\begin{align}
\label{eq:chp701}
H &=  \sum_{k= 1,2}\left[\int d\mathbf{r} \Psi_k^\dagger(\mathbf{r},t) {H}_0 \Psi_k(\mathbf{r},t) + \frac{U_k}{2}\int d\mathbf{r}  \Psi_k^\dagger(\mathbf{r},t)
\Psi_k^\dagger(\mathbf{r},t) \Psi_k(\mathbf{r},t) \Psi_k(\mathbf{r},t)\right]\nonumber\\
& +  U_{12} \int d\mathbf{r} \Psi_1^\dagger(\mathbf{r},t) \Psi_2^\dagger(\mathbf{r},t) \Psi_1(\mathbf{r},t) \Psi_2(\mathbf{r},t),
\end{align}
where $H_0$ is a single-particle Hamiltonian, $U_k  = 4 \pi\hbar^2 a_{\mathrm{sc}}^k/m$,  $m$ is the atomic mass, $a_{\mathrm{sc}}^{k}$ the $k$th species $s$-wave scattering length,  $U_{12}  = 4 \pi\hbar^2 a_{\mathrm{sc}}^{ab}/m$, \index{s-wave scattering length} and $a_\mathrm{sc}^{ab}$ is the inter-species scattering length\index{scattering length}. The single-particle Hamiltonian includes the confining potential that traps the atomic condensate, and includes the  effect of the environment that results in the different dynamics for condensates in different hyperfine states. 

The condensate wavefunction $\psi_k(\mathbf{r},t)$ for each\index{condensate wavefunction} component $k$ in the BEC are found by solving two coupled Gross-Pitaevskii equations~\index{Gross-Pitaevskii equation} 
\begin{align}
\label{eq:chp702}
\mu_1\psi_1(\mathbf{r},t) & = \left(H_0 + U_1n_1 |\psi_1(\mathbf{r},t) |^2  +  U_{12}n_2|\psi_2(\mathbf{r},t)|^2 \right) \psi_1(\mathbf{r},t),\\
\label{eq:chp703}
\mu_2\psi_2(\mathbf{r},t) & = \left(H_0 + U_2 n_2|\psi_2(\mathbf{r},t) |^2  +  U_{12}n_1|\psi_1(\mathbf{r},t)|^2 \right) \psi_2(\mathbf{r},t),
\end{align}
where $\mu_k$ is the chemical potential per particle\index{chemical potential}, and $n_k$ is the total number of particles in $k$th component. The condensate wavefunction $\psi_k(\mathbf{r},t)$ satisfies the normalization condition $\int d\mathbf{r}\, |\psi_k(\mathbf{r},t)|^2 = 1$.\index{condensate wavefunction}

Let $ a^\dagger_{1}$ and $ a^\dagger_{2}$ be operators that act on the vacuum state~\index{vacuum} to create an atom in components $1$ and $2$, respectively. They satisfy the commutation relation $[ a_k,a_j] =0$ and $ [ a_k, a^\dagger_j] = \delta_{kj}$. These operators are defined in (\ref{adaggerdef}). The field operators in terms of the $ a^\dagger_{1}$ and $ a^\dagger_{2}$ are 
\begin{equation}
\label{eq:chp704}
\Psi_1(\mathbf{r},t) = \psi_1(\mathbf{r},t) a_1, \qquad  \Psi_2(\mathbf{r},t) = \psi_2(\mathbf{r},t) a_2.
\end{equation}
Using (\ref{eq:chp702}), (\ref{eq:chp703}) and~(\ref{eq:chp704}) in 
 (\ref{eq:chp701}), we have
\begin{align}
\label{eq:chp705}
H & = \mu_+{\cal N} + \mu_- (n_1 - n_2)-g_+\frac{{\cal N}^2}{4} - g_- {\cal N}\left(\frac{n_1 - n_2}{2} \right)  \nonumber\\
& + \frac{1}{2}\Bigg[g_+ {\cal N} + g_-\left(n_1 - n_2\right) - 2\left(g_+ - g_{12} \right)\left(\frac{n_1 - n_2}{2} \right)^2
  + \frac{g_{12}{\cal N}^2}{2}\Bigg]
\end{align}
where ${\cal N} = n_1 + n_2$ is the total number of atoms in the two condensates, $n_1 =a^\dagger_1 a_1$, $n_2 = a^\dagger_2 a_2$, $\mu_\pm = (\mu_1 \pm \mu_2)/2$, $g_\pm = (g_{11}  \pm g_{22})/2 $, and 
\begin{align}
\label{eq:chp706}
g_{kk} & = U_k\int d\mathbf{r}\, \lvert\psi_k\rvert^4,  \qquad k = 1,2\\
\label{eq:chp707}
g_{12} & =  U_{12}\int d\mathbf{r}\, \lvert\psi_1\rvert^2 \lvert\psi_2\rvert^2.
\end{align}

The terms of the form $(a_1^\dagger a_1 - a_2^\dagger a_2)$ in  (\ref{eq:chp705})  imprints a linear phase on the atoms. The $g_-$, $g_+$, and $g_{12}$ are the two-body interactions terms. The $g_-$ imprints a linear phase that is enhanced by a factor ${\cal N}-1 $ on the atoms. Similarly, the chemical potential $\mu_-$ imprints a linear phase on the atoms. The term ($ g_+ - g_{12}$) imprints a nonlinear phase on the atoms. It is this phase that will determine if the distribution is squeezed and give the degree of the squeezing. For instance, if $g_{12}$ vanishes, then the two-component condensates\index{two-component Bose-Einstein condensates} are separated and do not overlap. Hence there is no inter-particle collisions between them. This is akin to the system studied in Sec. \ref{sec:chp6:BEC interferometry} where self interactions, that is the intra-particle collisions, dominate leading to phase diffusion.\index{phase diffusion} For $g_{12} \neq 0$ and $g_{12} <g_+$, the phase diffusion dominates the dynamics of the two-component condensates but \index{two-component Bose-Einstein condensates}some  fingerprints of the squeezing effect due to the non-negligible value of $g_{12}$ may be seen on the distribution of the atomic condensate. In the case where the phase diffusion term $g_{12} = g_+ $, the intra-particle collisions in each condensates cancels the \index{phase diffusion} inter-particle collisions between the two components. As a result, there is no non-linear phase imprinted on the atom. For $g_{12} > g_+$, the inter-particle collision between the two-component condensates dominates and drives the dynamics of the condensates. In this case, the atoms' distribution would show strong squeezing effects. We point out that none of these situations are static. The two-component states starting out with $g_{12} > g_+$ will evolve in such a manner that at some point in time $g_{12} \le g_{+}$, where the phase diffusion effect dominates the dynamics or the inter- and intra-particle collisions in the two-component condensates completely cancel out each other. Thus, generating a squeezed state would require optimizing the interactions such that the squeezing effect is not destroyed or washed out by phase diffusion. \index{two-component Bose-Einstein condensates}\index{phase diffusion}

From  (\ref{eq:chp705}), the Hamiltonian $H$, ${\cal N}$, $n_1 $, and $n_2 $ all commute. Hence, these four operators common set of eigenstates $\lvert N,n_1,n_2\rangle$. In what follows, we assume to work within the total particle sector of $ N $, and ignore all terms in the Hamiltonian only involving the total particle number $\cal N$ as they merely add a constant energy shift to the Hamiltonian.  We hence write  (\ref{eq:chp705}) as  
\begin{align}
\label{eq:chp708}
H_\mathrm{eff} & = \mu_- (n_1 - n_2)- g_- (\mathcal{N}-1)\left(\frac{n_1\ - n_2}{2} \right)  - \left(g_+ - g_{12} \right)\left(\frac{n_1 - n_2}{2} \right)^2 .
\end{align}

The operators $a_k^\dagger$ and $a_k$ may be related to the angular momentum operators in the same way as (\ref{schwingerboson}).  For our current notation this is
\begin{align}
\label{eq:chp709}
S_x &  = \frac{1}{2}\left(a^\dagger_{1}a_2 + a^\dagger_{2}a_1\right),\\
\label{eq:chp710}
S_y &  = \frac{i}{2}\left(a^\dagger_{2}a_1 - a^\dagger_{1}a_2\right),\\
\label{eq:chp711}
S_z &  = \frac{1}{2}\left(a^\dagger_{1}a_1 - a^\dagger_{2}a_2\right).
\end{align}
We note that to make the connection to standard conventions of angular momentum, we include here the factor of 1/2 in the definitions of the spin. The raising angular momentum operator is defined $S_+ = S_x + i S_y = a^\dagger_{1}a_2$ and lowering angular momentum operator is $ S_- = S_x - iS_y = a^\dagger_{2}a_1$. The Hamiltonian  (\ref{eq:chp708}) in terms of the angular momentum operators is 
\begin{equation}
\label{eq:chp712}
H_{\mathrm{eff}} = 2\mu_-S_z - g_-(\mathcal{N}-1) S_z - (g_+ - g_{12}) S_z^2.
\end{equation}
It can be shown that these operators obey the angular momentum commutation algebra with the Casimir invariant operator $S^2$ as
\begin{equation}
\label{eq:chp713}
S^2 = S_x^2  + S_y^2 + S_z^2 = \frac{{\cal N}}{2}\left(\frac{{\cal N}}{2}+ 1\right),
\end{equation}
The Hamiltonian  (\ref{eq:chp712}), $S^2$, $\cal N$ and $S_z$ all commute. Another set of orthogonal basis vectors for this  Hamiltonian is $\lvert N,s,m\rangle$. It is then evident that $S^2\lvert N,s,m\rangle = N/2(N/2 + 1)\lvert N,s,m\rangle = s(s+1)\lvert N,s,m\rangle$, from which we immediately conclude that $s = N/2$. For brevity  $\lvert N, s,  - s\rangle$ is henceforth written as $\lvert s, - s\rangle$.

The angular momentum states can be represented in terms of the Fock states of the Hamiltonian  (\ref{eq:chp708}) by expanding either of the angular momentum state $\lvert s, m= \pm s\rangle$ in terms of Fock states $\lvert N , n, N-n\rangle$ defined earlier.  Take for example the angular momentum state $\lvert  s,-s\rangle$ which is expanded in the Fock states as 
\begin{equation}
\label{eq:chp714}
\lvert  s,-s\rangle = \sum_{n =0} C_n \lvert N,n,N-n\rangle,
\end{equation}
where $C_n = \langle N, n, N - n\lvert s, - s\rangle$, $n_1 = n$, and $n_2 = N- n$.  To determine the coefficients $C_n$, the lowering angular momentum operator $S_- = a^\dagger_{2}a_1$ is applied to  (\ref{eq:chp714}) and set to zero, giving
\begin{align}
S_-  \lvert  s,-s\rangle & = 0  \nonumber \\
 & = \sum_{n} C_n  \sqrt{(N -n+1 ) n} \,\lvert N, n-1, N-n+1\rangle. \label{eq:chp715}
\end{align}
It is evident that except for $n= 0$, $C_n$ must vanish for every other term. Thus $\lvert  s, m=-s\rangle =  \lvert N, 0, N\rangle$ where $C_0 =1$  has been chosen to satisfy the normalization condition. The state corresponds to a situation where all the atoms are in the second component of the condensate $n_2 = N$. To get the rest of the angular momentum states in terms of the Fock state, the raising angular momentum operator $S_+ = a^\dagger_1a_2$ is applied repeatedly to $\lvert  s, -s\rangle$:
\begin{equation}
\label{eq:chp716}
\lvert  s, m\rangle = \sqrt{\frac{1}{(2s)!}\frac{(s-m)!}{(s+m)!}}\, S_+^{l+m}\,\lvert   s, -s\rangle,
\end{equation}
where  $\lvert  s, -s\rangle = \lvert N,  0, N\rangle$. This immediately gives 
\begin{equation}
\label{eq:chp717}
\lvert s, m\rangle = \lvert N, n, N-n\rangle.
\end{equation} 
Thus, there is a one-to-one correspondence between the angular momentum states and the Fock states. 

\begin{exerciselist} [Exercise]
    \item \label{q8-1}
    Using (\ref{eq:chp704}) in (\ref{eq:chp701}) verify  (\ref{eq:chp705}).
    \item \label{q8-2}
    By applying the lowering operator $S_-$ to (\ref{eq:chp714}), show that $C_{n\neq1} = 0$, and verify (\ref{eq:chp717}).
    \item \label{q8-3} 
    A three-level atom has three (energy) \index{three-level atom}levels such as the the three hyperfine energy levels of the $F=1$ of the alkali metals like \textsuperscript{$87$}Rb: $m_F = -1, 0, 1$. The operators $a^\dagger_{+1}$, $a^\dagger_0$, and $a^\dagger_{-1}$ act on vacuum to create an atom in the corresponding level, respectively. The operators describing the interactions in the system can be expressed in terms of angular momentum operators: $S_z = a^\dagger_{-1}a_{-1} - a^\dagger_{+1}a_{+1}$, $S_+ = \sqrt{2} (a^\dagger_{-1} a_0 + a^\dagger_0 a_{+1})$, and $S_- = \sqrt{2}(a^\dagger_{+1}a_0 + a^\dagger_0 a_{-1})$. Show that the operators $S_z$, $S_x$, and $S_+$ are angular momentum operators. Hint: verify that the operators satisfy angular momentum commutation relations.
    \item \label{q8-4}
    If the angular momentum state $\lvert s,-s\rangle$ are related to the three-level Fock states as $$\lvert s, -s\rangle = \sum_{k}^n C_k \lvert2k, n+s-k,n-k\rangle,$$
    determine $C_k$. How is the Fock state representation of the angular momentum state for a two level atom different from that of a three-level atom?
\end{exerciselist}

\subsection[Evolution of initial state]{Evolution of the initial state
	\label{sec:sec:chp7:InitialStateEvolution }}
The atoms are initially prepared with $N$ atoms in one of the internal states such as the $\lvert F = 1, m_f = -1\rangle $ of  \textsuperscript{87}Rb. This is represented by the state
\begin{equation}
\label{eq:chp718}
\lvert \Psi_{\mathrm{ini}} \rangle = \frac{\left(a^\dagger_{1}\right)^N}{\sqrt{N!}}\lvert 0 \rangle,
\end{equation} 
where $\lvert 0\rangle$ is the vacuum state. To create a linear superposition  of two-component atomic condensates, a $\pi/2$-pulse is used (see Sec. \ref{sec:prepscs}) to couple the bosonic operators $a^\dagger_{1}$ and $a^\dagger_{2}$ that act on vacuum state to create an atom in each state, respectively, according to the following rules\index{two-component Bose-Einstein condensates}
\begin{align}
\label{eq:chp719}
a^\dagger_{1} & \rightarrow \frac{1}{\sqrt{2}}\left(a^\dagger_{1} - ia^\dagger_{2}\right),\\
\label{eq:chp720}
a^\dagger_{2} & \rightarrow \frac{1}{\sqrt{2}}\left(-ia^\dagger_{1} + a^\dagger_{2}\right).
\end{align}
A single-atom state is transformed as 
\begin{equation}
\label{eq:chp721}
a^\dagger_{1}\lvert 0\rangle \rightarrow \frac{1}{\sqrt{2}}\left(a^\dagger_{1} - i a^\dagger_{2}\right) \lvert 0\rangle,
\end{equation}
so that the product state of $N$-particle system after the $\pi/2$-pulse has been applied becomes
\begin{align}
\label{eq:chp722}
\lvert \Psi_{\mathrm{split}}\rangle & = \frac{1}{\sqrt{2^N N!}} \left(a^\dagger_{1} -ia^\dagger_{2}\right)^{N} \lvert 0 \rangle\nonumber,\\
& = \frac{1}{\sqrt{2^N}} \sum_{n=0}^{N} \sqrt{\frac{N!}{n!(N-n)!}}(-i)^{N-n} \lvert N,n, N-n\rangle,
\end{align}
where
\begin{equation}
\label{eq:chp723}
\lvert N,n, N-n\rangle = \frac{(a^\dagger_{1})^n}{\sqrt{n!}} \frac{(a^\dagger_{2})^{N-n}}{\sqrt{(N-n)!}}\lvert 0 \rangle,
\end{equation}
is the state vector having $n_1 = n$ atoms in one of the hyperfine state $\lvert F = 1, m_f = -1\rangle $, and $n_2 = N-n$ atoms in the other hyperfine state  $\lvert F = 2, m_f = 1\rangle $.

The time evolution of the state vector is governed by the Hamiltonian  (\ref{eq:chp708}) 
\begin{equation}
\label{eq:chp724}
\lvert \Psi(t)\rangle = \exp\left[-\frac{i}{\hbar}\int_{0}^{t} H_\mathrm{eff} dt'\right] \lvert \Psi_{\mathrm{split}}\rangle .
\end{equation}
The states $\lvert N, n_1=n, n_2=N-n\rangle$ given by  (\ref{eq:chp723})  are eigenstates of the Hamiltonian with eigenvalues
\begin{equation}
\label{eq:chp725}
E(n_1,n_2) = \left(2\mu_- - g_-(N - 1)\right)\frac{(n_1 - n_2)}{2} - (g_+ - g_{12})\left(\frac{n_1 - n_2}{2}\right)^2.
\end{equation}
The state vector of the system at time $t=T$  is 
\begin{equation}
\label{eq:chp726}
\lvert \Psi(T)\rangle = \frac{1}{\sqrt{2^N N!}}\sum_{n=0}^{N}\frac{N!}{n!(N-n)!} e^{-i \phi_T  (n - N/2) + i\varphi (n - N/2)^2}\left(a^\dagger_{1}\right)^n \left(-ia^\dagger_{2}\right)^{N - n}\lvert 0 \rangle,
\end{equation}
where 
\begin{equation}
\label{eq:chp727}
\phi_T = \frac{1}{\hbar}\int_{0}^{T}\, dt (2\mu_- - g_-(N - 1)),
\end{equation}
is the linear phase difference accumulated due to the relative difference in the nonlinear self-interactions in each component and environment effects that are coupled to the condensate through the chemical potential.  Meanwhile,  \index{chemical potential}
\begin{equation}
\label{eq:chp727p}
\varphi = \frac{1}{\hbar} \int_{0}^{T}(g_+ - g_{12})\,dt
\end{equation}
is the nonlinear phase per atom due to self-interactions in each BEC component and the mutual interactions between the two-components.

\begin{exerciselist} [Exercise]
    \item \label{q8-5}
    Verify that the state $\lvert N,n,N-n\rangle$ is an eigenstate of (\ref{eq:chp712}) with eigenvalue (\ref{eq:chp725}).
    Verify that the state of the system at any time is as given in (\ref{eq:chp726}).
\end{exerciselist}

\section[Husimi \emph{Q}-function]{Husimi \emph{Q}-function
	\label{sec:chp7:Husimi}}

\begin{figure}[t]
	\begin{center}
		\includegraphics[angle=0,width= \textwidth]{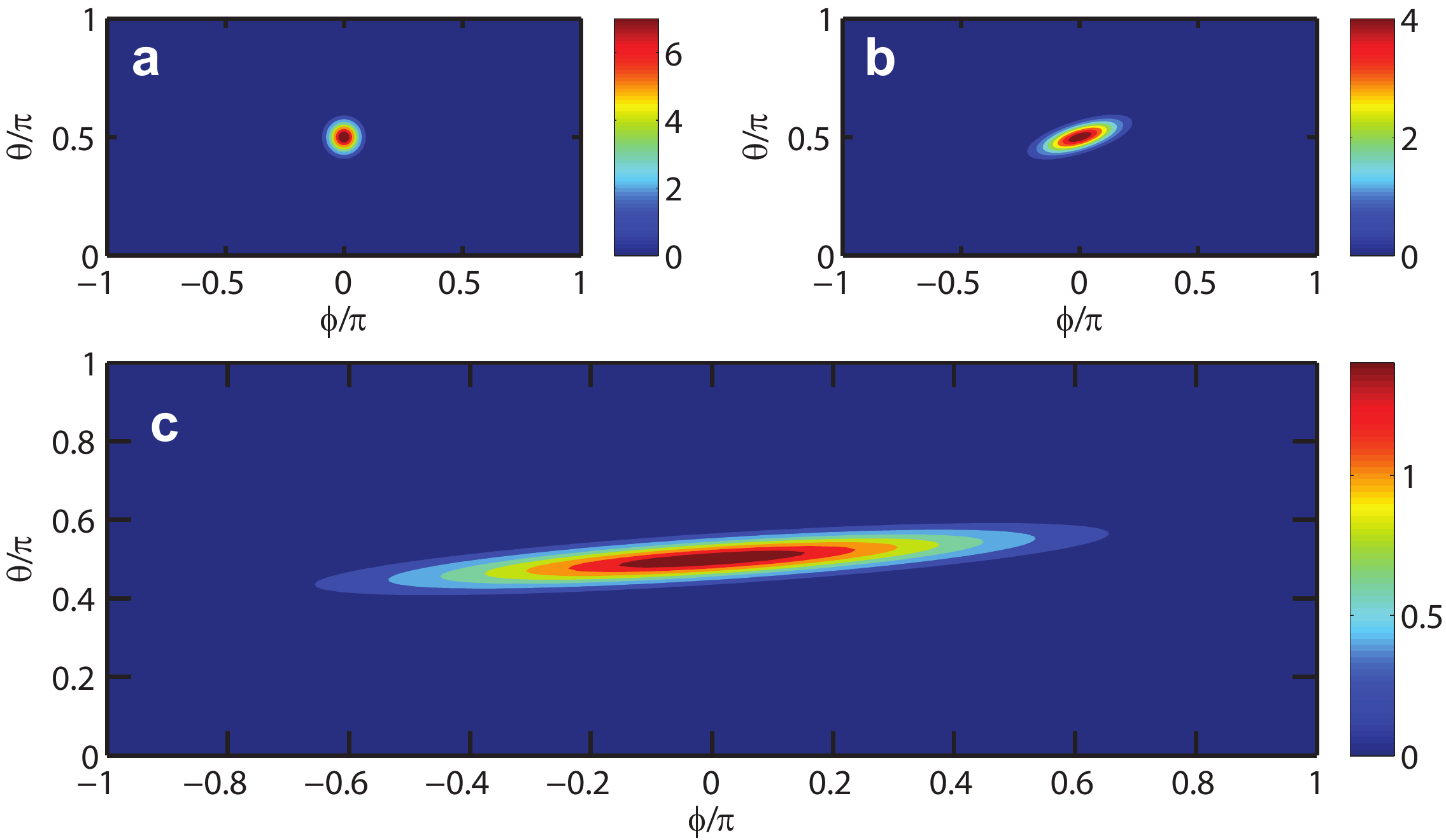}
		\caption{The $Q$-distribution as a function of $\phi$ and $\theta$ for varying interaction strengths (a) $\varphi = 0$; (b) $\varphi = 3/N$; and (c) $\varphi = 1/\sqrt{N}$. The parameters used are $N = 100$, $\phi_T = -3\pi/2$.}\label{fig:chp7-1}
	\end{center}
\end{figure}

To visualize the state (\ref{eq:chp726}), we plot the associated $Q$-function as defined in (\ref{qfuncdefinition}). The numerical calculations of the \emph{Q}-function are shown in Fig. ~\ref{fig:chp7-1}. It is seen that various strengths of the nonlinear interaction parameter $\varphi$ gave different shapes of the \emph{Q}-function. To make sense out this we begin analysis in the simplest case scenario that is $\varphi = 0$. The \emph{Q}-function in this case is given \index{Q-function}
\begin{equation}
\label{eq:chp731}
Q = \frac{N+1}{4\pi} \cos^{2N}\left(\frac{\theta}{2} - \frac{\theta_0}{2}\right) e^{-Nab(\phi - \frac{3}{2}\pi - \phi_T)^2},
\end{equation}
where 
\begin{align}
\label{eq:chp732}
a = \frac{\cos\theta/2\cos\theta_0/2}{\cos\left(\theta/2 - \theta_0/2\right)},\\
\label{eq:chp733}
b = \frac{\sin\theta/2\sin\theta_0/2}{\cos\left(\theta/2 - \theta_0/2\right)},
\end{align}
and $\theta_0 = \pi/2$ specifically for the state (\ref{eq:chp726}). It is clear from  (\ref{eq:chp731}) that the dominant term in the \emph{Q}-function comes from points $\theta = \theta_0$, and  $\phi =\tfrac{3}{2}\pi + \phi_T$.  Close to the maximum of the \emph{Q}-function, the width of the distribution in either the $\phi $ or $\theta$ directions is roughly the same and is of the order $1/\sqrt{N}$. Hence the \emph{Q}-function appears symmetric about its maximum, as expected since there are no non-linear interactions. 

As the nonlinear interaction parameter is increased, the width of the \emph{Q}-function changes. This can be seen from the analytical form of the \emph{Q}-function in the limit $N\gg1$,\index{Q-function}
\begin{align}
\label{eq:chp734}
Q &= \frac{N+1}{4\pi}\sqrt{\frac{1}{1 + 4N^2a^2b^2\varphi^2}}\cos^{2N}\left(\frac{\theta}{2} -\frac{\theta_0}{2}\right) \times\\
& \exp\left[-\frac{Nab}{1 + 4N^2a^2b^2\varphi^2}\left(\phi - \frac{3}{2}\pi - \phi_T - N\varphi(b - a)\right)^2\right].\nonumber
\end{align}
Notice that the location and width of the  maximum  along $\theta$ does not change. However, the location of width and the maximum along the $\phi$ axis changes with $\varphi$. At small values of $\varphi$ ($\varphi \ll 1/N$), the location of the distribution's peak along $\phi$ has dependence on both $\phi_T$ and $\varphi$. Also, its width along $\phi$ grows linearly  with the largest contribution coming from points around $\theta = \theta_0$. Since the width along the $\theta$ dimension remains fixed while the width along $\phi$ axis grows by some amount proportional to $\varphi$ as shown in Fig.~\ref{fig:chp7-1}(b)-(c), the distribution becomes rotated about  the location of its maximum such that it is tilted at an angle with the $\phi$ axis. Large values of $\varphi$ ($1/N <\varphi\ll 1/\sqrt{N}$) further increases in the width along $\phi$ and decreases the angle of inclination of the distribution with the $\phi$ axis, resulting in a tilted ellipse on the ($\theta$, $\phi$) plane. The net effect is that a measurement along the semi-minor axis of the ellipse results in a reduced error that is less than the shot noise. States with this property are refereed to as squeezed states and are exploited in metrology.

\begin{exerciselist} [Exercise]
    \item \label{q8-6}
    Verify that the $Q$-function for a coherent state is as given in (\ref{eq:chp731}). Also, verify (\ref{eq:chp734}). 
    \item \label{q8-6b} Verify that for $\varphi \ll 1/\sqrt{N}$ the width of the \emph{Q}-function along $\phi$ grows at a rate $4Nab\varphi^2$.
\end{exerciselist}

\section[Ramsey interferometry and Bloch vector]{Ramsey interferometry and Bloch vector\label{sec:chp5:Ramsey}}\index{Ramsey interferometry}

This section applies the results of the Sec.~\ref{sec:chp5:atomlight} to describe the working principles of Ramsey interferometry. This technique is often used to measure the transition frequency of atoms as done in magnetic resonance imaging. The process involves interacting an atom with electromagnetic radiation for a time $\tau_r$, which is very short in comparison with its decay time (or lifetime) $1/\gamma$. To illustrate the working principle, we will consider the interaction of two square pulses of electromagnetic wave with an atom.\index{Ramsey interferometry}

\subsection{Pure state evolution}

Consider a two-level atom with eigenstates $\lvert \psi_g\rangle$,$\lvert \psi_e \rangle$ of the Hamiltonian $H_0$, with energies $E_g$ and $E_e$ respectively. The two-level atom is interacting with two square pulses of an electromagnetic wave. Each square pulse of the radiation has a constant amplitude of duration $\tau_r$ and is separated by time interval $T$. During the interaction of the atom and the radiation, the total Hamiltonian of the atom and radiation is $H = H_0 + H_I$. If the interaction strength of the atom and radiation described by $H_I$ is sufficiently small, the eigenstates of $H$ can be approximated as a linear superposition of the eigenstate of the bare Hamiltonian $H_0$.  The state can then be written $\lvert \psi\rangle = a_g(t)\lvert \psi\rangle + a_e(t)\lvert \psi\rangle$, where $a_g(t)$ is the probability amplitude to populate the lower of the two eigenstates and $a_e(t)$ is the probability to populate the upper level of the two eigenstates. The evolution of the amplitudes $a_g(t)$ and $a_e(t)$  is described by the solution  (\ref{eq:chp512}). Moving into the frame that is rotating with eigenfrequencies of the bare atom states and assuming that the detuning $\Delta = \omega_e - \omega_g - \omega$ is large compared to the Rabi frequency $\Omega_{ge}$,  $\Delta \gg \Omega_{ge}$, as well as ignoring the phases of the radiation $\phi_L = 0$, the solution for  $a_e(t)$ in\index{rotating wave approximation} this limit becomes
\begin{align}
\label{eq:chp735}
c_e(t) & = -i\frac{\Omega_{ge}}{\Delta}e^{i t \Delta /2}\sin\left(\frac{t\Delta }{2}\right).
\end{align}
Applying  (\ref{eq:chp735}) to the square pulses for $t = 0$ to $t = \tau_r$, and from $t = T$ to $t = T + \tau_r$ gives
\begin{equation}
\label{eq:chp736}
c_e(t) = -2i\frac{\Omega_{ge}}{\Delta}\sin\left(\frac{\tau_r \Delta }{2}\right) \cos\left(\frac{T \Delta   }{2}\right)e^{i\tau_r\Delta /2}e^{iT \Delta /2}
\end{equation}
The probability that the atom is excited after applying the pulses is 
\begin{equation}
\label{eq:chp737}
|c_e(t)|^2 = \Omega_{ge}^2\tau_r^2  \left[\frac{\sin(\tau_r\Delta/2)}{\tau_r\Delta /2}\right]^2 \cos^2\left(\frac{T \Delta }{2} \right) .
\end{equation} 

\begin{figure}[t]
	\begin{center} 
		\includegraphics[angle=0,width= \textwidth]{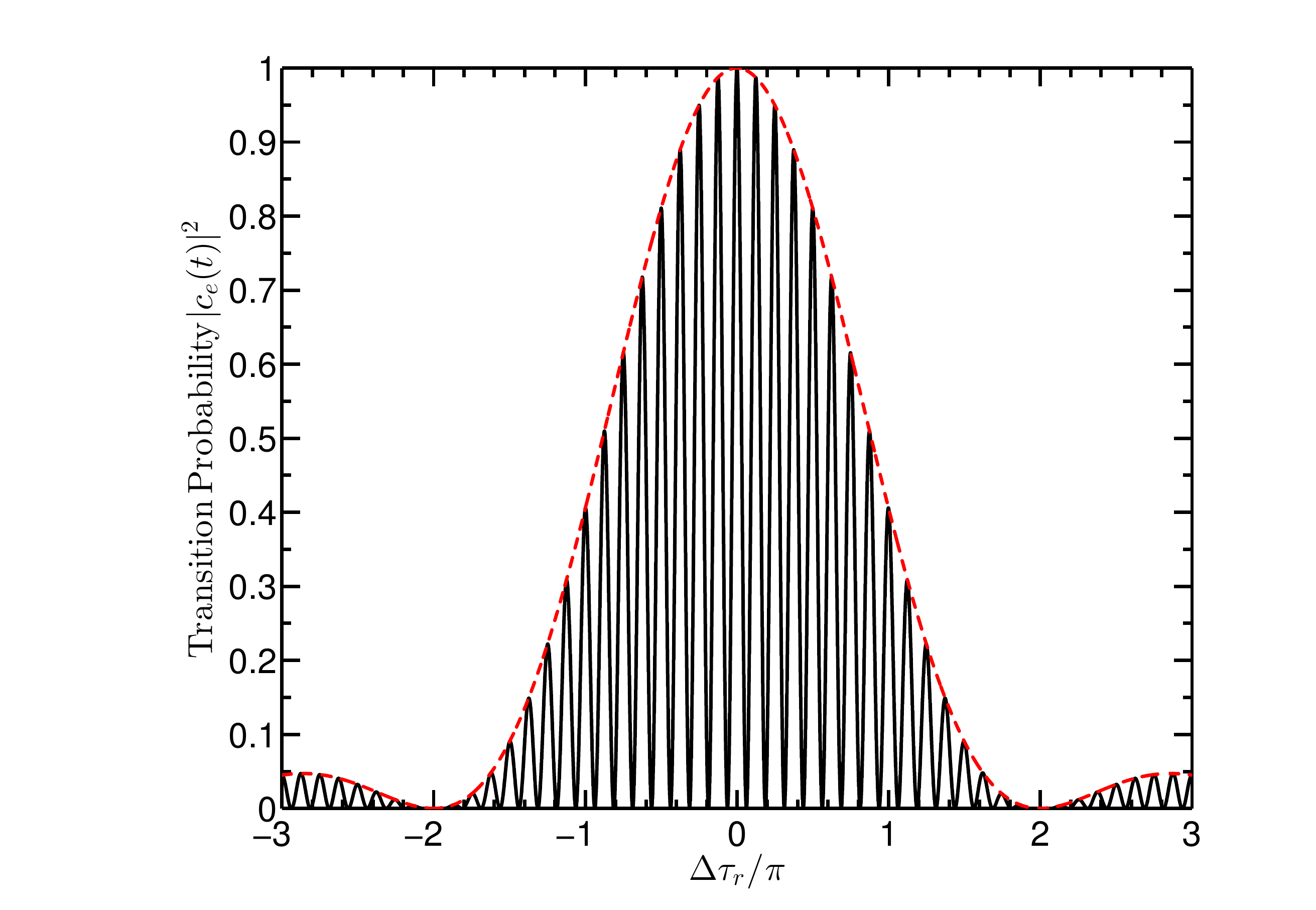}
		\caption{Ramsey interference. The red dashed line is the envelope from a single square pulse $ \Omega_{ge}^2\tau_r^2  \left[\frac{\sin(\tau_r \Delta/2)}{\tau_r \Delta  /2}\right]^2$, while the solid line is the interference pattern from two square pulses separated by a time interval $T$, (\ref{eq:chp737}). The parameters used are $\tau_r = \pi/4$, $T = 4\pi$, $\Omega_{ge} = 1/\tau_r$.}\label{fig:chp7-2}
	\end{center}
\end{figure}

Figure~\ref{fig:chp7-2} show the Ramsey fringes that are given in (\ref{eq:chp737}). \index{Ramsey fringes} The interference exhibited in the figure may be understood as follows. Each pulse taken independently (as in  (\ref{eq:chp735})) gives a $\mathrm{sinc}^2$ function with a maximum at $\tau_r \Delta  = 0$ and minimum occurring at $\tau_r \Delta  = 2\pi $. This gives the width of central maximum to be $\sigma_{\Delta,\mathrm{c}} = 2\pi/\tau_r$. However, for the two square pulses, there is a superposition of two effects;\index{diffraction} interference and diffraction as result of  applying the first pulse in the interval $t=0 $ to  $t= \tau_r$, and applying the pulse from $t = T$ to $t = T + \tau_r$.\index{interference} The maximum of a fringe is dictated by the cosine term, and occurs when the $T \Delta  $ is an integral multiple of $2\pi$. The central peak or maximum occurs for $\Delta = 0$, and goes to zero at $T \Delta  =  \pi$, giving the width of the central peak as $ \sigma_{\Delta,\mathrm{p}} = \pi/T$.  The $\mathrm{sinc}^2$ function is an envelope for the interference pattern and controls the amplitude of the  transition probability with the width of each band given by $\sigma_{\Delta,\mathrm{c}} = 2\pi/\tau_r$. Comparing with the single square pulse, one can see from (\ref{eq:chp735}) that the $\mathrm{sinc}$ term is being sliced by the $\cos^2$ term such that a better resolution is obtained. This is a  classic two-slit\index{diffraction} diffraction where the slits of size $\tau_r$ are separated in time by an interval $T$. Hence the benefit of the Ramsey spectroscopy is that it resolves the width of the frequency band $2\pi/\tau_r$ of a single square pulse into finer intervals of $\pi/T$ due to interference. \index{Ramsey spectroscopy}

\subsection[Mixed state evolution]{Mixed state evolution\label{sec:sec:blochvector}}
Thus far, the atom interaction with radiation has been treated using the wavefunction approach. In order to take into account the interaction with the environment, we must extend this to a  density matrix approach. Working in the rotating frame, \index{rotating frame}the wavefunction describing the evolution of an atom interacting with radiation, as described at the beginning of this section, is given by $\lvert \psi\rangle = c_g(t)\lvert \psi\rangle + c_e(t)\lvert \psi\rangle$. The evolution of the probability amplitudes $c_{g,e}(t)$ is as described in  (\ref{eq:chp510}). To calculate the evolution of the density matrix $\rho$, we need to know the elements of the matrix $\rho = \lvert \psi\rangle\langle \psi \rvert$,
\begin{equation}
\label{eq:chp738}
\rho = \left(\begin{array}{c}
c_g(t) \\
c_e(t)
\end{array}
\right)\left(\begin{array}{cc}
c_g^*(t)& c_e^*(t) 
\end{array}
\right) = 
\left(\begin{array}{cc}
c_g(t)c_g^*(t)& c_g(t) c_e^*(t)\\
c_e(t)c_g^*(t) & c_e(t) c_e^*(t)
\end{array}
\right) = \left(\begin{array}{cc}
\rho_{11} & \rho_{12}\\
\rho_{21} & \rho_{22}
\end{array}
\right).
\end{equation}
The diagonal elements $\rho_{11}$, $\rho_{22}$ of the density matrix are called populations\index{population} and are real. The off-diagonal elements, $\rho_{12}$ and $\rho_{21}$, are coherences\index{coherence} and are complex having a time dependent phase factor that describes the frequency response of the atom to radiation field. 

The evolution of the density matrix are calculated  using  (\ref{eq:chp510}). Noting for instance $\dot{\rho}_{11} = \dot{c}_{g}(t)c_g^*(t) + c_g(t)\dot{c}_{g}^*(t)$, $\dot{\rho}_{12} = \dot{c}_g(t)c^*_e(t) + c_g(t)\dot{c}_e^*(t)$, and assuming that $\Omega_{ge}$ is real as well as ignoring the phase of the radiation field, $\phi_L = 0$, we find from these that the evolution of the matrix elements are
\begin{align}
\label{eq:chp739}
\frac{d\rho_{11}}{dt} &= -i\frac{\Omega_{ge}}{2}(\rho_{21} - \rho_{12}) = -\frac{d\rho_{22}}{dt},\\
\label{eq:chp740}
\frac{d\rho_{12}}{dt} & = i\Delta\rho_{12} +i\frac{\Omega_{ge}}{2}(\rho_{11} - \rho_{22}),\\
\label{eq:chp741}
\frac{d\rho_{21}}{dt} & = -i\Delta\rho_{21} -i\frac{\Omega_{ge}}{2}(\rho_{11} - \rho_{22}).
\end{align}
From  (\ref{eq:chp739}), it is easy to deduce that $\tfrac{d\rho_{11}}{dt} + \tfrac{d\rho_{22}}{dt} = 0$ is a constant of motion and expresses the conservation of the probability, $c_g(t)c_g^*(t) + c_e(t)c_e^*(t) = 1$. The conservation of the probability allows three variables $u$, $v$, and $w$ to be defined as follows. Decomposing the coherences into their real and imaginary parts $\rho_{12} = u + i v$, $\rho_{21} = u - i v$,
\begin{align}
\label{eq;chp742}
u & =\frac{ \rho_{12} + \rho_{21}}{2} ,\\
\label{eq;chp743}
v & = i\frac{\rho_{21} - \rho_{12}}{2}.
\end{align}
Finally, due to the conservation of the probability, the population gives only one variable, the relative population difference which we define as
\begin{equation}
\label{eq:chp744}
w  = \rho _{11} - \rho_{22}.
\end{equation}
From (\ref{eq:chp739})-(\ref{eq:chp741}), the evolution of the  variables $u$, $ v$, and $w$ are
\begin{align}
\label{eq:chp745}
\frac{du}{dt} & = -\Delta v,\\
\label{eq:chp746}
\frac{dv}{dt} & = \Delta u + \Omega_{ge} w,\\
\label{eq:chp747}
\frac{d w}{dt} & = -\Omega_{ge} v.
\end{align}
Defining a Bloch vector $\mathbf{r} = u \hat{\mathbf{i}} + v\hat{\mathbf{j}} + w \hat{\mathbf{k}}$, and $\mathbf{Q} = -\Omega_{ge}\hat{\mathbf{i}} + \Delta\hat{\mathbf{k}}$, (\ref{eq:chp745})-(\ref{eq:chp747})  can be written in vector notation as 
\begin{equation}
\label{eq:chp748}
\frac{d\mathbf{r}}{dt} = \mathbf{Q} \times \mathbf{r}.
\end{equation}
The solution of  (\ref{eq:chp748}) may be obtained by using  (\ref{eq:chp509}) and (\ref{eq:chp512}) in the definitions of $u$, $v$ and $w$. Equation~(\ref{eq:chp748}) can be solved more generally under various conditions including damping and losses. Furthermore, the Bloch vector allows for the visualization of the Bloch vector \index{Bloch sphere} and gives an intuitive interpretation of the dynamics. For instance, $\dot{\mathbf{r} }\cdot\mathbf{r} =0$ immediately tells us that $\dot{\mathbf{r} }\cdot\mathbf{r} + \mathbf{r}\cdot\dot{\mathbf{r} } = 0$ so that $\mathbf{r}\cdot\mathbf{r} = u^2 + v^2 + w^2 = 1$ is a constant of motion. This is not surprising since $\mathbf{Q} \times \mathbf{r}$ gives a vector that is perpendicular to the plane of $\mathbf{Q}$ and $\mathbf{r}$, so the dot product of this vector $\mathbf{Q} \times \mathbf{r}$ with either $\mathbf{Q}$ or $\mathbf{r}$ must vanish. It is then evident that $\mathbf{Q} \cdot \tfrac{d\mathbf{r}}{dt} = 0$ or $\mathbf{Q}\cdot\mathbf{r}$ is also a constant of motion. This implies that motion of $\mathbf{r}$ must be such that the projection of $\mathbf{r}$ along $\mathbf{Q}$ must remain invariant. Suppose the vector $\mathbf{Q}$ is held fixed in space, then $ \mathbf{Q}\cdot\mathbf{r}= Q r \cos\theta$, where $\theta$ is some angle between $\mathbf{Q}$ and $\mathbf{r}$ measured from $\mathbf{Q}$. It then means that $\mathbf{r}$ is a vector on the surface of the cone traced out by $\theta$ measured from $ \mathbf{Q}$ (with $\mathbf{Q}$ at the center of the cone), that is $\mathbf{r}$ precesses around $\mathbf{Q}$. In interferometry as we shall see later, the goal is to be able to estimate this rotation angle $\theta$ with best precision possible. Our discussion so far does not account for loss mechanisms in the dynamics of the density matrix, such as spontaneous emission\index{spontaneous emission}. Such loss mechanisms and decoherence can be incorporated for example using master equation methods as seen in Secs. \ref{sec:spontaneous} and \ref{sec:loss}.   


\section[Error propagation formula and squeezing parameter]{Error propagation formula and squeezing parameter 
	\label{sec:chp7:errorpropagation}}\index{error propagation formula}
In the previous section, visual inspection  of the $Q$-function for the state  (\ref{eq:chp734}) showed that the noise in the state can be affected and controlled by the strength of interaction parameter. \index{Q-function} The amount of noise, which originates from quantum mechanical noise in this case, affects the accuracy of estimating particular parameters in an experiment.  
For example, it may be of interest to know the parameter $\phi$  that influences the relative population difference $S_z$. But the parameter $\phi$ cannot be directly measured in experiment. By measuring the relative population difference $S_z$, one hopes to get the best possible estimate of the unknown parameter $\phi$. The way that $S_z$ changes with $\phi$ may not necessarily by a linear function and may be different for various parameter choices of $\phi$.  That is, for a change from $ \phi $ to $\phi + \Delta \phi$, that is relative population difference changes from $ S_z(\phi) $ to $S_z(\phi + \Delta\phi)$. Assuming that the change in the parameter $\phi$ is very small, we can expand $S_z(\phi + \Delta\phi) \approx S_z(\phi) + \tfrac{dS_z(\phi)}{d\phi} \Delta\phi$ to get the error propagation formula, whose inversion gives the error in estimation of $\phi$ as
\begin{equation}
\label{eq:chp749}
\Delta \phi = \frac{\Delta S_z(\phi)}{\left|\frac{d S_z(\phi)}{d\phi}\right|},
\end{equation}
where $S_z(\phi + \Delta \phi) -S_z(\phi) = \Delta S_z(\phi)$ and  the modulus sign has been put on the derivative in recognition that it could sometimes be negative. 

To give a concrete example, consider using the state prepared in  (\ref{eq:chp726}) in sensing rotation or measure the an unknown field. The expectation values of the operators $S_x$, $S_y$ and $S_z$ are as follows
\begin{align}
\label{eq:chp750}
\langle\Psi(T)\lvert S_x \rvert\Psi(T) \rangle& =\frac{N}{2} \sin\phi_T \cos^{N-1}\varphi,\nonumber\\
\langle\Psi(T)\lvert S_y \rvert\Psi(T) \rangle& =-\frac{N}{2}  \cos\phi_T \cos^{N-1}\varphi,\\
\langle\Psi(T)\lvert S_z \rvert\Psi(T) \rangle& = 0.\nonumber
\end{align}
This implies that at time $T$, the state $\lvert \Psi(T)\rangle$ lies on the $x$-$y$ plane. Without loss of generality and setting $\phi_T = \pi$, the state $\lvert \Psi(T)\rangle$ points along $y$ axis at $t= T$. This state can be used to measure rotation a radiation field in any of the other two orthogonal directions, say the $x$ axis. The field tips the mean spin and causes it to rotate about the $x$ axis in the $y$-$z$ plane. The Hamiltonian describing the applied field within the rotating wave approximation is $H_I = \tfrac{\hbar\Delta}{2} S_z + \hbar\Omega_{ge} S_x/2$. The dynamics of the spin is governed by the Heisenberg\index{Heisenberg equation of motion} equations which are (\ref{eq:chp745})-(\ref{eq:chp747}), if $u$, $v$ and $w$ of the Bloch vectors\index{Bloch vector} are replaced with $S_x$, $S_y$ and $S_z$ respectively,  $\Delta = 0$ and $\Omega_{ge} = -\Omega_{ge}/2$. 

\begin{figure}[t]
	\begin{center}
		\includegraphics[angle=0,width= \textwidth]{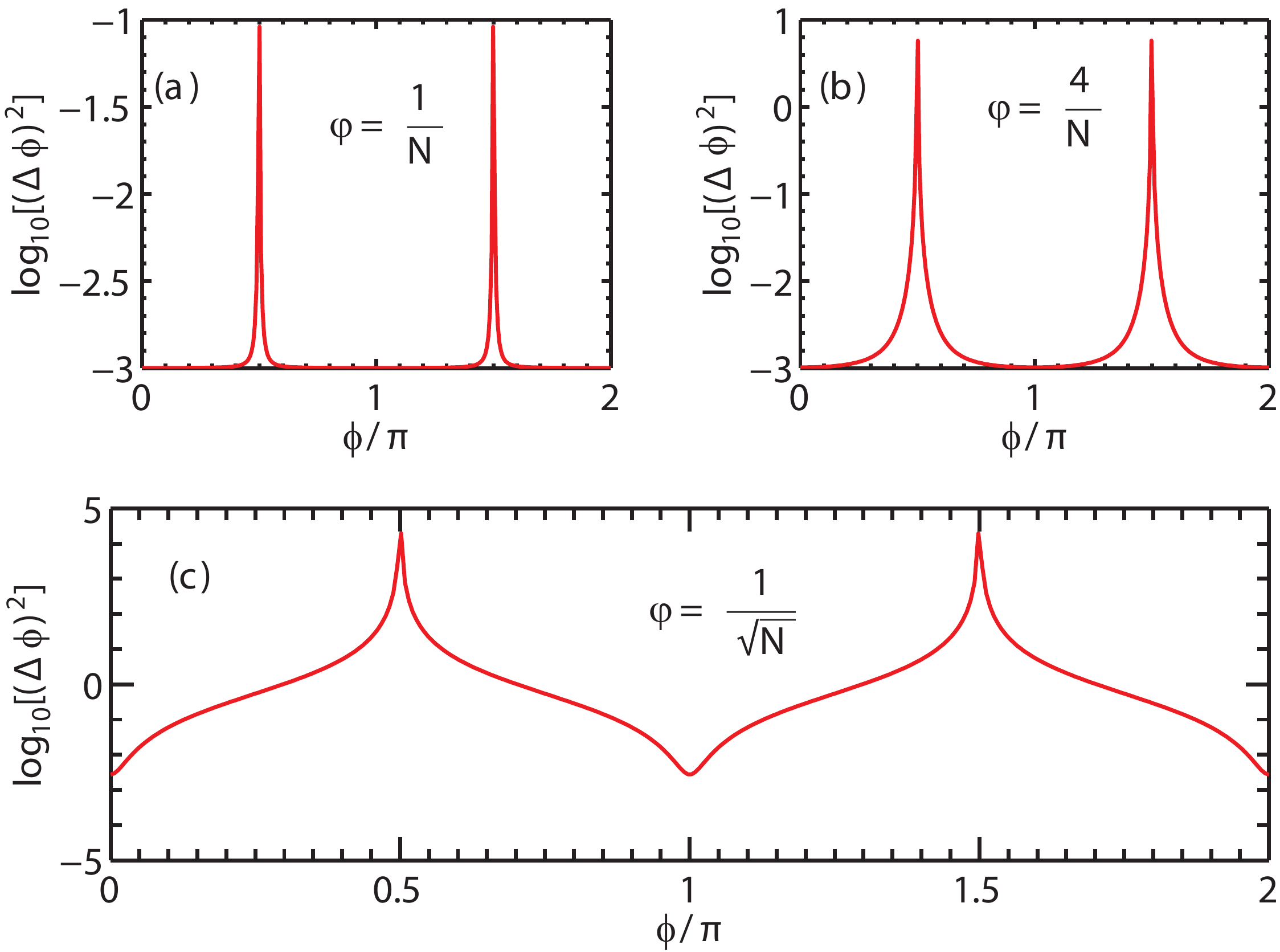}
		\caption{The error in estimating $\phi$ using the error-propagation formula  (\ref{eq:chp754}). For the state $\lvert \Psi(T)\rangle$, the following parameters were used, $N = 1000$,  and $\phi_T = 0$. The values of $\varphi$ are as shown in the figure.}\label{fig:chp7-3}
	\end{center}\index{error propagation formula}
\end{figure}

The time evolution of the spins is given by
\begin{align}
\label{eq:chp751}
S_x(t) & = S_x,\\
\label{eq:chp752}
S_y(t) & = S_y\cos\frac{\Omega_{ge}t}{2} + S_z\sin\frac{\Omega_{ge}t}{2},\\
\label{eq:chp753}
S_z(t) & = S_y\sin\frac{\Omega_{ge}t}{2} - S_z\cos\frac{\Omega_{ge}t}{2},
\end{align}
where $S_i(0) = S_i$, $i =x,y,z$. To estimate the angle $\phi = \Omega_{ge} t/2 $ at which the spin is rotating about the $x$-axis, one calculates the spin in $z$ direction (\ref{eq:chp753}), 
\begin{align}
\langle  S_z(t)\rangle = \langle  S_y\rangle\sin\phi - \langle  S_z\rangle \cos\phi = \langle  S_y\rangle\sin\phi,
\end{align} 
where $\langle S_i\rangle$ = $\langle\Psi(T)\lvert S_i\rvert\Psi(T)\rangle$, $i=x,y,z$. In the following discussion, we will approximate the change in $S_z$ (that is $\Delta S_z$) with the standard deviation $\sigma_{S_z}$ and similarly $\Delta \phi$ will be approximated with $\sigma_{\phi}$. Then, the error in estimating $\phi$ in accordance with  (\ref{eq:chp749}) then becomes
\begin{align}
\label{eq:chp754}
\sigma^2_{\phi} & = \frac{(\langle  (S_y)^2\rangle-\langle S_y\rangle^2)\sin^2\phi + \langle S_z^2\rangle\cos^2\phi }{\langle S_y \rangle^2\cos^2\phi },
\end{align}
where all the expectations are taken with respect to $ \Psi(T) $. The cross terms $\langle S_yS_z\rangle$ in the variance vanishes, hence are not present. \index{variance}  Note that the denominator of  (\ref{eq:chp754}) is the expectation of $\langle S_y(t)\rangle$, while the numerator is variance of $\langle S_z^2(T)\rangle$. The error $ \sigma^2_\phi $ is plotted in Fig.~\ref{fig:chp7-3}. As seen from the plots, as $\varphi$ remains small $0\leq \varphi \leq 1/N$, the error is relatively low except at $ \phi = n\pi/2$, $n$ being odd. The points $\phi = n\pi/2$ give value of $\phi$ where spin is aligned along the $z$ axis, thus $\langle S_y(t)\rangle$ is zero. Increasing value of $\varphi$ increases the error in estimating $\phi$. However, from Fig.~\ref{fig:chp7-3}(c) there exist error minima at values of $\phi = 0,m\pi$ ($m$ even), corresponding to the spins aligned along the $y$ axis. At these points, the variance of $S_z(t)$ is small\index{variance} while the mean value of $S_y(t)$ has its maximum value. Thus the error $\sigma_{\phi}$ is minimized for  $\phi = 0, m\pi$ giving
\begin{equation}
\label{eq:chp755}
\sigma^2_{\phi,\mathrm{min}} = \frac{\langle S_z^2\rangle}{\langle S_y\rangle^2}.
\end{equation}
The variance of $S_z$ is shot noise limited,\index{variance}  $\langle S_z^2\rangle = N/4$ while  ${\langle S_y\rangle^2}$ is given in  (\ref{eq:chp750}) with $\phi_T = \pi$. In summary, (\ref{eq:chp755}) states that if the direction of the mean spin $\langle \mathbf{S}\rangle$ of a state is along some axis, then the variance $\sigma_{S_\perp}$ that minimizes the error propagation formula is on a plane that is perpendicular to the mean spin $\langle \mathbf{S}\rangle$, $\sigma_{\phi} = \frac{\sigma_{ S_\perp}}{|\langle\mathbf{S}\rangle|}$. More so, when this error is benchmarked against the error obtained from a spin coherent state $\sigma_{\phi,\mathrm{s}} = 1/\sqrt{N}$, one gets a \index{spin coherent state} parameter $\xi_\mathrm{R} =\tfrac{\sigma_{\phi}}{\sigma_{\phi,\mathrm{s}}}$ that estimates the improvement of using the given state over the coherent state which is given in this case by
\begin{equation}
\label{eq:chp756}
\xi_\mathrm{R} = \frac{\sqrt{N}\sigma_{ S_\perp}}{|\langle\mathbf{S} \rangle| }.
\end{equation}

\begin{figure}[t]
	\begin{center}
		\includegraphics[angle=0,width= \textwidth]{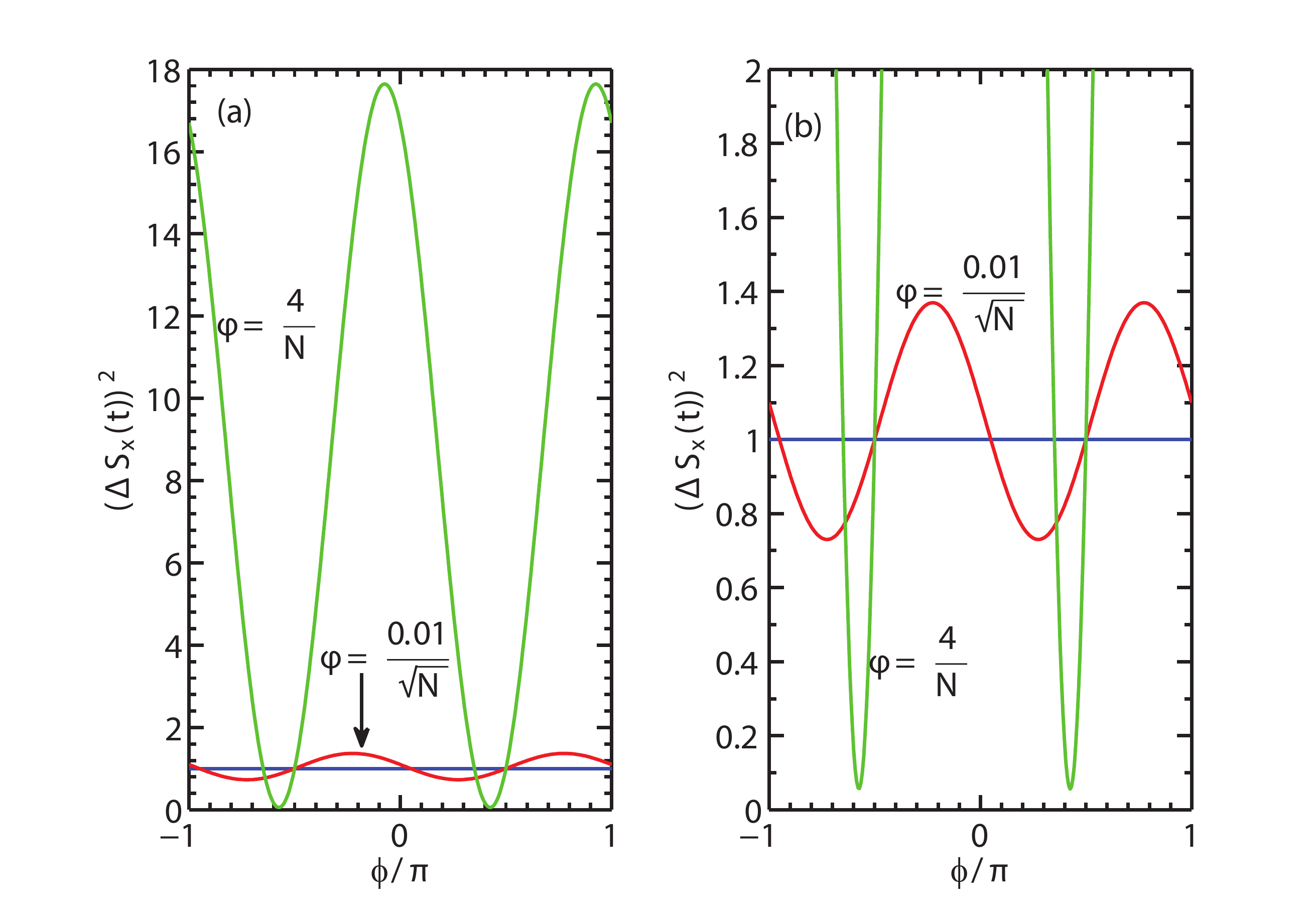}
		\caption{The normalized variance of $S_x(t)$,  (\ref{eq:chp760}). (b) shows a zoomed-in region of (a) to show the minimum values of variance. The parameters are $N = 1000$ and $\phi_T = 0$. The values of $\varphi$ are as shown in the figure, except for the straight line which is the shot-noise with $\varphi =0$.}\label{fig:chp7-4}
	\end{center}\index{variance}
\end{figure}

The failure of  (\ref{eq:chp755}) to obtain the desired improvement from using spin squeezing \index{spin squeezed state} stems from the fact that even though the error  $\sigma_{S_z}$ along the $z$-axis is perpendicular to the mean spin $S_y$ direction which is along $y$-axis, the direction $z$ is not that of the minimum variance. In order to utilize the squeezing effect contained in the state $\lvert \psi(T)\rangle$,  one can observe from the $Q$-function\index{Q-function} that the dominant spin direction of the atoms is in the $y$-axis. Any rotation that is not along the mean spin tends to shift the mean spin to another position. It then suggests that the best sensing will be obtained if the atoms are rotated about their dominant spin direction. For the phase of the laser light $\phi_L = \pi/2$ and detuning $\Delta=0$ in  (\ref{eq:chp510}), one obtains within the rotating wave approximation the Hamiltonian $H= \hbar\Omega_{ge}S_y$. This Hamiltonian rotates spin about the dominant spin direction $\langle\mathbf{S}\rangle$.  The evolution of the spin operators is
\begin{align}
\label{eq:chp757}
S_x(t) & = S_x\cos\Omega_{ge} t + S_z \sin\Omega_{ge} t  ,\\
\label{eq:chp758}
S_y(t) & = S_y,\\
\label{eq:chp759}
S_z(t) & = -S_x\sin\Omega_{ge} t + S_z\cos\Omega_{ge} t.
\end{align}
The mean spin $|\langle\mathbf{S}\rangle|$ direction has not changed, hence the variance on the plane perpendicular to the direction of the mean spin that points along $y$ can be found using either (\ref{eq:chp757}) or (\ref{eq:chp759}). Using  (\ref{eq:chp757})  gives that the variance along $x$ %
\begin{equation}
\label{eq:chp760}
\langle S_x^2(t)\rangle = \langle S_x^2\rangle\cos^2\phi + \langle S_zS_x + S_xS_z\rangle\sin\phi\cos\phi + \langle S_z^2\rangle\sin^2\phi,
\end{equation}
where $\phi = \Omega_{ge} t$. This result is shown in Fig.~\ref{fig:chp7-4}, normalized with respect to the shot-noise $N/4$. From Fig.~\ref{fig:chp7-4}(b), we see that there exists a $\phi$ value for which the variance is minimum and is well below the shot noise for a given $\varphi$ with $\varphi>0$. Minimizing $\langle S_x^2(t)\rangle$ with respect to $\phi$ gives the optimum $\phi$ value that optimizes the variance\index{variance}
\begin{equation}
\label{eq:chp761}
\tan(2\phi) = \frac{\langle S_x S_z + S_zS_x\rangle}{\langle S_x^2 \rangle - \langle S_z^2\rangle}.
\end{equation}
Using the property of the trigonometric functions, $\tan x = \sin x/\cos x$, then
\begin{align}
\label{eq:chp762}
\sin( 2\phi_\mathrm{opt})& = \frac{\langle S_xS_z\rangle + \langle S_zS_x\rangle}{\sqrt{(\langle S_xS_z\rangle + \langle S_zS_x\rangle)^2 + (\langle S_x^2\rangle - \langle S_y^2\rangle)^2}},\\
\label{eq:chp763}
\cos(2\phi_\mathrm{opt}) & = \frac{\langle S_x^2 \rangle - \langle S_z^2\rangle}{\sqrt{(\langle S_xS_z\rangle + \langle S_zS_x\rangle)^2 + (\langle S_x^2\rangle - \langle S_y^2\rangle)^2}}.
\end{align} 
Thus the variance\index{variance} is minimized if $\sin(2\phi) = -\sin(2\phi_\mathrm{opt})$, and $\cos(2\phi) = -\cos(2\phi_\mathrm{opt})$, which is  obtained for $\phi = \pi/2 + \phi_\mathrm{opt}$ in  (\ref{eq:chp760}), giving $\langle S_x^2(t)\rangle = \sigma_{ S_\perp}$ as
\begin{equation}
\label{eq:chp764}
\sigma^2_{ S_\perp,\mathrm{min}} = \frac{\langle S_x^2\rangle + \langle S_z^2\rangle}{2} - \frac{\sqrt{(\langle S_x^2\rangle - \langle S_z^2\rangle)^2 + (\langle S_xS_z\rangle + \langle S_zS_x\rangle)^2}}{2}.
\end{equation}
Similarly, for $\phi = \pi + \phi_\mathrm{opt}$ the variance is maximized $\sigma_{S_\perp,\mathrm{max}}$. This is often referred to the anti-squeezing, the opposite of squeezing effect. Other scenarios are also possible, such as having $\phi = \pi/2 - \phi_\mathrm{opt}$ or $\phi = \pi - \phi_\mathrm{opt}$. In this case, $\phi = \pi/2 - \phi_\mathrm{opt}$ will minimize the error $S_z(t)$ while $\phi = \pi - \phi_\mathrm{opt}$ will maximize the error in $S_z(t)$.

\begin{exerciselist} [Exercise]
    \item \label{q8-7}
    Calculate the variance of $S_z(t)$ (\ref{eq:chp753}) and expectation value of $S_y(t)$ (\ref{eq:chp752}), and verify (\ref{eq:chp754}).
    \item \label{q8-8}
    Calculate the variance of $S_z(t)$ (\ref{eq:chp759}) and find the angle that minimizes $\langle S_z^2\rangle$. For what value of $\phi$ is $\xi_\mathrm{R}$ anti-squeezed?
\end{exerciselist}

\section[Fisher information]{Fisher information
	\label{sec:chp7:fisherinformation}}\index{Fisher information}
In many practical situations the parameters of physical interest are not directly accessible due to experimental circumstances or from the fact that it can only be inferred from a measurement of a conjugate parameter. This scenario abounds in quantum mechanics where some parameters like phase do not have a defined operators. Thus, inference about the unknown parameter relies on indirect measurement of a conjugate observable, which  adds additional uncertainty for the estimated parameter, even for optimal measurements. The estimation of the parameter usually occurs in a two-step process, namely the parameter to be estimated is first encoded in the state of the system (to be probed), an then followed by a measurement of the system that reveals information about the parameter. 

Fisher information\index{Fisher information} gives the amount of information that can be extracted about an unknown parameter encoded in the state of any system  from a measurement of an observable of the system. The amount of information extracted depends of course on the procedure that is used in obtaining the information, and hence the procedure needs to be typically optimized for better and improved estimation. The best estimators are those that saturate the Cramer-Rao inequality\index{Cramer-Rao inequality}
\begin{equation}
\label{eq:chp765}
\sigma^2_{\theta}  \geq \frac{1}{MF(\theta)},
\end{equation}
where $M$ is the number of measurements. Equation (\ref{eq:chp765}) establishes a lower bound on the variance $\sigma^2_{\theta} $ of any estimator of the parameter $\theta$, which is given by the Fisher information\index{Fisher information}
\begin{equation}
\label{eq:chp766}
F(\theta) = \int dx p(x|\theta) \left(\frac{\partial\ln p(x|\theta)}{\partial\theta}\right)^2,
\end{equation}
where $p(x|\theta)$ is the conditional probability of obtaining the value of $x$ when the parameter has a value $\theta$. 

In quantum mechanics, the probability $p(x|\theta)$ of obtaining the value $x$ given that the quantum state of the system is parametrized by $\theta$ is $p(x|\theta) = \mathrm{Tr}\left[\Pi_x \rho_\theta\right]$, where $\Pi_x$ is a positive operator satisfying the normalization condition $\int dx \, \Pi_x = \mathbbm{1}$, and $\rho_\theta$ is the density matrix parametrized by the quantity $\theta$ that is to be estimated. Introducing the symmetric logarithmic derivative $L_\theta$ as a Hermitian operator satisfying the equation
\begin{equation}
\label{eq:chp767}
\frac{L_\theta \rho_\theta + \rho_\theta L_\theta}{2} = \frac{\partial \rho_\theta}{\partial \theta},
\end{equation}
then
\begin{equation}
\label{eq:chp768}
\frac{\partial p(x|\theta)}{\partial \theta} = \text{Re}\left(\mathrm{Tr}\left[\rho_\theta\Pi_xL_\theta\right]\right).
\end{equation}
The Fisher information  (\ref{eq:chp766}) is then written as\index{Fisher information} 
\begin{equation}
\label{eq:chp769}
F(\theta) = \int dx \frac{\text{Re}\left(\mathrm{Tr}\left[\rho_\theta\Pi_xL_\theta\right]\right)}{\mathrm{Tr}\left[\Pi_x\rho_\theta\right]}.
\end{equation}
For a given quantum measurement,  (\ref{eq:chp769}) gives a classical bound on the precision that can achieved through data processing. Optimizing the measurements $\Pi_x$ gives the ultimate bound on the Fisher information\index{Fisher information}
\begin{equation}
\label{eq:chp770}
F(\theta) \leq \mathrm{Tr}\left[\rho_\theta L^2_\theta\right]. 
\end{equation}
The quantum Fisher information $ F_{\mathrm{Q}}(\theta) = \mathrm{Tr}\left[\rho_\theta L^2_\theta\right]$\index{Fisher information} places a limit on the amount of information regarding $\theta$ that can be gained in any measurement. Using  (\ref{eq:chp770}) in  (\ref{eq:chp765}) gives the quantum Cramer-Rao bound \index{Cramer-Rao inequality} 
\begin{equation}
\label{eq:chp771}
\sigma^2_{\theta} \geq \frac{1}{MF_{\mathrm{Q}}(\theta)}
\end{equation}
for the variance of any estimator\index{variance}. This implies that any measurement can produce a minimum acceptable error that is determined by the quantum Fisher information, \index{Fisher information} and this bound  does not depend on the measurement.

For a pure state, the symmetric logarithmic derivative is $L_\theta = 2\tfrac{\partial \rho_\theta}{\partial \theta}$. Using the fact that $\rho_\theta= \rho_\theta^2$ then 
\begin{equation}
\label{eq:chp772}
\frac{\partial \rho_\theta}{\partial \theta } = \frac{\partial \rho_\theta}{\partial \theta}\rho_\theta + \rho_\theta\frac{\partial \rho_\theta}{\partial \theta}.
\end{equation}
From  (\ref{eq:chp772}) and the symmetric logarithmic derivative, we have that for a pure state the quantum Fisher information is
\begin{equation}
\label{eq:chp773}
F_Q = 4\left(\left\langle\frac{\partial \psi_\theta}{\partial \theta } \left\vert \frac{\partial \psi_\theta}{\partial \theta }\right.\right\rangle + \left(\left\langle\psi_\theta\left|\frac{\partial \psi_\theta}{\partial \theta}\right. \right\rangle\right)^2\right).
\end{equation} 
For a pure state $\lvert \psi_\theta\rangle$  parametrized by $\theta$, which is related to the initial state $\lvert \psi_0\rangle$ by the unitary transform $\lvert \psi_\theta\rangle = e^{-i\theta S}\lvert \psi_0\rangle$, the quantum Fisher information\index{Fisher information} is then given by
\begin{equation}
\label{eq:chp774}
F_Q = 4 \sigma^2_{S} = 4(\langle \psi_0 \lvert S^2 \rvert \psi_0\rangle - \langle \psi_0 \lvert S \rvert \psi_0\rangle^2 ),
\end{equation}
where $S$ is any of the spin operators $S_x,\,S_y,\,S_z$. Equation (\ref{eq:chp774}) immediately tells us that the quantum Fisher information\index{Fisher information} is independent of the parameter $\theta$ and is given by the variance\index{variance} in the operator $S$ that generates the parameter $\theta$. 

We now apply the above concepts to find the component of the total angular momentum operator $S_k$, $k=x,\,y\,,z$, that gives the maximum information on the estimation of rotation angle $\phi$ from a state $\lvert\psi_\phi\rangle$.  Here,  (\ref{eq:chp726}) is the initial state 
\begin{equation}
\label{eq:chp775}
\lvert\psi_\phi\rangle = e^{-i\phi S_k}\lvert \psi(T)\rangle,
\end{equation}
where $\phi_T =0$. According to  (\ref{eq:chp774}), finding the optimum operator amounts to calculating the variances\index{variance} which using  (\ref{eq:chp750}) are
\begin{align}
\label{eq:chp776}
\sigma^2_{S_x} & = \frac{N[(N+1) - (N-1)e^{-2(N-2)\varphi^2}]}{8},\\
\label{eq:chp777}
\sigma^2_{S_y} & =  \frac{N[(N+1) + (N-1)e^{-2(N-2)\varphi^2}]}{8} - \frac{N^2e^{-(N-1)\varphi^2}}{4},\\
\label{eq:chp778}
\sigma^2_{S_z} & = \frac{N}{4}.
\end{align}

\begin{figure}[t]
	\begin{center}
		\includegraphics[angle=0,width= \textwidth]{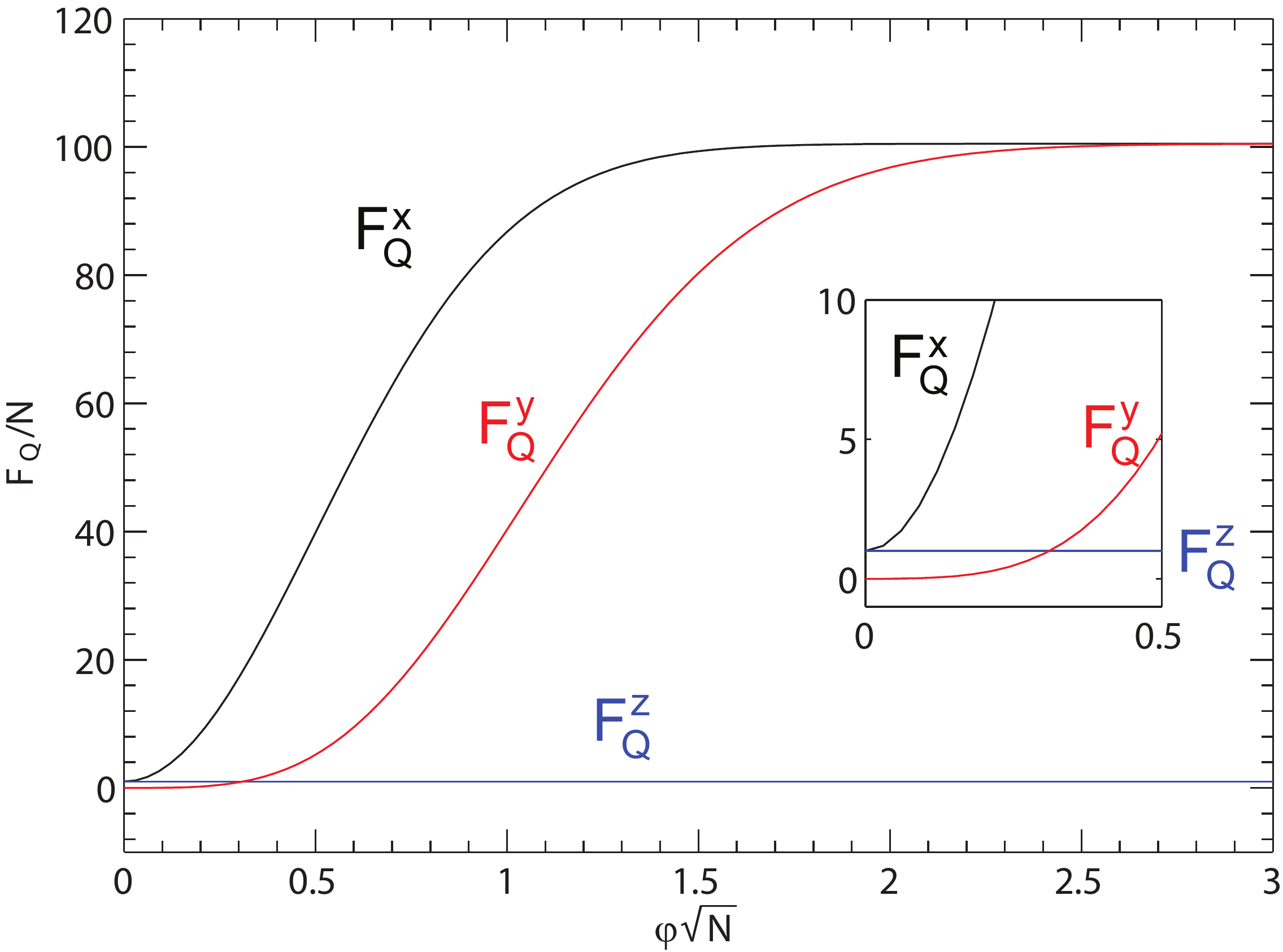}
		\caption{The quantum Fisher information as a function of the nonlinear phase per atom $\varphi$. The inset shows a zoomed-in region near origin. Parameters are  $N=200$.}\label{fig:chp7-6}\index{Fisher information}
	\end{center}
\end{figure}

The quantum Fisher information\index{Fisher information}  (\ref{eq:chp774}) is plotted in Fig.~\ref{fig:chp7-6}  using (\ref{eq:chp776})-(\ref{eq:chp778}). The Fisher information\index{Fisher information}  $F_Q^z$ for a rotation about the $z$-axis remains constant throughout the variation of atom-atom interactions $\varphi$. This is not surprising as the state $\lvert\psi_\phi\rangle$ is on the $x$-$y$ plane. Thus the fluctuations along the $z$-axis remain on average fixed and the rotations about the $z$-axis would not produce any information better that $N$. However, rotations about the $x$-axis and $y$-axis produce an increase in the amount of information that can be obtained. At $\varphi = 0$, we have $F_Q^x = N$ and $F_Q^y = 0$. This is because the $\lvert\psi_\phi\rangle$ is an eigenstate of $S_y$ with eigenvalue $N/2$. Thus, all the atoms in this state would simply produce an absolute phase for the rotation angle $\phi$. Since the atoms are known with certainty to be in the eigenstate of $S_y$ with eigenvalue $N/2$, there is no fluctuation and the variance is thus zero. As such, this state will not provide any useful information in sensing rotation as the $\sigma_{\theta=\phi}$ is infinite in (\ref{eq:chp765}). However, a rotation about the $x$-axis at $\varphi=0$ produces a variance\index{variance} of $N/4$, giving $F_Q^x = N$. As $\varphi $ increases the fluctuations about the $x$ and $y$ axes increase, and hence more information can be estimated about $\phi$. Also, at every $\varphi$ value $F^x_Q > F^y_Q$. Consequently, the variance in the estimation of $\phi$ (see (\ref{eq:chp765})), using a rotation about $x$-axis is smaller compared to the variance in estimation of $\phi$ using a rotation about the $y$-axis for the same $\varphi$ value. For large values of $\varphi$, the quantum Fisher information is approximately the same for a rotation about the $x$ and $y$ axes, such that $F^y_Q =F^x_Q\approx N(N+1)/2$. We thus see that $S_x$ is the operator that gives maximal information in estimating the rotation angle $\phi$.\index{Fisher information}

\begin{exerciselist} [Exercise]
    \item \label{q8-9}
    Using the initial state  $\lvert \Psi(T) \rangle $ with $\phi_T = 0$, verify (\ref{eq:chp776}), (\ref{eq:chp777}) and (\ref{eq:chp778}).
    \item \label{q8-10}
    The Fisher information is as defined in (\ref{eq:chp766}). Given the state (\ref{eq:chp726}), calculate the Fisher information. Hint: first calculate the probability density of having the $k$th atom in state say $\lvert k, N-k\rangle$. Next, take the derivative of the probability with the parameter $\phi_T$, and complete the calculation.
\end{exerciselist}

\section[Controlling the nonlinear phase per atom]{Controlling the nonlinear phase per atom
	\label{sec:chp7:controllingxi}}
From the preceding sections, we have seen that the nonlinear phase per atom $\varphi$ controls the amount of squeezing that is useful for measurement, and  dictates the  lower bound on the Fisher information. The parameter $\varphi$ is related to experimental parameters through the trap frequencies and the scattering lengths, the degree of overlap between the atomic BEC components in the trap,  and therefore could in principle be controlled. In experiments~\citep{matthews1998,mertes2007} where the scattering lengths $a_k$ of the different atomic species are close to the interspecies scattering length $a_{12}$, the nonlinear phase per atom becomes negligible. To see this, consider a two-component atomic BEC confined in a trapping potential of the form\index{two-component Bose-Einstein condensates}
\begin{equation}
\label{eq:chp779}
	V(x,y,z) = \frac{m_i}{2}(\omega^2_{x}x^2 + \omega_\perp^2 r^2_\perp).
\end{equation}
The $m\omega_\perp^2 r^2_\perp/2$ provides confinement in the transverse dimension $\mathbf{r}_\perp = (y,z)$. In many experiments, the atoms are more tightly confined in the transverse dimension than the axial dimension  $\omega_x\ll\omega_\perp$. Let us assume that the BEC occupies the lowest transverse mode in the trap
\begin{equation}
	\label{eq:chp780}
	\psi_\perp(r_\perp) = \frac{1}{\sqrt{\pi}a_\perp} \exp\left(-\frac{r_\perp^2}{2a^2_\perp}\right),
\end{equation}
where $a_\perp = \sqrt{\hbar/(m\omega_\perp)}$ is the transverse oscillator length. Writing the condensate wavefunction $\psi_i(\mathbf{r},t) = \psi_i(x,t)\psi_\perp(r_\perp)$, we find that (\ref{eq:chp702}) and (\ref{eq:chp703}) are reduced to a one-dimensional Gross-Pitaevskii equation in $\psi_i(x,t)$
\begin{align}
\label{eq:chp781}
	\mu_1\psi_1(x,t) & = \left[-\frac{\hbar^2}{2m}\frac{\partial^2}{\partial x^2} + \frac{1}{2}m\omega_x^2x^2 + \frac{U_1 n_1 }{2\pi  a_\perp^2}\lvert \psi_1(x,t)\rvert^2 + \frac{U_{12} n_2 }{2\pi a_\perp^2}\lvert \psi_2(x,t)\rvert^2 \right]\psi_{1}(x,t),\\
	\label{eq:chp782}
	\mu_2\psi_2(x,t) & = \left[-\frac{\hbar^2}{2m}\frac{\partial^2}{\partial x^2} + \frac{1}{2}m\omega_x^2x^2 + \frac{U_2 n_2 }{2\pi  a_\perp^2}\lvert \psi_2(x,t)\rvert^2 + \frac{U_{12} n_1 }{2\pi a_\perp^2}\lvert \psi_1(x,t)\rvert^2 \right]\psi_{1}(x,t).
\end{align}

Within the Thomas-Fermi approximation ($n_1,\, n_2 \gg 1$), the kinetic energy terms (\ref{eq:chp781}) and (\ref{eq:chp782}) can be neglected, and the density profile of the condensates becomes
\begin{align}
\label{eq:chp783}
	|\psi_{1}(x)|^2 &= 2\pi a_\perp^2 \frac{(\mu_1 - V(x))U_2 - (\mu_2 - V(x))U_{12}}{n_1(U_1U_2 - U^2_{12})},\\
	\label{eq:chp784}
	|\psi_{2}(x)|^2 &= 2\pi a_\perp^2 \frac{(\mu_{2} - V(x))U_1 - (\mu_1 -V(x))U_{12}}{n_2(U_1U_2 - U^2_{12})},
\end{align}
where $|\psi_{i}(x)|^2\geq 0$. The boundary of the cloud where the density vanishes $|\psi_{i}(x)|^2 = 0$ determines the spatial extent $R_{k,x}$, $k=1,\,2$, of the cloud. For  $\psi_{1}(x)$,
\begin{equation}
\label{eq:chp785}
	V(x= R_{1,x}) = \frac{1}{2} m\omega_x R_{1,x}^2 = \frac{U_2\mu_1 - U_{12}\mu_2}{U_2 - U_{12}}.
\end{equation}
The length $R_{1,x}$ of the cloud at the point where $|\psi_{1}(x)|^2$ vanishes then becomes
\begin{equation}
\label{eq:chp786}
	R_{1,x}^2 = \frac{2}{m\omega_x^2}\frac{U_2\mu_1 - U_{12}\mu_2}{U_2 - U_{12}}.
\end{equation}
A similar calculation for $|\psi_2(x)|^2$ gives 
\begin{equation}
\label{eq:chp787}
R_{2,x}^2 = \frac{2}{m\omega_x^2}\frac{U_1\mu_2 - U_{12}\mu_1}{U_1 - U_{12}}.
\end{equation}
Note that for non-interacting components $U_{12} = 0$, the length for a uniform one dimensional cloud is recovered. Hence the normalized density function becomes
\begin{align}
\label{eq:chp789}
	|\psi_1(x)|^2 & = \frac{3}{4R_{1,x}}\left(1 - \frac{x^2}{R_{1,x}^2}\right),\\
	\label{eq:chp790}
	|\psi_2(x)|^2 & = \frac{3}{4R_{2,x}}\left(1 - \frac{x^2}{R_{2,x}^2}\right).
\end{align}
Also, the normalization conditions 
\begin{align}
\label{eq:chp791}
\frac{3}{4} & = 2\pi a_\perp^2 R_{1,x} \frac{U_2\mu_1 - U_{12}\mu_2}{U_1U_2 - U_{12}^2},\\
\label{eq:chp792}
\frac{3}{4} & = 2\pi a_\perp^2 R_{2,x} \frac{U_1\mu_2 - U_{12}\mu_1}{U_1U_2 - U_{12}^2},
\end{align}
allow for the calculation of the chemical potentials. 

Consider a situation where the density of condensate in state $\lvert2\rangle$ is displaced  with respect to the condensate in state $\lvert 0\rangle$. Denoting the position of the center of the condensate in state $\lvert 2\rangle$ with respect to the condensate in $\lvert 0\rangle$ by $x_0$, the density of the condensate in state $\lvert 2\rangle$ is $|\psi_2(x)|^2 = \tfrac{3}{4R_{2,x}}\left(1 - \tfrac{(x - x_0)^2}{R_{2,x}^2}\right)$. The self-interaction energy terms $g_{kk}$  (\ref{eq:chp706}) is then
\begin{equation}
\label{eq:chp793}
g_{kk} = 3U_k/(10\pi a_\perp^2 R_{k,x}) .
\end{equation}
Meanwhile, the interspecies interaction energy is
\begin{equation}
\label{eq:chp794}
	g_{12} = \frac{3U_{12}(R_{1,x} + R_{2,x} - x_0)^3}{320 \pi a_\perp^2 R^3_{1,x}R^3_{2,x}}\left(x_0^2 + 3(R_{2,x} + R_{1,x})x_0 - 4(R^2_{1,x} + R^2_{2,x}) + 12 R_{1,x}R_{2,x}\right) .
\end{equation}
Equation (\ref{eq:chp794}) shows that $g_{12}$ depends on the separation $x_0$ between the centers of the condensate densities, and the ratio of condensate lengths $R_{2,x}/R_{1,x}$. The point $x_0 = R_{1,x} + R_{2,x}$ is where the condensate in state $\lvert1\rangle$ and $\lvert 2\rangle$ separate, and the interspecies interaction vanishes $g_{12}=0$. Thus $g_{12}=0$ for $x_0 \ge R_{1,x} + R_{2,x}$, and the dominant contribution to the nonlinear phase per atoms $\varphi$ then comes from the self-interaction terms $g_{kk}$. On the other hand, when $x_0 = 0$, $g_{12}$ becomes
\begin{equation}
\label{eq:chp795}
g_{12} = \frac{3U_{12}}{80\pi a_\perp^2 R_{1,x}}\left( 1 + \frac{R_{1,x}}{R_{2,x}}\right)^3\left(3\frac{R_{2,x}}{R_{1,x}} -  \left(\frac{R_{2,x}}{R_{1,x}}\right)^2 - 1 \right).
\end{equation}
If the ratio of the lengths of condensates $R_{2,x}/R_{1,x}$ are roughly the same, $R_{2,x}/R_{1,x}\approx 1$, then $g_{11}$, $g_{22}$ and $g_{12}$ all would have same form. Then both self-interaction terms and the intraspecies interaction term contribute to $\varphi$. In atomic species such as \textsuperscript{87}Rb BEC where the scattering lengths ($a_1$, $a_{2}$, and $a_{12}$) are all rather similar, then $\varphi \propto (a_1 + a_2 - 2a_{12})$ approximately vanishes.

One method to generate a nonlinear phase is to use spontaneous non-equilibrium dynamics. This is done by condensing the atoms into one of the hyperfine levels, say $\lvert 1 \rangle$. A combination of radio frequency and microwave pulses are then used to transfer some of the condensate in hyperfine level $\lvert 1 \rangle$ to another state $\lvert 2\rangle$, thus producing condensates that are in a linear superposition of state, $(\lvert1\rangle + \lvert2\rangle)/\sqrt{2}$, and interacting with each other. Immediately after the transfer of atomic population to $\lvert 2\rangle$, the density of the two atomic species containing different number of atoms $n_1$ and $n_2$ are not in equilibrium. This is because the nonlinear repulsive interaction energy due to the  difference in the scattering length of the two different hyperfine levels is no longer balanced by the confining potential. As such the equilibrium density distributions oscillate, driving each spin component to different regions of the trap. The decrease in overlap between the atomic densities at separation increases the nonlinear phase per atom which leads to squeezing~\citep{mertes2007,laudat2018}. 

Another approach is to use state-dependent potentials and Feshbach resonance~\citep{ockeloen2013,riedel2010} to achieve deterministic control of the interactions.  Different hyperfine levels have different magnetic moments and thus can be localized in different regions of a trap. This is exploited to produce state-dependent potentials, by physically separating the trap minima for the two-component species that coalesced during the production of the two component atomic condensate. As a result, deterministic control of the overlap and hence nonlinear phase per atom is achieved. Also, Feshbach resonance~\citep{gross2010} has been used to tune the scattering rates thereby controlling the nonlinear phase per atom.

\begin{exerciselist} [Exercise]
    \item \label{q8-11}
    Using (\ref{eq:chp780}) in (\ref{eq:chp702}) and (\ref{eq:chp703}), verify that the one-dimensional coupled Gross-Pitaevskii equation for a two-component atom is as given in (\ref{eq:chp781}) and (\ref{eq:chp782}).  Also verify using the Thomas-Fermi approximation that the normalized solutions are given by (\ref{eq:chp789}) and (\ref{eq:chp790}).
\end{exerciselist}

\section{References and further reading}
\begin{itemize}
    \item Sec.~\ref{sec:chp7:intro}: For precision interferometry beyond the standard quantum limit see~\cite{holland1993, bouyer1997, dowling1998correlated, bouchoule2002spin, dowling2008, kitagawa1993, urban2009observation, campos2003, saffman2010quantum, eto2013}. These references discuss spin-1  atom condensates and their components~\cite{ho1996, hall1998, hall1998b, mueller2002, ho1998, ohmi1998, stamper-kurn1998}. The first trapping of all the spin components of an  atomic BEC in a trap is found 
    here~\cite{barrett2001}.
    
    \item Sec.~\ref{sec:chp7:two-component}: These references discuss the dynamics of two-component atom Bose-Einstein condensates~\cite{matthews1998, hall1998, eto2018}. These references discuss the tunneling of atoms at a Josephson junction~\cite{milburn1997, smerzi1997, raghavan1999, steel1998,cataliotti2001}. The splitting of atom BEC in a trap is discussed here~\cite{javanainen1999}. The effect of two-body interactions on the operation of an atom interferometer are discussed in these references~\cite{javanainen1997, ilo-okeke2010}. The application of the spin-1 atoms to interferometry are discussed here~\cite{yurke1986}. Review articles on two-component atomic and spinor condensates by~\cite{kawaguchi2012,kasamatsu2005} are an excellent resource.
    
    \item Sec.~\ref{sec:chp7:Husimi}: These references are good resources for applications of the Husimi $Q$-Function~\cite{mandel1995, scully1999quantum, meystre2001atom}. Refs.~\cite{arecchi1972, gross2012} are review papers on the atomic coherent state. For an application of the $Q$-function on the atom BEC interferometry see~\cite{ilo2014theory, ilo-okeke2016}.
    
    \item Sec.~\ref{sec:chp5:Ramsey}: For an introduction to diffraction from a double slit see~\cite{halliday2013}.  Ramsey interferometry and the depiction of particle evolution on a  Bloch sphere may be found in these references~\cite{ramsey1950, ramsey1985, allen1975, shore1990, barnett1997}.
    
    \item Sec.~\ref{sec:chp7:errorpropagation}: Discussions on the error propagation formula and the comparison with the standard quantum limit are given in these references~\cite{wineland1994, wineland1992spin, kitagawa1993, caves1981}.
    
    \item Sec.~\ref{sec:chp7:fisherinformation}: The following materials give a detailed introduction to Fisher information~\cite{helstrom1974, braunstein1994, paris2009, pezze2014}.
    
    \item Sec.~\ref{sec:chp7:controllingxi}: For controlling the nonlinear phase per atom using state dependent potentials and Feshbach resonances see the following~\cite{riedel2010, gross2012, ockeloen2013}. The following references controlled the     nonlinear phase per atom by manipulating the spatial overlap between different spin components~\cite{mertes2007, laudat2018}. This paper~\cite{matthews1998} gives an experimental method to calculate the scattering lengths by controlling the nonlinear phase per atom. Ref.~\cite{myatt1997} is an experiment describing how to realize two atom BECs in different internal states with overlapping densities. Ref.~\cite{baym1996} is a theory paper detailing how to estimate parameters of an atomic BEC in a trap.
    
	
\end{itemize}

	\chapter[Quantum simulation]{Quantum simulation}\index{quantum simulation}

\label{ch:quantumsimulation}

\section{Introduction}
\label{sec:introsimulation}

It has long been appreciated that simulating quantum many-body problems is a difficult computational task.  Take for example an interacting system of $ N $ spin-1/2 particles.  The Hilbert space dimension grows exponentially as $ 2^N $, and therefore to find the ground state would generally require diagonalization of a Hamiltonian of matrix dimension $ 2^N \times 2^N $.  Even with the most powerful supercomputers available today, only about $ N = 40 $ spins can handled to perform exact calculations.  Typically in condensed matter physics one is interested in the behavior of large scale systems.  The number of atoms in a typical crystalline material could be more in the region of Avogradro's number\index{Avogradro's number}, $ N \approx 6 \times 10^{23} $.  This makes the study of realistic materials far beyond the reach of direct simulation. In condensed matter physics this has necessitated the development of sophisticated numerical techniques such as quantum Monte Carlo, density matrix renormalization group (DMRG), density functional theory, dynamical mean field theory, and series expansion methods to calculate the properties of the quantum many-body problem of interest. However, each of these methods have various shortcomings that prohibits certain quantum many-body problems to be applied with reliable accuracy, particularly those involving fermions in dimensions greater than one.  

The above problem is not one that is limited to condensed matter physics.  In high-energy physics, the fundamental theory of the strong force, quantum chromodynamics (QCD), is formulated as a SU(3) gauge-invariant quantum field theory, where the underlying fermions are quarks.  Particles such as protons and neutrons are emergent quasiparticle excitations of the underlying quantum field theory.  This again involves solving a quantum-many body problem involving fermions and gauge fields. Many decades of effort have been dedicated to verify that QCD is the correct theory of the strong force, particularly in the form of lattice gauge theories, which are discrete lattice versions of QCD are used to simulate the systems.  This again runs into the same issues that are raised above, where only a relatively small number of sites can be simulated, due to the exponential complexity.  Likewise, in quantum chemistry one aims to understand the nature of molecules which are fundamentally described by a Schrodinger equation for the many-electron system.  Due to the computational complexity various approximations are typically made, such as density functional methods.  Similarly, in atomic physics numerous approximations are made to understand the nature of heavy atoms as they involve large numbers of interacting electrons and nucleons.  

Richard Feynman originally proposed the quantum computer as a method of solving such quantum many-body problems.  The basic idea for this is rather simple. He noted that computers that we utilize today are based on classical logic, and do not utilize any aspect of quantum physics.  So the idea is that if the computers themselves used quantum physics then they might be better at solving quantum problems.  While it was shown later that indeed it is possible to gain an exponential speedup over classical computers using these methods, attaining this in practice is difficult due to all the same problems as 
realizing a quantum computer: decoherence, scalability, and errors occurring during qubit manipulations.  The trend recently has been to instead fabricate special-purpose systems that are manipulated so that they have the  Hamiltonian of interest. This approach is called {\it analogue quantum simulation} and is distinguished from the former algorithmic approach, which is termed {\it digital quantum simulation}. \index{quantum simulation!analogue}\index{quantum simulation!digital}

In this chapter, we will briefly describe the various techniques involved in performing quantum simulation. After introducing the problem in more detail, we will describe the difference between analogue and digital simulation.  We will focus more on analogue simulation, because digital simulation is more suited for performing quantum simulations when a quantum computer is available, which is out of the scope of this book.  There are already huge numbers of demonstrations for performing analogue quantum simulation on various quantum systems.  Rather than describe each of these, we will describe the main techniques that are used to perform such quantum simulations, which can be used together to construct various Hamiltonians as desired.  To illustrate these techniques, we also describe one of the classic experiments that demonstrates a quantum simulation of the Bose-Hubbard model\index{Bose-Hubbard model} with cold atoms.   For a comprehensive list of systems that have been simulated using these ideas, we refer the reader to the excellent review articles listed at the end of this chapter.

\section{Problem statement: What is quantum simulation?}
\label{sec:analogvsdigital}

The first ideas for quantum simulation originally were motivated by Richard Feynman, who proposed the quantum computer for simulating quantum many-body problems. Feynman made this conjecture in 1982, but it remained unproven until Lloyd gave an explicit argument that this is true in 1996.  Lloyd showed that it is possible to construct the Hamiltonian of a chosen physical system out of a limited set of controls, for example in a quantum computer.  While conceptually this was a breakthrough, this did not mean that one can easily perform quantum simulations.  Lloyd's method relies upon performing a series of gates to simulate the desired Hamiltonian.  This {\it digital quantum simulation} \index{digital quantum simulation}\index{quantum simulation!digital} approach requires the technological capability of something rather close --- or even equivalent --- to a full quantum computer.  

As the field of quantum technology developed since these early days, it became evident that quantum computers were a type of device that is one of the most challenging in terms of  engineering.  In a quantum computer, one needs to be able to perform a completely arbitrary unitary evolution and readout of a very large number of qubits.  Other quantum technological applications such as quantum cryptography are closer to being used in practice, since only few-qubit manipulations need to be performed. For this reason alternative directions were investigated where the technological demands are less. The approach was to construct purpose-built systems that have the same Hamiltonian as the system of interest.  This has been a popular approach since it is technologically more achievable, rather than first implementing a full programmable quantum computer.  In this {\it analogue quantum simulation}\index{quantum simulation!analogue} approach, one typically combines several techniques to build a specified Hamiltonian.  Then one experimentally studies the artificially created system, which often can be measured using advanced techniques which may not be available to other systems.  This gives a unique and practical way of studying complex quantum many-body systems.  

\begin{figure}[t]
\includegraphics[width=\textwidth]{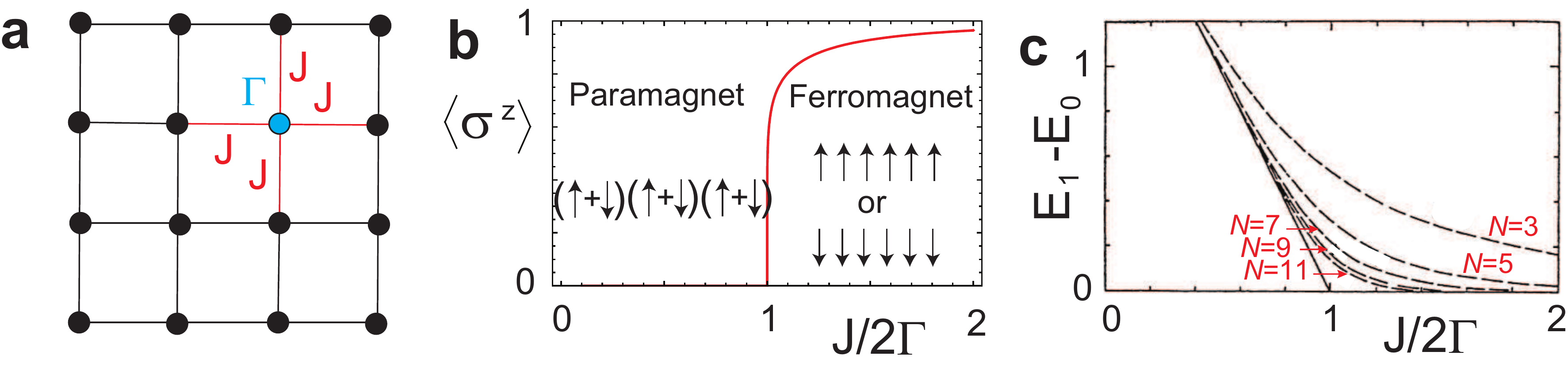}
\caption{The transverse Ising model.  (a) Schematic description of the transverse Ising model\index{Ising model} in two dimensions.  The couplings $ J $ indicate the interactions between nearest neighbor sites, and a magnetic field in the transverse direction of strength $ \Gamma $ is applied across all sites. (b) Phases of the one-dimensional transverse Ising model. The magnetization result of (\ref{pfeutyresult}) taken from \cite{pfeuty1970one} is shown.  (c) The energy gap reproduced from \cite{hamer1981finite}. Dashed lines show finite size results, the solid line is for the thermodynamic limit.  
}
\label{fig9-1}
\end{figure}

To understand the task of quantum simulation, it is illustrative to look at a simple example.  An archetypical quantum many-body problem from condensed matter physics is the transverse Ising model (see Fig. \ref{fig9-1}), which is specified by the Hamiltonian\index{Ising model}
\begin{align}
H = -\frac{J}{2} \sum_{j,k \in \text{n.n.}} \sigma^z_j \sigma^z_k - \Gamma \sum_j \sigma_j^x ,
\label{transverseisingham}
\end{align}
%
where $ J, \Gamma $ are some constants, and the sum runs over nearest neighbors (n.n.) of a lattice.  The lattice could have any geometry or dimensionality, we can consider it to be a square lattice for instance.  The Hamiltonian encodes any number of lattice sites, for example for $ N = 2 $ sites it can be equivalently represented as a matrix
\begin{align}
H_{N=2} = -\left(
\begin{array}{cccc}
J/2 & \Gamma &  \Gamma &  0  \\
\Gamma & - J/2 & 0 & \Gamma \\
\Gamma & 0 & - J/2 & \Gamma \\
0 & \Gamma &  \Gamma & J/2 
\end{array}
\right) .
\label{twositeham}
\end{align}
Here the basis vectors are taken as $ \{|00\rangle, |01\rangle, |10\rangle, |11\rangle \} $, where $ |0\rangle, |1\rangle $ are the eigenstates of the $  \sigma^z $ Pauli matrix\index{Pauli operators}.  A typical condensed matter physics problem would then involve diagonalizing the Hamiltonian to find the eigenstates.  In particular one is often interested in the ground state and low-lying excitations.  These states possess particular properties, which determine the physics of the model.  For example, in the one-dimensional transverse Ising model\index{Ising model}, there is a quantum phase transition\index{quantum phase transition} when $ J/2 = \Gamma$, where the ground state is a ferromagnet for  $ J/2 > \Gamma $ and a paramagnet for  $ J/2 < \Gamma $.  One can identify such phases using {\it order parameters}\index{order parameter} which gives characteristic behavior in the region of a quantum phase transition.  For instance, the magnetization has the relation
\begin{align}
\langle \sigma^z_i \rangle = \left\{
\begin{array}{cc}
(1- (\frac{2 \Gamma}{J})^2 )^{1/8} & \frac{J}{2 \Gamma} > 1 \\
0 & \frac{J}{2 \Gamma} \le 1
\end{array}
\right.
\label{pfeutyresult}
\end{align}
which is plotted in Fig. \ref{fig9-1}(b).  This shows a sharp transition at the quantum phase transition point.

While diagonalizing (\ref{twositeham}) is straightforward, unfortunately looking at such small systems do not usually exhibit the phase transition\index{phase transition} physics that one would like to observe.  The sharp transition occurs when the system size is large $ N \rightarrow \infty $\index{thermodynamic limit}, usually called the {\it thermodynamic limit}. \index{thermodynamic limit} In Fig. \ref{fig9-1}(c) the energy gap between the first excited state and the ground state is shown. The sharp transition only occurs for large systems, and for finite systems the precise transition is less visible.  In this way, one is usually more interested in larger systems.  However, this is a computationally expensive task, hence one typically performs smaller scale simulations and extrapolate to large scale systems.  Furthermore, there are often many generalizations of the above scenario that one would like to handle.  Looking at the ground state alone corresponds to a physical system at the absolute zero of temperature.  Realistically, one would like to look also at the effects of non-zero temperature. However this is another non-trivial computational task since one must also find the spectrum of states.  Another interesting but non-trivial task is to understand non-equilibrium physics of such systems, which do not follow a Boltzmann distribution\index{Boltzmann!distribution} of states.  This could involve time-dynamical problems, which are quite difficult from a computational point of view hence has been studied to a lesser extent.

\section{Digital quantum simulation}
\label{sec:digitalqs}\index{quantum simulation!digital}

We first describe the approach of digital quantum simulation, which is historically the first method that was found to have a quantum speedup over classical methods.  Suppose that we wish to study a Hamiltonian which takes the general form
\begin{align}
H = \sum_{j} H_j,
\label{hamgeneric}
\end{align}
where the $ H_j $ is a locally acting Hamiltonian.  This is a extremely common structure for physical Hamiltonians of interest, since the terms in a Hamiltonian are usually derived from reducing a particular forces which tend to be locally acting.  For example, in the transverse Ising model\index{Ising model} of (\ref{transverseisingham}), the $ \sigma^z_j \sigma^z_k $ originates from an exchange interaction which is typically short-ranged, and the $ \sigma_j^x  $ arises from an applied magnetic field which only acts one spin at a time.  The key point which makes (\ref{hamgeneric}) difficult to solve is that not all the terms in the Hamiltonian commute:
\begin{align}
[H_j, H_{j'}] \ne 0 . 
\end{align}
For example, for the transverse Ising model $ \sigma_j^x  $ does not commute with all $ \sigma^z_j \sigma^z_k $ terms which involve the same site $ j $.  

Now suppose that we prepare the initial state of the system in a particular state $ |\psi(0)\rangle $ and we want to find the time evolution according to the Hamiltonian.  At some later time $ t $ the state is given by
\begin{align}
|\psi(t)\rangle & = e^{-i H t/\hbar} |\psi(0)\rangle \nonumber \\
& = \exp(-\frac{it}{\hbar} \sum_{j} H_j ) |\psi(0)\rangle .
\label{timeevolvedwave}
\end{align}
Now further suppose that we have in our possession a quantum computer that can perform any combination of one and two qubit gates.  For example, such gates can be written respectively as
\begin{align}
U_l (\bm{n},\theta) & = \exp(-i\theta \bm{n} \cdot \bm{\sigma}_l /2 ) \nonumber \\
U_{l l'}(\phi)  & = \exp(-i \phi \sigma_l^z  \sigma_{l'} ^z)\end{align}
where $ \bm{n} $ is a unit vector specifying the axis of the single qubit gate, and $ \theta,\phi  $ are freely choosable angles.  A well-known result in quantum information states that an arbitrary unitary evolution can be made by combining the one and two qubit operations.  Thus the unitary evolution $ e^{-i Ht/\hbar} $ should also be written in terms of the elementary gates.  The question is then, how to perform this decomposition?

The decomposition can be achieved by a Suzuki-Trotter expansion\index{Suzuki-Trotter expansion}. \index{Suzuki-Trotter expansion} 
The first order Suzuki-Trotter expansion is written
\begin{align}
e^{(A+B)} = \lim_{n \rightarrow \infty} \left( e^{\frac{A}{n}} e^{\frac{B}{n}} \right)^n ,
\end{align}
where $ n $ is the Trotter number\index{Trotter number}.  In practice, the Trotter number is kept at a finite number so that one expands the unitary evolution in (\ref{timeevolvedwave}) according to 
\begin{align}
 e^{-i Ht/\hbar} & = \exp(-\frac{it}{\hbar} \sum_{j} H_j ) \nonumber \\ 
& = \left[ \exp(-\frac{i \Delta t }{\hbar} \sum_{j} H_j ) \right]^{t/\Delta t} \nonumber \\ 
& \approx \left[ \prod_j \exp(-\frac{i \Delta t }{\hbar} H_j )  \right]^{t/\Delta t}
\label{trotterexpanded}
\end{align}
This appears as a sequence of terms of the form $  \exp(-\frac{i \Delta t }{\hbar} H_j ) $ which is suitable to perform a gate decomposition.  For example, for the case of the transverse Ising model\index{Ising model} (\ref{transverseisingham}), the sequence of gates would involve a sequence of operations including
\begin{align}
U_l (\bm{x}, \lambda \Delta t/\hbar )  & = \exp(- \frac{i \lambda \Delta t}{\hbar} \sigma_l^x  ) \label{uxevol} \\
U_{l l'}(J \Delta t/\hbar ) & = \exp(- \frac{i J \Delta t}{\hbar} \sigma_l^z  \sigma_{l'} ^z)  . \label{uzzevol}
\end{align}
The error of the operation by performing the Trotter decomposition\index{Suzuki-Trotter expansion} (\ref{trotterexpanded}) for one of the time steps $ \Delta t $ (inside the square brackets) scales as $ O( (\Delta t)^2 ) $.  This means that the error of the whole operation (\ref{trotterexpanded}) scales as $ O( t \Delta t ) $.  

The key point here is that the number of terms in the decomposition (\ref{trotterexpanded}) is only proportional to the number of terms in the original Hamiltonian (\ref{hamgeneric}).  Taking the transverse Ising model\index{Ising model} as an example again, the total number of terms in a periodic one-dimensional chain with $ N $ sites is $ 2N $. The number of gates to evolve the system forward in time by $ \delta t $ is $N $ units of (\ref{uxevol}) and $ N $ units of (\ref{uzzevol}).  In fact many of these gates can be done in parallel, since terms that do not involve the same site commute.  Repeating this procedure $ t/\Delta t $ times, one achieves a time evolution of the state with a time $ t $.  In comparison,  time evolving a quantum state on a classical computer is a computationally intensive task.  On a classical computer, one way to perform this would be to find the eigenstates of the Hamiltonian $ | E_n \rangle $ and perform an expansion (\ref{timeevolvedwave}) according to
\begin{align}
| \psi(t) \rangle = \sum_n e^{-i E_n t/\hbar} \langle E_n | \psi (0) \rangle | E_n \rangle .  
\end{align}
The part that is computationally expensive is the evaluation of the eigenstates, which requires diagonalization of a matrix of dimension $ 2^N \times 2^N $.  This is far less efficient and hence using a quantum computer has an exponential speedup over classical methods. 

The above shows that time-evolution can be performed more efficiently. But what if one is interested in other aspects of a quantum many-body model?  We saw that in the previous section that often one is interested in finding the eigenstates and energies of the Hamiltonian, such as the ground and excited states.  For these tasks, one can also show that these can be performed with a quantum speedup.  For example, an eigenstate can be obtained by performing the quantum phase estimation algorithm\index{phase estimation algorithm!quantum} on the operator 
\begin{align}
e^{-iHt/\hbar} = \sum_n e^{-i E_n t/\hbar} | E_n \rangle  \langle E_n | .  
\end{align}
The time evolution operator applies a phase of $ \phi = E_n t/\hbar $ on the eigenstates, which is the phase to be estimated.  To perform the time evolution operator, we again use the Trotter expansion (\ref{trotterexpanded})\index{Suzuki-Trotter expansion}. Since the Trotter expansion can be performed efficiently, it also shows that the quantum phase estimation algorithm can be performed efficiently.  The eigenstates are found by a by-product of the same algorithm, where after a measurement of the energy is made, the initial state is projected into an eigenstate of the Hamiltonian. Numerous other methods to examine various quantities have been investigated in the past in the digital quantum simulation\index{quantum simulation!digital} approach.  We do not discuss them here, we refer the reader to several excellent review articles as given in Sec. \ref{sec:ch9refs}.

\begin{exerciselist}[Exercise]
\item \label{q9-1}
Expand $ \left( e^{\frac{A}{n}} e^{\frac{B}{n}} \right)^n $ to second order and verify that it agrees with the expansion of $ e^{(A+B)}  $.  What is the form of the error that occurs due to the Suzuki-Trotter expansion? What happens if $ [A, B ] = 0 $?   
\end{exerciselist}

\section{Toolbox for analogue quantum simulators}
\label{sec:toolbox}\index{quantum simulation!analogue}

The digital quantum simulation\index{quantum simulation!digital} approach described in the previous section assumes the availability of a highly controllable quantum system, consisting of a large number of qubits.  To perform the time evolution, one and two qubit gates were applied in sequence, in possible combination with quantum algorithms such as the phase estimation\index{phase estimation} algorithm.  This requires the capabilities of an experimental system that is close or equivalent to a full quantum computer.  While great progress has been made in recent years, at the time of writing only relatively small numbers of qubits can be fully controlled, hence the digital quantum simulation\index{quantum simulation!digital} approach can only be applied to small quantum systems. 

One of the features of many quantum-many problems is that there are usually symmetries that can be exploited.  For example, in the transverse Ising model\index{Ising model} (\ref{transverseisingham}), the couplings $ J, \Gamma $ are translationally invariant. This means that the gates such as (\ref{uxevol}) and (\ref{uzzevol}) are the same for all the qubits. This means that they can be applied in parallel across all qubits, as long as the gates commute.   In fact, depending upon the particular experimental system, it may be possible to even perform several Hamiltonians at the same time, even if they do not commute.  For example, one might be able to apply a one and two qubit operation
\begin{align}
H_j = - \frac{J}{4} \sum_{k \in \text{n.n.}} \sigma^z_j \sigma^z_k - \Gamma \sigma_j^x
\end{align}
on the $j$th qubit. Then performing this on all qubits $ H = \sum_j H_j $ reproduces (\ref{transverseisingham}),  directly implementing the desired Hamiltonian.  The only reason that we separated the two types of gates in (\ref{uxevol}) and (\ref{uzzevol}), is because of the assumption that these gates must be performed separately.  If experimentally it is in fact possible to perform them at the same time, then the procedure involving the Trotter decomposition\index{Suzuki-Trotter expansion} of the last section is clearly unnecessary and it is possible to directly perform the time evolution. 

What kind of Hamiltonians can be applied, and whether they can be done simultaneously, is largely an experimental question that depends upon the particular system.  Cold atomic systems are one of the most versatile systems with a variety of experimental techniques to construct various Hamiltonians. In addition to creating the Hamiltonian of interest, it is necessary to measure the system in an appropriate way to find the physics of interest.  For example, some observables such as the order parameter\index{order parameter} need to be extracted from the state of interest.  In this section, we outline the various techniques available for analogue quantum simulation with cold atoms. \index{quantum simulation!analogue}

\subsection{Optical lattices}
\label{sec:ch9optlat}\index{optical lattice}

As we have seen in Chapter \ref{ch:order}, the spatial wavefunction of BEC can be described by an order parameter\index{order parameter} in continuous space. Meanwhile, many of the models of interest that are discussed in condensed matter physics and other areas of physics are discrete lattice systems.  Therefore, if we are to use BECs to simulate the physics of a lattice model, some way of discretizing space is required.  This is performed experimentally using optical lattices, and has proved an immensely powerful tool to manipulate cold atom systems.

\begin{figure}[t]
\includegraphics[width=\textwidth]{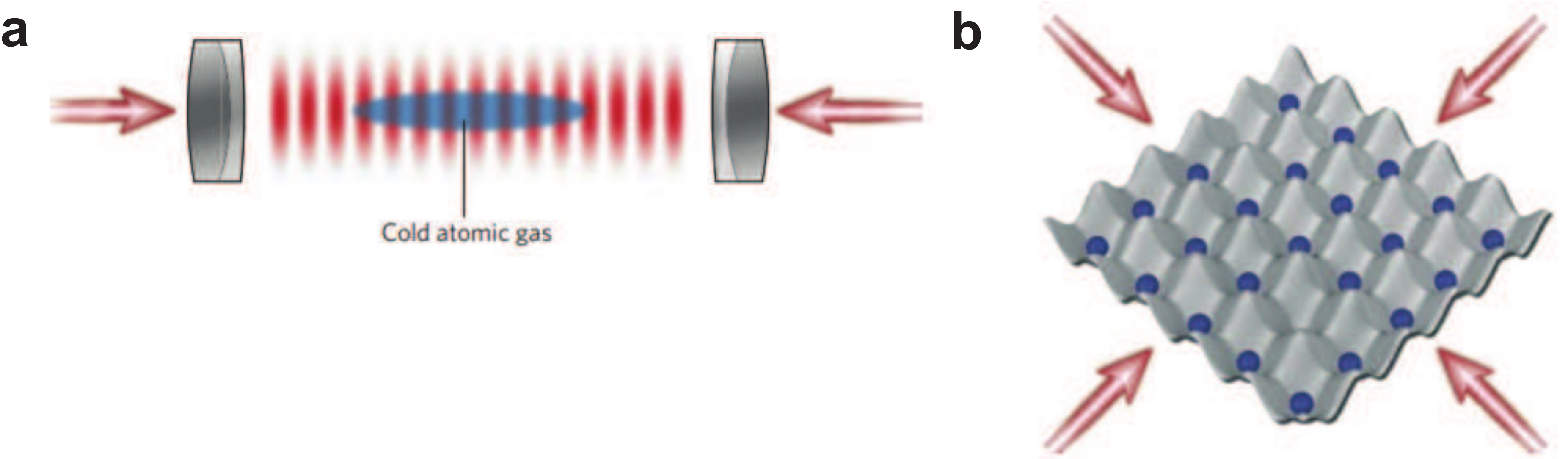}
\caption{
An optical lattice\index{optical lattice} for creating a periodic potential with cold atoms.  (a) Two counterpropagating laser beams produce a standing wave.  (b) The effective potential produced by the optical lattice. Diagrams reproduced from \cite{bloch2008quantum}.  }
\label{fig9-2}
\end{figure}

The simplest configuration of an optical lattice consists of two counter-propagating laser beams which produce a standing wave.  The counter-propagating lasers may be conveniently produced, for example, by using mirrors to reflect the laser light entering from one side (Fig. \ref{fig9-2}(a)).  This produces a light intensity that varies along the axis of the laser beams as
\begin{align}
I(x) = I_0 \cos^2 (\kappa x)
\label{intensitylaserbeam}
\end{align}
where $ \kappa = 2 \pi/\lambda $ and $ \lambda $ is the wavelength of the light.  

As we have already encountered in Sec. \ref{sec:acstark}, an off-resonant light field will create an ac Stark shift\index{ac Stark shift}.  From (\ref{rabifrequency}) we observe that the Rabi frequency\index{Rabi frequency} $ \Omega $ is proportional to the amplitude of the light, hence the energy shift (\ref{acstarkenergy}) is proportional to the intensity of the light.  Since the laser beam in the standing wave configuration is a spatially varying intensity, there will be a spatially varying energy due to the ac Stark shift according to
\begin{align}
V(x) \propto V_0 \cos (2 \kappa x) 
\label{opticallatticepotential}
\end{align}
where $ V_0  = - \frac{|\Omega|^2}{\Delta} $ and have removed an irrelevant constant, which contributes to a global energy offset.  This periodic potential is what is experienced by the atoms and is the effect of applying the optical lattice.\index{optical lattice}  
Generally the parameter $ V_0  $ is relatively easily controlled, for example by changing the laser intensity, so one can produce a potential depth as desired.  Although the above potential is only in one-dimension here, by applying more than one laser it is possible to produce two- and three-dimensional lattices by combining other lasers in standing wave configurations (Fig. \ref{fig9-2}(b)).  The simplest types of lattices are square lattices, but other lattice geometries are also possible, using combinations of laser beams in other orientations.

\subsection{Feshbach resonances}
\label{sec:feshbachbcs}\index{Feshbach resonance}

The optical lattice\index{optical lattice} manipulates the single particle Hamiltonian (\ref{multiham}), specifically the potential energy $ V(\bm{x}) $.  On the other hand, as discussed in Sec. \ref{sec:interactions}, there is another important component to the atomic Hamiltonian, which are the interactions between the atoms.  As we have already discussed in Sec. \ref{sec:feshbach}, by applying a magnetic field, it is possible to 
change the strength of the atom-atom interaction, and even change the sign between repulsive to attractive.  Feshbach resonances can therefore be used to manipulate the interaction part of the Hamiltonian (\ref{interactionham}) to a desired value.    

One of the best know experiments that demonstrated this technique was the observation of the transition between a BEC of molecules to a Bardeen-Cooper-Schrieffer (BCS) state\index{Bardeen-Cooper-Schrieffer state}. In the experiment performed by Regal, Greiner, and Jin, a trapped gas of fermionic $^{40}\text{K} $ atoms was cooled to quantum degeneracy and a magnetic field was applied to produce a Feshbach resonance.  In the regime where the scattering length corresponds to repulsive interactions, there exists a weakly bound molecular state, where the binding energy is controlled by the Feshbach resonance\index{Feshbach resonance}.  In this regime, the state is a BEC of bosonic diatomic molecules, and there are no fermion degrees of freedom.  When the Feshbach resonance is such that interactions are attractive, no molecules form, but there is still a condensation of Cooper pairs which can be described by a BCS theory.  Thus by controlling the Feshbach resonances it is possible to observe both phenomena in the same atomic system.

\subsection{Artificial gauge fields}
\label{sec:artgaugefield}\index{artificial gauge field}

In condensed matter physics, a common situation that one encounters is a charged particle in a magnetic field. The archetypical such experiment is the quantum Hall effect, where electrons in a two-dimensional plane have a magnetic field applied to it.   In terms of the Schrodinger equation this can be taken into account by a vector potential $ \bm{A}(\bm{x}) $, where the single particle Hamiltonian (\ref{singleparticleham}) becomes
\begin{align}
H_0 (\bm{x})  = -\frac{( \bm{p} - \bm{A}(\bm{x}) )^2}{2m} + V(\bm{x}).
\label{singleparticleham2}
\end{align}
where $ \bm{p} = - i \hbar \bm{\nabla} $ is the momentum operator.  One issue with using cold atoms is that they are charge neutral --- this means that applying a magnetic field will not produce such a vector potential.  How can one produce a magnetic, and more generally, a gauge field? Here we will introduce two methods that can be used to make an effective \index{vector potential}vector potential $\bm{A}(\bm{x}) $.  

The first method takes advantage of the fact that there is a very close relationship between rotation and a magnetic field. To illustrate this consider the single-particle Hamiltonian in a harmonic oscillator rotating around the $ z $-axis
\begin{align}
H_{\text{rot}} (\bm{x}) = \frac{p^2}{2m} + \frac{1}{2} m \omega^2 (x^2 + y^2 + z^2) - \Omega L^z
\end{align}
where  $ \Omega $ is the rotation frequency, and $ L^z = \hat{\bm{z}} \cdot (\bm{r} \times \bm{p}) $ is the angular momentum.  
The above can be rewritten as 
\begin{align}
H_{\text{rot}} (\bm{x}) = \frac{(\bm{p}-  \bm{A}(\bm{x}) )^2}{2m} + \frac{1}{2} m (\omega^2-\Omega^2) (x^2 + y^2 ) 
+\frac{1}{2} m \omega^2 z^2
\label{standardqhe}
\end{align}
where $  \bm{A}(\bm{x}) = m \Omega \hat{\bm{z}}  \times \bm{r} $.  Thus the rotation has the effect of producing an effective vector potential, as well as an anti-trapping in the perpendicular plane to the axis of rotation.

The second method takes advantage of Berry phases\index{Berry phase} that are produced in an adiabatic evolution.  The simplest configuration that shows the effect is when there is a spatially varying spin Hamiltonian.  Consider a similar situation to what was discussed in Sec. \ref{sec:transitions}, where a laser is applied to atoms such that it produces the effective Hamiltonian (\ref{interactionham2}).  The difference that we will consider here is that the Rabi frequency $ \Omega $ and detuning $ \Delta $ will have a  more general spatial dependence so that the overall Hamiltonian is
\begin{align}
H_{\text{Berry}} (\bm{x}) = \left( \frac{p^2}{2m} + V(\bm{x}) \right) 
\left(
\begin{array}{cc}
1 & 0 \\
0 & 1 
\end{array}
\right)
+ \hbar 
\left(
\begin{array}{cc}
0 & \Omega(\bm{x})/2 \\
\Omega^*(\bm{x}) /2 & \Delta(\bm{x})
\end{array}
\right) .
\label{berryham}
\end{align}
Here the first term is the spin-independent part of the Hamiltonian due to the kinetic and potential energy,  the second is due to the laser creating transitions between the spin states $ |1 \rangle $ and $ | 2 \rangle $.  The spin-dependent part of the Hamiltonian can be diagonalized and is given by
\begin{align}
| E^\pm_{\text{s}} (\bm{x}) \rangle = \Delta (\bm{x}) \mp \sqrt{\Delta^2 (\bm{x}) + |  \Omega (\bm{x}) |^2 } |1 \rangle - \Omega^* (\bm{x}) | 2 \rangle,
\label{chidiag}
\end{align}
where the spin-dependent energy being
\begin{align}
E^\pm_{\text{s}}(\bm{x}) = \frac{\Delta (\bm{x}) \pm \sqrt{\Delta^2 (\bm{x})  + |  \Omega (\bm{x}) |^2 } }{2} .
\end{align}

The above only considers the spin part of the Hamiltonian.  Let us now obtain an effective Hamiltonian including the spatial degrees of freedom. The most general form of the wavefunction including the spatial degrees of freedom takes the form
\begin{align}
| \Psi(\bm{x},t) \rangle = \psi^+ (\bm{x},t) | E^+_{\text{s}} (\bm{x}) \rangle + \psi^- (\bm{x},t) | E^-_{\text{s}} (\bm{x}) \rangle.
\end{align}
Consider that we prepare the state in the  lower energy state $ | E^-_{\text{s}} (\bm{x}) \rangle $, and all subsequent operations on the state will be sufficiently slow such that the adiabatic approximation\index{adiabatic approximation} holds.  In this case we can assume that the spin of the atoms will always remain in the state $ | E^-_{\text{s}}  (\bm{x}) \rangle $, and the wavefunction will be approximately
\begin{align}
| \Psi(\bm{x},t) \rangle \approx \psi^- (\bm{x},t) | E^-_{\text{s}} (\bm{x}) \rangle.
\label{approxpsiberry}
\end{align}

We now wish to obtain an equation of motion for the spatial wavefunction $ \psi^- (\bm{x},t) $.  First rewrite the Hamiltonian (\ref{berryham}) in the basis (\ref{chidiag}), yielding
\begin{align}
H_{\text{Berry}} (\bm{x}) | \Psi(\bm{x},t) \rangle  = & \left( \frac{p^2}{2m} + V(\bm{x}) \right)  \left( | E^+_{\text{s}}  (\bm{x})  \rangle  \langle E^+_{\text{s}}  (\bm{x}) | + | E^-_{\text{s}}  (\bm{x})  \rangle  \langle E^-_{\text{s}}  (\bm{x})  | \right) \nonumber \\
& + E^+_{\text{s}}  (\bm{x})  | E^+_{\text{s}}  (\bm{x})  \rangle  \langle E^+_{\text{s}}  (\bm{x}) | + E^-_{\text{s}} (\bm{x})   | E^-_{\text{s}} (\bm{x})  \rangle  \langle E^-_{\text{s}}  (\bm{x})  |  .
\label{rotatedberryham}
\end{align}
Applying (\ref{rotatedberryham}) to (\ref{approxpsiberry}) and projecting the result onto $ | E^-_{\text{s}} (\bm{x}) \rangle $ under the assumption that the spin component will always remain in this state, we have the effective Hamiltonian
\begin{align}
H_{\text{eff}}^{-} (\bm{x}) & = \langle E^-_{\text{s}}  (\bm{x})  | H_{\text{Berry}} (\bm{x})| \Psi(\bm{x},t) \rangle  \nonumber \\
& =  \langle E^-_{\text{s}} (\bm{x})  | \frac{p^2}{2m} \psi^- (\bm{x},t) | E^-_{\text{s}}  (\bm{x})  \rangle   + \left( V(\bm{x})  + E^-_{\text{s}} (\bm{x}) \right) \psi^- (\bm{x},t).
\end{align}
The first term above is the key difference to the more typical situation, where the spin eigenstates do not have a spatial dependence.  Due to the spatial dependence of $ | E^-_{\text{s}}  (\bm{x})  \rangle $, the derivative in the momentum $ \bm{p} = - i \hbar \bm{\nabla} $ will act on the spin eigenstates as well as the spatial wavefunction.  Evaluating this term we have
\begin{align}
H_{\text{eff}}^{-} (\bm{x})  = & \left( \frac{p^2}{2m} + V(\bm{x})  + E^-_{\text{s}} (\bm{x}) \right) \psi^- (\bm{x},t) \nonumber \\
& \frac{-i \hbar (  \bm{p} \psi^- (\bm{x},t) ) \cdot \langle E^-_{\text{s}} (\bm{x})  | \bm{\nabla} | E^-_{\text{s}}  (\bm{x})  \rangle  
- \hbar^2 \psi^- (\bm{x},t)  \langle E^-_{\text{s}} (\bm{x})  |\nabla^2 | E^-_{\text{s}}  (\bm{x})  \rangle}{2m} .
\end{align}
The last term may be written 
\begin{align}
\langle E^-_{\text{s}} (\bm{x})  |\nabla^2 | E^-_{\text{s}}  (\bm{x})  \rangle & = 
 \langle E^-_{\text{s}} (\bm{x})  | \bm{\nabla} \cdot \left( | E^+_{\text{s}}  (\bm{x})  \rangle  \langle E^+_{\text{s}}  (\bm{x}) | + | E^-_{\text{s}}  (\bm{x})  \rangle  \langle E^-_{\text{s}}  (\bm{x})  | \right) \bm{\nabla}  | E^-_{\text{s}}  (\bm{x})  \rangle \nonumber \\
& =  || \langle E^-_{\text{s}} (\bm{x})  | \bm{\nabla} | E^-_{\text{s}}  (\bm{x})  \rangle ||^2
+ || \langle E^-_{\text{s}} (\bm{x})  | \bm{\nabla} | E^+_{\text{s}}  (\bm{x})  \rangle  ||^2 
\end{align}
where the $ || \cdots || $ is the vector norm\index{vector norm} with respect to the spatial directions.  Introducing the effective vector and scalar potentials\index{vector potential}\index{scalar potential}
\begin{align}
\bm{A} ( \bm{x}) & = i \hbar \langle E^-_{\text{s}} (\bm{x})  | \bm{\nabla} | E^-_{\text{s}}  (\bm{x})  \rangle   \nonumber \\
W  ( \bm{x}) & = -\frac{\hbar^2}{2m} || \langle E^-_{\text{s}} (\bm{x})  | \bm{\nabla} | E^+_{\text{s}}  (\bm{x})  \rangle  ||^2,
\end{align}
we obtain the effective Hamiltonian 
%
\begin{align}
H_{\text{eff}}^{-} (\bm{x})  = & \left( \frac{( \bm{p} - \bm{A}(\bm{x}) )^2}{2m}  + V (\bm{x}) + E^-_{\text{s}} (\bm{x}) + W  ( \bm{x}) \right) \psi^- (\bm{x},t).
\end{align}
The above has the form of (\ref{singleparticleham2}) where the vector potential originates from the Berry phase\index{Berry phase} due to the spatial dependence of the spin eigenstates.  By suitably engineering the spatial dependence of the spin eigenstates (\ref{chidiag}), it is possible to produce a vector potential of the desired form.  The effective magnetic field corresponding to the vector potential\index{vector potential} is then given by 
\begin{align}
\bm{B}( \bm{x})  & = \bm{\nabla} \times \bm{A} ( \bm{x})  \nonumber \\
& = i \hbar  \bm{\nabla} \times  \langle E^-_{\text{s}} (\bm{x})  | \bm{\nabla} | E^-_{\text{s}}  (\bm{x})  \rangle   .
\end{align}

\subsection{Spin-orbit coupling}
\label{sec:spinorbitcoupling}
\index{spin-orbit coupling}

In some condensed matter systems, even more exotic Hamiltonians can arise beyond the relatively straightforward types of potentials and interactions that we have examined so far.  One class of interactions that has gained a lot of interest in recent years are spin-orbit coupled Hamiltonians.  Spin-orbit coupling can be one of the ingredients to form a topological insulator, where the bulk electronic band structure is that of a band insulator, but has topologically protected conductive states on the edge or surface.  Due to the great interest of such materials in condensed matter physics, a natural aim of quantum simulation has been to produce novel types of interactions which may display exotic properties.  The great advantage of such a quantum simulation approach --- where interactions are produced from the ``ground-up'' --- is that it is possible to engineer Hamiltonians that may be difficult, or even impossible, to produce in natural materials.  \index{topological insulator}

Firstly, what is spin-orbit coupling?  This is most familiar from traditional atomic physics, where the orbital angular momentum $ \bm{L} $ of an electron in an atom couples to its spin $\bm{S} $, and the interaction takes the form $ \propto\bm{L} \cdot \bm{S} $. The key point here is that the motion of the particle affects the spin, such that there is momentum-dependence to the energy of the spin.  Recalling that the energy of a spin in a magnetic field takes the form $ \propto \bm{B} \cdot \bm{S} $, we can view the spin-orbit coupling as an effective momentum-dependent Zeeman magnetic field.  

For electrons within an atom, it is angular momentum that couples to spin, but in condensed matter physics, regular linear momentum can also couple to the spin. For example, a coupling of the form 
$ \propto k_y S_x  - k_x S_y  $ is what is referred to as Rashba spin-orbit coupling\index{Rashba spin-orbit coupling} and can arise in two-dimensional semiconductor heterostructure systems.  Another type of spin-orbit coupling, referred to as the Dresselhaus effect\index{Dresselhaus effect}, results in a coupling of the form $ \propto k_x S_x  - k_y S_y $ up to linear terms.

\begin{figure}[t]
\includegraphics[width=\textwidth]{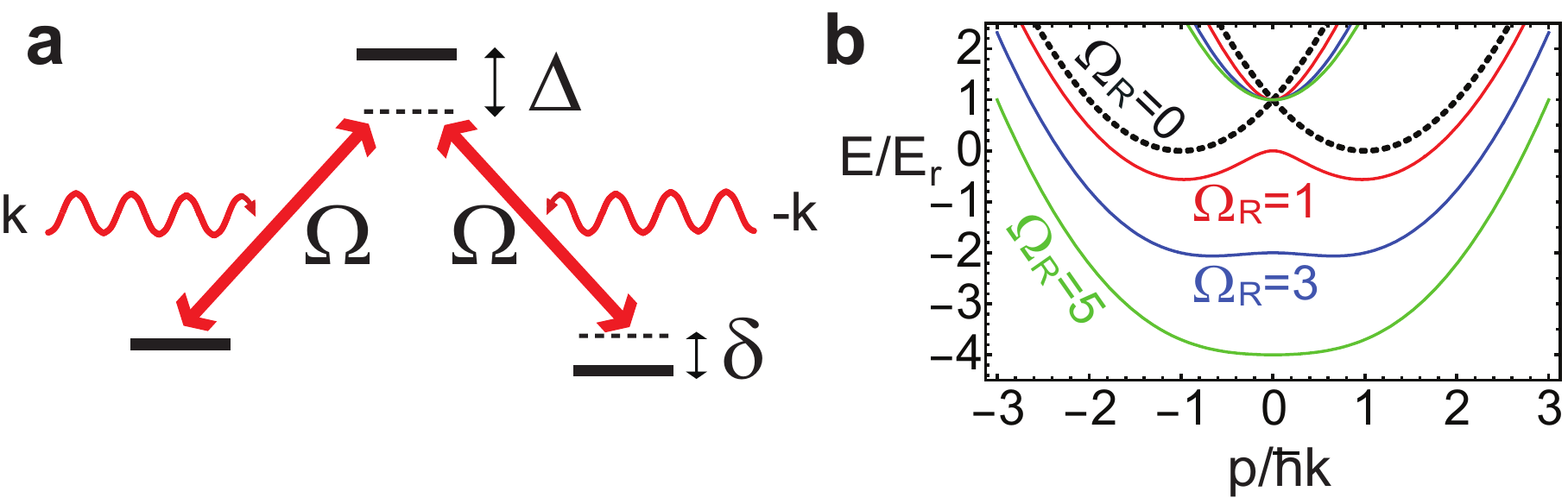}
\caption{Generating spin-orbit coupling in Bose-Einstein condensates.  (a) Raman scheme for coupling two atomic spin states.  The laser beams are counter-propagating with momenta $ \bm{k} $, which transfers a momentum $ 2\bm{k} $ during the spin transition.   (b) Energy spectrum of the spin-orbit coupled Hamiltonian (\ref{spinorbitham2}).  Here $ \Omega_R = \frac{|\Omega|^2}{2 \Delta} $ is the effective Raman coupling and $ E_r = \frac{\hbar^2 k^2}{2m} $ is the recoil energy.  }
\label{fig9-5}
\end{figure}

Such types of spin-orbit interaction can be produced in cold atoms by applying Raman fields between various internal states of atoms.  The basic idea is that the Raman fields are applied in such way that spin transfer between two states also results in a net momentum transfer due to the absorption and emission of a photon.  For example, suppose that in the Raman transition shown in Fig. \ref{fig9-5}(a) is produced by two counterpropagating laser beams each of wavenumber $ \bm{k} $.  On the first leg of the Raman transition a photon is absorbed, adding a momentum $ \bm{k} $ to the atom. On the second leg of the Raman transition, a photon is emitted with momentum $ - \bm{k} $, thus the atom loses momentum $ - \bm{k} $.  The total momentum change of the atom is $ 2 \bm{k} $.  For cold atoms this momentum change is not negligible in comparison to their average momenta, and causes a  spin-dependent shift to the dispersion relation. 

To show this in a concrete way, let us consider the situation of applying the two counterpropagating laser fields to a BEC.  The single particle Hamiltonian is
\begin{align}
H_{\text{spin-orbit}} ( \bm{x}) = 
\frac{p^2}{2m} 
\left( \begin{matrix}
1 & 0 & 0 \\
0 & 1 & 0 \\
0 & 0 & 1 \\
\end{matrix} \right) 
+ \hbar
\left( \begin{matrix}
0 & \Omega^* e^{-i \bm{k} \cdot \bm{x}}/2 & 0 \\
\Omega e^{i \bm{k} \cdot \bm{x}}/2 & \Delta & \Omega^* e^{-i \bm{k} \cdot \bm{x}}/2 \\
0 & \Omega e^{i \bm{k} \cdot \bm{x}}/2 & \delta   \\
\end{matrix} \right)  ,  
\label{spinorbithama}
\end{align}
where $ \delta $ is a two-photon detuning.  
Here we have ignored the trapping potential for simplicity and have included the spatial dependence of the Raman fields as in (\ref{transitionham}).  
Eliminating the intermediate level as in Sec. \ref{sec:acstark}, we obtain the effective 2-level Hamiltonian
\begin{align}
H_{\text{spin-orbit}} ( \bm{x}) &  \approx 
\frac{p^2}{2m} 
\left( \begin{matrix}
1 & 0  \\
0 & 1  \\
\end{matrix} \right) 
+ \hbar
\left( \begin{matrix}
- |\Omega|^2/4 \Delta & (\Omega^*)^2 e^{-2 i \bm{k} \cdot \bm{x}}/4 \Delta \\
\Omega^2 e^{2 i \bm{k} \cdot \bm{x}}/4 \Delta   & \delta -|\Omega|^2/4 \Delta  \\
\end{matrix} \right)   \nonumber \\
 & =   \frac{p^2}{2m}  I - \frac{|\Omega|^2}{4 \Delta } I + \frac{\delta}{2} ( I -  \sigma^z)  + \frac{ (\Omega^*)^2 e^{-2 i \bm{k} \cdot \bm{x}}}{ 4 \Delta } \sigma^+  + \frac{\Omega^2 e^{2 i \bm{k} \cdot \bm{x}}}{4 \Delta } \sigma^- , 
\label{effectivespinorbit}
\end{align}
where we wrote the various terms in terms of Pauli operators\index{Pauli operators}. We can rewrite this Hamiltonian to remove the spatially dependent Raman terms by applying a unitary transform
\begin{align}
U = e^{i \bm{k} \cdot \bm{x} \sigma^z} .  
\end{align}
This removes the spatial dependence because the spin operators transform as
\begin{align}
U \sigma^+ U^\dagger & =  e^{2 i \bm{k} \cdot \bm{x}} \sigma^+ \nonumber \\
U \sigma^- U^\dagger & =  e^{-2 i \bm{k} \cdot \bm{x}} \sigma^-  \nonumber \\
U \sigma^z U^\dagger & = \sigma^z .
\label{spinunitarytransspinorb}
\end{align}
Meanwhile the momentum operator is shifted according to 
\begin{align}
U \bm{p} U^\dagger = \bm{p} - \hbar \bm{k} \sigma^z
\label{unitarymomentum}
\end{align}
due to the fact that $ \bm{p} = -i \hbar \bm{\nabla} $.  The transformed Hamiltonian then reads
\begin{align}
U H_{\text{spin-orbit}} ( \bm{x}) U^\dagger & = 
 \frac{ (\bm{p}- \hbar \bm{k} \sigma^z  )^2 }{2m}  I - \frac{|\Omega|^2}{4 \Delta } I + \frac{\delta}{2} ( I -  \sigma^z)  + 
\frac{ (\Omega^*)^2 }{ 4 \Delta } \sigma^+  + \frac{\Omega^2 }{4 \Delta } \sigma^-  \label{spinorbitham} \\
& = \left(
\begin{matrix}
\frac{ (\bm{p}- \hbar \bm{k})^2 }{2m}  - \frac{|\Omega|^2}{4 \Delta } & \frac{ (\Omega^*)^2 }{ 4 \Delta } \\
\frac{\Omega^2 }{4 \Delta }  &  \frac{ (\bm{p} + \hbar \bm{k} )^2 }{2m}    - \frac{|\Omega|^2}{4 \Delta } + \delta 
\end{matrix}
\right) . \label{spinorbitham2} 
\end{align}
Expanding out the first term in (\ref{spinorbitham}), we have
\begin{align}
 \frac{ (\bm{p}- \hbar \bm{k} \sigma^z  )^2 }{2m}  =  \frac{p^2 }{2m} - \frac{\hbar}{m}  (\bm{k} \cdot \bm{p} )\sigma^z
+ \frac{\hbar k^2}{2m} .
\end{align}
The component of the momentum parallel to the Raman fields is coupled to the spin of the particle, realizing the spin-orbit interaction.\index{spin-orbit interaction}

Let us briefly analyze the Hamiltonian (\ref{spinorbitham2}) to understand what the effect of the spin-orbit interaction is.   First, consider the limit where $ |\Omega | \rightarrow 0  $.  In this case, the two spin states are uncoupled and the dispersion corresponds to two parabola displaced by a momentum $ \pm \hbar \bm{k} $.  The two-photon detuning $ \delta $ creates an energy offset between the two spin states.  Turning the Raman fields on, the two spin states couple, producing an anti-crossing of the levels and a double-well structure forms in momentum space  (see Fig. \ref{fig9-5}(b)). For larger values of $ |\Omega | $, there is a point where the double-wells merge into a single minimum. Such a phase transition was experimentally observed by \cite{lin2011spin} where spin-orbit coupling was produced in a BEC for the first time. We note that similar methods can be used to form artificial gauge fields, providing another method to that described in the previous section.

\begin{exerciselist}[Exercise]
\item \label{q9-6}
Examining the $ p= 0 $ state, verify that the effective Hamiltonian (\ref{effectivespinorbit}) has the same spectrum as (\ref{spinorbithama}) for the lowest two levels, as long as $ | \Omega |, \delta \ll \Delta $.  
\item \label{q9-7}
Verify the unitary transforms (\ref{spinunitarytransspinorb}) on the Pauli operators\index{Pauli operators}.   
\item \label{q9-8}
Show that  
\begin{align}
e^{ i k x} p_x e^{-i k x} \psi(x) = (p_x- \hbar k) \psi(x)
\end{align}
where $ p_x = -i \hbar \frac{\partial}{\partial x} $.  Use this to verify (\ref{unitarymomentum}).  
\item \label{q9-9}
Perform the unitary transformation to derive the spin-orbit Hamiltonian (\ref{spinorbitham2}).  
\item \label{q9-10}
Diagonalize the Hamiltonian (\ref{spinorbitham2}) and show that the spectrum follows the form shown in Fig. \ref{fig9-5}(b).  
\end{exerciselist}

\subsection{Time-of-flight measurements}
\label{sec:timeofflight}\index{time-of-flight measurement}

So far, we have discussed methods of creating various potentials and interactions for the trapped atoms.  However in order to understand the nature of the quantum many-body state of the system that has been created with the atoms, the atoms need to be measured in an appropriate way to extract information that characterizes the state.  One of the most important measurement techniques in the context of quantum simulation is the {\it time-of-flight measurement}, which allows one to measure the momentum distribution of the atoms.  \index{time-of-flight measurement}  

\begin{figure}[t]
\includegraphics[width=\textwidth]{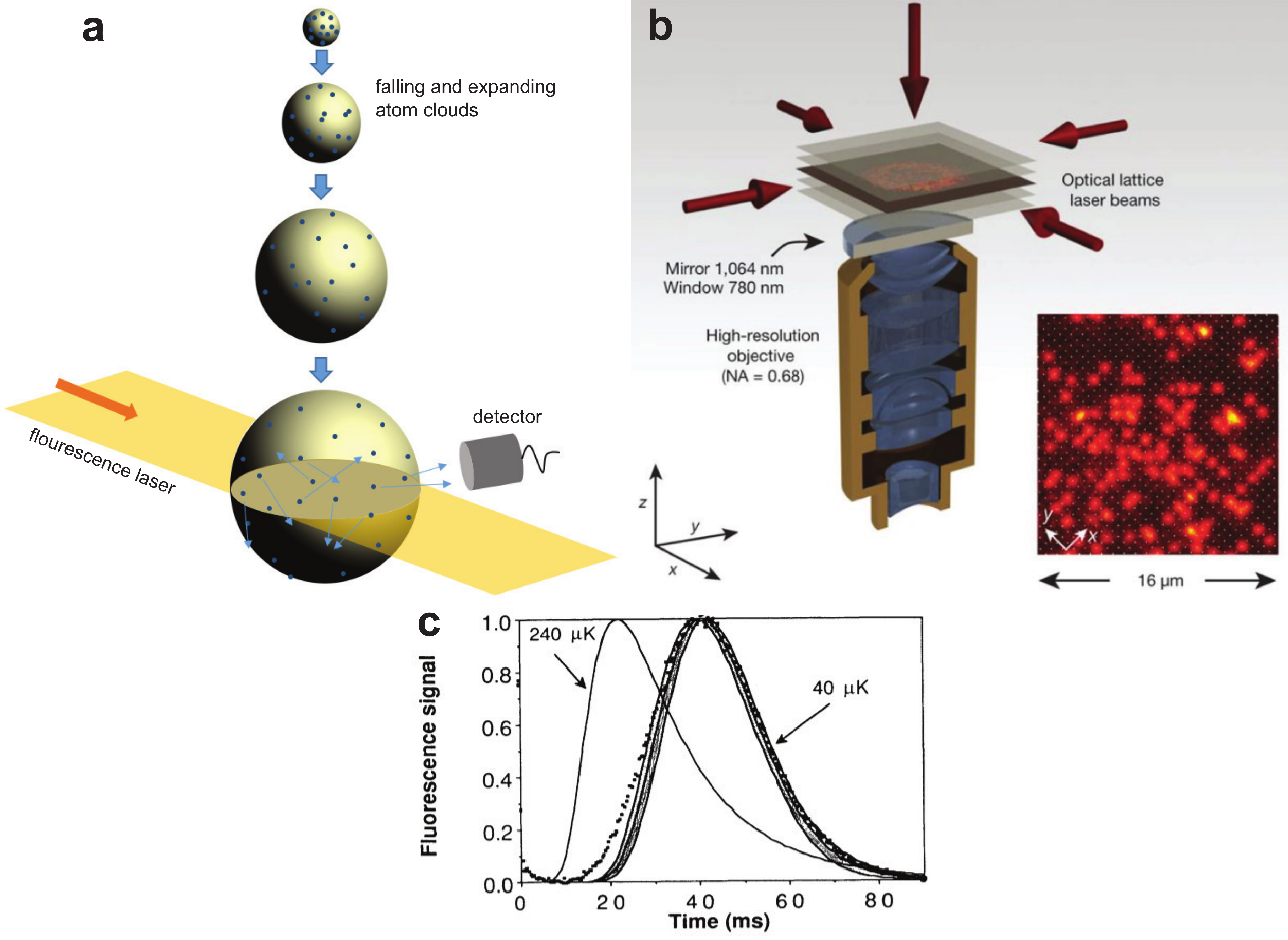}
\caption{
Methods for measuring the quantum many-body state of cold atoms for quantum simulation.  (a) Momentum space imaging using time-of-flight measurements  (adapted from \cite{phillips1998nobel}).   (b) Spatial imaging using the quantum gas microscope (reproduced from \cite{sherson2010single}).  (c)  Measurement data for cooled atoms above the BEC transition temperature showing a Maxwell-Boltzmann distribution at $ T = 40 \mu$K (reproduced from \cite{lett1988observation}).  }\index{time-of-flight measurement}\index{Maxwell-Boltzmann distribution}
\label{fig9-3}
\end{figure}

The way that this is performed is illustrated in Fig. \ref{fig9-3}(a).  After the suitable quantum state has been prepared with the atoms (by applying a suitable Hamiltonian for example), they are released from the trap, and they fall under the influence of gravity. Typically the trap size is much smaller than the distance that the atoms fall, so to first approximation we may neglect the size of the atom cloud when it is first released from the trap.   A given state of the atoms will generally consist of a variety of different momenta, and hence velocities of the atoms.  Below the trap, laser light illuminates the atoms and is tuned at the frequency such that the fluorescence of the atoms can be measured.  Due to the distribution of velocities of the atoms, the atom cloud will expand while falling under the influence of gravity.  Suppose an atom is initially moving with a velocity $ \bm{v} = (v_x, v_y, v_z) $ when it is just released from the trap.  Then the coordinates of the atom after a time $ t $ is 
\begin{align}
x & = v_x t \nonumber \\
y & = v_y t \nonumber \\
z & = v_z t - \frac{1}{2} g t^2 ,
\end{align}
where $ g = 9.8 \text{m}/\text{s}^2 $ is the acceleration due to gravity taken to be in the $ z $-direction. A measurement of the positions of the atoms $ (x,y,z)$ and the time $ t $ can thus be used to infer the velocity $ \bm{v} $ (and hence the momentum) of the atoms.  Despite the classical nature of the argument above, even for a quantum state being released from the trap, the momentum distribution of the atoms can be measured.  In this case, the atoms undergo interference while expanding, such that the fluorescence image corresponds to a measurement in momentum space.  An example of this is discussed later in Fig. \ref{fig9-4}. 

Time-of-flight measurements can also be used to infer the temperature of an atomic cloud, assuming a Maxwell-Boltzmann velocity distribution. An example of a time-of-flight measurement is shown in the inset of Fig. \ref{fig9-3}(c).  Here the fluorescence of the atoms are detected at a fixed distance $ z = - l_0 $ below the trap.  The center of the fluorescence peak corresponds to atoms which had an initial velocity of $ v_z = 0 $, and arrive at a time $ t = \sqrt{2l_0/g} $.  The atoms which happened to be moving downwards $ v_z < 0  $ will fluoresce earlier than the atoms which were moving upwards $ v_z > 0 $, and thus we obtain a distribution to the fluorescence signal.  By looking at the time distribution of the fluorescence, one can deduce the velocity   distribution of the atoms, from which the temperature can be inferred.   \index{fluorescence imaging}

%
%

\subsection{Quantum gas microscope}
\label{sec:quantumgasmicro}\index{quantum gas microscope}

Time-of-flight measurements\index{time-of-flight measurement} allow one to examine the momentum space distribution of the atoms. It is then a natural question to ask whether one can also measure the spatial distribution of the atoms.   Measuring the spatial distribution with high resolution is in fact a more technically demanding task than the momentum measurement.  The reason is that in order to measure the spatial distribution, one requires the measurement of the BEC {\it in situ}, i.e. while the atoms are still in the trap.  When the atoms are within the trap, they have a high optical density and the size of the condensate is very small, making absorption imaging (the most common method of measuring BECs) difficult to apply.  In the first quantum simulation\index{quantum simulation} experiments with cold atoms, such as with the Bose-Hubbard model\index{Bose-Hubbard model} performed in 2002 (we will discuss this in more detail in Sec. \ref{sec:bosehubbard}), no methods were available to image BECs at the resolution of the optical lattice period, and most of the results were inferred using time-of-flight measurements\index{time-of-flight measurement}. However, from 2008 and onwards several methods were developed to image BECs at the resolution of the optical lattice period.

The most well-known method of performing spatial measurements is using fluorescence imaging\index{fluorescence imaging}. In fluorescence imaging, photons originating from spontaneous emission during the laser cooling cycle are collected and imaged. Using a high-resolution microscope (see Fig. \ref{fig9-3}) it is possible to detect the presence of atoms at the level of a single optical lattice site.  In one of the first experiments to perform this by Sherson {\it et al.}, the optical lattice period was 532nm, while the wavelength of the detected light was 780nm, and the numerical aperture was $ \text{NA} = 0.68 $.  This gives a optical resolution of about 700nm, which is comparable to the optical lattice period. By collecting a sufficiently high number of photons (several thousands per pixel) one can accurately deduce the presence of an atom at a optical lattice site with fidelity over $ 99 \% $.  

\begin{exerciselist}[Exercise]
\item \label{q9-4}
Derive  the effective Hamiltonian (\ref{standardqhe}).   
\end{exerciselist}

\section{Example: The Bose-Hubbard model}
\label{sec:bosehubbard}\index{Bose-Hubbard model}

One of the best known examples of analogue quantum simulation\index{quantum simulation!analogue} was the realization of the Bose-Hubbard model using cold atoms realized by Greiner, Bloch and co-workers (\cite{greiner2002m}).  Since this experiment there has been numerous examples of quantum simulation using a variety of different physical systems.  Several reviews are available which summarize these realizations of quantum simulation, we refer the reader to these for more details.  The realization Bose-Hubbard model remains one of the most beautiful experiments of the field and we explain in more detail the realization and results.

\subsection{The experiment}

The experiment involved first producing a BEC of $ ^{87} \text{Rb} $ atoms in the state $ F = 2, m_F = 2 $ with $ 2 \times 10^5 $ atoms. An optical lattice\index{optical lattice} in three dimensions was then applied to the BEC, using three standing waves of light in orthogonal directions in a similar way as shown in Fig. \ref{fig9-2}.  This produces a periodic potential of the form 
\begin{align}
V(\bm{x}) = \frac{V_0}{2} \left[ \cos ( 2 \kappa x) + \cos ( 2 \kappa y) + \cos ( 2 \kappa z) \right] ,
\label{bosehubbardpotential}
\end{align}
where we have generalized (\ref{opticallatticepotential}) to three dimensions. When the optical lattice is applied to the BEC, the desired potential $ V_0 $ is gradually increased such that the system remains in the quantum many-body ground state of the whole system. Once the lattice is applied, the atoms occupy approximately 65 lattice sites in a single direction, with approximately 2.5 atoms per site at the center.  After the optical lattice is prepared, the momentum distribution of the atoms was measured using a time-of-flight measurement.\index{time-of-flight measurement}

\subsection{Effective Hamiltonian}
\label{sec:effectivehamiltonian}

In this section we derive the effective Hamiltonian produced by the optical lattice\index{optical lattice}. The derivation of an effective Hamiltonian is an essential step in many quantum simulation experiments since several approximations are required to obtain the simple models that are often studied from a theoretical perspective.  The same techniques can be used in this section can be applied to derive other related lattice models. 

We start by substituting the potential  (\ref{opticallatticepotential}) into (\ref{multiham}), such that we have 
\begin{align}
{\cal H}_0 = \int d \bm{x} a^\dagger (\bm{x}) \left( -\frac{\hbar^2}{2m} \nabla^2 + \frac{V_0}{2} \left[ \cos ( 2 \kappa x) + \cos ( 2 \kappa y) + \cos ( 2 \kappa z) \right]  \right)  a (\bm{x})  ,
\label{opticallatticeham}
\end{align}
where the potential is (\ref{bosehubbardpotential}). Due to the rectangular symmetry, the equations for the $ x, y, z $ directions are separable. First examining the $ x $-direction, the periodic potential implies that the eigenstates of the single particle Hamiltonian obey Bloch's theorem and takes a form \index{Bloch's theorem} 
\begin{align}
\psi_k^{(m)} (x) = u_k^{(m)} (x) e^{ikx} .
\label{blochwavefunction}
\end{align}
where $ u_k(x) = u_k(x + \lambda/2) $  is a periodic function with periodicity of the optical lattice, and the $ m $ labels the $m$th band of the band structure that arises from the periodic potential. We take the lowest band to be the ground state and labeled as $ m = 0 $. In the $ y $- and $ z $- directions the solutions also obey Bloch's theorem and we can write the full eigenstate as 
\begin{align}
\Psi_{\bm{k}}^{(\bm{m})} (\bm{x}) = \psi_{k_x}^{(m_x)} (x) \psi_{k_y}^{(m_y)} (y) \psi_{k_z}^{(m_z)} (z)  .
\label{totalblochstate}
\end{align}
where $ \bm{m} = (m_x,m_y,m_z) $ is the band index. 

The Bloch functions\index{Bloch function} can be in turn transformed to give Wannier functions, defined as \index{Wannier functions}
\begin{align}
w_{\bm{m}} (\bm{x}-\bm{x}_j) = \int_{B_{\bm{m}}} d\bm{k} e^{-i \bm{k} \bm{x}_j} \Psi_{\bm{k}}^{(\bm{m})} (\bm{x}) ,
\end{align}
where the integration is carried out over the momenta for the $ \bm{m} $th band and $ \bm{x}_j $ denotes the location of the $j$th 
lattice site.  For example, for the lowest energy band, the range of $ \bm{k} $ is $ k_{x,y,z} \in [-\kappa, \kappa] $.  The Wannier functions can be considered to be a localized version of Bloch functions, and obey orthogonality properties
\begin{align}
\int d\bm{x} w_{\bm{m}}^* (\bm{x}-\bm{x}_j) w_{\bm{m}'} (\bm{x}-\bm{x}_{j'}) = \delta_{j j'} \delta_{\bm{m} \bm{m}'}.
\label{orthogonalitywannier}
\end{align}

The full set of Wannier functions form a complete set which serve to expand the bosonic operators $ a(x) $ in a similar way to (\ref{basisaninv}). Expanding the operators according to 
\begin{align}
a(\bm{x}) = \sum_{j} \sum_{\bm{m}} w_{{\bm{m}}} (\bm{x}-\bm{x}_j) a_{j\bm{m}}.  
\label{wanniertransform}
\end{align}
Substituting this into (\ref{opticallatticeham}), we obtain
\begin{align}
{\cal H}_0 = \sum_{j j'} \sum_{\bm{m} \bm{m}'} T_{j, j'}^{\bm{m}, \bm{m}'} a_{j\bm{m}}^\dagger a_{j'\bm{m}'}   ,
\label{fullsingleparticleham}
\end{align}
where the hopping matrix elements are defined
\begin{align}
T_{j, j'}^{\bm{m}, \bm{m}'} & =  \int d \bm{x}  w_{{\bm{m}}}^* (\bm{x}-\bm{x}_j)   [-\frac{\hbar^2}{2m} \nabla^2 + V(\bm{x}) ]  
w_{{\bm{m}}'} (\bm{x}-\bm{x}_{j'}) \label{energyformhoppingorig}  \\
& = \frac{\delta_{\bm{m} \bm{m}'} }{2 \kappa}  \int_{B_{\bm{m}}} d \bm{k} E_{\bm{k}} e^{i \bm{k} (\bm{x}_j-\bm{x}_{j'})}.
\label{energyformhopping}
\end{align}
Here we substituted the definition of the Wannier functions in the first line and used the fact that the Bloch functions are eigenstates of the Hamiltonian $ H_0  \Psi_{\bm{k}}^{(\bm{m})} (\bm{x} ) = E_{\bm{k}}  \Psi_{\bm{k}}^{(\bm{m})} (\bm{x} )  $.  

The expression (\ref{fullsingleparticleham}) constitutes a change of basis from the position basis of the bosons to the Wannier basis, and no approximations have been made so far.  Typically one makes further assumptions to simplify the Hamiltonian.  For example, for sufficiently low temperatures, the bosons will not occupy the higher energy bands and one can consider the lowest band alone.  Furthermore, since Wannier integrals (\ref{energyformhoppingorig}) involve the overlap of two localized functions, we expect the magnitude of the integrals to decrease with $ |\bm{x}_j- \bm{x}_j'| $.  Taking only nearest neighbor (NN) terms, the Hamiltonian (\ref{fullsingleparticleham}) is
\begin{align}
{\cal H}_0 \approx - t \sum_{j, j' \in \text{NN}} \left(  a_{j}^\dagger a_{j'} + a_{j'}^\dagger a_{j} \right)  + \epsilon \sum_{j} a_{j}^\dagger a_{j}  ,
\label{approxsinglepar}
\end{align}
where 
\begin{align}
t & \equiv -T_{j,j'}^{\bm{0},\bm{0}}  \nonumber \\
& = -\int d \bm{x}  w_{{\bm{m}}}^* (\bm{x}-(\bm{x}_j-\bm{x}_{j'}))   [-\frac{\hbar^2}{2m} \nabla^2 + V(\bm{x}) ]  
w_{{\bm{m}}'} (\bm{x})
\label{hoppingt}
\end{align}
is the nearest neighbor hopping and $ \epsilon \equiv T_{j,j}^{\bm{0},\bm{0}} $. In (\ref{hoppingt}) we use the fact that there is a translational symmetry such that $ t $ are the same for all lattice sites, which is also true of $ \epsilon $.  Such a Hamiltonian takes the form of a typical lattice model that might be considered in many strongly correlated condensed matter problems.  Although we have used bosons to derive the Hamiltonian (\ref{fullsingleparticleham}), we note that the same steps can be used to derive a fermionic Hamiltonian, starting from fermion operators (\ref{fermioncomm1}).

In the above, we have only considered the kinetic energy component of the Hamiltonian.  We must also consider the interaction Hamiltonian, as defined by (\ref{interactionham}). In (\ref{interactionmanybodyham}) a transformation was made into the basis of the single particle Hamiltonian that trap the atoms.  Due to the presence of the optical lattice, the relevant single particle basis is now the Bloch wavefunction, or equivalently the Wannier basis.  As such, we substitute (\ref{wanniertransform}) into (\ref{interactionham}).  In a similar way to (\ref{interactionmanybodyham}), we obtain
\begin{align}
 {\cal H}_I & =  \frac{1}{2} \sum_{j_1 j_2 j_3 j_4}  \sum_{\bm{m}_1 \bm{m}_2 \bm{m}_3 \bm{m}_4} g^{\bm{m}_1, \bm{m}_2, \bm{m}_3, \bm{m}_4}_{j_1, j_2, j_3, j_4}  a^\dagger_{j_1 \bm{m}_1} a^\dagger_{j_2 \bm{m}_2} a_{j_3 \bm{m}_3} a_{j_4 \bm{m}_4}
\label{interactionmanybodyham2}
\end{align}
where
\begin{align}
g^{\bm{m}_1, \bm{m}_2, \bm{m}_3, \bm{m}_4}_{j_1, j_2, j_3, j_4} = & \int d \bm{x} d \bm{x}' w_{\bm{m}_1}^* ( \bm{x}- \bm{x}_{j_1}) 
w_{\bm{m}_2}^* ( \bm{x}'- \bm{x}_{j_2}) \nonumber \\
& \times U(\bm{x},\bm{x}') w_{\bm{m}_3} ( \bm{x}- \bm{x}_{j_3})  w_{\bm{m}_4} ( \bm{x}'- \bm{x}_{j_4})  .
\label{interactionmatrixwannier}
\end{align}
For the case of the contact potential (\ref{swaveint}), we obtain
\begin{align}
g^{\bm{m}_1, \bm{m}_2, \bm{m}_3, \bm{m}_4}_{j_1, j_2, j_3, j_4}  = \frac{4 \pi \hbar^2 a_s}{m} \int d \bm{x} 
 w_{\bm{m}_1}^* ( \bm{x}- \bm{x}_{j_1}) 
w_{\bm{m}_2}^* ( \bm{x}- \bm{x}_{j_2}) w_{\bm{m}_3} ( \bm{x}- \bm{x}_{j_3})  w_{\bm{m}_4} ( \bm{x}- \bm{x}_{j_4})  .
\label{interactionmatrix2wannier}
\end{align}
Again, due to similar assumptions as above, often one considers only the lowest band 
$ \bm{m}_1 = \bm{m}_2 = \bm{m}_3 = \bm{m}_4 = \bm{0} $.  Due to the localized nature of the Wannier functions, the largest matrix element will correspond to the case when $ j_1 =j_2 = j_3 = j_4 $.  This largest term is called the on-site interaction, \index{on-site interaction} and can be defined
\begin{align}
U = g^{\bm{0},\bm{0},\bm{0},\bm{0}}_{j,j,j,j} = \frac{4 \pi \hbar^2 a_s}{m} \int d \bm{x} | w_{\bm{0}} ( \bm{x}) |^4 . \label{onsiteinter}
\end{align}
Since $ | w_{\bm{0}} ( \bm{x}) |^2 $ is a normalized probability function for the lowest energy band, it has dimensions of inverse length.  This means that the smaller the spatial extent of the Wannier function, the large the integral will be. A Wannier function with a small spatial extent can be achieved by increasing the laser intensity, thereby producing a larger potential $ V_0 $, which has a tighter potential for each site. 

To lowest order the interaction part of the Hamiltonian can be written
\begin{align}
 {\cal H}_I \approx \frac{U}{2}  \sum_{j} a_{j}^\dagger a_{j}^\dagger a_{j} a_{j} .
\end{align}
Combined with (\ref{approxsinglepar}) this Hamiltonian corresponds to a model of bosons that are hopping on a lattice.  The total Hamiltonian 
\begin{align}
 {\cal H} &  =  {\cal H}_0 +  {\cal H}_I \nonumber \\
& = - t \sum_{j,j' \in \text{NN}} \left(   a_{j}^\dagger a_{j'} + a_{j'}^\dagger a_{j}  \right)  + \epsilon \sum_{j} a_{j}^\dagger a_{j}  + \frac{U}{2} \sum_{j} a_{j}^\dagger a_{j}^\dagger a_{j} a_{j}
\label{bosehubbardhamil}
\end{align}
is called the Bose-Hubbard model. \index{Bose-Hubbard model}

\begin{figure}[t]
\includegraphics[width=\textwidth]{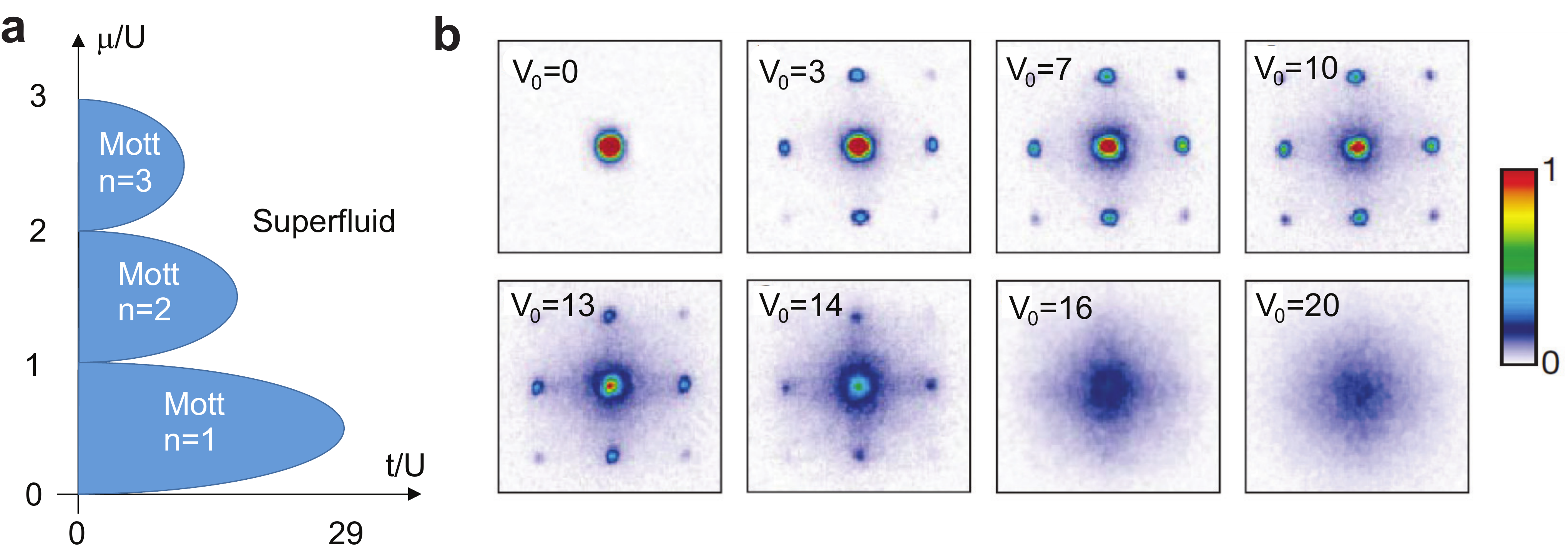}
\caption{
Quantum simulation of the Bose-Hubbard model.\index{Bose-Hubbard model} (a) The theoretical zero-temperature phase diagram of the three-dimensional Bose-Hubbard model on a cubic lattice with $ \epsilon = 0 $.  (b) Experimental results showing the time-of-flight measurements for the optical lattice potential amplitudes as indicated (reproduced from  \cite{greiner2002m}).  The potentials are measured in units of the recoil energy  $ \frac{\hbar^2 \kappa^2}{2m} $, where $ \kappa = 2 \pi/\lambda $ is the wavenumber of the optical lattice and $ m $ is the mass of an $^{87} \text{Rb} $ atom. }
\label{fig9-4}
\end{figure}

\subsection{Experimental observation of the phase transition}

The Bose-Hubbard model\index{Bose-Hubbard model} has been well-studied in condensed matter physics.  The model has primarily two phases, a Mott insulating phase\index{Mott insulating phase} and a superfluid phase\index{superfluid phase}.  The superfluid phase is characterized by long-range phase coherence, as is present in a BEC. The prototypical superfluid\index{superfluid} state occurs in the limit of $ U = 0 $ and takes the form
\begin{align}
| \Phi(U=0) \rangle = \frac{1}{\sqrt{N!}} \left( \frac{1}{\sqrt{M}} \sum_{j=1}^M a_{j}^\dagger \right)^N | 0 \rangle ,
\label{uzerolimit}
\end{align}
where $N $ is the total number bosons in the system. 
In this state, each boson is in a superposition across all lattice sites, and all $ N $ bosons occupy this same state. In the reverse limit where the interactions are large and the kinetic energy is small, the interactions cause the bosons to localize and there are a fixed number of atoms per lattice site.  The prototypical state occurs when $ t = 0 $, and the Mott insulator\index{Mott insulator} state take the form
\begin{align}
| \Phi(t=0) \rangle =  \prod_{j} \frac{\left( a_{j}^\dagger \right)^n}{\sqrt{n!}} | 0 \rangle .
\label{tzerolimit}
\end{align}
where we have assumed there are $ n $ bosons per site, also called the filling factor.   \index{filling factor}

The schematic phase diagram of the model is shown in Fig. \ref{fig9-4}(a).  The phase diagram consists of several separated regions (usually called ``lobes'') corresponding to different filling factors $ n $.  For $ t = 0 $ the ground state is always in a Mott insulating phase\index{Mott insulating phase}. The chemical potential $ \mu $ controls the number of particles in the system.  As the chemical potential\index{chemical potential} increases, the lattice fills up in a stepwise fashion, where the number of bosons per site increases by one per lobe. To understand this, set $ t = 0 $ in (\ref{bosehubbardhamil}), and add a chemical potential term
\begin{align}
 {\cal H}(t=0) - \mu N = \epsilon  \sum_j n_j  + \frac{U}{2}  \sum_j  n_j (n_j-1) - \mu \sum_j n_j  ,
\end{align}
where we have written pairs of bosonic operators as number operators $ n_j = a_j^\dagger a_j $. Assuming $ M $ lattice sites, the energy can be written
\begin{align}
E_0(t=0, n) = M \left( n (\epsilon-\mu)  + \frac{U}{2} n (n-1) \right) .
\end{align}
Since the chemical potential is the energy that is available per particle on a lattice site, let us calculate what additional energy is required to add an extra boson from $ n $ to $ n+ 1 $ per lattice site.  We can evaluate
\begin{align}
\frac{E_0(t=0, n+1) - E_0(t=0, n)}{M}  & =  \epsilon -\mu  + U n  .
\label{energydiff}
\end{align}
The filling corresponds to when (\ref{energydiff}) gives zero, hence we obtain the filling factor to particle number relation\index{filling factor}
\begin{align}
\mu = \epsilon + U n ,
\end{align}
which increases in units of $ U $ as shown in the phase diagram Fig. \ref{fig9-4}(a). 

For a particular filling factor $ n $, starting from a Mott insulator phase\index{Mott insulator phase} with $ t = 0 $, as $ U $ is decreased and/or $ t $ is increased, at some point there is a phase transition to a superfluid state\index{superfluid}.  The precise value of $ t/U $ depends upon the chemical potential\index{chemical potential} as shown in Fig. \ref{fig9-4}(a).  For the $ n = 1 $ Mott lobe, the phase transition occurs at 
$ U/t \approx 29 $ for a three dimensional cubic lattice.  For larger filling factors the phase transitions occur at smaller values of $ U/t $.  

In the experiment, the ratio of $ t/U $ can be changed by changing the lattice potential $ V_0 $. As the potential $ V_0 $ is increased, this has the effect of making the Wannier functions $ w_{{\bm{m}}} (\bm{x})  $  more localized.   Examining the formulas for $ t $ and $ U $, (\ref{hoppingt}) and (\ref{onsiteinter}) respectively, we can deduce that as $ V_0 $ is increased, $ t $ will decrease, while $ U $ will increase.   The hopping energy $ t $ decreases because the overlap between the Wannier functions $ w_{{\bm{m}}} (\bm{x}-(\bm{x}_j-\bm{x}_{j'}))  $ and $ w_{{\bm{m}}} (\bm{x})  $ decrease due to the increased localization.  Meanwhile, (\ref{onsiteinter}) increases because the square of a normalized probability distribution will increase with localization. For example, the square of a normalized Gaussian distribution  is
\begin{align}
\int p_{\text{Gauss}}^2 (x) & = \frac{1}{2\sigma \sqrt{\pi}},
\end{align}
where
\begin{align}
p_{\text{Gauss}}(x) =  \frac{e^{-x^2/2\sigma^2}}{\sqrt{2\pi}\sigma } ,
\label{gaussiandist}
\end{align}
thus is inversely proportional to the standard deviation. 
In this way, by increasing $ V_0 $, it has the effect of traversing the phase diagram of Fig. \ref{fig9-4}(a) from left to right.  

Returning to the experiment, the time-of-flight\index{time-of-flight measurement} images with various amplitudes of the potentials are shown in Fig. \ref{fig9-4}(b).
For $ V_0 = 0 $,  no optical lattice\index{optical lattice} is applied, and the momentum distribution corresponds to a single peak at $ \bm{k} = 0 $.  As the periodic potential amplitude is applied, the state corresponds to the lattice superfluid \index{superfluid}state (\ref{tzerolimit}), with a coherent superposition across lattice sites.  This produces an interference pattern much like that observed from a diffraction grating, where the sources are the lattice sites and the interference occurs due to matter wave interference. Another way to view this is that the momentum representation of the Bloch states\index{Bloch state} (\ref{totalblochstate}) contain components of integer multiples of the frequency $ \kappa $ of the lattice.  As the potential is further increased, the characteristic interference pattern is lost, due to the Mott insulator transition\index{Mott insulator}.  At this point there is no longer any diffraction grating-like interference, since there is no phase coherence between sites in a state like (\ref{tzerolimit}).  The pattern changes to the momentum distribution of atoms which are localized in position space with the dimensions of the optical lattice spacing.  Taking the spatial distribution to be a Gaussian (\ref{gaussiandist}) with $ \sigma = \lambda = 2 \pi/\kappa  $, the momentum space distribution should be another Gaussian with approximate standard deviation $ \propto \pi^2/\sigma = \kappa/2 \pi $, which is the observed result.

\begin{exerciselist}[Exercise]
\item \label{q9-2}
Verify  the orthogonality relations of the  Wannier functions (\ref{orthogonalitywannier}).  
\item \label{q9-3}
Verify  the expression for the hopping energies (\ref{energyformhopping}). 
\item \label{q9-5}
Verify that states (\ref{uzerolimit}) and (\ref{tzerolimit}) are the ground states of the Bose-Hubbard Hamiltonian\index{Bose-Hubbard Hamiltonian} (\ref{bosehubbardhamil}) in their respective limits.   
\end{exerciselist}

\section{References and further reading}
\label{sec:ch9refs}

\begin{itemize}
\item Sec. \ref{sec:introsimulation}: Review articles and commentaries on quantum simulation \cite{georgescu2014quantum,gross2017quantum,buluta2009quantum,bloch2012quantum,bloch2008quantum,schaetz2013focus,cirac2012goals,bloch2008many,paivi2014quantum,lewenstein2007ultracold,carr2009cold}.  Feynman's conjecture first introducing the idea of simulating quantum systems with quantum computers \cite{feynman1982simulating}.  Review article on quantum Monte Carlo   \cite{foulkes2001quantum},  Density Matrix Renormalization Group (DMRG)  \cite{white1992density,schollwock2011density}, density functional theory \cite{jones2015density}, dynamical mean field theory \cite{georges1996dynamical}, series expansion methods \cite{oitmaa2006series}, lattice QCD \cite{ratti2018lattice}, and quantum chemistry \cite{levine1991quantum,mcardle2018quantum}. 
\item Sec. \ref{sec:analogvsdigital}: Original works showing the efficiency of quantum simulation  \cite{lloyd1996universal,feynman1982simulating}. Original work exactly solving the 1D transverse Ising model  \cite{pfeuty1970one}.  Finite size analysis of the transverse Ising model \cite{hamer1981finite}. A textbook introduction to spin systems \cite{caspers1989spin}. 
\item Sec. \ref{sec:digitalqs}: Original works introducing digital quantum simulation \cite{lloyd1996universal,wiesner1996simulations}. Review articles discussing digital quantum simulation \cite{georgescu2014quantum,cirac2012goals}.  Experimental demonstrations of digital quantum simulation \cite{lanyon2011universal,barends2015digital,las2014digital,schindler2013quantum}.  Theoretical work showing method of performing digital quantum simulation of lattice gauge theory \cite{byrnes2006simulating}. Original works introducing the Suzuki-Trotter expansion \cite{trotter1959product,suzuki1976generalized,suzuki1976relationship}. 
\item Sec. \ref{sec:ch9optlat}: Original works experimentally implementing optical lattices \cite{verkerk1992dynamics,jessen1992observation,hemmerich1993two,grynberg1993quantized,hemmerich1993sub,kastberg1995adiabatic}.  Review articles and commentaries discussing optical lattices \cite{grimm2000optical,bloch2005ultracold,bloch2008quantum,bloch2008many,bloch2012quantum,paivi2014quantum}.  Optical lattices have found a variety of different uses in addition to quantum simulation, including atom sorting and transport \cite{miroshnychenko2006quantum,mandel2003coherent}, investigating physics of Anderson localization \cite{billy2008direct,schreiber2015observation} and Bose-Fermi mixtures \cite{lewenstein2004atomic}, optical lattice clocks \cite{takamoto2005optical,lemke2009spin,ushijima2015cryogenic,bloom2014optical}, lattice solitons \cite{efremidis2003two,yang2003fundamental}, molecular formation \cite{stoferle2006molecules}.   Producing 3D atomic structures with tweezers \cite{barredo2018synthetic}. 
\item Sec. \ref{sec:feshbachbcs}: For references relating to Feshbach resonances please see Sec. \ref{sec:feshbach}. Experimental realization of the BEC-BCS crossover with cold fermionic atoms  \cite{regal2004observation,bourdel2004experimental,greiner2005probing,regal2003creation,zirbel2008collisional}.  Theoretical proposal and analysis of the atomic BEC-BCS crossover \cite{ohashi2002bcs,gong2011bcs}.  
Review articles on the BEC-BCS crossover \cite{chen2005bcs,zwerger2011bcs,strinati2018bcs}.  
Experimental observation of condensation of molecules using a Feshbach resonance \cite{zwierlein2003observation}. 
\item Sec. \ref{sec:artgaugefield}: Review articles discussing artificial gauge fields induced by rotation \cite{fetter2009rotating,cooper2008rapidly,cohen1977quantum}. Review article discussing methods of producing artificial gauge fields for neutral atoms \cite{dalibard2011colloquium,goldman2014light}.
Experimental works demonstrating synthetic fields for cold atoms \cite{lin2009synthetic,lin2011synthetic}. Original works introducing Berry's phase \cite{pancharatnam1956generalized,longuet1958studies,berry1984quantal}. Further works discussing 
Berry phases in adiabatic evolution \cite{berry1989,dum1996gauge,mead1979determination,jackiw1987berry}. 
\item Sec. \ref{sec:spinorbitcoupling}: Experiments realizing spin-orbit coupling in a BEC \cite{lin2011spin,anderson2012synthetic,wu2016realization,li2017stripe}. 
Theoretical works analyzing spin-orbit coupling in BECs \cite{osterloh2005cold,ruseckas2005non,dudarev2004spin,stanescu2008spin,hu2011probing,ho2011bose,wang2010spin}. 
Review articles on spin-orbit coupling \cite{hasan2010colloquium,galitski2013spin,zhai2012spin,zhang2018spin}. 
\item Sec. \ref{sec:timeofflight}: Experiment demonstrating the time-of-flight method   \cite{lett1988observation}. Theoretical calculations modelling the time-of-flight method \cite{yavin2002calculation,brzozowski2002time}. Review articles discussing time-of-flight measurements \cite{phillips1998nobel,bloch2005ultracold}.  
\item Sec. \ref{sec:quantumgasmicro}: Experiments demonstrating the quantum gas microscope \cite{bakr2009quantum,sherson2010single,cheuk2015quantum,haller2015single,yamamoto2016ytterbium}.  Review articles and books on the quantum gas microscope \cite{kuhr2016quantum,paivi2014quantum}. Experimental realization of scanning electron microscopy of a cold atomic gas \cite{gericke2008high}.
\item Sec. \ref{sec:bosehubbard}: Experimental realizations of the Bose-Hubbard model with cold atoms \cite{greiner2002m,campbell2006imaging,white2009strongly}.  Theoretical proposals of Bose-Hubbard model quantum simulator with cold atoms 
\cite{jaksch1998cold,mekhov2007probing}. 
Theoretical investigations of the physics of Bose-Hubbard model \cite{capogrosso2007phase,melko2005supersolid,scarola2005quantum,gadway2010superfluidity,alon2005zoo}.
Other examples of theoretical proposals for quantum simulation \cite{romans2004quantum,damski2003atomic,damski2003creation,byrnes2010mott}.
Other examples of experimental realizations of quantum simulation \cite{o2002observation,paredes2004tonks,stoferle2004transition,yan2013observation,schneider2008metallic,chabe2008experimental,li2010measurement,langen2015experimental,de2013nonequilibrium,jendrzejewski2012three,hensinger2001dynamical,leblanc2013direct,preiss2015strongly,labuhn2016tunable,hart2015observation,deng2016observation,tang2018thermalization,bernien2017probing,weimer2010rydberg}.   
Experimental realization of slow light \cite{hau1999light,liu2001observation,labeyrie2003slow,boyer2007ultraslow}.
Review articles relating to cold atom quantum simulators \cite{dulieu2009formation,tarruell2018quantum}. 

\end{itemize}

  \chapter[Entanglement between atom ensembles]{Entanglement between atom ensembles}

\label{ch:entanglement}

\section{Introduction}

Up to this point, we have primarily dealt with entanglement only within a single atomic ensemble or Bose-Einstein condensate.  For example, for spin squeezed Bose-Einstein condensates we introduced a criterion for detecting entanglement between the particles in Sec. \ref{sec:entanglementsqueezing}.  However, one of the distinctive features of entanglement is the non-local aspect which Einstein famously called a ``spooky action at a distance''. Furthermore, in fact that notion of entanglement introduced in Sec. \ref{sec:entanglementsqueezing} is a rather ambiguous one, since the bosons within a Bose-Einstein condensate are one composite system.  A more conventional notion of entanglement is between the states of two spatially distinct particles.  In this chapter we will explore entanglement defined between two atomic ensembles.  Due to the large number of degrees of freedom, we will see that the types of entanglement that can be created are much more complex than for qubits.  We will also describe various ways that such entanglement can be quantified and detected.

\section{Inseparability and quantifying entanglement}
\label{sec:inseparability}

First let us define precisely what it means for two systems to be entangled. Once it is defined, we will want to find measures of entanglement such that we can put a numerical value on the amount of entanglement that are possessed by a quantum state.   We will treat pure states and mixed states on a different footing since each have a different approach for quantifying entanglement.

\subsection{Pure states}

\label{sec:purestateent}\index{pure state}

Consider that the wavefunction consists of two subsystems, labeled by $ A $ and $ B $ with basis states $ | n \rangle_A $ and $ | m \rangle_B $ in each. The Hilbert space dimension of each of the subsystems separately is $ D_{A,B} $, such that the total dimension is $ D = D_A D_B $. 
The most general pure state that can be written in this case is
\begin{align}
|\Psi \rangle = \sum_{nm} \Psi_{nm}  | n \rangle_A   | m \rangle_B .
\label{generalabstate}
\end{align}
A subclass of the general pure state is the {\it product state} \index{product state}, which takes the form
\begin{align}
| \psi\rangle_A \otimes | \phi\rangle_A  = \sum_{nm} \psi_n   \phi_m | n \rangle_A  | m \rangle_B 
\label{productstate}
\end{align}
where we have expanded the product state wavefunction in terms of the basis states with coefficients $  \psi_n, \phi_m $.  
Clearly it is possible to choose $  \Psi_{nm} = \psi_n   \phi_m  $, but there are states of the form (\ref{generalabstate}) that cannot be written in the form (\ref{productstate}).  Examples of such states are the Bell states\index{Bell state}
\begin{align}
\frac{1}{\sqrt{2}} \left( |0 \rangle_A |0 \rangle_B  \pm |1 \rangle_A |1 \rangle_B \right) \nonumber \\
\frac{1}{\sqrt{2}} \left( |0 \rangle_A |1 \rangle_B  \pm |1 \rangle_A |0 \rangle_B \right) .
\end{align}
The product states are a special class of states which by definition have no entanglement.  Thus a definition of an entangled pure state is any state that cannot be written in the form of\index{pure state} (\ref{productstate}).

For a pure bipartite system\index{bipartite system} as we consider here, a convenient and complete way of quantifying the amount of entanglement is using the {\it von Neumann entropy}. \index{von Neumann entropy} This is defined by
\begin{align}
E = - \text{Tr} ( \rho_A \log \rho_A ) ,
\label{entropydef}
\end{align}
where 
\begin{align}
\rho_A & = \text{Tr}_B (  |\Psi \rangle \langle \Psi | ) 
\label{reduceddensity}
\end{align}
is the reduced density matrix over subsystem $ B $.  The operation $ \text{Tr}_B $ denotes a partial trace, where the trace is only taken over the variable in subsystem $ B $.  For our state (\ref{generalabstate}), the reduced density matrix is
\begin{align}
\rho_A & = \sum_{n n' m} \Psi_{nm} \Psi_{n'm}^* |n \rangle_A \langle n'|_A .
\end{align}

In (\ref{entropydef}), the logarithm denotes the matrix logarithm, and the base of the logarithm can be chosen by convenience.  Common choices are base 2, which are used for qubit systems.  To evaluate (\ref{entropydef}) it is most convenient to work in the diagonalized basis of the reduced density matrix $ \rho_A $.  Suppose (\ref{reduceddensity}) is diagonalized according to
\begin{align}
\rho_A = \sum_n \lambda_n | \lambda_n \rangle \langle  \lambda_n | .
\end{align}
Then the von Neumann entropy can be written\index{von Neumann entropy}
\begin{align}
E = - \sum_n \lambda_n \log \lambda_n .
\end{align}
This is a non-negative quantity which takes a maximum value
\begin{align}
E_{\text{max}} = \log D_{\min} ,
\label{smaxentropy}
\end{align}
with $ D_{\min} = \min(D_A, D_B) $. Then the normalized von Neumann entropy is
\begin{align}
\frac{E}{E_{\text{max}}} = - \text{Tr} ( \rho_A \log_{D_{\min}} \rho_A ) .
\end{align}
Thus taking the base of the logarithm to be the minimum dimension of $ A $ and $ B $, we obtain a quantity that ranges from 0 to 1.

\subsection{Mixed states}

\label{sec:mixedstateent}

For pure states\index{pure state}, defining and quantifying entanglement is relatively straightforward using product states and the von Neumann entropy\index{von Neumann entropy}.  More generally, a quantum state is described by a mixed state\index{mixed state}.  A mixed state consists of a statistical mixture of different quantum states, and can arise in practice due to non-zero temperature, or the presence of decoherence or noise.   Thus any quantum state in a realistic experimental system is usually described by a mixed state since it is not possible to perfectly prepare a pure state.  

The most general bipartite mixed state\index{bipartite state} is written as 
\begin{align}
\rho = \sum_i P_i | \Psi_i \rangle \langle \Psi_i |
\label{generalmixed}
\end{align}
where $ | \Psi_i \rangle  $ are a set of orthogonal bipartite states on $ A $ and $ B $ as before, and $ P_i $ are the probabilities of obtaining them. The state (\ref{generalmixed}) could describe an ensemble of bipartite quantum states in several ways.  The state could for example describe an ensemble\index{ensemble} of bipartite systems\index{bipartite system} where there are numerous copies of the quantum system. The states $ | \Psi_i \rangle  $ occur with a relative population $ P_i $.   An example of this would be an NMR experiment where there are many molecules which all constitute the bipartite system.  Another way (\ref{generalmixed}) could be realized is that there is only one quantum system, but the experiment is run numerous times. Each run of the experiment gives a different quantum state probabilistically, and the $ P_i $ describes the statistics of obtaining each state.  

A separable state \index{separable state} is a mixed state consisting of the product states (\ref{productstate}).  The general form of a separable state is
\begin{align}
\rho_{\text{sep}} = \sum_i p_i \rho_i^A \otimes \rho_i^B .
\label{separable}
\end{align}
We may now give a more general definition of entanglement. Any state that cannot be written in the form (\ref{separable}) is then defined to be an entangled state.  This is a more general definition than that of the previous section because pure states are a specific example of the mixed states shown here --- (\ref{generalabstate}) is an instance of (\ref{generalmixed}), and (\ref{productstate}) is an instance of (\ref{separable}).  

Unlike the pure state case where there is a relatively simple formula (\ref{entropydef}) which quantifies the amount of entanglement, for mixed states\index{mixed state} it is more difficult to even give a test for finding whether a state is entangled or not.  The most general way to do this is to search the full space of separable states and check whether it can be written in the form (\ref{separable}).  This approach is however quite numerically intensive due to the large number of parameters which specify the general separable state\index{separable state}.  A more convenient method would be to have a formula, much like the von Neumann entropy\index{von Neumann entropy}, that depends only on the density matrix and one may find the amount of entanglement. 

One of the most popular methods to achieve this is the Peres-Horodecki criterion. \index{Peres-Horodecki criterion} To define this criterion, first write a general bipartite density matrix as
\begin{align}
\rho = \sum_{n'm'nm} \rho_{n'm'nm}|n '\rangle_A | m' \rangle_B \langle n |_A \rangle m |_B
\end{align}
where we have simply expanded all the matrix elements in terms of the basis states $ |n \rangle_A $ and $ |m \rangle_B$.  One of the properties of a density matrix is that its eigenvalues are always positive, since it can always be written in the form (\ref{generalmixed}).  This property can be stated as the positive semi-definiteness of the density matrix:
\begin{align}
\rho \ge 0  .
\end{align}
Now define the partial transpose of the density matrix as
\begin{align}
\rho^{T_B} & = \sum_{n'm'nm} \rho_{n'm'nm}|n '\rangle_A | m \rangle_B \langle n |_A \langle m' |_B \nonumber \\
& = \sum_{n'm'nm} \rho_{n'mnm'}|n '\rangle_A | m' \rangle_B \langle n |_A \langle m |_B \nonumber.
\end{align}
This involves taking the transpose only on subsystem $ B $.  The Peres-Horodecki criterion\index{Peres-Horodecki criterion} then states that if $ \rho $ is separable then its partial transpose is also a density matrix and is positive semi-definite
\begin{align}
\rho^{T_B} \ge 0    \hspace{1cm} \text{(if $ \rho $ is separable)}
\end{align}
Hence if $ \rho^{T_B} $ has any negative eigenvalues, $ \rho $ must be entangled. Due to this test involving the positivity of the partial transpose (PPT), it is also often called the {\it PPT criterion}. \index{PPT criterion}  We note that although we took the partial transpose with respect to $ B $ above, we could equally have done so with $ A $ instead, the same results would be obtained. 

The above procedure can be extended to quantify the amount of entanglement in a straightforward way.  Since the signature of entanglement is the presence of negative eigenvalues of $ \rho^{T_B} $, one would expect that the more entangled the state is, the more negative the eigenvalues are.  A suitable measure of entanglement is the sum of only the negative eigenvalues, which is called the {\it negativity}, defined as \index{negativity}
\begin{align}
{\cal N} =\frac{1}{2}\sum_i\left( |\lambda_i | - \lambda_i \right) .
\end{align}
Here, $ \lambda_i $ are the eigenvalues of  $ \rho^{T_B} $.  In matrix invariant form this can be equivalently be written as 
\begin{align}
{\cal N} & = \frac{||\rho^{T_B} ||_1 - 1}{2}
\end{align}
where $ || X ||_1 =  \text{Tr} \sqrt{X^\dagger X } $ is the trace norm of a matrix. We used the fact that the sum of the eigenvalues of the partial transposed density matrix still adds to 1
\begin{align}
\text{Tr} ( \rho^{T_B} ) = 1.
\end{align}
The negativity\index{negativity} is a non-negative number between 
\begin{align}
0 \le {\cal N} \le \frac{D_{\min} -1}{2} ,
\end{align}
where $ D_{\min} = \min(D_A, D_B) $. 

It is sometimes more natural to use a slightly different form of the negativity with the same basic idea so that it has a form more similar to the von Neumann entropy (\ref{entropydef})\index{von Neumann entropy}.  In this case we define the {\it logarithmic negativity} \index{logarithmic negativity} as
\begin{align}
E_N = \log ||\rho^{T_B} ||_1
\end{align}
The logarithmic negativity\index{logarithmic negativity} is also a non-negative number in the range
\begin{align}
0 \le E_N \le \log D_{\min} ,
\end{align}
which has the same range as the von Neumann entropy \index{von Neumann entropy} (\ref{smaxentropy}). 

The above shows that the PPT criterion\index{PPT criterion} is simple to calculate and gives a quantification of  entanglement for mixed states\index{mixed state} in the form of negativity\index{negativity} and logarithmic negativity\index{logarithmic negativity}.  It has unfortunately one problem, that it is only a {\it necessary} condition for separability for systems larger than $ 2 \times 2 $ or $ 2 \times 3 $. This means that there are states that are in fact entangled, but do not violate the PPT criterion\index{PPT criterion}.  In other words, there are states that $ {\cal N } = E_N = 0 $, despite the fact that they are in fact entangled.  However, if the PPT criterion is violated and $ {\cal N }>0 $ or $ E_N > 0 $, one can know for sure that the state is entangled.  For system dimensions that are $ 2 \times 2 $ or $ 2 \times 3 $, the PPT criterion is necessary and sufficient, meaning that there is no ambiguity for states that do not violate the PPT criterion.  In practice, the PPT criterion can detect many entangled states, hence it provides a powerful and convenient way of detecting entanglement for mixed states.

\begin{exerciselist}[Exercise]
\item \label{q10-1}
Why doesn't the von Neumann entropy\index{von Neumann entropy} (\ref{entropydef}) work as an entanglement measure for mixed state? Show that it cannot work by contradiction by evaluating the entropy for a separable state (\ref{separable}).  
\item \label{q10-2}
Evaluate the von Neumann entropy and logarithmic negativity\index{logarithmic negativity} for the state
\begin{align}
| \Psi \rangle =  \cos \theta |0 \rangle_A |0 \rangle_B  + \sin \theta |1 \rangle_A |1 \rangle_B  .
\end{align}
Does the von Neumann entropy and logarithmic negativity have the same value for all $ \theta $?
\end{exerciselist}

\section{Correlation based entanglement criteria}

The entanglement measures of the previous section give a straightforward way of evaluating the amount of entanglement between two systems.  For systems with relatively small Hilbert space dimensions (e.g. two qubits), it is within the capabilities of modern experiments to perform measurements on a given quantum state such that the density matrix is reconstructed.  But when 
the system dimension is much larger, such as with an atomic gas or BEC, this becomes impractical very quickly.  To see this, recall that for a $ D $-dimensional density matrix, there are $ D^2 - 1 $ free parameters.  Thus to tomographically reconstruct a two qubit density matrix one requires $ 4^2 - 1 = 15 $ independent measurements.  On the other hand, to fully reconstruct the density matrix of 10 qubits, we require 1,048,575 independent measurements!

It is therefore more practical for larger systems to find other ways of detecting entanglement that do not require a complete set of measurements.  Correlation based entanglement criteria provide exactly such a way of detecting entanglement.  Typically the way that these are formulated is in the following way.  First, a mathematical statement about any separable state is made, often in the form of an inequality.  Then, this expression is evaluated for an entangled state, and if it is violated it is concluded that the state is entangled.  Several such criteria have been formulated in the literature, here we provide some of the more well-known methods.

\subsection{Variance based criteria}
\label{sec:dgcz}

The first criterion we introduce was derived first for quantum optical states, where the natural observables are position and momentum operators as seen in (\ref{positionmomentum}).  Entanglement occurs in such systems between two modes labeled by $ i \in \{ A, B\} $, for which we define the observables as 
\begin{align}
x_i & = \frac{1}{\sqrt{2}} ( a_i + a^\dagger_i) \nonumber \\
p_i & = -\frac{i}{\sqrt{2}}   ( a_i - a^\dagger_i) .
\label{positionmomentum2mode}
\end{align}
The {\it Duan-Giedke-Cirac-Zoller criterion} \index{Duan-Giedke-Cirac-Zoller criterion} then states that for separable states
\begin{align}
\sigma_u^2 \sigma_v^2 &  \ge \xi^2 | \langle [x_A,p_A] \rangle| + \frac{ | \langle [x_B,p_B] \rangle|}{\xi^2} . \hspace{1cm}  \text{(for separable states)}
\end{align}
In the case of the position and momentum operators we can immediately evaluate the right hand side to give
\begin{align}
\sigma_u^2 \sigma_v^2 &  \ge \xi^2 + \frac{1}{\xi^2} , \hspace{1cm}  \text{(for separable states)}
\label{duancrit}
\end{align}
where $ \sigma_u^2, \sigma_v^2 $ are the variances of the operators
\begin{align}
u & = | \xi | x_A + \frac{x_B}{\xi}  \nonumber \\
v & = | \xi | p_A - \frac{p_B}{\xi}
\label{uvdefs}
\end{align}
and $ \xi $ is an arbitrary non-zero parameter. If it is found that (\ref{duancrit}) is violated, then the state is entangled.  The criterion is derived using a combination of the uncertainty relations between $ x_i $ and $ p_i $, and the Cauchy-Schwarz inequality\index{Cauchy-Schwarz inequality}. 

The above is defined in terms of observables of quantum optical states.  How can we apply it to spin systems? As we saw in Sec. \ref{sec:holstein}, it is possible to transform spin variables to bosonic operators using the Holstein-Primakoff transformation\index{Holstein-Primakoff transformation}.  Due to the square root factors in (\ref{holstein}), this can't be done perfectly.  In the case that 
$ N \gg 1 $ we can approximate the square root 
\begin{align}
\sqrt{N - a^\dagger a} \approx \sqrt{N} - \frac{a^\dagger a}{2} \dots 
\end{align}
using a Taylor expansion\index{Taylor series}.  Taking just the leading order term  we have
\begin{align}
S_+ & \approx \sqrt{N}  a^\dagger \nonumber \\
S_- & = \sqrt{N}   a  .
\label{holsteinapprox}
\end{align}
To ensure that this is a reasonable approximation, we should assume that $ \langle a^\dagger a \rangle $ is small, which implies that 
\begin{align}
\langle S_z \rangle \approx - N 
\end{align}
from (\ref{holsteinz}).  We can then approximate the position and momentum operators as 
\begin{align}
x & \approx \frac{S_- + S_+}{\sqrt{2N}} =  \frac{S_x}{\sqrt{2N}} \nonumber \\
p & \approx -i \frac{(S_- - S_+)}{\sqrt{2N}} =  - \frac{S_y}{\sqrt{2N}} 
\label{approxhpvar}
\end{align}
for the ensembles $ A $ and $ B $. These variables can be used in (\ref{uvdefs}) to evaluate the criterion. To evaluate  (\ref{uvdefs}), we only require second order correlations in the spin operators, a much easier task than tomographically reproducing the full density matrix.  

In the above we obtained minus signs in the polarization direction and the momentum.  
We can equally polarize the spins in the reverse direction and redefine the operators in different way.  For $ S_z $ operators polarized in the positive direction we have
\begin{align}
[S_x, S_y ] = 2i S_z \approx 2i N  .
\end{align}
Defining the position and momentum variables in the way (\ref{approxhpvar}) without the minus sign, we have variables that obey canonical commutation relations\index{commutation relations} $ [x, p ] = i $.  Similar types of inequalities to (\ref{duancrit}) were derived with tighter bounds which can be used as more sensitive detectors of entanglement \cite{giovannetti2003characterizing,hofmann2003violation}.  

%
%
%


\subsection{Hillery-Zubairy criteria}
\label{sec:hz}

Another correlation based entanglement criterion was originally discovered in the context of quantum optics.  The first {\it Hillery-Zubairy criterion} \index{Hillery-Zubairy criteria} states that for any separable state 
\begin{align}
| \langle a_A a_B \rangle |^2 \le \langle a_A^\dagger a_A \rangle \langle a_B^\dagger a_B \rangle \hspace{1cm}  \text{(for separable states)} .
\label{hzcrit1}
\end{align}
In order to conclude that entanglement is present, one requires a violation of the inequality, which requires that the left hand side is as large as possible.  Since the expectation value on the left hand side destroys two bosons simultaneously, one typically requires a state with quantum correlations that are a superposition of states that differ by an $ a_A $ and a $ a_B $ photon.  A two-mode squeezed state takes exactly this form
\begin{align}
\sqrt{1-\tanh^2 r} \sum_{n=0}^\infty \tanh r |n\rangle_A | n\rangle_B  , 
\label{tmss}
\end{align}
where $ r $ is the squeezing parameter and  the states $ |n \rangle $ are photonic Fock states\index{Fock states} (\ref{fockstatephoton}).   

A similar type of inequality where the correlations on the left hand side is of the form where the total boson number is conserved can be written. The second  Hillery-Zubairy criterion states that 
\begin{align}
| \langle a_A a_B^\dagger \rangle |^2 \le \langle a_A^\dagger a_A a_B^\dagger a_B \rangle . \hspace{1cm}  \text{(for separable states)\index{separable state}} 
\label{hzcrit2}
\end{align}
Again to conclude that entanglement is present one must have a larger correlation on the left hand side of the inequality than the right. 
In order to apply this to spin systems, one may apply of the Holstein-Primakoff approximation\index{Holstein-Primakoff transformation} as in the last section.  Alternatively it can be directly applied to systems involving two spatial modes of a BEC, for example in a double well potential.

\subsection{Entanglement witness}
\label{sec:entanglementwitness}

In the previous sections several methods were given that relied upon calculating particular correlations of specific operators.  In a general experiment the specific operators and correlations may not be available, in which case they cannot be used.  One may require the construction of an entanglement witness\index{entanglement witness} based on some arbitrary correlations of observables.  The {\it entanglement witness} approach provides such a general approach of determining whether entanglement is present from a set of observables of the form
\begin{align}
 \langle \xi^A_i  \otimes \xi^B_j  \rangle ,
\end{align}
where the $  \xi^A_i $ and $ \xi^B_j $ label a set of operators on subsystems $ A $ and $ B $ respectively. In the approach, one constructs a witness operator
\begin{align}
W = \sum_{i=1}^{M_A}  \sum_{j=1}^{M_B} c_{ij} \xi^A_i  \otimes \xi^B_j
\label{woperator}
\end{align}
where $ c_{ij} $ are real coefficients to be determined and $ M_A $ and $ M_B $ are the number of operators in $ A $ and $ B $ respectively.  The procedure is then to optimize the coefficients such that:
\begin{align}
\text{Minimize } \langle W \rangle \hspace{3mm} & \nonumber \\
 \text{Subject to: (1) } & W = P + Q^{T_A} \nonumber \\
 \text{(2) } & P \ge 0  \nonumber \\
 \text{(3) } &Q \ge 0 \nonumber \\
 \text{(4) } & \text{Tr}(W) = 1 .
\end{align}
If it is found that $  \langle W \rangle < 0 $, then the state is entangled.  

Generally the optimization is performed numerically such that the coefficients are obtained iteratively. Given a set of measurement operators $  \xi^A_i  $ and $  \xi^B_j $ (which may involve the identity), one randomly chooses a set of coefficients $ c_{ij} $. One then normalizes the coefficients such as to satisfy condition (4).  If it is found that $ W $ satisfies the condition (1), then one evaluates $ \langle W \rangle $ to see if it is negative.  If it is negative then the state is entangled, if it is positive, another trial version of $W$ is generated and the process is repeated.

\subsection{Covariance matrix}\index{covariance matrix}

\label{sec:covariance}

Another useful correlation based method is based on constructing a covariance matrix. \index{covariance matrix} A general covariance matrix is defined with respect to a set of Hermitian operators  $ \xi_i $ where $ i \in [1,M] $.  The $ M \times M $  covariance matrix is defined as  
\begin{align}
V_{jk} \equiv  \frac{1}{2} \langle \{ \xi_j, \xi_k \} \rangle - \langle \xi_j \rangle  \langle \xi_k \rangle , 
\end{align}
which is a real symmetric matrix.  We may also define the commutation relations between these operators by the $ M \times M $ commutation matrix defined as \index{commutation matrix}
\begin{align}
\Omega_{jk} \equiv  -i\langle [\xi_j,\xi_k] \rangle,
\label{omega}
\end{align}
which is a real antisymmetric matrix.  

The covariance matrix and commutation matrix can be put together to give a generalized statement of the uncertainty relation. The matrix inequation 
\begin{align}
 V+\frac{i}{2}\Omega \ge 0 
\label{Simon1}
\end{align}
succinctly summarizes the uncertainty relation between the operators $ \xi_j$.  The meaning of (\ref{Simon1}) is in terms of the semi-positive nature of the matrix, i.e. that it has no negative eigenvalues.  Eq. (\ref{Simon1}) is true for any set of operators, and is never violated for any kind of quantum state, separable or entangled.

\begin{framed}
{\centering \bf Covariance matrix formulation of the uncertainty relation \\\index{covariance matrix}
}
\bigskip

Eq. (\ref{Simon1}) summarizes the generalized uncertainty relations between an arbitrary number of operators.  In Sec. \ref{sec:uncertainty} we encountered the Schrodinger uncertainty relation which gave an inequality between two operators.  This same idea can be generalized to any number of operators, forming increasingly complex inequalities.  

For example, the Schrodinger uncertainty relation\index{Schrodinger uncertainty relation} gives the relationship bounding the product of the variances between two operators $ \xi_1, \xi_2 $
\begin{align}
{\cal I}_{12} \equiv & \sigma^2_{\xi_1}\sigma^2_{\xi_2} - \left| \frac{\langle\{\xi_1,\xi_2\}\rangle}{2} -\langle \xi_1 \rangle \langle \xi_2 \rangle \right|^2   - \left|  \frac{\langle[\xi_1 ,\xi_2]\rangle}{2i} \right|^2 \ge 0  , \nonumber
\end{align} 
where $ \sigma^2_\xi \equiv \langle \xi^2 \rangle - \langle \xi \rangle^2 $. Following the same procedure, one can straightforwardly derive the Schrodinger uncertainty relations\index{Schrodinger uncertainty relation} for $ M $ operators  $ \xi_1, \dots, \xi_M $.  For example, for $ M =1 $ one obtains 
\begin{align}
{\cal I}_{1} \equiv \sigma^2_{\xi_1} \ge 0
\end{align}
and for $ M = 3 $ we obtain
\begin{align}
{\cal I}_{123} \equiv & \sigma^2_{\xi_1}\sigma^2_{\xi_2}\sigma^2_{\xi_3} - \langle{f_1}|f_1\rangle|\langle{f_2}|f_3\rangle|^2 -\langle{f_2}|f_2\rangle|\langle{f_3}|f_1\rangle|^2  \nonumber \\ 
&  - \langle{f_1}|f_2\rangle\langle{f_2}|f_3\rangle\langle{f_3}|f_1\rangle   -\langle{f_2}|f_1\rangle\langle{f_3}|f_2\rangle\langle{f_1}|f_3\rangle   &\nonumber \\
&- \langle{f_3}|f_3\rangle|\langle{f_1}|f_2\rangle|^2  \ge 0 
  \label{ee}
\end{align} 
where  $ |f_i\rangle=(\xi_i -\langle \xi_i \rangle)|\Psi\rangle $. 

The remarkable feature of (\ref{Simon1}) is that it contains information about all the Schrodinger uncertainty 
relations as described above. Taking for example the $ M = 3 $ case, the first order invariant of (\ref{Simon1}) yields $  {\cal I}_{1} +  {\cal I}_{2}  +  {\cal I}_{3} \ge 0 $, i.e. the sum of the variances of the operators is non-negative.  The second order invariant (sum of principle minors) yields $  {\cal I}_{12} +  {\cal I}_{23}  +  {\cal I}_{13} \ge 0 $, which is the sum of the standard Schrodinger uncertainty relation between all operator pairs.  Finally, the third order invariant (i.e. the determinant) yields $ {\cal I}_{123} \ge 0  $, which is the three operator Schrodinger uncertainty relation\index{Schrodinger uncertainty relation}.  In the $ M $-operator case,  (\ref{Simon1}) summarizes the $ 1, 2, \dots, M $-operator Schrodinger uncertainty relation via the matrix invariants in a highly succinct way.  

\end{framed}

The covariance matrix\index{covariance matrix} criterion then says that for any separable state
\begin{align}
\tilde{V} + \frac{i}{2} \tilde{\Omega} \ge 0  \hspace{1cm}  \text{(for separable states)}
\label{covariancematrixcrit}
\end{align}
where
\begin{align}
\tilde{V}_{jk} &=  \frac{1}{2} \langle \{ \xi_j, \xi_k \}^{T_B} \rangle - \langle \xi_j^{T_B} \rangle  \langle \xi_k^{T_B} \rangle  \nonumber  \\
\tilde{\Omega}_{jk} &=-i\langle [\xi_j,\xi_k]^{T_B} \rangle .
\label{pttransformed}
\end{align}
Again, if (\ref{covariancematrixcrit}) is violated, then the state is entangled, otherwise the test is inconclusive. The meaning of the inequality is again in relation to the positivity of the $ M \times M $ matrix. For separable states\index{separable state} (\ref{covariancematrixcrit}) possesses only positive or zero eigenvalues, for entangled states (\ref{covariancematrixcrit}) can possess negative eigenvalues. The matrix elements of 
(\ref{covariancematrixcrit}) can be written equally as
\begin{align}
[\tilde{V} + \frac{i}{2} \tilde{\Omega}]_{jk} = \langle ( \xi_j \xi_k )^{T_B} \rangle - \langle \xi_j^{T_B} \rangle  \langle \xi_k^{T_B} \rangle  .
\label{pttransformed2}
\end{align}

The most well-known example of the covariance matrix\index{covariance matrix} entanglement criterion is again in the context of quantum optical systems.  The operators of the covariance matrix are taken to be
\begin{align}
\xi = (x_A, p_A, x_B, p_B ) .
\label{opticaloperators}
\end{align}
The commutation matrix can be written in this case as
\begin{align}
\Omega = \left(
\begin{array}{cccc}
0& -1 & 0 & 0 \\
1 & 0 & 0 & 0 \\
0 & 0 & 0 & -1 \\
0 & 0 & 1 & 0 
\end{array}
\right)
= \left(
\begin{array}{cc}
- J & 0 \\
0 & - J 
\end{array}
\right)
\end{align}
where we defined
\begin{align}
J = \left(
\begin{array}{cc}
0 & 1 \\
-1 & 0
\end{array}
\right) .
\end{align}

To evaluate the entanglement criterion we require evaluating the covariance and commutation matrices involving the partial transposed operators (\ref{covariancematrixcrit}).  For the choice (\ref{opticaloperators}), the partial transposed operators have the simple relation
\begin{align}
\xi^{T_B} = (x_A, p_A, x_B, -p_B ) ,
\end{align}
when working in the $ x $-basis.  The matrix elements of the partial transposed operators are
\begin{align}
\tilde{V} = 
\left(
\begin{array}{cccc}
V_{11} & V_{12} & V_{13} & -V_{14} \\
V_{12} & V_{22} & V_{23} & -V_{24} \\
V_{13} & V_{23} & V_{33} & -V_{34} \\
-V_{14} & -V_{24} & -V_{34} & V_{44} 
\end{array}
\right) .
\label{transposedcovar}
\end{align}
That is, the matrix elements involving $ p_B $ are inverted in sign. For the commutation matrix all elements are unchanged
\begin{align}
\tilde{\Omega}_{jk} = \Omega_{jk}
\label{transposedcommu}
\end{align}
where we used the fact that $ (x_B p_B)^{T_B} = p_B^{T_B} x_B^{T_B} = - p_B x_B $.  One can then substitute (\ref{transposedcovar}) and (\ref{transposedcommu}) into (\ref{covariancematrixcrit}) and check for positivity of the matrix.  

Checking the positivity of (\ref{covariancematrixcrit}) typically involves finding the eigenvalues of the matrix $ \tilde{V} + \frac{i}{2} \tilde{\Omega} $.  One can bypass the diagonalization by evaluating the determinant 
\begin{align}
\text{det} ( \tilde{V} + \frac{i}{2} \tilde{\Omega} ) \ge 0  \hspace{1cm}  \text{(for separable states)\index{separable state}},
\end{align}
since the determinant is the product of all the eigenvalues. As long as the number of negative eigenvalues is an odd number, entanglement will still be detected using the determinant approach. 
 For the example of the operators (\ref{opticaloperators}), this can be written in the form
\begin{align}
\det A \det B + \left( \frac{1}{4} - | \det C | \right)^2 - \text{Tr} \left( AJCJBJ C^T J \right) \nonumber \\
\ge \frac{ \det A +  \det B}{4}  \hspace{1cm}  \text{(for separable states)\index{separable state}},
\end{align}
where we have defined the covariance submatrix operators
\begin{align}
V & =  
\left(
\begin{array}{cc}
A & C \\
C^T & B
\end{array}
\right)  \nonumber \\
A & = 
\left(
\begin{array}{cc}
V_{11} & V_{12} \\
V_{12} & V_{22}
\end{array}
\right)  \nonumber \\
B & = 
\left(
\begin{array}{cc}
V_{33} & V_{34} \\
V_{34} & V_{44}
\end{array}
\right)  \nonumber \\
C & = 
\left(
\begin{array}{cc}
V_{13} & V_{14} \\
V_{23} & V_{24}
\end{array}
\right) .
\end{align}

In the case of total spin operators the procedure is similar.  For example one can instead use the operators
\begin{align}
\xi = (S^A_x, S^A_y, S^A_z, S^B_x, S^B_y, S^B_z ) .
\label{xispinopers}
\end{align}
The partial transpose of the operators gives in the $ S_z $-basis
\begin{align}
\xi^{T_B} =  (S^A_x, S^A_y, S^A_z, S^B_x, -S^B_y, S^B_z ) .
\label{partialtransposespin}
\end{align}
In the previous case the commutation matrix\index{covariance matrix} was independent of the state because the commutator $ [x, p ] = i $ evaluates to a constant.  Here, the commutation matrix will depend upon the expectation value with respect to the state in question.  When evaluating (\ref{pttransformed}) or (\ref{pttransformed2}) one must be careful to interchange the order of the operators if both the operators are on subsystem $ B $, e.g. $ ( S^B_x S^B_y )^{T_B} = (S^B_y)^{T_B} ( S^B_x )^{T_B} = 
-S^B_y S^B_x  $.

\begin{exerciselist}[Exercise]
\item \label{q10-3}
Explain why there are $ D^2 - 1 $ free parameters in a $ D $-dimensional density matrix.  Hint: An arbitrary density matrix can be written
\begin{align}
\rho_{nn} & = p_n \hspace{1cm} (n<D) \nonumber \\
\rho_{DD} & = 1 - \sum_{n=1}^{D-1} \nonumber \\
\rho_{n'n} & = \rho_{nn'}^* = a_{n' n}
\end{align}
where $ p_n $ are real numbers and $ a_{n' n} $ are complex numbers. 
\item \label{q10-4}
Evaluate the Hillery-Zubairy criterion\index{Hillery-Zubairy criterion} (\ref{hzcrit1}) for the two mode squeezed state\index{squeezed state} (\ref{tmss}).  What values of $ r $ does the criterion detect entanglement for?
\item \label{q10-5}
Verify that the covariance and commutation matrices\index{covariance matrix} for the transposed operators satisfy (\ref{transposedcovar}) and (\ref{transposedcommu}) for (\ref{opticaloperators}).  
\item \label{q10-6}
Verify that the partial transpose of the operators (\ref{xispinopers}) gives (\ref{partialtransposespin}).  Hint: Use the definition of the spin operators in the form (\ref{fockstateop}) and directly take the transpose.  
\end{exerciselist}

\section{One-axis two-spin squeezed states}
\label{sec:szsz}
\index{one-axis two-spin squeezed states}
\index{squeezed state!one-axis two-spin}

Up to this point we have discussed entanglement from a rather general perspective, without considering specific examples.  Due to the large Hilbert space available to spin ensembles, the types of entangled states that can be generated between spins are rather complex, much more so than for qubits.  In this section we discuss an example of a type of entangled state that can be generated between two atomic ensembles.   

\begin{figure}[t]
\includegraphics[width=\textwidth]{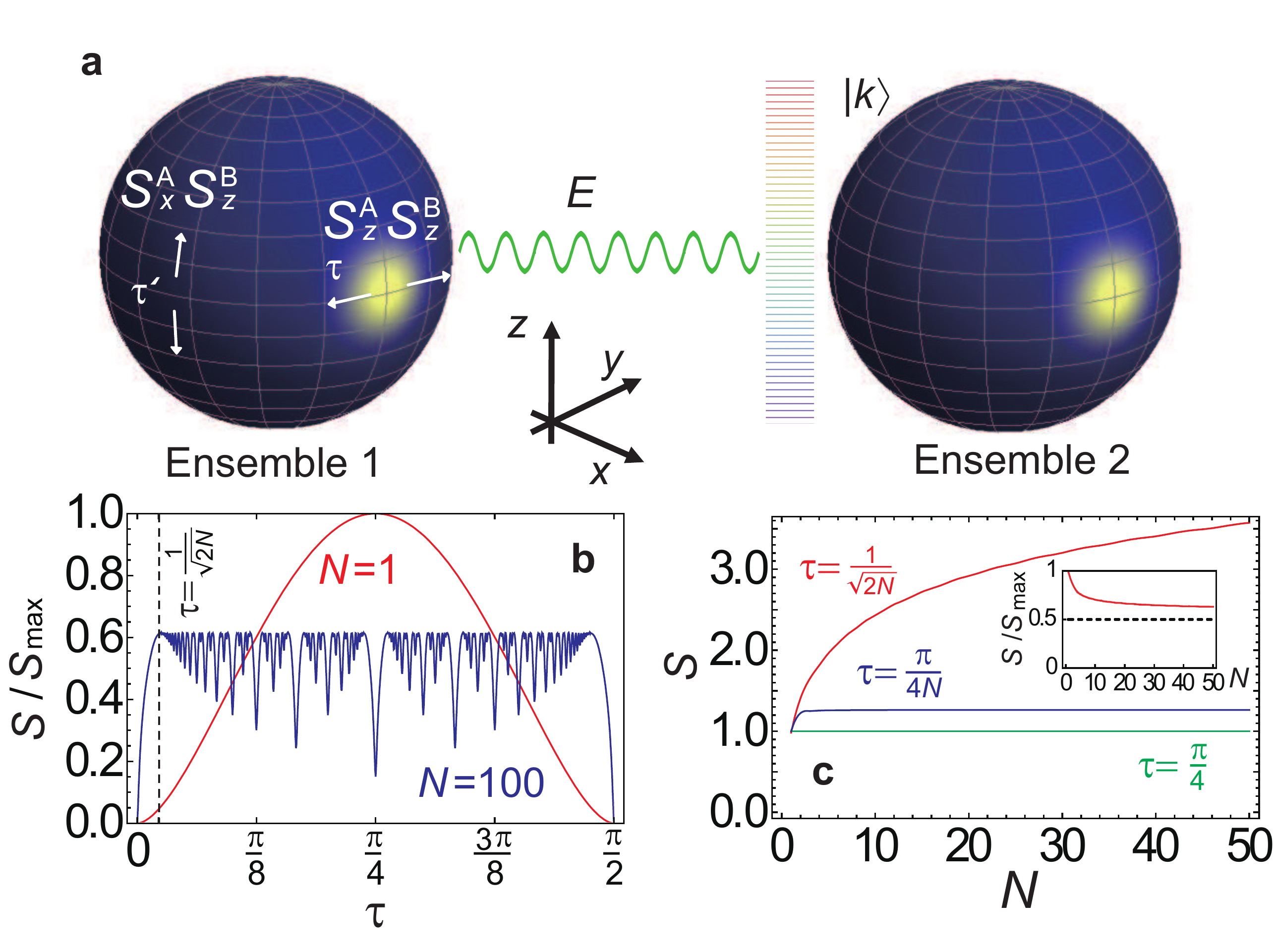}
\caption{
Entangling operation between two atomic ensembles.  (a) The schematic operation considered in this paper. The $Q$-functions for the initial spin coherent state $ | \frac{1}{\sqrt{2}},\frac{1}{\sqrt{2}} \rangle \rangle $ on each BEC are shown. The $ S_z^A S_z^B $  interactions induces a fanning of coherent states\index{coheren state} in the direction shown by the arrows.  
(b) The von Neumann entropy\index{von Neumann entropy} normalized to the maximum entanglement ($E_{\mbox{\tiny max}} = \log_2 (N+1) $) between two BEC qubits for  $ N = 1 $ and $ N =100 $ after operation of $ S_z^A S_z^B $ for a time $ \tau $. (c) Entanglement at times $  \tau = \frac{\pi}{4N}, \frac{1}{\sqrt{2N}}, \frac{\pi}{4} $ for various boson numbers $ N $. Inset: The same data for $ \tau = \frac{1}{\sqrt{2N}} $, but normalized to $E_{\mbox{\tiny max}} $. }
\label{fig10-1}
\end{figure}

In (\ref{squeezingham}) we saw an example of a squeezing Hamiltonian which can be produced by interactions between the bosons.  This was an interaction on the same ensemble, but it is possible to also have interactions of the same form between two ensembles\index{ensemble}.  The Hamiltonian in this case is
\begin{align}
H_{\text{1A2S}} = \hbar \kappa S_z^A S_z^B  ,
\end{align}
where the total $S_z $ spin for the two ensembles are labeled by $A$ and $B$. We call this the one-axis two-spin (1A2S) squeezed state since it is a two-spin generalization of the one-axis squeezed state on a single ensemble.  Applying such a Hamiltonian on two ensembles that are in 
 maximal $ S_x $-eigenstates we obtain (see Fig. \ref{fig10-1}(a))
\begin{align}
& e^{-i S_z^A S_z^B \tau} | \frac{1}{\sqrt{2}}, \frac{1}{\sqrt{2}} \rangle \rangle_A 
| \frac{1}{\sqrt{2}}, \frac{1}{\sqrt{2}} \rangle \rangle_B \nonumber \\
& = \frac{1}{\sqrt{2^N}} \sum_k  \sqrt{N \choose k} | \frac{e^{i(N-2k)\tau}}{\sqrt{2}} , \frac{e^{-i(N-2k)\tau}}{\sqrt{2}} \rangle \rangle_A | k \rangle_B ,
\label{entangledstate}
\end{align}
where the subscripts $A$ and $B$ on the states label the two ensembles and $ \tau = \kappa t $ is the dimensionless time.

Using the techniques shown in the previous sections we can directly quantify that entanglement is present.  Since (\ref{entangledstate}) is a pure state and the full quantum state is available, it is straightforward to evaluate the entanglement using the von Neumann entropy (\ref{entropydef}).  We compare the entanglement generated between the two ensembles by comparing to the standard qubit case ($ N= 1$) in Fig. \ref{fig10-1}(b).  The entanglement $  E $ is normalized to $E_{\mbox{\tiny max}} = \log_2 (N+1) $, the maximum entanglement possible between two $ N $ particle systems. For $ N = 1 $ we see the expected behavior, where a maximally entangled state is reached at $ \tau = \pi/4 $. For $ N = 100 $ the entanglement shows a repeating structure with period $ \pi/2 $, but otherwise no apparent periodicity in between. 
The fluctuations become increasingly at finer timescales as $ N $ grows, where the timescale of the fluctuations occur at
an order of $ \sim 1/N $. The basic behavior can be summarized as follows. The entanglement increases monotonically until a characteristic time $ \tau = 1/\sqrt{2N} $, after which 
very fast fluctuations occur, bounded from above by a ``ceiling'' in the entanglement.

\begin{figure}[t]
\includegraphics[width=\textwidth]{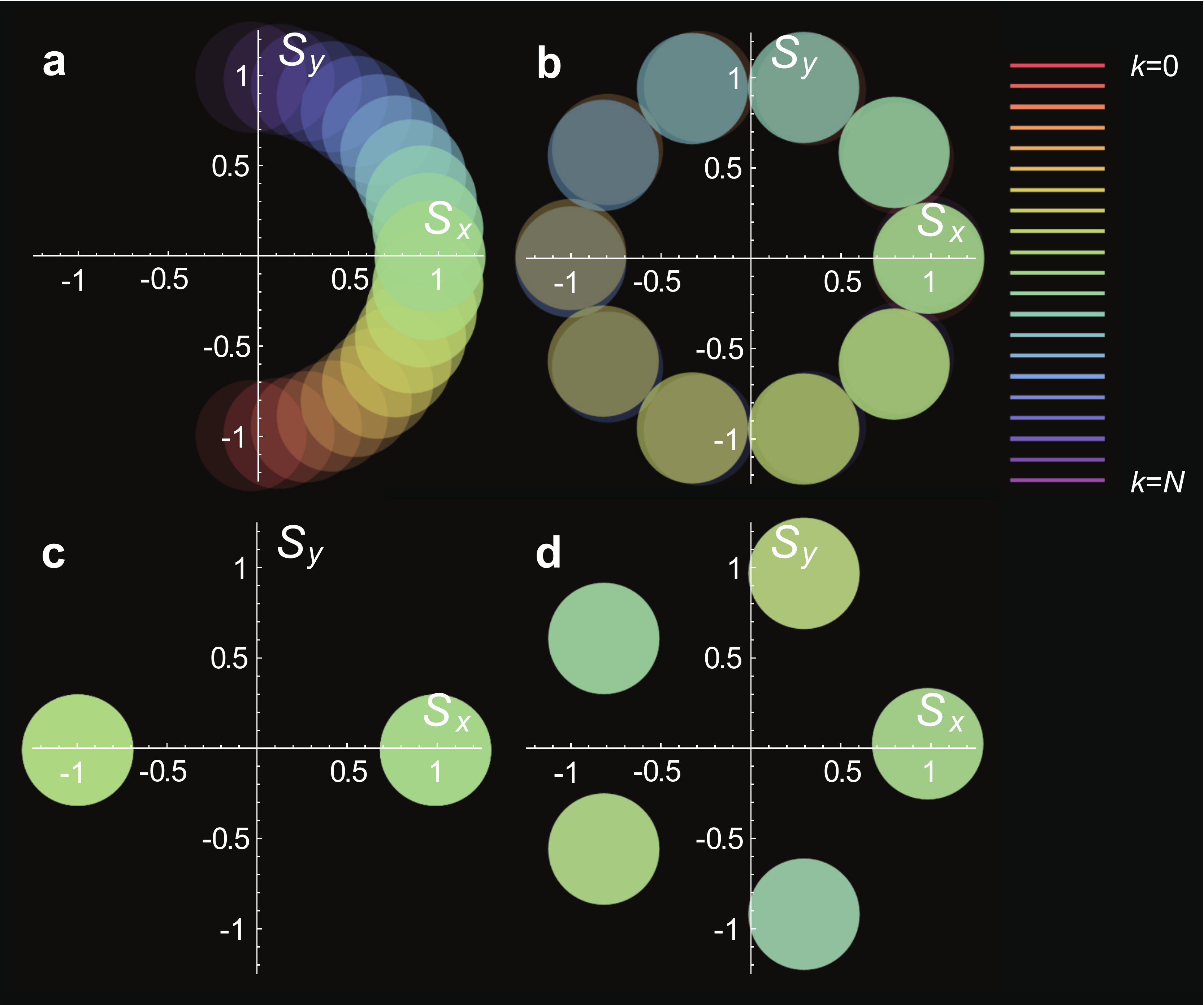}
\caption{
Visualization of the entangled state (\ref{entangledstate}) for gate times (a) $ \tau = \frac{\pi}{4 N} $ (b) $ \tau = \frac{1}{\sqrt{2N}} $  (c) $ \tau = \frac{\pi}{4} $ (d) $ \tau = \frac{\pi}{5} $ (adapted from Ref. \cite{byrnes2013}).  Eigenstates of the $ S_z $ operator, $ | k \rangle $,  are entangled with spin coherent states as marked in the key (far right).  The radii of the circles correspond to approximate distances for diminishing overlap of the spin coherent states. All plots correspond to $ N = 20 $.  }
\label{fig10-2}
\end{figure}

What is the origin of the complex behavior of the entanglement?  Figure \ref{fig10-1}(a) shows the 
$ Q $-function\index{Q-function}  for the initial state $ \tau = 0 $ of one of the BECs  on the surface of the normalized Bloch sphere\index{Bloch sphere}. Evolving the system now in $ \tau $, a visualization of the states is shown in Fig. \ref{fig10-2}. Keeping in mind that the spin coherent states\index{coherent state!spin} form a quasi-orthogonal set of states (\ref{expansionoverlap}), we represent various spin coherent states in the sum (\ref{entangledstate}) each as a circle of radius $ r = \sqrt{\frac{2}{N}} $, which corresponds to the distance where the
overlap between two spin coherent states start to diminish exponentially.  As all spin coherent states in (\ref{entangledstate}) are along the equator of the Bloch sphere $ S_z=0 $, we flatten the Bloch sphere along the $ S_z $-direction, such that each spin coherent state is located at an angle $ \phi $ in the $S_x$-$S_y$ plane. Each circle is marked such as to show the  correlation a particular $ | k \rangle $ state that the spin coherent state is entangled with (see Fig. \ref{fig10-1}(a)).  

There are several characteristic times which we consider separately. As $ \tau $ is increased, the circles fan out in 
both clockwise and counterclockwise directions. At $ \tau = \frac{\pi}{4N} $, the extremal states $ k=0,N  $ reach the $ \pm S_y $ directions.   For $ N = 1 $ (the standard qubit case) this gives a maximally entangled state. Figure \ref{fig10-1}(c) shows the amount of entanglement for times $ \tau = \frac{\pi}{4N} $, which shows that it quickly approaches the asymptotic value of $ E \approx 1.26 $. A maximally entangled state would have entanglement $ E_{\mbox{\tiny max}} = \log_2 (N+1) $, and thus gate times of $ \tau = \frac{\pi}{4N} $ correspond to relatively small amounts of entanglement, equivalent to approximately one pair of entangled qubits. 

The next characteristic time is $ \tau = \frac{1}{\sqrt{2N}}$.  This is the time when the spin coherent states\index{coherent state!spin} are separated far enough such that adjacent circles have no overlap (see Fig. \ref{fig10-2}(b)). The entanglement, as can be seen in Fig. \ref{fig10-1}(b), reaches its maximal 
value at this time, and beyond this executes fast fluctuations briefly returning to this value.  It is also a characteristic time due to 
the weight factors in (\ref{entangledstate}).  Approximating  
\begin{align}
\sqrt{\frac{1}{2^N}{N \choose k}} \approx (\frac{2}{\pi N})^{1/4} \exp  [-\frac{1}{N}(k-N/2)^2 ],
\end{align}
we see that the only the terms between $ k = N/2 \pm \sqrt{N}  $ have a significant weight in the summation and the other terms only contribute and exponentially small amount. At the time $ \tau = \frac{1}{\sqrt{2N}}$, these states are spread out over the unit circle, since the angular positions of the spin coherent states in (\ref{entangledstate}) reach order unity for the first time.  These two observations give us a simple way of estimating the asymptotic value of the entanglement in the limit $ N \rightarrow \infty $. 
Starting from (\ref{entangledstate}) and discarding small weight contributing states, we have 
\begin{align}
\frac{1}{\sqrt{2N}} \sum_{k=N/2-\sqrt{N}}^{k=N/2+\sqrt{N}} | \frac{e^{i(N-2k)\tau}}{\sqrt{2}} , \frac{e^{-i(N-2k)\tau}}{\sqrt{2}} \rangle \rangle_A | k \rangle_B
\end{align}
where we have approximated the binomial factor within the range to be constant.  Assuming that the spin coherent states are orthogonal, this
gives an entanglement of $ E \approx \log_2 \sqrt{N} $.  Relative to the maximal value this gives 
\begin{align}
\lim_{N\rightarrow \infty} E/ E_{\mbox{\tiny max}} = 1/2  .
\label{asymptoticent}
\end{align}
This shows that even in the limit of $ N \rightarrow \infty $ entanglement survives. Thus although the limit $ N \rightarrow \infty $ is considered to be a classical limit under certain situations, in this case it is clear that quantum effects are present at all $ N $. The inset of Fig. \ref{fig10-1}(c) shows results consistent with the asymptotic result (\ref{asymptoticent}).

\begin{figure}[t]
\includegraphics[width=\textwidth]{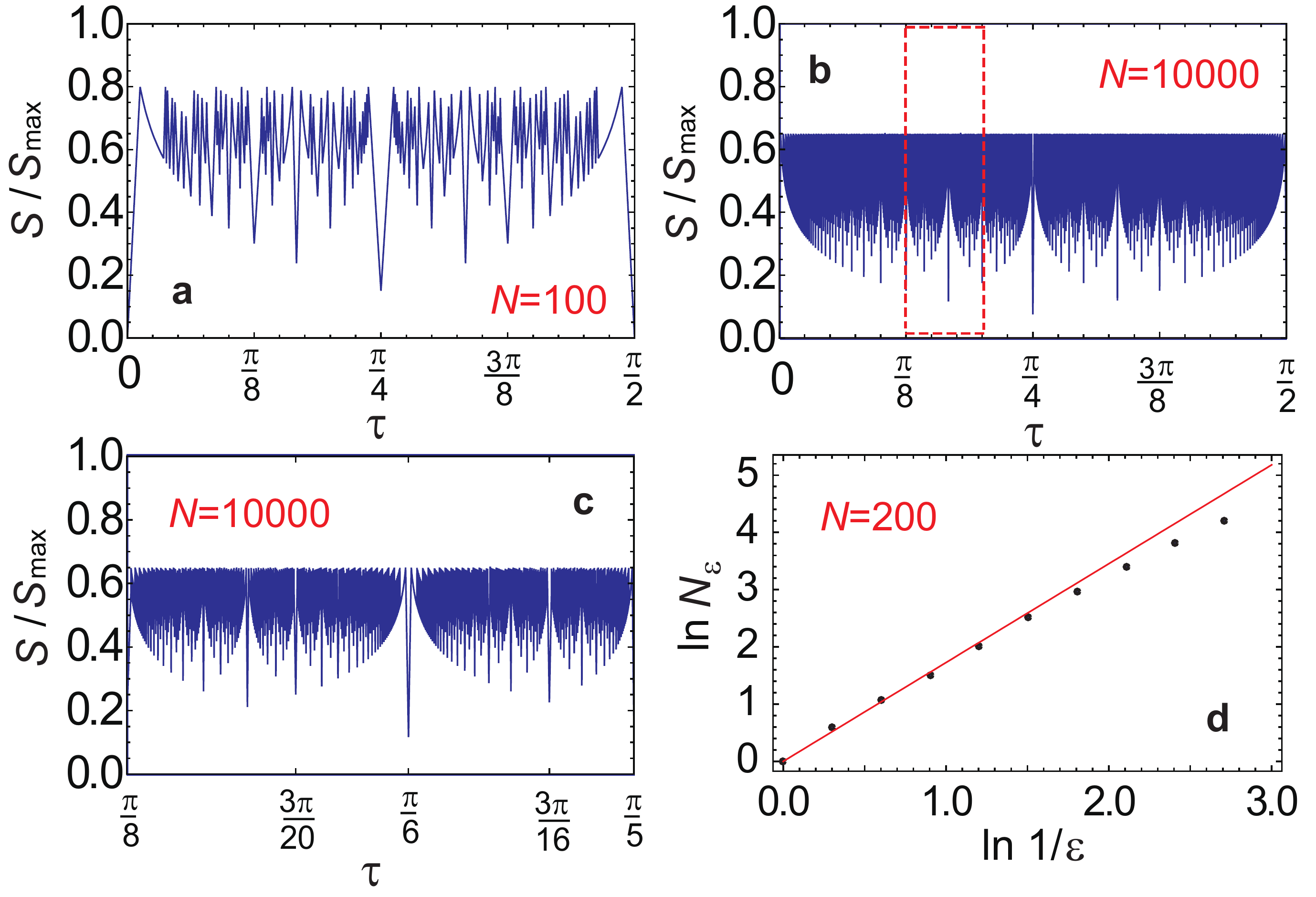}
\caption{\label{fig10-3}
The entanglement as given by the approximate formula Eq. (\ref{rationalent}) (valid for times $ \tau $ that are a rational multiple of $ \pi/4 $) for  (a) $ N =100 $ and (b) $ N = 10000$. (c) The self-similar behavior of the entanglement can be observed in a zoomed in view of (b).  (d) Logarithmic plot of the number of boxes $ N_{\varepsilon} $ versus 
the size of the box $ \varepsilon $ in the box-counting method to determine the Haussdorff (fractal) dimension\index{fractal}\index{Haussdorff dimension} of the exact entanglement curve for $ N = 200$. }
\end{figure}

We now consider the origin of the fast fluctuations in the entanglement in Fig. \ref{fig10-1}(b).  Due to the periodicity of the spin coherent states\index{coherent state!spin}, the circles align with high symmetry at particular $ \tau $ times as seen in Figs. \ref{fig10-2}(c) and \ref{fig10-2}(d).  These points correspond to sharp dips in the entanglement as seen in Fig. \ref{fig10-1}(b). For example, at $ \tau = \pi/4 $, (\ref{entangledstate}) can be written for even $ N $
\begin{align}
& \frac{1}{2} \left( (-1)^{N/2} | \frac{1}{\sqrt{2}} , \frac{1}{\sqrt{2}} \rangle \rangle_A  + | \frac{1}{\sqrt{2}} , -\frac{1}{\sqrt{2}} \rangle \rangle_A  \right) | \frac{1}{\sqrt{2}} , \frac{1}{\sqrt{2}} \rangle \rangle_B \nonumber \\
& + \frac{1}{2} \left( | \frac{1}{\sqrt{2}} , \frac{1}{\sqrt{2}} \rangle \rangle_A  - (-1)^{N/2} | \frac{1}{\sqrt{2}} , -\frac{1}{\sqrt{2}} \rangle \rangle_A  \right) | \frac{1}{\sqrt{2}} , -\frac{1}{\sqrt{2}} \rangle \rangle_B .
\label{schrodingercat}
\end{align}
For large $ N $, the states $ | \frac{1}{\sqrt{2}} , \frac{1}{\sqrt{2}} \rangle \rangle $ and $ | \frac{1}{\sqrt{2}} , - \frac{1}{\sqrt{2}} \rangle \rangle $ require flipping $ N $ bosons from $ +S_x $ to $ -S_x $ eigenstates.  Thus the terms in the brackets are Schrodinger cat states. The positive parity Schrodinger cat state\index{Schrodinger cat state} contains only even number Fock states\index{Fock states}, while the negative parity Schrodinger cat state contains only odd number Fock states\index{Fock states}.  Taking into account that 
\begin{align}
\langle \langle \frac{1}{\sqrt{2}} , -\frac{1}{\sqrt{2}}  | \frac{1}{\sqrt{2}} , \frac{1}{\sqrt{2}} \rangle \rangle = 0 ,
\end{align}
(\ref{schrodingercat}) is precisely equivalent to a Bell state\index{Bell state}, which has an 
entanglement $ E = \log_2 2 = 1 $.  For odd $N$ and $ \tau = \pi/4 $, the states can be written in terms of the eigenstates of $ S_y$
\begin{align}
& \frac{1}{2} \left( (-1)^{(N+1)/2} | \frac{e^{-i\pi/4}}{\sqrt{2}} , \frac{e^{i\pi/4}}{\sqrt{2}} \rangle \rangle_A  + | \frac{e^{-i\pi/4}}{\sqrt{2}} , -\frac{e^{i\pi/4}}{\sqrt{2}} \rangle \rangle_A  \right) | \frac{e^{-i\pi/4}}{\sqrt{2}} , \frac{e^{i\pi/4}}{\sqrt{2}} \rangle \rangle_B \nonumber \\
& + \frac{1}{2} \left( | \frac{e^{-i\pi/4}}{\sqrt{2}} , \frac{e^{i\pi/4}}{\sqrt{2}} \rangle \rangle_A  - (-1)^{(N+1)/2} | \frac{e^{-i\pi/4}}{\sqrt{2}} , -\frac{e^{i\pi/4}}{\sqrt{2}} \rangle \rangle_A  \right) | \frac{e^{-i\pi/4}}{\sqrt{2}} , -\frac{e^{i\pi/4}}{\sqrt{2}} \rangle \rangle_B .
\label{schrodingercatybasis}
\end{align}
This state also has entanglement $ E = 1 $, hence for all $ N $ the entanglement is unity, in agreement with the numerical results of Figure \ref{fig10-1}(c).

Similar arguments may be made for any $\tau$  that is a rational multiple of $ \pi/4 $. Consider a general time $ \tau = \frac{m \pi}{4d} $, where $ m,d $ are integers. The angular difference between adjacent circles in Figure \ref{fig10-2} may be deduced to be $ \Delta \phi = 4 \tau = \frac{m\pi}{d} $. The number of circles must then be the first integer multiple of $ \Delta \phi $ that gives a multiple of $ 2 \pi $.  This is $  \mbox{LCM} (m/d,2) d /m $, where $ \mbox{LCM} $ is the least common multiple\index{lowest common multiple}.  The amount of entanglement, valid for $ d \lesssim \sqrt{N} $,  can then be written
\begin{align}
E = \log_2 \left[ \frac{ \mbox{LCM} (m/d,2) d }{m} \right] .
\label{rationalent}
\end{align}
The entanglement at times that are an irrational multiple of $ \pi/4 $ do not have any particular alignment of the circles, 
and thus can be considered to be approximately randomly arranged over $ \phi $.  Similar arguments to (\ref{asymptoticent}) may then 
be applied, and at these points the entanglement jumps back up to $ E/E_{\mbox{\tiny max}} = 1/2 $ for $ N \rightarrow \infty $. We see that (\ref{rationalent}) correctly reproduces the 
result for $ \tau = \pi/4 $ in Fig. \ref{fig10-1}(c), with $ E = 1 $ for all $ N $.  

The formula (\ref{rationalent}) is directly plotted in Fig. \ref{fig10-3}(a).  We see that the dips in the entanglement are well reproduced in comparison to the exact result of Fig. \ref{fig10-1}(b). The ceiling value of the entanglement does not agree exactly since this depends upon the cutoff imposed on $ d $, a quantity which does not have a precise cutoff.  Neglecting the ceiling value, this allows us to obtain the structure for large $ N $  as seen in Fig. \ref{fig10-3}(b)(c).  We now more clearly see the self-similar repeating structure, characteristic of a fractal\index{fractal}.  Since the set of rational numbers is everywhere dense, in the limit of $ N \rightarrow \infty $, the entanglement has a pathological behavior where there are an infinite number of
dips between any two values of $ \tau $.  Such a dependence of whether a parameter is either rational or irrational is familiar in physics contexts from models such as the long-range antiferromagnetic Ising model\index{Ising model}, where the filling factors equal to a rational number occupy a finite extent of the 
chemical potential\index{chemical potential}, forming the so-called ``devil's staircase''.  In analogy to this, 
the behavior of the entanglement here is described by a {\it devil's crevasse}, \index{devil's crevasse} where every rational multiple of $ \pi/4 $
in the gate time gives a sharp dip in the entanglement.  The fractal dimension of the curve may be estimated using the box counting method (Figure \ref{fig10-3}d), applied to the exact entanglement curve (i.e. not Eq. (\ref{rationalent})).  Using several different $ N $ values and extrapolating to $ N \rightarrow \infty$, the Haussdorff dimension\index{Haussdorff dimension} is found to be $ d \approx 1.7 $, showing clearly the fractal\index{fractal} behavior of the entanglement.

\section{Two-axis two-spin squeezed states}
\label{sec:2a2s}
\index{two-axis two-spin squeezed states}
\index{squeezed state!two-axis two-spin}

In Sec. \ref{sec:squeezedstates} we saw the one-axis (1A1S) and two-axis  (2A1S) squeezed states on single ensembles. In the previous section, we generalized the one-axis squeezed state to its two-spin version, the one-axis two-spin (1A2S) squeezed state.  It is thus only natural to ask what the two-spin version of the two-axis squeezed state is.  Accordingly, we define the the two-axis two-spin (2A2S) Hamiltonian according to
\begin{align}
H_{\text{2A2S}} = \frac{\hbar \kappa}{2} (S^A_x S_x^B - S_y^A S_y^B) = \hbar \kappa(S_+^A S_+^B + S_-^A   S_-^B ) ,  
\label{eq:Hamiltonian}
\end{align}
where $  \kappa $ is an interaction constant. 
The 2A2S squeezed states are produced by a unitary evolution according to the Hamiltonian (\ref{eq:Hamiltonian}) for a time $ t $, 
\begin{align}
|\psi_{\text{2A2S}} (t) \rangle & = e^{-i H_{\text{2A2S}}  t/\hbar} | 1 ,0 \rangle \rangle_A | 1,0 \rangle \rangle_2
\nonumber \\
& = e^{-i (S_+^A S_+^B + S_-^A   S_-^B) \tau } | 1,0 \rangle \rangle_A | 1,0 \rangle    \rangle_2 ,
\label{state}
\end{align}
where we have defined a dimensionless time $ \tau =  \kappa t $.  The initial states are maximally polarized states in the $ S_z $-direction, the same as for the 2A1S Hamiltonian. In a similar way to the 2A1S Hamiltonian, the 2A2S Hamiltonian cannot be diagonalized using a linear transformation of the bosonic operators. Thus generally it must be studied using numerical methods. For small times we may expand the exponential in (\ref{state}) to second order, where the first terms in the expansion are
\begin{align}
|\psi(t)\rangle \approx & ( 1 - \tau^2 N^2) | N \rangle_1 | N \rangle_2 -i\tau N |N-1 \rangle_1 |N-1 \rangle_2 \nonumber \\
& - \tau^2 N (N-1)  |N-2 \rangle_1 |N-2 \rangle_2 + \dots ,
\label{expexpan}
\end{align}
where the states we have used are Fock states (\ref{fockstates}). All terms in the superposition have the same Fock numbers for the two BECs, hence the first thing to notice is that the state is perfectly correlated in $ S^z_j $.


To obtain some intuition about the state (\ref{state}), let us first examine the state for small evolution times. For a system with fixed particle number $ N $, the Holstein-Primakoff approximation (\ref{holsteinapprox2}) can be used, where we approximate the 2A2S Hamiltonian by
\begin{align}
H_{\text{2M}} \approx \hbar \kappa N (a_A a_B + a_A^\dagger a_B^\dagger) . 
\label{twomodesqueezeham}
\end{align}
The Hamiltonian (\ref{twomodesqueezeham}) is exactly the two-mode squeezing Hamiltonian  considered in quantum optics, with an multiplicative factor of $ N $.  The transformation of the mode operators is 
\begin{align}
e^{i H_{\text{2M}} t/\hbar  } a_A e^{-i H_{\text{2M}}  t/\hbar} = a_A \cosh N \tau  - i a_B^\dagger \sinh N \tau  \nonumber \\
e^{iH_{\text{2M}} t /\hbar} a_B e^{-iH_{\text{2M}}  t/\hbar} = a_B \cosh N \tau  - i a_A^\dagger \sinh N \tau .
\end{align}
We can deduce the time for which the Holstein-Primakoff approximation is valid by evaluating the population of the $a_1$ and $a_2$ states.  The population of the two states are always equal $ \langle a_1^{\dagger}  a_1 \rangle = \langle a_2^{\dagger}  a_2 \rangle  $ and we obtain
\begin{align}
\langle a_1^{\dagger}(t)  a_1(t) \rangle &= \langle 0 | (a_1^\dagger \cosh N \tau  + i a_2 \sinh N \tau) (a_1 \cosh N \tau  - i a_2^\dagger \sinh N \tau) | 0\rangle  \nonumber \\ 
&= \sinh ^2 N\tau \approx e^{2N\tau}/4 ,
\end{align}
where in the last step we assumed $ N\tau \gg 1 $.  Demanding that $ \langle a_1^\dagger a_1 \rangle \ll N $, we have the criterion for the validity of the Holstein-Primakoff approximation as
\begin{align}
\tau \ll \frac{\ln(4N)}{2N} .
\label{hpvalidtime}
\end{align}

Next let us define the canonical position and momentum operators as 
\begin{align}
x_j & = \frac{a_j + a^\dagger_j}{\sqrt{2}} \approx \frac{S_+^j + S_-^j}{\sqrt{2N}} = \frac{S_x^j}{\sqrt{2N}}  \nonumber \\
p_j & = \frac{-i a_j + i a^\dagger_j}{\sqrt{2}}  \approx \frac{-i S_+^j + i S_-^j}{\sqrt{2N}}  = \frac{S_y^j}{\sqrt{2N}} ,
\end{align}
where $ j \in \{A, B \}$.  For the choice of phase between the two terms in (\ref{twomodesqueezeham}), the relevant operators are those that are rotated by $45^\circ$ with respect to the quadrature axes
\begin{align}
\tilde{x}_j & = \frac{x_j + p_j}{\sqrt{2}} \approx \frac{\tilde{S}_x^j}{\sqrt{2N}} \nonumber \\
\tilde{p}_j & = \frac{p_j-x_j}{\sqrt{2}} \approx \frac{\tilde{S}_y^j}{\sqrt{2N}} ,
\end{align}
where we used the definitions (\ref{tildevariables}). The correlations for which the quantum noise is suppressed are then $ \tilde{x}_A + \tilde{x}_B  $ and $ \tilde{p}_A - \tilde{p}_B $. This can be seen by evaluating
\begin{align}
e^{i H_{\text{2M}} t/\hbar} (\tilde{x}_A + \tilde{x}_B) e^{- iH_{\text{2M}} t/\hbar} & = e^{-N \tau} (\tilde{x}_A + \tilde{x}_B) \nonumber \\
e^{i H_{\text{2M}} t/\hbar} (\tilde{p}_A - \tilde{p}_B) e^{- iH_{\text{2M}} t/\hbar} & = e^{-N \tau} (\tilde{p}_A - \tilde{p}_B) , 
\label{cvlimitsqueezed}
\end{align}
which become suppressed for large squeezing times. The corresponding anti-squeezed variables are
\begin{align}
e^{i H_{\text{2M}} t/\hbar} (\tilde{x}_A - \tilde{x}_B) e^{- iH_{\text{2M}} t/\hbar} & = e^{N \tau} (\tilde{x}_A - \tilde{x}_B) \nonumber \\
e^{i H_{\text{2M}} t/\hbar} (\tilde{p}_A + \tilde{p}_B) e^{- iH_{\text{2M}} t/\hbar} & = e^{N \tau} (\tilde{p}_A + \tilde{p}_B) .
\label{cvlimitantisqueezed}
\end{align}
%


We now directly evaluate the correlations produced by the 2A2S Hamiltonian numerically, without applying the Holstein-Primakoff approximation. From (\ref{cvlimitsqueezed}) we expect that the variances of the observables 
\begin{align}
O_{\text{sq}} \in \{ \tilde{S}_x^A + \tilde{S}_x^B, \tilde{S}_y^A - \tilde{S}_y^B \} 
\label{eprquants}
\end{align}
become suppressed, for short times when the Holstein-Primakoff approximation holds. The observables in the perpendicular directions (\ref{cvlimitantisqueezed}) 
\begin{align}
O_{\text{asq}} \in \{  \tilde{S}_x^A - \tilde{S}_x^B, \tilde{S}_y^A + \tilde{S}_y^B \}
\label{eprquantsanti}
\end{align}
are the anti-squeezed variables.   

\begin{figure}[t]%
\includegraphics[width=\linewidth]{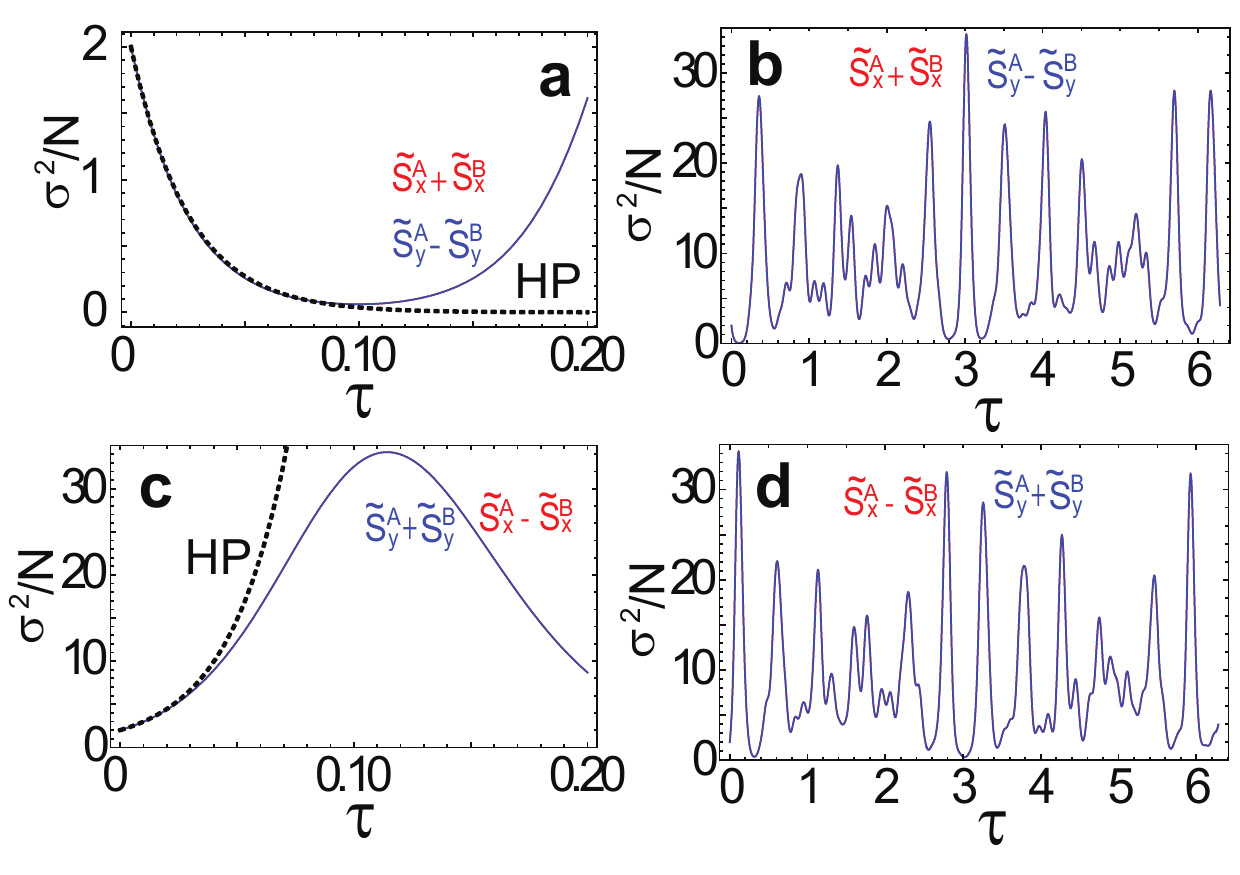}
\caption{Variances of EPR-like observables in the two-axis two-spin squeezed state.  The variances of the (a)(b) squeezed variables $ \tilde{S}_{x}^{A}+ \tilde{S}_{x}^{B}$ and $ \tilde{S}_{x}^{A}- \tilde{S}_{x}^{B}$; (c)(d) anti-squeezed variables $ \tilde{S}_{x}^{A}- \tilde{S}_{x}^{B}$ and $ \tilde{S}_{x}^{A}+ \tilde{S}_{x}^{B}$ are plotted as a function of the dimensionless interaction time $ \tau $.  The Holstein-Primakoff (HP) approximated variances are shown by the dotted lines. (a)(c) show timescales in the range $ \tau \sim 1/N $ and (b)(d) show longer timescales $ \tau \sim 1 $.  The number of atoms per ensembles is taken as  $N = 20$. }
\label{fig10-4}%
\end{figure}

In Fig. \ref{fig10-4}(a), the variances of the observables (\ref{eprquants}) are plotted for short timescales $  \tau \sim 1/N $. We see that the two variances have exactly the same time dependence and take a minimum at a time which we define as the optimal squeezing time $ \tau_{\text{opt}}$. 
In the time region $ 0 \le \tau \le \tau_{\text{opt}}  $, the variance agrees well with Holstein-Primakoff approximation, giving
\begin{align}
\sigma^2_{\tilde{S}_x^A + \tilde{S}_x^B} = \sigma^2_{\tilde{S}_y^A - \tilde{S}_y^B}  \approx 2N e^{-2N \tau},
\label{squeezedvarhp}
\end{align}
which follows from the relations (\ref{cvlimitsqueezed}) and the fact that $ \text{Var} (O_{\text{sq}}, \tau=0 ) = 2N $. Beyond these times the variance increases and no longer follows (\ref{squeezedvarhp}).  For longer timescales, as shown in Fig. \ref{fig10-4}(b), the variance follows aperiodic oscillations between low and high variance states.  Some relatively low variance states are achieved (e.g., particularly around $ \tau \approx 3 $), although the minimum variance at the times $ \tau_{\text{opt}} $ is not attained again.  The anti-squeezed variables (\ref{eprquantsanti}) are shown in  Fig. \ref{fig10-4}(c) for short timescales $  \tau \sim 1/N $.  Again the two variables (\ref{eprquantsanti}) have exactly the same time dependence, and initially increase according to
\begin{align}
\sigma^2_{  \tilde{S}_x^A - \tilde{S}_x^B} = \sigma^2_{ \tilde{S}_y^A + \tilde{S}_y^B} \approx 2N e^{2N\tau},
\end{align}
which follows from (\ref{cvlimitantisqueezed}).  In contrast to genuine two-mode squeezing, the variance does not increase unboundedly but reaches a maximum.

 \begin{figure}[ht]%
\includegraphics[width=\linewidth]{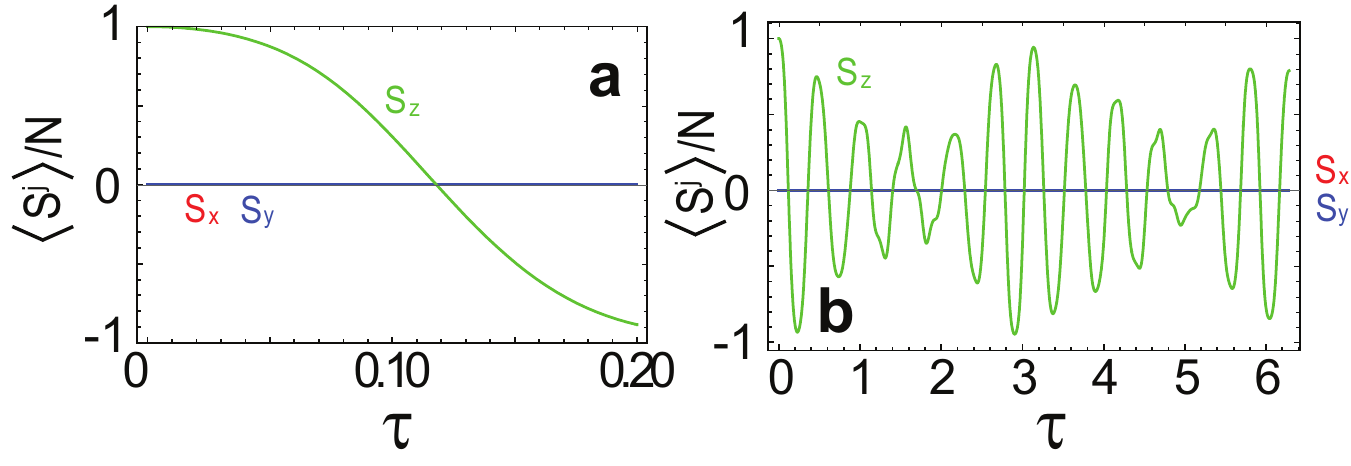}
\caption{Expectation values of spin operators for the two-axis two-spin squeezed state for (a) short timescales $ \tau \sim 1/N $; (b) long timescales  $ \tau \sim 1$.  The number of atoms per ensembles is taken as  $N = 20$. }
\label{fig10-5}%
\end{figure}


It is also instructive to examine the expectation values of the spin operators in the 2A2S squeezed state.  Figure \ref{fig10-5} shows the expectation values of the operators $ S_x^j$, $S_y^j$, $ S_z^j $.  Due to the symmetry between the initial state of the two ensembles and the 2A2S Hamiltonian, identical values are obtained for the two ensembles $ j \in \{A, B \} $.  Furthermore, the expectation values of two of the operators are always zero:
\begin{align}
\langle S_x^j (\tau) \rangle = \langle S_y^j (\tau) \rangle = 0  .
\end{align}
This can be seen from (\ref{expexpan}), where the Hamiltonian creates pairs of equal number Fock states.  Since the $S_x $ and $S_y $ operators shift the Fock states by one unit, the expectation values of $ S_x $ and $ S_y $ are zero for all time. Meanwhile, the expectation value of  $ S_z $ undergoes aperiodic oscillations and flips sign numerous times during the evolution. In particular, a sign change is observed in the vicinity of $ \tau_{\text{opt}} $, although it is not exactly at this time.  This can be understood from (\ref{expexpan}), where at $ \tau \sim 1/N $ the sum contains all terms with a similar magnitude.


The optimal squeezing time, obtained by minimizing the variance of the squeezed variable $ \tilde{S}_{x}^A+ \tilde{S}_{x}^B$ has a dependence approximated by 
\begin{align}
\tau_{\text{opt}} \approx \frac{p_0 + p_1 \ln N }{N}
\label{newfitform}
\end{align}
Where for large $N $ the parameters are $ p_0 = 0.467, p_1 = 0.508 $.  Using the optimal squeezing times, the maximal squeezing that can be attained can be estimated from the Holstein-Primakoff relation according to
\begin{align}
\min_\tau  \text{Var} (\tilde{S}^x_1 + \tilde{S}^x_2, \tau) & = 
\min_\tau  \text{Var} (\tilde{S}^y_1 - \tilde{S}^y_2, \tau)  \nonumber \\
& \approx 2Ne^{-2N \tau_{\text{opt}}^{(\text{sq})} } \nonumber \\
& \approx  \frac{2 N}{e^{2p_0} N^{2 p_1}} , 
\label{minvaluesqueezed}
\end{align}
where in the last line we substituted the expression (\ref{newfitform}). The minimal squeezing level tends to improve approximately as $ 1/N $ for larger ensemble sizes due to the factor of $ N^{2 p_1} $ in the denominator, relative to the spin coherent state variance $ 2N $.


\begin{figure}
\includegraphics[width=\columnwidth]{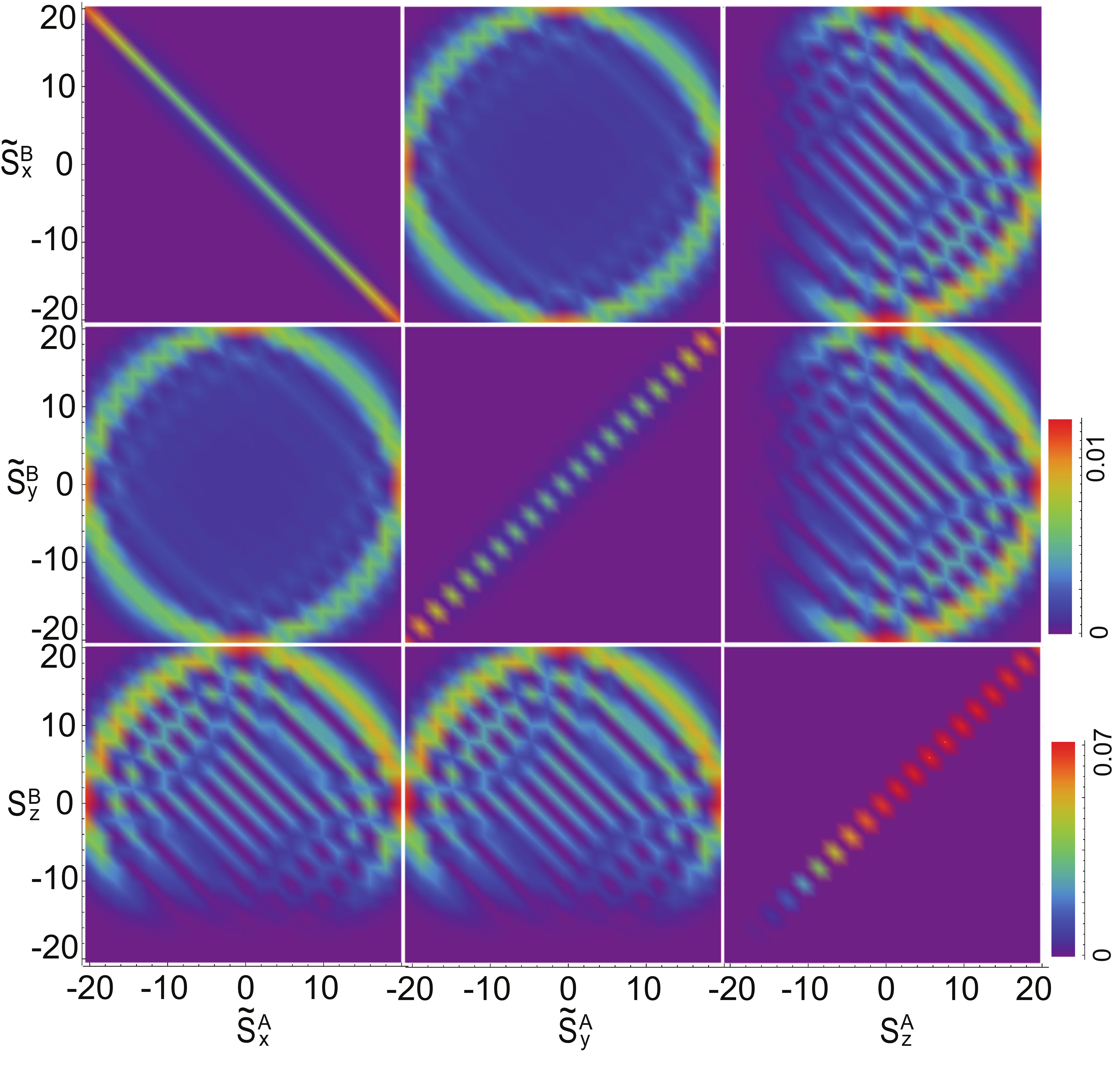}
\caption{Probability distributions of the two-axis two-spin state measured in various bases for $ \tau = 0.1 \approx \tau_{\text{opt}} $ and $ N = 20 $. The density plot legends for similar shaped distributions are the same. The legend for the 
$ (\tilde{S}_y^A,\tilde{S}_x^B) $ and $ (S_z^A,\tilde{S}_x^B) $ is the same as $ (S_z^A,\tilde{S}_y^B )$.}
\label{fig10-6}
\end{figure}

Another way to visualize the correlations is to plot the probability distributions when the state (\ref{state}) is measured in various bases. Specifically the probability of a measurement outcome $ k_A, k_B $ for various Fock states is then 
\begin{align}
p_{l_A l_B} (k_A, k_B ) = | \langle \psi (t) | 
\left( |k_A \rangle_{l_A} \otimes | k_B \rangle_{l_B}  \right)  |^2 ,
\end{align}
where $ l_A, l_B \in \{ x,y,z \} $ and are defined in Sec. \ref{sec:kxkzconv}.  The probabilities for two evolution times  near $\tau_{\text{opt}} $ are shown in Fig. \ref{fig10-6}. The effect of the correlations are seen in the $ (\tilde{S}_x^A, \tilde{S}_x^B )$ and $ (\tilde{S}_y^A, \tilde{S}_y^B )$ measurement combinations, where the most likely probabilities occur when $ \tilde{S}_x^A = -\tilde{S}_x^B $ and $  \tilde{S}_y^A = \tilde{S}_y^B $ respectively. This means that the quantities $ \tilde{S}_x^A + \tilde{S}_x^B $ and $  \tilde{S}_y^A - \tilde{S}_y^B $ always take small values and hence are squeezed. The probability distribution for the measurements for the four cases $ (\tilde{S}_{x,y}^A, \tilde{S}_{x,y}^B )$ initially starts as a Gaussian centered around $ \tilde{S}_{x,y}   = 0 $ and becomes increasingly squeezed.  For the  $ (S_z^A, S_z^B )$ measurement, we see Fock state correlations arising from the fact that the 2A2S Hamiltonian always produces Fock states in pairs, as shown in (\ref{expexpan}).  For the remaining correlation pairs, the distributions are always symmetrical in the variables $ \tilde{S}_x $ and $ \tilde{S}_y $, hence give zero when averaged. Thus there is no correlation between the remaining variables. The lack of correlations in the off-diagonal combinations in Fig. \ref{fig10-6} is also a feature of standard Bell states.

\begin{figure}
\includegraphics[width=\columnwidth]{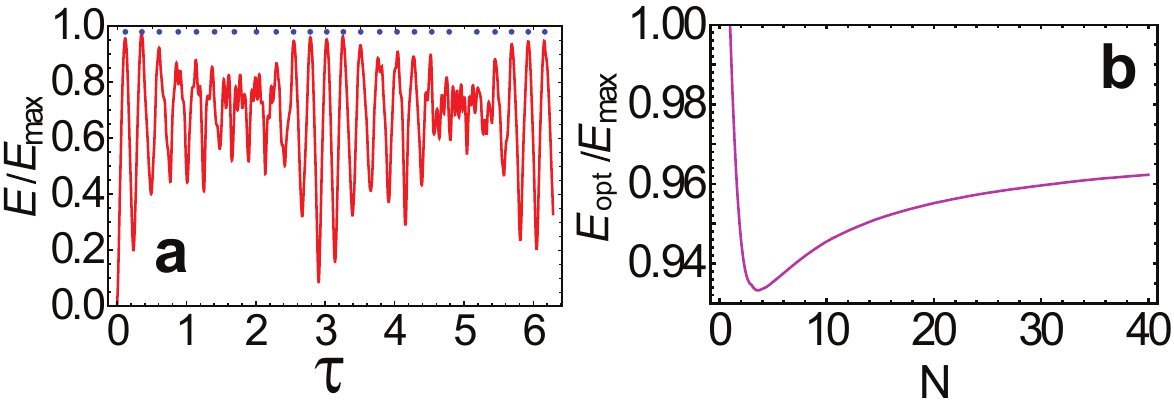}
\caption{Entanglement of the two-axis two-spin squeezed state, as measured by the von Neumann entropy $E$  normalized to the maximum value $E_{\text{max}} = \text{log}_2 (N+1)$.  (a) As a function of the interaction time $\tau$ for $ N = 20 $. The dots in each figure denote the times when $ \langle S_z^A \rangle = 0 $ for each $N $.  (b) The maximum value of the von Neumann entropy as a function of $ N $.  \label{fig10-7} }
\end{figure}

We now turn to the entanglement that is generated in 2A2S state. The entanglement that we consider is that present between the two BECs, which forms a natural bipartition in the system.  Figure \ref{fig10-7}(a) shows the von Neumann entropy normalized to the maximum value $E_{\text{max}} = \text{log}_2(N+1)$ for two $N+1$ level systems.  We see that the entanglement first reaches a maximum at a similar time to the optimal squeezing time $ \tau_{\text{opt}} $, and reaches nearly the maximum possible entanglement between the two BECs.  For larger values of $N $, the oscillations have a higher frequency, with a period that is $ \sim 2 \tau_{\text{opt}} $.  Figure \ref{fig10-7}(b) shows the maximal entanglement as a function of $ N $.  For each $ N $, we find the maximum value of the entanglement by optimizing the time in the vicinity of the first maximum.   We see that the optimized entanglement approaches the maximum possible entanglement $E_{\text{max}} $ for large $N$. 
The entanglement oscillates between large and small values and tends to occur at the values corresponding to $ \langle S_z \rangle = 0 $. This is true not only in the vicinity of the first maximum in the entanglement, but for all $ \tau $.  Figure \ref{fig10-7}(a) marks all the times (with a dot in the figure) where $ \langle S_z \rangle =0 $. We see that each peak in the entanglement occurs when $ \langle S_z \rangle =0 $. 

%


We have seen in Fig. \ref{fig10-7}(b) that near-maximal entanglement can be obtained at optimized evolution times of the 2A2S Hamiltonian.  We have also seen in Fig. \ref{fig10-6} that at the optimized squeezing times, very flat distributions of the correlations can be obtained.  These facts suggest that a good approximation for the state in the large $N $ regime is
\begin{align}
|\psi (\tau_{\text{opt}}) \rangle & \approx | \text{EPR}_{-} \rangle ,
\end{align}
where we defined the state
\begin{align}
| \text{EPR}_{-} \rangle = \frac{1}{\sqrt{N+1}} \sum_{k=0}^N | k \rangle^A_{\tilde{x}} | N-k \rangle^B_{\tilde{x}}  .
\label{spinEPRxbasis}
\end{align}
This state has the maximum possible entanglement $E_{\text{max}} $ between the two BECs, and exhibits squeezing in the variable $ \tilde{S}_x^A + \tilde{S}_x^B $. Algebraic manipulation allows one to rewrite this state equally as
\begin{align}
| \text{EPR}_{-} \rangle & = \frac{1}{\sqrt{N+1}} \sum_{k=0}^N (-1)^k | k \rangle^A_{\tilde{y}} | k \rangle^B_{\tilde{y}}  \label{spinEPRybasis}  \\
& = \frac{1}{\sqrt{N+1}} \sum_{k=0}^N (-1)^k | k \rangle^A_{z} | k \rangle^B_{z}  ,
\label{spinEPRzbasis}
\end{align}
which have the correct $ \tilde{S}_y^A - \tilde{S}_y^B $ and $ \tilde{S}_z^A - \tilde{S}_z^B $ correlations, in agreement with Fig. \ref{fig10-6}. Such a state is a type of spin-EPR state which exhibits correlations in a similar way to Bell states and continuous variable two-mode squeezed states, in all possible bases.  In fact, the correlations are for any choice of basis such that 
\begin{align}
| \text{EPR}_{-} \rangle  = \frac{1}{\sqrt{N+1}} \sum_{k=0}^N  | k \rangle_{\bm{n}}^A | k \rangle_{\bar{\bm{n}}}^B ,
\label{prototypeepr}
\end{align}
where $ \bar{\bm{n}} = (- \sin \theta \cos \phi,\sin \theta \sin \phi,\cos \theta ) $.  

\begin{figure}
\includegraphics[width=\columnwidth]{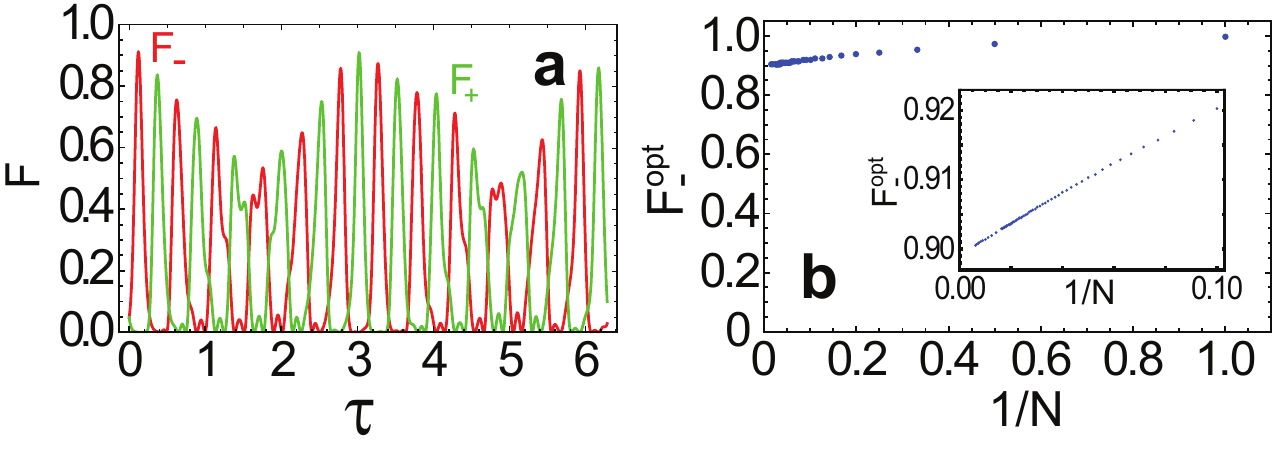}
\caption{(a) Fidelities of the two-axis two-spin squeezed state (\ref{state}) with respect to the spin-EPR states (\ref{spinEPRzbasis}) and (\ref{eprplus}).   Time dependence of the fidelity for $ N = 20 $.  (b) Optimized fidelity (\ref{fminusdef}) as a function of $ 1/N$.  Inset shows zoomed in region for large $ N $. \label{fig10-8}}
\end{figure}

In Fig. \ref{fig10-8} show the fidelity of the 2A2S squeezed state with reference to the spin-EPR state, defined as 
\begin{align}
F_{-} = | \langle \text{EPR}_{-} | \psi(\tau) \rangle |^2 .  
\label{fminusdef}
\end{align}
We also plot the fidelity with respect to another spin-EPR state defined without phases
\begin{align}
F_+ = | \langle \text{EPR}_{+} | \psi(\tau) \rangle |^2 ,  
\end{align}
where 
\begin{align}
| \text{EPR}_{+} \rangle = \frac{1}{\sqrt{N+1}} \sum_{k=0}^N | k \rangle^A_{z} | k \rangle^B_{z} .  
\label{eprplus}
\end{align}
We see that the state attains high overlap with the $ | \text{EPR}_{-} \rangle  $ at a time $ \tau_{\text{opt}} $ as expected, and oscillates with peaks at similar times as the peaks in the anti-squeezing parameters seen in Fig. \ref{fig10-4}(d). On comparison with Fig. \ref{fig10-7}(b), we see that every second peak in the entanglement corresponds to the peaks for the fidelity $ F_- $.  The remaining peaks occur for the fidelity $F_+ $.  The timing of the peaks in $F_+ $ also match the peaks in squeezing parameters in Fig. \ref{fig10-4}(b). This makes it clear that the effect of the 2A2S Hamiltonian is to first generate a state closely approximating $ | \text{EPR}_{-} \rangle  $, which 
then subsequently evolves to $ | \text{EPR}_{+} \rangle  $, and this cycle repeats itself in an aperiodic fashion.  In Fig. \ref{fig10-8}(b) we examine the scaling of the fidelity with $ N $.  We optimize the interaction time $ \tau $ such as to maximize $ F_- $ in the region of the optimal squeezing time.  The optimal time in terms of fidelity is again found to be most similar to times when $ \langle S_z \rangle = 0 $, but not precisely the same. The state approaches $ F_- \approx 0.9 $, showing that the 2A2S squeezed state has a high overlap with the $ | \text{EPR}_{-} \rangle  $ state.

\begin{exerciselist}[Exercise]
\item \label{q10-7}
Show that the form of the spin-EPR state (\ref{prototypeepr}) is algebraically equivalent to (\ref{spinEPRxbasis}).  Hint: Use the definition (\ref{expansionkykz}) of the rotated Fock states and show that the rotation factors can be eliminated.  
\end{exerciselist}

\section{References and further reading}

\begin{itemize}
\item Sec. \ref{sec:inseparability}: General reviews on entanglement \cite{guhne2009entanglement,horodecki2009quantum,plenio2014introduction,amico2008entanglement,friis2019entanglement}.
\item Sec. \ref{sec:purestateent}: Original reference on von Neumann entropy \cite{vonveumann1927}. For further reviews and books on entropy \cite{nielsen2000,wehrl1978general}.
\item Sec. \ref{sec:mixedstateent}: Original works for the PPT criterion \cite{peres,Horodecki19961,horodecki1997separability}. 
Negativity entanglement measure for mixed states \cite{PhysRevLett.95.090503,vidal2002computable}. 
\item Sec. \ref{sec:dgcz}: The Duan-Giedke-Cirac-Zoller 
criterion \cite{duan2000inseparability}. Other variance based entanglement criteria \cite{giovannetti2003characterizing,hofmann2003violation}.
\item Sec. \ref{sec:hz}: The Hillery-Zubairy criterion \cite{hillery06}.
\item Sec. \ref{sec:entanglementwitness}: Methods of constructing entanglement witnesses \cite{horedecki1996separability,lewenstein2000optimization,terhal2002detecting,szangolies2015detecting}.
\item Sec. \ref{sec:covariance}:  The connection between  covariance matrices and uncertainty relations \cite{robertson1934indeterminacy}.  The covariance matrix entanglement criterion for optical quadratures was introduced in \cite{simon00}.  This was then generalized to arbitrary operators \cite{tripathi18}.
\item Sec. \ref{sec:szsz}: In an atomic context the one-axis two-spin squeezed states were theoretically investigated in \cite{byrnes2013,kurkjian2013spin}.  Various proposals for producing this in cold atoms \cite{pyrkov2013entanglement,treutlein2006microwave,jing2019split,rosseau2014entanglement}.
\item Sec. \ref{sec:2a2s}: Theoretical investigation of the two-axis two-spin squeezed states  \cite{kitzinger2020two}.  
\end{itemize}

	\chapter[Quantum information processing with atomic ensembles]{Quantum information processing with atomic ensembles}

\label{ch:quantuminfo}

\section{Introduction}

\index{quantum metrology}
\index{quantum information processing}

In this chapter we will give an overview of how atomic ensembles can be used for quantum information processing.  What we mean here by ``quantum information processing'' is in a different sense to that used to describe the whole field of quantum information.  We have seen up to this point that atomic ensembles can be used for a variety of different applications, ranging from matter wave and spin interferometry, which can be applied to precision measurements (the field of ``quantum metrology''), and quantum simulations of various many-body systems.  The {\it field} of quantum information encompasses these types of applications in the sense that they are all applications of quantum mechanics.  On the other hand, ``quantum information processing'' refers to tasks that are more connected to {\it information} and how  quantum operations can be performed on them.  The tasks that we will discuss are simple protocols such as quantum teleportation which has a fundamental place in the field as a simple yet non-trivial operation. 
 Other tasks include quantum computation where the aim is to perform algorithmic tasks that cannot be performed using a conventional classical computer.  In particular, we will examine Deutsch's algorithm\index{Deutsch's algorithm} as a simple example showing quantum mechanical speedup, and how it is implemented with atomic ensembles. Another example shows how adiabatic quantum computing\index{adiabatic quantum computing} can be performed with atomic ensembles.  \index{ensemble}

\section{Continuous variables quantum information processing}

\label{sec:cv}

\subsection{Mapping between spin and photonic variables}

\label{sec:spinphoton}

The first and simplest way that atomic ensembles can be used for quantum information processing is to use the Holstein-Primakoff transformation\index{Holstein-Primakoff transformation} (see Sec. \ref{sec:holstein}) to approximate spin 
variables as quadrature operators.  Consider an atomic ensemble that contains a large number of atoms $ N \gg 1 $, and is strongly polarized in the $ + S_x $-direction (for example the maximally polarized state $ | 1/\sqrt{2}, 1/\sqrt{2} \rangle \rangle $).  Writing the spin operators in the $ x $ basis using the transformation (\ref{axtransform}) we have
\begin{align}
S_x & = a_x^\dagger a_x - b_x^\dagger b_x  \nonumber \\
S_y & = -i b_x^\dagger a_x + i a_x^\dagger b_x \nonumber \\
S_z & = b_x^\dagger a_x +  a_x^\dagger b_x .
\end{align}
Since the state is polarized in the $ + S_x $-direction, we expect that the number of atoms in the $ b_x $ state will be small $ \langle b_x^\dagger b_x \rangle \approx 0 $. Applying the Holstein-Primakoff transformation\index{Holstein-Primakoff transformation} (\ref{holstein}) we may write the spin operators as 
\begin{align}
S_x & \approx N \nonumber \\
S_y & \approx \sqrt{N}(-ib_x + i b_x^\dagger) = \sqrt{2N} p \nonumber \\
S_z & \approx \sqrt{N}(b_x + b_x^\dagger) = \sqrt{2N} x  ,
\label{spinholstein}
\end{align}
where we used the definitions of the quadrature operators (\ref{positionmomentum}). 

One of the main techniques to visualize photonic states is to plot the (optical) Wigner function\index{Wigner function} of a quantum optical state in phase space $ (x,p) $.  We can visualize the state in a similar way using the (spin) Wigner functions as discussed in Sec. \ref{sec:wignerreps}.  For example, the maximally polarized state $ | 1/\sqrt{2}, 1/\sqrt{2} \rangle \rangle $, which corresponds to the photonic vacuum state, has a similar Wigner function to the optical case as shown in Fig. \ref{fig5-5}(a).  Performing a one-axis twisting squeezing operation on such a state produces a squeezed state as shown in Fig. \ref{fig5-5}(e) and \ref{fig5-5}(f).   In addition to squeezing operations, one can perform displacement operations that rotate the maximally polarized state to a spin coherent state\index{coherent state!spin} centered at a different point using (\ref{generalrotation}).  This would be performed by applying a Hamiltonian $ \cos \theta S_z + \sin \theta S_y $ which rotates the state away from the maximally $ S_x $-polarized state.  As long as the displacement is not too large, so that $ \langle b_x^\dagger b_x \rangle $ is small, this corresponds to an optical coherent state\index{coherent state!optical} as discussed in (\ref{coherentfock}). Such states are all dominantly polarized in the $ + S_x $-direction, which is why they have an equivalence to the photonic case. 

States that deviate significantly from the maximally $ + S_x $-polarized state cannot be mapped exactly to photonic states since the Holstein-Primakoff approximation\index{Holstein-Primakoff transformation} breaks down.  As a simple example, for a spin coherent state that is rotated too far from the $ + S_x $-polarized state will have different properties to a genuine optical state.  An optical coherent state\index{coherent state!optical} can be displaced in any direction an arbitrary amount
\begin{align}
e^{\beta a^\dagger - \beta^* a} | \alpha \rangle = | \alpha + \beta \rangle 
\end{align}
where the states are optical coherent states (\ref{coherentfock}).  In contrast, a displacement of a spin coherent state (\ref{generalrotation}) will eventually result in the coherent state returning to the original position after a rotation angle of $ 2 \pi $ (see Fig. \ref{fig5-4}).  One can visualize that the Holstein-Primakoff approximation\index{Holstein-Primakoff transformation} as performing quantum operations such that it is on a locally flat region of the Bloch sphere.\index{Bloch sphere}

For measurements, two common types of optical measurements that are performed are photon counting\index{photon counting} and homodyne measurements\index{homodyne measurement}.  For the former, according to the transformation (\ref{spinholstein}), it is evident that counting the number of atoms in the $ b_x $ state  corresponds to a projection to the photon Fock basis. For homodyne measurements, one would like to measure the expectation value of the generalized quadrature operator
\begin{align}
x(\theta) = \frac{1}{\sqrt{2}} ( e^{-i \theta}  a + e^{i \theta}  a^\dagger) \leftrightarrow 
\frac{\cos \theta S_z + \sin \theta S_z}{\sqrt{2N}} .
\end{align}
The measurement of such an operator can be made by measurement of spin operators in the basis $ \cos \theta S_z + \sin \theta S_z $.  

The above shows that one atomic spin ensemble can be used as an effective photonic mode. 
This can be extended to the multi-mode case by simply having more atomic ensembles.  One important operation that is required between modes is to perform entangling operations between them.  We have discussed in Sec. \ref{sec:szsz} one type of entangling operation that can be produced between atomic ensembles. In the Holstein-Primakoff approximation\index{Holstein-Primakoff transformation}, this corresponds to a $ x_A x_B $ type of Hamiltonian.  Other types of quantum states can also be produced which are analogous to two-mode squeezing, and have been produced by shining common light beam through two atomic ensembles (see \cite{julsgaard2001experimental}). 

We thus see that many of the operations that are possible in the photonic case have exact analogues to the atomic case. One operation which is somewhat less natural in the atomic case in comparison to the photonic case is the beam splitter operation, which mixes two photonic modes
\begin{align}
a & \rightarrow \cos \theta a + \sin \theta a'  \leftrightarrow \frac{\cos \theta }{2\sqrt{N}} ( S_z + i S_y )
+ \frac{\sin \theta }{2\sqrt{N'}} ( S_z' + i S_y' )
\nonumber \\
a' & \rightarrow \sin \theta a - \cos \theta a' \leftrightarrow \frac{\sin \theta }{2\sqrt{N}} ( S_z + i S_y )
- \frac{\cos \theta }{2\sqrt{N'}} ( S_z' + i S_y' ),
\label{modeinterference}
\end{align}
where the dashes denote the second photonic mode or second atomic ensemble. Physically this would correspond to mixing two atomic ensembles and interfering the spins on one with another.  While it is not impossible to perform such an operation using matter interferometry techniques, this is more experimentally challenging than performing rotations and squeezing operations between atomic levels on the same ensemble. \index{ensemble}

\subsection{Example: Quantum teleportation}
\label{sec:teleportationcv}\index{quantum teleportation}

\begin{figure}[t]
\includegraphics[width=\textwidth]{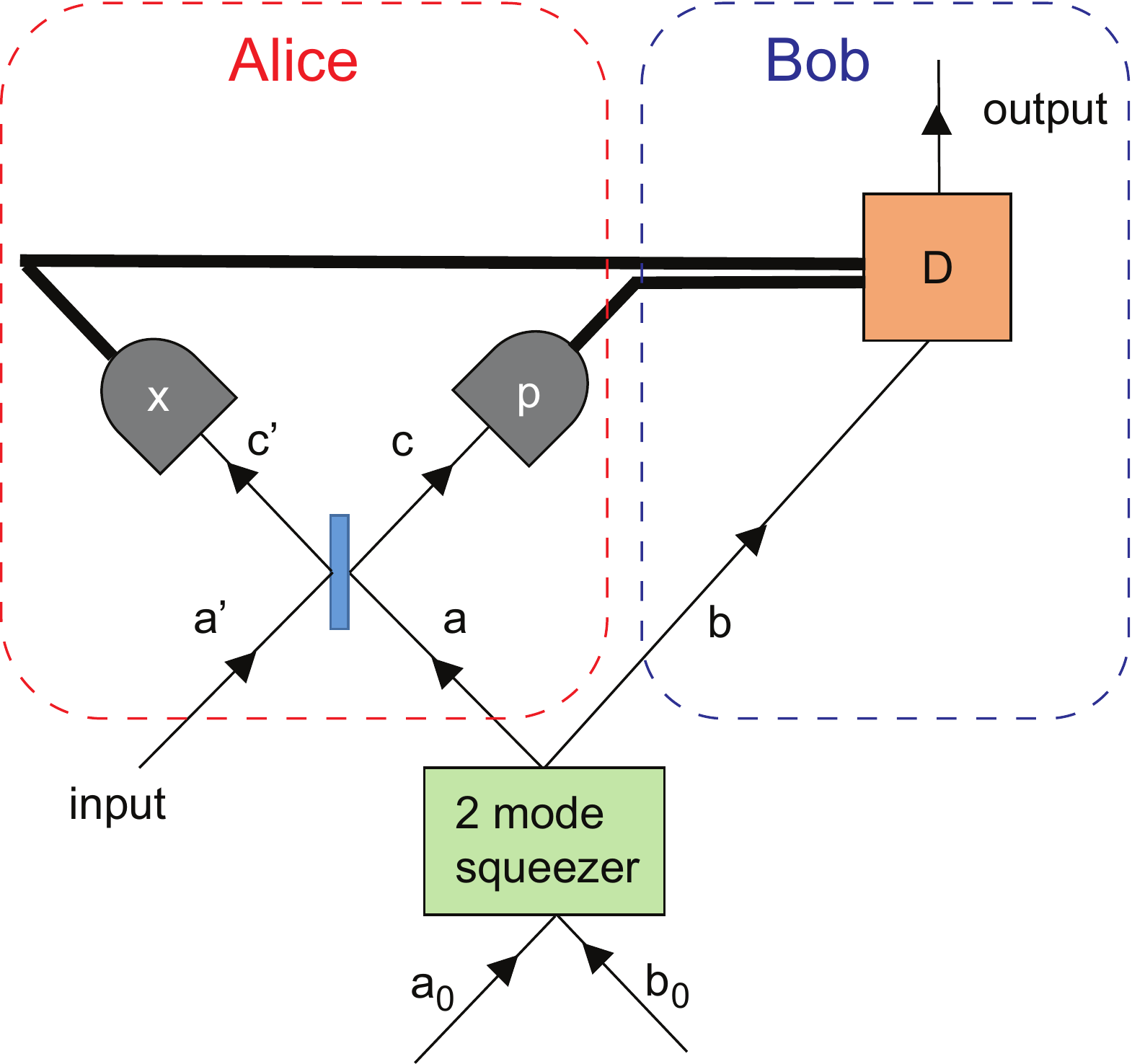}
\caption{The quantum teleportation\index{quantum teleportation} protocol for continuous variable optics.  Thin lines denote the optical modes and thick lines denote classical communication.  A two-mode squeezed state is produced in modes $ a $ and $ b $.  The state to be teleported in mode $ a' $ enters a beam splitter interfering with mode $ a $. The modes exiting the beam splitter are measured using a homodyne detectors, measuring the $ x $ and $ p $ quadratures.  The results of the measurements are sent to a displacement operator $ D $ which makes a displacement of the mode $ b $.  
}
\label{fig11-1}
\end{figure}

We now give an example of a quantum information protocol that can be implemented using the Holstein-Primakoff approximation\index{Holstein-Primakoff transformation}. In quantum teleportation an unknown quantum state is transferred from one mode to another with the use of entanglement and classical communication.  The teleportation protocol is most easily represented as a optical quantum circuit diagram as shown in Fig. \ref{fig11-1}.  The sequence of operations consist of the following steps.
\begin{enumerate}
\item Produce a two mode squeezed state on modes $ a $ and $ b $.
\item Prepare a state in mode $ a' $ to be teleported.
\item Interfere modes $ a $ and $ a' $ with a 50/50 beam splitter.
\item Measure the output modes $ c $ and $ c' $ with respect to the $ p $ and $ x $ quadratures respectively and send the results to Bob.
\item Perform a conditional displacement on mode $ b $ based on the measurement outcomes.
\end{enumerate}
The simplest way that the protocol can be described is in the Heisenberg picture\index{Heisenberg picture}.  Let us work through each of the steps in the protocol above, and obtain the transformations in the mode operators.  

The two-mode squeezing operation corresponds to applying the operator $ U = e^{-r( a_0 b_0 -  a_0^\dagger b_0^\dagger)} $.  This produces a transformation
\begin{align}
a & = a_0 \cosh r + b_0^\dagger \sinh r  \nonumber \\
b & = b_0 \cosh r + a_0^\dagger \sinh r , 
\end{align}
where $ r $ is the squeezing parameter. 
In terms of quadrature operators, using (\ref{positionmomentum}) we have
\begin{align}
x_a & = \frac{1}{\sqrt{2}} ( e^r x_{a_0} + e^{-r} x_{b_0}) \nonumber \\
p_a & = \frac{1}{\sqrt{2}} ( e^{-r} p_{a_0} + e^{r} p_{b_0}) \nonumber \\
x_b & = \frac{1}{\sqrt{2}} ( e^r x_{a_0} - e^{-r} x_{b_0}) \nonumber \\
p_b & = \frac{1}{\sqrt{2}} ( e^{-r} p_{a_0} - e^{r} p_{b_0}),
\end{align}
where $ x_a = ( a+ a^\dagger)/\sqrt{2} $ and $ p_a = -i( a- a^\dagger)/\sqrt{2} $ and similarly for the other modes. This state correlations in the variables $ x_a - x_b  $ and $p_a + p_b $ because these are exponentially suppressed according to 
\begin{align}
x_a - x_b & = \sqrt{2} e^{-r} x_{b_0} \nonumber \\
p_a + p_b & =  \sqrt{2} e^{-r} p_{b_0} .
\label{eprcorres}
\end{align}

The next step is to prepare a quantum state in mode $ a $.  In the Heisenberg picture\index{Heisenberg picture} we evolve the operators, hence no calculation is required since the state remains time-invariant. Next, we interfere the modes $ a $ and  $ a' $, giving the transformation 
\begin{align}
c' & = \frac{1}{\sqrt{2}} ( a' - a) \nonumber \\
c & = \frac{1}{\sqrt{2}} ( a' + a) .
\end{align}
In terms of quadrature operators this is 
\begin{align}
x_{c'} = \frac{1}{\sqrt{2}} ( x_{a'} - x_a ) \label{xcp} \\
p_{c'} = \frac{1}{\sqrt{2}} ( p_{a'} - p_a ) \label{pcp} \\
x_{c} = \frac{1}{\sqrt{2}} ( x_{a'} + x_a ) \label{xc} \\
p_{c} = \frac{1}{\sqrt{2}} ( p_{a'} + p_a ) \label{pc} .
\end{align}

Next, the modes $ c $ and $ c' $ are measured, and collapses the $ x_{c'} $ and $ p_{c} $ quantum variables to a classical random variables $ x_{c'}^{M} $ and $ p_{c}^{M}  $ respectively.  These two variables are at this point classical c-numbers and are classically communicated to Bob. 

Now let us look at the quadratures of Bob's mode before it enters the displacement operation $ D $. We can rearrange the correlations (\ref{eprcorres}) to write 
\begin{align}
x_b & = x_a - \sqrt{2} e^{-r} x_{b_0} = x_{a'} - \sqrt{2} x_{c'}^M - \sqrt{2} e^{-r} x_{b_0} \nonumber \\
p_b & = -p_a + \sqrt{2} e^{-r} p_{b_0} = p_{a'} - \sqrt{2} p_c^M + \sqrt{2} e^{-r} p_{b_0} ,
\end{align}
where we used (\ref{xcp}) and (\ref{pc}) to rewrite $ x_a  $ and $ p_a $.  For an infinitely squeezed state\index{squeezed state} $ r \rightarrow \infty $, the last term in the above quadratures become zero.  This shows that the quadratures of the mode $ b $ are the same as for the mode $ a' $, up to a constant offset.  Since the offsets  $ x_{c'}^M $ and $p_{c}^M $ are classical numbers, these can be canceled by a displacement operation
\begin{align}
D(x_{c'}^M, p_{c}^M)  = e^{ - \sqrt{2} (x_{c'}^M + i p_{c}^M) b^\dagger + \sqrt{2}  (x_{c'}^M - i p_{c}^M ) b } .
\end{align}
This completes the teleportation operation, and the mode emerging from the displacement operation has the same state as the input mode $ a' $.  

The above protocol was first performed experimentally with optical continuous variable modes in 1998, following the above optical circuit.  In the context of atomic ensembles, the above protocol was first performed by Krauter, Polzik, and co-workers using two atomic ensembles and optical modes in 2013.  In their experiment, the atomic ensembles were trapped in paraffin coated glass cells, where the effect of the paraffin is to prolong the spin coherence time despite the numerous collisions the atoms undergo with the glass walls.  Polarized atomic ensembles were used to realize the modes $ a' $ and $ b $ in the Holstein-Primakoff approximation\index{Holstein-Primakoff transformation}. Meanwhile mode $ a $ is implemented by an optical mode.  A light mode was sent through to the atomic ensemble corresponding to mode $ b $, producing a two-mode squeezed state\index{squeezed state} between the ensemble and the light.  The emerging light mode $ a $ then interacts with the second atomic ensemble corresponding to the mode $ a' $, this time with a beam-splitter type of interaction.  The light mode is then measured, and the classical results are fed back to the atomic ensemble corresponding to mode $ b $.  For further details we refer the reader to the original experiment \cite{krauter2013deterministic}. 

The above example shows the way in which atomic ensembles can be used to perform an experiment which is initially designed for optical modes. 
Numerous other examples of continuous variables quantum information processing is possible, which we do not reproduce here.  We refer the reader to the excellent review for other protocols that can be implemented \cite{braunstein2005}.

\section{Spinor quantum computing}

\label{sec:beyondholstein}

In the previous section, we saw that it is possible to use spin ensembles as effective bosonic modes that can be used for continuous variables quantum information processing. In order to perform this mapping, the Holstein-Primakoff approximation\index{Holstein-Primakoff transformation} was used, which requires that the spin ensembles are dominantly polarized in the $ S_x $ direction. This means that one must always work with states that are close to the fully polarized $ S_x $ eigenstate, or in terms of the Wigner\index{Wigner function} or $Q$-distribution\index{Q-function}, it must be primarily centered around $ \theta = \phi = 0 $ (see Figs. \ref{fig5-5}(a) and \ref{fig5-6}(a)).  This is a restriction in terms of the types of states that can be used, and experimentally there is no reason to only look at such a small class of states.  We may ask whether it is possible to use more of the Hilbert space available to the ensembles.  Due to the strong analogy between quantum optical states and spin ensemble states that we have seen in Chapter \ref{ch:spinorbec}, this suggests that it might be possible to find a way of performing quantum information processing that is more naturally suited to the structure of states that one encounters with spin ensembles. 

The most natural way to encode a single qubit with ensembles is to use the spin coherent states\index{coherent state!spin} that we encountered in Sec. \ref{sec:spincoherentstates}.  Suppose that one wishes to encode the quantum information associated with a single qubit.  This could be part of a quantum memory in a quantum computer for example, that could be used as part of an quantum algorithm.  Physically, this could be implemented using thermal atomic ensembles in glass cells, as described in the previous section, or with atom chips which can be patterned to have multiple traps.  The same quantum information could be stored in a qubit or a spin coherent state, according to the mapping
\begin{align}
\alpha |0 \rangle + \beta | 1 \rangle \rightarrow | \alpha, \beta \rangle \rangle = \frac{1}{\sqrt{N!}}( \alpha a^\dagger + \beta b^\dagger )^N | 0 \rangle .
\end{align}
Both of these states contain exactly the same amount of quantum information, since they involve the same parameters $ \alpha, \beta $.  Of course, the spin ensemble could potentially have many other types of state, not only spin coherent states, since the Hilbert space dimension is in fact much larger.  However, for the set of all spin coherent states, there is a one-to-one mapping between the qubit states and spin coherent states. 

Single qubit gates also have a perfect analogy with spin coherent states.  As we saw in (\ref{generalrotation}), applying a Hamiltonian $ H = \bm{n} \cdot \bm{S} $ on a spin coherent state has exactly the corresponding effect for qubits, in terms of the mapping of the parameters $ (\alpha, \beta) \rightarrow (\alpha', \beta') $.  For example, common gates such as as the $ Z $ phase flip gate\index{Z phase flip gate},  $ X $ bit flip gate\index{X bit flip gate}, and Hadamard gate\index{Hadamard gate} are performed by the operations
\begin{align}
e^{-i \pi S_z /2} | \alpha, \beta \rangle \rangle & = e^{-i\pi N/2} | \alpha, - \beta \rangle \rangle \hspace{1cm} &(Z\text{-gate})
 \label{zgateens} \\
e^{-i \pi S_x /2} | \alpha, \beta \rangle \rangle & = (-i)^N | \beta, \alpha \rangle \rangle  
\hspace{1cm} &(X\text{-gate})
 \label{xgateens}  \\
e^{-i 3\pi S_y/4} | \alpha, \beta \rangle \rangle & = (-1)^N| \frac{\alpha + \beta}{\sqrt{2}},  \frac{\alpha -\beta}{\sqrt{2}} \rangle \rangle  .
\hspace{1cm} & (\text{Hadamard gate})
\label{hgateens}
\end{align}
We have written the global phase factors to match the expressions in (\ref{spinrotationsxyz}), but they are physically irrelevant. 

For two qubit gates, the types of states that are produced with ensembles are analogous, but do not have the perfect equivalence as single qubits.  A typical type of interaction that might be present between ensembles is the $ H = S_z^{A} S_z^B $ interaction, where $ A $ and $ B $ label the two ensembles. For the qubit case, the analogous Hamiltonian is $ H = \sigma_z^A \sigma_z^B $ and can produce an entangled state 
\begin{align}
& e^{-i \sigma_z^A \sigma_z^B \tau}     \left(\frac{|0 \rangle+ |1 \rangle}{\sqrt{2}} \right) \left(\frac{|0 \rangle+ |1 \rangle}{\sqrt{2}} \right)  = \left( \frac{e^{i \tau} | 0 \rangle + e^{-i \tau} | 1 \rangle}{\sqrt{2}} \right) | 0 \rangle + 
\left( \frac{ e^{-i \tau} | 0 \rangle + e^{i \tau} | 1 \rangle }{\sqrt{2}} \right) | 1 \rangle   .
\label{qubitszsz}
\end{align}
For a time $ \tau = \pi/4 $, the above state becomes a maximally entangled state
\begin{align}
 e^{-i \sigma_z^A \sigma_z^B \pi/4 }  \left( \frac{|0 \rangle+ |1 \rangle}{\sqrt{2}} \right) \left( \frac{|0 \rangle+ |1 \rangle}{\sqrt{2}} \right)   = \frac{1}{\sqrt{2}} \left(  | - y \rangle  | 0 \rangle +  | + y  \rangle  | 1 \rangle \right) ,
\end{align}
where the eigenstates of the $ \sigma^y $ Pauli operator are 
\begin{align}
| + y \rangle & = \frac{| 0 \rangle + i | 1 \rangle}{\sqrt{2}} \nonumber \\
| - y \rangle & = \frac{i | 0 \rangle + | 1 \rangle}{\sqrt{2}} .
\end{align}
Such a Hamiltonian can be used as the basis of a CNOT gate, which produces a maximally entangled state. 

For ensembles, the $ H = S_z^{A} S_z^B $ interaction produces the state (\ref{entangledstate}) 
\begin{align}
& e^{-i S_z^{A} S_z^B \tau} | \frac{1}{\sqrt{2}}, \frac{1}{\sqrt{2}} \rangle \rangle_A 
| \frac{1}{\sqrt{2}}, \frac{1}{\sqrt{2}} \rangle \rangle_B  = \frac{1}{\sqrt{2^N}} \sum_k  \sqrt{N \choose k} | \frac{e^{i(N-2k)\tau}}{\sqrt{2}} , \frac{e^{-i(N-2k)\tau}}{\sqrt{2}} \rangle \rangle_A | k \rangle_B \nonumber \\
& = \frac{1}{\sqrt{2^N}} \Big[ 
| \frac{e^{i N \tau}}{\sqrt{2}} , \frac{e^{-i N\tau}}{\sqrt{2}} \rangle \rangle_A | 0 \rangle_B 
+ \sqrt{N} | \frac{e^{i (N-2) \tau}}{\sqrt{2}} , \frac{e^{-i (N-2) \tau}}{\sqrt{2}} \rangle \rangle_A | 1 \rangle_B + \dots  \nonumber \\
& +  \sqrt{N \choose N/2}   | \frac{1}{\sqrt{2}} , \frac{1}{\sqrt{2}} \rangle \rangle_A | N/2  \rangle_B
+ \dots +
| \frac{e^{-i N \tau}}{\sqrt{2}} , \frac{e^{i N\tau}}{\sqrt{2}} \rangle \rangle_A | N \rangle_B  \Big] .
\label{twobecszsz}
\end{align}
Comparing (\ref{qubitszsz}) and (\ref{twobecszsz}), we see that a similar type of state is produced where the first ensemble\index{ensemble} is rotated by an angle $ 2 (N - 2k) \tau $ around the $ S_z $ axis, for a Fock state\index{Fock states} $ |k \rangle $ on the second ensemble (see Fig. \ref{fig10-2}).  The types of correlations are similar for both the qubit and ensemble cases, but there are also differences.  The first most obvious difference is that the ensemble consists of $ N +1 $ terms, whereas the qubit version only has two terms. Thus although the same type of correlations are present to the qubit entangled state, it is a higher dimensional generalization.  The other difference is the presence of the binomial factor which weight the terms. The binomial function approximately follows a Gaussian distribution with a standard deviation $ \sim \sqrt{N} $.  Thus the most important terms are in the range $k \in [N/2-\sqrt{N}, N/2+\sqrt{N}] $.  

Using a sequence of single and two ensemble Hamiltonians, it is possible to produce a large range of effective Hamiltonians.  A well-known quantum information theorem states that if it is possible to perform an operation with Hamiltonians $ H_A $ and $ H_B $, then it is also possible to perform the operation corresponding to $ H_C = i [H_A,H_B] $ \cite{lloyd1995almost}. Therefore, the combination of single and two ensemble Hamiltonians  may be combined to form more complex effective Hamiltonians.  Using this theorem one can show that for a $ M $ ensemble system, it is possible to produce any Hamiltonian of the form
\begin{align}
\label{prodspinham}
H_{\text{eff}} \propto \prod_{n=1}^M S^{(n)}_{j(n)} .
\end{align}
where $ j(n) \in \{ 0, x, y, z \} $ and $ S_0 = I $.  An arbitrary sum of such Hamiltonian may also be produced by performing a Trotter expansion\index{Suzuki-Trotter expansion} \cite{lloyd1995almost}. For BEC qubits in general higher order operators can also be constructed (e.g. $ (S_j)^l$ with $ l\ge 2 $). 

The above results suggest that it should be possible to use the one and two ensemble Hamiltonians $ H_n = \bm{n} \cdot \bm{S}^{(n)} $ and  $ H_{nm} = S_z^{(n)} S_z^{(m)} $ together to perform various quantum information processing tasks.  We show several examples that demonstrate this explicitly in the following sections.  When constructing a quantum algorithm\index{quantum algorithm} using ensembles, often there is no unique mapping since the Hilbert space of the ensemble case is much larger. Thus there is a lot of freedom in implementing a given quantum algorithm using ensembles, with some mapping being more favorable than others.  We now discuss what factors should be included in a good mapping from qubits to ensembles.  

\paragraph{Equivalence of the circuit} The first most obvious requirement is that the ensemble version of the algorithm in fact does perform the same quantum computation as its qubit counterpart. Once the quantum algorithm is complete, one should be able to read off the result of quantum computation by a readout of the BEC qubits, which may involve some simple encoding rule to obtain the standard qubit version.  As we show in the example for Deutsch's algorithm\index{Deutsch's algorithm} below, this may also include the requirement that certain circuit elements such as the oracle behave in the same way as the qubit counterparts when operated outside of the quantum circuit.  

\paragraph{Elementary gates} Generally it is assumed that the only operations that are available are the one and two ensemble Hamiltonians $ H_n = \bm{n} \cdot \bm{S}^{(n)} $ and  $ H_{nm} = S_z^{(n)} S_z^{(m)} $.  This is because these operations are readily producible experimentally. The same goes for the measurements that are performed, these should be in a basis that can be implemented in a realistic way in the lab.  This might involve measuring ensembles in the $ S_z $ basis for example.  

\paragraph{Decoherence} The algorithm should be robust against decoherence.  This is a particularly important issue since the spin ensembles that may use a macroscopic number of atoms.  Encoding of the quantum information using particularly sensitive states such as Schrodinger cat states will generally not be practical as in the presence of decoherence the quantum algorithm will not complete with high fidelity. 

\paragraph{Algorithmic complexity} The complexity of the mapped algorithm should still retain the quantum speedup of the original qubit version of the algorithm.   Here, the way complexity is calculated is with respect to the number of gate operations of the Hamiltonians $ H_n $ and $ H_{nm} $ are executed.  Thus gates such as (\ref{zgateens})-(\ref{hgateens}) would count as 1 gate.  Although each ensemble may include a large number of atoms $ N $, there is little dependence to the experimental complexity with $ N $, justifying this way of counting resources.

\begin{figure}[t]
\includegraphics[width=\textwidth]{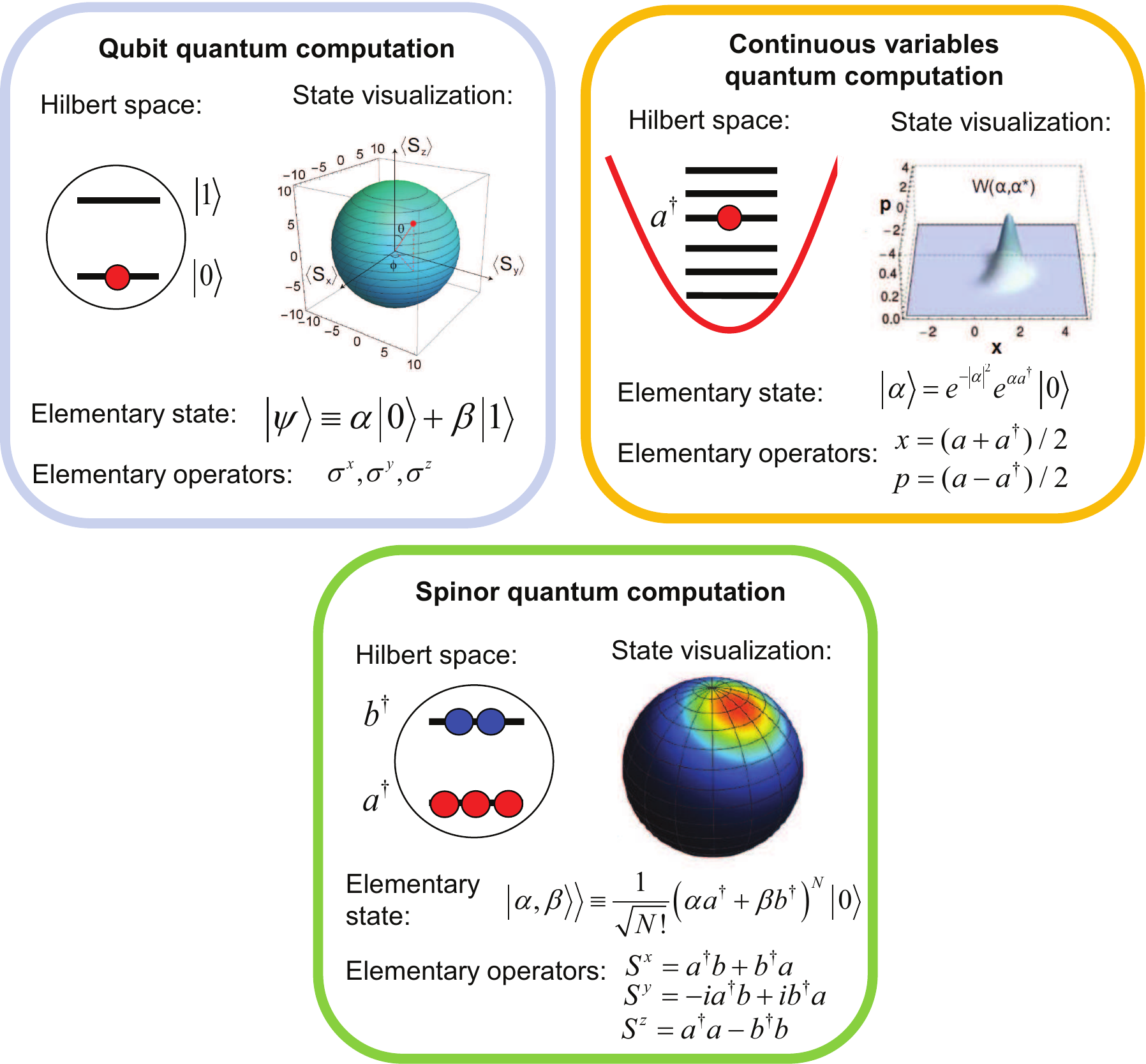}
\caption{Relationship between qubit, continuous variables, and spinor quantum computation. For each case, the Fock states\index{Fock state} spanning the Hilbert space, a typical state visualization using Wigner\index{Wigner function} or $Q$-functions\index{Q-function}, an elementary quantum state, and the Hamiltonian operators used to manipulate them are shown. }
\label{fig11-3}
\end{figure}

In Fig. \ref{fig11-3} we summarize the differences between qubit, continuous variables, and spinor quantum computing.  We can see that spinor quantum computing is in many ways a hybrid approach of the qubit and continuous variables approaches. Spin ensembles have a similar structure of states in terms of the Bloch sphere\index{Bloch sphere} and its operators have a similar commutation relations.  Meanwhile, spin ensembles also are similar to continuous variables in that large Hilbert spaces are used to store the quantum information.

\section{Deutsch's algorithm}
\label{sec:deutschalg}\index{Deutsch's algorithm}

An example of a simple quantum algorithm where we see a quantum advantage is Deutsch's algorithm.  This is a straightforward example where it is possible to implement the same algorithm on ensembles, in a regime beyond the Holstein-Primakoff approximation.\index{Holstein-Primakoff transformation}  We first briefly explain how the algorithm works for standard qubits and then show an explicit translation to spin ensembles.

\begin{figure}[t]
\includegraphics[width=\textwidth]{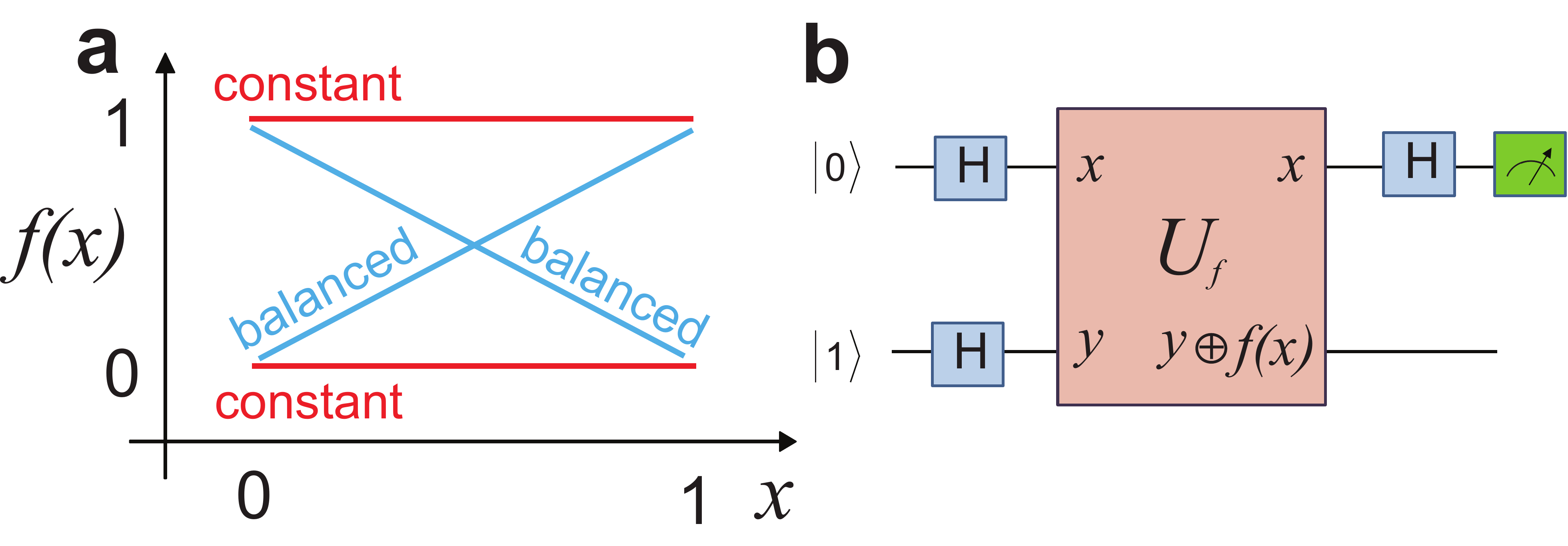}
\caption{Deutsch's algorithm. (a) The four constant and balanced algorithms as implemented by the oracle. (b) The quantum circuit implementing Deutsch's algorithm.  }
\label{fig11-2}
\end{figure}

\subsection{Standard qubit version}
\label{sec:deutschqubit}

Consider the binary function $ f(x) $ which takes a binary variable $ x $ as its argument and returns another binary number.  There are in fact only four types of such functions, they are shown in Fig. \ref{fig11-2}(a). Two of these functions always give the same output regardless of the input $ x $.  These are called {\it constant} functions. The other two functions, which do have a dependence on $ x $ are called  {\it balanced} functions. Now say that someone gives you a black box which gives the output $ f(x) $ given the input $ x $.  In order to identify whether a particular black box was in the constant or balanced categories, classically, at least two evaluations would be necessary. For instance if it is found that $ f(0) = 0 $, this rules out two of the four functions, but one cannot tell which of the remaining two it is, of which there is one constant and one balanced.  

Now let us see if we can achieve the same task in a quantum mechanical setting. The quantum version of the black box is the oracle, which performs the same operation as the black box when given classical inputs. The main difference with a quantum mechanical operation is that we must make sure that any operation we perform is a unitary operation, which is always a reversible operation. At the moment, the black box is not necessarily reversible.  For example, for the constant function $ f(x) = 1 $, 
you would not be able to work out what $ x $ was from the output, since in both cases the same result of $ 1 $ is the result.  The way to make the oracle reversible is to have an oracle  with two inputs and outputs, following the operation
\begin{align}
U_f |x\rangle | y \rangle = | x \rangle | y  \oplus f(x) \rangle ,
\end{align}
where $ x, y \in \{0,1 \} $ and $\oplus $ is addition modulo 2. It is clear that this is a reversible operation since one can perform the same gate twice and recover the original state $ U_f^2 = I $.  The oracle works in essentially the same way as the black box mentioned above.  For example,  if one sets $ y =0 $ and  chooses an input $ x $, the output of the second register is the same as the output of the black box.  

Quantum mechanically, it is possible to work out whether a given oracle is constant or balanced with only one evaluation. The quantum circuit for this is shown in Fig. \ref{fig11-2}(b). Following the circuit we can show that the output of the circuit is
\begin{align}
\left\{
\begin{array}{cc}
|0 \rangle |- \rangle & \hspace{1cm} \text{if} f(0) = f(1) \\
|1 \rangle |- \rangle & \hspace{1cm}  \text{if} f(0) \ne f(1) 
\end{array} \right. ,
\label{deutschqubitout}
\end{align}
where we have discarded irrelevant global phases. Therefore, the output of the first qubit is dependent only upon whether the oracle is balanced or constant.  Since in the quantum circuit the oracle is only called once, this gives a speedup of a factor of 2 compared to the classical case.  
While a factor of 2 speedup does not sound too impressive, it is possible to extend Deutsch's algorithm\index{Deutsch's algorithm} to a $n$-qubit oracle, and is called the Deutsch-Jozsa algorithm\index{Deutsch-Jozsa algorithm}.  In this case the speedup compared to the classical case is $ 2^n $, which is a huge speedup when when $ n $ is large.

\begin{exerciselist}[Exercise]
\item \label{q11-1}
Work through the quantum circuit in Fig. \ref{fig11-2}(b) and show that the output is (\ref{deutschqubitout}).  
\end{exerciselist}

\subsection{Spinor quantum computing version}

We now show that the above logical operations can also be performed using atomic ensembles.  Following Sec. \ref{sec:beyondholstein}, we can map the input states such that the input state is
\begin{align}
| 0, 1 \rangle \rangle |1,0\rangle \rangle  .
\end{align}
The Hadamard gates then operate on this state such that the state before the oracle is
\begin{align}
| \frac{1}{\sqrt{2}}, \frac{1}{\sqrt{2}}  \rangle \rangle |\frac{1}{\sqrt{2}}, - \frac{1}{\sqrt{2}} \rangle \rangle .
\label{deutschinput}
\end{align}

Now we must make sure that the oracle can be mapped in a way that has an analogous operation to the qubit example.  To show that the ensemble implementation of the oracle has the same effect as the qubit implementation, we distinguish two cases  when the input states $ |x \rangle $  and $ | y \rangle $ are   (i)  in the computational basis $ \{|0 \rangle, | 1 \rangle \}$; and (ii) involve superpositions such as the $\{  | + \rangle ,  | - \rangle \} $ states.  We call the former case when the oracle is working in ``classical mode'', and the second as working in ``quantum mode''.  Of course for genuine qubits, the quantum mode is simply the superposition of the classical cases, so this never needs to be distinguished.  However for the ensemble case, since a state 
\begin{align}
| \frac{1}{\sqrt{2}}, \frac{1}{\sqrt{2}}  \rangle \rangle  \ne \frac{| 1,0 \rangle \rangle 
+ | 0,1 \rangle \rangle}{\sqrt{2}} ,
\end{align}
we must ensure that the oracle works in the desired way for both cases. 

It is possible to implement the oracle in many different ways, but a particularly simple choice corresponds to taking the Hamiltonians for the four cases
\begin{align}
H_{f=0} & = 0 \\
H_{f=1} & = S^x_2-N \\
H_{f=\{1,0\}} & = \frac{1}{2} \left( 1- S^z_1/N \right) (S^x_2-N)  \\
H_{f=\{0,1 \}} & = \frac{1}{2} \left( 1+ S^z_1/N \right) (S^x_2-N)  ,
\label{deutschhams}
\end{align}
and it is implied that we evolve these for a time $ t = \pi/2 $.  First let us verify that in classical mode, the Hamiltonian produce the correct outputs. 
\begin{align}
e^{-i H_{f=0} t} |x ,1-x\rangle \rangle | y,1-y \rangle \rangle & = |x,1-x \rangle \rangle | y,1-y \rangle \rangle \nonumber \\
e^{-i H_{f=1} t} |x,1-x \rangle \rangle | y,1-y \rangle \rangle & =  |x,1-x \rangle \rangle | 1-y,y \rangle \rangle 
\nonumber \\
e^{-i H_{f=\{1,0\}} t} |0,1 \rangle \rangle | y,1-y \rangle \rangle & =  |0,1 \rangle \rangle | 1-y,y \rangle \rangle 
\nonumber \\
e^{-i H_{f=\{1,0\}} t} |1,0 \rangle \rangle | y,1-y \rangle \rangle & =  |1,0 \rangle \rangle | y,1-y \rangle \rangle 
\nonumber \\
e^{-i H_{f=\{0,1\}} t} |0,1 \rangle \rangle | y,1-y \rangle \rangle & =  |0,1 \rangle \rangle | y,1-y \rangle \rangle 
\nonumber \\
e^{-i H_{f=\{0,1\}} t} |1,0 \rangle \rangle | y,1-y \rangle \rangle & =  |1,0 \rangle \rangle | 1-y,y \rangle \rangle  ,
\end{align}
where $ x,y \in \{0,1 \} $ and we have discarded any irrelevant global phase factors. The above shows that the Hamiltonians for the oracles (\ref{deutschhams}) give the same results as for qubits when operated in classical mode.

Operating in quantum mode, after the initial Hadamard gates, the Hamiltonian is applied on the state (\ref{deutschinput}).  For the constant cases, the resulting state is
\begin{align}
e^{-i H_{f=0} t} | \frac{1}{\sqrt{2}}, \frac{1}{\sqrt{2}}  \rangle \rangle |\frac{1}{\sqrt{2}}, - \frac{1}{\sqrt{2}} \rangle \rangle = | \frac{1}{\sqrt{2}}, \frac{1}{\sqrt{2}}  \rangle \rangle |\frac{1}{\sqrt{2}}, - \frac{1}{\sqrt{2}} \rangle \rangle \\
e^{-i H_{f=1} t} | \frac{1}{\sqrt{2}}, \frac{1}{\sqrt{2}}  \rangle \rangle |\frac{1}{\sqrt{2}}, - \frac{1}{\sqrt{2}} \rangle \rangle = | \frac{1}{\sqrt{2}}, \frac{1}{\sqrt{2}}  \rangle \rangle |\frac{1}{\sqrt{2}}, - \frac{1}{\sqrt{2}} \rangle \rangle, 
\end{align}
which is the same state up to a global phase. For the balanced cases, the resulting state is 
\begin{align}
e^{-i H_{f=\{1,0\}} t} | \frac{1}{\sqrt{2}}, \frac{1}{\sqrt{2}}  \rangle \rangle |\frac{1}{\sqrt{2}}, - \frac{1}{\sqrt{2}} \rangle \rangle = | \frac{1}{\sqrt{2}}, -\frac{1}{\sqrt{2}}  \rangle \rangle
 |\frac{1}{\sqrt{2}},  -\frac{1}{\sqrt{2}} \rangle \rangle \\
e^{-i H_{f=\{1,0\}} t} | \frac{1}{\sqrt{2}}, \frac{1}{\sqrt{2}}  \rangle \rangle |\frac{1}{\sqrt{2}}, - \frac{1}{\sqrt{2}} \rangle \rangle = | \frac{1}{\sqrt{2}}, -\frac{1}{\sqrt{2}}  \rangle \rangle 
|\frac{1}{\sqrt{2}}, - \frac{1}{\sqrt{2}} \rangle \rangle , 
\end{align}
which is again the same state up to a global phase.

Finally, after the Hadamard gate, the state for constant cases become
\begin{align}
| 0, 1 \rangle \rangle |\frac{1}{\sqrt{2}}, - \frac{1}{\sqrt{2}} \rangle \rangle  .
\end{align}
For the balanced cases the state is
\begin{align}
|1,0 \rangle \rangle |\frac{1}{\sqrt{2}}, - \frac{1}{\sqrt{2}} \rangle  \rangle  .
\end{align}
This is the equivalent result as the qubit case (\ref{deutschqubitout}).  Since the quantum circuit for the ensemble case is the same as the qubit case, and the oracle is called only once, this has the same quantum speedup over the classical case by a factor of two.  

The Hamiltonians (\ref{deutschhams}) involve either single ensemble rotations (\ref{spinrotationsxyz}) or two ensemble interactions of the form (\ref{generalabstate}).  Furthermore, the time evolution of the interaction term is $ \tau = \pi/4N $ in dimensionless units.  From the discussion in Sec. \ref{sec:szsz} we know that the types of states that are generated with this interaction do not involve fragile Schrodinger cat-like states.  This means that the ensemble version of the quantum algorithm satisfies the requirements discussed in
 Sec. \ref{sec:beyondholstein} and is a satisfactory mapping to spinor quantum computing.

\begin{figure}[t]
\includegraphics[width=\textwidth]{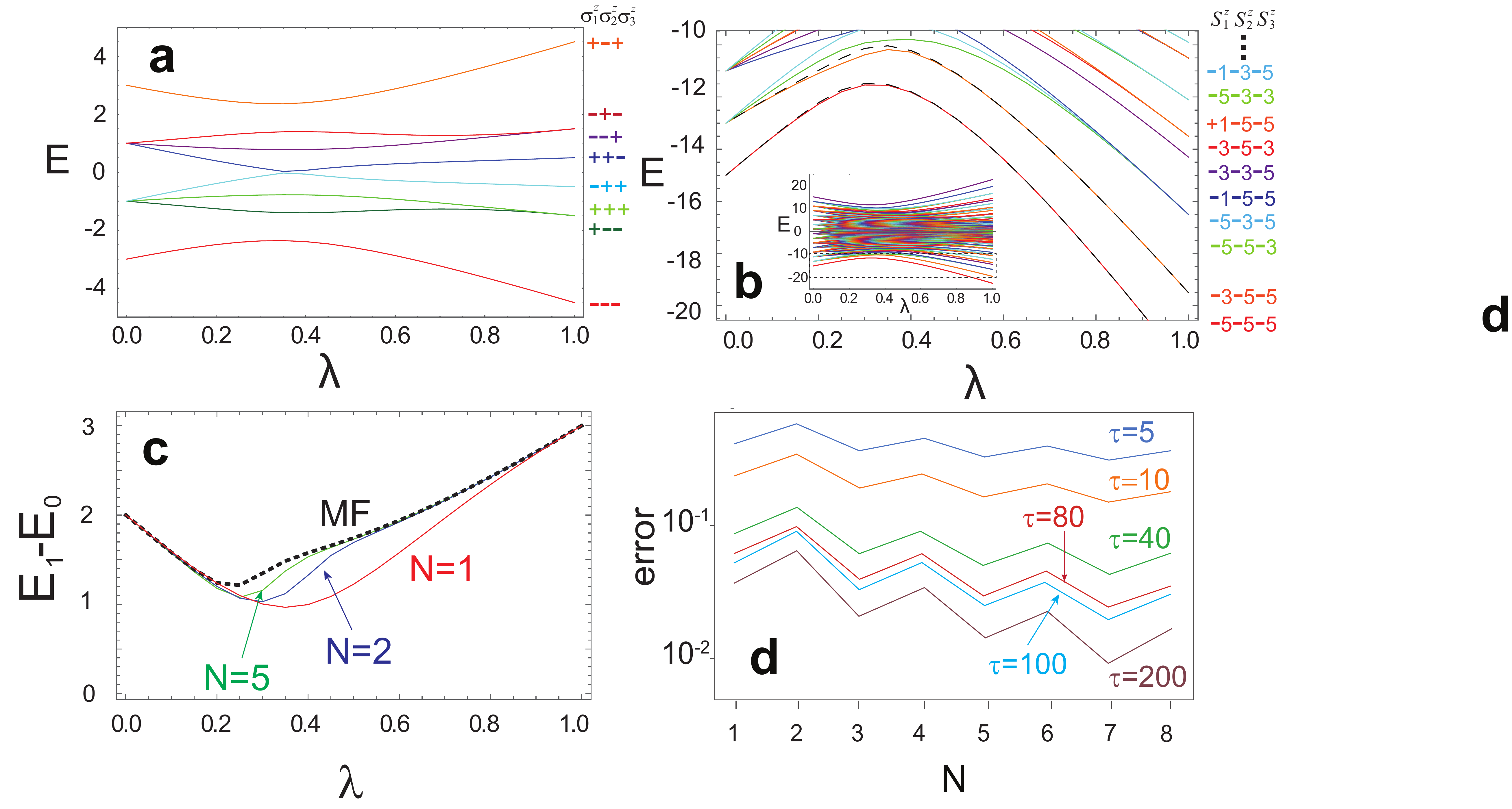}
\caption{Energy spectrum and gap energies of the adiabatic quantum computing Hamiltonian. Spectrum of  (\ref{aqcham}) with $ M =3 $ with  (a)  $ N = 1 $ and  (b) $ N = 5 $ for parameters  $J_{12}=-0.5, J_{13}=0, J_{23}=-1, K_1=0.5, K_2=0, K_3=1 $.  The mean field approximation for the $N = 5 $ is shown as the dashed lines for the ground and first excited state.  (c) The gap energy for the ensemble qubit numbers as shown.   (d) The final error probability for 60 instances averaged versus $N$ for various $\tau$ and $ M=3 $.  Here $ \tau $ is the time for changing the adiabatic parameter $ \lambda = t/\tau $, and the Hamiltonian (\ref{aqcham}) is evolved for a time $ \tau $.  }
\label{fig11-4}
\end{figure}

\section{Adiabatic quantum computing}
\label{sec:adiabatic}

Adiabatic quantum computing (AQC)\index{adiabatic quantum computing} is an alternative approach to traditional gate-based quantum computing where quantum adiabatic evolution is performed in order to achieve a computation.  In the scheme, the aim is to find the ground state of a Hamiltonian $ H_Z $ which encodes the problem to be solved. In addition, an initial Hamiltonian $ H_X $ which does not commute with the problem Hamiltonian is prepared, where the ground state is known.  For example, in the qubit formulation, a common choice of these Hamiltonians are
\begin{align}
H_Z & = \sum_{i=1}^M \sum_{j=1}^M J_{ij} \sigma^{(i)}_z \sigma^{(j)}_z +  \sum_{i=1}^M K_i \sigma^{(i)}_z  \label{hzham} \\
H_X & = - \sum_{i=1}^M \sigma^{(i)}_x \label{hxham}
\end{align}
where $ \sigma^{(i)}_{x,z} $ are Pauli matrices\index{Pauli operators} on site $ i $, and $ J_{ij} $ and $ K_i $ are coefficients which determine the problem to be solved, and there are $ M $ qubits.  We take $ J_{ij} = J_{ji} $ and $ J_{ii} = 0 $. The form of (\ref{hzham}) as chosen is rather general and can encode a wide variety of optimization problems.  For example, MAX-2-SAT and MAXCUT can be directly encoded in (\ref{hzham}), which is a NP-complete problem, meaning that any other NP-complete problem can be mapped to it in polynomial time.  AQC then proceeds by preparing the initial state of the quantum computer in the ground state of $ H_X $ (in the case of the above is a superposition of all states), then applying the time-varying Hamiltonian
\begin{align}
H = (1-\lambda) H_X + \lambda H_Z, 
\label{aqcham}
\end{align}
is prepared where  $ \lambda $ is a time-varying parameter that is swept from $ 0 $ to $ 1 $. 

In the AQC framework, the speed of the computation is given by how fast the adiabatic sweep is performed. To maintain adiabaticity, one must perform the sweep sufficiently slowly, such that the system remains in the ground state throughout the evolution. The sweep time is known to be proportional to the inverse square of the minimum energy gap of the Hamiltonian (\ref{aqcham}). One of the issues with AQC is that the minimum gap energy is typically unknown prior to a computation, and it is difficult to make general statements of the size of this gap.  One of the attractive features of AQC is that time-sequenced gates do not need to be applied, but is nevertheless known to be equivalent to gate-based quantum computation. 

The qubit Hamiltonians (\ref{hzham}) and (\ref{hxham}) can be mapped onto spin ensembles in a straightforward way. The equivalent Hamiltonian is
\begin{align}
H_Z & = \frac{1}{N} \sum_{i=1}^M \sum_{j=1}^M   J_{ij} S^{(i)}_z S^{(j)}_z +   \sum_{i=1}^M  K_i S^{(i)}_z   \label{hzhamens}  \\
H_X & = -  \sum_{i=1}^M  S^{(i)}_x ,\label{hxhamens}
\end{align}
where the $ S^{(i)}_{x,z} $ are the total spin operators as usual. Each of the ensembles are first prepared in a fully polarized state of  $ S_i^{x} $, according to
\begin{align}
|\psi(0) \rangle = \prod_{j=1}^M | \frac{1}{\sqrt{2}},\frac{1}{\sqrt{2}} \rangle \rangle_j ,
\end{align}
then adiabatically evolved to the ground state of $H_Z $. After the adiabatic evolution, measurements of the spin ensembles are made in the $ S^z $ basis.  If the evolution is successful, one should find that 
\begin{align}
\text{sgn} ( \langle S^{(j)}_z \rangle_{\text{ens}} ) = \text{sgn} ( \langle \sigma^{(j)}_z \rangle_{\text{qubit}} ),
\label{equivalentham}
\end{align}
where the right hand side of the expression are the spins for the original qubit operators as in (\ref{hzham}) and the expectation value is taken with respect to an adiabatic evolution of the qubit problem.  

The ground state of (\ref{hzhamens}) and (\ref{hzham}) are equivalent in the sense of (\ref{equivalentham}).  This is easy to see for states such that $ \langle S_z^{(j)} \rangle   = \pm N $ since substitution of this into (\ref{hzhamens}) gives exactly the same energy up to a factor of $ N $ as (\ref{hzham}). The state of the ground state can be written 
\begin{align}
\langle S_z^{(j)} \rangle_0 = \sigma_0^{(j)} N 
\label{groundstateense}
\end{align}
where $ \sigma_0^{(j)} \in \{ -1,1 \} $ is the ground state configuration.  The energetic ordering of the states do not change for these states, hence the lowest energy state are equivalent in terms of (\ref{equivalentham}).  The ensemble Hamiltonian however also possesses many more states that lie in the range $  \langle S_z^{(j)} \rangle \in [-N, N ] $. Is it possible that such states have a lower energy than those with (\ref{groundstateense})?  We do not show the arguments here but it can be shown that these intermediate states always have an energy exceeding the ground state (see \cite{mohseni2019} for further details).  This means that as long as it is possible to maintain adiabaticity throughout the evolution, evolving (\ref{aqcham}) with  (\ref{hzhamens}) and (\ref{hxhamens}) gives an equivalent way of solving the qubit problem. 

While the above shows it is possible to use the ensemble version of the Hamiltonian in place of the qubits, an important question is whether the performance of the adiabatic procedure changes.  One crucial question is whether the minimum gap increases or decreases due to the replacement with ensembles.  Here we can distinguish two cases depending upon the specific instance of the problem Hamiltonian (\ref{hzhamens}).  The first is when the first excited state is a perturbation of the original ground state.  For a specific ground state configuration (\ref{groundstateense}), the first excited state involves a single spin flip on one of the ensembles $ k $ such that one of the ensembles is 
\begin{align}
\langle S_z^{(j)} \rangle_1 = \left\{
\begin{array}{cc}
\sigma_0^{(j)} N & j \ne k \\
\sigma_0^{(j)} (N -2) & j = k 
\end{array}
\right. .
\label{excitedstateense1}
\end{align}
The second case is when the first excited state has a completely different character to the ground state.  In this case the first excited state is
\begin{align}
\langle S_z^{(j)} \rangle_1 = \sigma_1^{(j)} N 
\label{excitedstateense2}
\end{align}
where $ \sigma_1^{(j)} \in \{ -1, 1 \} $ and denotes the excited qubit state configuration.  It can be easily shown that the energy difference between the ground and excited state (\ref{excitedstateense2}) is proportional to $ N $, while for (\ref{excitedstateense1}) is independent of $ N $.  Thus for large $ N $ as with ensembles, the state (\ref{excitedstateense1}) is the more typical case.  

Figure \ref{fig11-4} shows energy spectrum, energy gap, and error probabilities for randomly generated problem instances. In Fig. \ref{fig11-4}(a)(b) the energy spectrum of the Hamiltonian (\ref{aqcham}) is plotted for the same problem instance.  The use of ensembles gives many more states, but the energy scale of the Hamiltonian is also increased by a factor of $ N $.  The combination of these effects means that the gap energy stays approximately the same, as can be seen in Fig. \ref{fig11-4}(c).  In fact for the case where the first excited state takes the form (\ref{excitedstateense1}), the gap energy slightly increases. More importantly, due to the ensemble mapping (\ref{hxhamens}) the low energy states are all logically equivalent states according to (\ref{equivalentham}), as can be seen in Fig. \ref{fig11-4}(b).  This means that even if the system is diabatically excited, or decoherence produces some excitations, it does not contribute to a logical error, as long as the number of excitations is small.    This translates to a reduced error for the adiabatic evolution as shown in Fig. \ref{fig11-4}(d), which improves with the ensemble size $ N $.  We note that the improvement with $ N $ comes only when the first excited state is of the form (\ref{excitedstateense1}), which occurs at a critical ensemble size $N_c $, and below this the performance can initially degrade with $ N $.  In this way, the ensemble encoding achieves quantum error suppression of logical errors in adiabatic quantum computing.

\section{References and further reading}

\begin{itemize}
\item Sec. \ref{sec:cv}: Reviews articles and books on continuous variables quantum information processing \cite{braunstein2005,gerd2007quantum,weedbrook2012gaussian,braunstein2012quantum,adesso2014continuous,loock2011}. 
\item Sec. \ref{sec:spinphoton}: Original work on Holstein-Primakoff transformation \cite{holstein1940field}. Review article on atomic ensemble continuous variables using the mapping \cite{hammerer2010quantum}.  
\item Sec. \ref{sec:teleportationcv}: Experimental demonstrations of continuous variable quantum teleportation \cite{furusawa1998unconditional,krauter2013deterministic,sherson2006quantum}. Theoretical proposals of discrete \cite{bennett1993,vaidman1994teleportation} and continuous variable quantum teleportation \cite{braunstein1998teleportation}. Review of continuous variable quantum teleportation \cite{furusawa2007quantum}.  Entanglement between atomic ensembles: \cite{julsgaard2001experimental,krauter2011entanglement}. 
\item Sec. \ref{sec:beyondholstein}: Original theoretical proposals of spinor quantum computing \cite{byrnes2012macroscopic,byrnes2015macroscopic,adcock2016quantum}. Experimental demonstrations of atom chips \cite{folman2000controlling,hansel2001bose,bohi2009,riedel2010}. Review articles and books reviewing atom chips \cite{reichel2011atom,keil2016fifteen}. Proof of universality of gate decompositions \cite{lloyd1995almost}. Other theoretical proposals  quantum information schemes using atomic ensembles \cite{lukin2001dipole,rabl2006hybrid,brion2007quantum,ebert2015coherence,calarco2000quantum}. 
\item Sec.~\ref{sec:deutschalg}: Original theoretical work introducing Deutsch's algorithm \cite{deutsch1992rapid}.  Reviews and books further discussing the algorithm \cite{cleve1998quantum,nielsen2000}.  Spinor quantum computing mapping of Deutsch's algorithm \cite{semenenko2016implementing}. 
\item Sec.~\ref{sec:adiabatic}: Original theoretical work introducing adiabatic quantum computing \cite{farhi2001quantum}.  Review articles on adiabatic quantum computing \cite{RevModPhys.80.1061,RevModPhys.90.015002}.  Equivalence between adiabatic quantum computing and standard quantum computing \cite{mizel2007simple,aharonov2008adiabatic}.  Spinor quantum computing mapping of adiabatic quantum computing \cite{mohseni2019}. 
\end{itemize}

  \backmatter


  \addtocontents{toc}{\vspace{\baselineskip}}
  \theendnotes

  \bibliography{references}\label{refs}
  \bibliographystyle{cambridgeauthordate}

  \cleardoublepage


 \printindex



\end{document}